\documentclass[12pt]{article}

\usepackage{star}
\usepackage{tikzit}
\usepackage{enumitem}
\usepackage{natbib}
\usepackage[toc,page]{appendix}
\usepackage{rotating}
\graphicspath{{img/}{tikz/}{img/tikz/}}

\def\spacingset#1{\renewcommand{\baselinestretch}%
{#1}\small\normalsize} \spacingset{1}
\newcommand{\blind}{1}

\def\deg{{\sf deg}}
\def\neig{{\sf nei}}

\def\adj{{\sf adj}}

\newcommand{\nodea}{v}
\newcommand{\nodeb}{u}
\newcommand{\nseta}{V}
\newcommand{\nsetb}{U}

\newcommand{\edge}[1]{(#1)}
\newcommand{\Tbd}[2]{\cL_{{#2} }}
\newcommand{\Tn}[2]{\cN_{{#2} }}
\newcommand{\Ts}[2]{\cP_{{#2}}}
\newcommand{\press}[1]{\texttt{reduce}(#1)}
\newcommand{\tq}{\tilde {q}}

\begin{document}

\def\spacingset#1{\renewcommand{\baselinestretch}%
{#1}\small\normalsize} \spacingset{1}


\if1\blind
{
  \title{Parallel sampling of decomposable graphs using Markov chain on junction trees}     
\author{Mohamad Elmasri\thanks{Department of Statistical Sciences, University of Toronto, 100 St. George Street, Toronto, ON  M5S 3G3, Canada; E-mail: \texttt{mohamad.elmasri@utoronto.ca}.}}
  \maketitle
} \fi

\if0\blind
{
  \bigskip
  \bigskip
  \bigskip
  \begin{center}
    {\LARGE\bf Parallel sampling of decomposable graphs using Markov chain on junction trees}
\end{center}
  \medskip
} \fi

\bigskip
\begin{abstract}
  Bayesian inference for undirected graphical models is mostly restricted to the class of decomposable graphs, as they enjoy a rich set of properties making them amenable to high-dimensional problems. While parameter inference is straightforward in this setup, inferring the underlying graph is a challenge driven by the computational difficulty in exploring the space of decomposable graphs. This work makes two contributions to address this problem. First, we provide sufficient and necessary conditions for when multi-edge perturbations maintain decomposability of the graph. Using these, we characterize a simple class of partitions that efficiently classify all edge perturbations by whether they maintain decomposability. Second, we propose a novel parallel non-reversible Markov chain Monte Carlo sampler for distributions over junction tree representations of the graph. At every step, the parallel sampler executes simultaneously all edge perturbations within a partition. Through simulations, we demonstrate the efficiency of our new edge perturbation conditions and class of partitions. We find that our parallel sampler yields improved mixing properties in comparison to the single-move variate, and outperforms current state-of-the-arts methods in terms of accuracy and computational efficiency. The implementation of our work is available in the Python package \textsf{parallelDG}.
\end{abstract}

\noindent%
\vfill

\newpage
\addtolength{\textheight}{0.5in}

\section{Introduction}
A graphical model represents a collection of joint probability distributions for a vector of random variables \(Y = (Y_{1}, \ldots, Y_{p})\). In these models, the distributions are subject to conditional independence constraints specified by a graph, comprised of vertices \(\cbr{1,\ldots, p}\). A notable class of graphical models focuses on cases where the underlying graph $G$ is undirected and decomposable (chordal). Bayesian structure learning, or model determination, involves inferring the underlying conditional independence graph and the model parameters concurrently. This process is based on observed data and predefined prior specifications. This work develops Markov chain Monte Carlo methods for computational inference in this settings.

The decomposability assumption is a severe restriction on the space of possible graphs. Less than \(8\times 10^{-5}\) of graphs with 12 vertices are decomposable, and exact enumeration exists only for graphs with up to 15 vertices~\citep{olsson2022sequential,wormald1985counting}. It is therefore impractical to use Bayesian methods that require quantification of the prior normalization constant. While general Bayesian structure learning methods can take minutes to converge for medium-sized problems~\citep{mohammadi2021accelerating}, assuming decomposability shrinks this convergence time to seconds. In sparse high-dimensional problems, this inferential efficiency is vital.

The computational advantage stems from the unique clique-separator factorization property of decomposable graphs. Specifically, when a decomposable graph $G$ represents the conditional independence constraints of a random vector $Y$, the joint distribution of $Y$ can be expressed as:
\begin{equation}
\label{eq:joint-csf}
p(Y) = \frac{\prod_{C \in \clq{G}}p(Y_{C})}{\prod_{S \in \sep{G}}p(Y_{S})},
\end{equation}
where \(Y_{A}\) is a subvector of \(Y\) indexed by the set $A$, $\clq{G} = \cbr{C_1, \ldots, C_c}$ is a set of complete subgraphs of $G$, known as maximal cliques, and $\sep{G}$ is a set of intersections between elements in~$\clq{G}$~\cite[Ch.~4.4]{lauritzen1996}. This factorization facilitates statistical inference by permitting operations on the marginals of \(Y\) that are specified by \(\clq{G}\).

The line of work by~\cite{frydenberg1989decomposition, giudici1999decomposable, thomas2009enumerating, Green01032013} has led to the development of an efficient Metropolis--Hastings (MH) algorithm for decomposable graphical models~\citep{hastings1970monte}. This method leverages the \emph{junction tree} representation of decomposable graphs to create efficient proposals that preserve decomposability throughout the states of the chain. The junction tree sampler introduced in~\cite{Green01032013} is particularly fast, even for large-dimensional problems. However, it exhibits a high degree of within-sample correlation, a topic explored in Section~\ref{sec:numerical} and by~\cite{olsson2019bayesian}. The class of multi-edge perturbations in~\cite{Green01032013} only sufficiently maintains decomposability across the chain, which results in suboptimal proposals. These proposals not only increase the rejection rate of the sampler but also impact its overall efficiency and accuracy.

Our first contribution delineates the sufficient and necessary conditions for multi-edge updates to preserve the decomposability of the underlying graph. We derive this understanding from a simple algorithm that generates decomposable graphs via random walks on trees (Sec.~\ref{sec:decomp-graphs-random-walks}), which exhibits junction tree-like properties. Leveraging these properties, we define a partitioning scheme for junction trees (Sec.~\ref{sec:graph-perturbations}). Specifically, for any given graph vertex, all tree elements are divided into three disjoint sets: those the vertex can connect to, disconnect from, and all others. These sets enumerate all possible edge perturbations related to the vertex that maintain decomposability.

Our second contribution (Sec.~\ref{sec:single-move-sampler}) introduces a hierarchical latent sampler for junction trees. This sampler projects junction trees into higher dimensions while retaining their desirable factorization properties (Sec.~\ref{sec:graph-prior-post}). While updates in this latent space do not directly correspond to graph updates, when combined with our multi-edge partitioning sets, they enable simultaneous parallel updates on the graph. To demonstrate the benefits of parallelism, we first implement a single-move reversible MH Markov chain (Mc) (Sec.~\ref{sec:graph-prior-post}) and compare it with a parallel non-reversible MH-Mc (Sec.~\ref{sec:jt-sampler}).
\begin{figure}[!ht]
  \centering
  \includegraphics[width=0.43\textwidth]{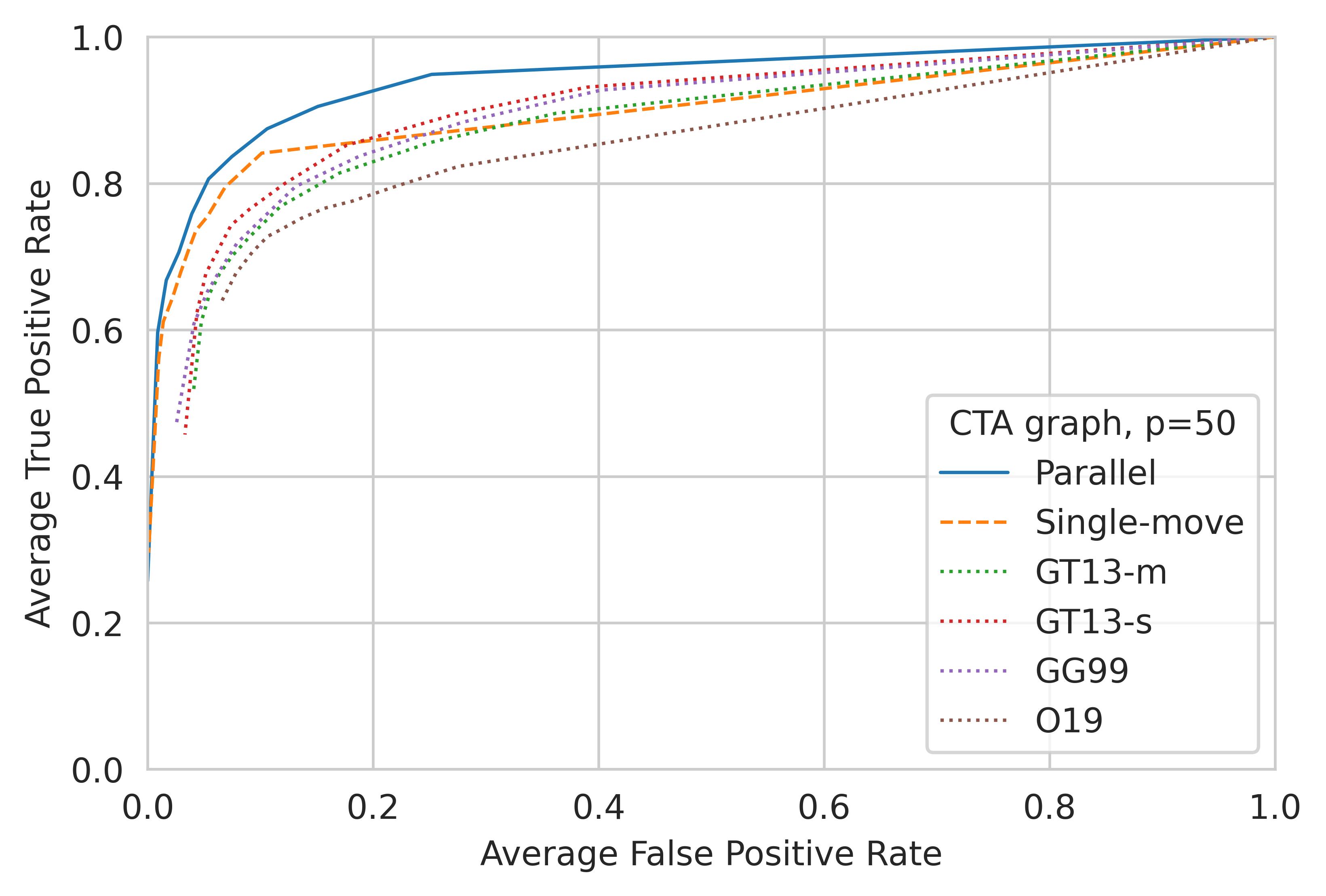}
  \includegraphics[width=0.43\textwidth]{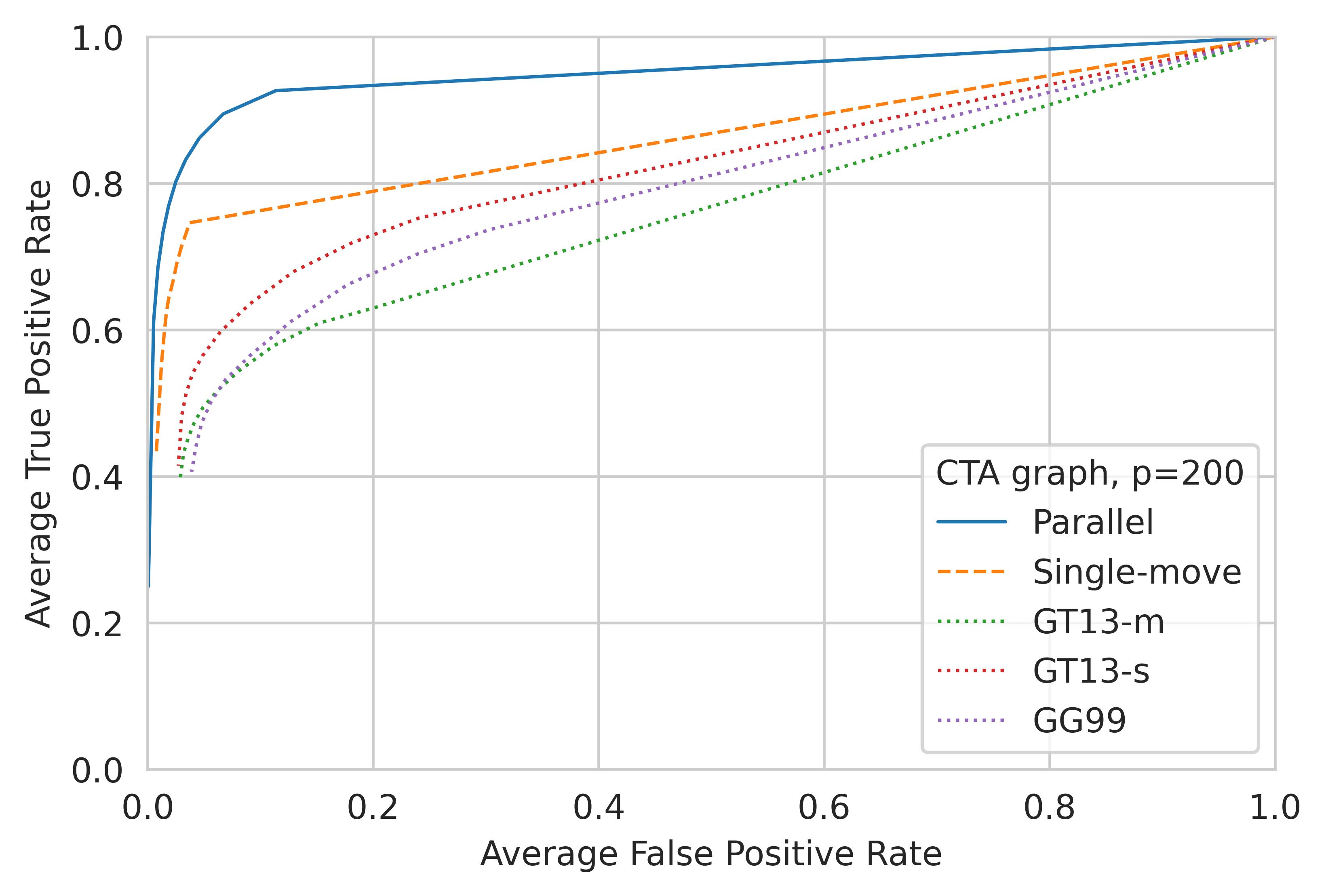}
  
  \caption{ROC curves from the simulation study (Sec.~\ref{sec:numerical}) over 10 replications. Each replication involves a random decomposable graph, with $p$-vertices, generated via the Christmas tree algorithm (CTA). The plot compares our parallel and single-move samplers to state-of-the-art MH-Mc (GT13, GG99)~\citep{Green01032013, giudici1999decomposable} and Gibbs (O19)~\citep{olsson2019bayesian} samplers, for $n=100$, $p=100$ (left) and 200 (right).}
      \label{fig:ggm-roc-cta}
\end{figure}

It is now well-understood that non-reversible chains offer significant advantages over their reversible counterparts, as documented in various studies~\citet{neal1998suppressing,rey2015irreversible,bierkens2016non,duncan2016variance}. To immediately showcase these benefits and the effectiveness of our decomposability-preserving conditions, we present an average Receiver Operating Characteristic (ROC) plot across 10 replications in Figure~\ref{fig:ggm-roc-cta}. In each replication, a random decomposable graph $G$ is generated using the Christmas tree algorithm ~\citep{olsson2022sequential}, with $p=\cbr{50, 200}$ vertices and a sample size of 100 from a Gaussian graphical model under a $G$-Wishart prior~\citep{roverato2002hyper}. Our proposed parallel and single-move samplers achieve impressive accuracy in the general case $(p=50)$ and, more so, in the high-dimensional case $(p=200)$, than the state-of-the-art MH-Mc junction tree-based samplers (GT13-\{m,s\}) by~\citet{Green01032013}, the MH-Mc graph-based sampler (GG99) by~\citet{giudici1999decomposable}, and the Gibbs sampler (O19) by~\citet{olsson2019bayesian}. Remarkably, our samplers converge in less than 3 minutes for all cases. Competing MH-Mc samplers converge in 1--3 for $p=50$ and over an hour for $p=200$. The Gibbs sampler (O19) demonstrates limited feasibility for larger $p$ values, requiring as much as 5 hours to converge at $p=50$. To our knowledge, this problem has not been addressed by any frequentist methods, nor are there other fully Bayesian approaches.

Our simulation study (Sec.~\ref{sec:numerical}) further reveals a substantial reduction in within-sample correlation, accelerated and more stable convergence, and more than five-fold increase in the acceptance rate of our sampler relative to state-of-the-arts methods. For more details, refer to Section~\ref{sec:numerical} and Supplementary Material (SM) Sections~\ref{app:sec:ggm} and~\ref{app:sec:ggm:sparse}.

The multi-edge partitioning sets and parallel sampling approach we've characterized have potential applications in various statistical problems, extending beyond the scope of this work. This includes the sampling of directed acyclical graphs and potential applications in computational graph theory. Similar to the approaches in~\cite{giudici1999decomposable, Green01032013, olsson2019bayesian}, our latent junction tree sampler is compatible with proper parameter-inference methods across different statistical models. A notable application is in multinomial models for discrete data~\citep{tarantola2004mcmc}.

\section{Decomposable graphs and junction trees} \label{sec:expand-junct-trees}
We begin by reviewing some theoretical underpinnings and properties of decomposable graphs and junction trees. We frame this work in the classic graphical modelling literature, where \cite{lauritzen1996} and \cite{cowell2006probabilistic} are excellent references on the topic.

An \emph{undirected graph} \(G=(\cV,\cE)\) is composed of a set of \emph{vertices} \(\cV\) connected by a set of undirected \emph{edges} \(\cE\). For any subset \(V \in \cV\), \(G_{V}\) denotes the induced subgraph with the vertex set \(V\). A path \(\nodea_{0} \sim \nodea_{k} \subseteq G\) consists of a sequence of vertices \(\rbr{\nodea_{0}, \nodea_{1}, \ldots, \nodea_{k}}\), such that \(\edge{\nodea_{i}, \nodea_{i+1}} \in \cE\). A graph is \emph{connected} when there is a path between every pair of vertices, and it is \emph{complete} (a clique) if an edge exists between every pair of vertices. A complete subgraph is called a \emph{maximal clique} if it is not a subgraph of any other clique. A graph is called a \emph{tree} if there is a unique path between any pair of nodes.

A graph \(G\) is \emph{decomposable} if and only if the set of maximal cliques of \(G\) can be ordered as \((C_{1}, \ldots, C_{c})\), such that, for each \(i=1, \ldots, c\), if
\begin{equation}
\label{eq:running-intersection-property}
S_{i} = C_{i} \cap \bigcup_{j=1}^{{i-1}}C_{j} \quad \text{then} \quad S_{i} \subset C_{k}, \text{ for some } k < i; 
\end{equation}
\(S_{i}\) may be empty. The relation in~\eqref{eq:running-intersection-property} is called the running intersection property, and the sequence $(C_1, \ldots, C_c)$ is called the \emph{perfect ordering sequence}. The set \(\sep{G} = \cbr{S_{1}, \ldots ,S_{c}}\), formed by~\eqref{eq:running-intersection-property}, is known as the \emph{minimal separators} of \(G\), and define \(\clq{G} = \cbr{C_{1}, \ldots, C_{c}}\). Maximal cliques are unique to a decomposable graph, while separators can repeat in~\eqref{eq:running-intersection-property}. The edge-relation \(\edge{C_{i}, C_{k}}\) that \(S_{i} \) forms between \(C_{i}\) and \(C_{k}\) in~\eqref{eq:running-intersection-property} leads to a tree representation of \(G\). A (reduced) \emph{junction tree} of \(G\) is a tree with a vertex set as the maximal cliques and an edge set as the minimal separators of \(G\) written as \(J = (\clq{G}, \sep{G})\). It follows that \(G\) is decomposable if and only if it admits a junction tree representation. A decomposable graph admits multiple (reduced) junction tree representations. We denote \(\jtree{G}\) to be the set of (reduced) junction trees of \(G\). The maximal cardinality search algorithm of~\cite{Tarjan:1984:SLA:1169.1179} allows a junction tree representation to be found in time of order \(|\cV| + |\cE|\). Refer to SM Table~\ref{table:notations} for all notations.

As in \cite{Green01032013}, we adopt the convention of allowing separators to be empty, ensuring that every junction tree is connected. We use the terms \emph{vertex} and \emph{edge} specifically for the elements of $G$, and reserve \emph{cliques} and \emph{separators} for the elements of junction trees. This distinction is important: a vertex represents a single element, while a clique or a separator can represent multiple vertices.

Junction trees are general graphical objects defined independently of decomposable graphs. A tree $T$, with vertices as subsets of $\cV$, is termed a \emph{junction tree} if for any pair of vertices $C_{1}, C_{2}$ in $T$, and any vertex $C$ on the unique path $C_{1} \sim C_{2} \subseteq T$, 
\begin{equation}
  \label{eq:junction-property}
  C_{1} \cap C_{2} \subseteq C. 
\end{equation}

The relationship described by~\eqref{eq:junction-property} is known as the \emph{junction property}. This property can also be expressed by stating that for any subset $\nseta \subseteq \cV$, the vertices in $T$ containing $\nseta$ form a connected subtree. Therefore, any tree $T = (\cC, \cS)$ comprising subsets of cliques $\cC$ from $G$ that satisfy~\eqref{eq:junction-property} qualifies as a junction tree. In light of this, we can construct junction trees for a decomposable graph $G$, where $\cC$ may include but is not limited to the maximal cliques $\clq{G}$.
\begin{figure}[!ht]
  \centering
  \includegraphics[width=0.6\textwidth]{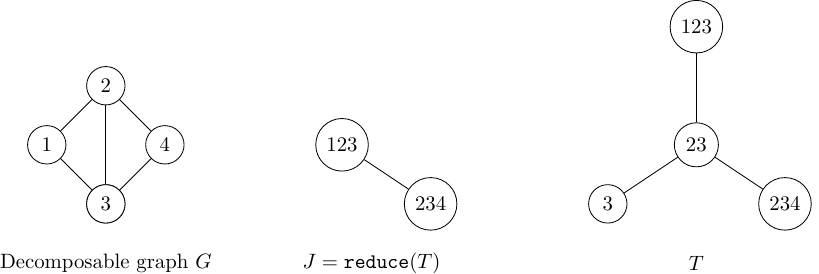}
  \caption{A decomposable graph $G$, its unique junction tree $J = (\clq{G}, \sep{G})$, and an expanded tree $T = (\cC, \cS)$, where $\clq{G} \subset \cC$.}
\label{fig:example_random_walk_graph}
\end{figure}

The decomposable graph $G$ in Figure~\ref{fig:example_random_walk_graph} admits the junction tree $J = (\clq{G}, \sep{G})$, as well as the expanded junction tree $T$ that arises from the junction property. This property allows for the inclusion of additional, though probabilistically superfluous, relations in $T$. In this case, $T$ comprises four cliques: two maximal, $\cbr{1,2,3}$ and $\cbr{2,3,4}$, and two non-maximal, $\cbr{3}$ and $\cbr{2,3}$.

Moving forward, we will consistently refer to junction trees in the context of their junction property, i.e., having $\clq{G} \subseteq \cC$, and denote them as $T$, unless specified otherwise. In cases where we are discussing reduced junction trees, these will be denoted as $J$. The elements of junction trees will be referred to as cliques and separators, instead of vertices and edges, reflecting their representation as subsets of vertices. A junction tree $T$ can always be compressed into a reduced junction tree $J = \press{T} = (\clq{G}, \sep{G})$. This compression involves iteratively removing cliques in $T$ that are subsets of an adjacent clique, rewiring their associated edges to the adjacent superset clique, and eliminating self-loops. We denote this compression operation as $\press{T}$.

We simplify notations by using $C \in T$ to denote a clique in $T$, and $\edge{C, C'} \in T$ to represent a separator in $T$. The notation $\cC$ and $\cS$ are reserved for sets of cliques and separators in $T$, respectively, i.e., $T = (\cC, \cS)$. For a graph vertex $\nodea \in \cV$, $T_{\nodea}$ denotes the induced subtree of $T$, comprising every $C \in T_{\nodea}$ that includes $\nodea$, indicated as $\nodea \in C$. For a specific junction tree $T$, we define $g(T)$ as the unique decomposable graph represented by $T$. This graph is constructed by connecting vertices within cliques of $T$, namely by adding an edge $\edge{\nodea,\nodeb}$ to $G$ for every $\nodea, \nodeb \in C$, with $C \in T$. We denote $\deg(C, T) = \abr{\neig(C, T)}$ as the number of cliques adjacent to $C$ in $T$. These notations are applicable to $J$ as well.

\section{Decomposable graphs from random-walks on trees}\label{sec:decomp-graphs-random-walks}

In this section, we introduce an algorithm that generates decomposable graphs from random walks on trees, beginning with a basic hierarchical model. Let $\pi$ be a probability measure on the space of the specified random variable. Generate an arbitrary tree skeleton $t$, and conditional on $t$, sample a junction tree $T$ in the hierarchical scheme
\begin{equation}
\label{eq:hierarchical-model}
 t  \sim \pi(t), \qquad T \given t \sim \pi(T\given t),
\end{equation}

while ensuring that the sampling of $T$ adheres to the junction property~\eqref{eq:junction-property}. One such sampling method initiates $p$ random walks on a tree $t$ of $p$ vertices. Each walk starts at an arbitrary vertex and continues exploring $t$, visiting new vertices with a probability $\rho \in \rbr{0,1}$. Each walk forms a connected subtree $t_i$ of $t$. When combined, these $p$ subtrees constitute a junction tree of a $p$-vertex decomposable graph. We detail this sampling process in Algorithm~\ref{alg:random-walk}.
\begin{algorithm}
\KwIn{An arbitrary tree $t$ with \(\cbr{1,\ldots, p}\) vertices. $\rho \in \rbr{0,1}$.}
    \For{ $i\gets1$ \KwTo $p$}{
      draw \(j \sim \mathsf{Uniform}(\cbr{1,\ldots, p})\)\;
      \(w = \cbr{j}\); \tcp{visited vertices}
      \(a = \cbr{}\); \tcp{attempted vertices}
      \While{\(\neig(t_{w}, t) \setminus \cbr{w \cup a}\not=\emptyset\)}{
        \(k \sim\textsf{Uniform}(\neig(t_{w}, t) \setminus \cbr{w \cup a})\)\;
        \lIf{\(\mathsf{Uniform}([0,1])\leq \rho\)}{ add \(k\) to \(w\) {\bf else} add \(k\) to \(a\)}
      }
      \(t_{i}\) $\gets$ a copy of \(t\), where the label of each vertex \(k \in w\) is set to \(i\).
    }
    \Return $\bigcup_{i \leq n} t_{i}$ \tcp{union over labels.}
\caption{Random walks on a tree.}
\label{alg:random-walk}
\end{algorithm}

\begin{figure}[!h]
  \centering
  \includegraphics[width=0.7\textwidth]{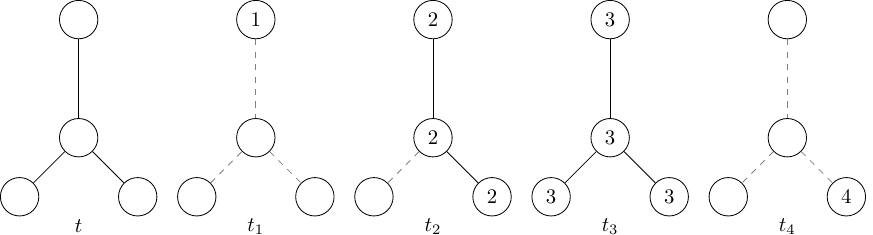}
  \caption{Example of a 4-vertex skeleton tree $t$ and 4 random walks $\cbr{t_i}$, $i = 1, \dots, 4$, generated via Algorithm~\ref{alg:random-walk}. Visited vertices are labeled by the walk's count ($i$) with solid-edge subtrees. The resulting $T$, $J=\press{T}$ and decomposable graph $G$ are in Fig.~\ref{fig:example_random_walk_graph}.}
\label{fig:example_random_walk}
\end{figure}

Figure~\ref{fig:example_random_walk} showcases an example from Algorithm~\ref{alg:random-walk} with four random walks independently initiated on a skeleton $t$, with $\rho = 1/2$. The resulting junction tree $T$, the reduced junction tree $J = \press{T}$, and the corresponding decomposable graph are illustrated in Fig.~\ref{fig:example_random_walk_graph}.

The parameter $\rho$ in Algorithm~\ref{alg:random-walk} (line 7) controls the sparsity of the generated subtrees. Lower values of $\rho$ result in smaller subtrees, while higher values tend to produce more saturated subtrees. This parameter is utilized as a prior in SM Section~\ref{sec:decomp-graphs-seven}.

From the junction property~\eqref{eq:junction-property}, we understand that for a given junction tree $T$ and a graph vertex $\nodea \in \cV$, $T_{\nodea}$ is a connected subtree. In fact, if we remove all vertices in $T_{\nodea}$ except for $\nodea$, the resulting structure is the subtree skeleton $t_{\nodea}$, since $t_{\nodea} = T_{\nodea} \setminus \cbr{\cV\setminus\cbr{\nodea}}$. Therefore, updating $t_{\nodea}$ as described in Algorithm~\ref{alg:random-walk} ensures that $T$ maintains its status as a junction tree, as is substantiated in the following theorem.
\begin{theorem} \label{thm:1} A tree \(T\) generated by Algorithm~\ref{alg:random-walk} is a junction tree.
\end{theorem}
\begin{proof} By construction, every \(C, C' \in T\) are connected via a unique path \(C\sim C' \subseteq T\). If \(\nodea \in C\cap C'\), then \(C, C' \in T_{\nodea}\), and so is every clique on the path \(C\sim C'\).
\end{proof}

Algorithm~\ref{alg:random-walk} is both simple and insightful in demonstrating how decomposable graphs can be constructed. It highlights three essential properties that underpin our work: \((i)\) a junction tree can be updated solely through vertex-induced subtrees \(\cbr{T_{\nodea}}\) for each \(\nodea \in \cV\), particularly when conditional independence is not the primary focus; \((ii)\) a random decomposable graph can be constructed from the union of independently generated subtrees \(\cbr{t_{i}}\) as shown in Fig.~\ref{fig:example_random_walk}, useful when considering conditional independence; \((iii)\) each subtree \(T_{\nodea}\) can be expanded and contracted in multiple directions simultaneously by manipulating \(t_{\nodea}\). Specifically, \((iiia)\) expanding \(T_{\nodea}\) involves enlarging the \(\nodea\)th random-walk \(t_{\nodea}\) in parallel to adjacent vertices in \(t\) (lines 5-7 in Alg.~\ref{alg:random-walk}), and \((iiib)\) contracting \(T_{\nodea}\) is effectively a reversal of its expansion, achieved by reducing \(t_{\nodea}\) from its leaf nodes.

Observations \((i)-(iii)\) form the cornerstone of our analysis, facilitating a parallelized approach for manipulating junction trees through their junction property. The forthcoming section establishes a connection between the expansion and contraction methods detailed in \((iiia)-(iiib)\) and updates to decomposable graphs.
\section{Decomposable graph updates}\label{sec:graph-perturbations}
In this section, we delineate how the random-walk process, as outlined in Algorithm~\ref{alg:random-walk}, leads to the effective partitioning of a junction tree. This partitioning streamlines the process of updating a decomposable graph and enables us to establish the precise conditions necessary to preserve decomposability under multi-edge perturbations.

For a given junction tree \(T=(\cC, \cS)\) and graph vertex \(\nodea \in \cV\), we partition \(\cC\) into three distinct clique sets: cliques that \(\nodea\) can be added (\emph{connect}) to, referred to as  \(\Tn{}{\nodea}\), cliques that \(\nodea\) can be removed from (\emph{disconnect}), referred to as \(\Tbd{}{\nodea}\), and everything else. Graphically, $\Tn{}{\nodea}$ consists of cliques in $T$ that are adjacent to $T_{\nodea}$, and $\Tbd{}{\nodea}$ are the leaf cliques of $T_{\nodea}$, as
\begin{equation}\label{eq:nei-bound-clique-basic-no-cond}
\begin{aligned}
& \Tn{}{\nodea} (T):= \cbr{C \in T: \edge{C , C'} \in  T\,, C' \in T_{\nodea}\,, C \not \in T_{\nodea}},\\
  & \Tbd{}{\nodea}(T) := \cbr{C  \in T_{\nodea}:  \deg(C, T_{\nodea})= 1}.
\end{aligned}
\end{equation}

The terms \emph{add} and \emph{remove} are used when referring to updates on $T$, and \emph{(dis)connect} when referring to updates on the underlying graph $G$. In $T$, adding vertex $\nodea$ to an adjacent clique $C \in \Tn{}{\nodea}$ implies updating $C$ to $C \cup \cbr{\nodea}$. Removing $\nodea$ from a leaf clique $C \in \Tbd{}{\nodea}$ implies updating $C$ to $C\setminus\cbr{\nodea}$. While we primarily work on $T$, in $G = g(T)$, these operations correspond to the following: Adding $\nodea$ to $C$ entails connecting edges $\edge{\nodea, \nodeb}$ in $G$, where $\nodeb \in C\setminus C_{\adj}$; here, $C_{\adj}$ is the unique clique in $T_{\nodea}$ adjacent to $C$ (i.e., $\edge{C, C_{\adj}} \in T$). Removing $\nodea$ from a clique $C \in \Tbd{}{\nodea}$ involves disconnecting all graph edges $\edge{\nodea, \nodeb}$ from $\nodeb \in C \setminus C_{\adj}$, where $\edge{C, C_{\adj}} \in T_{\nodea}$. In both cases, there is a unique clique $C_{\adj} \in T_{v}$ adjacent to the relevant clique. Example~\ref{example:1} illustrates these sets graphically. 
\begin{figure}[h!]
  \centering
  \includegraphics[width=0.5\textwidth]{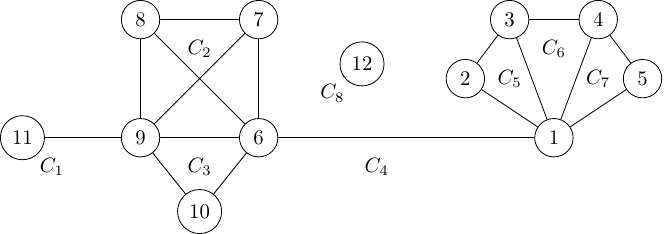}
  \caption{12-vertex decomposable graph of 8 maximal cliques \(\clq{G} = \cbr{C_{1}, \ldots, C_{8}}\).}
\label{fig:example1_graph}
\end{figure}

\begin{example} \label{example:1} Figure~\ref{fig:example1_graph} illustrates a 12-vertex, 8-maximal-clique decomposable graph $\cbr{C_{1}, \ldots, C_{8}}$, with its junction tree $T$ shown in Fig.~\ref{fig:example1_junction} (left). The tree $T$ includes one clique without any graph vertices, represented as $\emptyset$, and two non-empty non-maximal cliques labeled as $(8,7)$ and $3$ to indicate the vertices they contain. The induction of $T$ to vertex $6$ ($T_{6}$), depicted in Fig.~\ref{fig:example1_junction} (right), consists of 3 cliques $\cbr{C_{2}, C_{3}, C_{4}}$ connected with solid edges. The set $\Tn{}{6}(T)$ is highlighted in blue and $\Tbd{T}{6}(T)$ in green.
\begin{figure}[h!]
  \centering
  \includegraphics[width=0.25\textwidth]{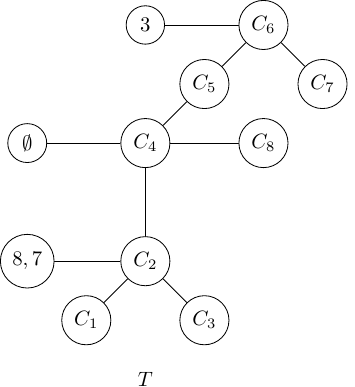}\qquad \qquad
  \includegraphics[width=0.25\textwidth]{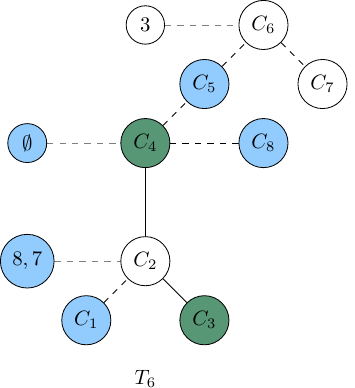}
   \caption{A junction tree \(T\) (left) of decomposable graph in Fig.~\ref{fig:example1_graph}, its induction by vertex \(6\) (right) in solid lines, with neighboring \(\Tn{}{6}\) (blue) and leaf \(\Tbd{}{6}\) (green) cliques. None-maximal cliques are labeled with vertices contained in them, as \(\emptyset, \cbr{8,7}\) and \(\cbr{3}\).}
\label{fig:example1_junction}
\end{figure}
\end{example}
Cliques in \(\Tn{}{6}\) and \(\Tbd{}{6}\) are nonadjacent in \(T\), except \(C_{4} \in \Tbd{}{6}\), which is adjacent to \(\emptyset, C_{5}, C_{8} \in \Tn{}{6}\). Removing \(6\) from \(C_{4}\) would hence remove \(\emptyset, C_{5}\) and \(C_{8}\) from \(\Tn{}{6}\). Similarly, adding \(6\) to any of \(\emptyset, C_{5}\) or \(C_{8}\), would remove \(C_{4}\) from \(\Tbd{}{6}\). Therefore, we term a clique \(C \in \Tn{}{\nodea}\cup \Tbd{}{\nodea}\) a {\it partition-divisor} if it is adjacent to other cliques in those sets. A partition-divisor occurs only when a leaf clique \(C \in \Tbd{}{\nodea}\) is adjacent to a clique in \(\Tn{}{\nodea}\).

As implicitly mentioned above, the sets \(\Tn{}{\nodea}\) and \(\Tbd{}{\nodea}\) exhibit a congruence relation across updates, where leaf cliques become neighboring cliques, and vice versa, if updated. In Example~\ref{example:1}, adding vertex \(6\) to the neighboring clique \(C_{5} \in \Tn{}{6}(T)\) to form the clique \(C'_{5} = C_{5} \cup \cbr{6}\) in a new junction \(T'\) renders \(C'_{5} \in \Tbd{}{6}(T')\). Similarly, if 6 is removed from \(C_{4}\), the new \(C'_{4}\) becomes a neighboring clique to \(T'_{6}\). This congruence relation, demonstrated in~\eqref{eq:congruence}, will come in handy in carrying graph updates across a Markov chain, as shown in subsequent sections.
\begin{equation}
\label{eq:congruence}
C \in \Tn{}{\nodea} \Longleftrightarrow C\cup\cbr{\nodea} \in \Tbd{}{\nodea}.
\end{equation}

It is evident from Fig.~\ref{fig:example1_junction} that $\Tbd{}{\nodea}$ and $\Tn{}{\nodea}$ are both vertex-wise symmetric in relation to graph updates. Specifically, for an unconnected vertex pair $\nodea, \nodeb \in \cV$, $\nodeb \in C$ for some $C \in \Tn{}{\nodea}$ if and only if $\nodea \in C'$ for some $C' \in \Tn{}{\nodeb}$. If $\nodea$ and $\nodeb$ are in a single clique of $G$, then $\nodeb \in C$ for some $C \in \Tbd{}{\nodea}$ if and only if $\nodea \in C'$ for some $C' \in \Tbd{}{\nodeb}$. These findings align with the single-edge decomposability-preserving conditions illustrated in~\cite{frydenberg1989decomposition} and \cite{giudici1999decomposable}.

Extending this concept to a multi-vertex set $\nseta \subseteq \cV$, where $\nseta$ forms a clique in $G$, is straightforward by utilizing $T_{\nseta}$ in~\eqref{eq:nei-bound-clique-basic-no-cond}. For enhanced readability, the partition sets in~\eqref{eq:nei-bound-clique-basic-no-cond} are primarily described for singleton vertices.

Having analytically and visually demonstrated the class of partitions generated by~\eqref{eq:nei-bound-clique-basic-no-cond}, we now present theorems that illustrate the necessary and sufficient conditions for a perturbation in the edge set of a decomposable graph to maintain decomposability.
\begin{theorem}\label{lem:connect} Let \(G = (\cV, \cE)\) be a decomposable graph, and let \(\nseta, \nsetb\) be two disjoint subsets of \(\cV\), both forming complete graphs in \(G\). Suppose we form a graph \(G' = (\cV, \cE')\) by adding edges from every vertex in \(\nseta\) to every vertex in \(\nsetb\). Then, \(G'\) is decomposable if and only if \(\nsetb\) can be partitioned into mutually exclusive subsets of vertices ordered as \(\rbr{\nsetb^{(1)}, \ldots \nsetb^{(K)}}\), where $\nsetb^{(i)} \cap U^{(j)} = \emptyset$ if $i\neq j$, such that, there exists a path of maximal cliques $\rbr{C_1, \ldots, C_K}$ in some $J \in \jtree{G}$ satisfying the recursive relation that \(U^{(1)} \subseteq C_1\) for \(C_1 \in \Tn{}{V}(J)\), and \(U^{{(k)}} \subseteq C_{k}\) for \(C_{k} \in \Tn{}{U^{(k-1)}}(J)\) where $k=2, \ldots, K$.
\end{theorem}

\begin{theorem}\label{lem:disconnect} Let \(G = (\cV, \cE)\) be a decomposable graph, and let \(\nseta, \nsetb\) be two disjoint subsets of \(\cV\) that are completely connected, i.e.~\(\nseta\cup \nsetb\) is complete in \(G\). Suppose we form a graph \(G' = (\cV, \cE')\) by disconnecting all edges \(\edge{\nodea, \nodeb}\), where \(\nodea \in \nseta\) and \(\nodeb \in \nsetb\). Then, \(G' = (\cV, \cE')\) is decomposable if and only if \(\nsetb\) can be partitioned into mutually exclusive subsets of vertices ordered as \(\rbr{\nsetb^{(1)}, \ldots \nsetb^{(K)}}\), where $\nsetb^{(i)} \cap U^{(j)} = \emptyset$ if $i\neq j$, such that, there exists a path of maximal cliques $\rbr{C_1, \ldots, C_K}$ in  some $J \in \jtree{G}$ satisfying the recursive relation that \(U^{(1)} \subseteq C_1\) for \(C_1 \in \Tbd{}{V}(J)\), and \(U^{{(k)}} \subseteq C_{k}\)  for \(C_{k} \in \Tbd{}{U^{(k-1)}}(J)\) where $k=2, \ldots, K$.
\end{theorem}

A special case in Theorems~\ref{lem:connect} and~\ref{lem:disconnect} occurs when the entire vertex set $\nsetb$ is a subset of $C_1$, implying $\nsetb = \nsetb^{(1)}$ and the sets $\cbr{\nsetb^{(k)}}$ are empty for $k = 2, \ldots, K$. Consequently, $C_1$ is identified as a neighboring clique ($C_1 \in \Tn{}{\nseta}(J)$) in Theorem~\ref{lem:connect}, or as a leaf clique ($C_1 \in \Tbd{}{\nseta}(J)$) in Theorem~\ref{lem:disconnect}. When the sets $\cbr{\nsetb^{(k)}}$ for $k = 1, \ldots, K$ are non-empty, it indicates that $\nsetb$ spans multiple adjacent cliques in $J$, forming a path. Metaphorically, a disconnect move can be likened to peeling a bandage, where $\nseta$ is sequentially disconnected from $\nsetb^{(1)}$, then $\nsetb^{(2)}$, and so on. In this context, $\nsetb^{(k)}$ is always part of a leaf clique in the junction tree created by disconnecting $\nseta$ from $\cup_{j < k} \nsetb^{(j)}$. Conversely, a connect move is akin to a reverse operation, where $\nsetb^{(k)}$ is always in a neighboring clique to the induced junction tree $J_\nseta$, formed by connecting $\nseta$ to $\cup_{j < k} \nsetb^{(j)}$. These processes are consistent with the junction property~\eqref{eq:junction-property}.

The main consequence of Theorems~\ref{lem:connect} and~\ref{lem:disconnect} is that they provide a comprehensive enumeration of all decomposability-preserving updates related to a vertex, as defined by the partition sets in~\eqref{eq:nei-bound-clique-basic-no-cond}. Although Theorems~\ref{lem:connect} and~\ref{lem:disconnect} are framed in the context of graph updates, pertinent to graphical models, analogous principles for updates on general junction trees can be directly derived from the junction property~\eqref{eq:junction-property}. The next section introduces a single-move sampler and extends it to a parallel sampler over junction trees.

\section{Single-move sampler}\label{sec:single-move-sampler}
\subsection{Proposal probability}
This section introduces a single-move sampler based on the partition sets in~\eqref{eq:nei-bound-clique-basic-no-cond}, highlighting its key properties. For a given state of the chain $T$, the update of clique $C$ to $C'$, resulting in the new state $T'$, occurs with a probability 
\begin{equation}
  \label{eq:proposal-basic}   
  q(T, T')  = \frac{1}{2}\frac{1}{|\cV|}\frac{1}{\abr{\Ts{}{\nodea}(T)}} \one\{C \in {\Ts{}{\nodea}(T)}\}.
\end{equation}

The ratio $1/\abr{\Ts{}{\nodea}(T)}$ accounts for the probability of selecting clique $C$ from the set $\Ts{}{\nodea}$, where $\Ts{}{\nodea} = \Tn{}{\nodea}$ applies to an addition, and $\Ts{}{\nodea} = \Tbd{}{\nodea}$ to a removal move. The factor of $1/2$ accounts for the choice between update types, and $1/\abr{\cV}$ for the selection of a vertex.

For the addition move, the reverse operation involves enumerating $\Tbd{}{\nodea}(T')$ of the new junction tree $T'$, which can be computed from $\Tbd{}{\nodea}(T)$ due to the congruence relation between the sets defined in~\eqref{eq:nei-bound-clique-basic-no-cond} and~\eqref{eq:congruence}. If $\nodea$ is added to $C \in \Tn{}{\nodea}(T)$ to form $C'$, then $C'$ becomes a leaf clique in $\Tbd{}{\nodea}(T')$. The status of other cliques in $\Tbd{}{\nodea}(T)$ remains unchanged, except when the clique adjacent to $C$ in $T_{\nodea}$ is a partition-divisor. In Example~\ref{example:1}, adding $6$ to $C_{1}$ increases the number of leaves in $T'_{6}$ by one, as $C_{2}$ is not a partition-divisor. However, adding $6$ to $C_{5}$ or any neighbor of the partition-divisor $C_{4}$ would result in $C_{4}$ losing its status as a leaf clique in $T'_{6}$, thereby making $\Tbd{}{6}(T') = \Tbd{}{6}(T)$. Let $C_{\adj}$ be the unique clique in $T_{\nodea}$ adjacent to $C$, $\edge{C_{\adj}, C} \in T$, then 
\begin{equation}
  \label{eq:reverse-proposal-connect}
  \abr{ \Tbd{}{\nodea}(T')} = \abr{\Tbd{}{\nodea}(T)} + \one\{C_{\adj} \not\in \Tbd{}{\nodea}(T)\}.
\end{equation}
For the removal move, the reverse operation involves enumerating $\Tn{}{\nodea}(T')$, which can similarly be computed from $\Tn{}{\nodea}(T)$. If $\nodea$ is removed from $C$, and $C$ is a partition-divisor, all neighbors of $C$ in $\Tn{}{\nodea}(T)$ lose their status as neighboring cliques to $T_{\nodea}$. For instance, if $6$ is removed from $C_{4}$ (Fig.~\ref{fig:example1_junction}), all neighbors (in blue) of $T_{6}$ ($\cbr{\emptyset, C_{5}, C_{8}}$) are not in $\Tbd{}{6}(T')$, and $C_{4}$ would become a leaf node of $T'_{6}$. Otherwise, $\abr{\Tn{}{\nodea}(T')} = \abr{\Tn{}{\nodea}(T)} + 1$. Resulting in 
\begin{equation}
  \label{eq:reverse-proposal-disconnect}
    \abr{ \Tn{}{\nodea}(T')  } = \abr{\Tn{}{\nodea}(T)}  + 2 - \deg(C, T). 
\end{equation}

In all cases, quantifying the reverse operation ratio $q(T', T)$ requires only the enumeration of the sets $\Tbd{}{\nodea}(T)$ and $\Tn{}{\nodea}(T)$. This enumeration can be efficiently carried out with a single pass over $T_{\nodea}$ and its neighbors, without necessitating an actual modification to

\subsection{Prior and posterior probabilities}\label{sec:graph-prior-post}

Following~\eqref{eq:hierarchical-model}, we propose a hierarchical sampler that iteratively samples a junction tree $T$ and the underlying skeleton $t$ from the posteriors $\pi(T \given t)$ and $\pi(t \given T)$, respectively. Assuming a skeleton tree with $p$ vertices (as a decomposable graph with $p$ vertices has at most $p$ maximal cliques), Cayley's theorem from~\cite{cayley1889theorem} tells us there are $p^{p-2}$ possible trees of size $p$. Thus, a uniform prior on $t$ would be $\pi(t) \propto p^{-(p-2)}$.

Turning to prior specification over the space of junction trees.~\cite{byrne2015structural,green2018structural} have illustrated an important family of graph laws through algebraic characterization, known as the~\emph{clique-separator} factorization laws. Laws in this family have densities \(\pi(T)\) that factorize as
\begin{equation}
    \label{eq:csf}
    \pi(T) \propto \frac{\prod_{C \in \cC}\phi(C)}{\prod_{S \in \cS} \psi(S)},
  \end{equation}
Over the class of decomposable graphs, and consequently over the class of junction trees, this applies for positive functions $\phi, \psi$. As noted by~\citet[Thm. 1]{green2018structural}, a graph law over the class of decomposable graphs, which encompasses all such graphs, is weakly structural Markov if and only if it adheres to a clique-separator factorization law. Specifying decomposability through junction trees does not modify the clique-separator factorization of the joint distribution in~\eqref{eq:joint-csf}, as elucidated in the following proposition.
\begin{proposition} \label{lem:jtree-csf}
For a random vector \(Y\) that has a decomposable conditional independence graph \(G\), where \(T = (\cC, \cS)\) is a junction tree of \(G\) having \(\clq{G} \subseteq \cC\) and \(\sep{G}\subseteq \cS\), the distribution of \(Y\) factorizes as
\begin{equation}
\label{eq:joint-jtree-csf}
p(Y) = \frac{\prod_{C \in \cC}p(Y_{C})}{\prod_{S \in \cS}p(Y_{S})} = \frac{\prod_{C \in \clq{G}}p(Y_{C})}{\prod_{S \in \sep{G}}p(Y_{S})}.
\end{equation}
\end{proposition}
\begin{proof} Let \(C \in \cC\) be any non-maximal clique in \(T\), i.e.,~there exists a \(\edge{C', C} \in \cS\) such that \(C \subseteq C'\). The terms in~\eqref{eq:joint-jtree-csf} that are specific to \(C\) and \(C'\) are
  \[\frac{p(Y_{C})p(Y_{C'})}{p(Y_{C\cap C'})} = \frac{p(Y_{C})p(Y_{C'})}{p(Y_{C})} = p(Y_{C'}).\]
  
  Carrying such cancellations over non-maximal cliques leads to the last term in~\eqref{eq:joint-jtree-csf}.
\end{proof}

Following a similar argument to Proposition~\ref{lem:jtree-csf}, the density \(\pi(T)\) in~\eqref{eq:csf} factorizes in terms of \((\clq{G}, \sep{G})\), where \(G = g(T)\), if \(\psi=\phi\), as
\begin{equation}
\label{eq:junction-prior-cfs}
\pi(T) \propto \frac{\prod_{C \in \cC}\phi(C)}{\prod_{S \in \cS} \psi(S)} = \frac{\prod_{C \in \clq{G}}\phi(C)}{\prod_{S \in \sep{G}} \psi(S)},
\end{equation}
otherwise, a multiplicative factor of \(\prod_{C \in \cbr{\cC \setminus\clq{G}}}\phi(C)/\prod_{S \in \cbr{\cS \setminus \sep{G}}}\psi(S)\) exists on the right-hand side of~\eqref{eq:junction-prior-cfs}.

Turning to posterior specification, we have~\(\pi(t\given T) \propto 1\), as the likelihood and prior terms cancel out, leaving a symmetric proposal, which follows exactly as the tree randomization step performed in~\cite{Green01032013}, detailed in~\cite[Sec.~5]{thomas2009enumerating}. For a posterior sample from $\pi(T \given t)$, we employ a standard Metropolis--Hastings step as outlined by~\citep{hastings1970monte}. Given a junction tree $T$, we first uniformly sample a vertex $\nodea \in \cV$, then compute the partitions $\Tbd{}{\nodea}(T)$ and $\Tn{}{\nodea}(T)$. Subsequently, we uniformly select an update-type and a partition $C$ for the update. Let $C_{\adj}$ be the unique clique in $T_{\nodea}$ adjacent to $C$ (i.e., $\edge{C_{\adj}, C} \in T$). The acceptance probability of the proposed update of $C$ to $C'$ and the new junction tree $T'$ is 
\begin{equation}
  \label{eq:acceptance-prop-single-move}
  \begin{aligned}
    \alpha(T, T'\given t) &= \min \cbr{1, \frac{\pi(T'\given t)q(T, T')}{\pi(T \given t)q(T', T)}} \\
    &= \min \cbr{1,\frac{p(Y_{C'})p(Y_{C\cap C_{\adj}})}{p(Y_{C'\cap C_{\adj}})p(Y_{C})}  \frac{\phi(C')\psi(C\cap C_{\adj})}{\psi(C'\cap C_{\adj})\phi(C)}\frac{\abr{\Ts{}{\nodea}(T')}}{\abr{\Ts{}{\nodea}(T)}}}, 
  \end{aligned}
\end{equation}
if \(C \in {\Ts{}{\nodea}(T)}\), and zero otherwise. If the update-type is an addition, $\Ts{}{\nodea}(T) = \Tn{}{\nodea}(T)$ and $\Ts{}{\nodea}(T') = \Tbd{}{\nodea}(T')$. Conversely, if the update-type is a removal, $\Ts{}{\nodea}(T) = \Tbd{}{\nodea}(T)$ and $\Ts{}{\nodea}(T') = \Tn{}{\nodea}(T')$. Both quantities can be computed using~\eqref{eq:reverse-proposal-connect} and~\eqref{eq:reverse-proposal-disconnect}. In~\eqref{eq:acceptance-prop-single-move}, it is sufficient to use only the indicator function associated with $q(T, T')$ because the validity of the reverse operation depends on the congruence relation~\eqref{eq:congruence}, which in turn is contingent on the initial move. Another consequence of this congruence relation is that the step in~\eqref{eq:acceptance-prop-single-move} is reversible across the chain states. Furthermore, given that $\pi(T \given t) > 0$, there is always a sequence of junction trees from any state of the chain to the single-clique junction tree. The finiteness of the chain's state space implies irreducibility, thus ensuring the ergodicity of the chain.

Computing the acceptance probability as specified in~\eqref{eq:acceptance-prop-single-move}, while not necessitating an actual modification of $T$, does require the enumeration of the two state-dependent sets defined in~\eqref{eq:nei-bound-clique-basic-no-cond}. Section~\ref{sec:jt-sampler} introduces a parallel strategy for sampling over junction trees, with localized proposal that bypass the need to consider the entire state $T$. Consequently, the sampler only needs to enumerate a single partition set for each proposal, greatly streamlining the sampling process.

\section{Parallel sampler}\label{sec:jt-sampler}
 \subsection{Local proposal probability}
In the single-move sampler (Sec.~\ref{sec:single-move-sampler}), a transition from state $T$ to $T'$ involves updating the clique $C$ to $C'$. In this process, only $C$ is modified, while all other cliques in $T'$ remain unchanged from their state in $T$. This aspect is captured in the posterior ratio in~\eqref{eq:acceptance-prop-single-move}, which depends only on the marginals of $C, C'$, and $C_\adj$. It is the proposal probability in~\eqref{eq:proposal-basic} that ties the acceptance ratio~\eqref{eq:acceptance-prop-single-move} to the overall state of $T$. This connection is specifically through the set $\Ts{}{\nodea}(T)$, which determines (i) the likelihood of selecting $C$ from $\Ts{}{\nodea}(T)$, and (ii) the validity of the move, denoted as $\one\cbr{C \in \Ts{}{\nodea}(T)}$.

To develop a parallel sampler, it is essential that all proposals within a partition be mutually exclusive. Specifically, updating a clique in $\Ts{}{\nodea}(T)$ should not alter the state or validity of any other clique within the same set. Under this condition, it becomes feasible to update all cliques in $\Ts{}{\nodea}(T)$ concurrently, this is formalized in the following proposition.
\begin{proposition} \label{prop:exclusive} Let $T = (\cC, \cS)$ be a junction tree over the vertex set $\cV$. For $\nodea \in \cV$, define $\Tn{}{\nodea}(T)$ and $\Tbd{}{\nodea}(T)$ as in~\eqref{eq:nei-bound-clique-basic-no-cond}. For any distinct cliques $C_1, C_2 \in T$:
\begin{enumerate}[label=({\Roman*})]
    \item If $C_1, C_2 \in \Tn{}{\nodea}(T)$, when $\nodea$ is added to $C_1$ to arrive at state $T_1$, then  $C_2 \in \Tn{}{\nodea}(T_1)$.\label{prop:case:1}
    \item If $C_1, C_2 \in \Tbd{}{\nodea}(T)$, and \(\abr{\cbr{C: C \in T_{\nodea}}}>2\), when $\nodea$ is removed from $C_1$ to arrive at state $T_1$, then $C_2 \in \Tbd{}{\nodea}(T_1)$. \label{prop:case:2}
\end{enumerate}
\end{proposition}
\begin{proof} The proof follows by construction. By~\eqref{eq:nei-bound-clique-basic-no-cond}, in both cases, there exists a clique $C_\adj \in T_\nodea$ such that $\rbr{C_2, C_\adj} \in T$ and $C_\adj \neq C_1$. In~\ref{prop:case:1}, adding $\nodea$ to $C_1$, neither alters $C_\adj$ nor the edge $\rbr{C_2, C_\adj}$ $\implies \rbr{C_2, C_\adj} \in T_1 \implies C_2 \in \Tn{}{\nodea}(T_1)$. In~\ref{prop:case:2}, removing $\nodea$ from $C_1$ does not affect the edge $\rbr{C_2, C_\adj}$ $\implies$ $C_2 \in \Tbd{}{\nodea}(T_1)$.
\end{proof}

Under the conditions of Proposition~\ref{prop:exclusive}, consider a sequence of updates $C_1,\ldots, C_K \in \Ts{}{\nodea}(T)$, starting at state \(T\) and resulting in states $T_1, \ldots, T_K$. The validity of the $k$th update can be established by induction as
\[\cbr{C_k \in \Ts{}{\nodea}(T)} \Longleftrightarrow  \cbr{C_k \in \Ts{}{\nodea}(T_{k-1})}.\]

Therefore, the execution order does not affect the validity of updates in $\Ts{}{\nodea}(T)$, as long as these updates are valid in the initial state $T$. To visualize this, all (blue) cliques in $\Tn{}{6}$ (Fig.~\ref{fig:example1_junction}) can be updated simultaneously, as their validity relies on cliques in $T_6$ (the initial state). In this case, $T$ can be transformed into one of $2^{5}$ possible states (since $\abr{\Tn{}{6}}=5$). The choice among these potential states is made independently based on the acceptance probability of each proposal. Similarly, the (green) cliques in $\Tbd{}{6}$ can also be updated concurrently, as their validity is contingent upon $C_2$, a non-leaf clique of $T_6$.

Proposition~\ref{prop:exclusive}~\ref{prop:case:2} highlights a scenario where proposals in $\Ts{}{\nodea}(T)$ are not mutually independent. Initially, if $\abr{\cbr{C: C \in T_{\nodea}}}<2$, indicating that $T_{\nodea}$ is a single-clique tree, it follows that $\Tbd{}{\nodea}(T) = \emptyset$. Secondly, if $\abr{\cbr{C: C \in T_{\nodea}}}=2$, meaning $T_{\nodea}$ comprises only two cliques, then removing $\nodea$ from $C_{1}$ results in $C_{2} \not \in \Tbd{}{\nodea}(T_{1})$, as $C_{2}$ now has degree 0 in $T_{\nodea}$. Given the limitation of this scenario to just two possible moves, we revert to using the single-move sampler. This is achieved by adjusting the proposal mechanism in the parallel sampler. Specifically, at state $T$, the proposal probability for updating $C_{i} \in \Ts{}{\nodea}(T)$ to $C'_{i}$ is 
\begin{equation}
  \label{eq:proposal} 
  \begin{aligned}
  q(C_{i}, C'_{i}; T) &=  \frac{1}{2}\frac{1}{\abr{\cV}}\frac{1}{2^{U(C_{i},C_{\adj(i)})}} \one\cbr{C_{i} \in \Ts{}{\nodea}(T)},\\
  U(C_{i}, C_{\adj(i)}) &=\one\cbr{\Ts{}{\nodea}(T) = \cbr{C_{i}, C_{\adj(i)}}}.
  \end{aligned}
\end{equation}

  Here, $C_{\adj(i)} \in T_{\nodea}$ is the unique clique adjacent to $C_{i}$, such that $\edge{C_{\adj(i)}, C_{i}} \in T$. The condition $U(C_{i}, C_{\adj(i)})=1$ corresponds to the scenario where $\abr{\cbr{C: C \in T_{\nodea}}}=2$, and the move is a removal, namely, $\Ts{}{\nodea}(T)=\Tbd{}{\nodea}(T)=\cbr{C_{i}, C_{\adj(i)}}$. Consequently, the factor $2^{-U(C_{i}, C_{\adj(i)})}$ accounts for the selection choice $1/\abr{\Tbd{}{\nodea}(T)}$ when $\abr{\cbr{C: C \in T_{\nodea}}}=2$, and the single-move proposal is employed. In all other cases,~\eqref{eq:proposal} does not incorporate the selection probability $1/\abr{\Ts{}{\nodea}(T)}$, as specified in~\eqref{eq:proposal-basic}, since all proposals are executed concurrently. By the congruence relation~\eqref{eq:congruence}, the validity of the reverse operation hinges on the event \(\cbr{C_{i} \in \Ts{}{\nodea}(T)}\). Resulting in a proposal ratio of \(q(C_{i}, C'_{i}; T)/q(C'_{i}, C_{i}; T') =2^{U(C'_{i}, C'_{\adj(i)})-U(C_{i}, C_{\adj(i)})} \one\cbr{C_{i} \in \Ts{}{\nodea}(T)}  \).
  
  By eliminating the need to decide which clique to update, the parallel sampler only requires the enumeration of a single partition set, either $\Tn{}{\nodea}$ or $\Tbd{}{\nodea}$, based on the chosen update-type. In contrast, the single-move sampler requires the enumeration of both sets.
  \subsection{Prior and posterior probabilities}
  In this section, we construct a parallel sampler over the space of junction trees. We adhere to the iterative sampling method and prior setup used in the single-move sampler (Sec.~\ref{sec:single-move-sampler}), employing a parallelized Metropolis--Hastings step, as per \citep{hastings1970monte}, to sample from the posterior $\pi(T \given t)$. Our sampler is detailed in Algorithm~\ref{alg:parallel-sampler}.
\begin{algorithm}
  \SetKwBlock{DoParallel}{do in parallel:}{}
  \KwIn{Initiate an arbitrary tree $t$ with \(p=\abr{\cV}\) vertices, and set $T = t$. For \(M\) steps, let \(N<M\) be the frequency at which \(t\) is updated.}
  \For{ $k\gets1$ \KwTo $M$}{
    draw \(\nodea\sim\text{Uniform}\cbr{1, \ldots, p}\), choose an update-type uniformly\;
    \DoParallel{update every $C_{i} \in {\Ts{}{\nodea}(T)}$ to $C'_{i}$, and tree \(T'_{i}\), with probability
\begin{equation}
  \label{eq:acceptance-prop}
  \begin{aligned}
     \alpha_{i}(C_{i}, C_{i}'\given t) = &\min \cbr{1, \frac{\pi(C_{i}, C'_{i} \given t)q(C_{i}, C'_{i}; T)}{\pi(C'_{i}, C \given t)q(C'_{i}, C ; T'_{i})}} \\
    = &\min \cbr{1,\frac{p(Y_{C'_{i}})p(Y_{C_{\adj(i)}\cap C_{i}})}{p(Y_{C_{\adj(i)}\cap C'_{i}})p(Y_{C_{i}})}  \frac{\phi(C'_{i})\psi(C_{\adj(i)}\cap C_{i})}{\psi(C_{\adj(i)}\cap C'_{i})\phi(C_{i})}\frac{2^{U(C'_{i}, C'_{\adj(i)})}}{2^{U(C_{i}, C_{\adj(i)})}}}.
  \end{aligned}
\end{equation}
      }
    
      let $T'$ be the resulting junction tree, and set \(T = T'\)\;
    \lIf{$k \mod N = 0$}{draw $t' \sim \pi(t\given T) \propto 1$}
  }

\caption{Parallel sampler over junction trees.}
\label{alg:parallel-sampler}
\end{algorithm}

In~\eqref{eq:acceptance-prop}, $\phi$ and $\psi$ are as defined in~\eqref{eq:csf}, $q(C_{i}, C'_{i};T)$ is as specified in~\eqref{eq:proposal}, and $C_{\adj(i)} \in T_{\nodea}$ is the unique clique adjacent to $C_{i}$, such that $\edge{C_{\adj(i)}, C_{i}} \in T$.
Although the first equality in~\eqref{eq:acceptance-prop} is defined in terms of \(T\), the final ratio only involves terms associated with \(C_{i}\), its update \(C'_{i}\), and its neighbor \(C_{\adj(i)}\), since here \(C'_{\adj(i)} = C_{\adj(i)}\) by construction. As a consequence, all updates are localized to the clique of interest and its neighbors in \(T\). Moreover, since we do not update the skeleton \(t\) after every update, all updates in \(\Ts{}{\nodea}(T)\) are carried out simultaneously. The proposed parallel sampling scheme is straightforward to implement in practice. Some computational considerations are discussed in Section~\ref{sec:comp-cons}.

The chain governed by~\eqref{eq:acceptance-prop} is irreducible and aperiodic over a finite state space, since it is possible to reach the single-clique junction tree with a finite number of steps, from any state, and backward, and return to the single-clique junction tree in any number of steps. Moreover, \(\pi(T_{i}\given t) >0\). Hence, a unique stationary distribution exists. However, the chain is not reversible, as it only satisfies the partial detail balance equations~\citep{whittle1985partial}.

Conditional on the skeleton $t$, the marginal state space of clique $C_{i}$ is reversible by the congruence relation in~\eqref{eq:congruence} across chain states. This relation entails that leaf cliques become neighboring cliques, and vice versa, upon being updated. Consequently, the congruence relation indicates that all updated cliques in $\Ts{}{\nodea}(T)$ are now included in $\Ts{}{\nodea}(T')$. However, the latter set may contain additional cliques. If all cliques in $\Ts{}{\nodea}(T')$ were updated in a reverse move, the resulting state $T''$ might not revert to the original state $T$. Therefore, reversibility is partially ensured within the set $\Ts{}{\nodea}(T)$, or when $\cbr{C\cup \cbr{\nodea}: C \in \Ts{}{\nodea}(T)} = \Ts{}{\nodea}(T')$ in an addition move and $\cbr{C\setminus\cbr{\nodea}: C \in \Ts{}{\nodea}(T)} = \Ts{}{\nodea}(T')$ in a removal move.

\subsection{Computational considerations}\label{sec:comp-cons}
Our proposed single-move and parallel samplers are based on the junction property of trees, as discussed in Section~\ref{sec:expand-junct-trees}. In practice, \(T = (\cC, \cS)\) can be larger than \(\press{T} = (\clq{G}, \sep{G})\), where \(G = g(T)\). Two factors influence the number of non-empty cliques in \(\cC\), the size of the underlying skeleton \(t\), and the prior over \(T\), as specified in~\eqref{eq:csf}.

The proposed samplers can be initialized with a skeleton \(t\) of an arbitrary number of vertices \(p\). This setup can lead to \(\cC\) having up to \(p-1\) non-empty cliques, as a complete graph is a single clique. However, a very large \(p\) can result in a slowdown in convergence due to potentially redundant updates in the graph space. Conversely, a conservatively chosen \(p\) might lead to a limited number of accepted updates. We suggest initializing \(t\) with \(p = |\cV|\) vertices, as the no-edge graph yields the maximum number of cliques over the vertex set.

Prior specification plays a pivotal role in controlling the number of non-empty cliques in \(\cC\). Very diffuse priors lead to \(\cC\) being saturated with many non-empty and non-maximal cliques. In contrast, tighter priors are more conservative in proposing updates, effectively reducing the size of \(\cC\) throughout the chain iterations, even with a large \(t\). This trade-off is formalized in the following example. Consider the \(i\)th acceptance ratio in~\eqref{eq:acceptance-prop}. Proposition~\ref{lem:jtree-csf} allows us to factor out likelihood ratios corresponding to non-maximal cliques, effectively computing the likelihood ratio as if over \(\press{T} = (\clq{G}, \sep{G})\). If \(\phi=\psi\), the prior ratio cancels all terms associated with non-maximal cliques. Therefore, for our discussion, we assume that \(\phi\neq\psi\).

If \(C \in \cC\) is a non-maximal clique, then \(C\) is contained in one of its neighbors in \(T\), say \(C_{\adj}\). Consequently, $C$'s contribution to the prior ratio in the acceptance probability simplifies to the multiplicative factor $\phi(C)/\psi(C_{\adj}\cap C)=\phi(C)/\psi(C)$. Updating $C$ to the proposed $C'$ can alter its maximal status, such as $C'$ being maximal while $C$ is non-maximal, and vice versa. This scenario leads to four cases, (a)-(d), as outlined in Table~\ref{tb:update-cases}. Each case in this table denotes the contribution of the non-maximal clique to the acceptance ratio in the form of a multiplicative factor. For instance, in (a), when both $C$ and $C'$ are maximal cliques, the non-maximal clique's contribution to the acceptance probability ratio is 1. In (b), if $C$ is maximal while $C'$ is non-maximal, aside from $C$'s contribution, $C'$ contributes with the prior ratio $\psi(C')/\phi(C')$. In the special case (d), the acceptance probability is entirely influenced by the prior ratio, as no other clique in $T$ undergoes a change in status.
\begin{table}[ht]
  \centering
  \caption{Multiplicative contribution of non-maximal cliques to the acceptance ratio}
  \label{tb:update-cases}
\begin{tabular}{ l | l l}
  &   \(C' \) maximal &   \(C'\) non-maximal \\
  \hline 
  \(C\) maximal        & (a)\;:\; 1  & (b)\;:\; \(\frac{\phi(C')}{\psi(C')}\) \\
  \(C\) non-maximal    & (c)\;:\; \({\frac{\psi(C)}{{\phi(C)}}}\)  & (d)\;:\; \({\frac{\phi(C')\psi(C)}{\psi(C')\phi(C)}}\) 
\end{tabular}
\end{table}

Adjusting the contribution factors in Table~\ref{tb:update-cases} can effectively control the number of non-maximal cliques in \(\cC\). For instance, a tight prior that penalizes separators, by setting \(\psi(C) > \phi(C)\), can favor proposals involving maximal cliques. In such a configuration, the likelihood of accepting a non-maximal proposal in case (b) is reduced, while acceptance of the maximal proposal in case (a) is facilitated. The outcome in case (d) is contingent upon the specific prior used. For instance, under the clique exponential family priors~\citep{bornn2011,green2018structural}, where \(\phi(C) = \exp(\alpha|C|)\) and \(\psi(S) =\exp(\beta|S|)\) for constants \(\alpha, \beta > 0\), the ratio \(\phi(C)/\psi(C)\) becomes \(\exp\cbr{|C|(\alpha - \beta)}\). Thus, in case (d), the factor translates to \(\exp\cbr{\xi(\alpha - \beta)}\), where \(\xi = 1\) if \(|C'| > |C|\) and \(-1\) otherwise. If \(\beta > \alpha\), case (d) then tends to favor smaller non-maximal cliques. This phenomenon is explored numerically in SM Section~\ref{app:prior-effect}.

\section{Numerical performance of the new samplers}\label{sec:numerical}
Our samplers, designed to target a posterior over junction trees with $p$ vertices ($\pi(T \given Y, t)$), map this posterior onto decomposable graphs using the operator $g$. This mapping influences the frequency of graph samples, as more junction tree representations lead to higher sample frequency~\citep{thomas2009enumerating}. Achieving uniform sampling across decomposable graphs necessitates a prior considering these representations, a strategy implemented by~\cite{Green01032013}. In SM Section~\ref{sec:decomp-graphs-seven}, we demonstrate uniform sampling over reduced junction trees and approximate this uniformity across broader junction trees using Algorithm~\ref{alg:random-walk}'s visit probability $\rho$. Despite computational challenges in enumerating these broader spaces, our approach effectively approximates uniform sampling across decomposable graph spaces, as shown for $p=7$.

In the next subsections, we compare our samplers in the context of a Gaussian decomposable graphical model. The conjugate posterior of a Gaussian graphical model with a Wishart prior for the covariance matrix is derived in SM Section~\ref{sec:infer-setup-gauss}, following the classic approaches of~\citet{dawid1993} and \citet{giudici1999decomposable}. For alternative setups, such as under a log-linear model, refer to works like~\cite{tarantola2004mcmc}.

\subsection{Gaussian graphical model simulation setup}\label{sec:simul-setup-gauss}

To illustrate the advantages of our methods, we consider the following graph setups: an autoregressive (AR) graph with lag varying between 1 and 5, and a random decomposable graph sampled following the Christmas tree algorithm (CTA) of~\cite{olsson2022sequential} with parameters set to 0.5. 

\begin{figure}[!ht]
  \centering
  \includegraphics[width=0.19\textwidth]{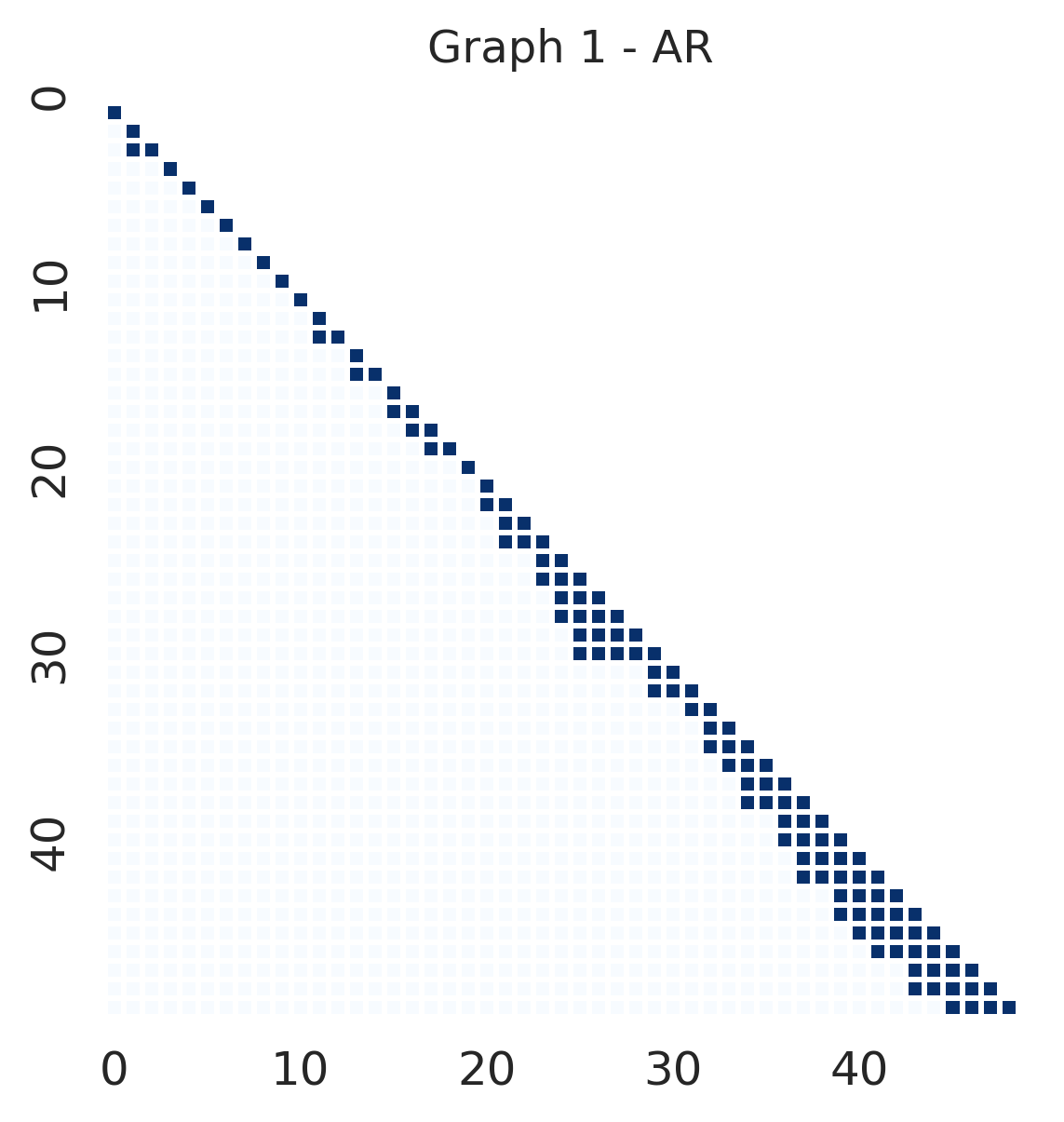}
  \includegraphics[width=0.29\textwidth]{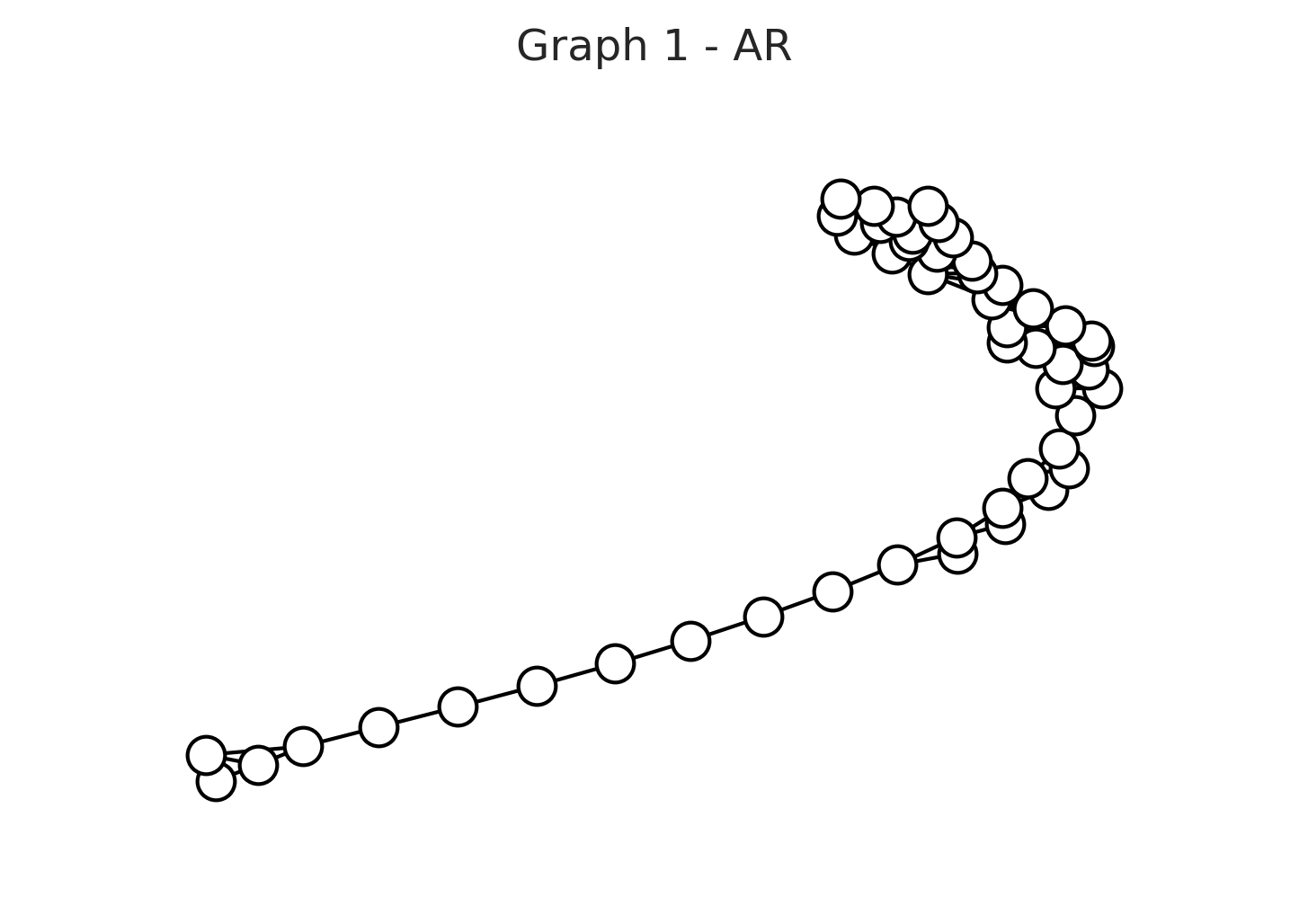}
  \includegraphics[width=0.19\textwidth]{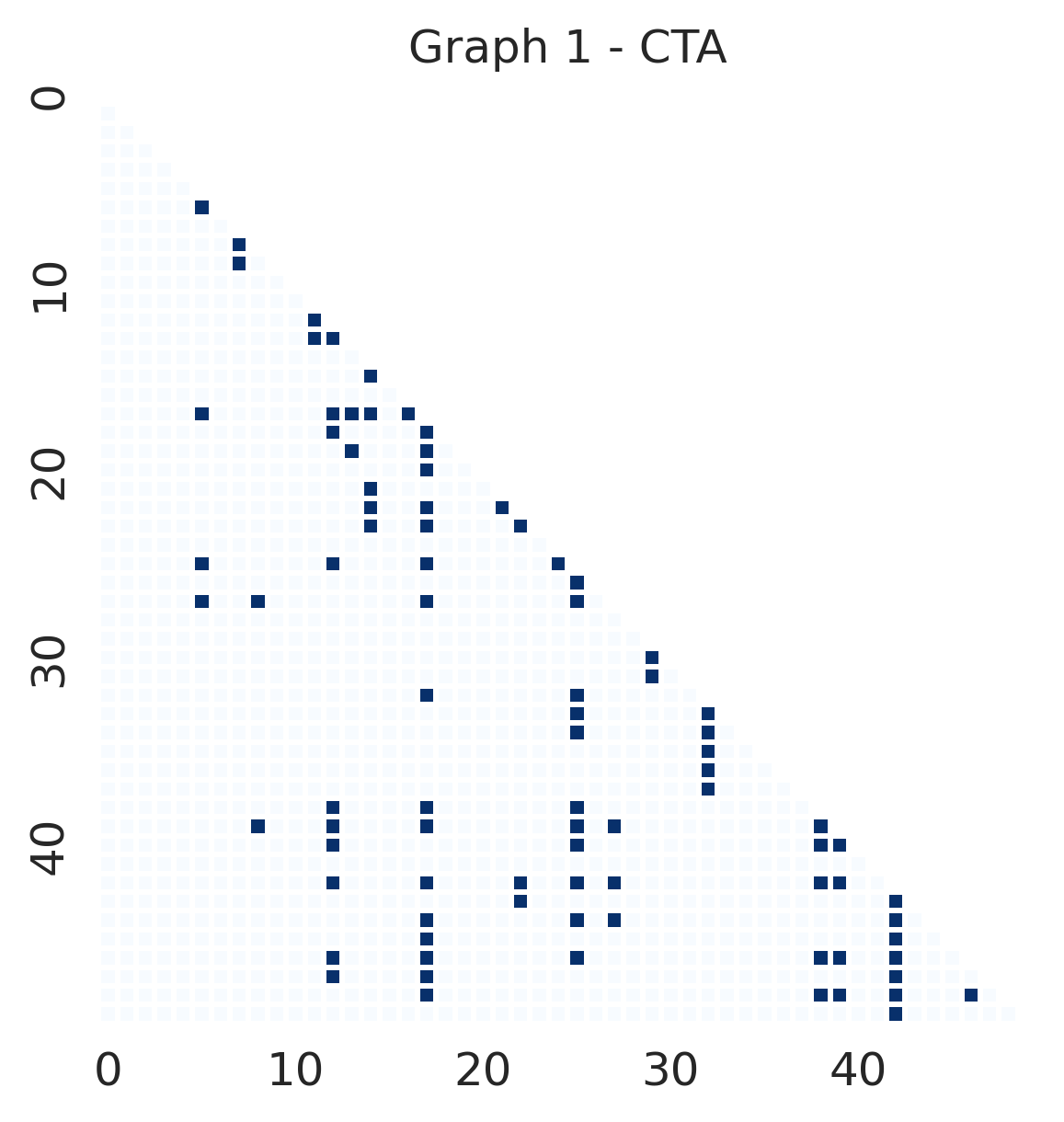}
  \includegraphics[width=0.29\textwidth]{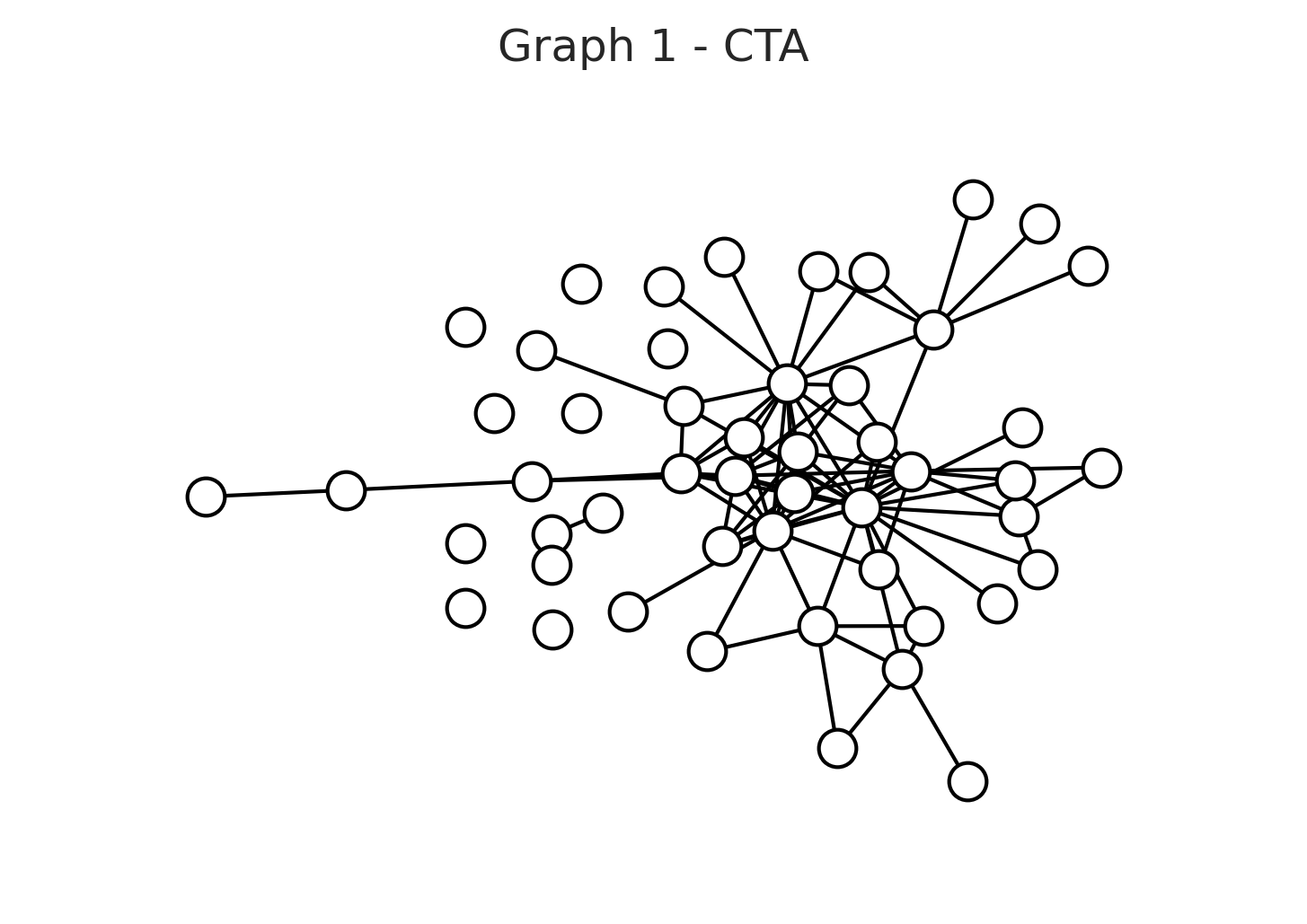}
  \caption{Simulated 50-vertex autoregressive (AR) and Christmas tree algorithm (CTA)-generated decomposable graphs, alongside their adjacency matrices.}
  \label{fig:ggm-ture-G}
\end{figure}

For each setup, we sample a graph \(G\) and generate a dataset of 100 samples following a \(N_{p}(0, \Theta^{-1})\) distribution, where \(\Theta \sim \text{inverse-Wishart}(\delta, \mathbb{I}_{p} | G)\) and \(\mathbb{I}_{p}\) is the identity matrix. This process is repeated 10 times, setting \(p=\cbr{50, 200}\) and \(\delta=3\), resulting in 10 unique datasets per dimensions $p$, each associated with a distinct conditional independence graph. Figure~\ref{fig:ggm-ture-G} illustrates an example of two sampled decomposable graphs \(G\) alongside their adjacency matrices, for each graph setup and $p=50$. Other simulated graphs are presented in SM Fig~\ref{app:fig:ture-graph-auto} and~\ref{app:fig:ture-graph-cta}, for $p=50$.

To simulate real-world conditions and avoid bias from prior knowledge of the generative model, in the posterior sampling, we assign a hyper Wishart prior to \(\Theta\). For each clique \(C\), the degree of freedom is set at \(\delta=1\), with the \(\abr{C}\)-dimensional identity matrix as the scale matrix. This choice of prior leads to a conjugate posterior, as elaborated in SM Section~\ref{sec:infer-setup-gauss}. Unless stated otherwise, we employ a uniform prior over \(T\), setting \(\pi(T) \propto 1\).

The next sections compare our single-move sampler (Sec.~\ref{sec:single-move-sampler}) and parallel sampler (Sec.~\ref{sec:jt-sampler}) with four existing Bayesian samplers for decomposable graphs: the multi-edge (GT13-m) and single-edge (GT13-s) junction tree samplers of~\cite{Green01032013}, the single-edge (GG99) sampler of~\cite{giudici1999decomposable}, and the particle Gibbs sequential Monte Carlo sampler (O19) by~\cite{olsson2019bayesian}. In contrast to our junction-based sampler, the samplers by~\cite{Green01032013} are graph-based, meaning at every step they reduce \(T\) to \( J=\press{T}\). Their GT13-m sampler updates an arbitrary number of edges in line with their decomposability-preserving proposals. This differs from our all-or-none sampling method, where \(\nodea\) is either added to or removed from a clique. Our single-move sampler effectively operates as a multi-edge sampler, except in cases involving only two vertices. This contrasts with GT13-s, which consistently selects two vertices at every step.

We execute the single move sampler for one million steps, and the parallel sampler for 500,000 steps when $p=50$ and one million when $p=200$. We initiate the samplers with a random $p$-vertex skeleton \(t\) and set it as the junction tree \(T\), which corresponds to the junction tree of a no-edge $p$-vertex graph. We sample a posterior of the skeleton \(\pi(t \given T)\) every 100 Metropolis--Hastings samples of \(\pi(T \given t)\), following the recommendation in~\cite{Green01032013}. The sampling of \(\pi(t \given T)\) follows the randomization approach in~\cite{thomas2009enumerating}.

We execute one million steps for all MH-Mc competing samplers when $p=50$, and 30 million when $p=200$. In the latter case, this large number of steps was required for convergence. We incorporate junction tree randomization in the samplers of~\cite{Green01032013} every 100 steps. While~\cite{Green01032013} defines this as junction tree randomization, we refer to it as a hierarchical sampling scheme, as outlined in~\eqref{eq:hierarchical-model}. The Gibbs sampler of~\cite{olsson2019bayesian} is analytically more complex than a Metropolis--Hastings, and more computationally intensive. Therefore, we executed it for 10,000 steps, with each step comprising 50 particles when $p=50$. This sampler is computationally unfeasible for $p=200$. All competing samplers are assigned a uniform graph prior. Simulations were performed using the {\it Benchpress} framework~\citep{rios2021benchpress} with code outlined in SM Section~\ref{app:sec:simcode}.

Updates in both our parallel and single-move samplers are performed on the junction tree \(T\), yet not every update on \(T\) corresponds to a change in the underlying decomposable graph \(G = g(T)\). To evaluate the performance of these samplers in the graph space, we process each chain post-sampling to generate a new chain with the graph \(G\) as the state variable. Metrics calculated on this processed chain are termed as graph-updates metrics, distinguishing them from junction-updates metrics.

\subsection{Simulation results for the general case}\label{sec:general-ggm}
\begin{figure}[ht!]
  \centering
  \includegraphics[width=0.32\textwidth]{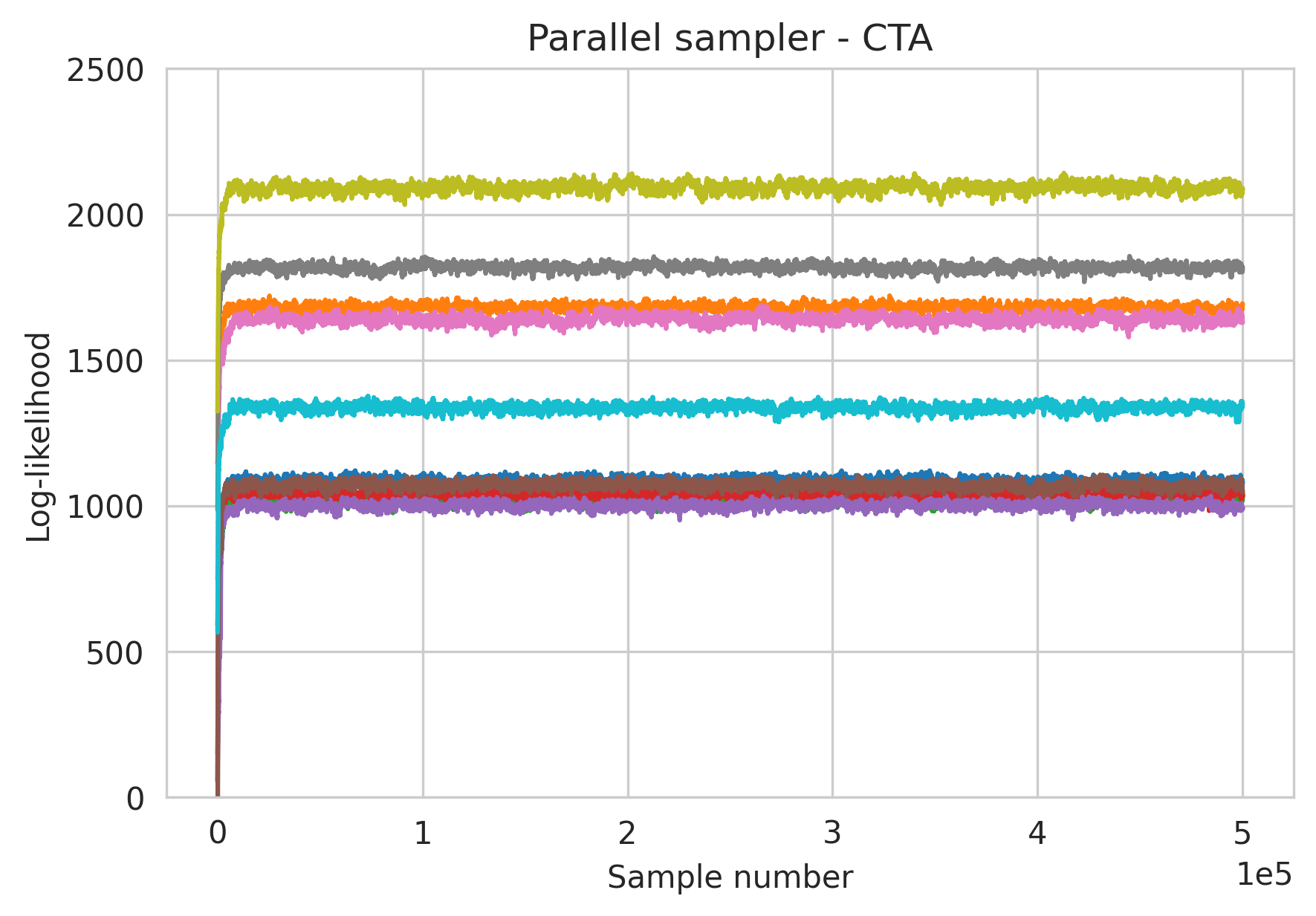}
  \includegraphics[width=0.32\textwidth]{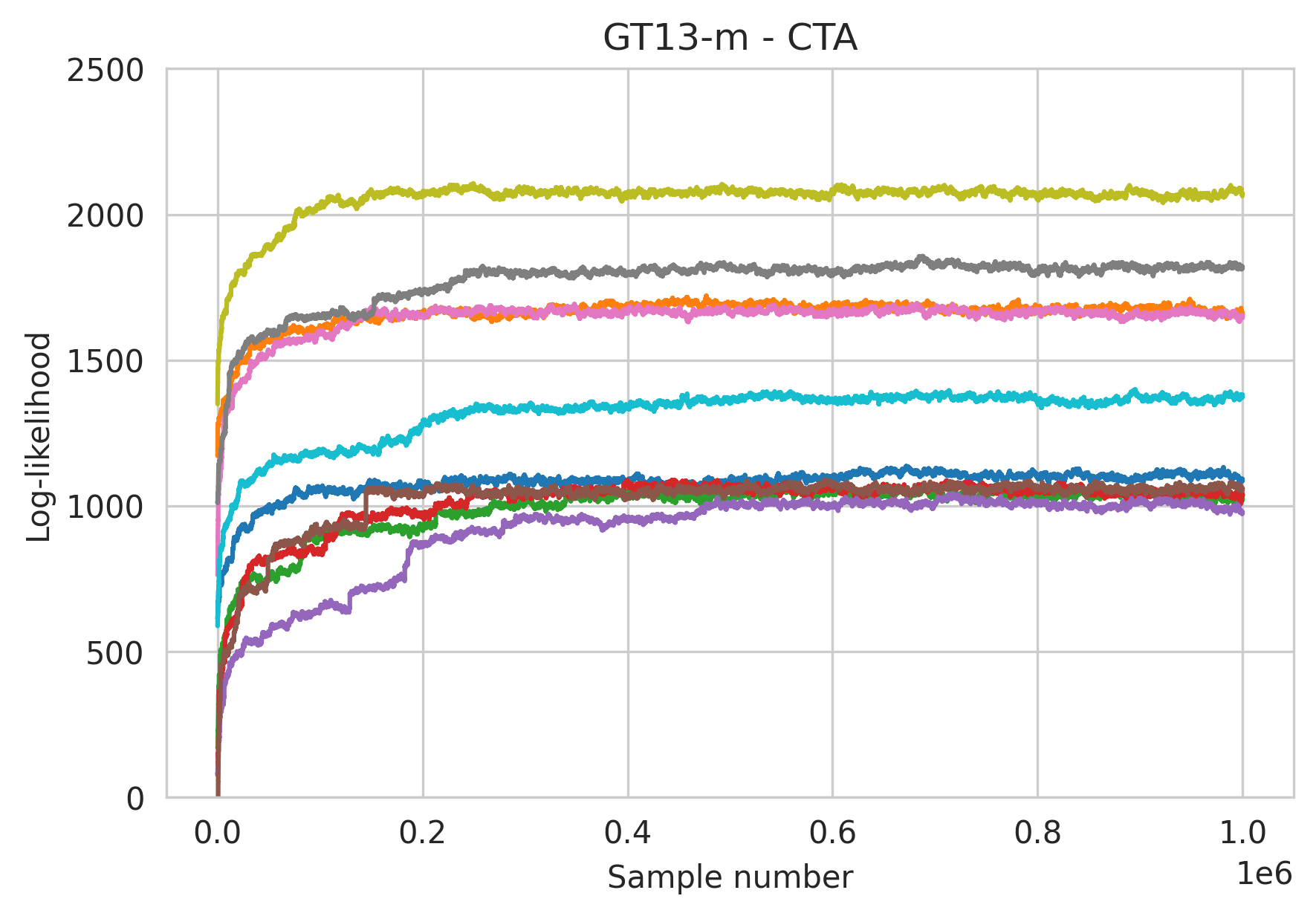}
  \includegraphics[width=0.32\textwidth]{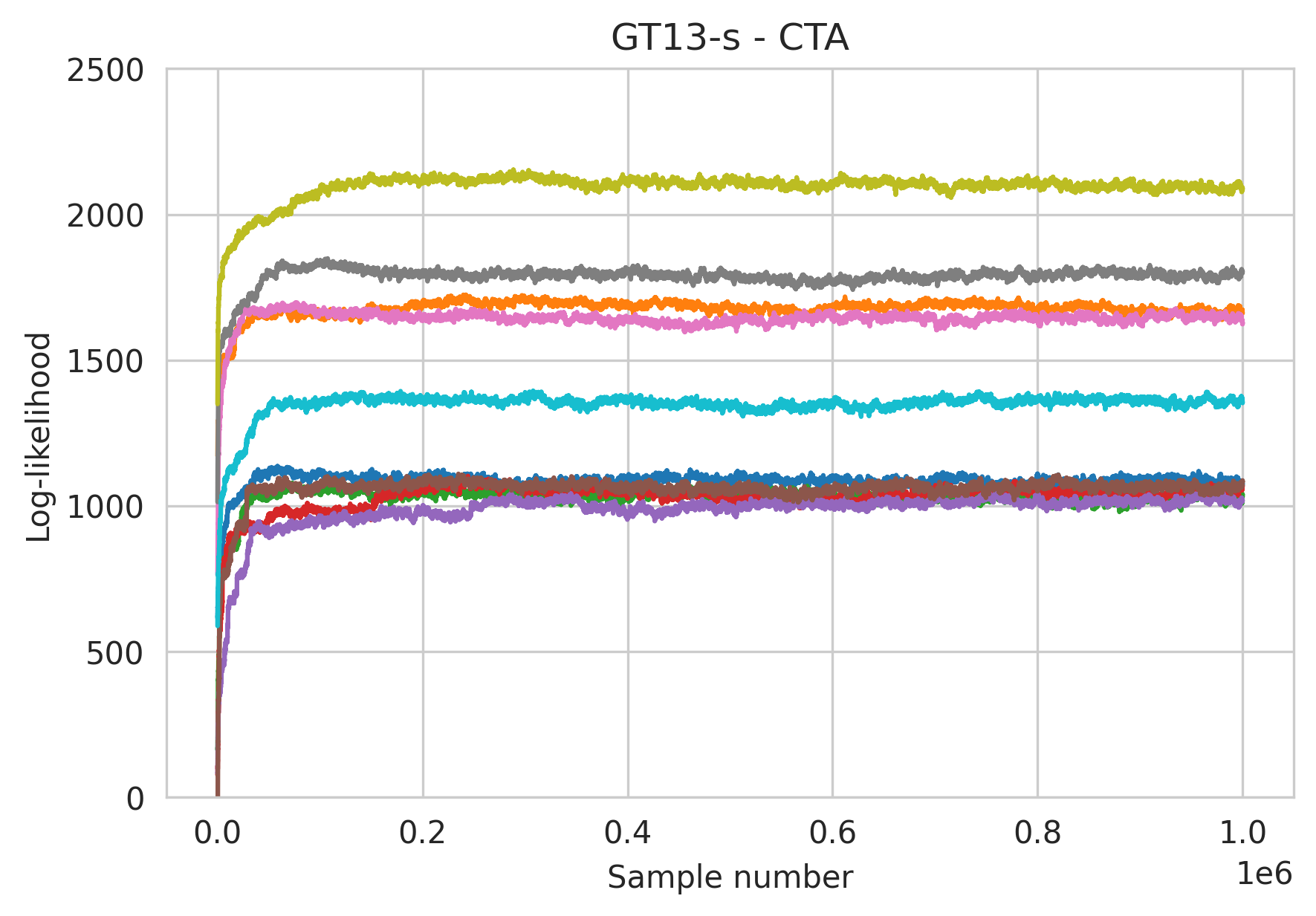}
  
  \includegraphics[width=0.32\textwidth]{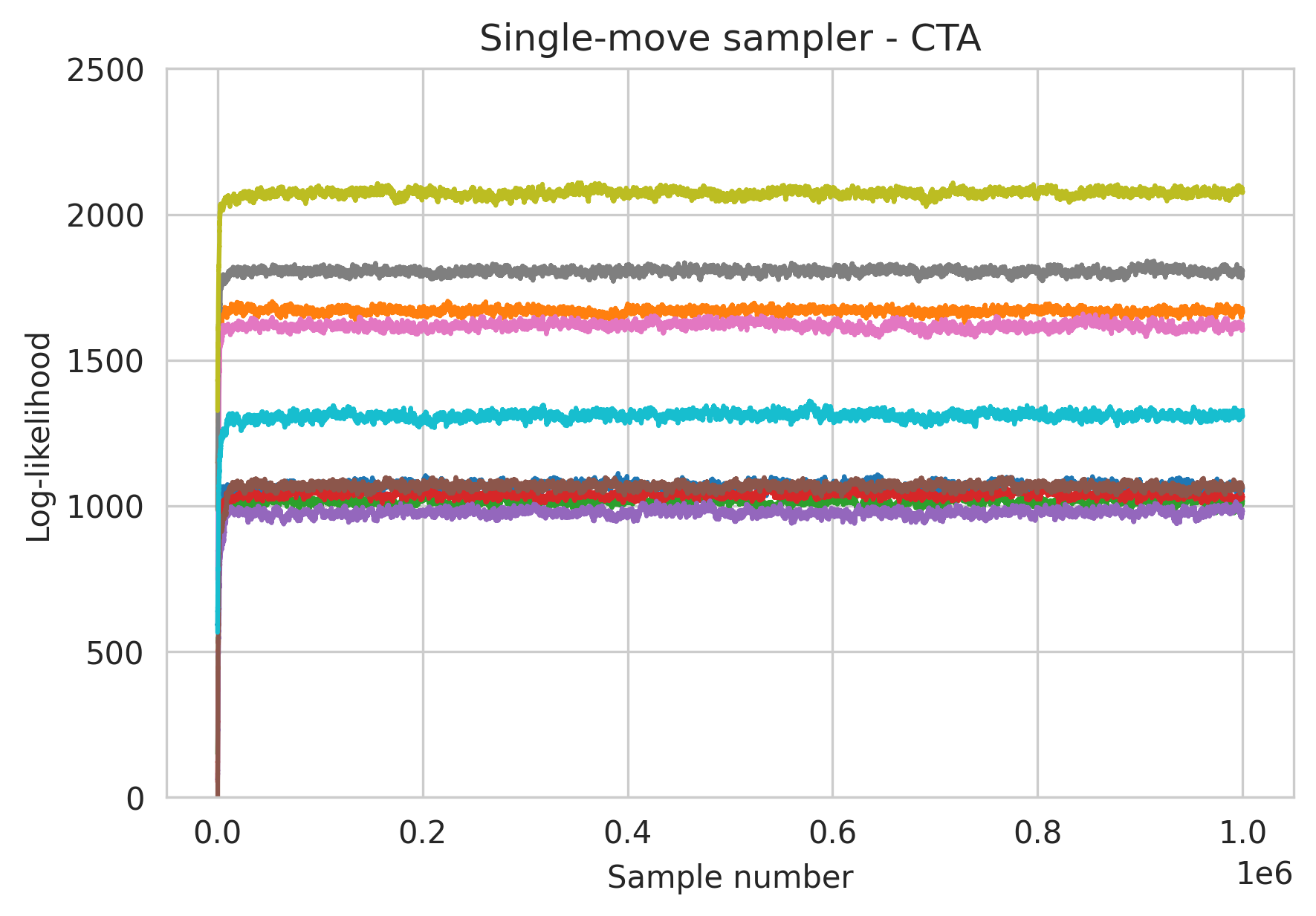}
  \includegraphics[width=0.32\textwidth]{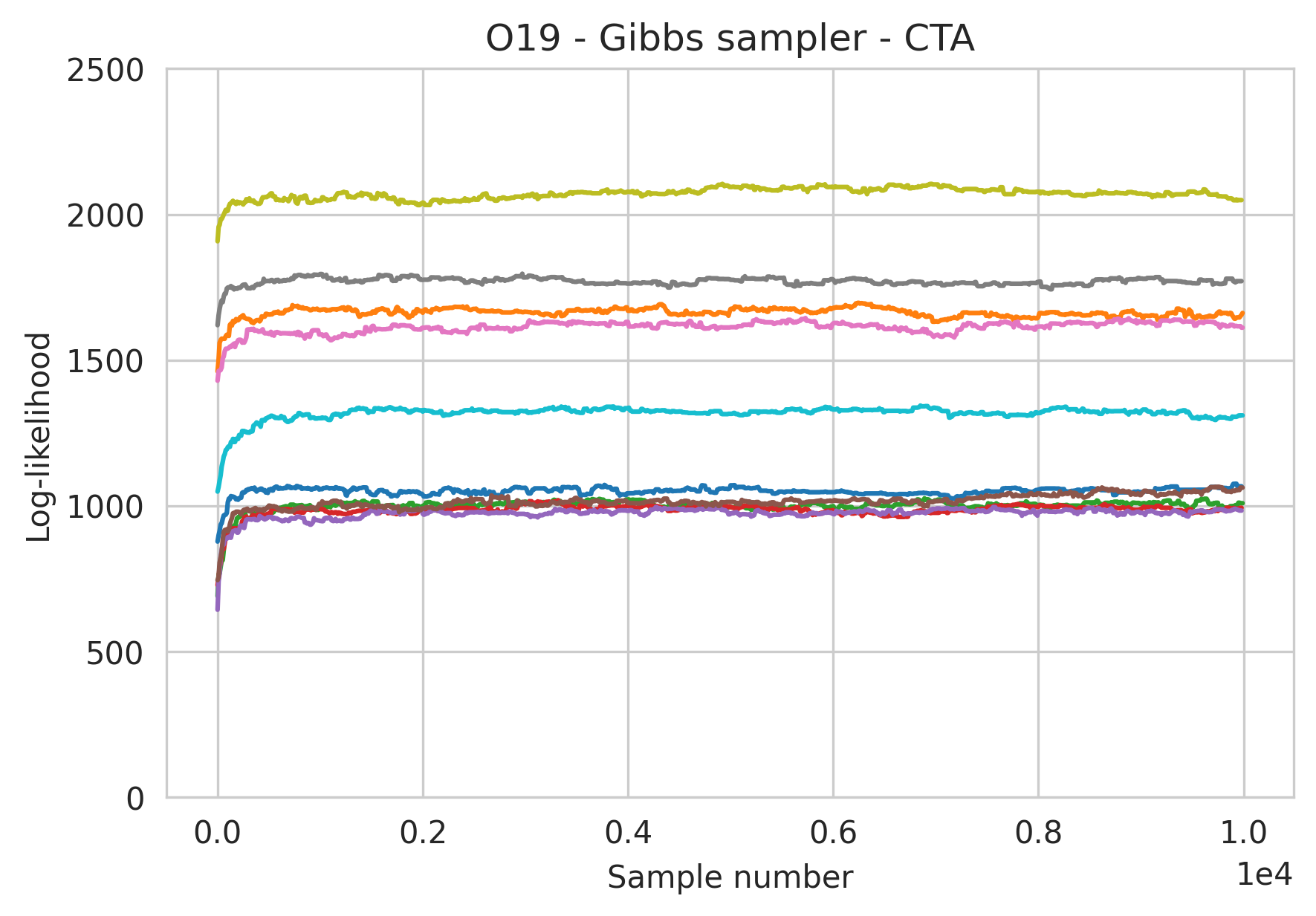}
  \includegraphics[width=0.32\textwidth]{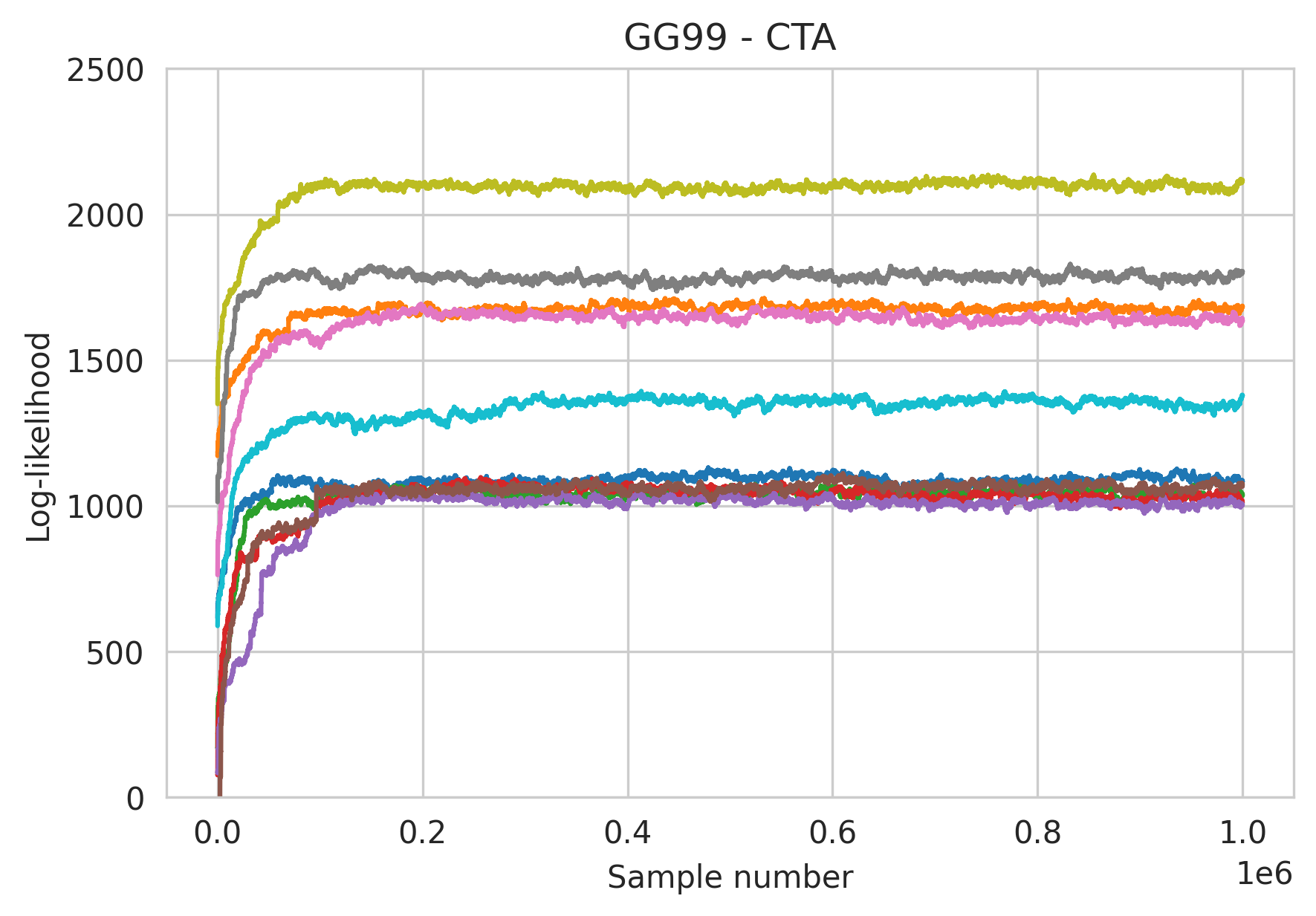}
  \caption{Log-likelihood traceplots of 10 replications (color-matched) for our samplers and competing ones, under the Christmas tree algorithm (CTA) and $(n,p)=(100,50).$}
  \label{fig:gt-parallel-traceplot}
\end{figure}

\begin{figure}[ht]
  \centering
  \includegraphics[width=0.32\textwidth]{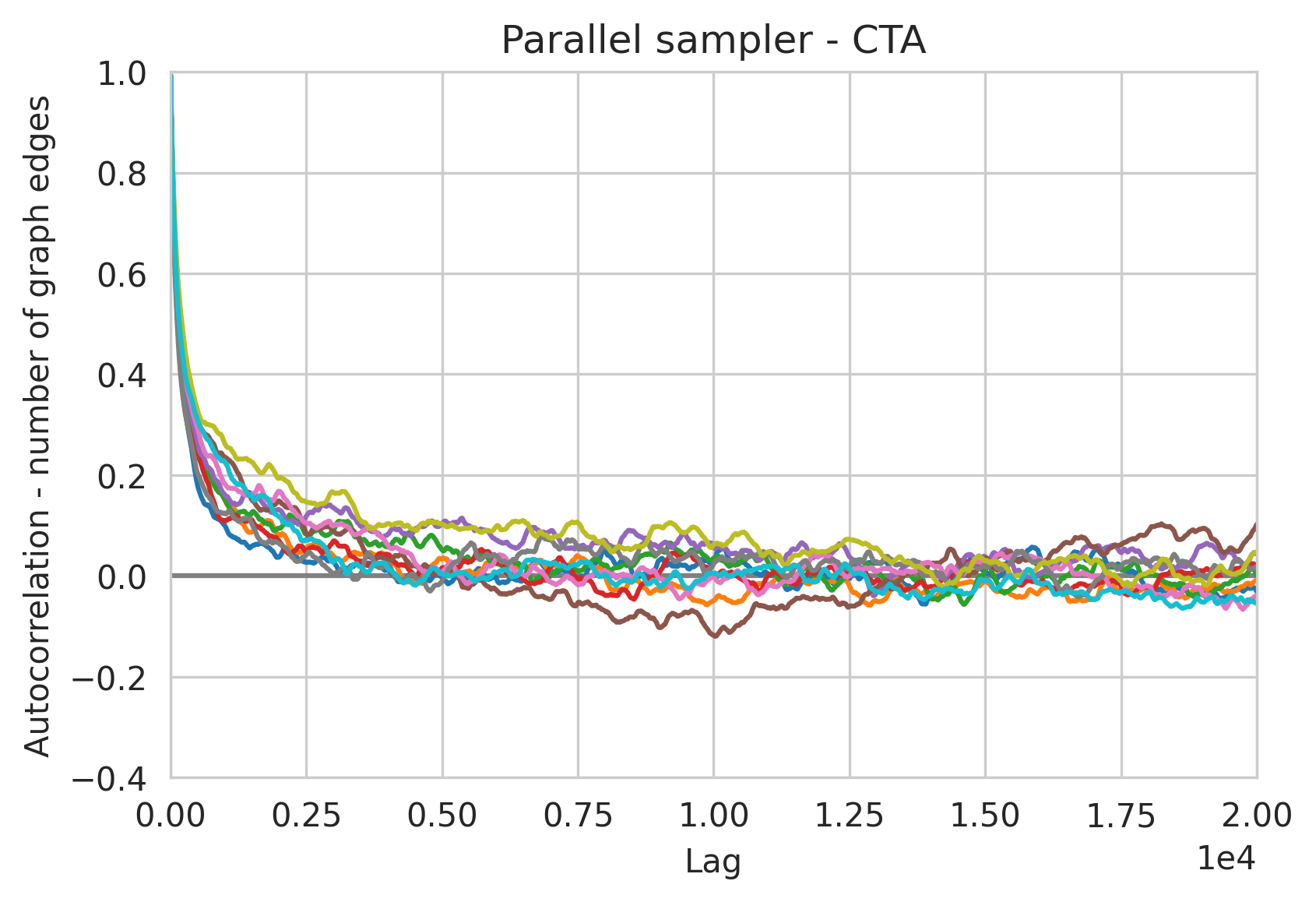}
  \includegraphics[width=0.32\textwidth]{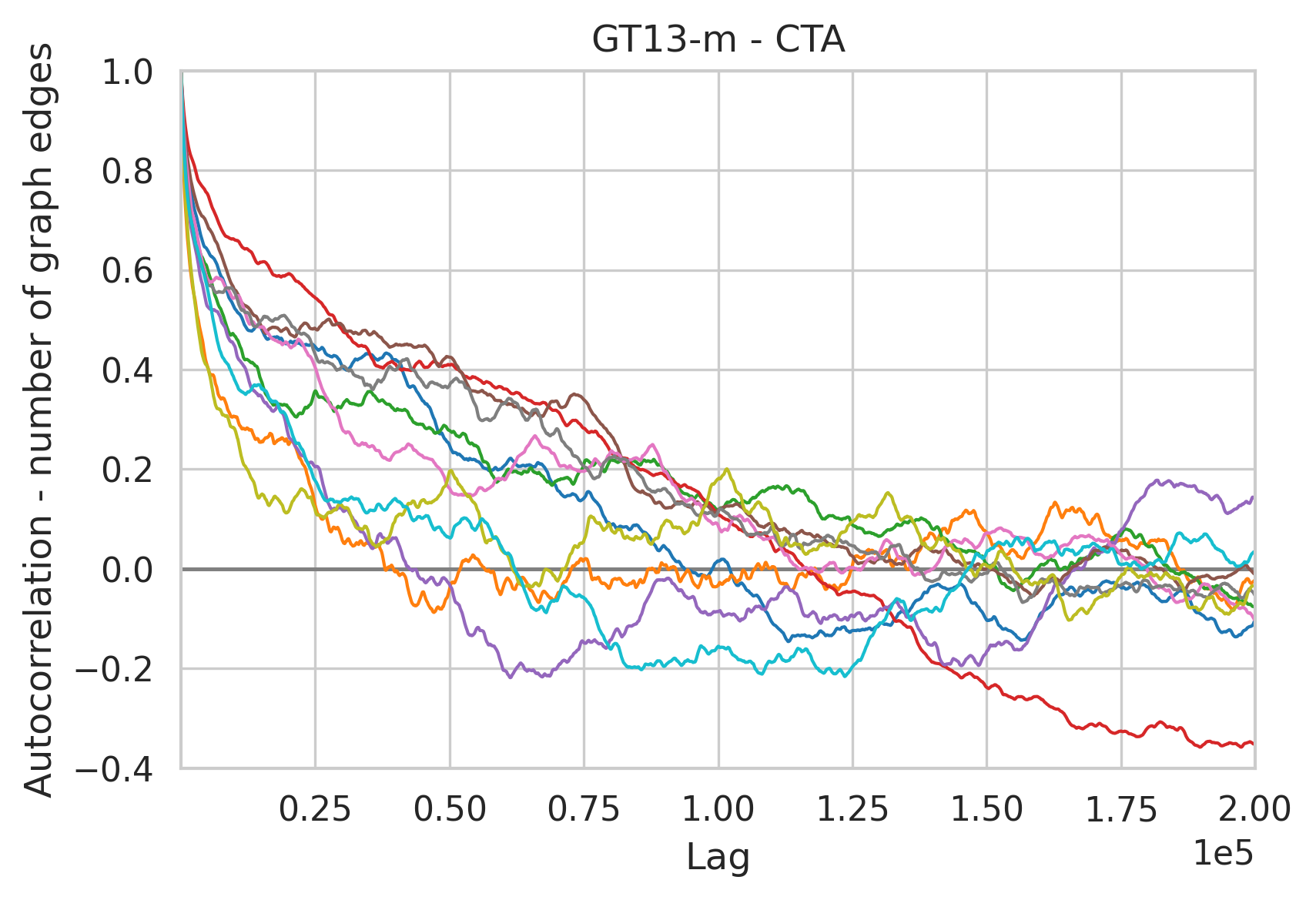}
  \includegraphics[width=0.32\textwidth]{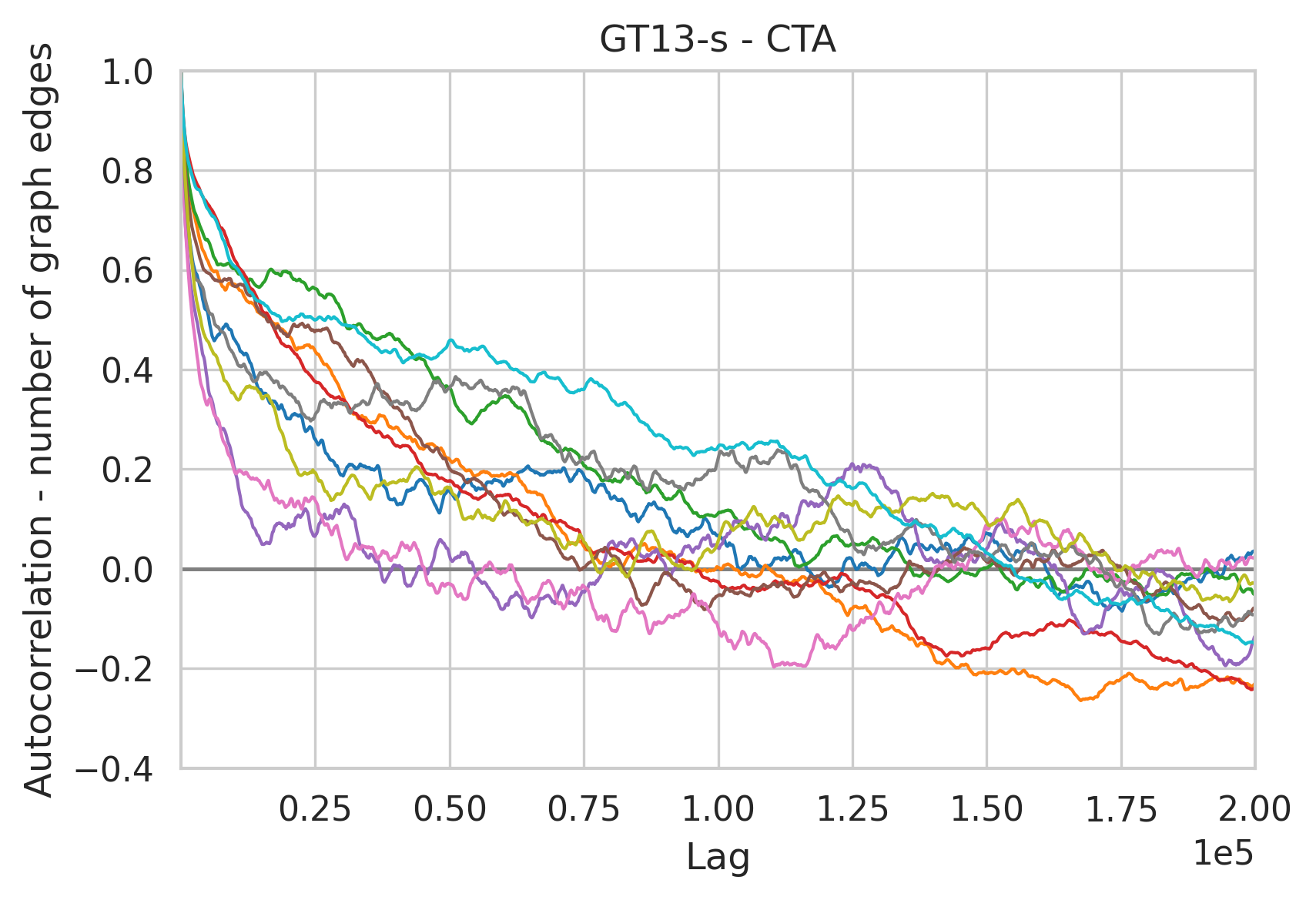}

  \includegraphics[width=0.32\textwidth]{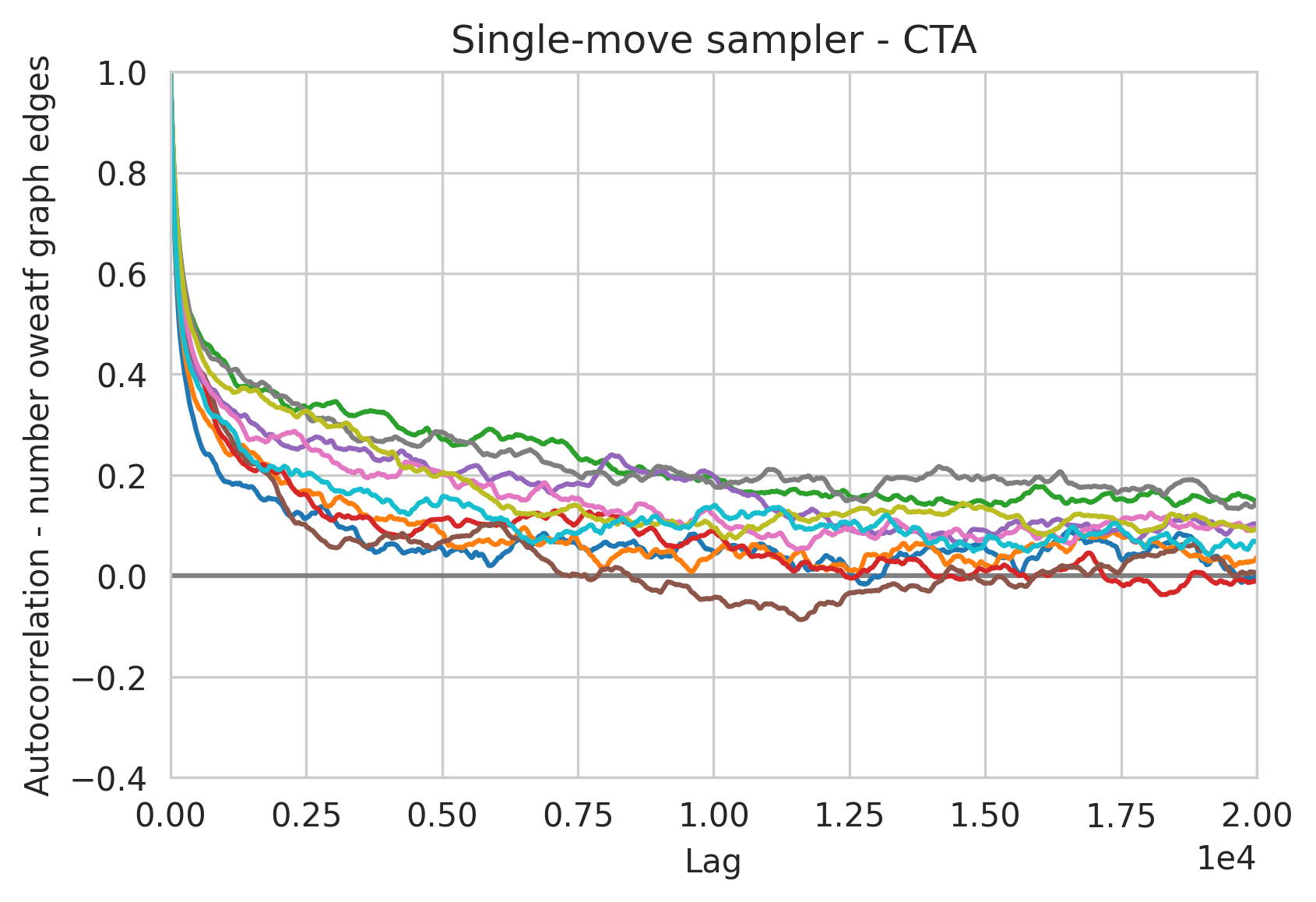}
  \includegraphics[width=0.32\textwidth]{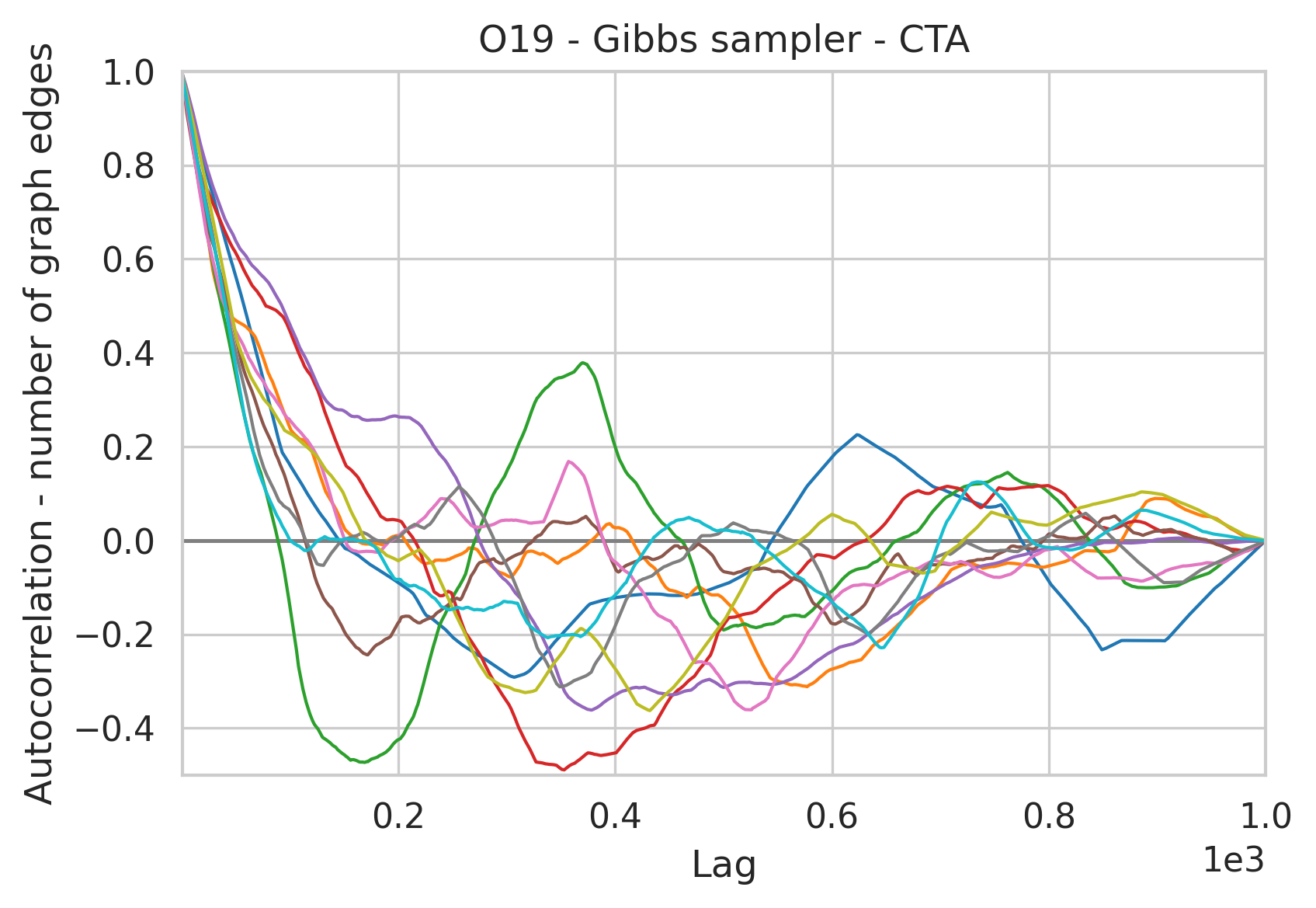}
  \includegraphics[width=0.32\textwidth]{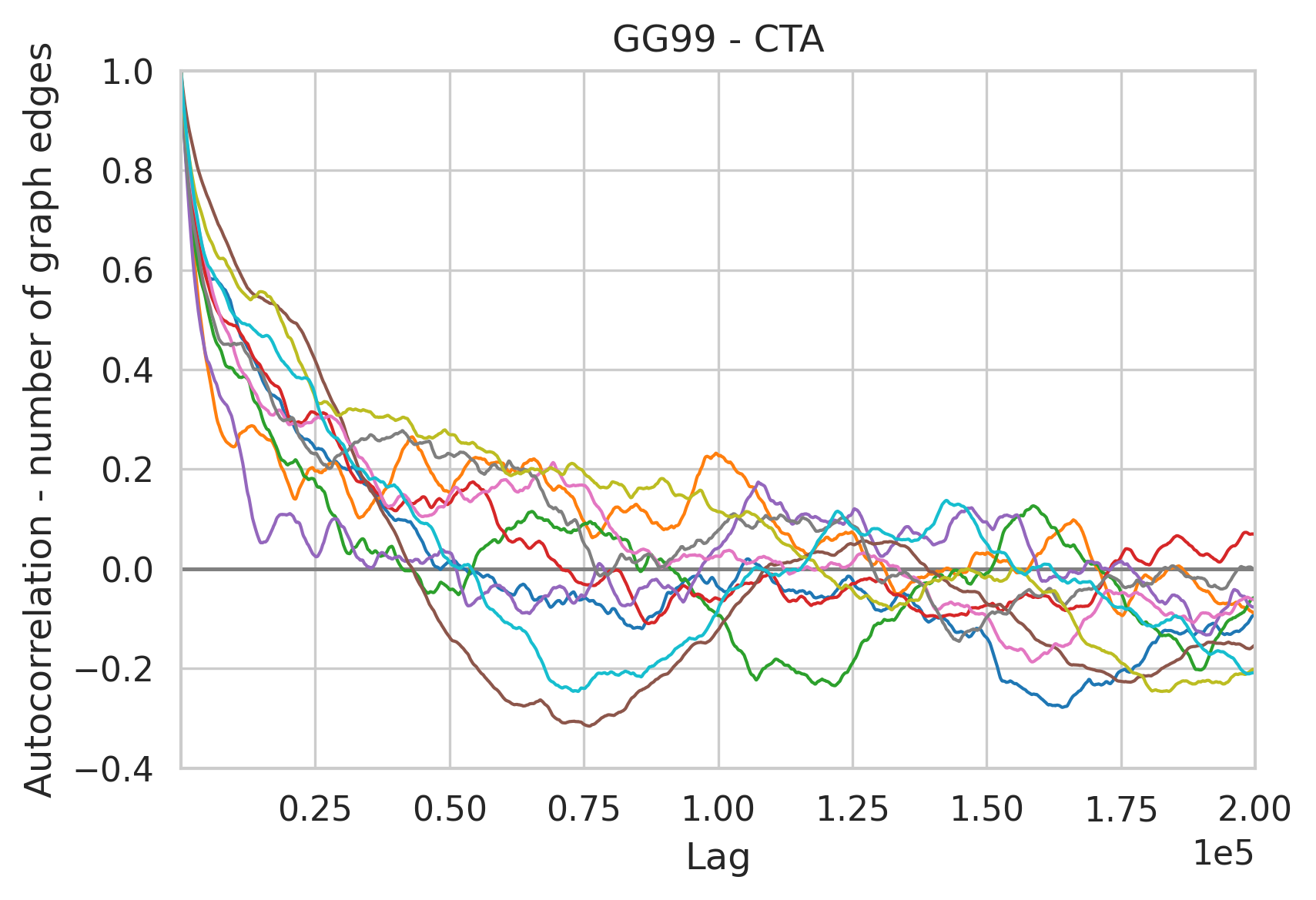}

  \caption{Within-sample correlation over the last 300,000 samples of each of the 10 replications (color-matched), under the Christmas tree algorithm and $(n,p)=(100,50)$.}
  \label{fig:ggm-auto-correlation}
\end{figure}

Our parallel and single-move chains have achieved convergence within the initial 50,000 steps (Fig.~\ref{fig:gt-parallel-traceplot}, left panel). The within-sample correlation concerning the number of edges in the underlying graph for each chain is almost negligible for these samplers. Specifically, the correlation drops below 0.2 for all parallel chains by lag 2500 and by lag 10000 for the single-move chains. They maintain a level close to zero subsequently (Fig.~\ref{fig:ggm-auto-correlation}, left panel). 

Overall, we observed consistently faster convergence to the chain's stationary state, less within-sample correlation and improved computation speed with the parallel sampler versus the single-move sampler. The computational speed improvement is attributable to parallelization; the parallel sampler executed approximately 1,500,000 total updates for 500,000 steps, achieving about 300\% greater efficiency. Each update for the parallel sampler requires fewer computations than the single-move sampler, since the former does not require a proposal ratio computation as in ~\eqref{eq:acceptance-prop-single-move}

\cite{Green01032013}'s chains required approximately 600,000 steps to converge, with at least two multi-edge (GT13-m) chains failing to do so. The within-sample correlation for these samplers remains high, only dropping below 0.2 at a lag 100,000.~\cite{giudici1999decomposable}'s chains (G99) performed similarly to GT13-s.~\cite{olsson2019bayesian}'s chains (O19) showed improved within-sample correlation, falling below 0.2 at lag 1000, however, thier run-time exceeded 4 hours, in contrast to the sub-3-minute for other samplers. The relative computational efficiency for 1000 steps is about 0.2 seconds for all samplers, except O19, which is approximately 1 hour. Simulations are run on an Intel Xeon E5-2698 v4 2.20GHz processor, for detailed analysis of computation time see SM Figure~\ref{fig:computation_time}. Computational speed of our samplers scales linearly with $p$ (SM Fig.~\ref{fig:time-samplers}). 

GT13-m median acceptance rate is $\approx$4\%, while our parallel sampler exceeds 26\% acceptance rate. The GT13-s aligns with our decomposability conditions, showing comparable rates to our single-move sampler and double that of the non-junction tree based GG99 sampler. Further demonstrating the advantage of junction-tree based samplers. 

\begin{figure}[!ht]
  \centering
  \includegraphics[width=0.44\textwidth]{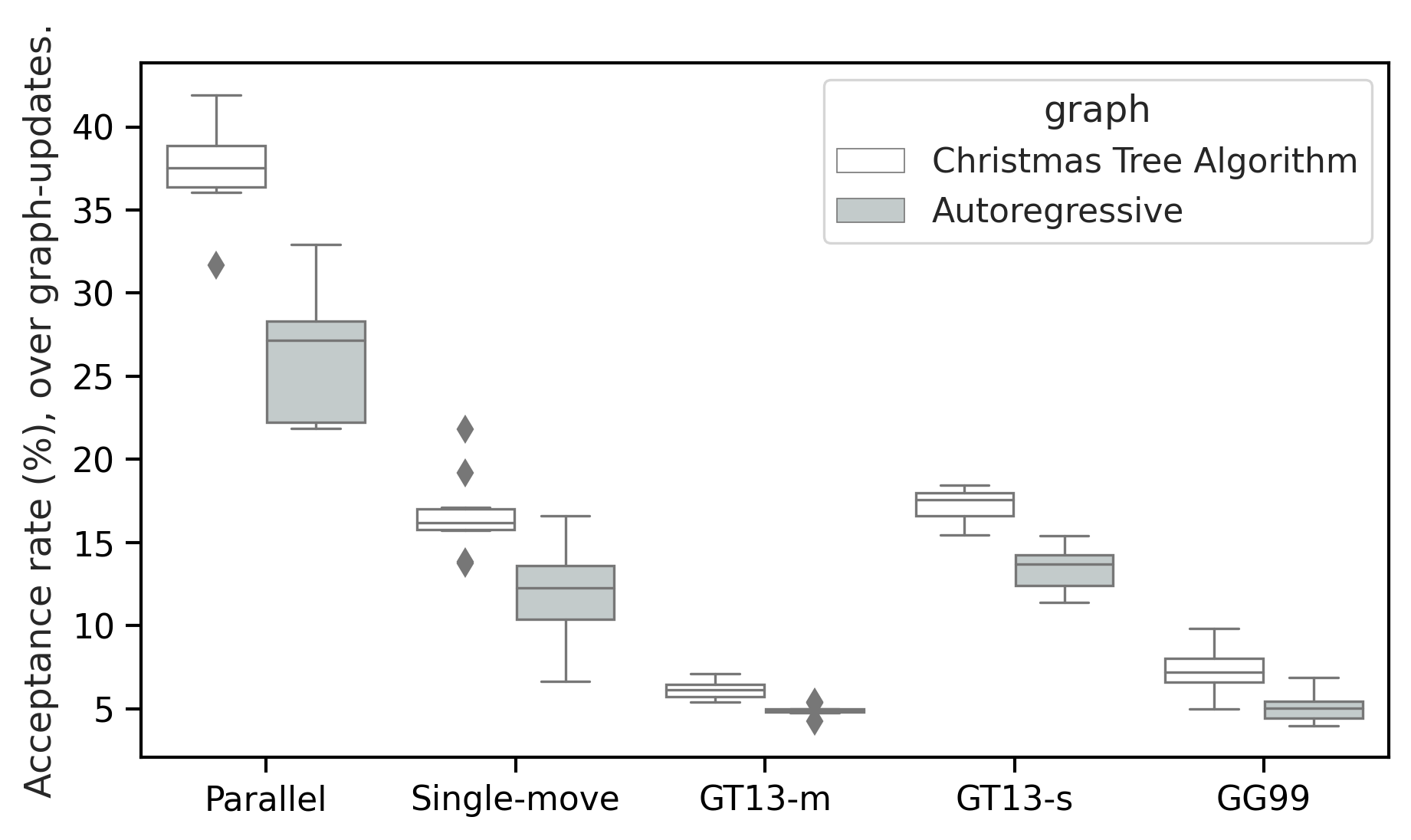}
  \includegraphics[width=0.38\textwidth]{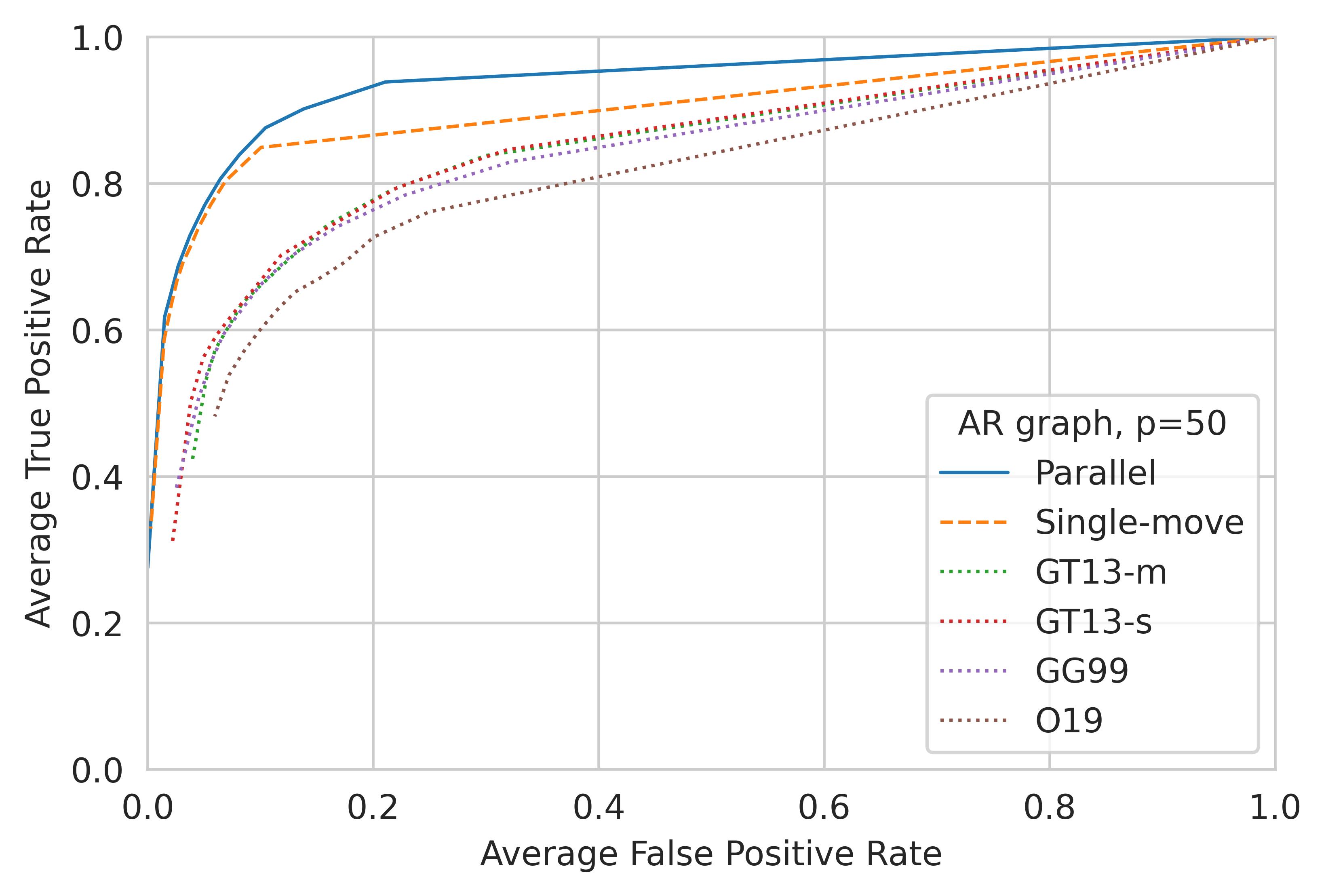}
  \caption{(Left) Boxplot of acceptance rates over graph-updates comparing our samplers and competitors, across two graph structures and 10 replications. (Right) ROC curves comparing the same samplers under the autoregressive graph structure and $(n,p)=(100,50)$.}
      \label{fig:ggm-roc}
\end{figure}

Alongside improved mixing properties, our samplers demonstrate higher accuracy than competing samplers. This is evidenced by the ROC curves in Figures~\ref{fig:ggm-roc} (right) and~\ref{fig:ggm-roc-cta} (left), computed from the last 300,000 samples of each chain, where the true and false positive rates are computed and then averaged these across all replications. 

Our samplers aim at a distinct distribution (Sec.~\ref{sec:numerical}). Nonetheless, empirically, our chains converged close to competing methods (Fig.~\ref{app:fig:loglikeli-traceplots}). Analysis of the autoregressive graph structure show similar results to the Christmas tree algorithm, and thus differed to SM Section~\ref{app:sec:ggm}, alongside more diagnostic plots on the simulations of this section.

\subsection{Simulation results for the high-dimensional case}\label{sec:sparse-ggm}
In the high-dimensional case ($p=200$), our samplers demonstrated distinct convergence properties. While similar in acceptance rates and computational time to the general case ($p=50$, Sec.~\ref{sec:general-ggm}), our samplers reached stationarity within the first 600,000 steps (SM Fig.~\ref{app:fig:corr-traceplots-sparse}), in 2--3 minutes. In contrast, some GT13-s and GG99 chains required around 20 million steps to converge, lasting about 40 minutes. Most GT13-m chains did not converge even after 30 million steps, exceeding an hour in computational time. This discrepancy is evident in the traceplots showing the number of graph edges (SM Fig.~\ref{fig:gt_parallel_size_traceplot-sparse}). The accuracy gains of our samplers far exceeded competing ones, where the gap is much larger than the case of $p=50$, as shown by the ROC curves in Figure~\ref{fig:ggm-roc-cta} (right) and SM Figure~\ref{fig:boxpot-roc-sparse} (right). The ROC curves were computed from the last 500,000 samples of each chain. For additional diagnostic plots of the simulations in this section, please refer to SM Section~\ref{app:sec:ggm:sparse}.

\section{Discussion}
Computational bottlenecks still remain in the quest to design Markov chain samplers for high-dimensional decomposable graphs. Exploiting the junction tree representation led to the largest gain in sampling efficiency, as established in the work of~\cite{Green01032013,giudici1999decomposable,olsson2019bayesian, olsson2022sequential}, and herein. A greater part of this efficiency gain is driven mainly by an improvement in graph proposals that the junction property explicitly exposed. Nonetheless, the junction representation is not unique. Counting the number of junction tree representations for a fixed decomposable graph is known to be computationally expensive~\citep{thomas2009enumerating}, which led~\cite{Green01032013} to propose a modified Metropolis--Hastings chain that is inferior to the classical one, to avoid such computation. Our proposed method trades the need to quantify the space of junction trees with a form of latent dimensional expansion coupled with hierarchical sampling of junction trees. Here, the latent expansion represents non-maximal cliques of the underlying decomposable graph.

While some updates in this latent space are probabilistically superfluous, as they involve non-maximal cliques, this work has demonstrated that such latent expansion is promising for two reasons. First, controlling the size of this expansion can be done efficiently and intuitively by restricting the size of the latent junction trees to the number of graph vertices; it is intuitive since a no-edge graph has the maximum number of maximal cliques over the set of vertices. Second, parallel sampling, which improves mixing properties, seamlessly integrates into our framework as it does not necessitate junction tree modification post-update—a requirement in other junction tree-based samplers.

Parallelism is a key benefit of our sampler over competitors, but not the only one. All proposals from our samplers inherently preserve decomposability, with rejections occurring due to likelihood ratio. A consequence of  Theorems~\ref{lem:connect} and~\ref{lem:disconnect}. By contrast, the disconnect move in the multi-edge sampler of~\citet[Sec.~3.2 conditions (a)-(d)]{Green01032013} can cause decomposability-based rejections, leading to less optimal mixing. These rejections occur at the proposal stage without a full evaluation of the data likelihood~\citep[Sec.~3.4]{Green01032013}. This leads to faster computational time (SM Fig.~\ref{fig:time-samplers}), yet reduced acceptance rates (Fig.~\ref{fig:ggm-roc} left), for the multi-edge sampler when compared to others.

\cite{Green01032013} implemented a multi-edge update per proposal, choosing edge sets uniformly from a valid superset with probability proportional to \(2^{m} -1\), where \(m\) is the cardinality of the superset. This approach appears less efficient compared to alternatives. The average acceptance rate for their multi-edge proposal is lower than the single-edge proposal, as shown in~\citep[Supp.~Fig.~3]{Green01032013} and Figure~\ref{fig:ggm-roc} (left). This work opted for an all-or-none multi-edge update, since it is more intuitive with respect to the junction property, where it does not require any modification to the tree skeleton. Our proposal outperformed competing ones in efficiency and accuracy.

Recent advances by~\cite{uhler2018exact} in providing exact formulas for normalizing constants of Wishart priors have addressed a major analytical bottleneck in nondecomposable graphical models, previously a significant challenge in the field as noted by~\cite{jones2005, wang2010simulation, mitsakakis2011metropolis, roverato2002hyper, atay2005monte, dellaportas2003bayesian}. However, computational barriers persist. In moderate-dimensional problems, state-of-the-art Bayesian structure learning methods for nondecomposable graphical models executes 1000 steps in 15 minutes, as shown by~\cite{mohammadi2021accelerating}, compared to less than a second for similar tasks in decomposable models, as shown here and in~\cite{Green01032013,giudici1999decomposable}. In fact, under a Gaussian graphical model of 150 vertices, our samplers execute 1000 steps in 0.3 seconds. This dramatic reduction in computational time highlights the efficiency advantage of assuming decomposability in the conditional independence graph \(G\). 

It still remains unclear whether Bayesian inference for graphical models can be achieved for large-dimensional problems, beyond 1000 vertices for example, with no reliance on the decomposability framework.

\section*{Acknowledgements} The author wishes to acknowledge Felix Rios (Royal Institute of Technology in Stockholm) for providing necessary code through the {\sf trilearn} Python package, and the structure learning benchmarking workflow {\sf Benchpress}, and for his comments, which  improved the quality of this work. The author was supported by an NSERC postdoctoral fellowship.

\bibliographystyle{ims}
\bibliography{references}
\bigskip

\newpage
\renewcommand{\appendixpagename}{Supplementary material}
\renewcommand{\appendixtocname}{Supplementary material}
\renewcommand{\thesection}{S\arabic{section}}  
\renewcommand{\thetable}{S\arabic{table}}  
\renewcommand{\thefigure}{S\arabic{figure}}
\begin{appendices}
\section{Table of notations}
\begin{table}[ht!]
\caption{Table of Notations}
\label{table:notations}
\small
\begin{tabular}{ll}
\hline
Symbol & Description \\
\hline
\(G = (\cV, \cE)\) & An undirected decomposable graph with vertex set $\cV$ and edge set $\cE$\\
\(V \subseteq \cV\) & A subset of vertices \(\cV\) \\
\(\clq{G}\) & The set of unique maximal cliques of \(G\) \\
\(\sep{G}\) & The set of unique minimal separators of \(G\) \\
\(J = (\clq{G}, \sep{G})\) & A junction tree of \(G\) with vertex set $\clq{G}$ and edge set $\sep{G}$\\
\(\jtree{G}\) & The set of junction trees of \(G\) \\
\(\cC\) & A superset of cliques of $G$\\
$\cS$ & A superset of separators of $G$ \\
\(T = (\cC, \cS)\) & A junction tree with vertex set $\cC$ and edge set $\cS$\\
\(C \in \cC\) & Any clique in the set $\cC$\\
\(G_{V}\) & The induced subgraph with the vertex set \(V\) \\
\(T_{\nodea}\) & The induced subtree of \(T\), where every \(C\in T_{\nodea}\) includes \(\nodea\) \\
\(u \sim v \subseteq G\) & A path between $v$ and $u$ in $G$ consisting of a sequence of vertices \\
\(C_1 \sim C_2 \subseteq T\) & A path between cliques $C_1$ and $C_2$ in $T$ consisting of a sequence of cliques in $\cC$\\
$\abr{A}$ & The size of the set $A$ \\
\(\neig(C, T)\) & The set of cliques in \(T\) adjacent to \(C\) \\
\(\deg(C, T)\) & The number of adjacent cliques to \(C\) in \(T\), that is $\abr{\neig(C, T)}$\\
$\press{T}$ & A reduction of $T$ from relations over $\rbr{\cC, \cS}$ to one over $\rbr{\clq{G}, \sep{G}}$\\
\(g(T)\) & The unique decomposable graph represented by \(T\) \\
\hline
\end{tabular}
\end{table}

\section{Proof of Theorem~\ref{lem:connect} and~\ref{lem:disconnect}} \label{proof:lem:connect-disconnect}

\subsection{Theorem~\ref{lem:connect}}\label{proof:lem:connect}
We consider the simplest case, where \(\nseta = \cbr{\nodea}\)  and $\nsetb \subset \cV$ is a subset of vertices that form a complete graph in $G = (\cV, \cE)$. There is no edge between $\nseta$ and any node in $\nsetb$, and $G$ is connected. Let $J = (\clq{G}, \sep{G})$, be a junction tree of $G$. Let $G' = (\cV, \cE')$ be a graph such that $\cE \subset \cE'$, and $G'$ is formed by connecting edges $\edge{\nodea, \nodeb}$, for every $\nodeb \in \nsetb$. Let $C_\nseta \in \clq{G}$ be a maximal clique containing $\nseta$. Similarly, let $C_\nsetb \in \clq{G}$ such that $\nsetb \subseteq C_\nsetb$. When $\nsetb$ is partitioned into mutually exclusive sets $\cbr{\nsetb^{(1)}, \ldots, \nsetb^{(K)}}$, then none of those sets are empty, as it contradicts the fact that $\nsetb$ is completely connected.  It is sufficient to prove Theorem~\ref{lem:connect} for the simplest case when $\nsetb = \nsetb^{(1)}$, since the same argument can be applied to connect to $\nsetb^{(2)}$, once connected to $\nsetb^{(1)}$, and so forth. 

\begin{figure}[!ht]
  \centering
  \includegraphics[width=1\textwidth]{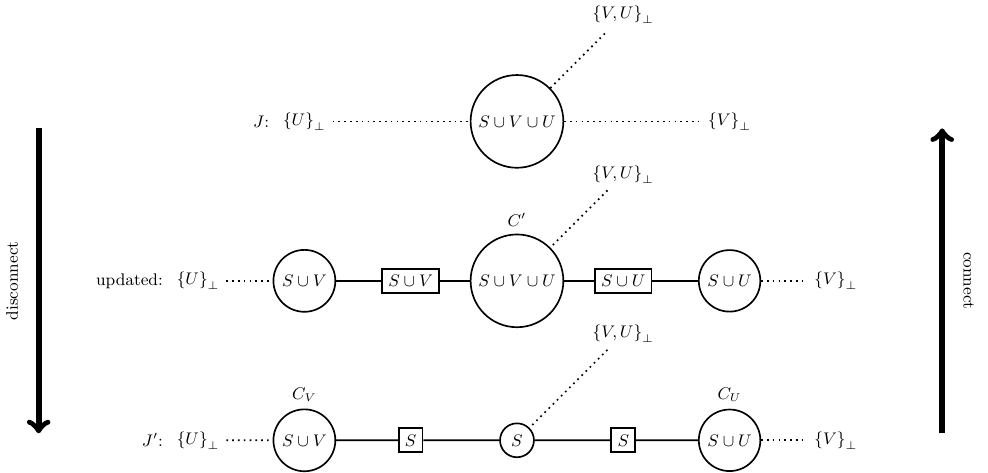}
  \caption{Updating junction tree \(J\) to \(J'\) (and backward) by (dis)connecting vertices in \(\nseta, \nsetb \in \cV\). The notation \(\cbr{A}_{\perp}\) indicates a subtree branch that does not contain \(A \subset \cV\). For a disconnect, modify \(J\) to include two dummy cliques \(S\cup \cbr{\nseta}\) and \(S\cup \cbr{\nsetb}\) in the junction between \(S\cup \nseta\cup\nsetb\) and both branches \(\cbr{\nsetb}_{\perp}\) and \(\cbr{\nseta}_{\perp}\). The final tree \(J'\) is formed by removing \(\nseta\cup\nsetb\) from the \(S\cup \nseta\cup\nsetb\). \(J'\) adheres to the running intersection property and forms a decomposable graph that does not include edges between \(\nseta\) and \(\nsetb\). A connect-move is the opposite. Circles are cliques, boxes are separators.}
      \label{fig:disconnect-x-y}
\end{figure}

\begin{itemize}
\item If  $C_\nsetb  \in \Tn{}{\nseta}(J)$, then $G'$ is decomposable. This case follows directly from~\cite[Prop.~1]{Green01032013}. Here we present a different proof. The steps are illustrated graphically in Fig.~\ref{fig:disconnect-x-y}.  We know that $\edge{C_\nseta, C_\nsetb} \in J$. Let $C' = \nseta \cup S \cup\nsetb $, where $S = C_\nseta\cap C_\nsetb$. Create $J'$ as follows, add $C'$ between $C_\nseta$ and $C_\nsetb$ by replacing the edge $\edge{C_\nseta, C_\nsetb}$ with $\edge{C_\nseta, C'}$ and $\edge{C', C_\nsetb}$. $J'$ satisfies the junction property~\eqref{eq:junction-property}. Since for any clique $C_i, C_j, \in J$, if their path intersects the edge $\edge{C_\nseta, C_\nsetb}$, then $C_i\cap C_j \subseteq S \subset C' \in J'$. Also, it is easily verified that $\rbr{C_{1}, \ldots, C_\nseta, C', C_\nsetb, \ldots, C_{c}}$ is a perfect ordering sequence that generates $J'$, when $\rbr{C_{1}, \ldots, C_\nseta, C_\nsetb, \ldots, C_{c}}$ is the perfect ordering sequence of $G$, as defined in~\eqref{eq:running-intersection-property}. Now $C'$ or $C_\nsetb$ might not be maximal in $J'$, but then we can reduce $J'$ to $\press{J'} =(\clq{G'}, \sep{G'})$, where \(G' = g(J')\). 

\item  We will show the necessary part of the condition by showing that $G'$ contains a non-chordal cycle of 4 vertices. We will pick those vertices from the recursive simplicial subsets over a path in $J$. Assume $G'$ is decomposable, and there does not exist a junction tree $J$ of $G$ such that $\edge{C_\nseta, C_\nsetb} \in J$. Assume that there exists a $J =(\clq{G}, \sep{G}) $ such that the path $C_\nseta\sim C_\nsetb$ contains a single maximal clique $C$, i.e.~$ C_\nseta\sim C_\nsetb = (C_\nseta, C, C_\nsetb)$. This path, by definition, is part of the perfect ordering sequence of \(G\). Define the following:  $S_\nseta = C_\nseta \cap C$ and $S_\nsetb = C \cap C_\nsetb$. Now, we know that: (i) $S_\nsetb \setminus C_\nseta \neq \emptyset$, otherwise $S_\nsetb \subseteq C_\nseta$ meaning that there exists a junction tree that has the edge $\edge{C_\nseta, C_\nsetb}$, which is a contradiction; and similarly (ii)  $S_\nseta \setminus C_\nsetb \neq \emptyset$. Then, let $b \in S_\nsetb \setminus C_\nseta$ and $a \in S_\nseta \setminus C_\nsetb$. Then all the following edges exits in $G'$, $\edge{\nodea, a}, \edge{a,b} \edge{b,\nodeb}, \edge{\nodea, \nodeb}$, where $\nodea \in \nseta, \nodeb \in \nsetb$. However, neither edges $\edge{\nodea, b}$, nor $\edge{a, \nodeb}$ exists in $G'$, hence $G'$ has a non-chordal cycle of the 4 vertices $\edge{\nodea, a, b, \nodeb, \nodea}$, leading to a contradiction. 

Now if the path $C_\nseta \sim C_\nsetb = (C_\nseta, C, C', C_\nsetb)$, one can find a vertex $d$ in $C\cap C'\setminus C_\nseta$ such that the non-chordal cycle  $\edge{\nodea, a,d, b, \nodeb}$ exist in $G'$. Here, $b \in C_\nsetb \cap C' \setminus C$. It is possible that $d$ equals $a$ or $b$. In all cases, $G'$ is not chordal. 
\end{itemize}
By induction, the results of Theorem~\ref{lem:connect} follows. 

\subsection{Theorem~\ref{lem:disconnect}}
Under similar setup in proof of Theorem~\ref{lem:connect}, in Section~\ref{proof:lem:connect}. Here $\nsetb$ is fully connected to $\nseta$ in $G$ and $G'$ is a graph formed from $G$ by removing edges $\edge{\nodea, \nodeb},$ where $ \nodea \in \nseta, \nodeb \in \nsetb$. Again, it is sufficient to prove Theorem~\ref{lem:disconnect} for the simplest case when $\nsetb = \nsetb^{(1)}$, since the same argument can be applied to disconnect from $\nsetb^{(2)}$, once disconnected from $\nsetb^{(1)}$, and so forth.
\begin{itemize}
    \item If $\nsetb \in C$ for some $C \in \Tbd{}{\nseta}(T)$, then $\nseta \cup \nsetb$ are contained in only one maximal clique $C$, and $G'$ is decomposable. This follows from~\cite[Prop.~2]{Green01032013}. We present a proof based on the running intersection property~\eqref{eq:running-intersection-property}. This process is demonstrated visually in Fig.~\ref{fig:disconnect-x-y}.  Define $C' = \nseta \cup S \cup \nsetb$, where $S \subset \cV$. Let $C'_\nseta, C'_\nsetb \in \clq{G}$ be such that $(C_1, \ldots, C'_\nseta, C', C'_{\nsetb}, \ldots, C_c)$ is the perfect ordering sequence of $T$ formed by~\eqref{eq:running-intersection-property}. Here $\edge{C'_\nseta, C'} \in T$ and $\edge{C', C'_\nsetb} \in T$, by definition of junction trees edges with relation to~\eqref{eq:running-intersection-property}.  Now remove $C'$ from the ordering and add $S$ instead, alongside the cliques $C_\nseta = S \cup \nseta$  and $C_\nsetb = S\cup \nsetb$, to form the new ordering $(C_1, \ldots, C'_\nseta, C_\nseta, S, C_\nsetb, C'_{\nsetb}, \ldots, C_c)$ which respects~\eqref{eq:running-intersection-property} and generates $G'$ and $T' = (\clq{G'}, \sep{G'})$. To see this, let $C_i,  C_j \in \clq{G}$ if the path $C_i \sim C_j \subseteq T$ intersects $\edge{C'_\nseta, C'}$ or $\edge{C', C'_\nsetb}$, then $C_i \cap C_j \subseteq S$, where $S$ is contained in $C_\nseta$ and $C_\nsetb$ as well. From~\eqref{eq:running-intersection-property}, we have $C_\nseta \cap \cbr{C_1, \ldots, C'_\nseta} \subseteq C'_\nseta $, $S \cap \cbr{C_1, \ldots, C'_\nseta, C_\nseta} \subseteq C_\nseta $, and   $C_\nsetb \cap \cbr{C_1, \ldots, C'_\nseta, S} \subseteq S$, while everything else remains the same.  If $T'$ contains non-maximal cliques, reduce it as $\press{T'}$.  

    \item To show the necessary part of the conditions. Assume that $G'$ is decomposable, but there does not exist a $C \in \clq{G}$ such that $C \in \Tbd{}{\nseta}(T)$, for any junction tree $T$ of $G$. We will show by contradiction, that $G'$ has a non-chordal cycle of 4 vertices, by picking vertices of recursively simplical subsets of a path in $T$ that contains $\nseta\cup \nsetb$. For simplicity, let $\nseta\cup \nsetb$ be contained in only two maximal cliques $C_1, C_2 \in \clq{G}$, then by the run intersection property~\eqref{eq:running-intersection-property}, there exists a junction tree \(T\), of \(G\), such that the edge $\edge{C_1, C_2} \in T$. Let $C_1 = A \cup \nseta \cup \nsetb \cup S$ and $C_2 = S \cup \nseta \cup \nsetb \cup B$, such that \(C_{1}\cap C_{2} = \nseta \cup \nsetb \cup S\), for $A, B, S \subset \cV$. Here $A$ and $B$ are no connected, and not empty, i.e.~\(A\neq \emptyset\) and \(B \neq \emptyset\). Otherwise, either \(C_{1}\) or \(C_{2}\) would be non-maximal. Disconnecting edges $\edge{\nodea, \nodeb}, \nodea \in \nseta, \nodeb \in \nsetb$, in $G$ would result in the cycle $(a, \nodea, b , \nodeb, a)$ for $a \in A$ and $b \in B$. Since the edges in this cycle associated with $a$ and $b$ still exist. This contradicts the fact that $G'$ is decomposable. 
\end{itemize}
This completes the proof of Theorem~\ref{lem:disconnect}.

\section{Inference setup for Gaussian graphical model}\label{sec:infer-setup-gauss}
Consider a \(p\)-dimensional zero-mean (for simplicity) Gaussian random vector \(Y \in \RR^{n}\) that is globally Markov with respect to a decomposable graph \(G\). Its precision matrix (inverse covariance) \(\Theta\) belongs to the set
\[A_{G} =\cbr{\Theta \in \cM_{p}^{+}: \Theta_{{ij}} = 0 \text{ for all} \cbr{i,j} \not \in G},\]
where \(\cM_{p}^{{+}}\) is the space \(p\times p\) of positive definite matrices. Conditional independence between \(i\)th and \(j\)th variable in Gaussian models is equivalent to having \(\Theta_{ij}=0\) \citep{speed1986gaussian}. Let \(y= \cbr{y_{i}}\), for \(i=1, \ldots, n\), be \(n\) observations from this model. Following the factorization law in~\eqref{eq:joint-csf}, the full likelihood of \(y\) can be specified by the clique marginals \(\clq{G}\) of \(G\), and their separators \(\sep{G}\). For clique \(C \in \cC\), the likelihood marginal is
\[f(y_{C} \mid \Theta_{C}) = \frac{1}{\rbr{2\pi}^{\abr{C}}} \abr{\Theta_{C}}^{{n/2}} \exp \cbr{-\tr\rbr{\Theta_{C}D_{C}}/2},\]
where \(D\) is the sample covariance matrix, as \(D = n^{{-1}}\sum_{i=1}^{n}y_{i}y_{i}^{\top}\), \(\abr{\Theta}\) is the determinant of \(\Theta\), \(\abr{C}\) is the cardinality of \(C\), and \(y_{C}\) is the subvector \(y\) indexes by \(C\), similarly for all other elements. Similarly, for \(f(y_{S} \given \Theta_{S})\), where \(S \in \cS\). Following \cite{dawid1993}, a conjugate prior for \(\Theta\) is a hyper-Wishart distribution that is also hyper Markov with respect to \(G\), having the form
\[\eta\rbr{\Theta \given G, \varphi} = \frac{\prod_{C \in \cC} \eta\rbr{\Theta_{C}\given \varphi_{C}}}{\prod_{S \in \cS}\eta\rbr{\Theta_{S}\given \varphi_{S}}}. \]

Each prior marginal \(\eta\rbr{\Theta_{C}\given \varphi_{C}}\) is of the form 
\[\eta\rbr{\Theta_{C}\given \varphi_{C}} \propto {\abr{\Theta_{C}}^{\beta_{C}}\exp \cbr{-\tr\rbr{\Theta_{C}Q_{C}}/2}}, \]
with normalization constant \(\int \eta\rbr{\Theta_{C}\given \varphi_{C}} d\Theta_{C} = 2^{\delta\abr{C}/2}{\Gamma_{\abr{C}}(\beta_{C})}/{\abr{Q_{C}}^{\beta_{C}}}\), where \(\Gamma_{k}\) denotes the multivariate gamma function. Here \(\beta_{C} = (\delta + \abr{C} - 1)\), \(\varphi_{C} = (\delta,Q_{C})\), \(Q \in \cM_{p}^{+}\) a scale positive definite matrix and \(\delta > \max_{C \in \cC}\cbr{\abr{C}-1}\) the number of degrees of freedom.
By~\citet[Thm. 3.9]{dawid1993}, the collection of priors \({\cbr{\eta\rbr{\Theta_{C}\given \varphi_{C}}}}\) are pairwise hyper consistent and there exists a unique hyper Wishart joint distribution of the form
\[f(y_{C} \mid \Theta_{C})\eta\rbr{\Theta_{C}\given \varphi_{C}} \propto \frac{1}{\rbr{2\pi}^{\abr{C}}} \abr{\Theta_{C}}^{\alpha_{Q}} \exp \cbr{-\tr{\rbr{\Theta^{}_{C}\rbr{D_{C} + Q_{c}}}/2}},\]
where \(\alpha_{C} = \rbr{\delta + n + \abr{C}-1}/2\). By conjugacy, it is possible to integrate out \(\Theta\) of the term above. With a junction prior of the form~\eqref{eq:csf}, the junction posterior is
\begin{equation}
  \label{eq:posterior-ggm}
  \pi(T\given t) \propto \frac{\prod_{C \in \cC}\phi(C)\rho(C)}{\prod_{S \in \cS} \psi(S) \rho(S)},
\end{equation}
with
\[ \rho(C) = \frac{\abr{Q_{C}}^{\alpha_{C}}}{\abr{Q_{C} + D_{C}}^{\beta_{C}}}\frac{\Gamma_{\abr{C}}\rbr{\alpha_{C}}}{\Gamma_{\abr{C}}\rbr{\beta_{C}}},\]
and \(\rho(S), S \in \cS\) is defined similarly. As indicated by~\cite{grone1984positive}, given that \(G\) is decomposable and \(\Theta\) is positive definite for each clique, \(\Theta\) is both existent and unique for each graph.
\clearpage

\section{Simulation code}\label{app:sec:simcode}
To run the simulation code, please follow these steps:

\begin{enumerate}
    \item \textbf{Install Benchpress:} Download and install Benchpress from the official documentation. For detailed instructions, visit the Benchpress documentation website at \url{https://benchpressdocs.readthedocs.io/en/latest/}.

    \item \textbf{Clone the GitHub Repository:} Use the following commands to clone the specific branch of the Benchpress repository:
    \begin{verbatim}
git clone https://github.com/felixleopoldo/benchpress.git
cd benchpress
git checkout paralleldg
git pull origin paralleldg
    \end{verbatim}

    \item \textbf{Run the Simulation:} Execute the simulation using Snakemake as:
    {\footnotesize 
    \begin{verbatim}
    # A single repication of (n,p)=(100,50)  (~2hrs)
    snakemake --cores all --use-singularity 
        --configfile config/elmasri_parallel_single_run.json
    
        --configfile config/elmasri_parallel.json

    # A single repications of (n,p)=(100,200) (~4hrs)
    snakemake --cores all --use-singularity 
        --configfile config/elmasri_parallel_high-dim_single_rung.json
    
    # 10 repications of (n,p)=(100,200) (~48hrs)
    snakemake --cores all --use-singularity 
        --configfile config/elmasri_parallel_high-dim.json
    \end{verbatim}
    }
    (compute times are based on an Intel i7 with 4 cores)
    \item \textbf{Results:} will be available at \textsf{benchpress/results/outputs/}.
\end{enumerate}
\clearpage

\section{Decomposable graphs of seven vertices}\label{sec:decomp-graphs-seven}

We analyzed all 2,097,152 undirected graphs on seven labeled vertices, identifying 61,675 decomposable ones. We indexed each graph's cliques into a counter table and used the algorithm from~\cite{thomas2009enumerating} to calculate the number of possible reduced junction tree representations for each graph, denoted as \(\abr{\jtree{G}}\). The no-edge graph is represented by 16,807 reduced junction trees, while there are 18,447 graphs, each represented by a single reduced junction tree.

First, we restricted the sampler to the reduced junction tree space, ensuring that every update led to a change in the underlying graph. We set the junction tree prior as~\(\pi(T\given t) \propto 1 \), indicating uniform sampling over the space of reduced junction trees. In a subsequent run, we defined the prior as~\(\pi(T\given t) \propto 1/\abr{\jtree{g(T)}}\), aiming for uniform sampling over the space of decomposable graphs.

We initiated the single-move sampler with a no-edge graph and ran it for 1,000,000 steps. This produced a trajectory of junction trees, which we then converted into a trajectory of decomposable graphs. For each decomposable graph, we computed empirical frequencies, cumulatively forming an estimated distribution function. As depicted in SM Figure~\ref{app:fig:single-move-empirical-CDF}, there is a strong match between the expected and estimated cumulative distribution functions for the two priors. This alignment suggests that our single-move sampler effectively and uniformly traverses both the reduced junction tree space and the decomposable graph space.

When not limited to the space of reduced junction trees, our single-move sampler uniformly samples from the broader space of junction trees with \(p\)-vertices. Enumerating this space is a significant challenge, requiring a brute force approach that's only viable for very small values of \(p\) since no analytical expression exists. Drawing inspiration from Algorithm~\ref{alg:random-walk} and the visit probability~\(\rho \in \rbr{0,1}\), we devised a proposal over the update-type. This approach aims to approximate uniform sampling across the space of decomposable graphs.

Let \(\tq(\text{add})\) be the modified proposal that integrate the probability that the random walk on~\(T_{\nodea}\) visits a neighboring clique~\(C \in \Tn{\nodea}{}\), that is \(\nodea\) is added to \(C\). We express this as
\begin{equation} \label{eq:prior-update-type}
  \tq(\text{add}) = \frac{e^{-\rho}}{1 + e^{-\rho}}, \qquad \rho \in (-\infty,\infty),
 \end{equation}

 As \(\rho\) deviates from 0, it skews the proposal probability of an addition move. Positive values of \(\rho\) discourage such additions, while negative values encourage them. This framework allows us to modulate our sampling behavior based on the desired balance between addition and removal of nodes.

\begin{figure}[ht]
  \centering
  \includegraphics[width=0.8\textwidth]{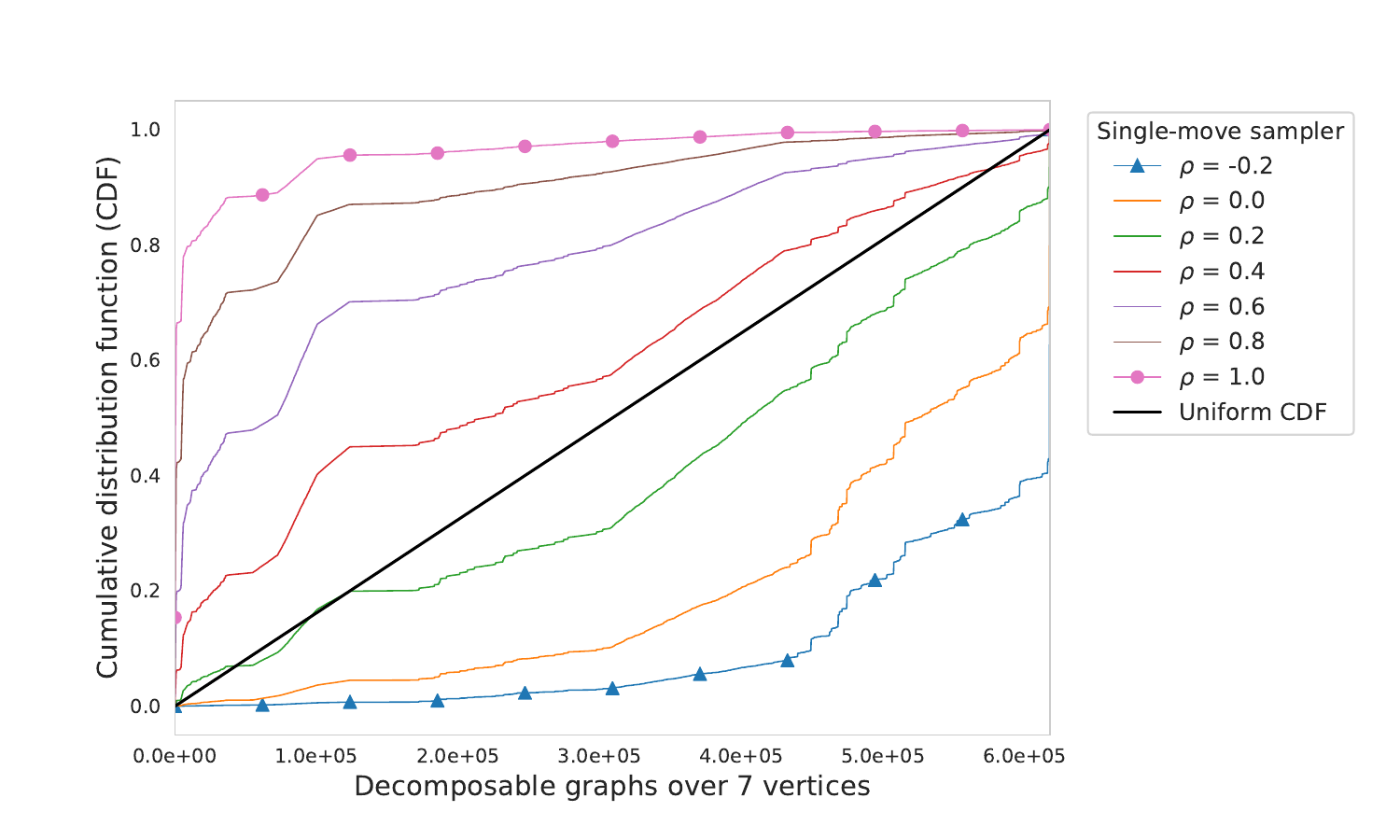}
  \caption{Cumulative distribution functions for decomposable graphs over seven vertices, estimated from samples drawn from the single-move sampler with added proposal specified in~\eqref{eq:prior-update-type}. The x-axis enumerates the graphs in decreasing order based on the number of reduced junction tree representations.}
\label{fig:parallel-prior-sensitivity}
\end{figure}

 Following the simulation setup above, we equip the single-move and parallel samplers with the proposal~\eqref{eq:prior-update-type}, where the acceptance ratios in~\eqref{eq:acceptance-prop} and~\eqref{eq:acceptance-prop} are multiplied by the factor ~\(\tq(\text{add})/(1-\tq(\text{add}))\), or its reciprocal, accordingly. Figure~\ref{fig:parallel-prior-sensitivity} shows that, when \(\rho = 0.4\) (or~\(\pi(\text{add}) = 0.4\)), our single-move sampler is able to traverse the space of decomposable graphs over seven vertices approximately uniformly. Varying \(\rho\) can lead to sampling decomposable graphs more or less frequently, according to the number of reduced junction tree representations. The parallel sampler behaves similarly (SM Fig.~\ref{app:fig:single-move-prior-sensitivity}). Performing 1,000,000 updates took around 60 seconds for the single-move sampler, and half that time for the parallel sampler.

\section{Effect of graph prior on acceptance rate}\label{app:prior-effect}
We observed a notable influence of the graph prior on the acceptance rate in our single-move sampler. With a uniform prior and an autoregressive graph structure, the average acceptance rate for junction-updates is about 48\%. However, when using a modified exponential family prior, defined as \(\phi(C) = \exp\left(\alpha \left(|C|-1\right)\right)\) and \(\psi(S) = \exp\left(\beta |S|\right)\) with \(\alpha=2\) and \(\beta=4\), this rate dropped significantly to 22\%, as shown in the following table. This substantial decrease in acceptance rate illustrates the impact of more conservative priors in reducing the number of non-maximal cliques in the junction tree \(T\), as detailed in Section~\ref{sec:comp-cons}
\begin{table}[ht!]
    \centering
    \caption{Average acceptance rate (\%) on different priors and update types}
    \begin{tabular}{lrr}
    \toprule
                       &  Graph-updates &  Junction-updates \\
    \midrule
         Uniform prior &          28.17 &             48.05 \\
    Exponential prior &          12.38 &             22.52 \\
    \bottomrule
\end{tabular}
\end{table}

\section{Extra simulation results for the general case} \label{app:sec:ggm}
A step in our parallel sampler can encompass multiple proposals executed concurrently. Specifically, updating the chain from \(T\) to \(T'\) may entail \(k > 0\) parallel updates, which are order-invariant, resulting in a sequence such as \(T_{i}, T_{i1}, \ldots, T_{ik}=T'\). To elucidate the impact of parallelism, we post-process each chain after sampling to derive a corresponding serial chain, where each proposal is individually indexed. Consequently, a step consisting of \(k\) parallel updates in the parallel sampler is represented as \(k\) individual steps in the serial chain.

\begin{figure}[ht]
  \centering
  \includegraphics[width=0.7\textwidth]{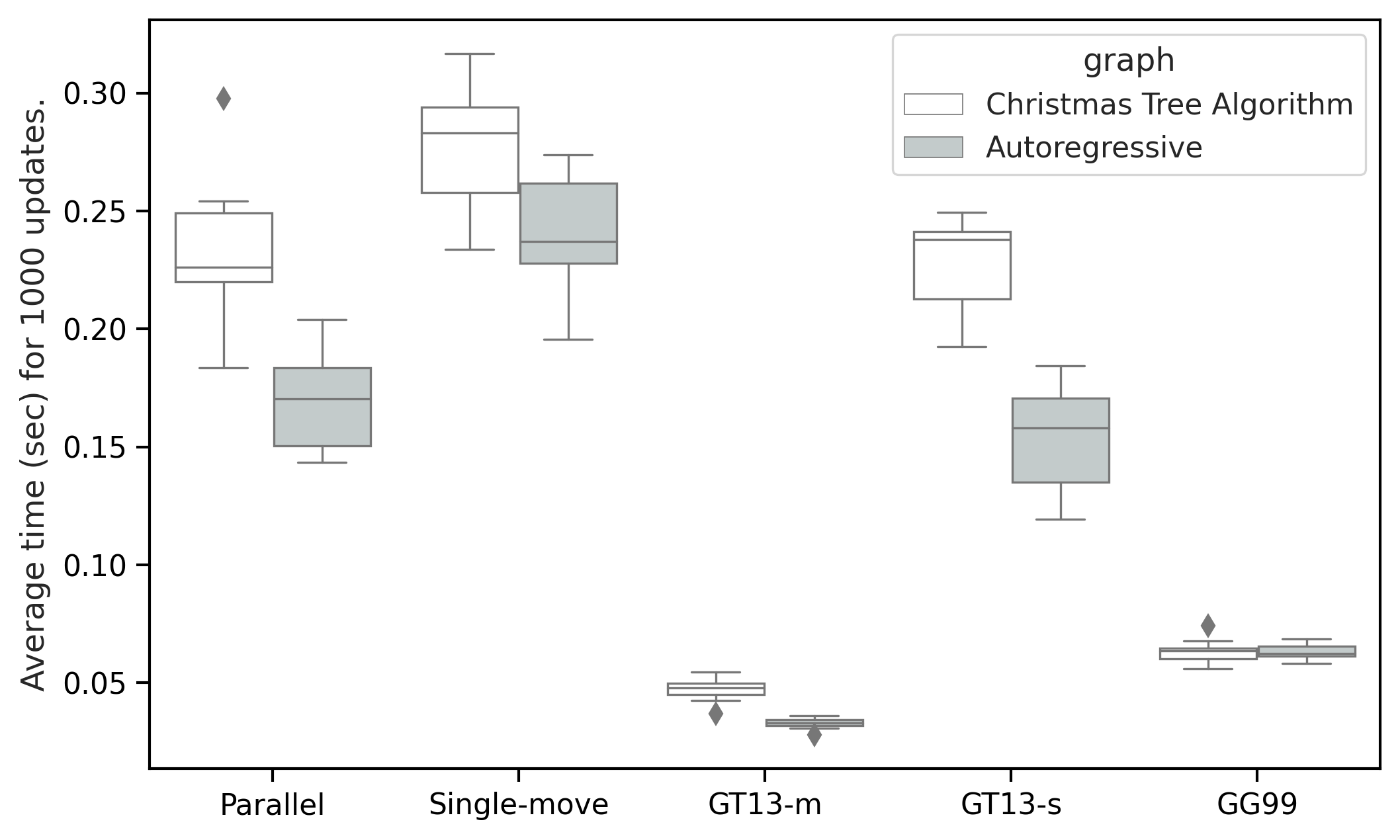}
  \caption{Boxplot of the average computation time in seconds for 1000 steps across two graph structures, detailed in Section~\ref{sec:simul-setup-gauss}, over 10 replications, on an Intel Xeon E5-2698 v4 2.20GHz processor, when $(n,p)=(100,50)$.}
  \label{fig:computation_time}
\end{figure}

\begin{table}[ht!]
\centering
 \caption{Mean and standard deviation of compute time (in seconds) for 1000 steps of the Gaussian decomposable graphical models discussed in Section~\ref{sec:numerical}, when $(n,p)=(100,50)$}
\begin{tabular}{lrrrrrr}
\toprule
{} &  Parallel &  Single-move &  GT13-m &  GT13-s &     O19 &  GG99 \\
\midrule
mean  &      0.20 &         0.26 &    0.04 &    0.19 & 3810.16 &  0.06 \\
std   &      0.04 &         0.03 &    0.01 &    0.04 &  122.58 &  0.01 \\
\bottomrule
\end{tabular}

    \label{app:tb:compute_time}
\end{table}

\begin{figure}[ht!]
  \centering
  \includegraphics[width=0.45\textwidth]{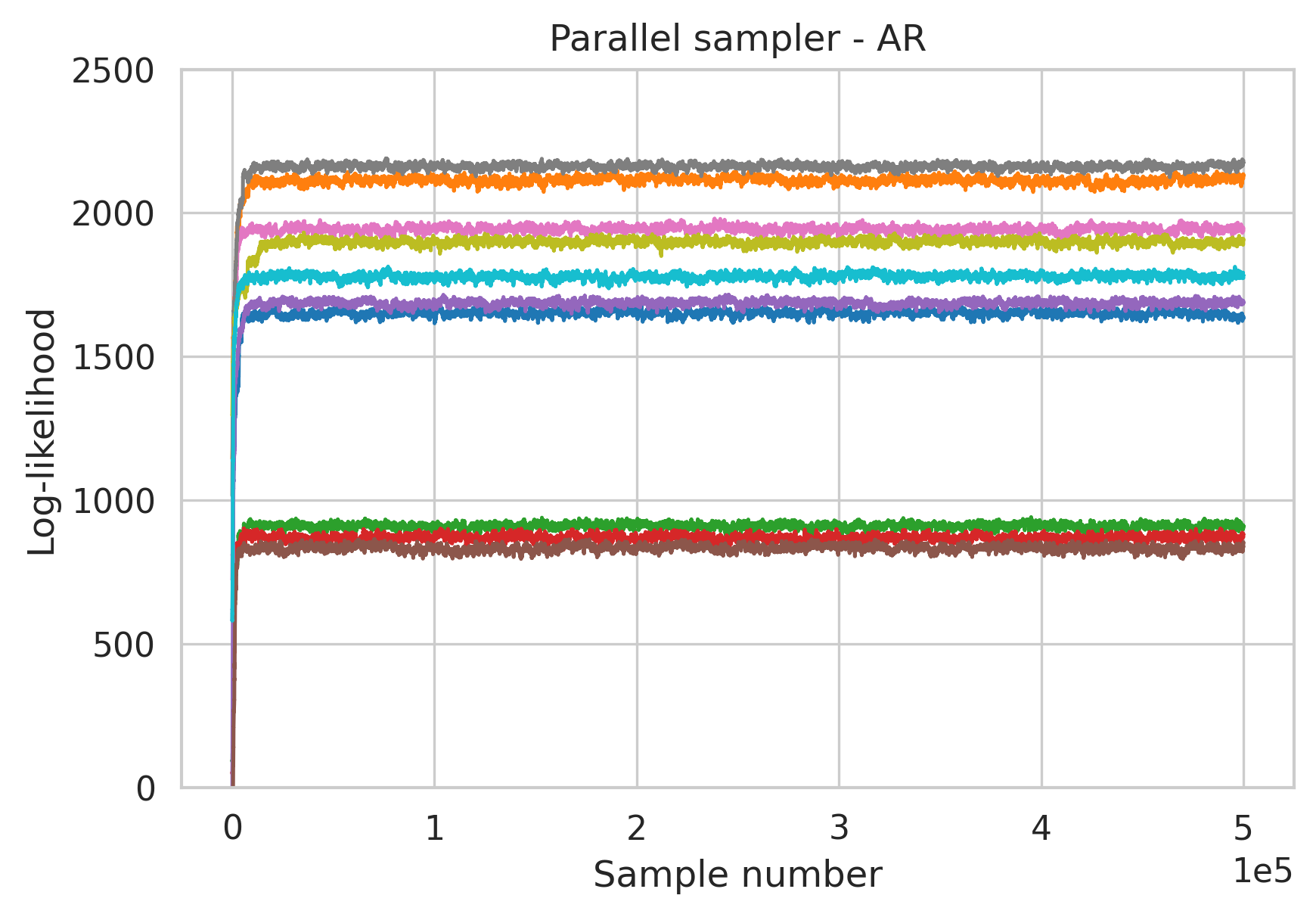}
  \includegraphics[width=0.45\textwidth]{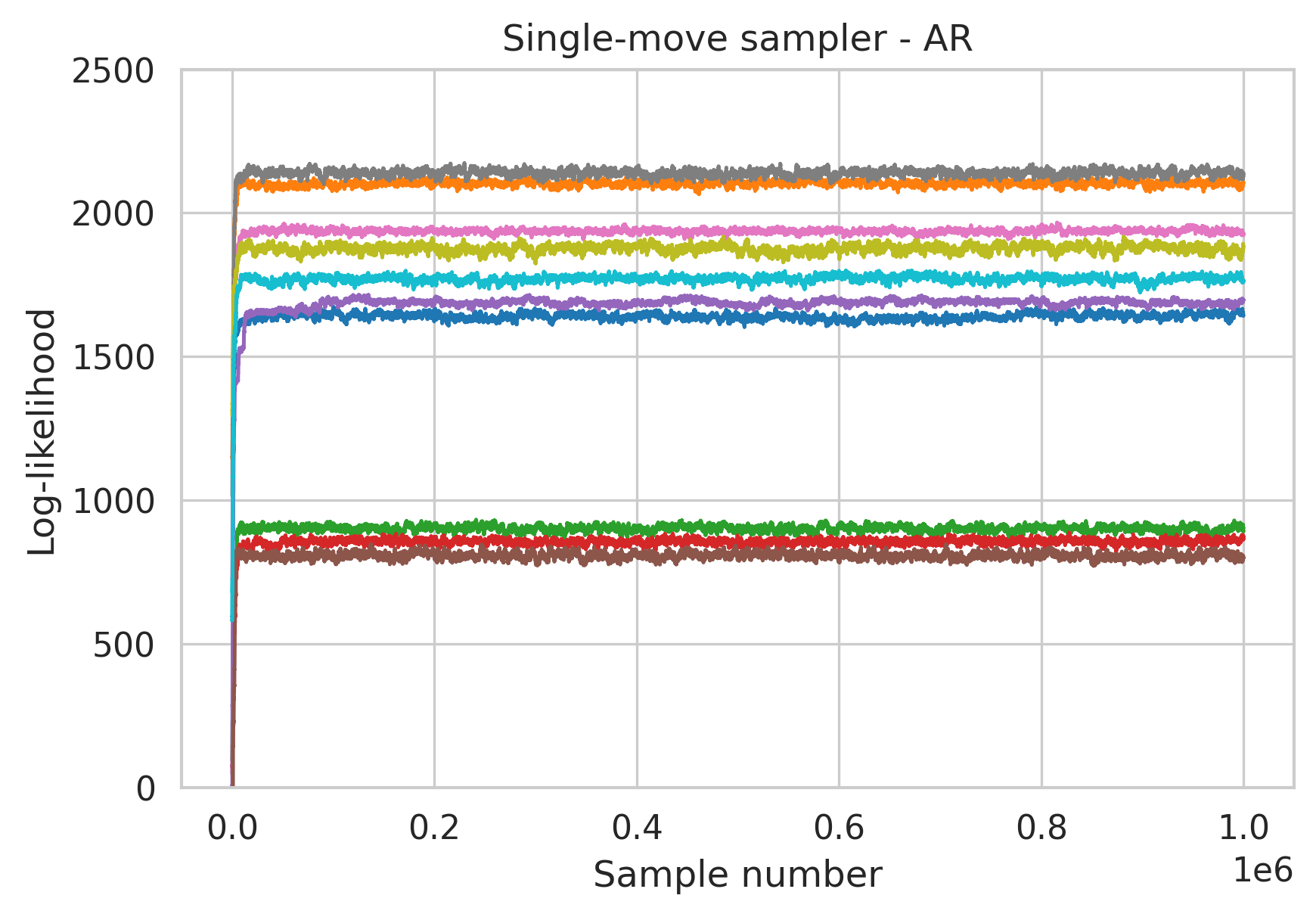}
  
  \includegraphics[width=0.45\textwidth]{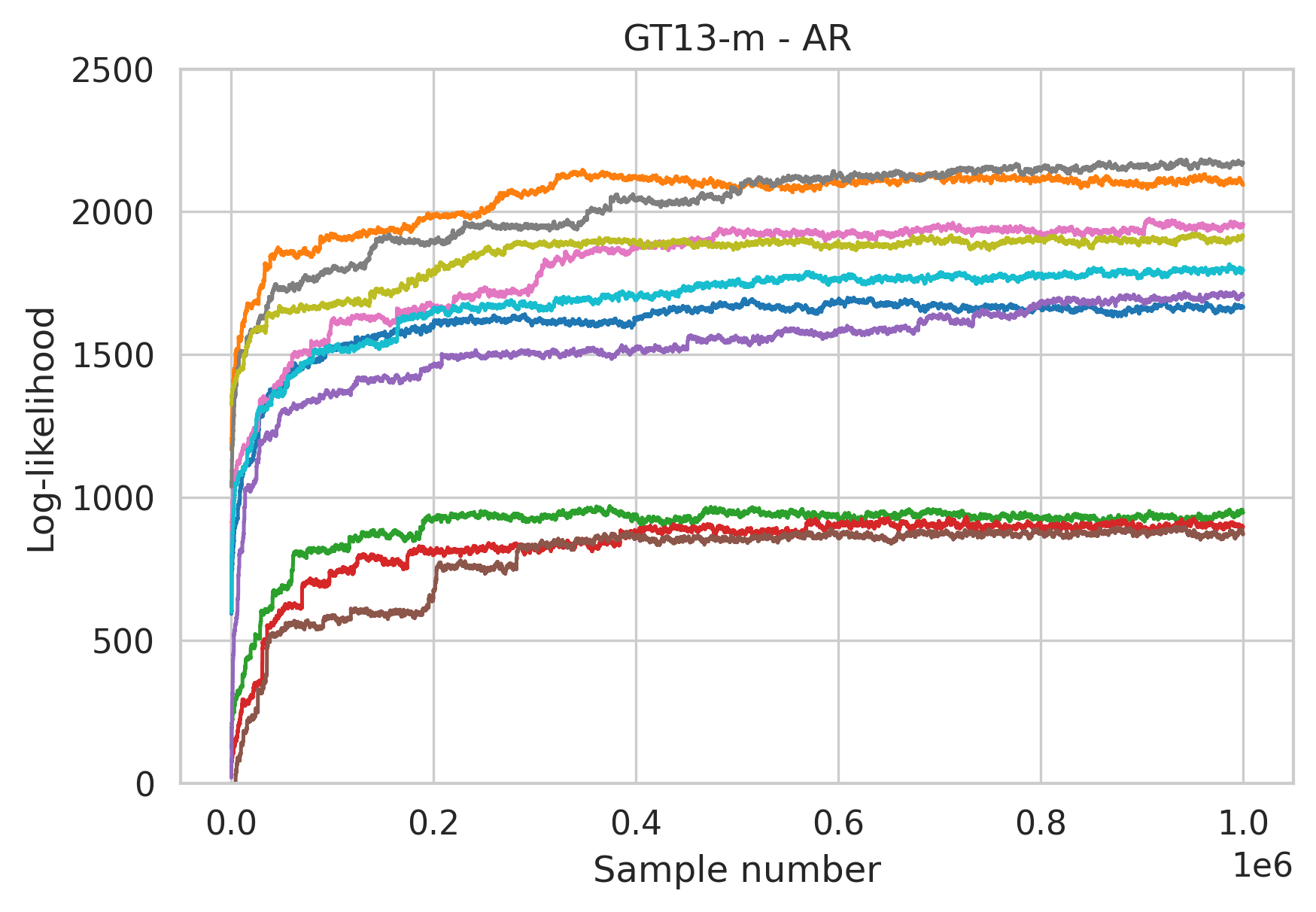}
  \includegraphics[width=0.45\textwidth]{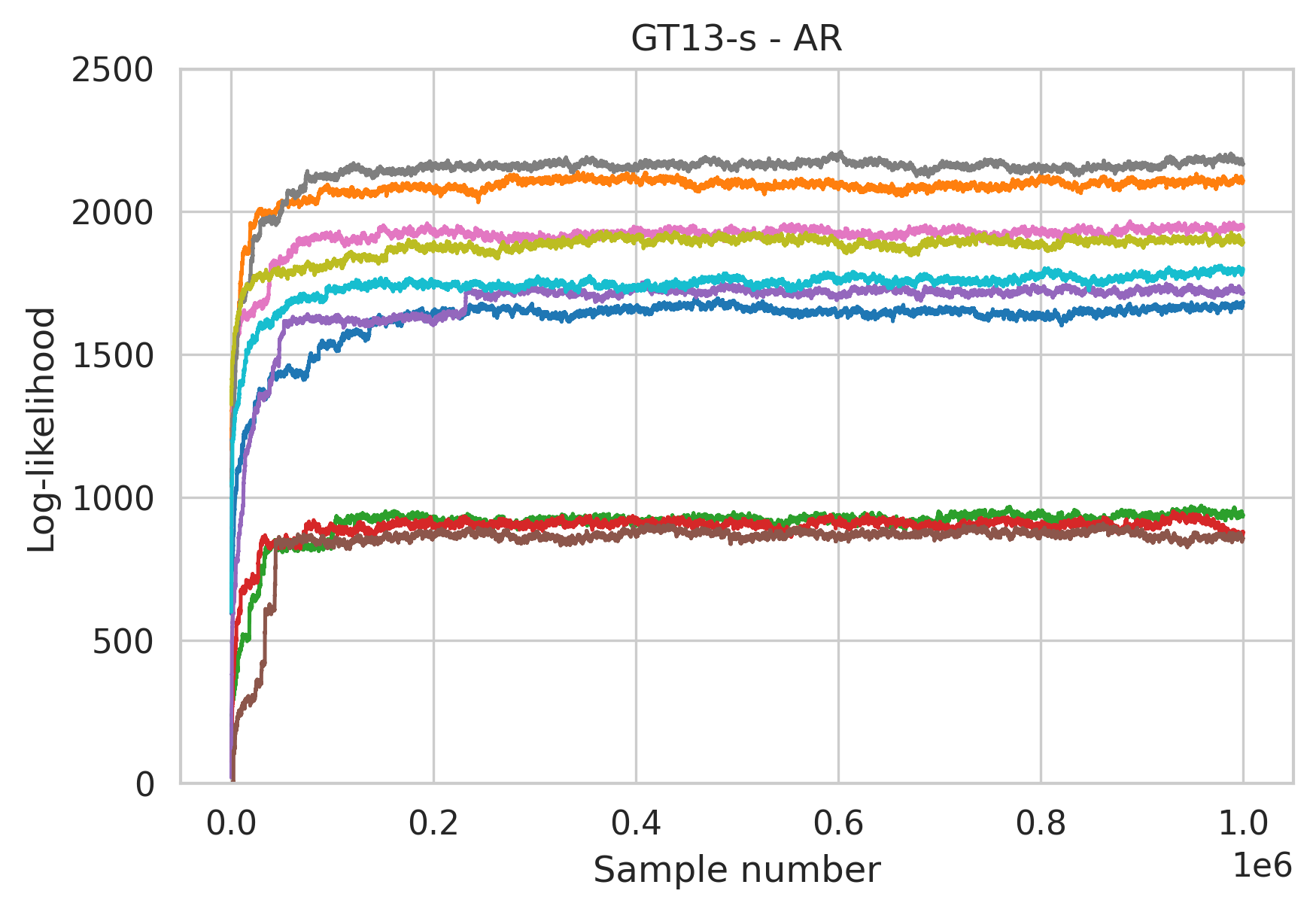}
  
  \includegraphics[width=0.45\textwidth]{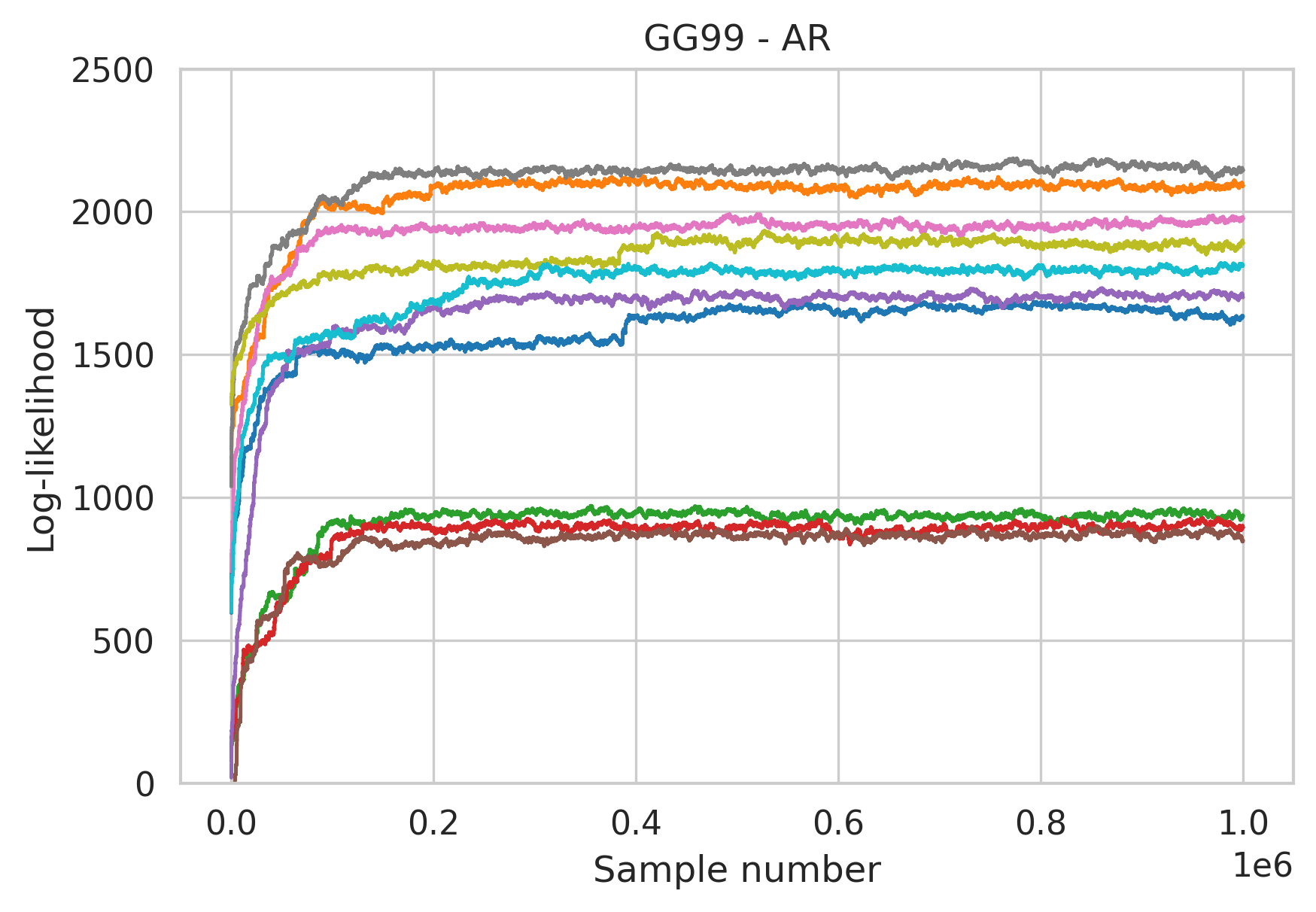}
  \includegraphics[width=0.45\textwidth]{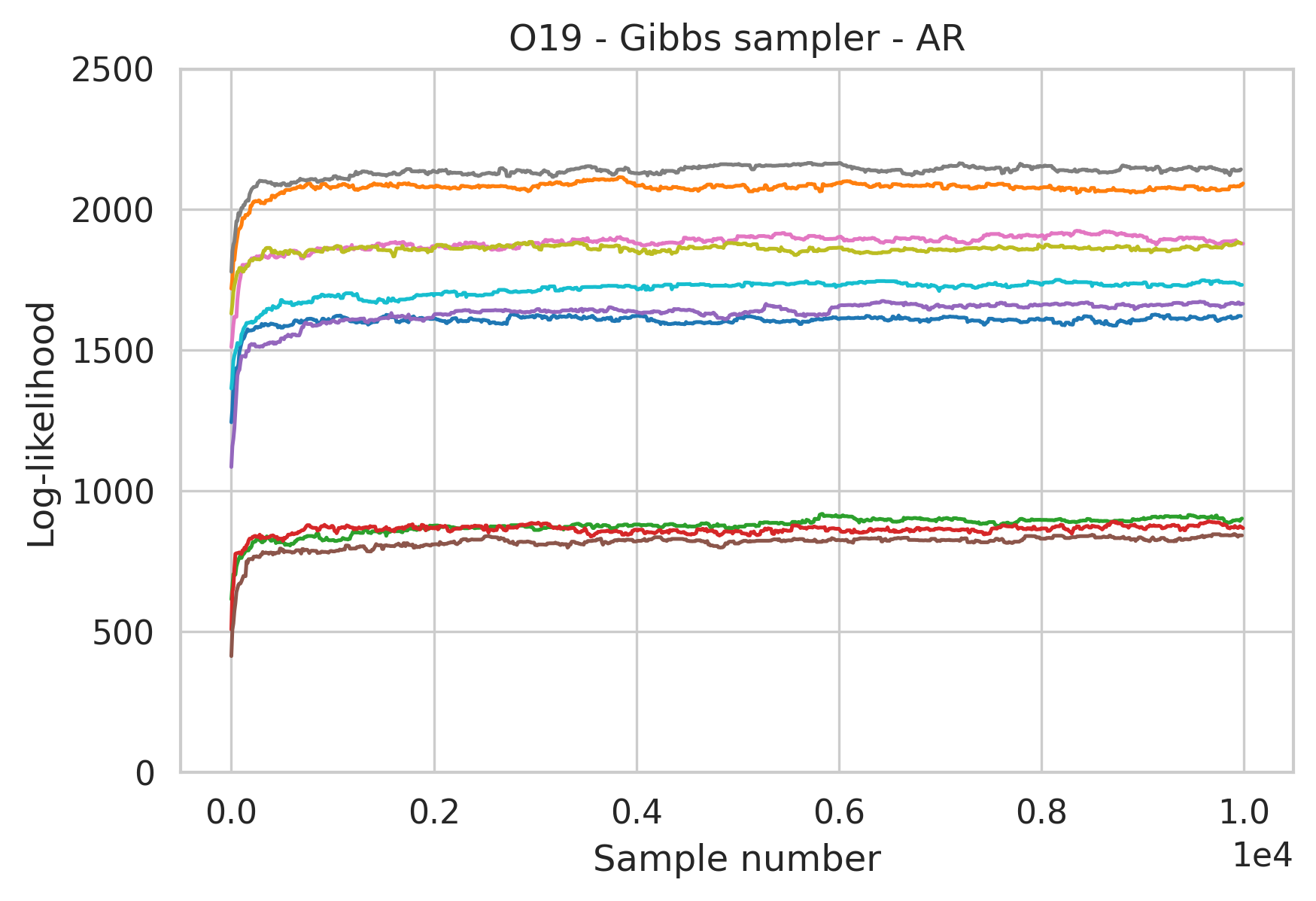}
\caption{Likelihood traceplots for 10 replications (color-matched) for samplers discussed in Section~\ref{sec:numerical}, under the autoregressive graph structure and $(n,p)=(100,50)$.}
   \label{app:fig:loglikeli-traceplots}
 \end{figure}

\begin{figure}[ht!]
  \centering

  \includegraphics[width=0.45\textwidth]{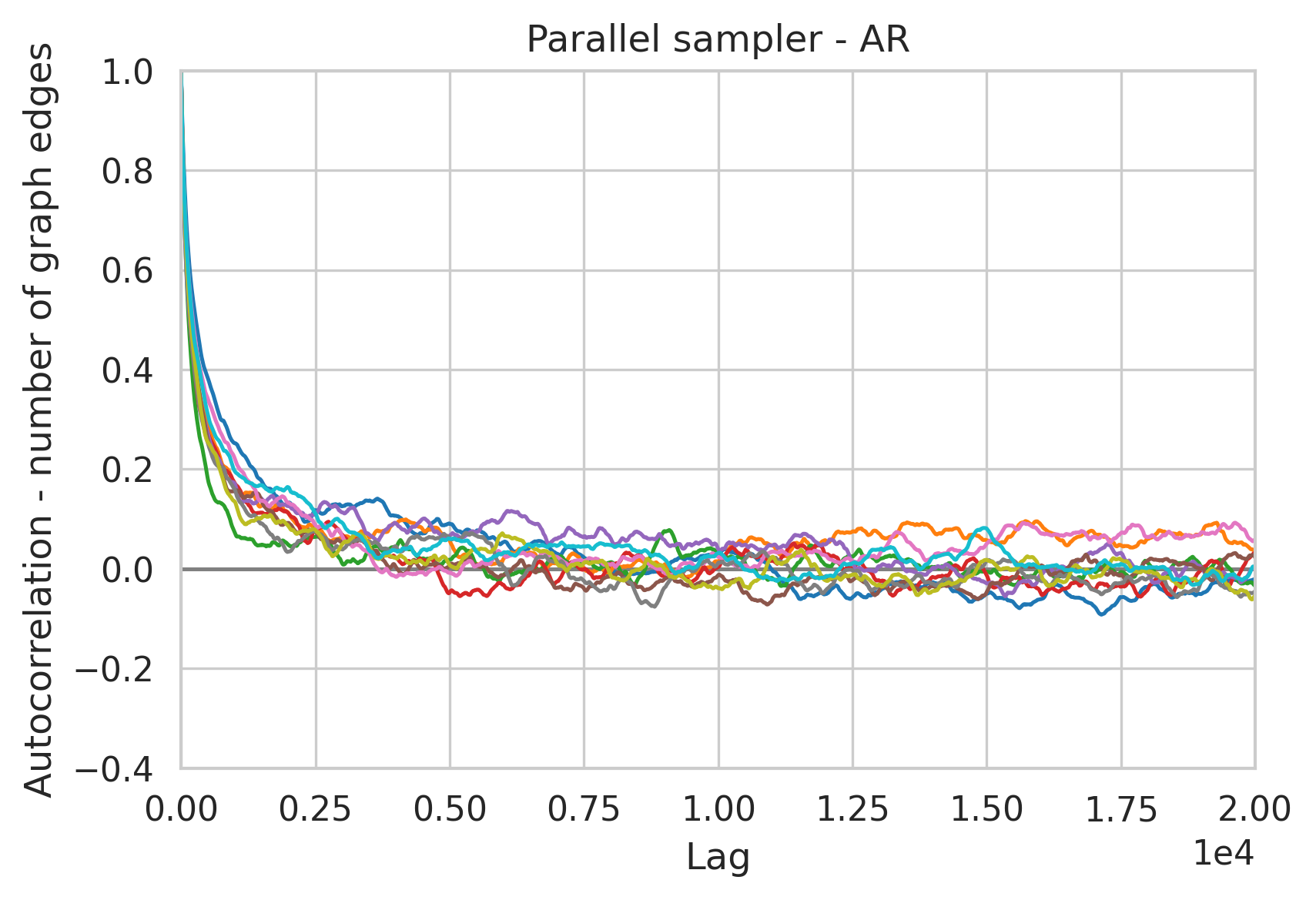}
  \includegraphics[width=0.45\textwidth]{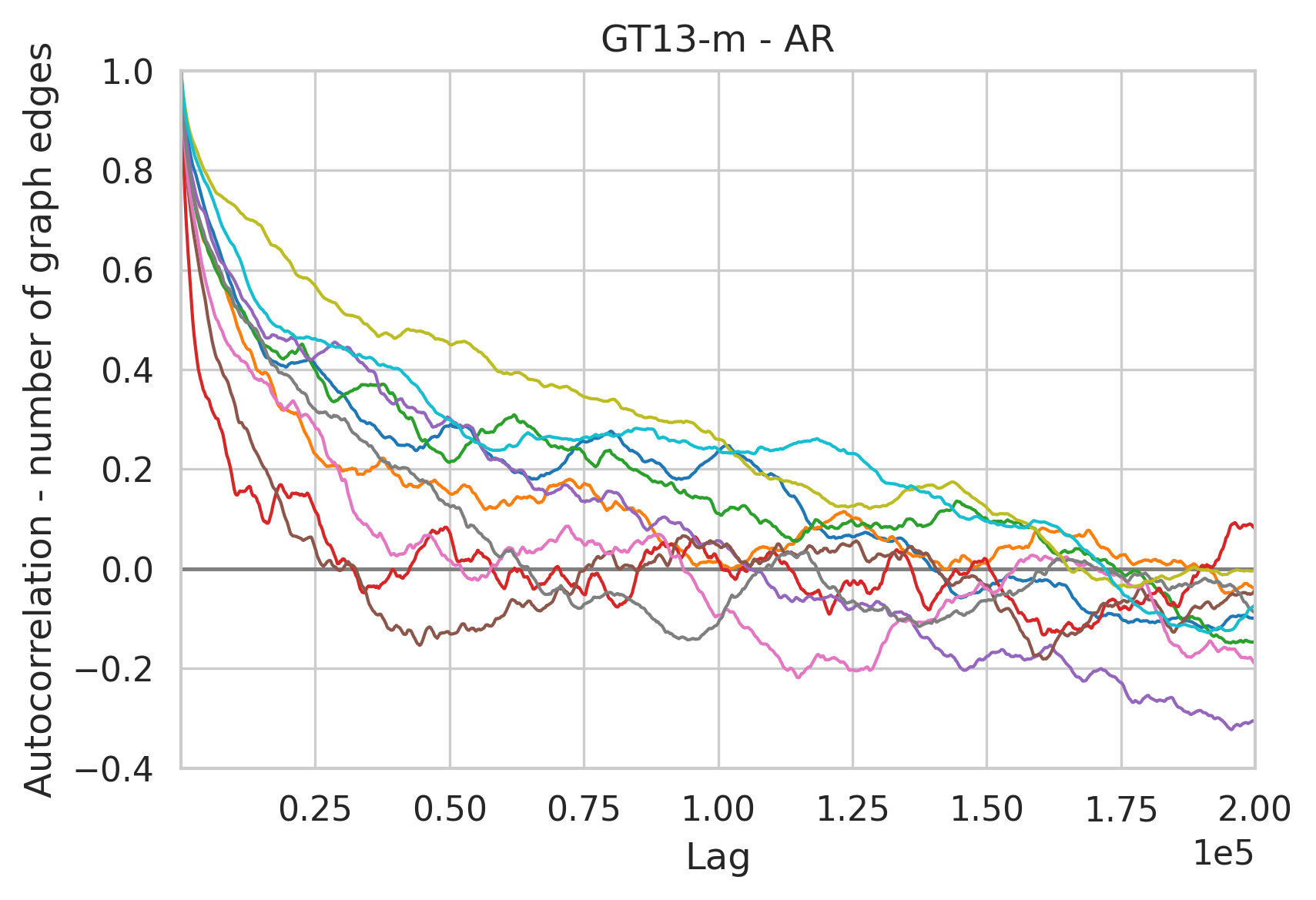}

  \includegraphics[width=0.45\textwidth]{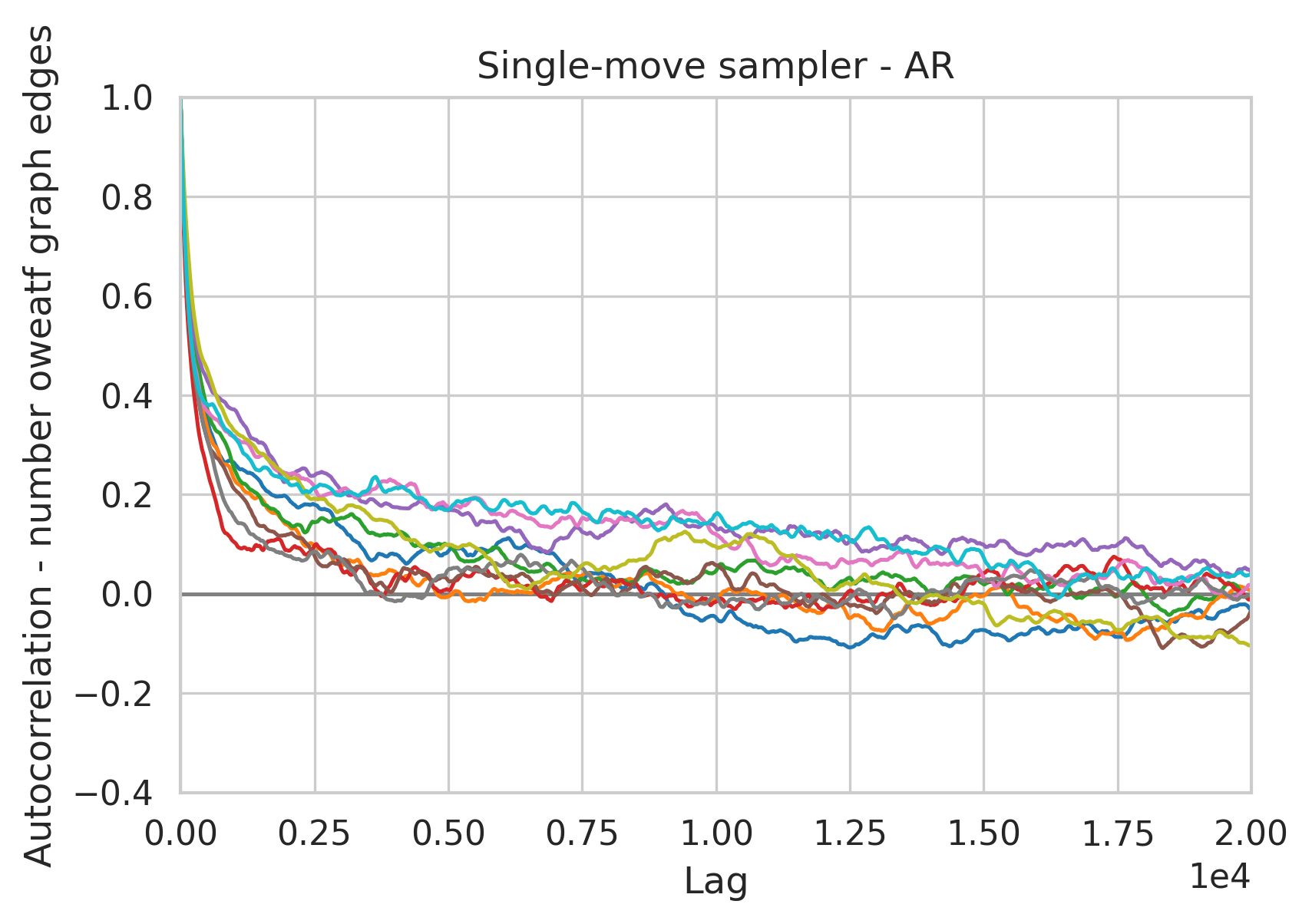}
  \includegraphics[width=0.45\textwidth]{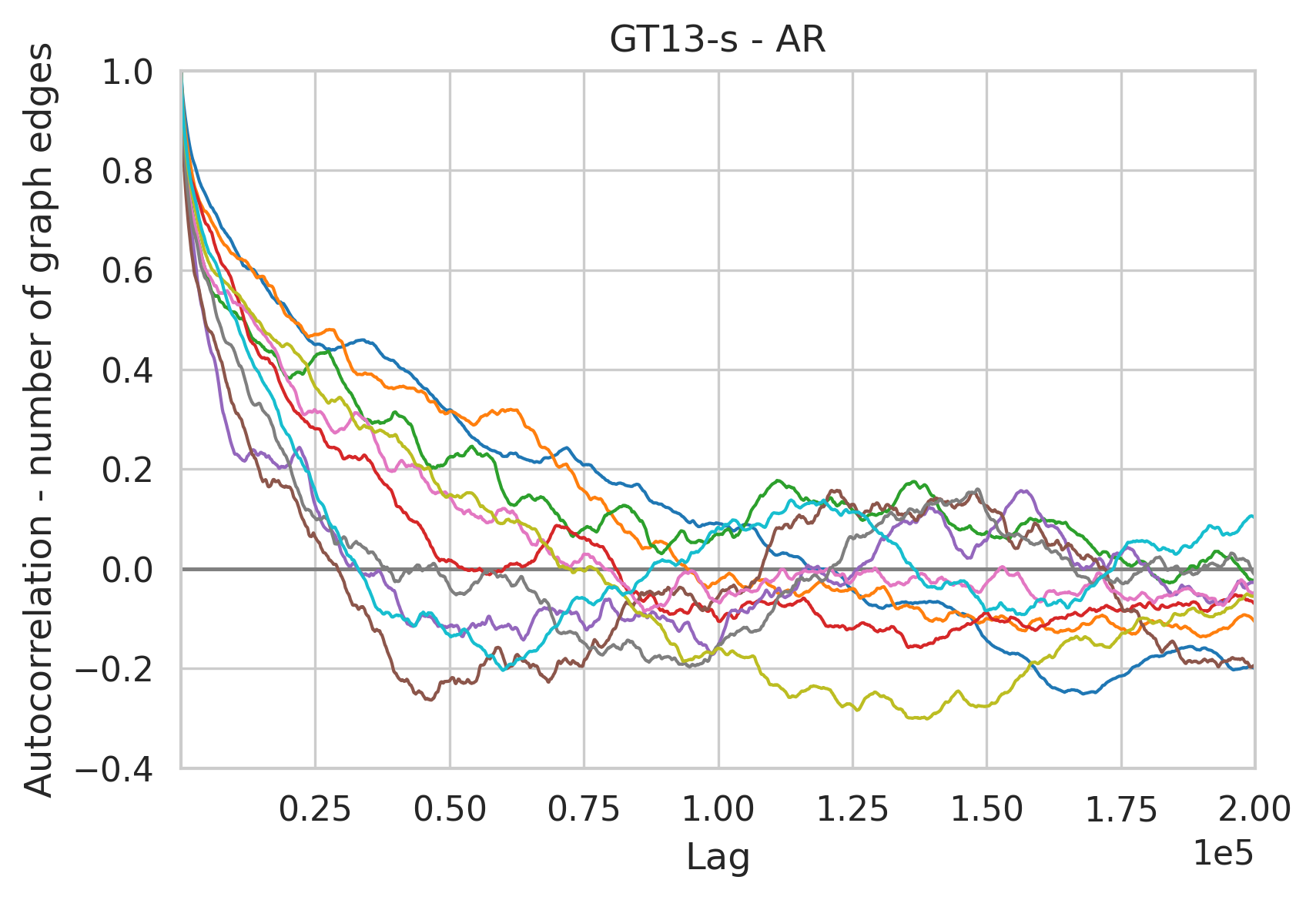}
  
  \includegraphics[width=0.45\textwidth]{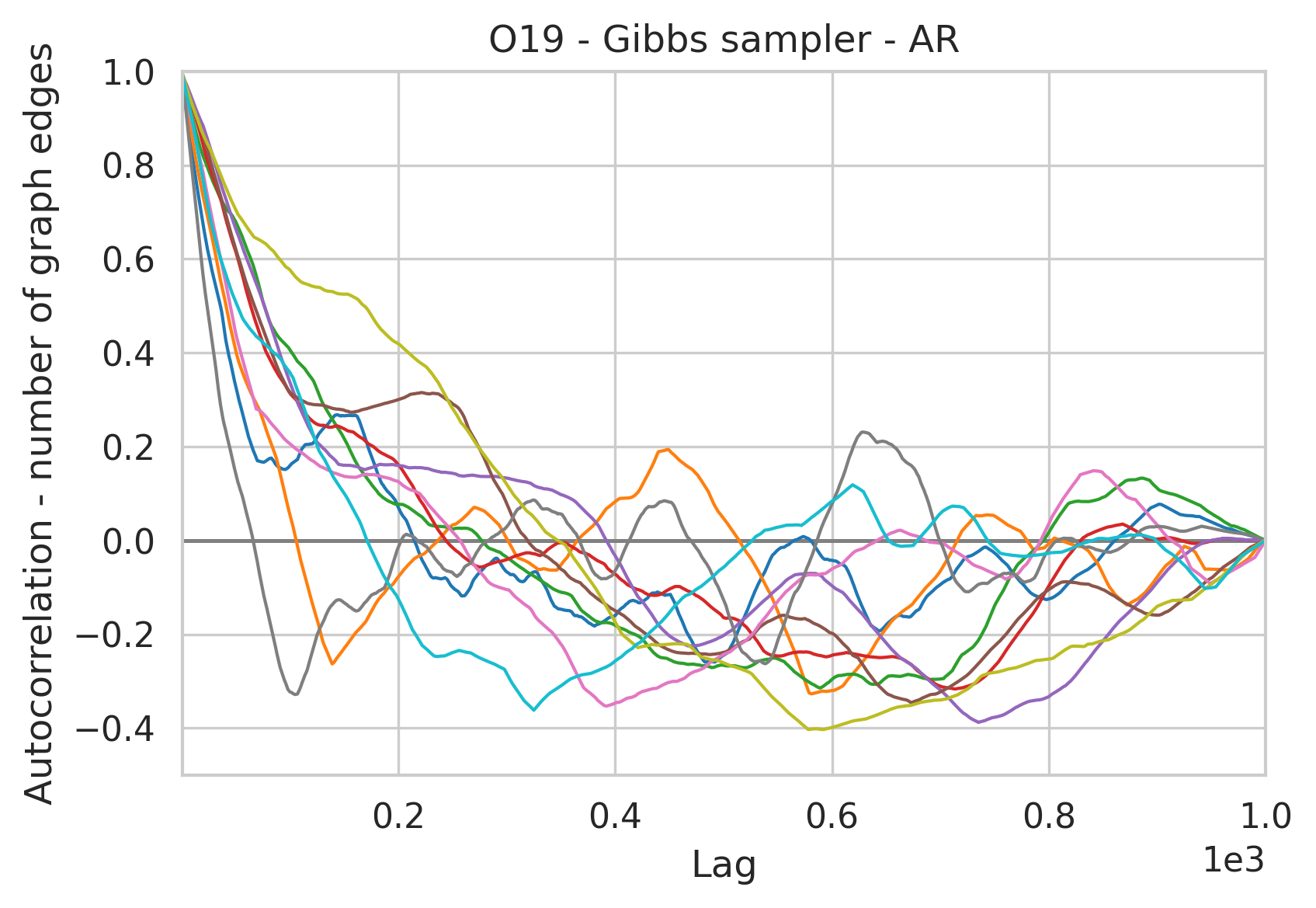}
  \includegraphics[width=0.45\textwidth]{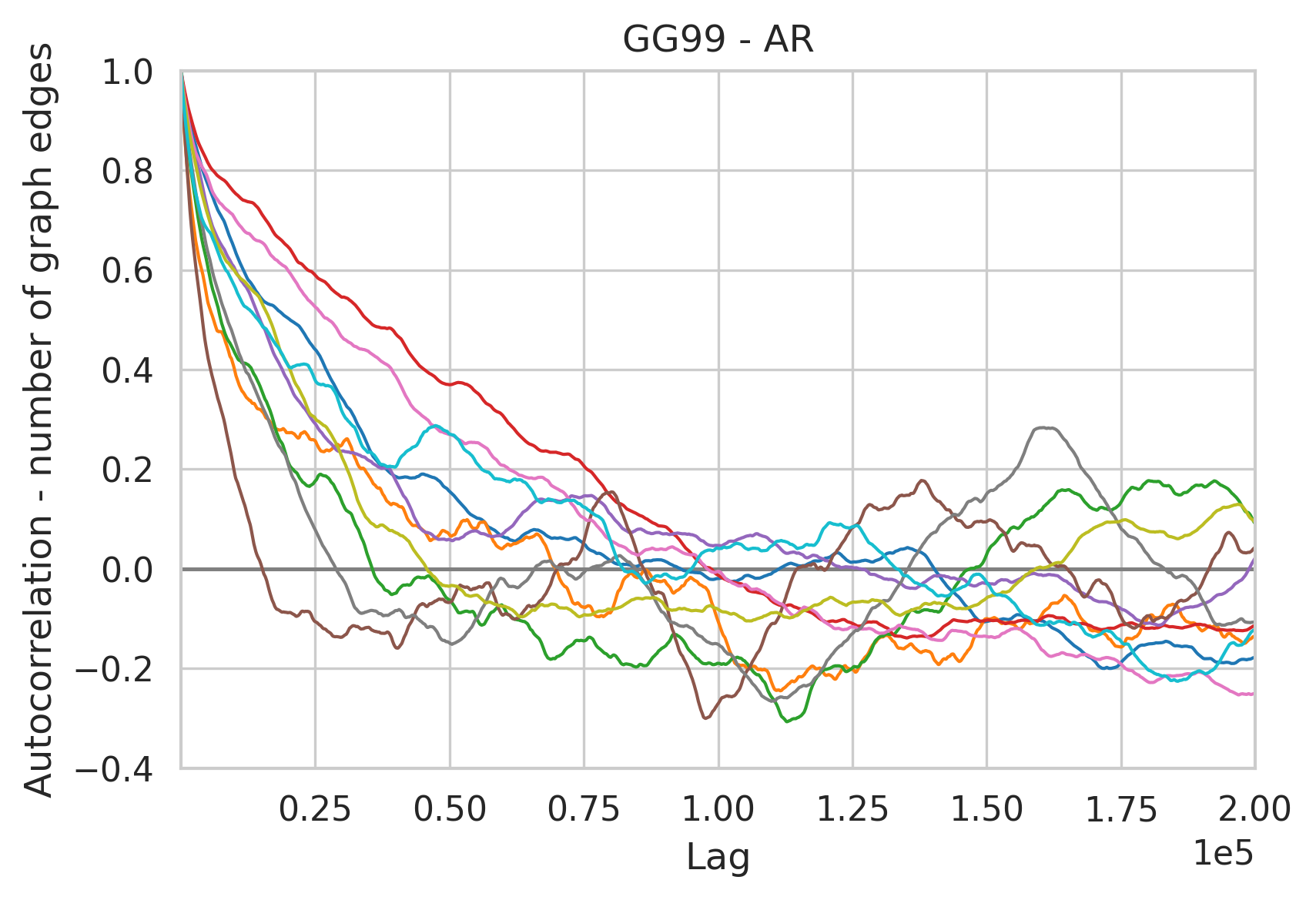}

\caption{Autocorrelation plots for 10 replications (color-matched) for samplers discussed in Section~\ref{sec:numerical}, under the autoregressive graph structure and $(n,p)=(100,50)$.}
   \label{app:fig:corr-traceplots}
 \end{figure}

\begin{figure}[ht]
  \centering
  \includegraphics[width=0.32\textwidth]{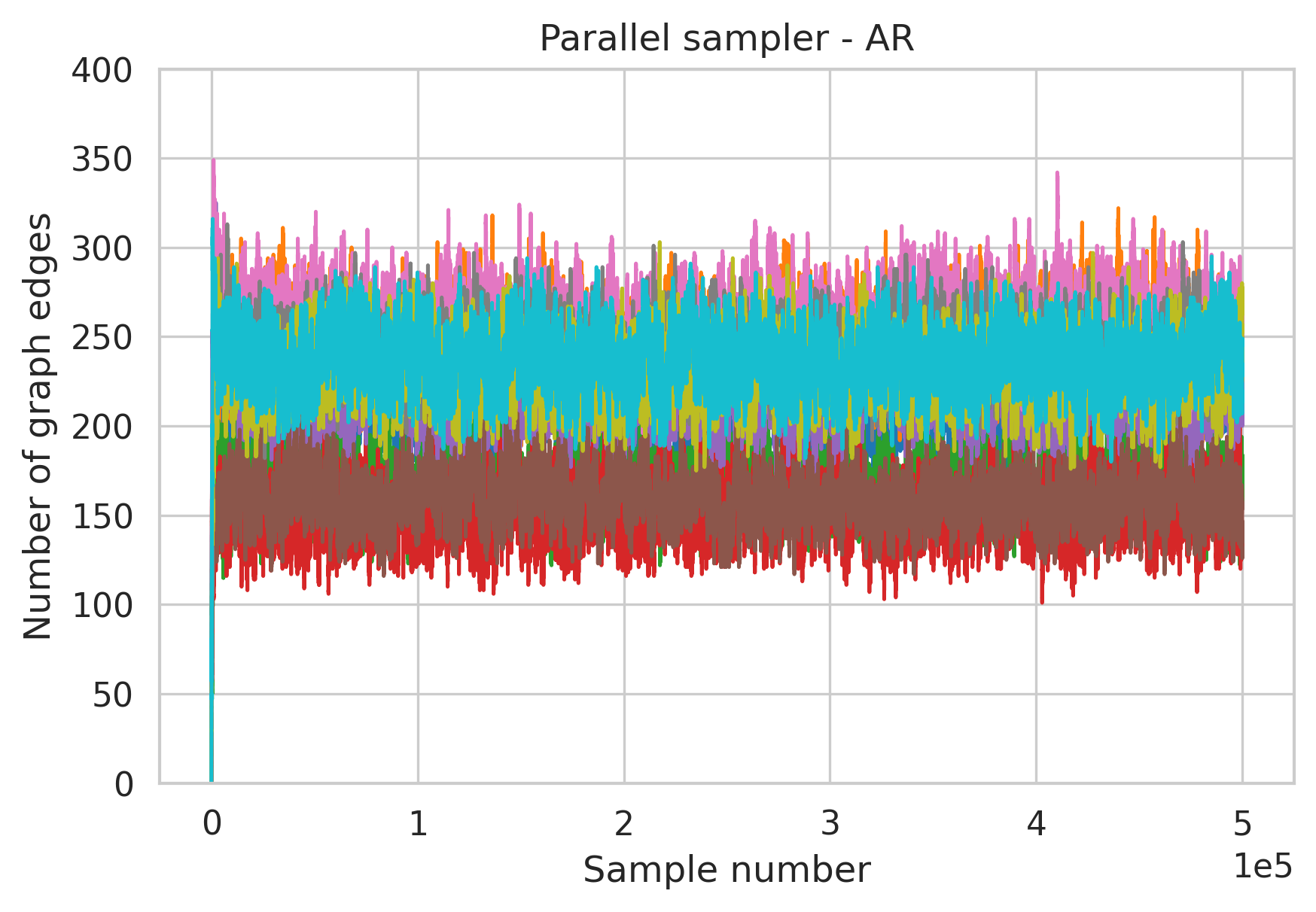}
  \includegraphics[width=0.32\textwidth]{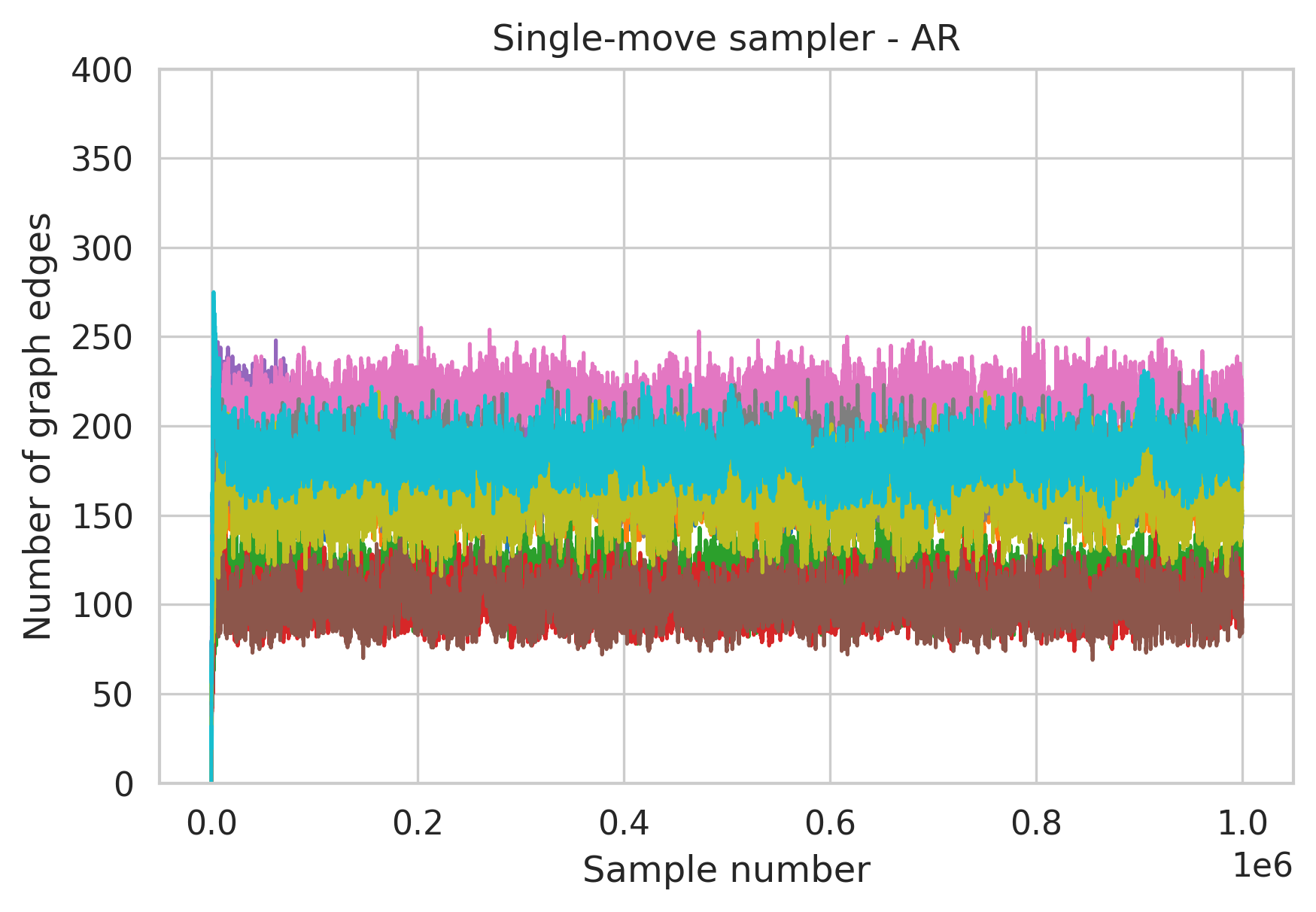}
  \includegraphics[width=0.32\textwidth]{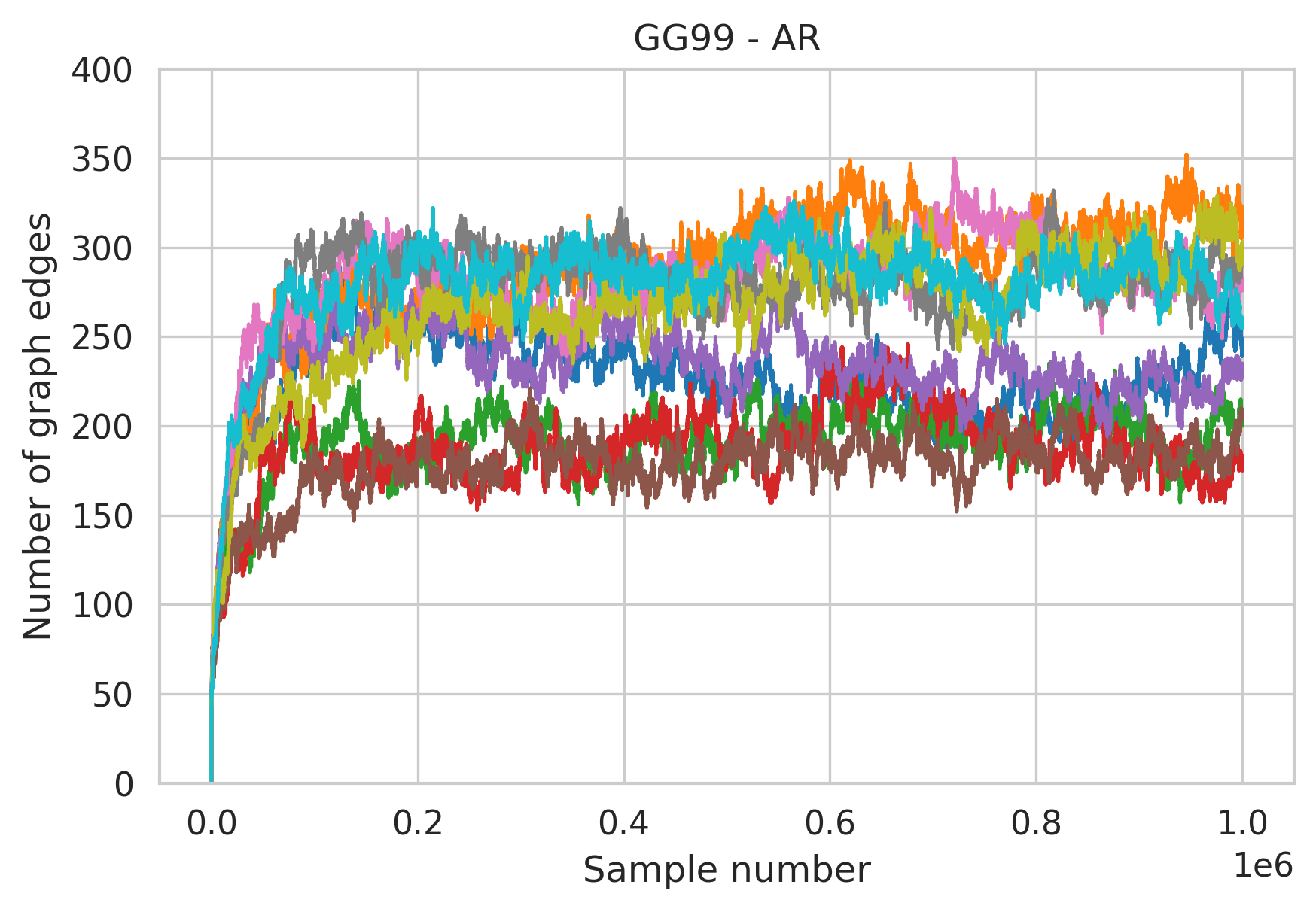}
  
  \includegraphics[width=0.32\textwidth]{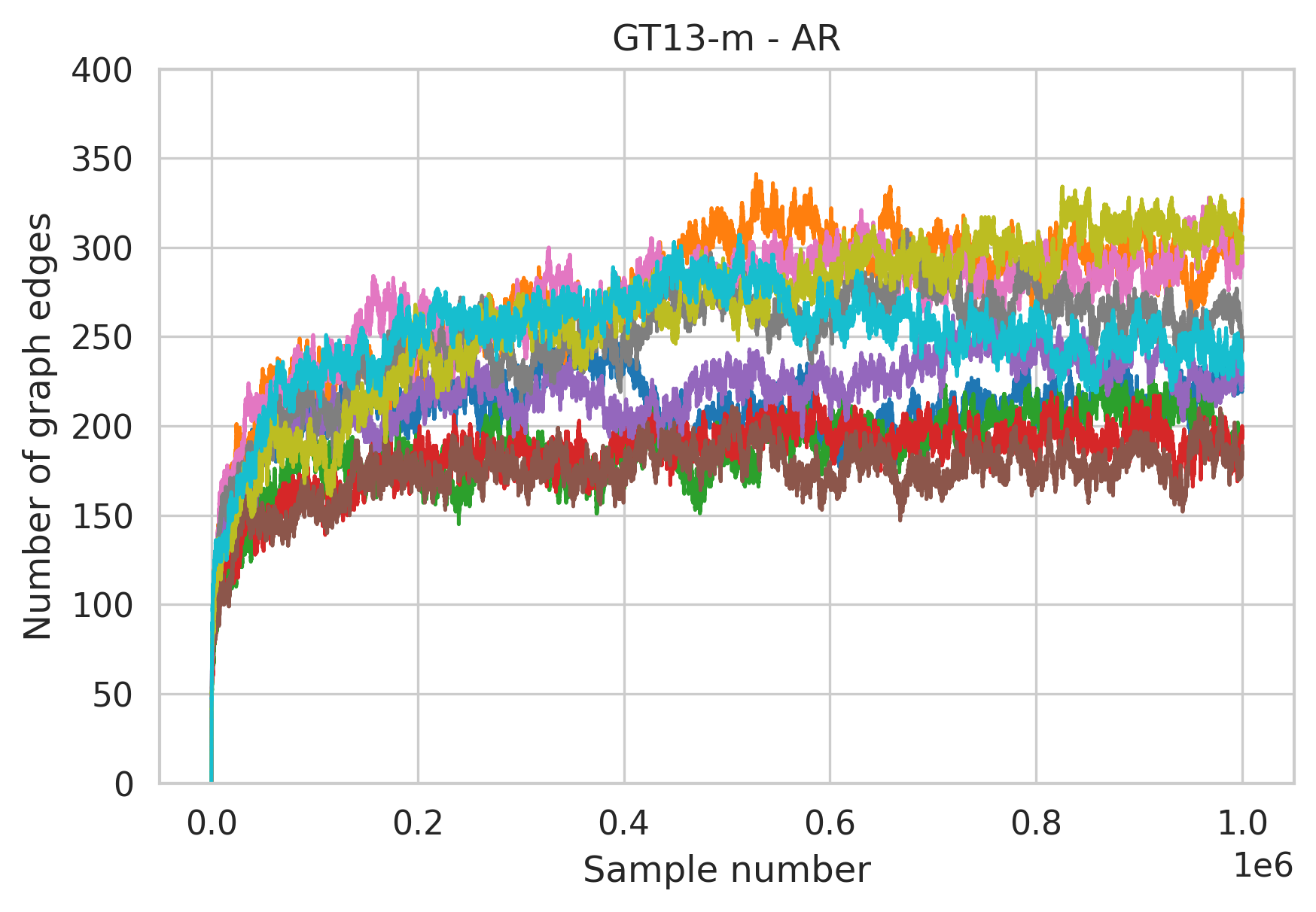}
  \includegraphics[width=0.32\textwidth]{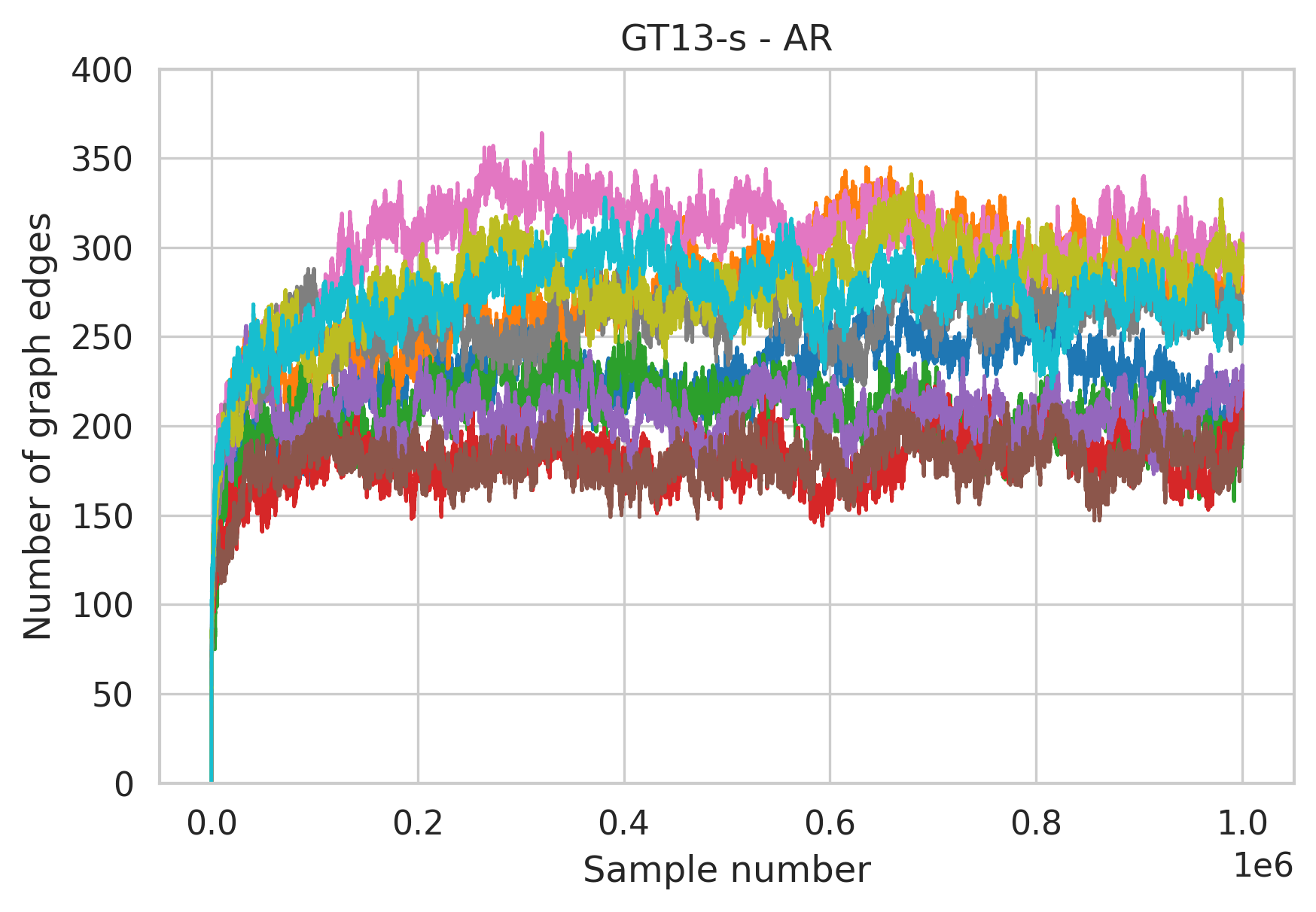}
  \includegraphics[width=0.32\textwidth]{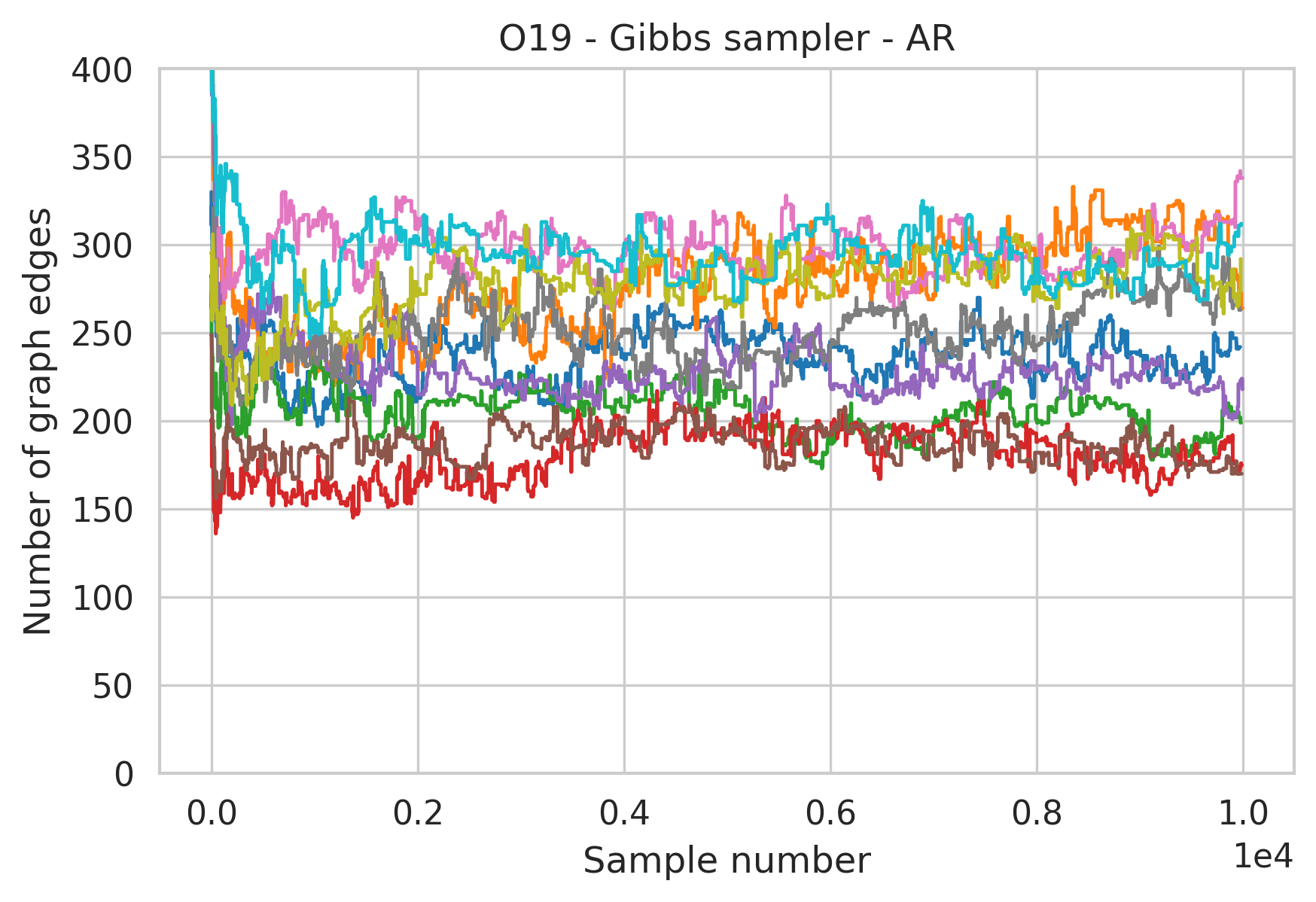}

  \includegraphics[width=0.32\textwidth]{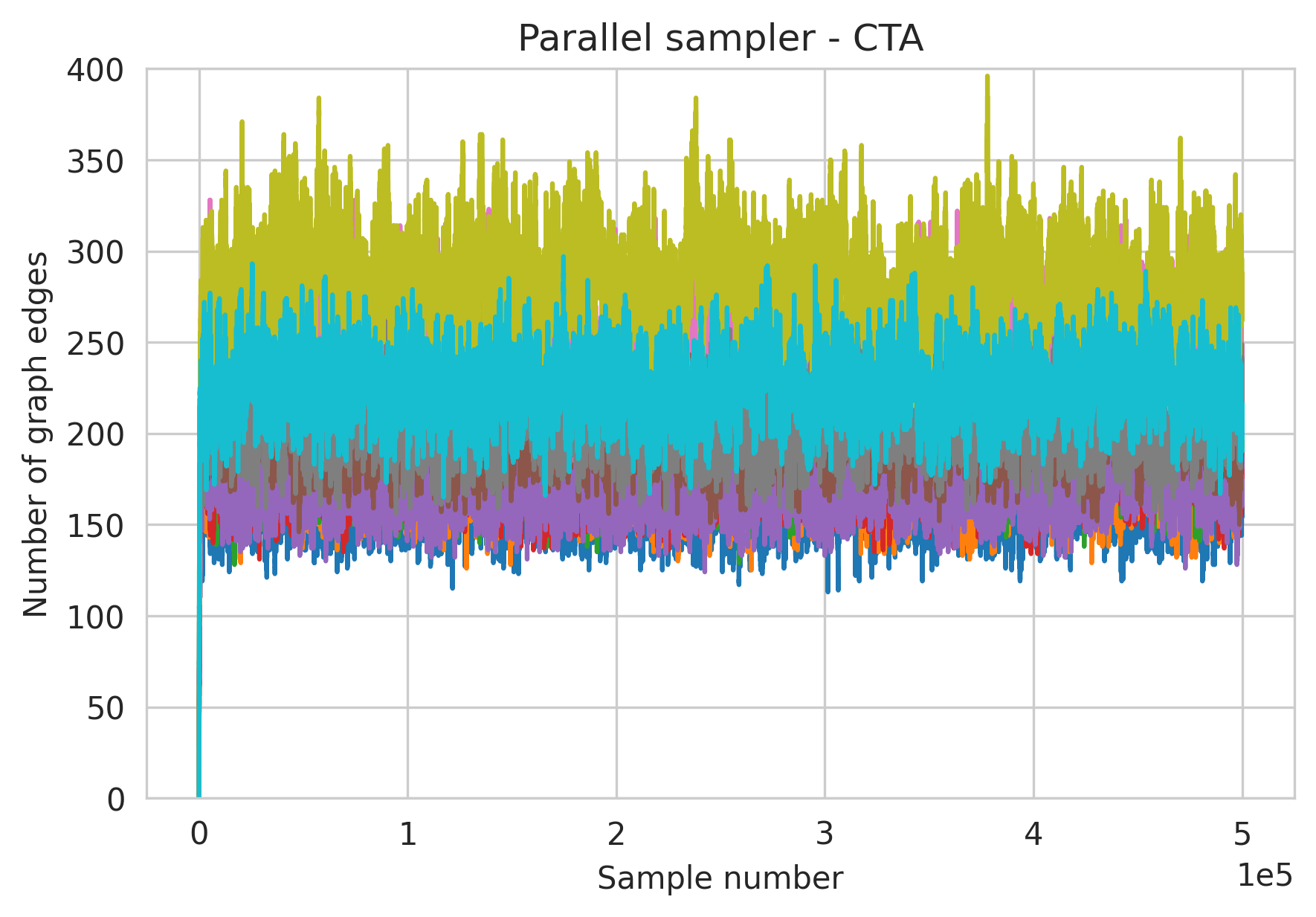}
  \includegraphics[width=0.32\textwidth]{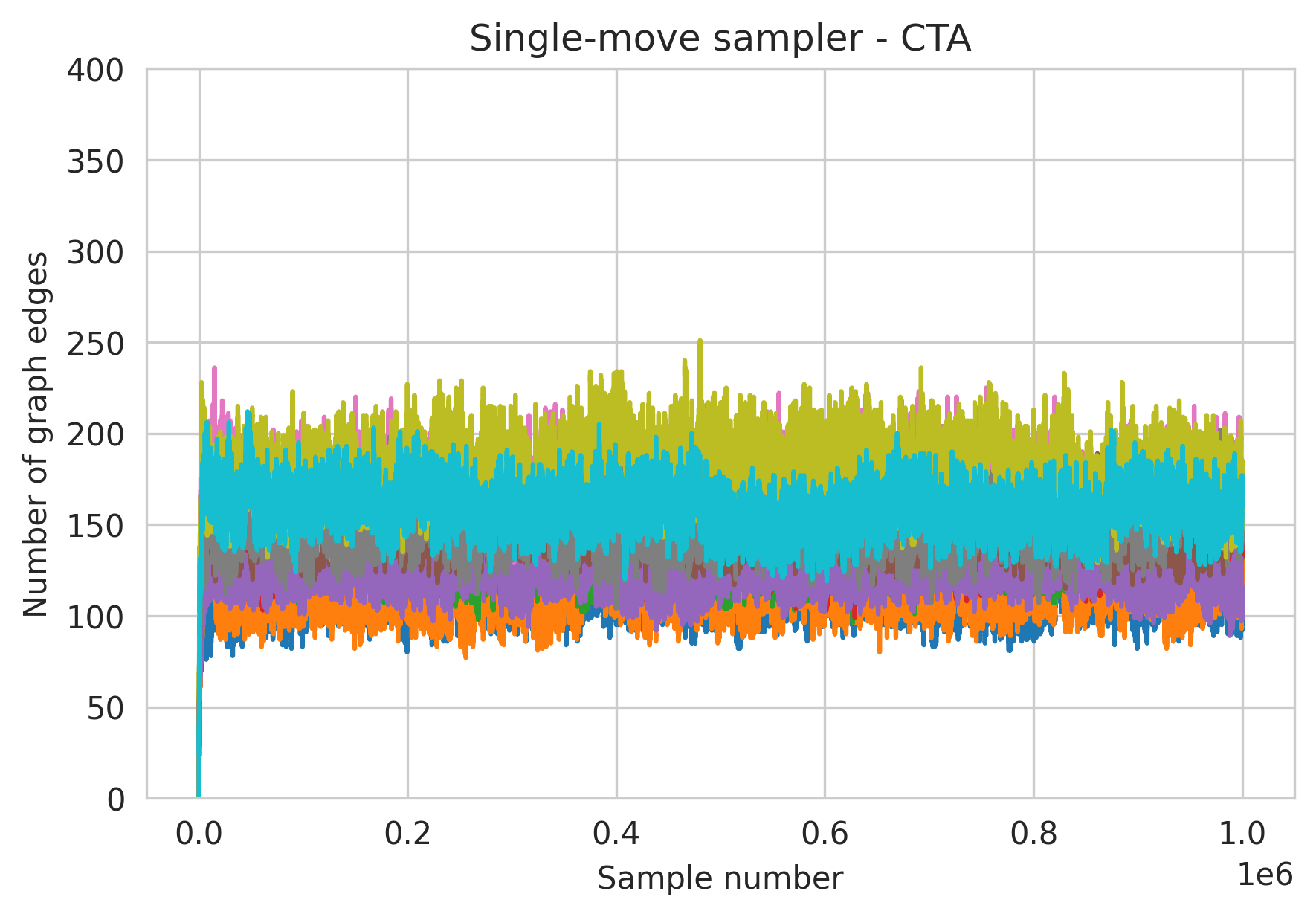}
  \includegraphics[width=0.32\textwidth]{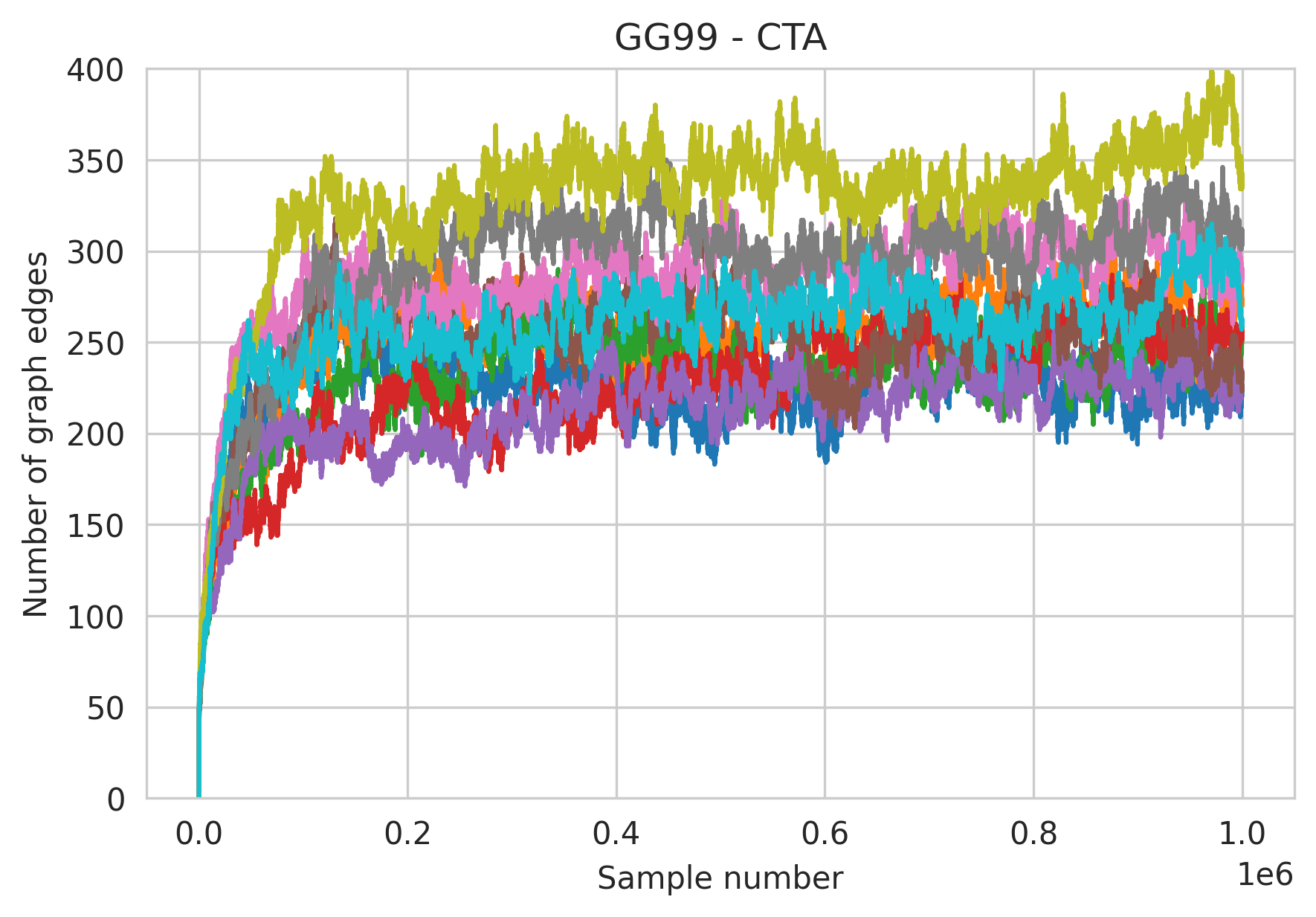}
  
  \includegraphics[width=0.32\textwidth]{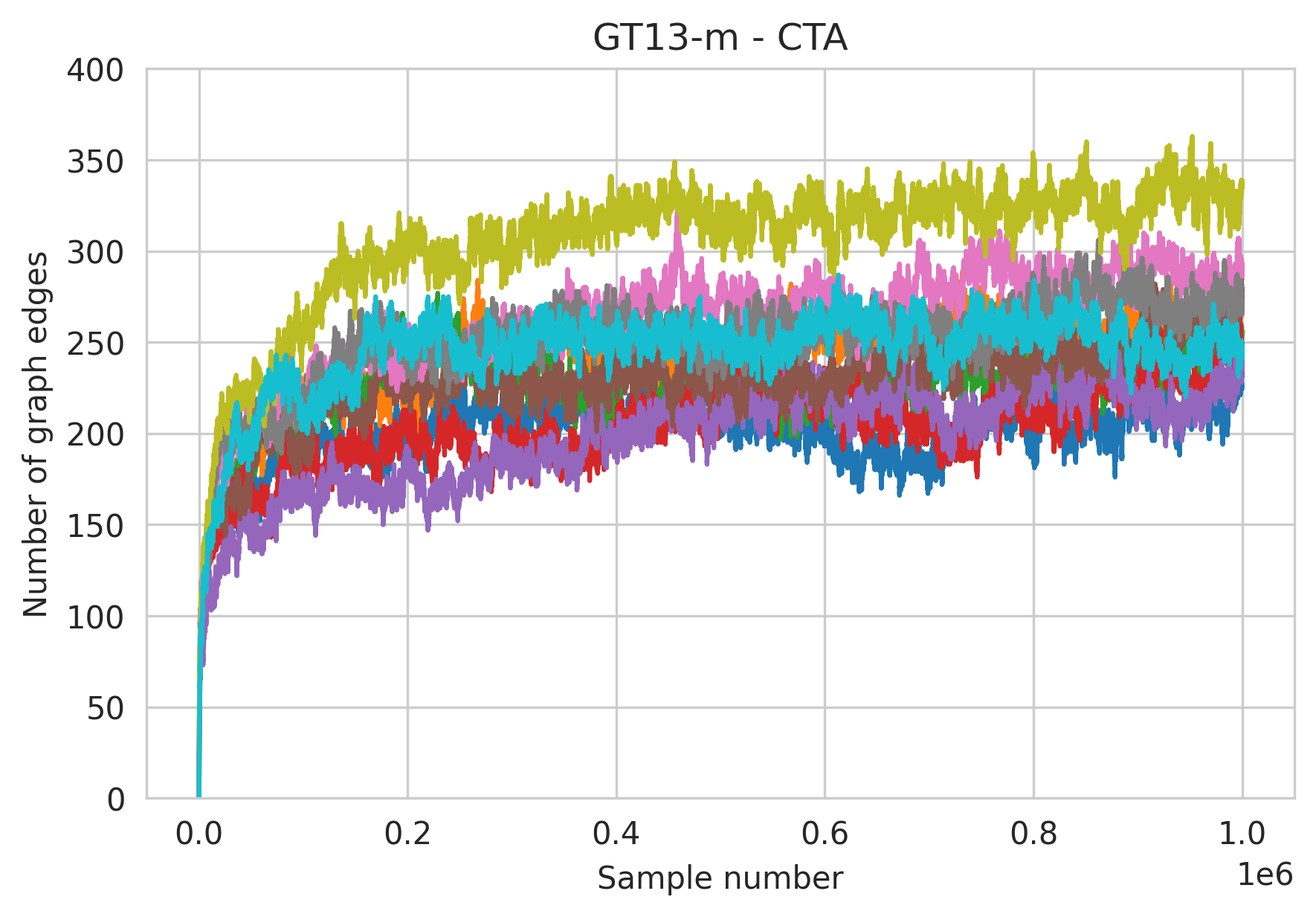}
  \includegraphics[width=0.32\textwidth]{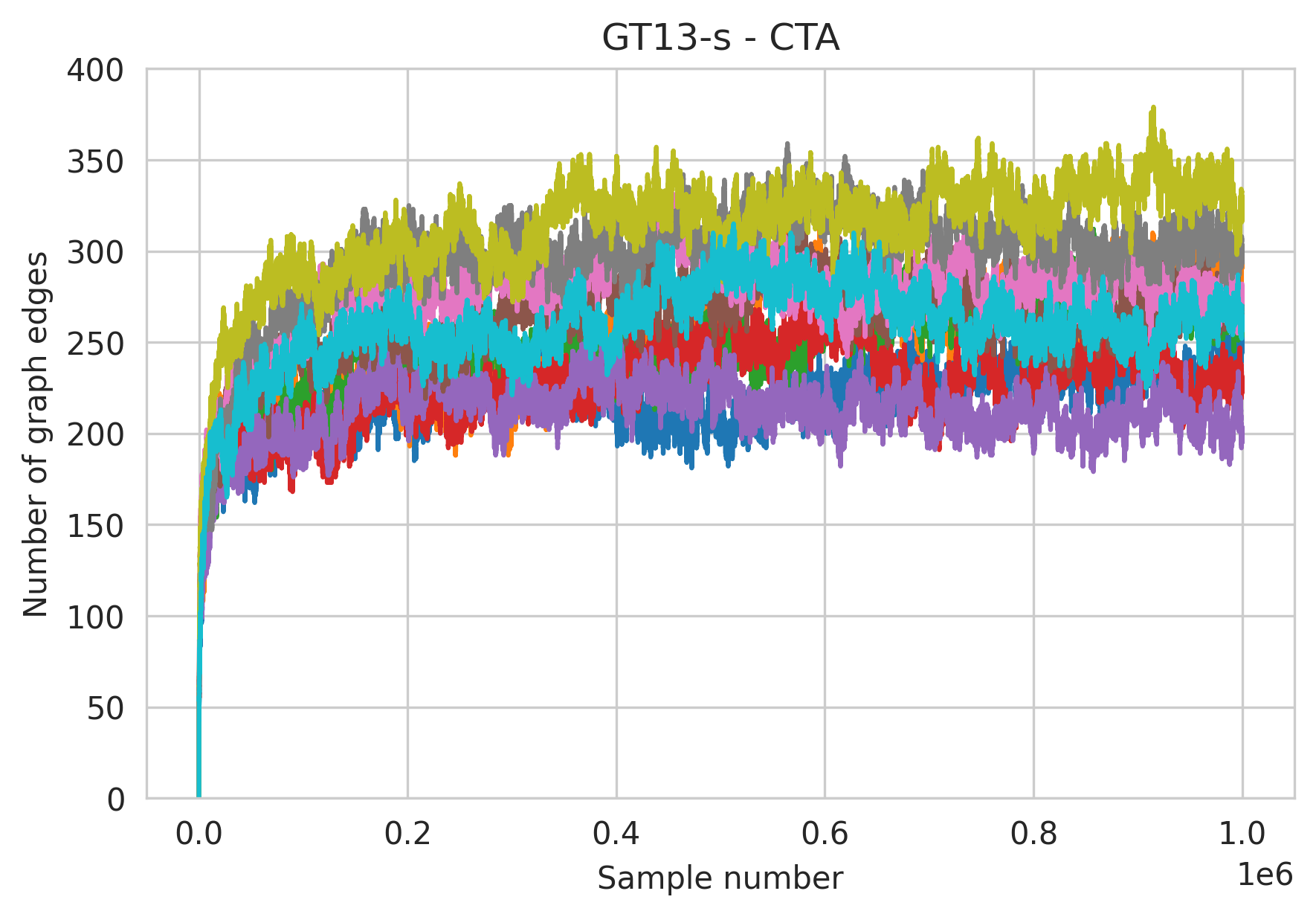}
  \includegraphics[width=0.32\textwidth]{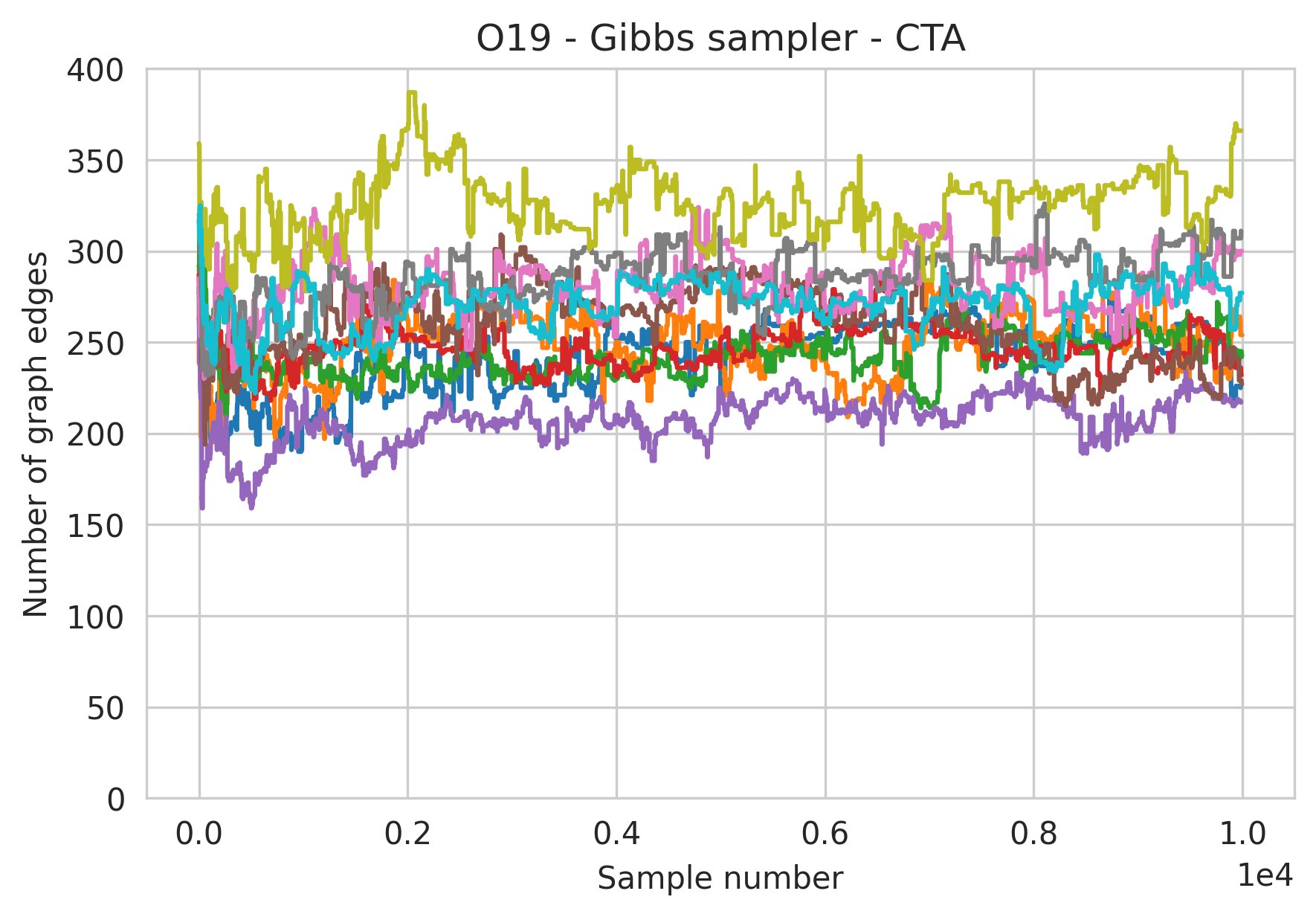}
  
    \caption{Traceplots showing the number of graph edges for 10 replications (color-matched), as outlined in Section~\ref{sec:simul-setup-gauss} and $(n,p)=(100,50)$.}
  \label{fig:gt_parallel_size_traceplot}
\end{figure}

\begin{figure}[ht]
  \centering
  \includegraphics[width=0.19\textwidth]{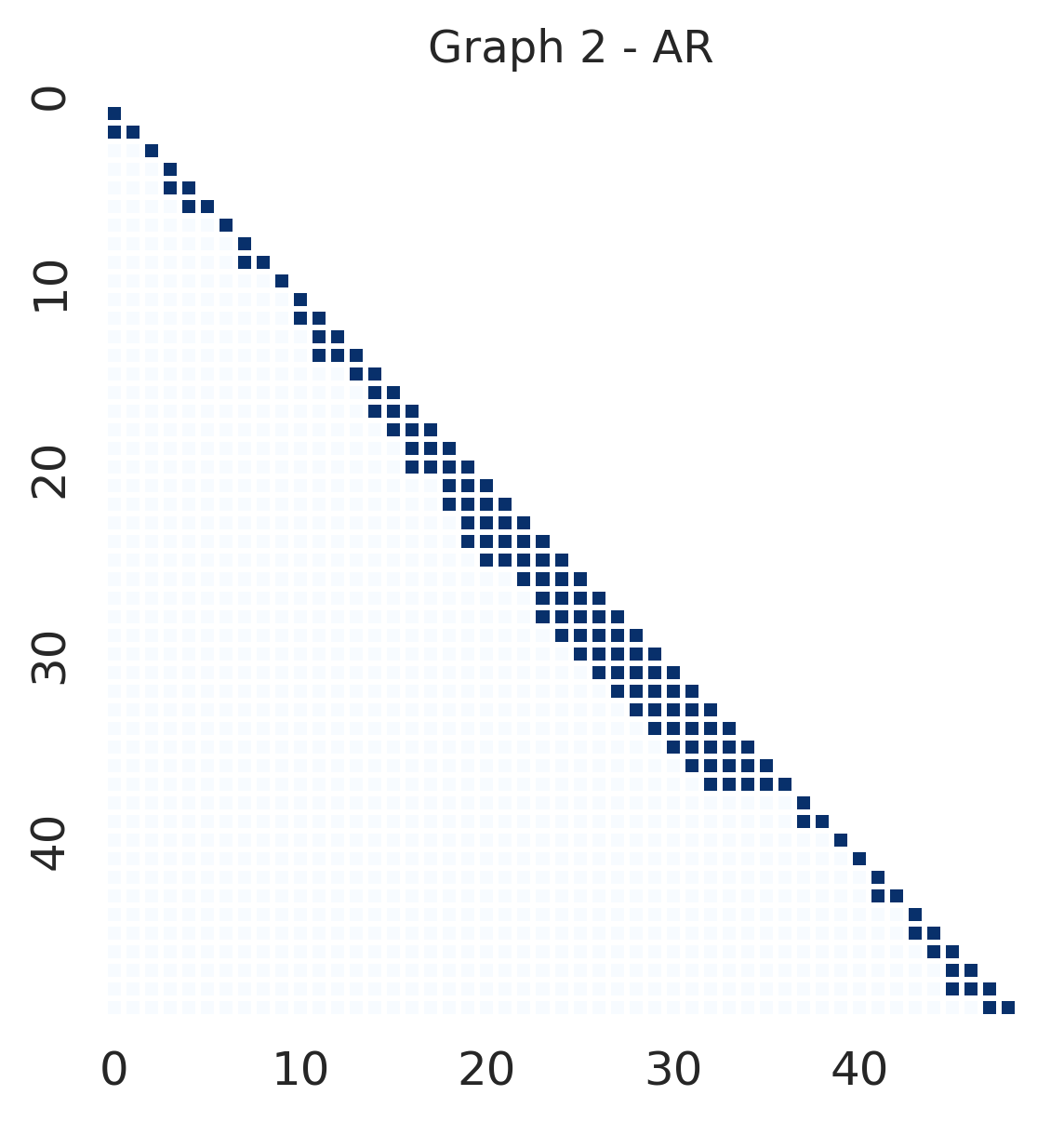}
  \includegraphics[width=0.29\textwidth]{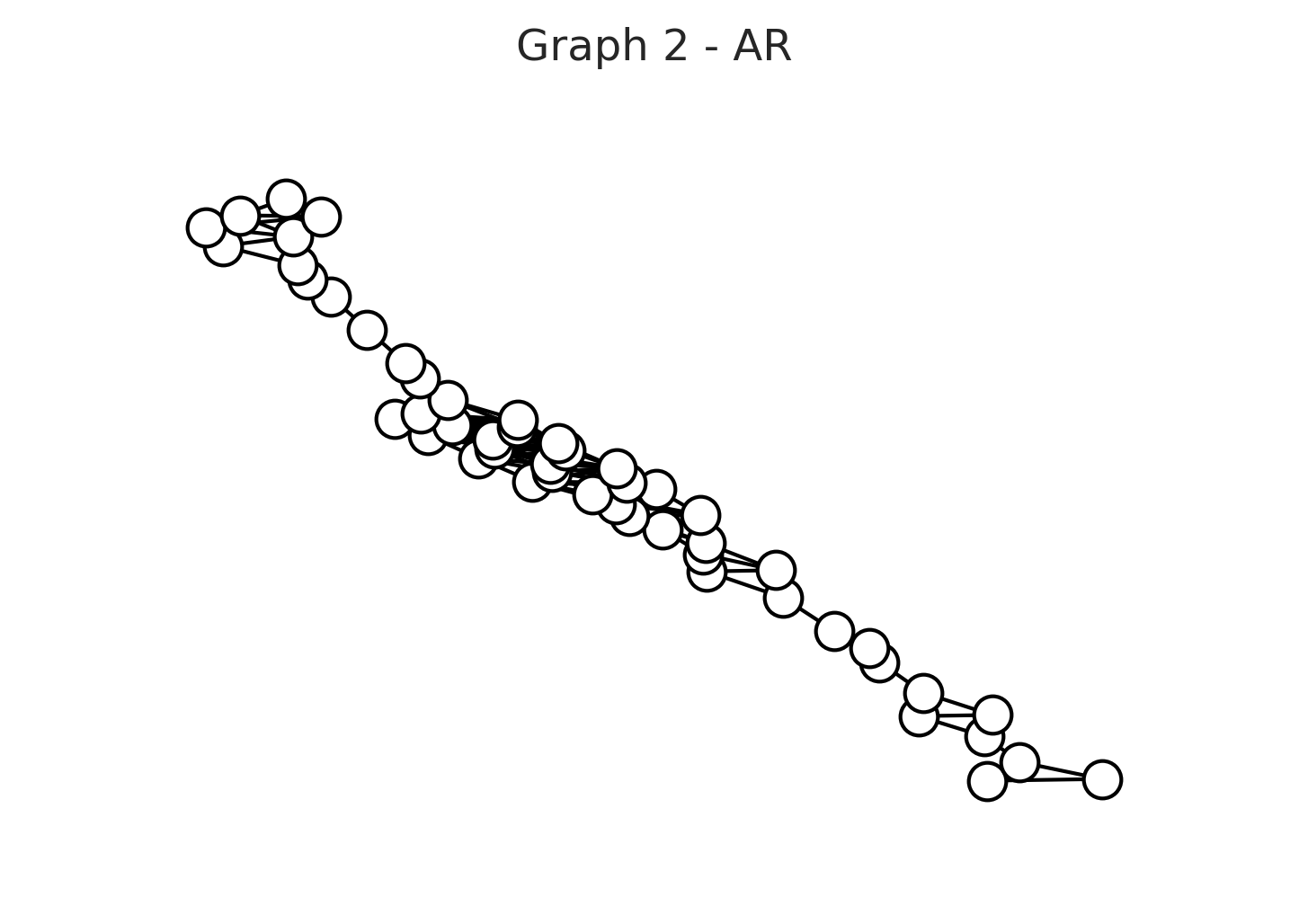}
   \includegraphics[width=0.19\textwidth]{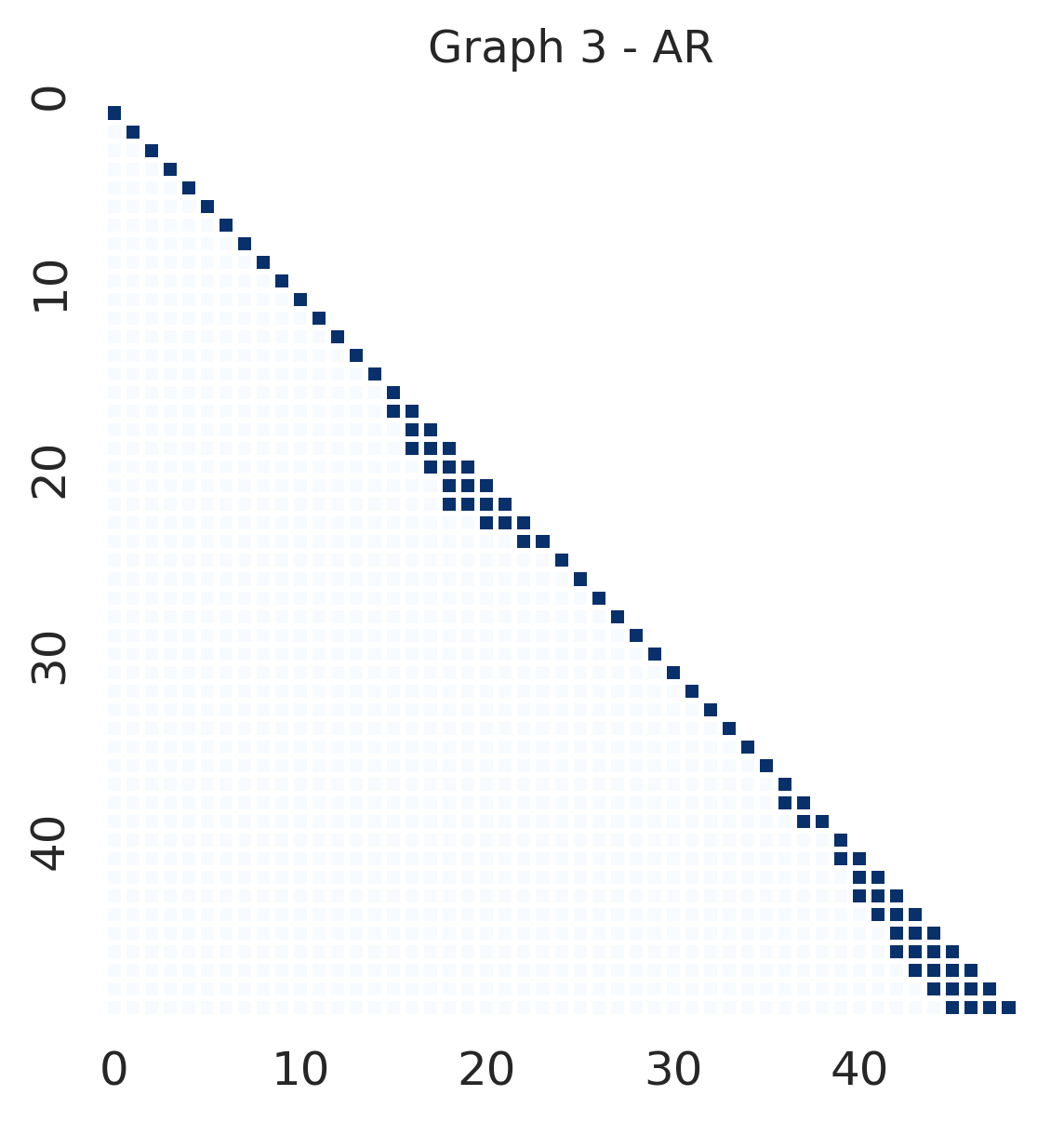}
  \includegraphics[width=0.29\textwidth]{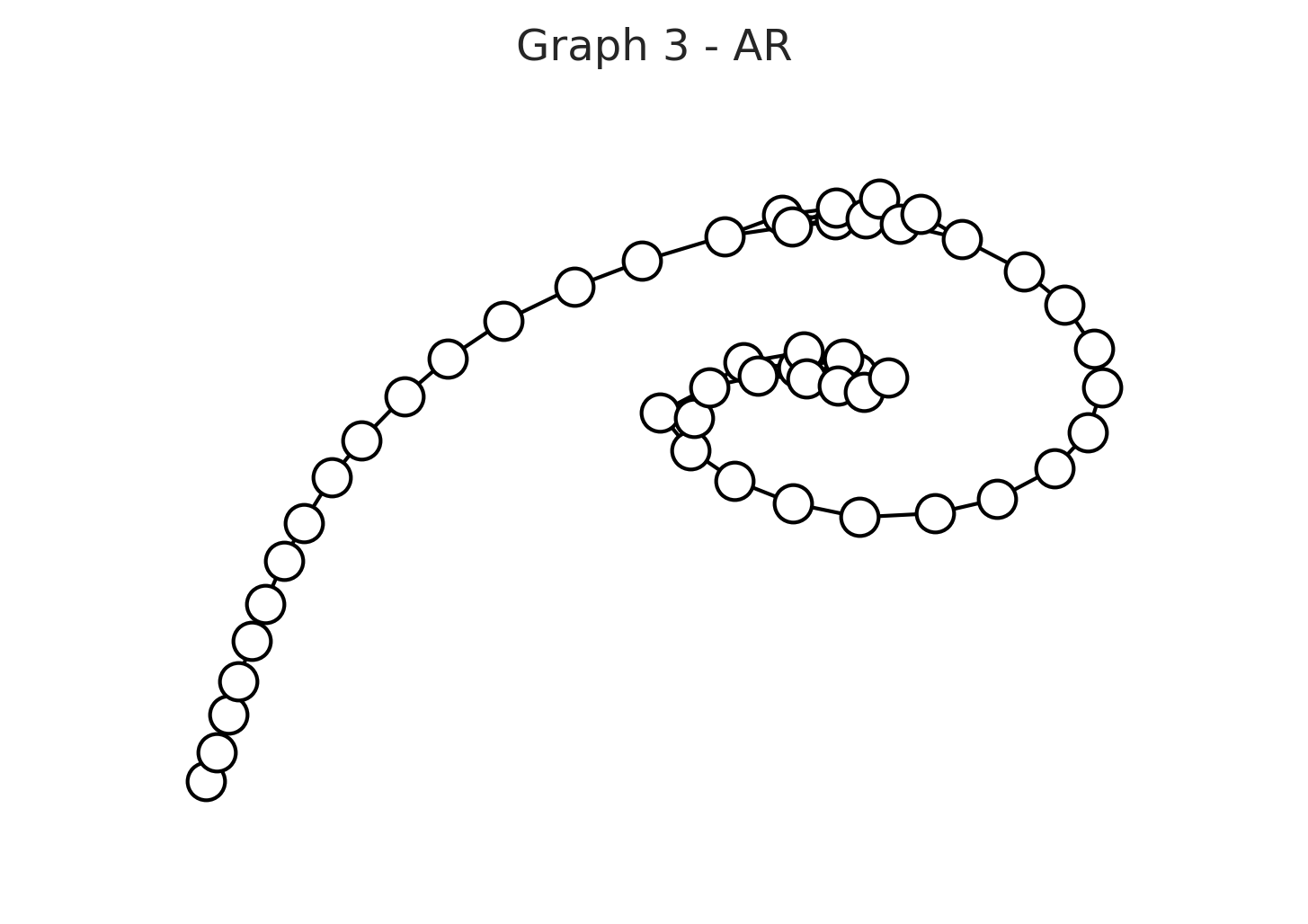}

  \includegraphics[width=0.19\textwidth]{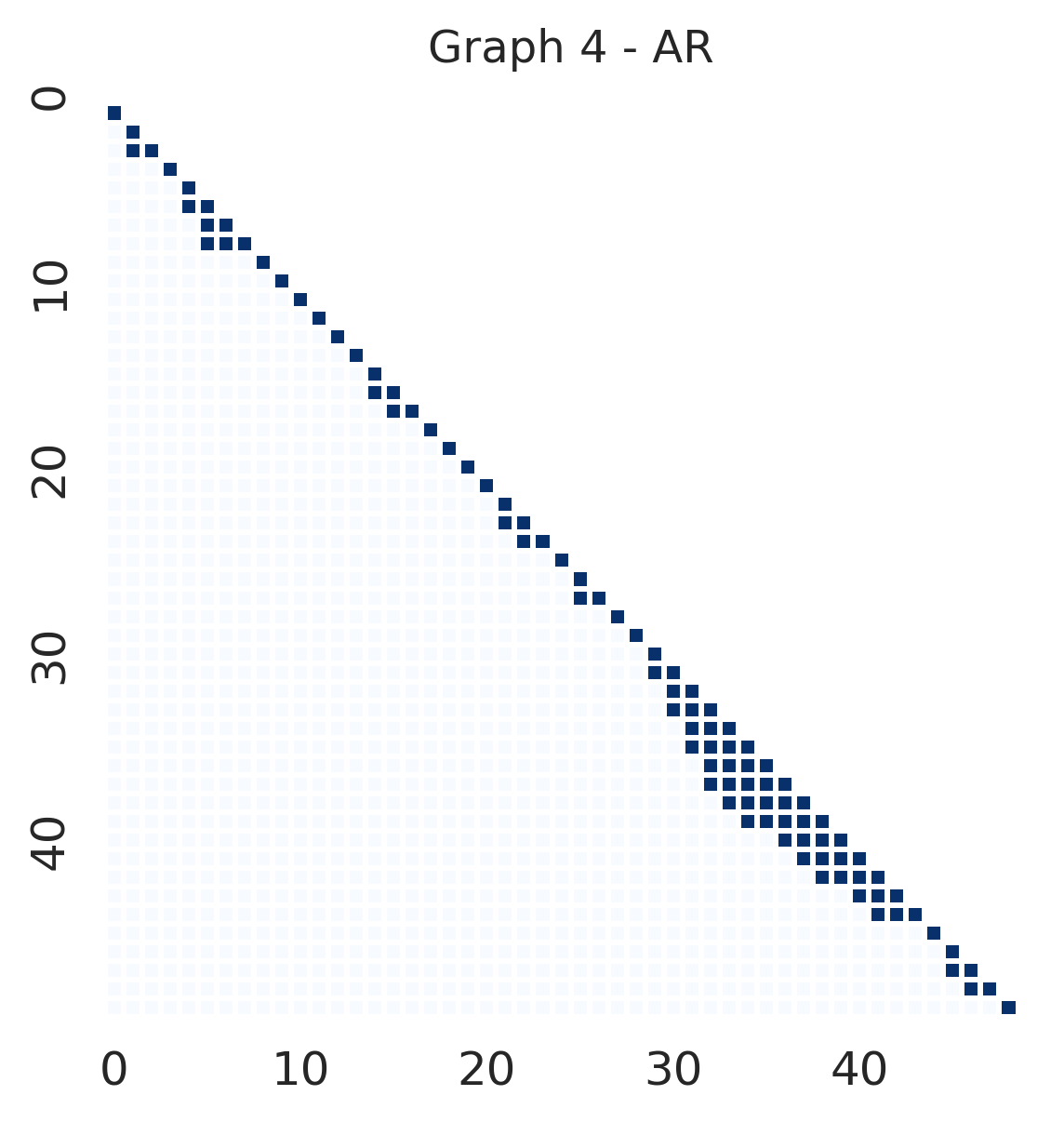}
  \includegraphics[width=0.29\textwidth]{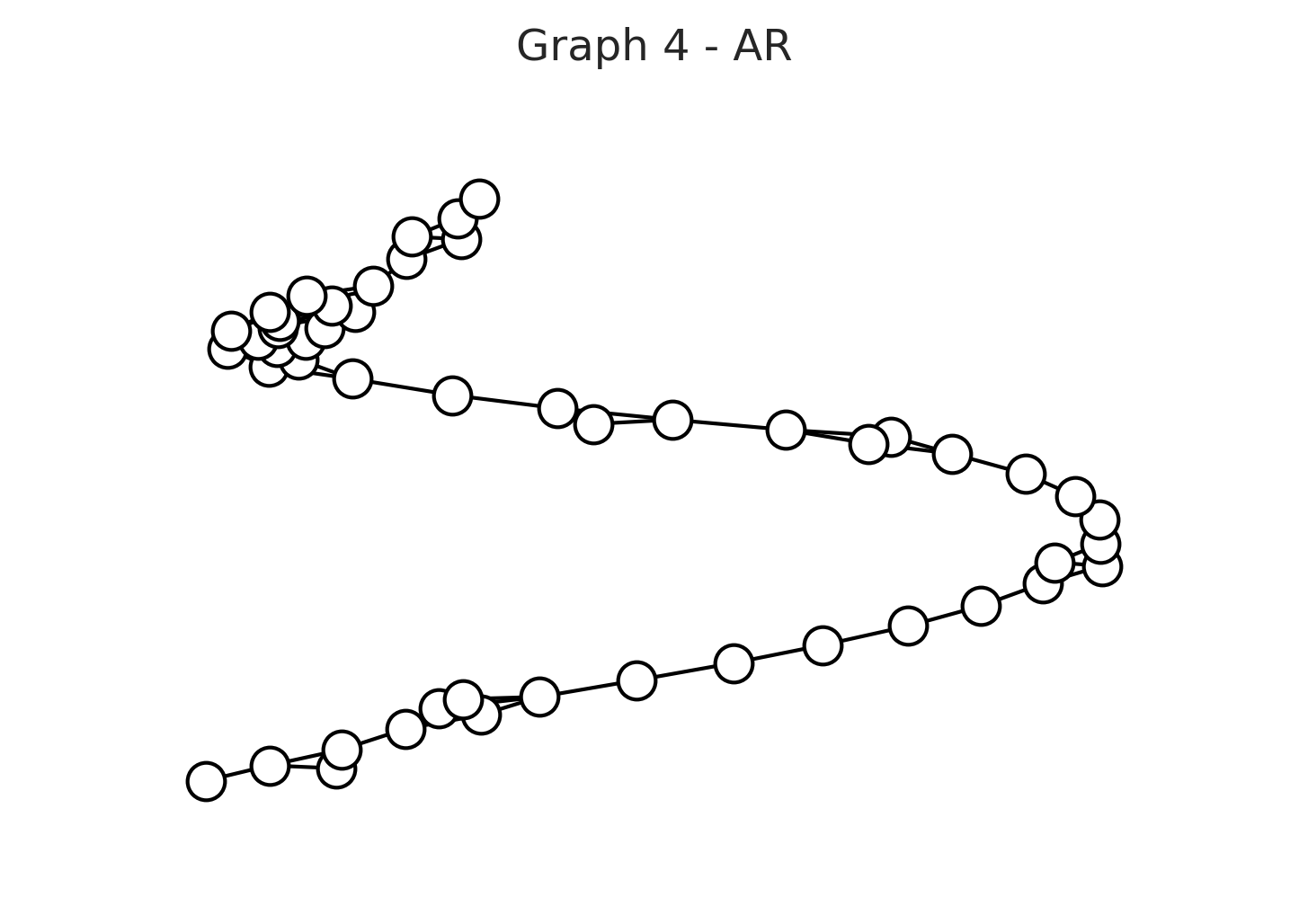}
   \includegraphics[width=0.19\textwidth]{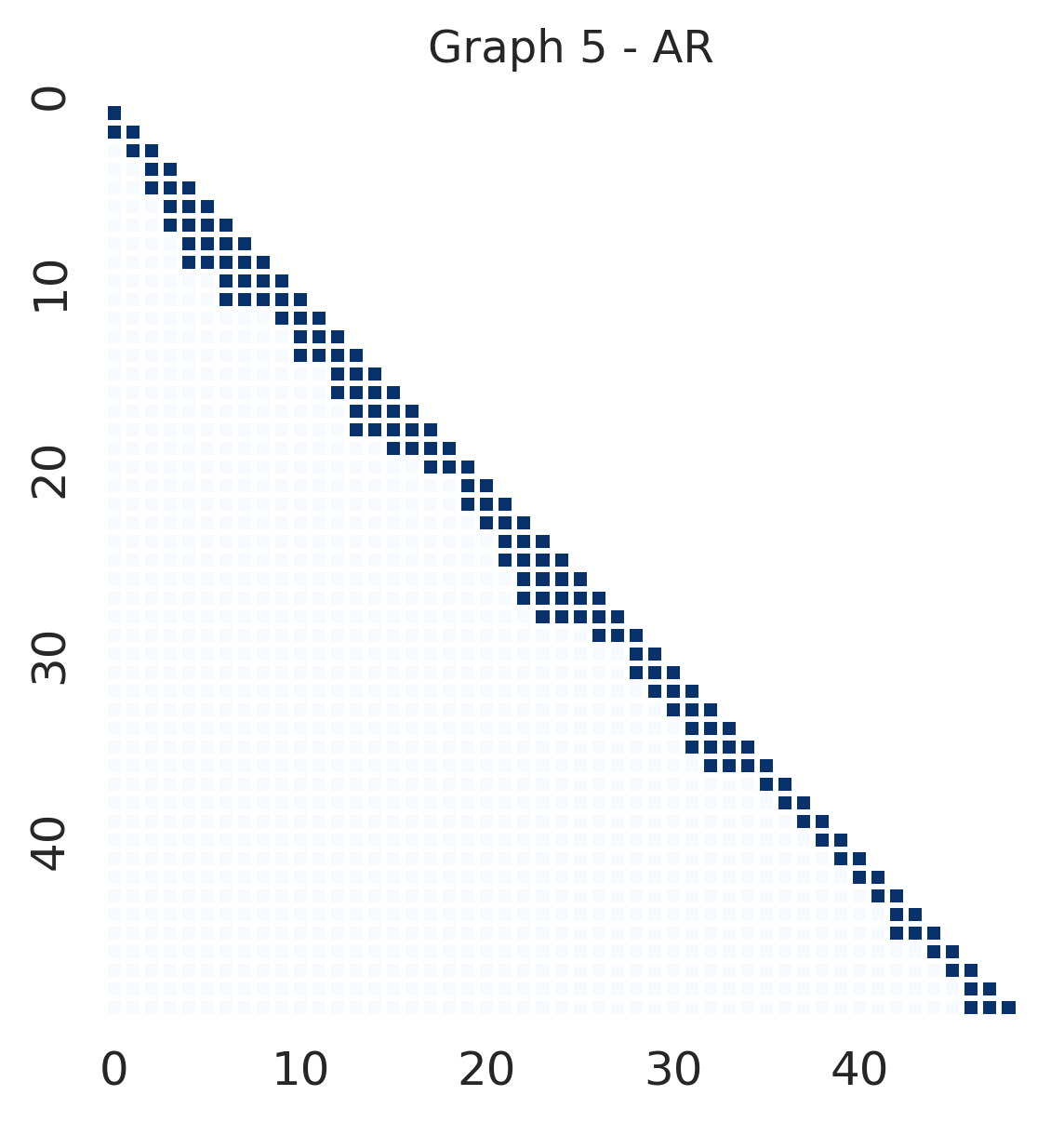}
   \includegraphics[width=0.29\textwidth]{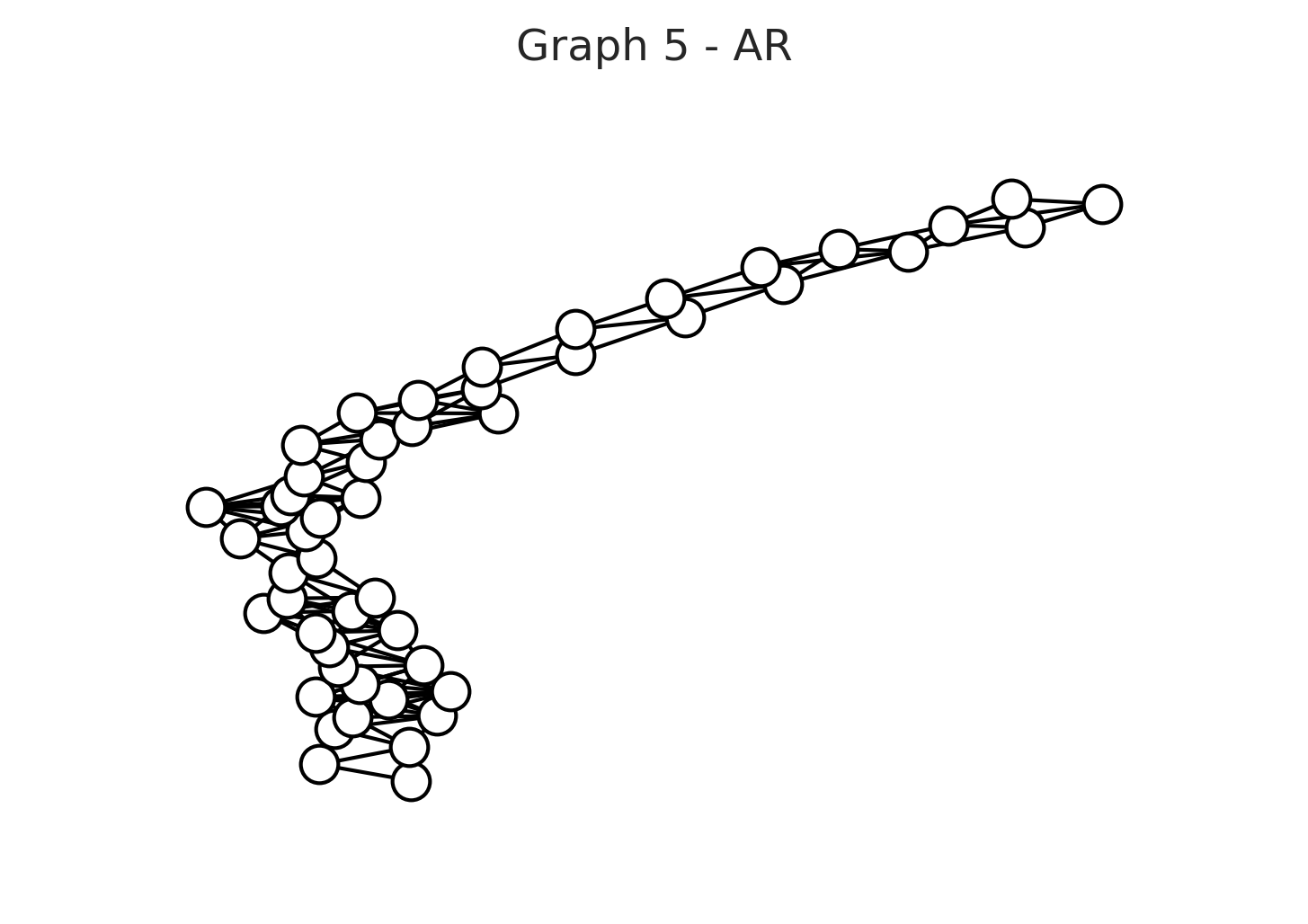}

   \includegraphics[width=0.19\textwidth]{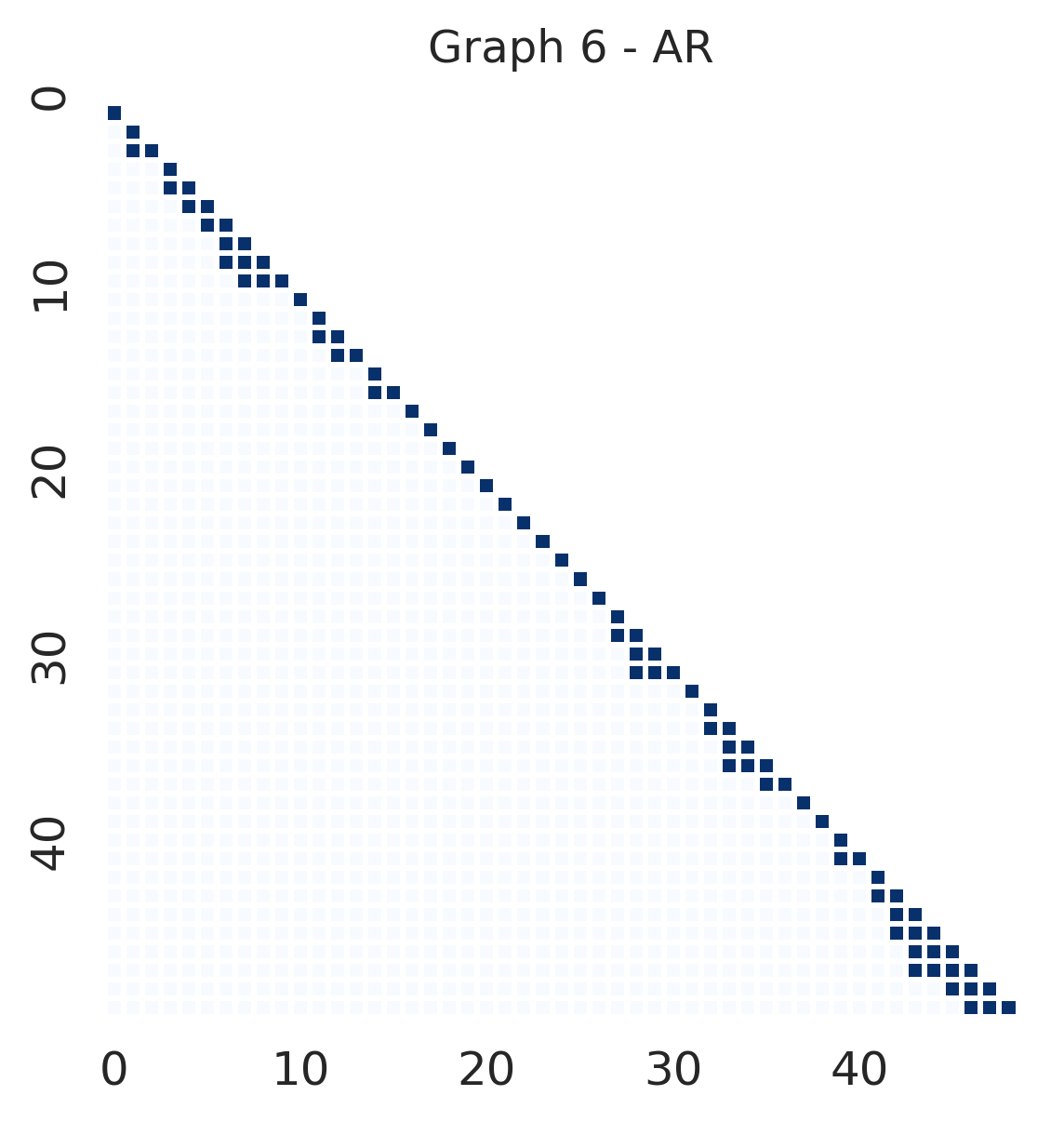}
  \includegraphics[width=0.29\textwidth]{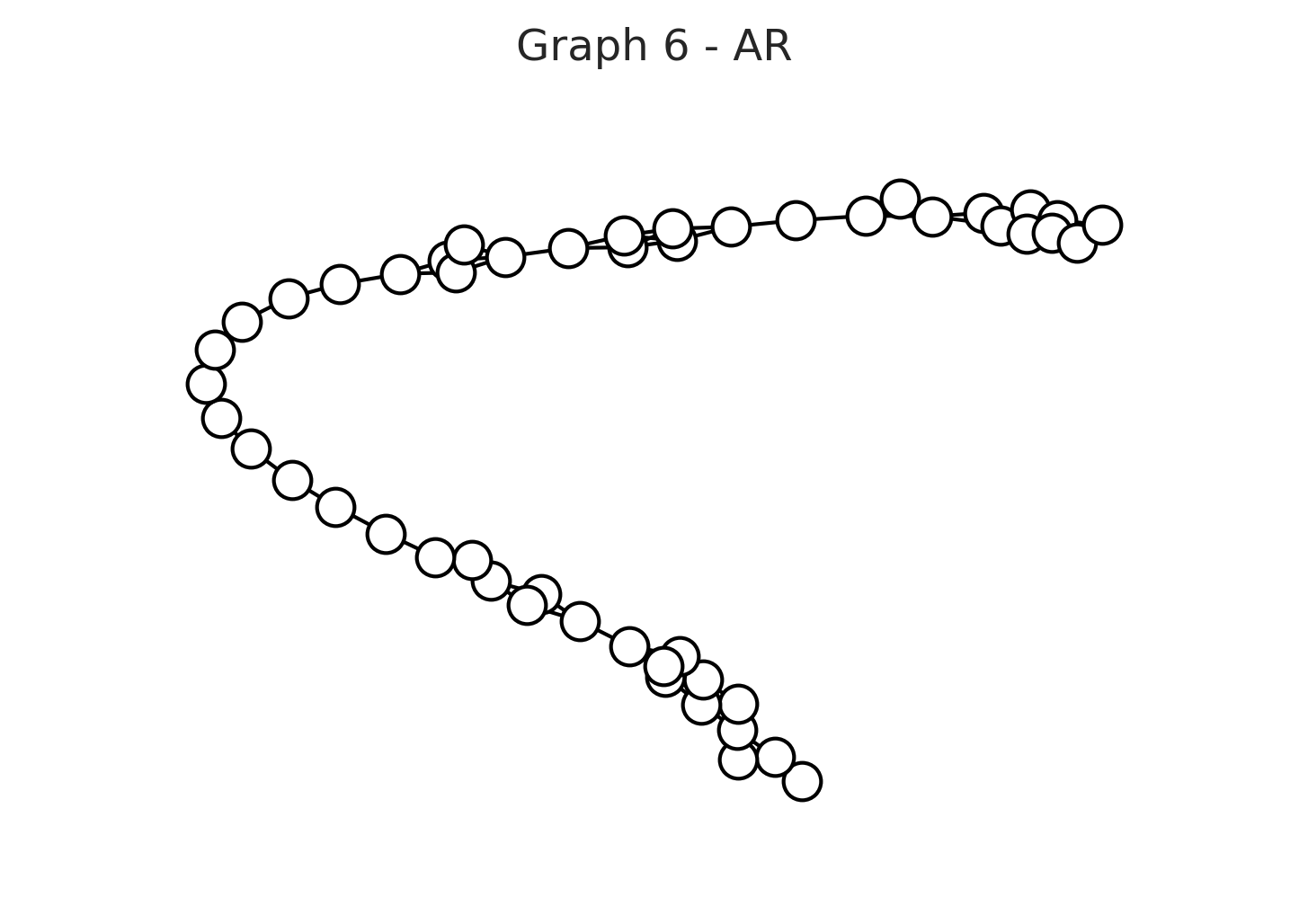}
   \includegraphics[width=0.19\textwidth]{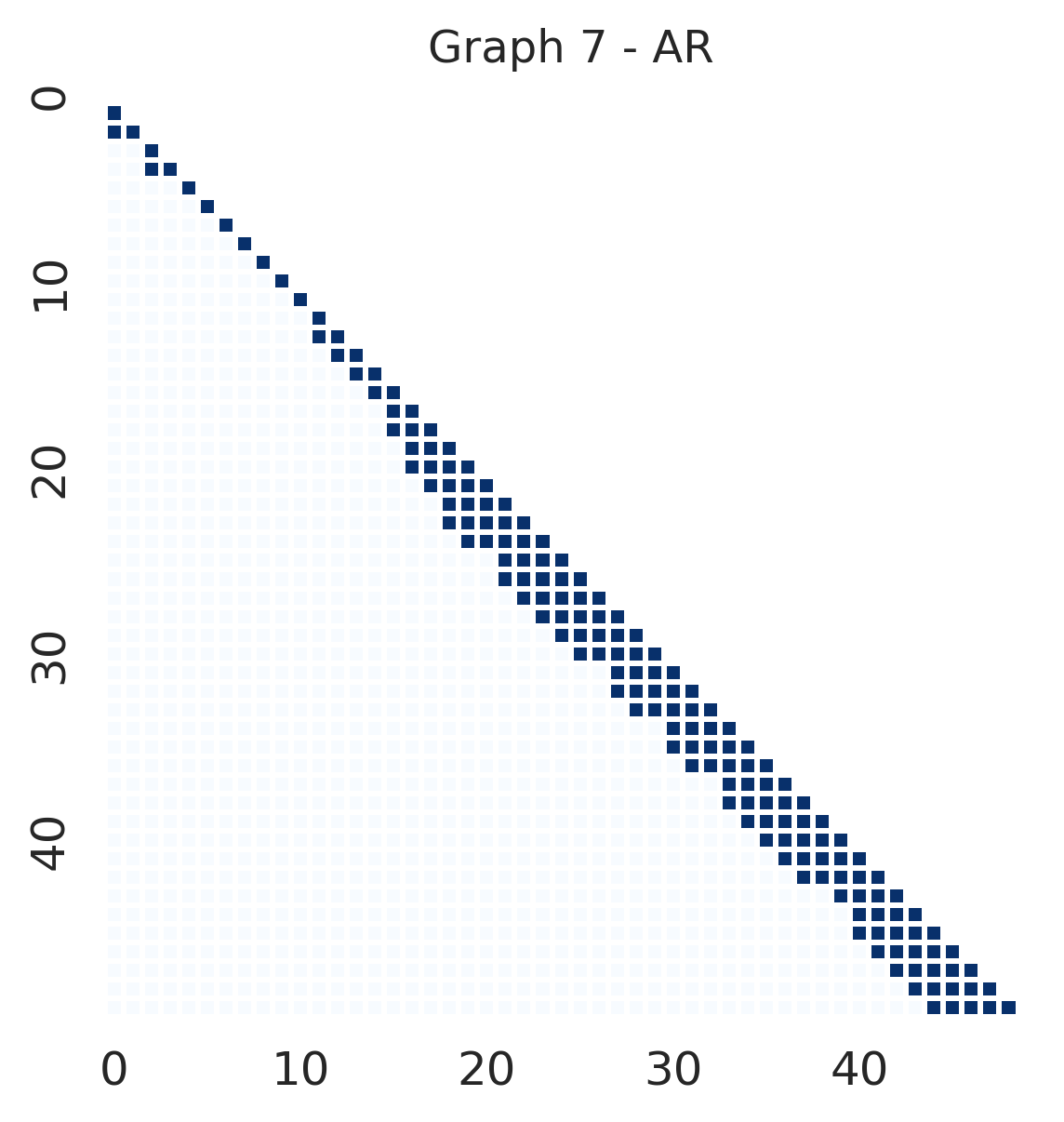}
  \includegraphics[width=0.29\textwidth]{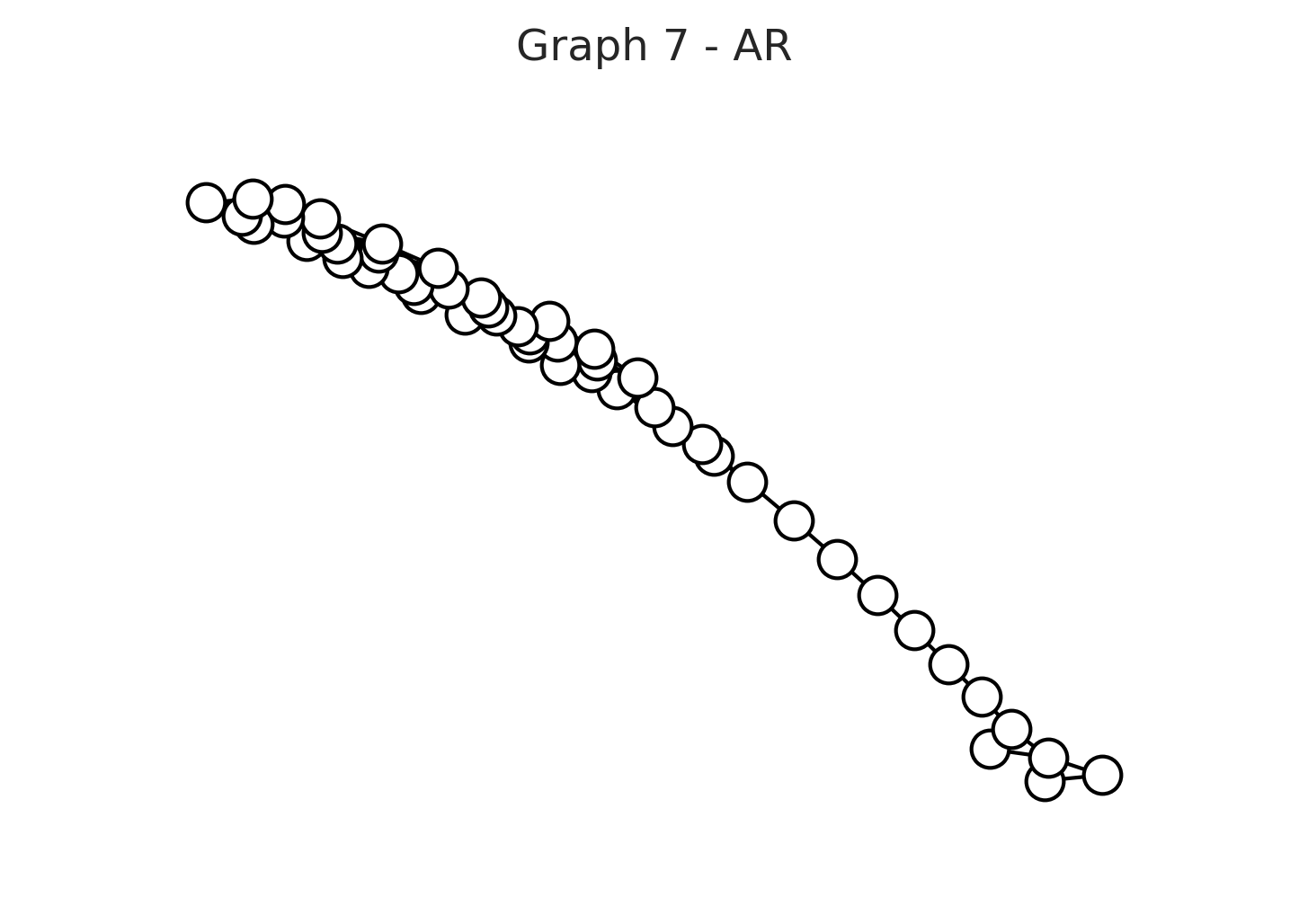}

  \includegraphics[width=0.19\textwidth]{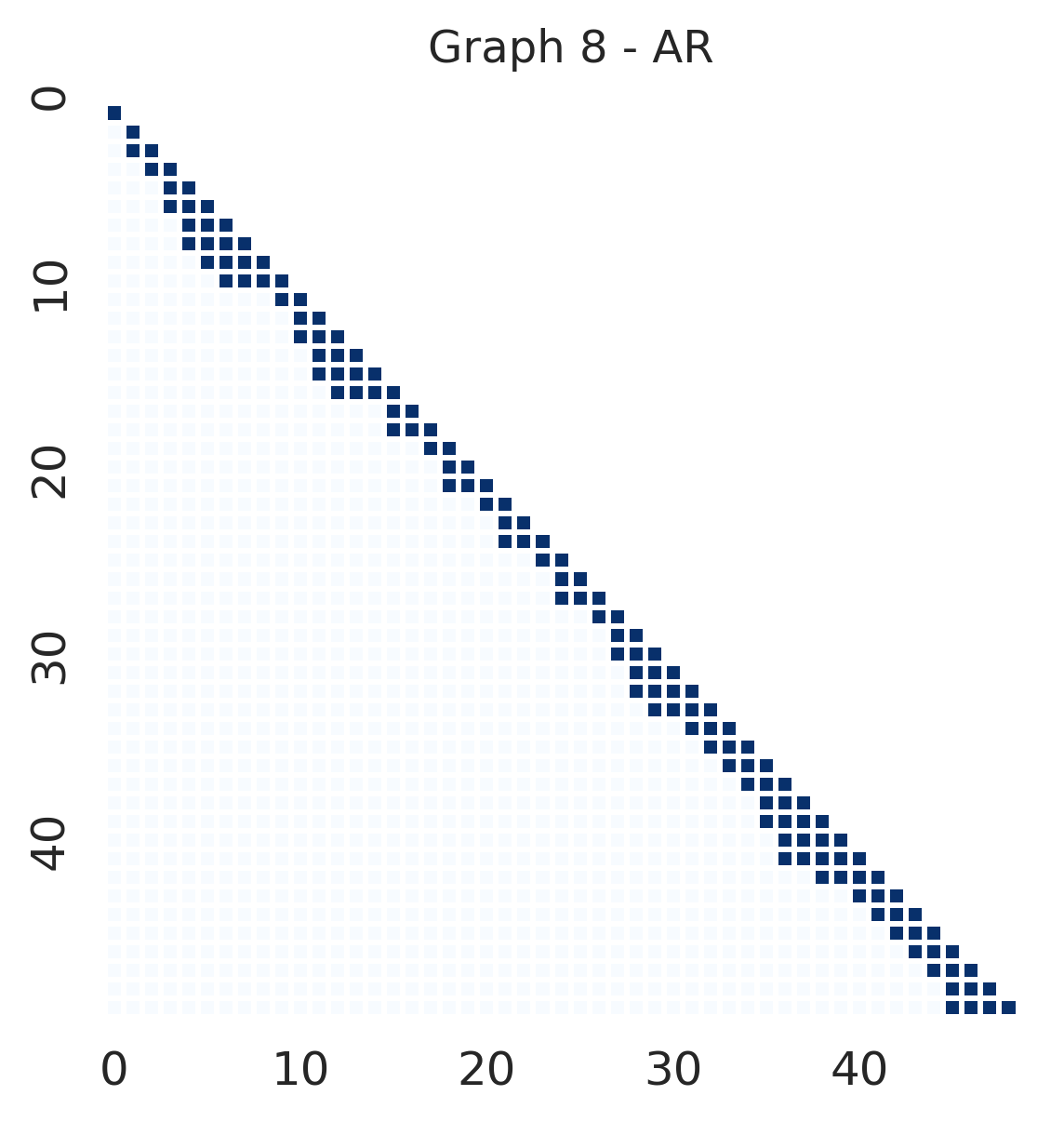}
  \includegraphics[width=0.29\textwidth]{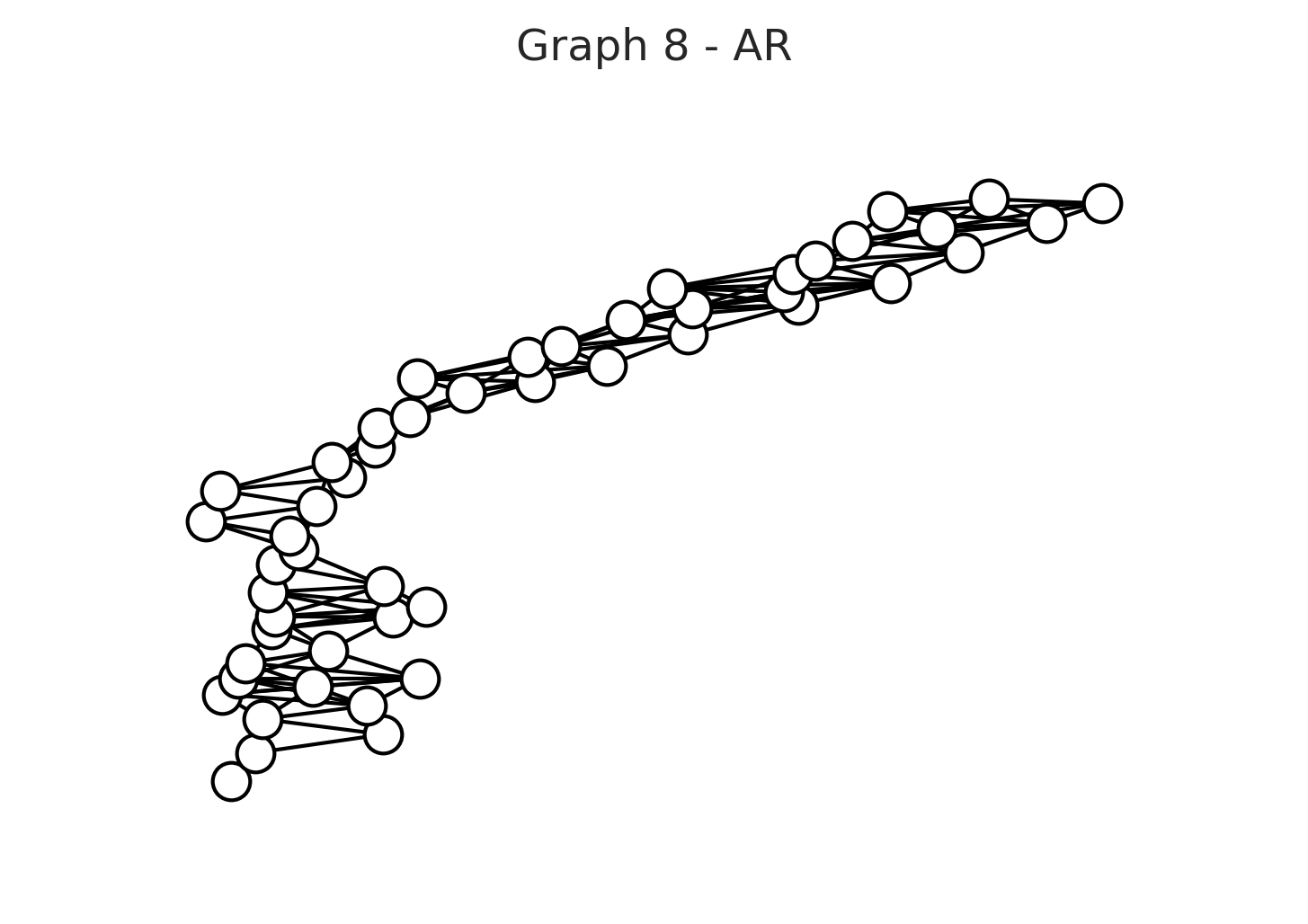}
  \includegraphics[width=0.24\textwidth]{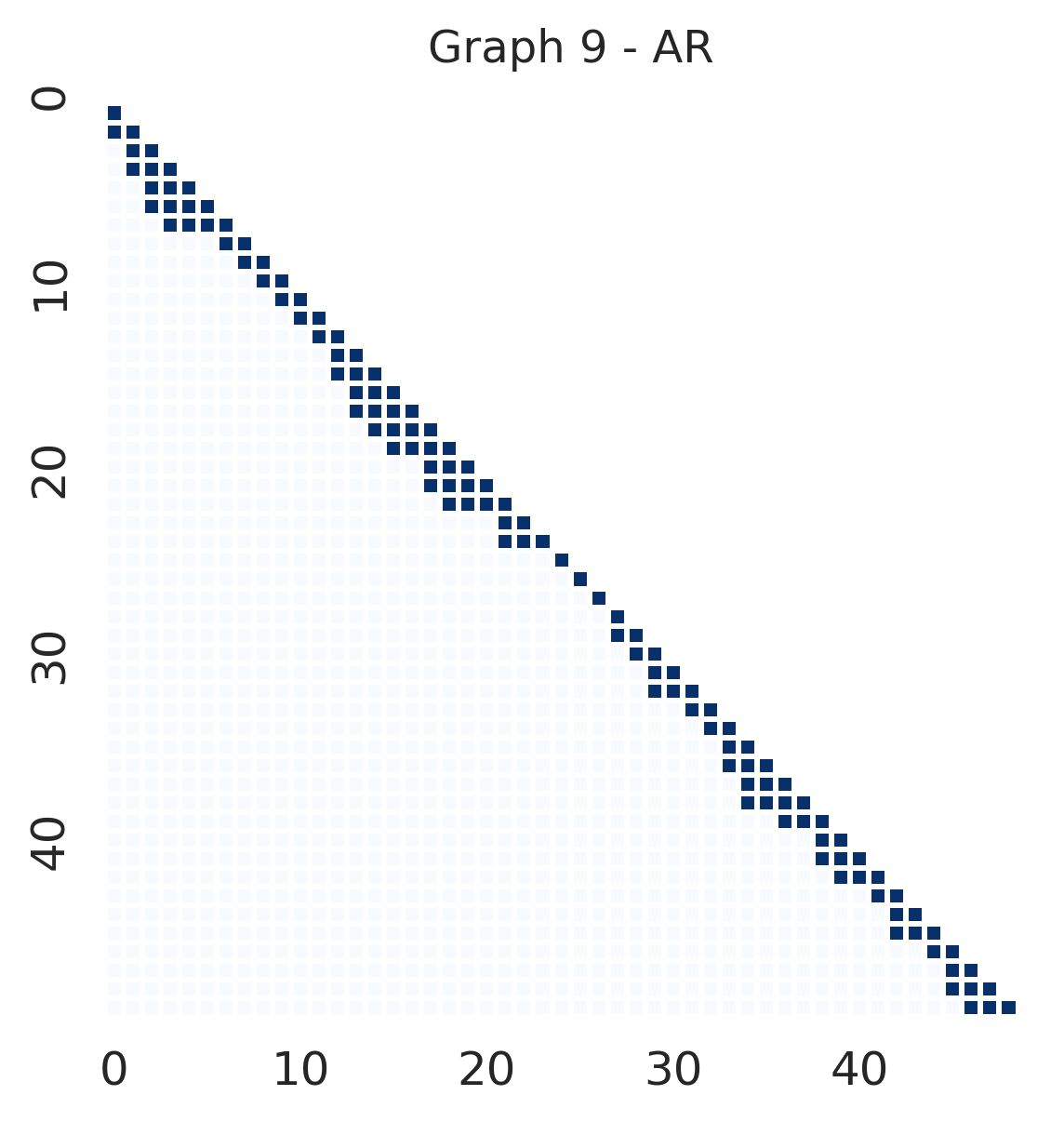}
  \includegraphics[width=0.24\textwidth]{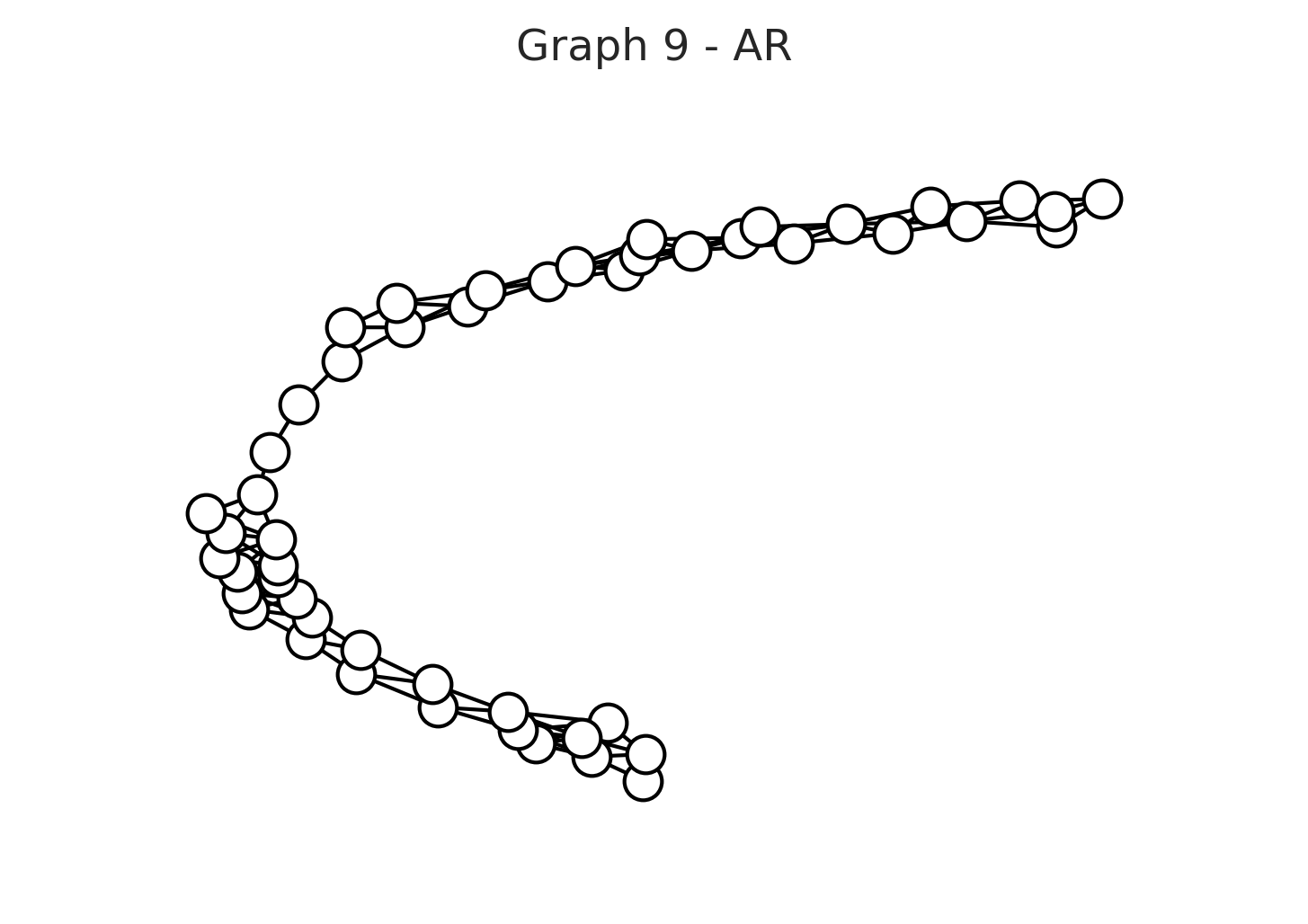}

  \includegraphics[width=0.24\textwidth]{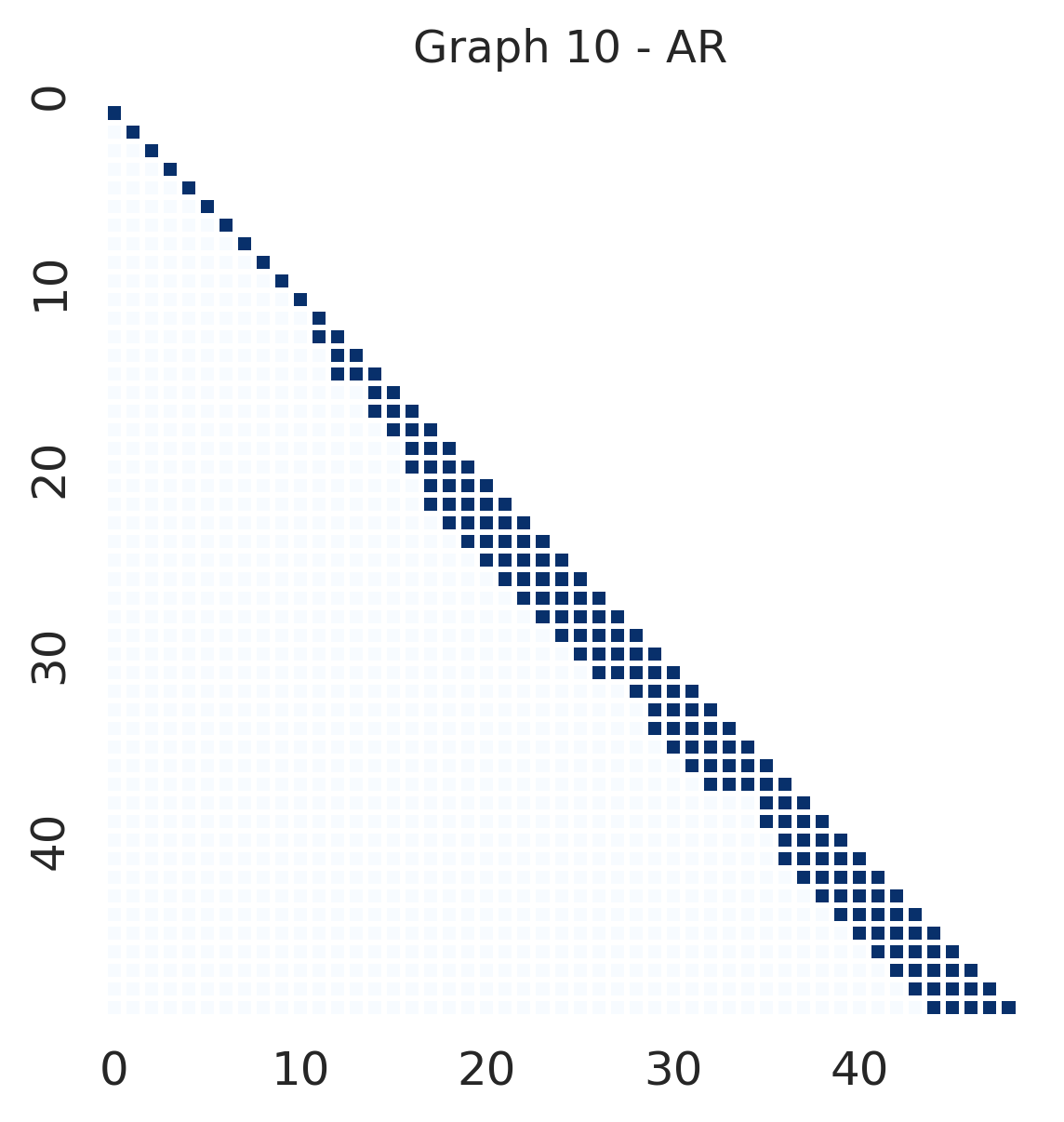}
  \includegraphics[width=0.24\textwidth]{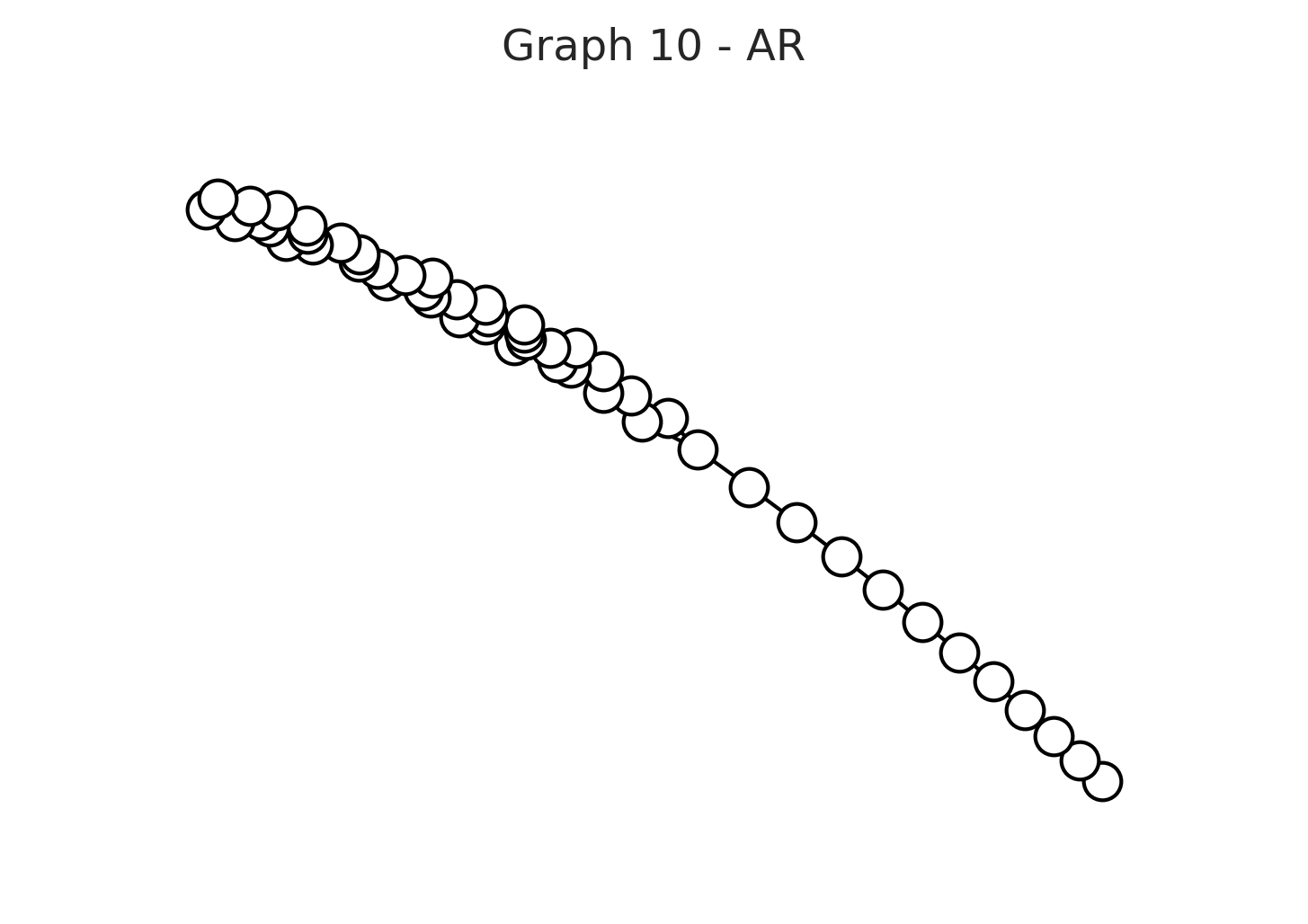}

  \caption{Graph and adjacency matrix plots discussed in Section~\ref{sec:simul-setup-gauss}, generated using the autoregressive structure and $p=50$.}  
  \label{app:fig:ture-graph-auto}
\end{figure}

\begin{figure}[ht]
  \centering
  \includegraphics[width=0.19\textwidth]{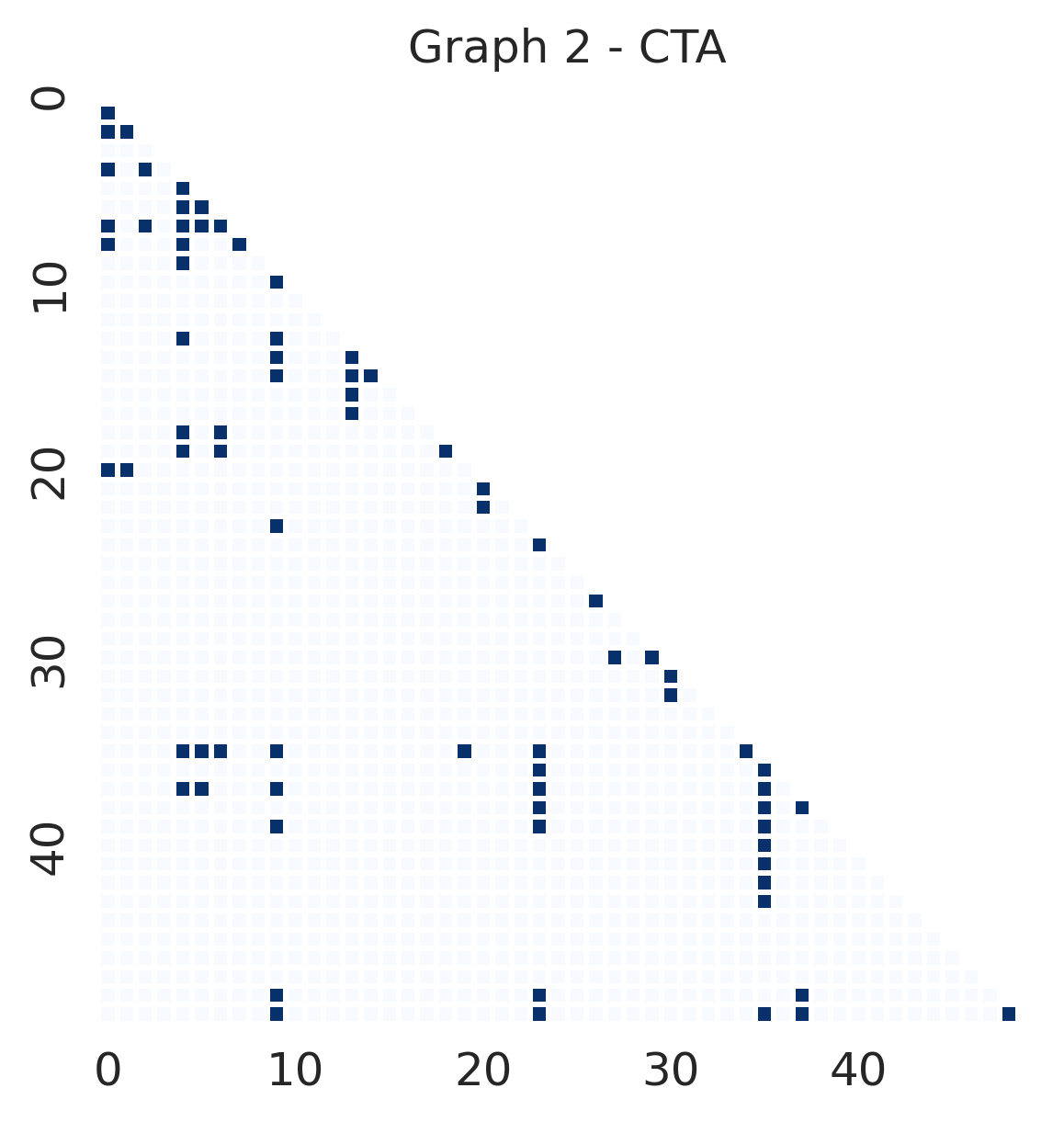}
  \includegraphics[width=0.29\textwidth]{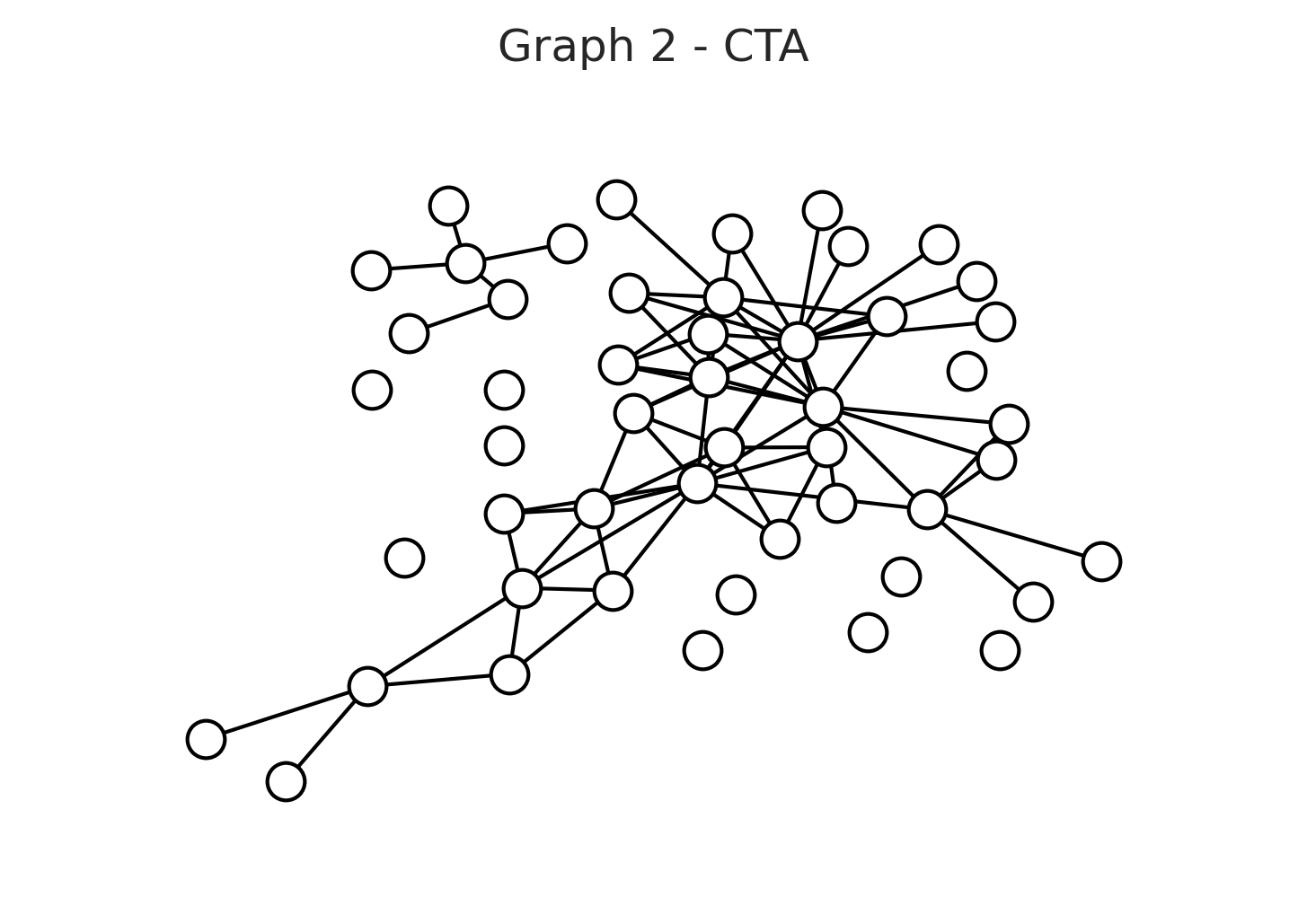}
    \includegraphics[width=0.19\textwidth]{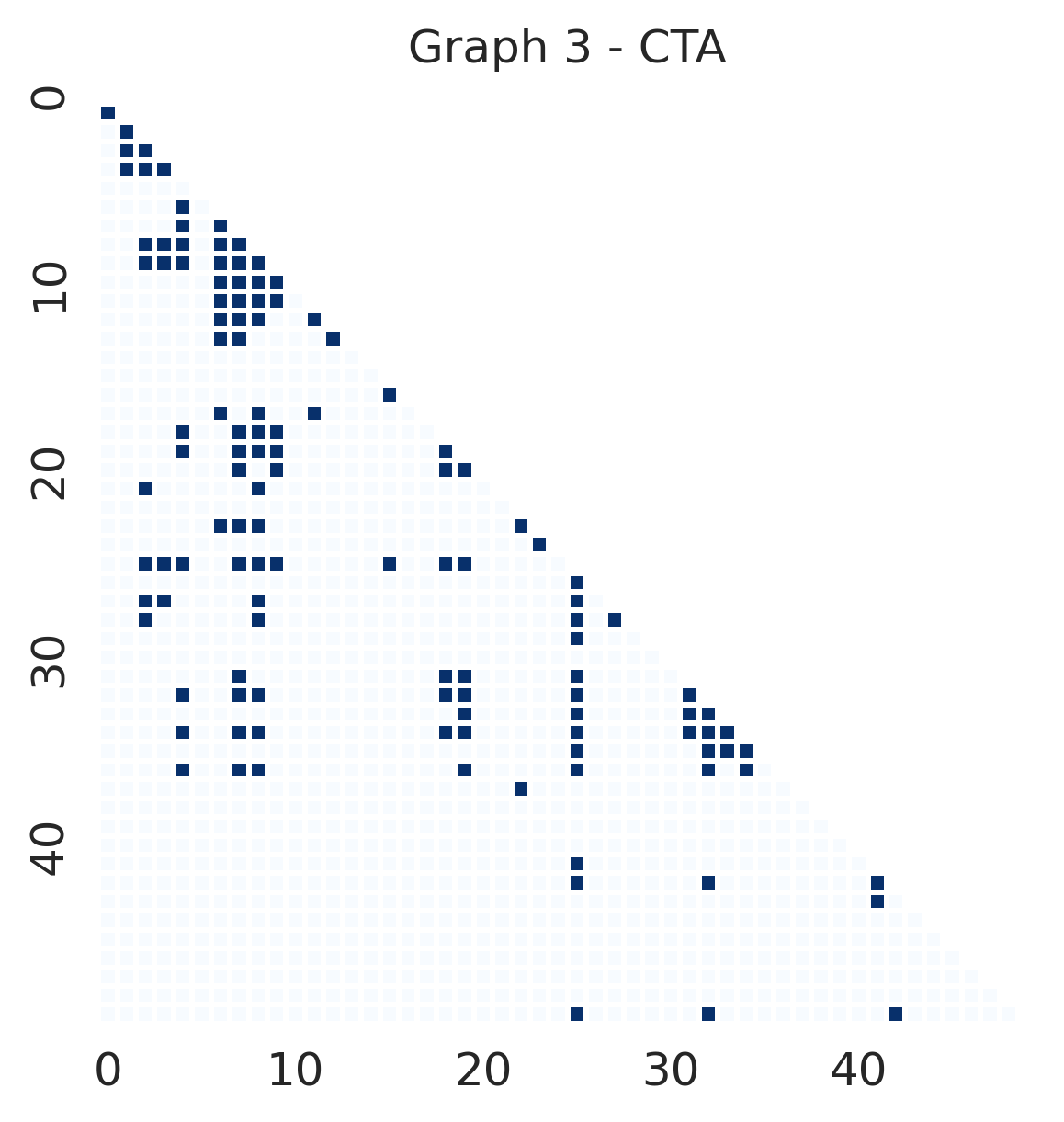}
  \includegraphics[width=0.29\textwidth]{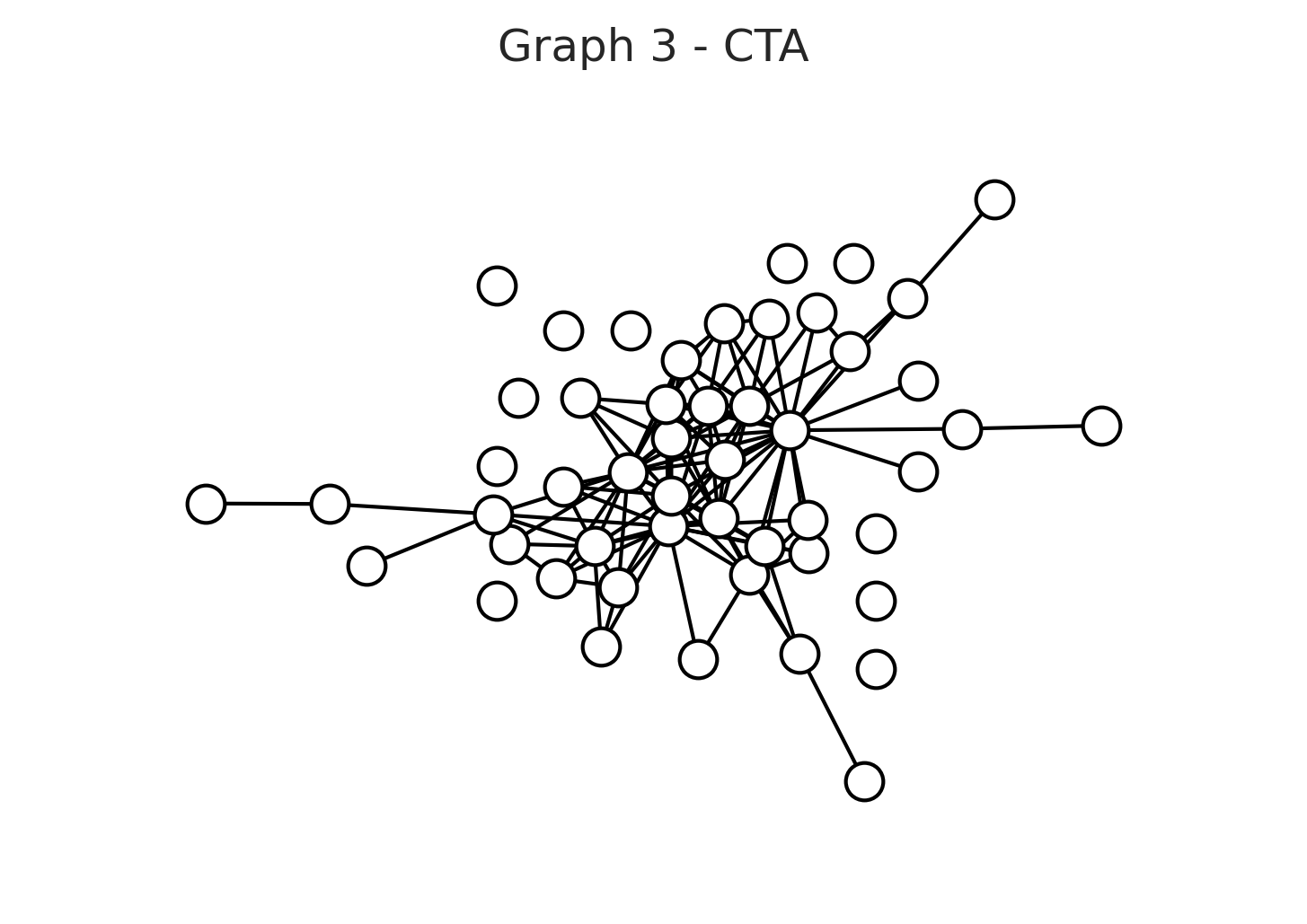}

  \includegraphics[width=0.19\textwidth]{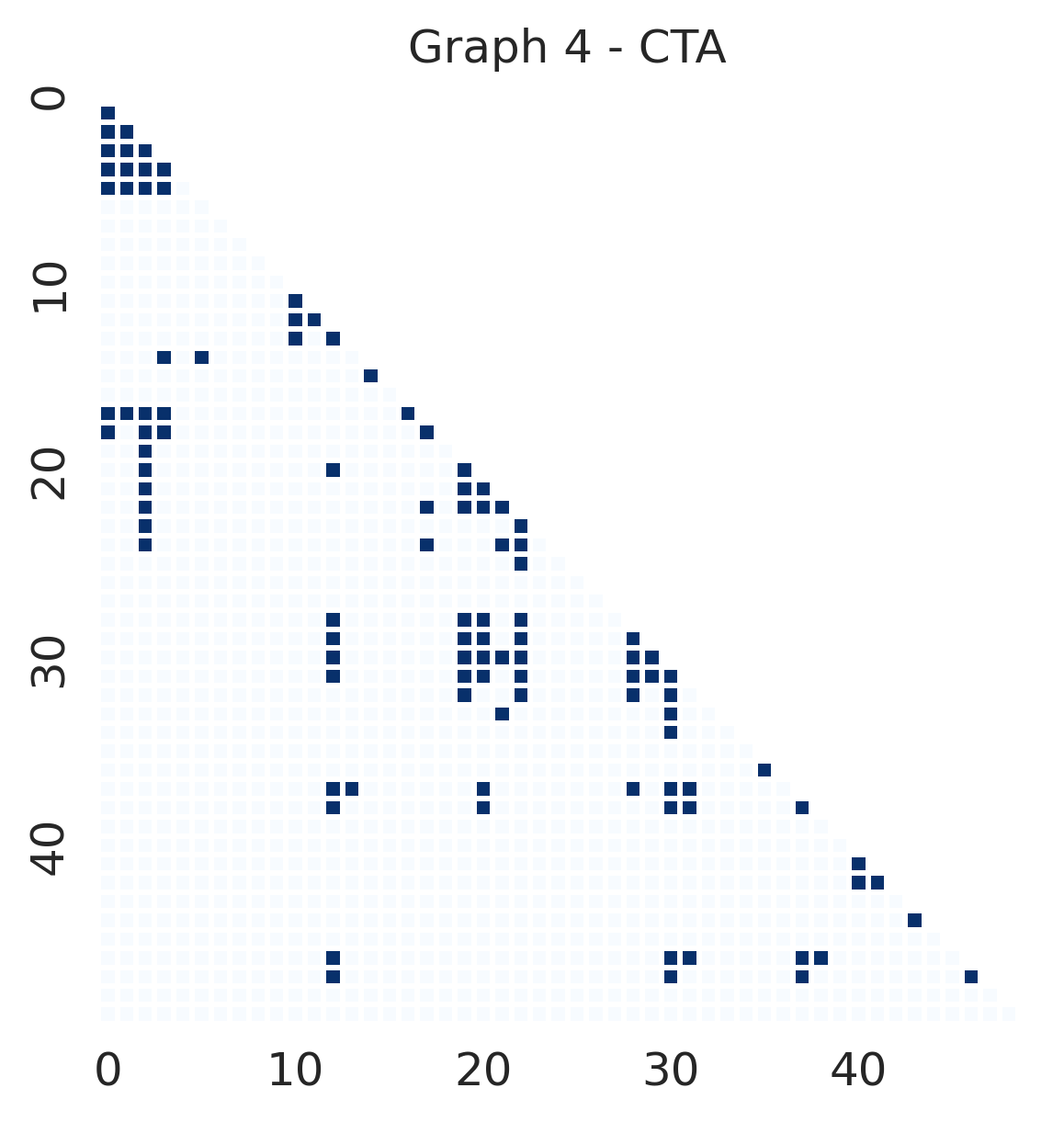}
  \includegraphics[width=0.29\textwidth]{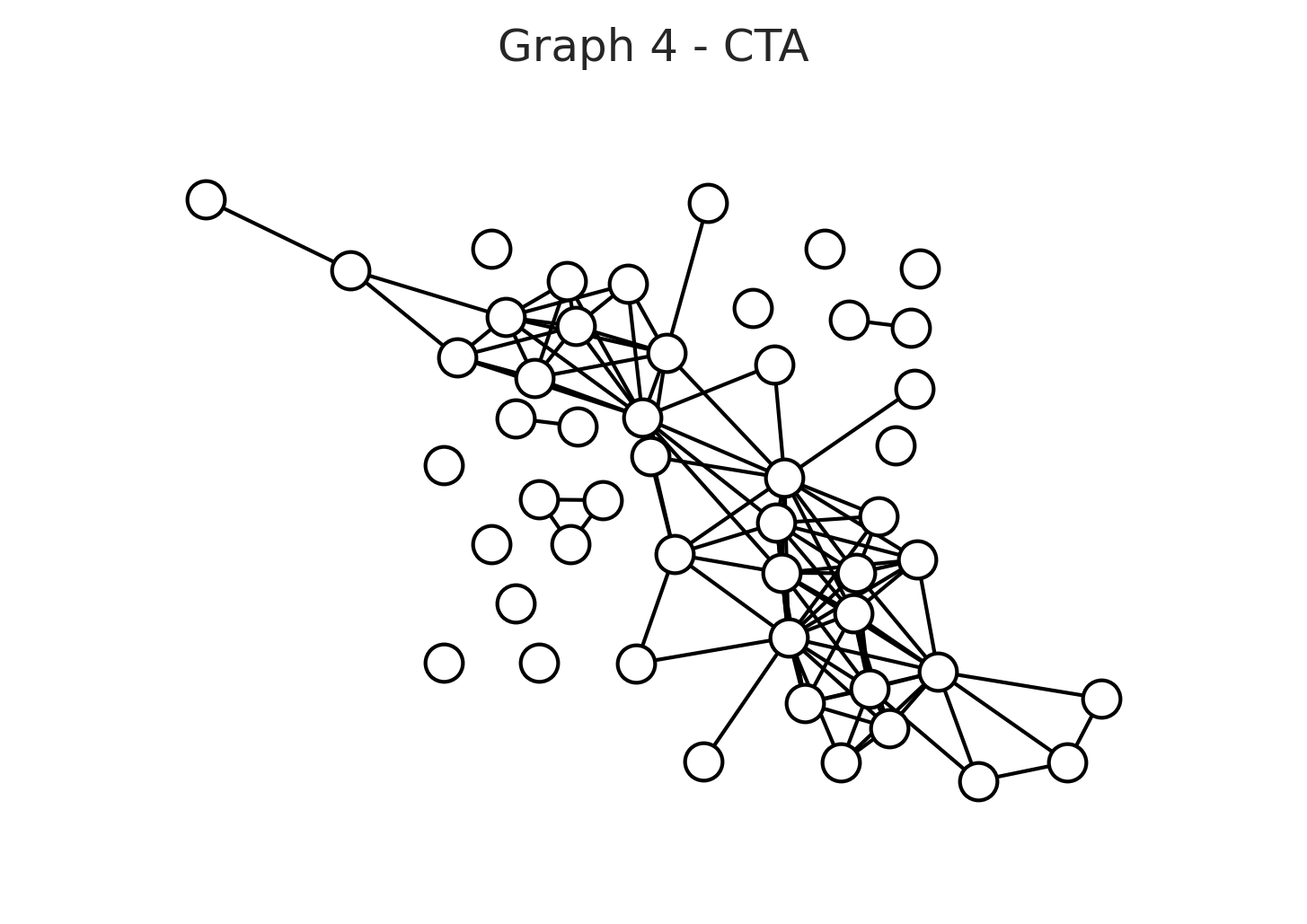}
    \includegraphics[width=0.19\textwidth]{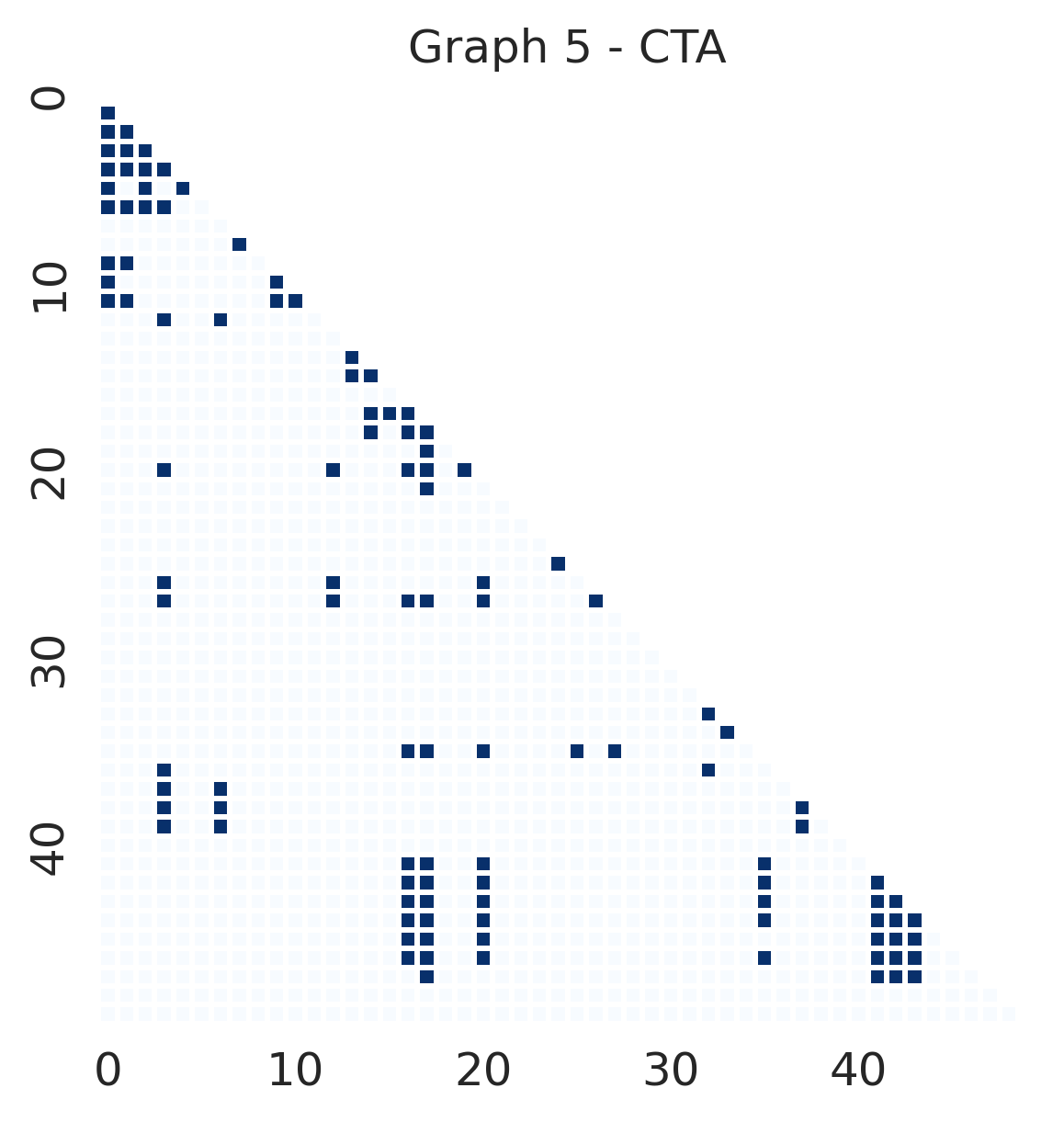}
  \includegraphics[width=0.29\textwidth]{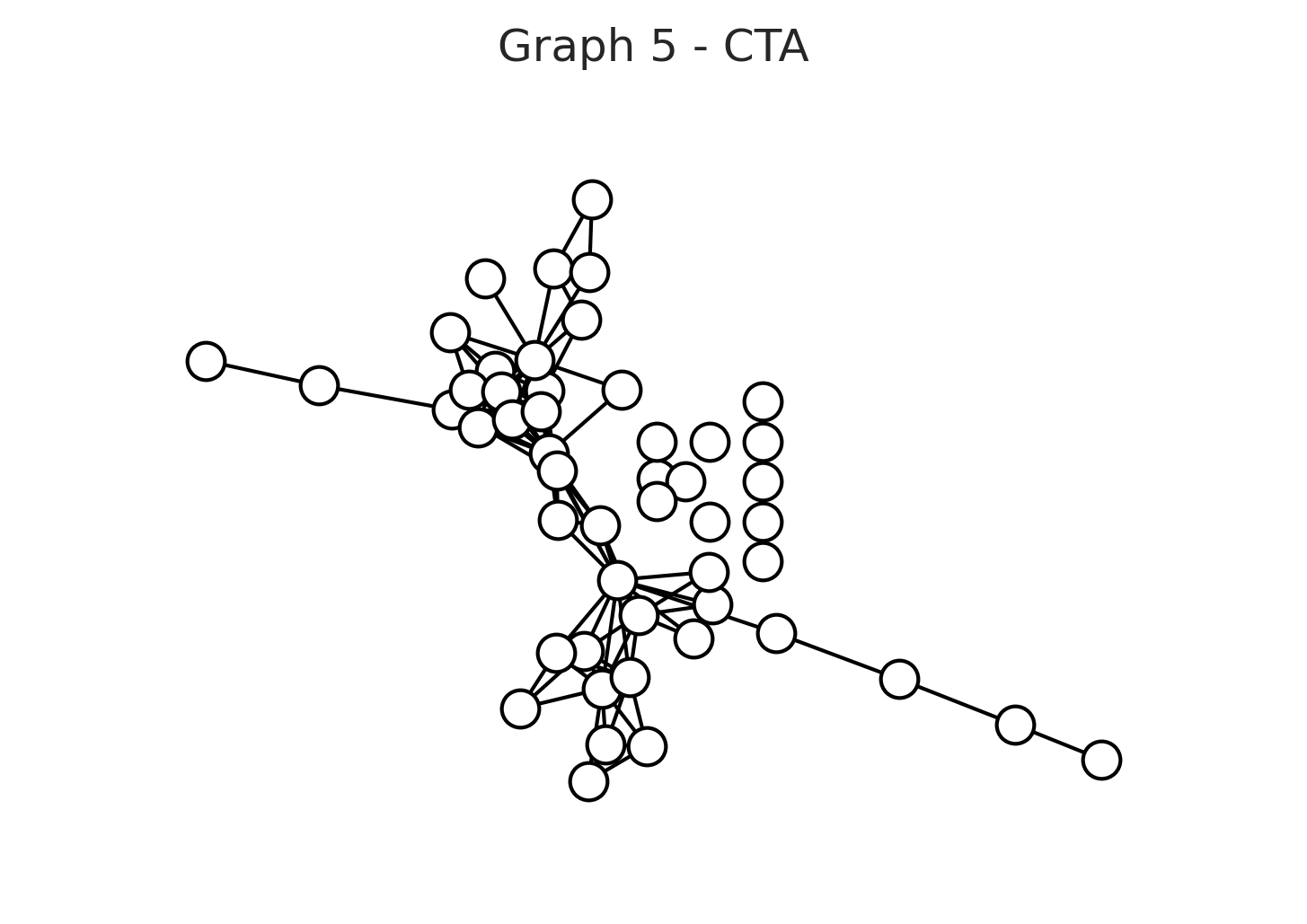}

  \includegraphics[width=0.19\textwidth]{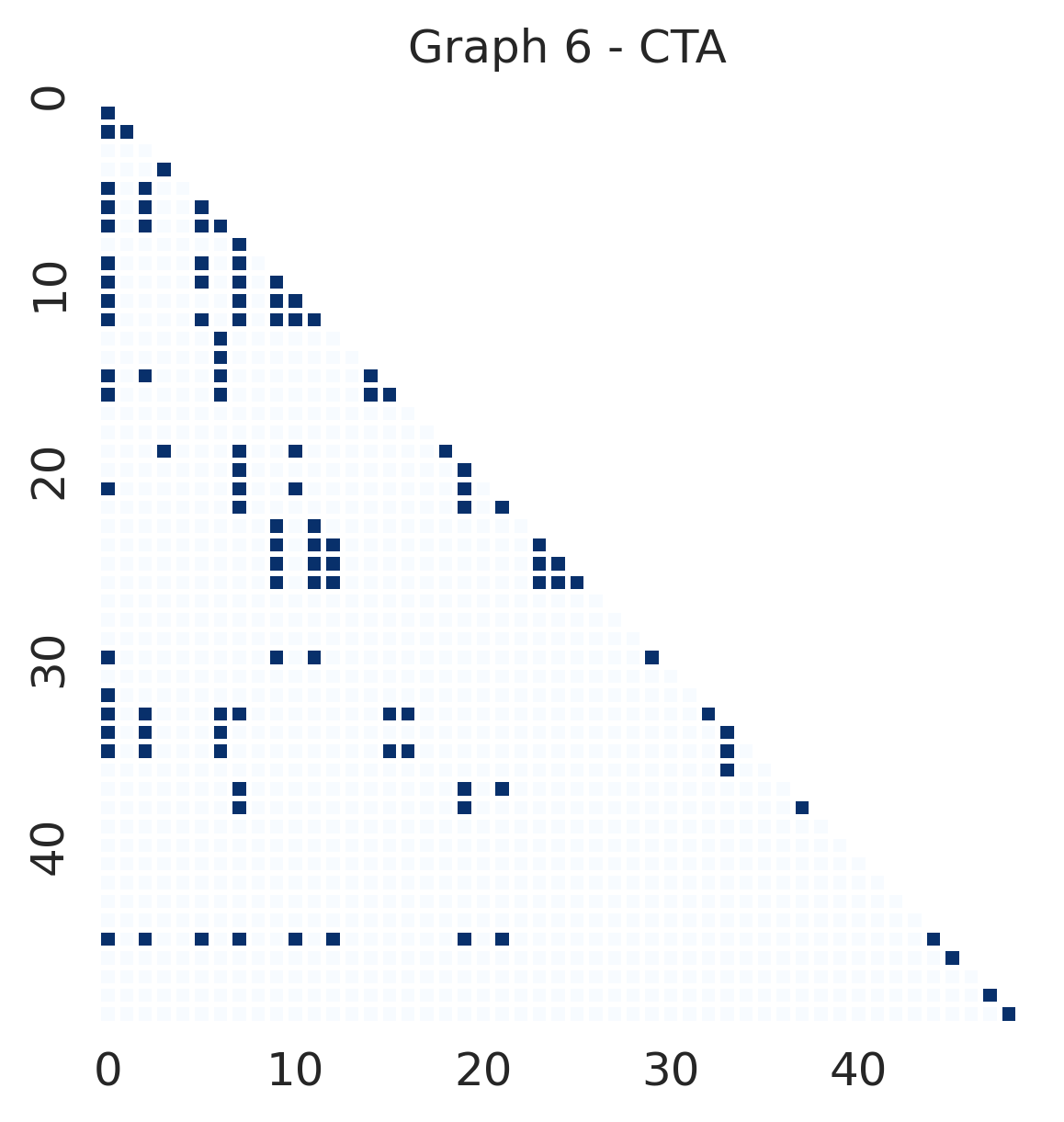}
  \includegraphics[width=0.29\textwidth]{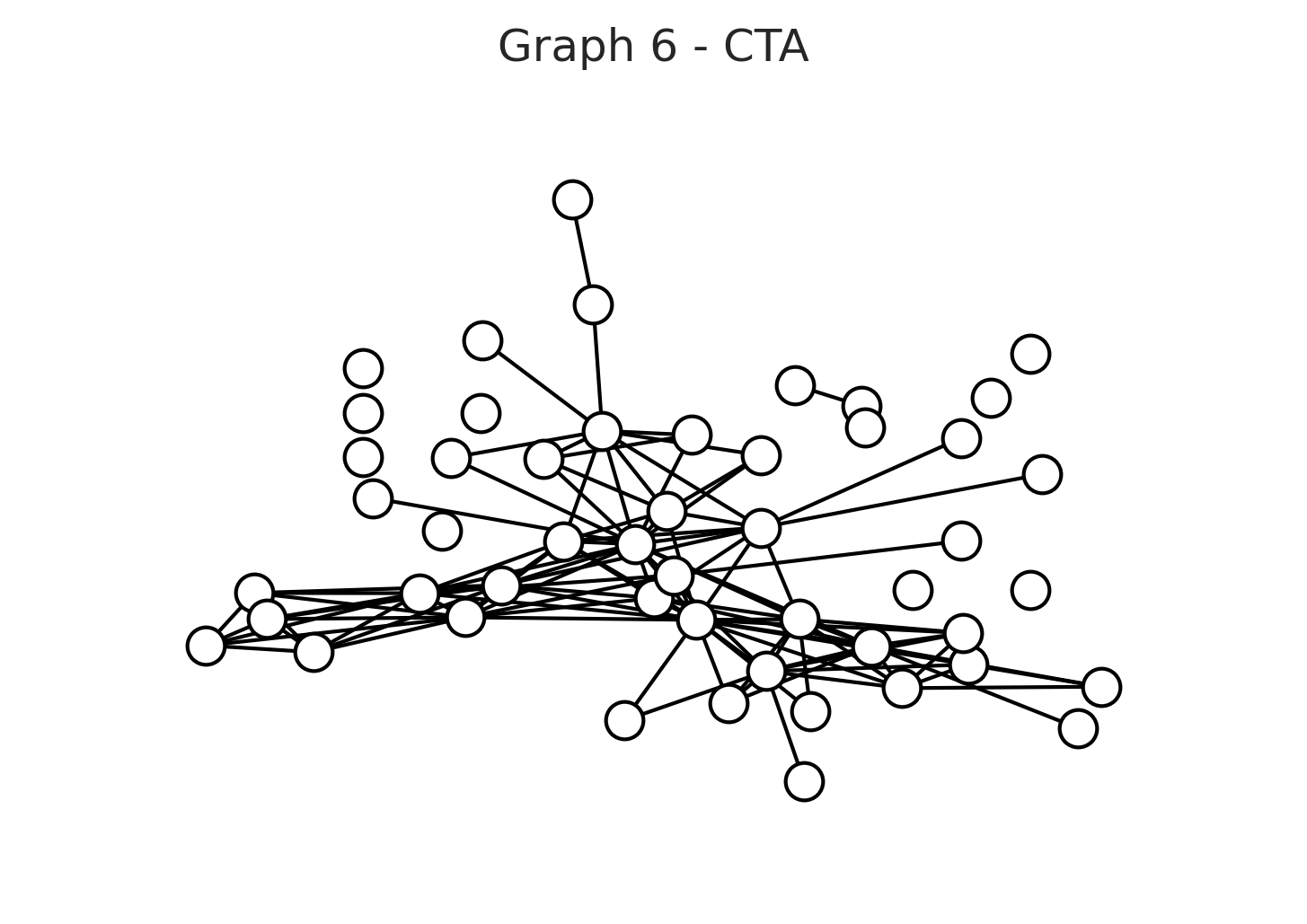}
    \includegraphics[width=0.19\textwidth]{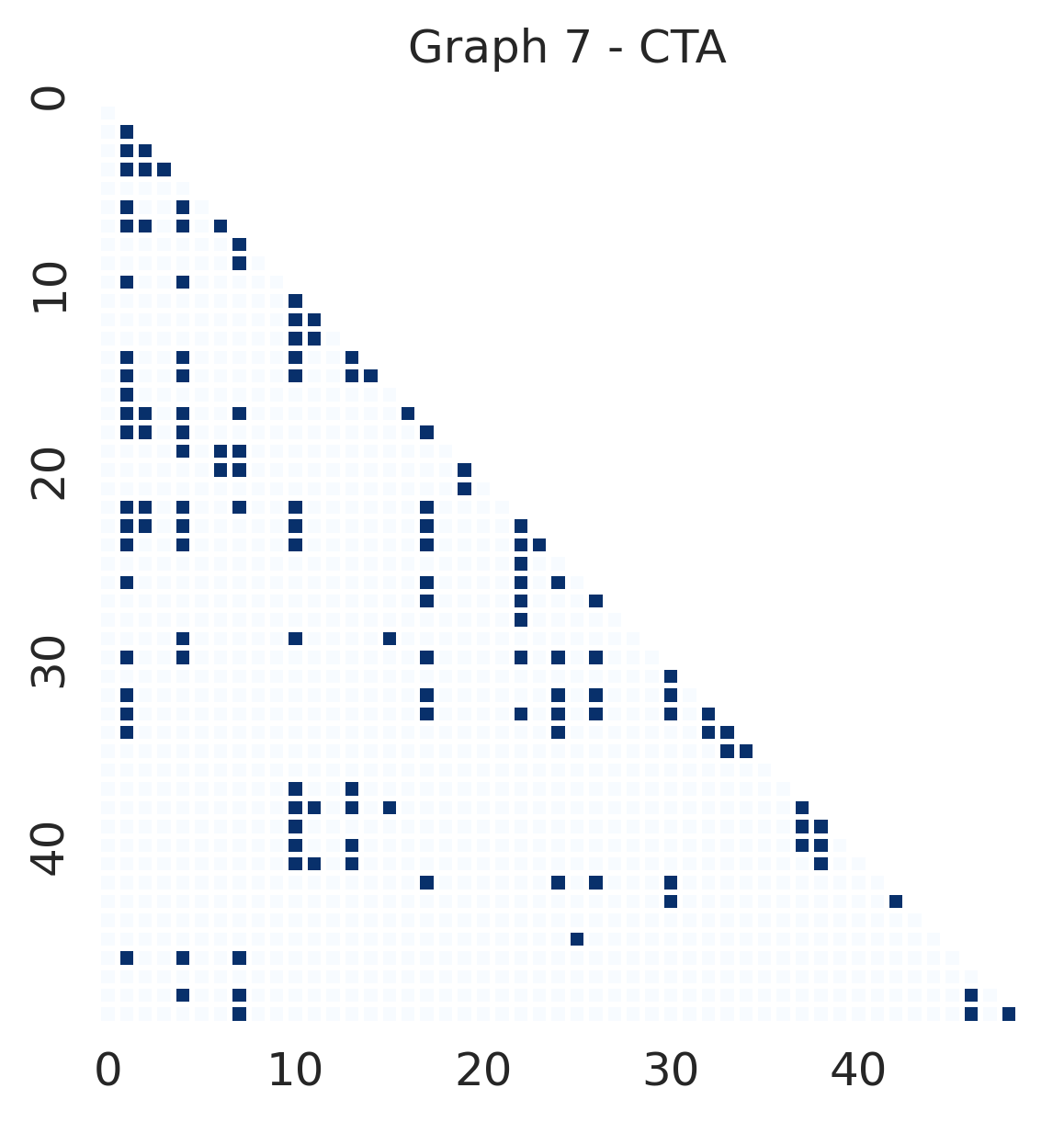}
  \includegraphics[width=0.29\textwidth]{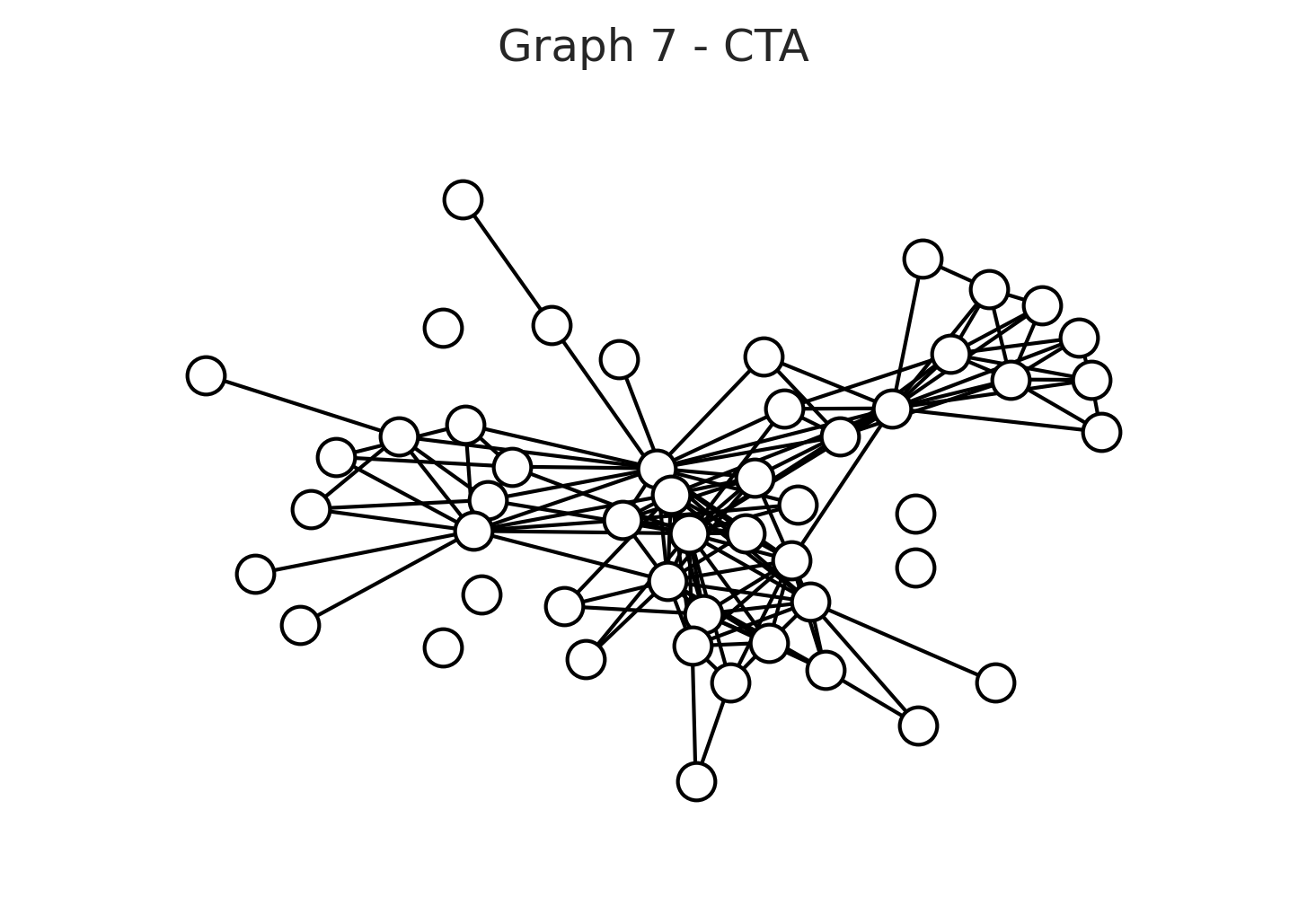}

  \includegraphics[width=0.19\textwidth]{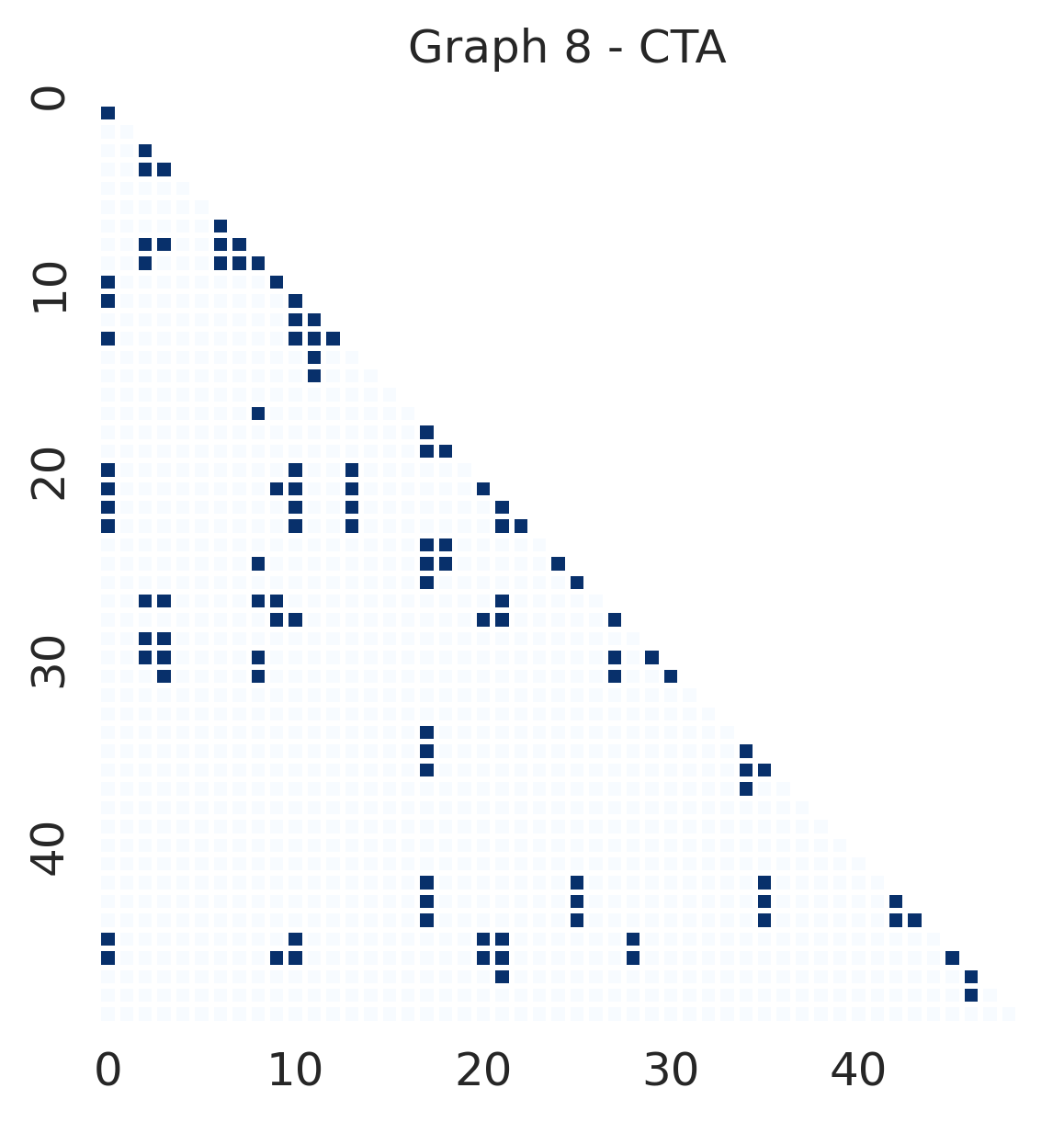}
  \includegraphics[width=0.29\textwidth]{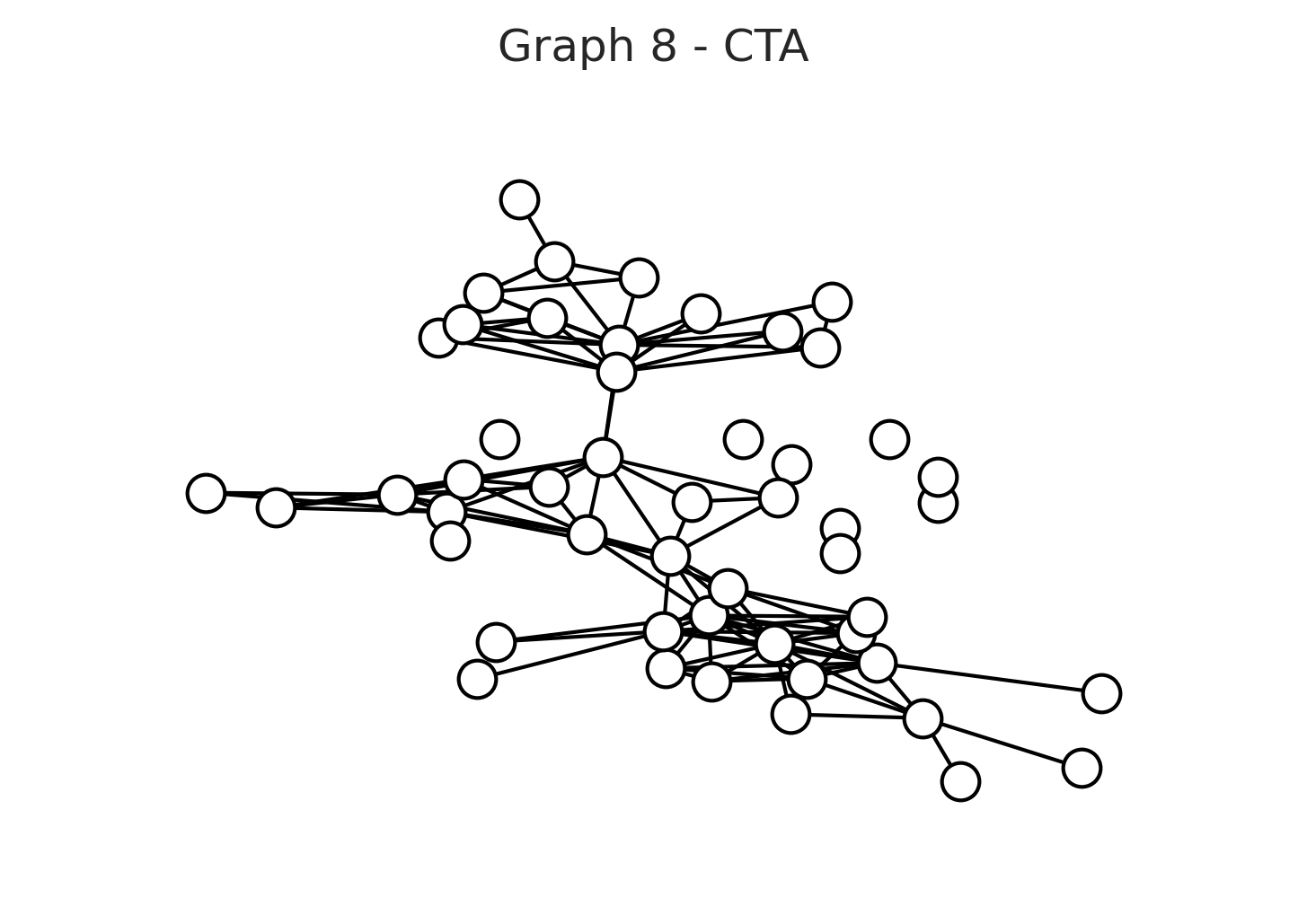}
  \includegraphics[width=0.19\textwidth]{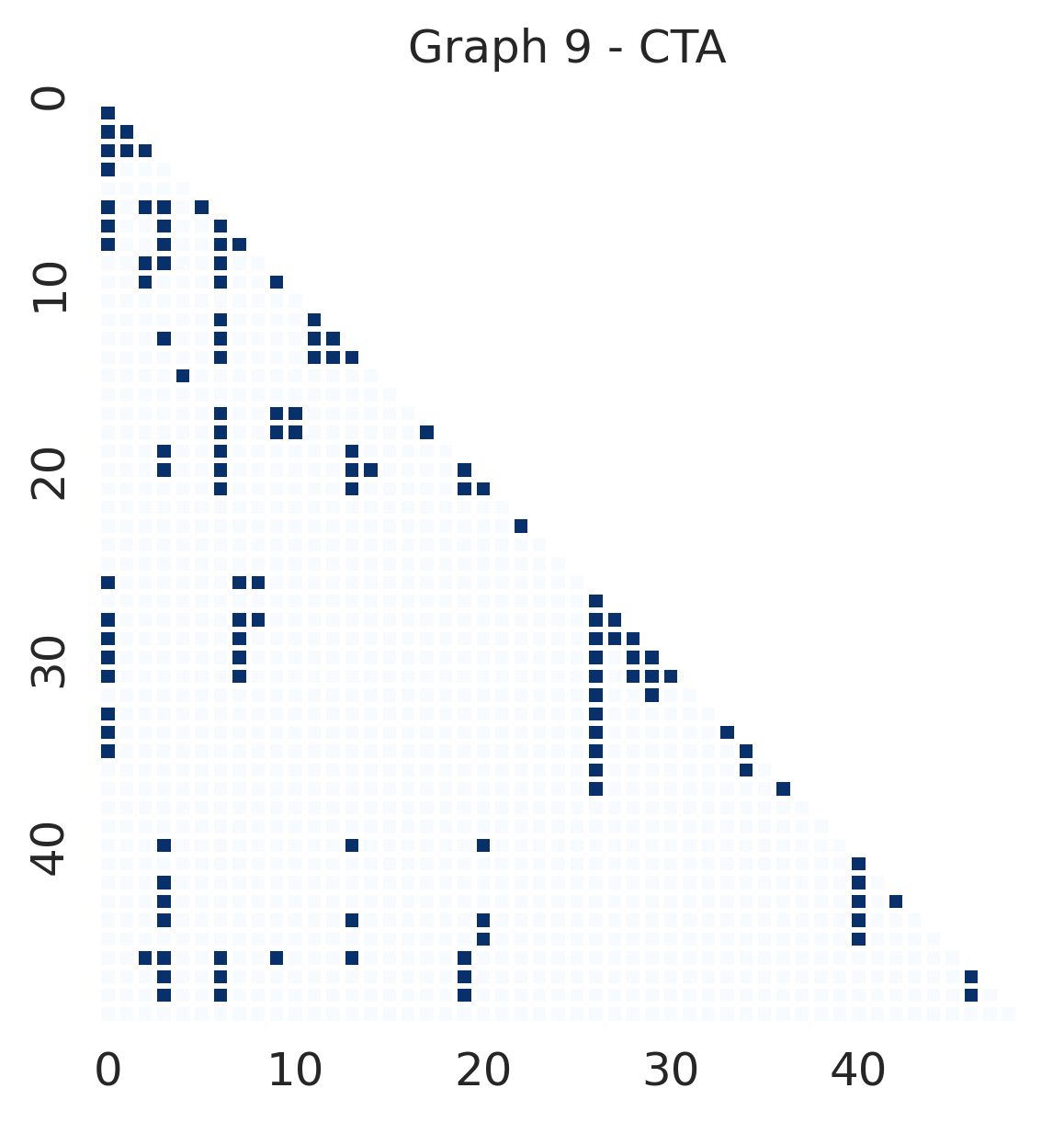}
  \includegraphics[width=0.29\textwidth]{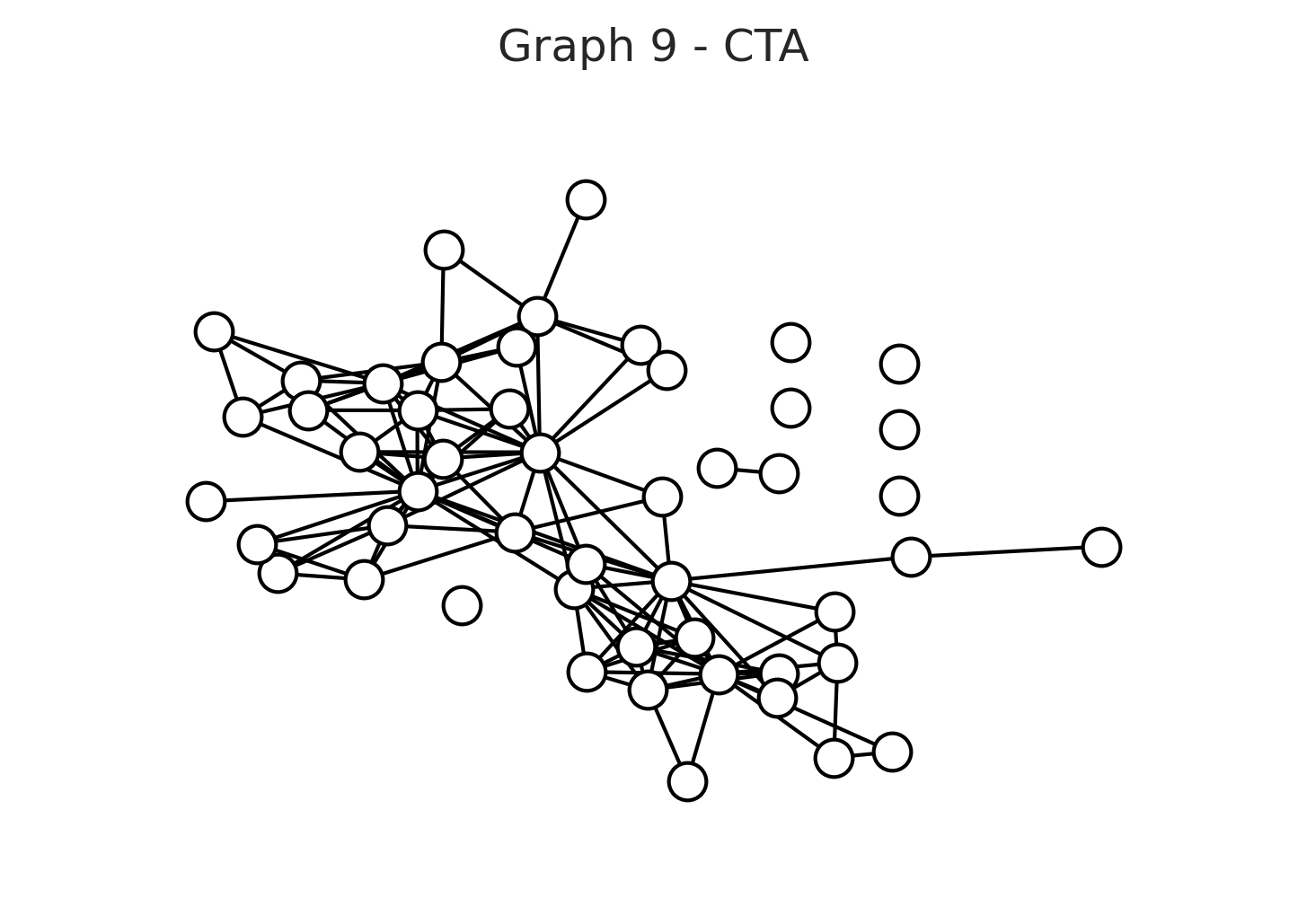}

  \includegraphics[width=0.19\textwidth]{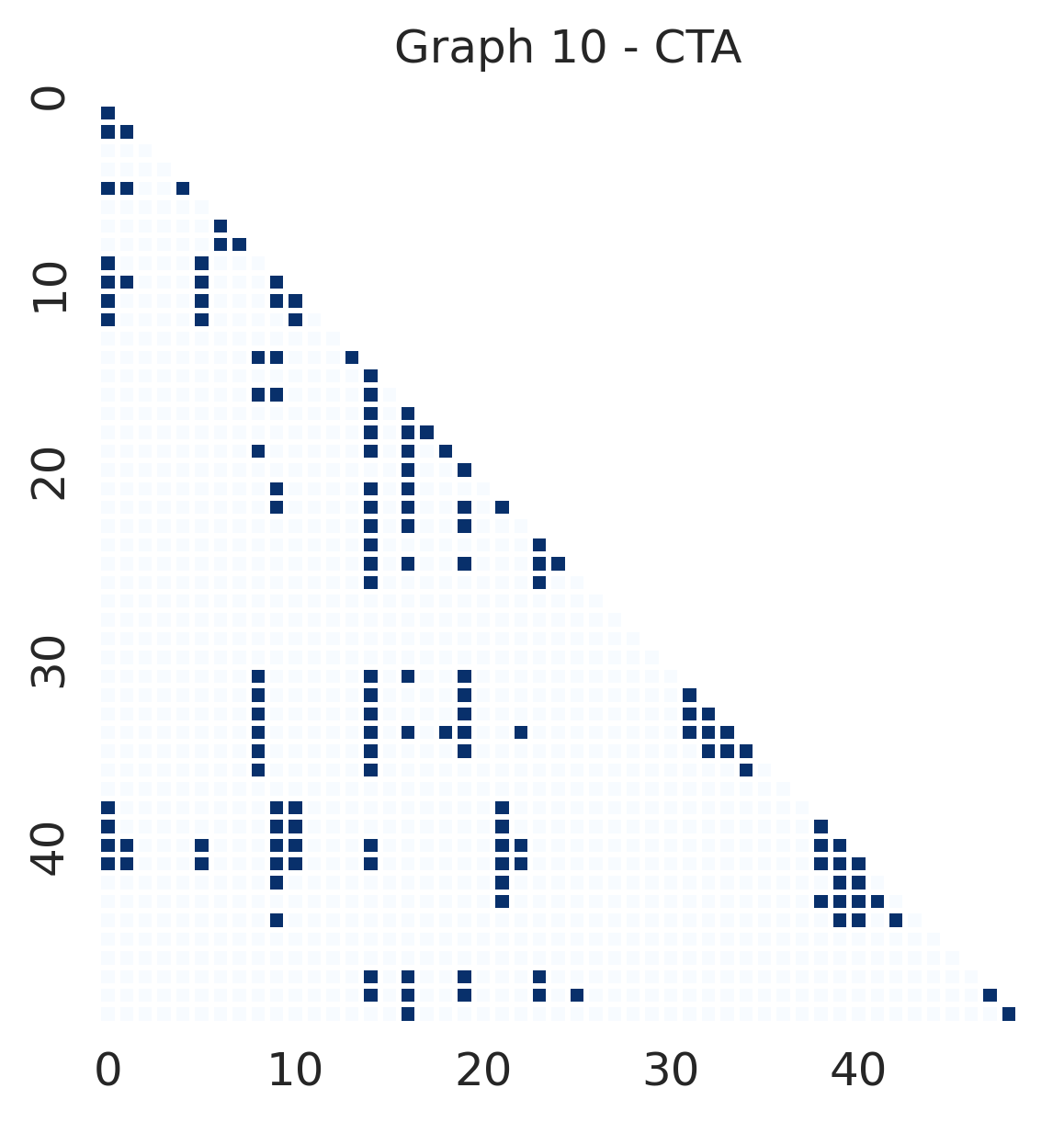}
  \includegraphics[width=0.29\textwidth]{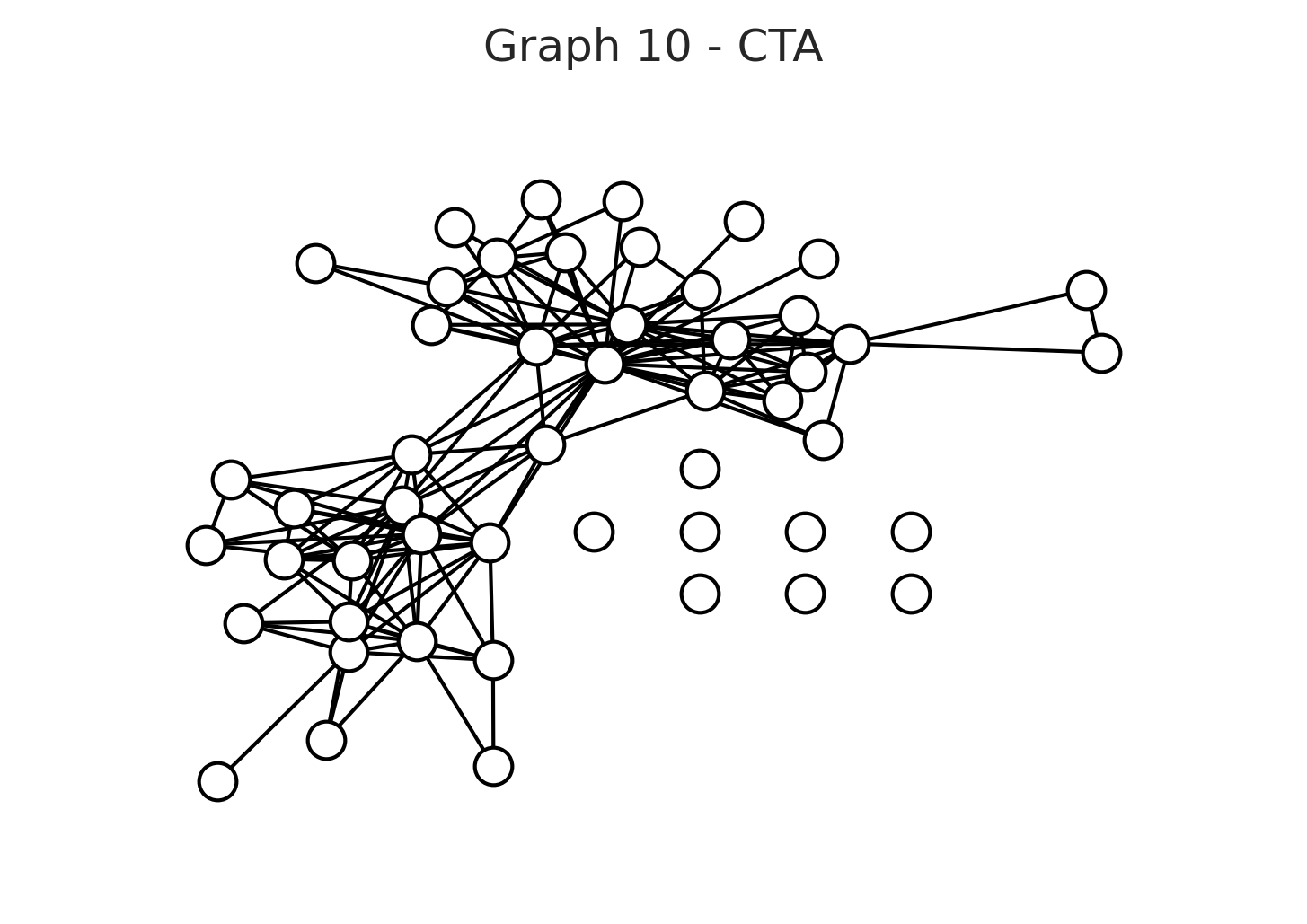}
  
  \caption{Graph and adjacency matrix plots discussed in Section~\ref{sec:simul-setup-gauss}, generated using the Christmas tree algorithm~\citep{olsson2022sequential} and $p=50$.}
  \label{app:fig:ture-graph-cta}
\end{figure}

\begin{figure}[ht]
  \centering
  \includegraphics[width=0.19\textwidth]{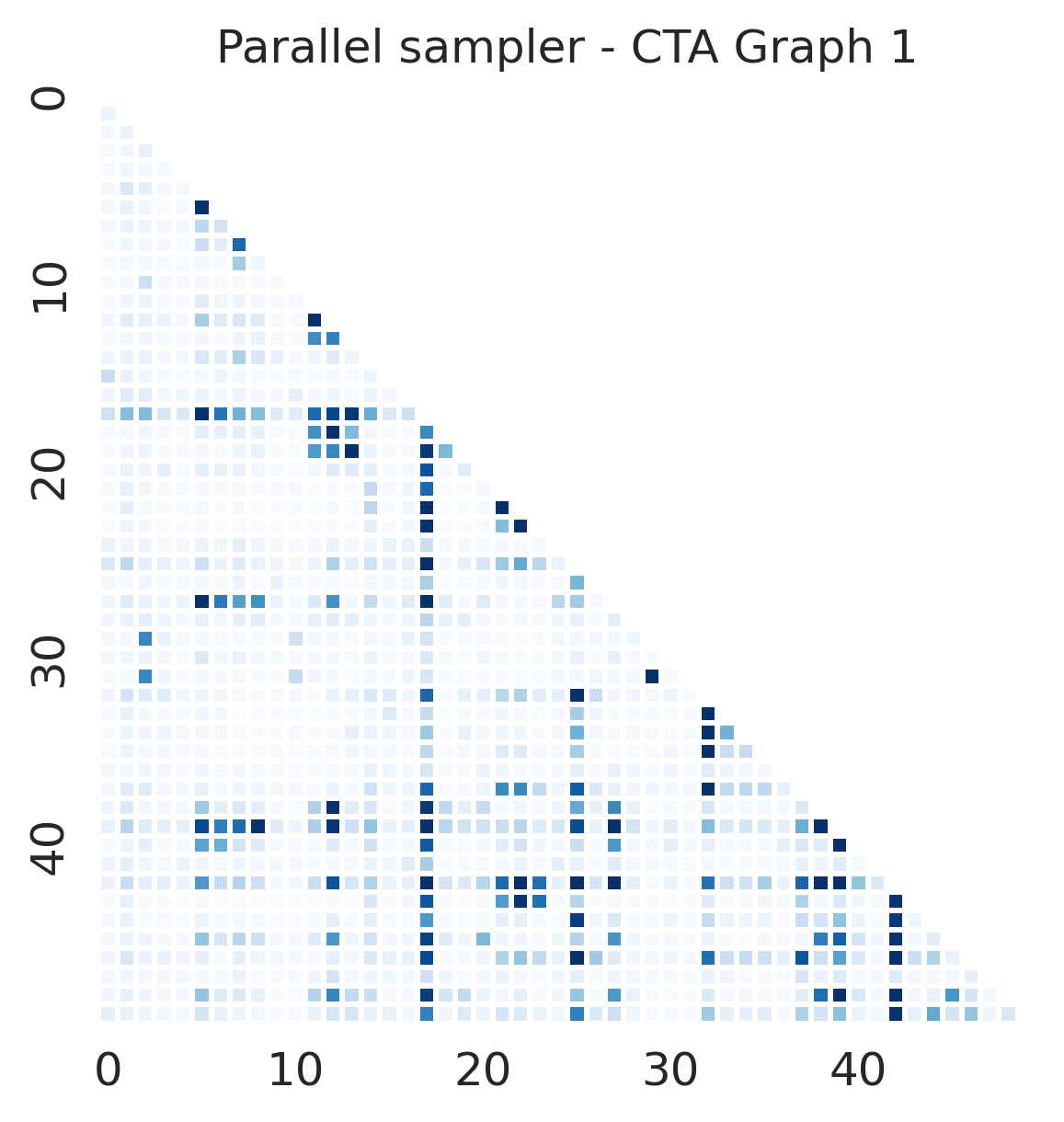}
  \includegraphics[width=0.19\textwidth]{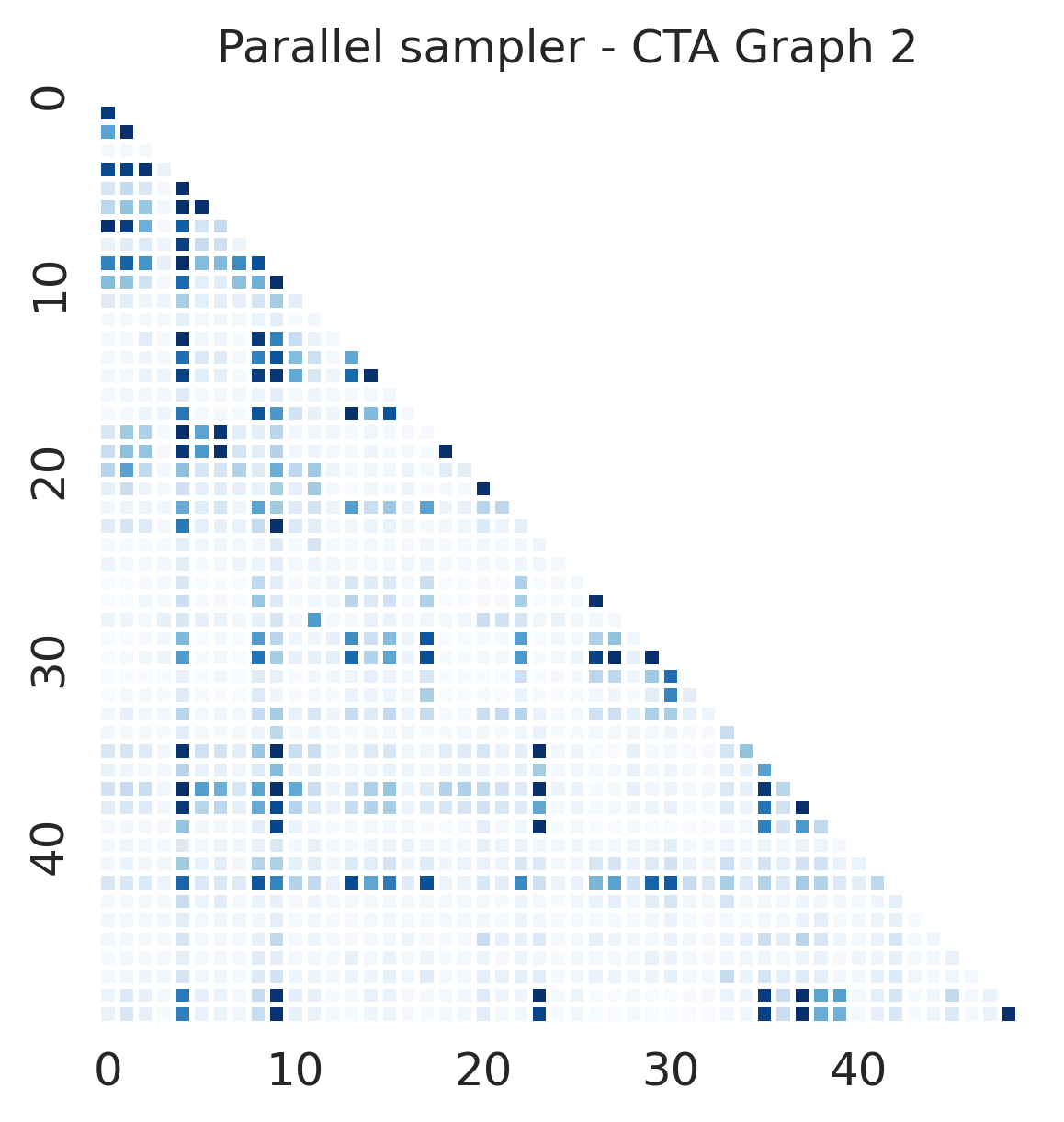}
  \includegraphics[width=0.19\textwidth]{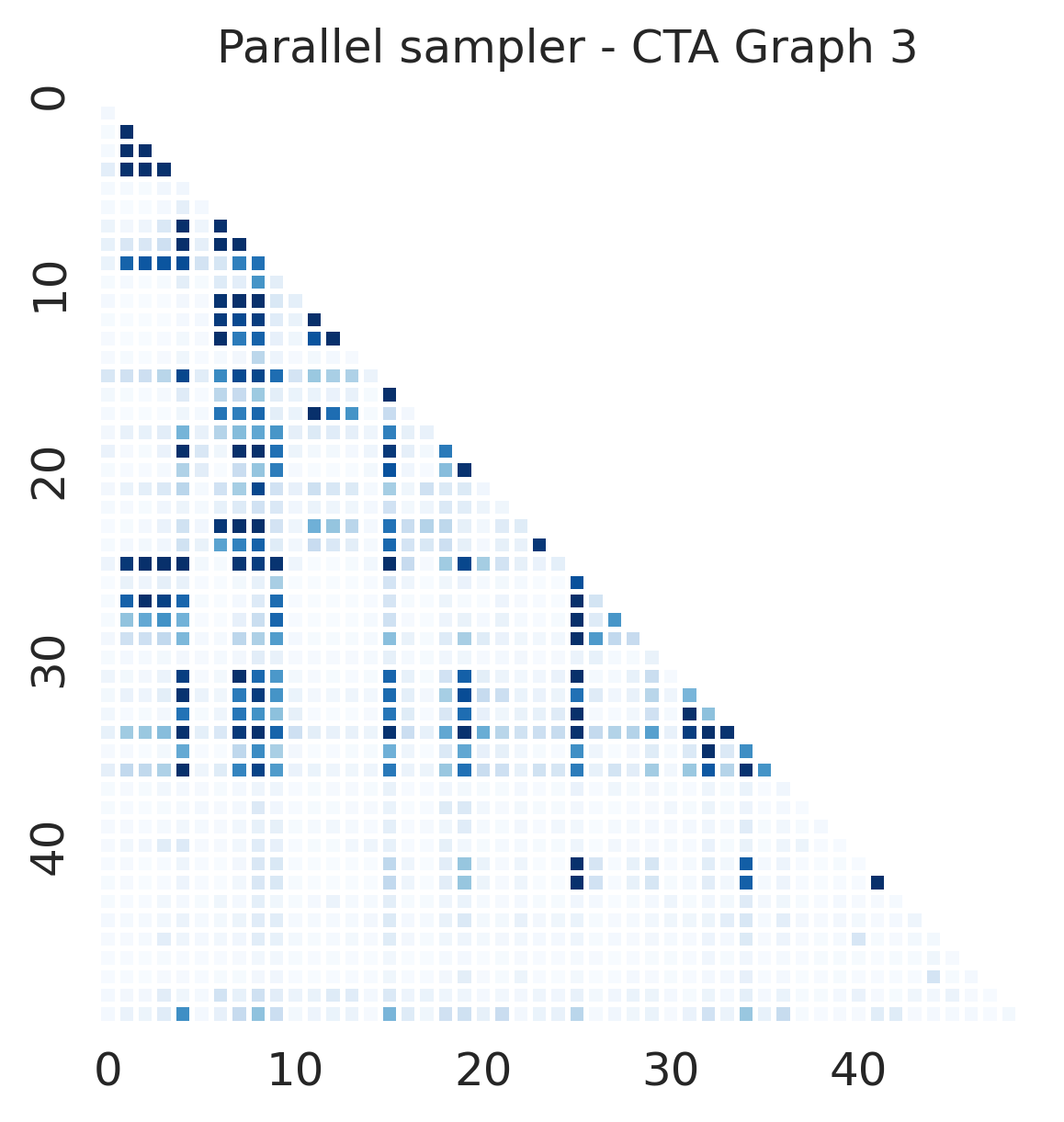}
  \includegraphics[width=0.19\textwidth]{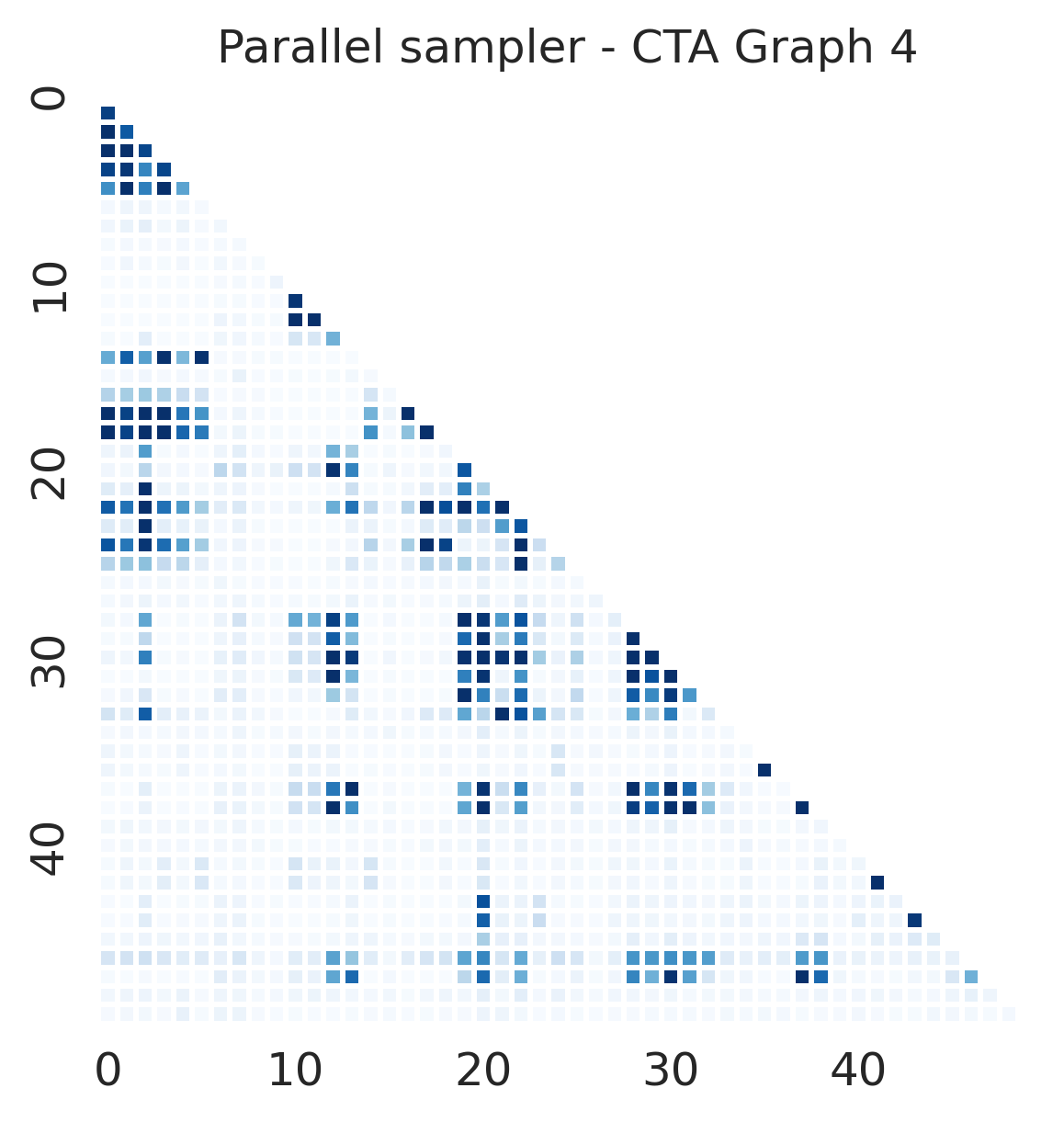}
  \includegraphics[width=0.19\textwidth]{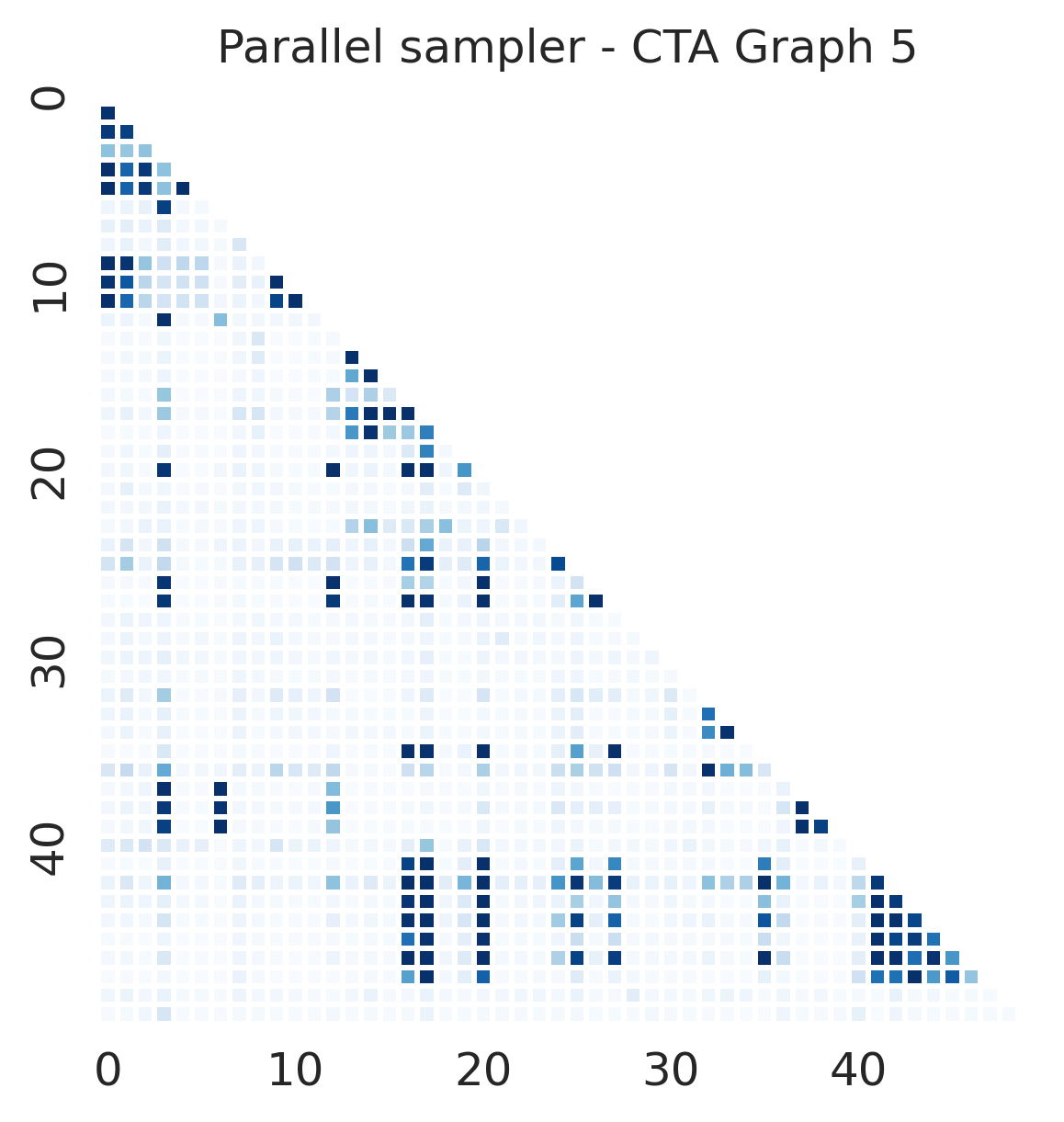}

  \includegraphics[width=0.19\textwidth]{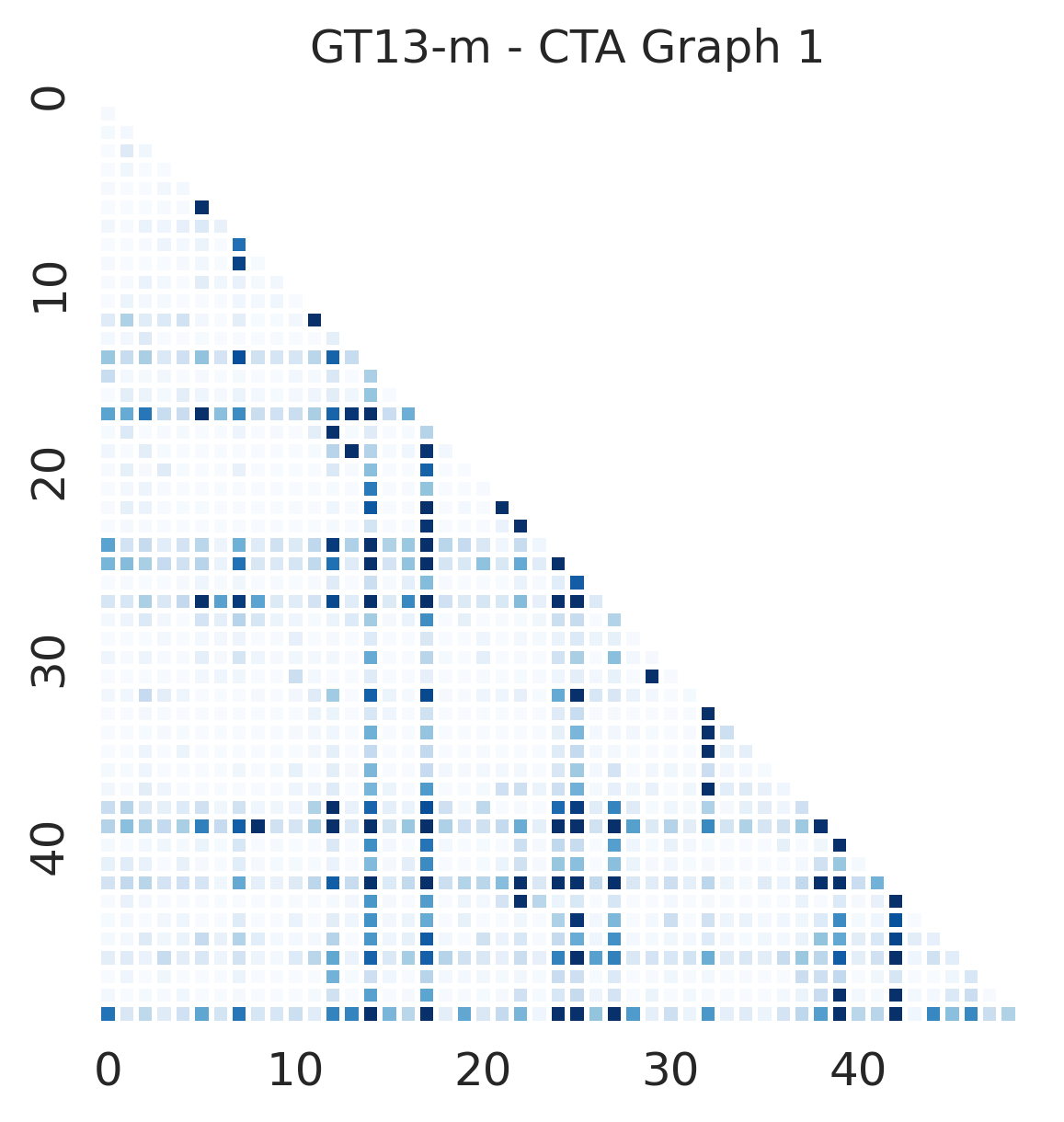}
  \includegraphics[width=0.19\textwidth]{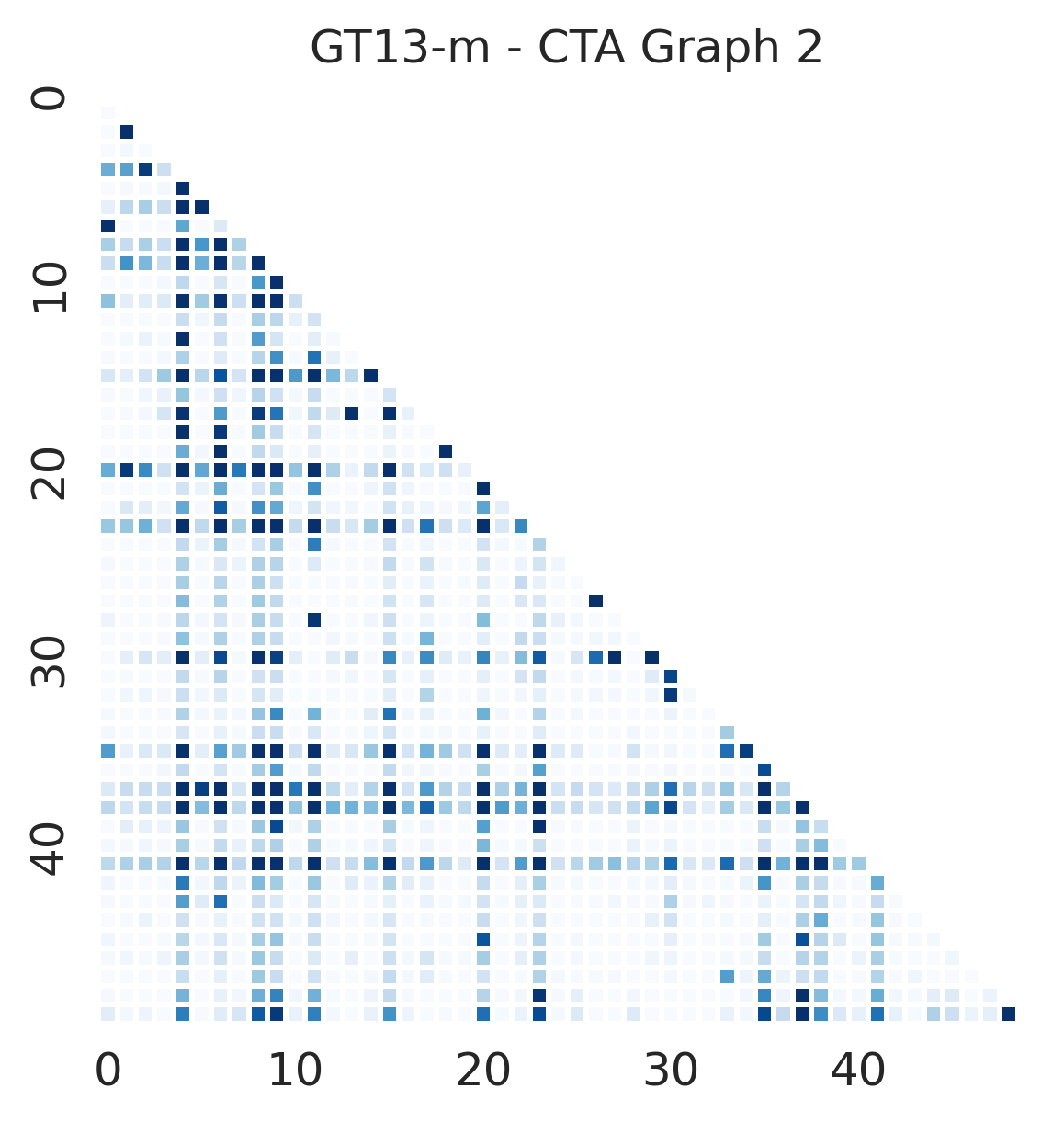}
  \includegraphics[width=0.19\textwidth]{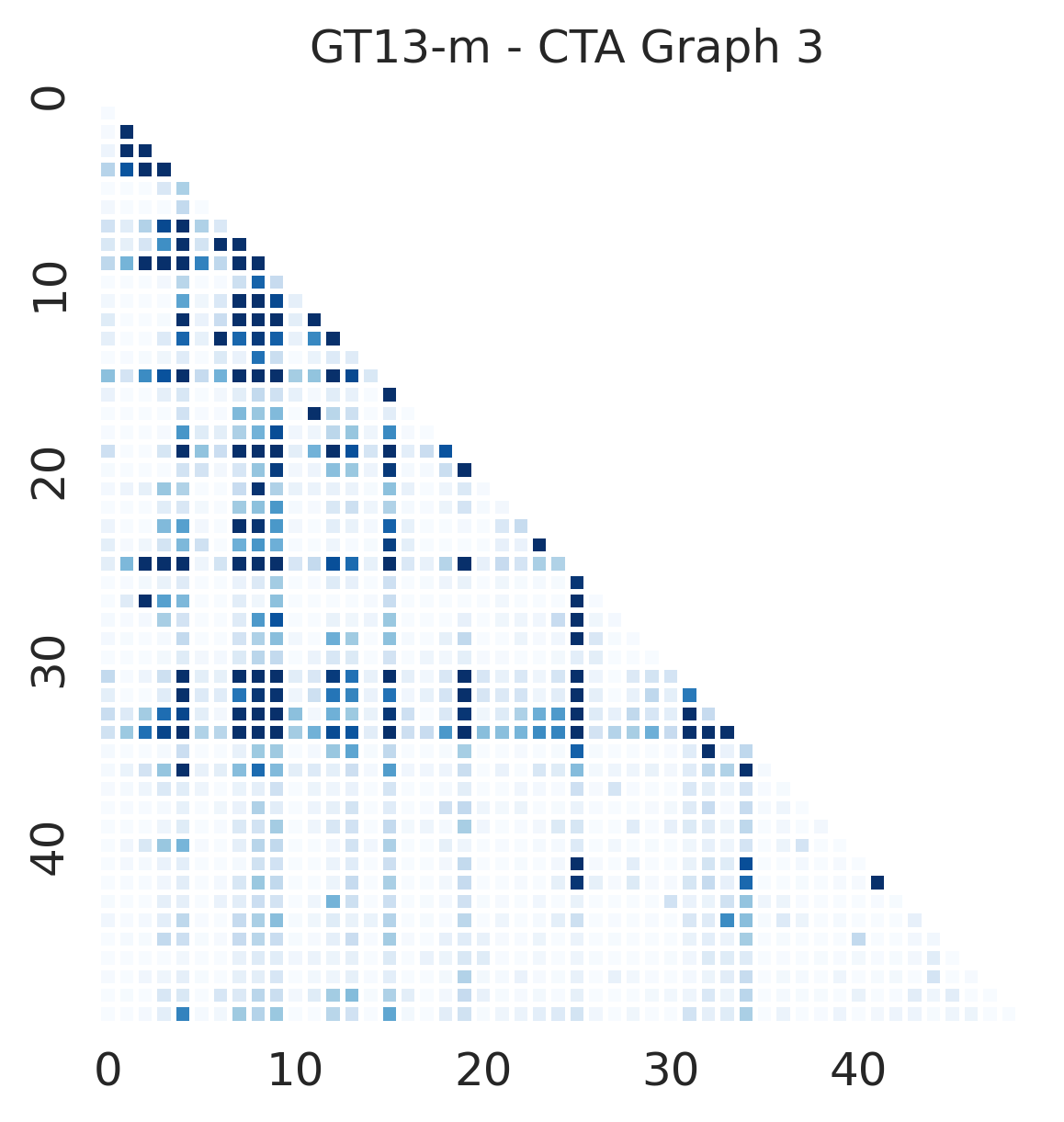}
  \includegraphics[width=0.19\textwidth]{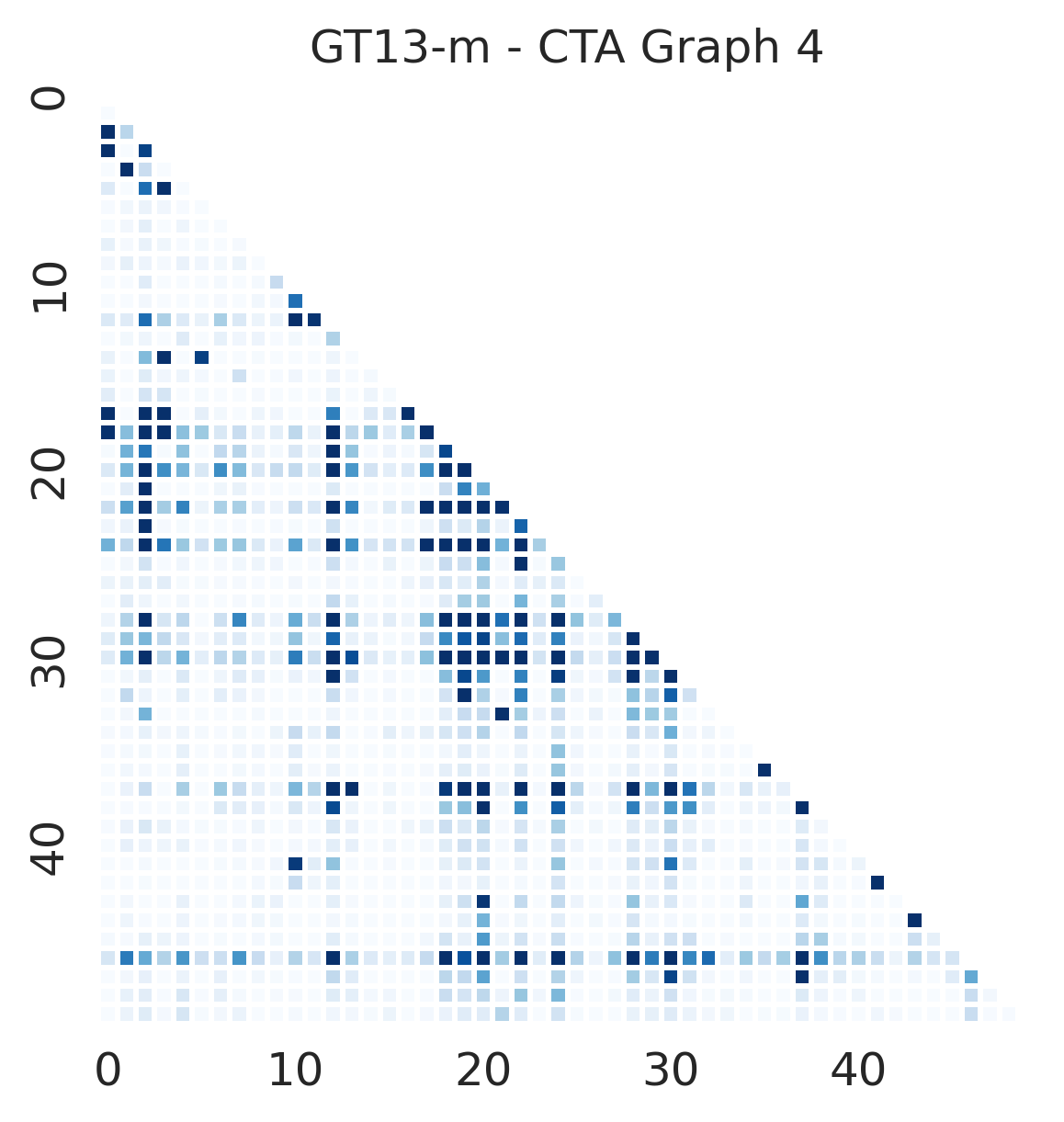}
  \includegraphics[width=0.19\textwidth]{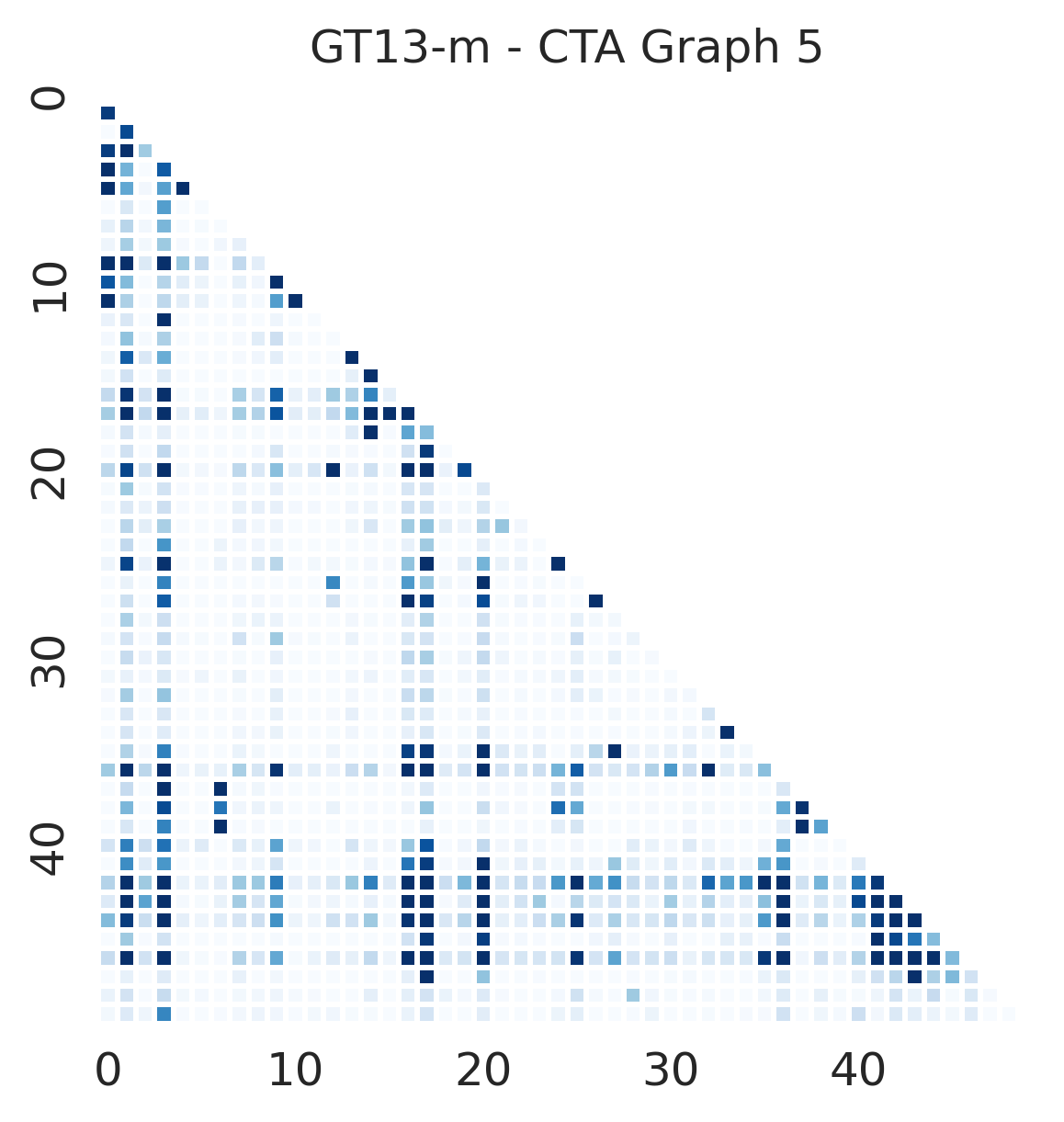}

  \includegraphics[width=0.19\textwidth]{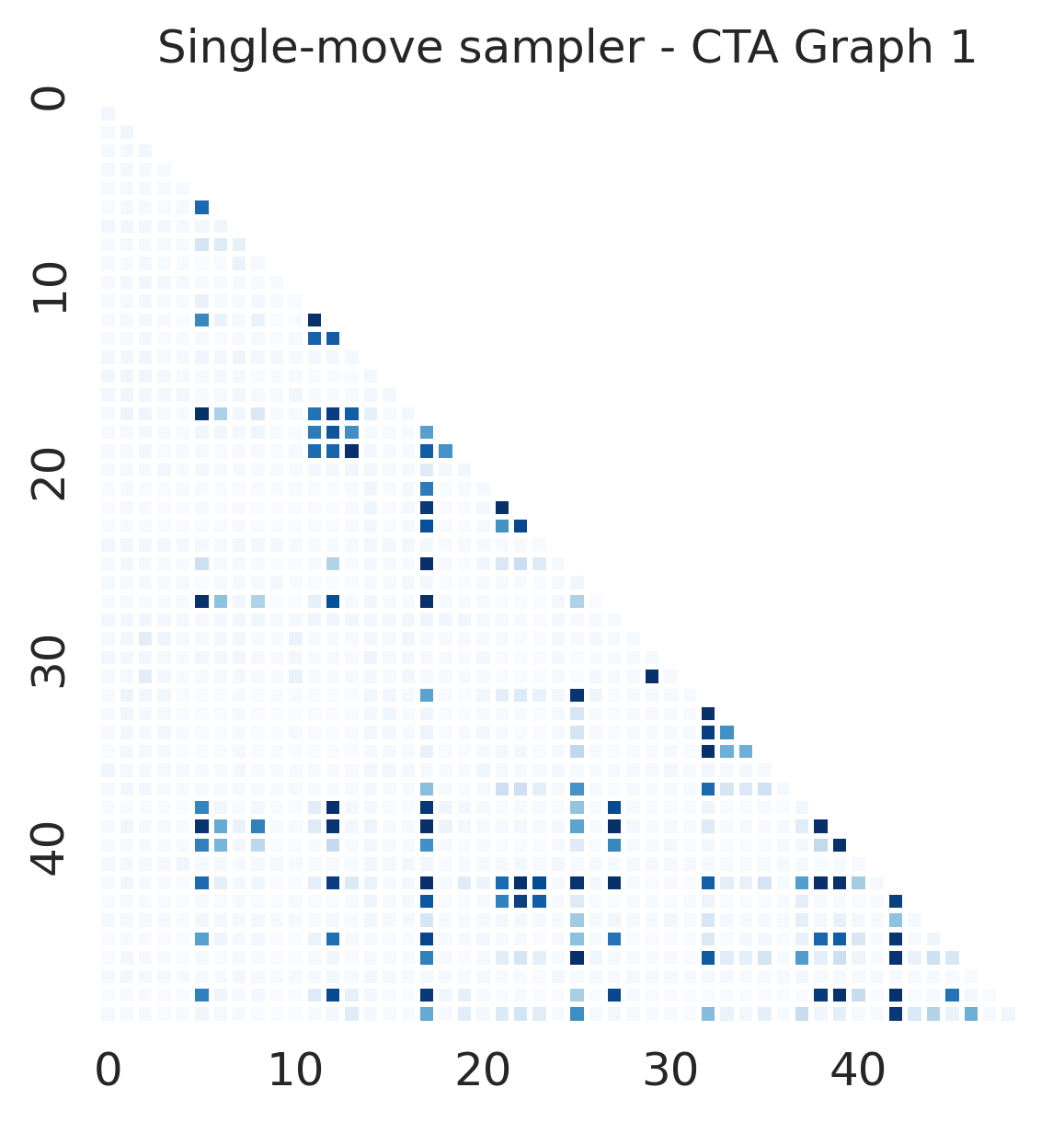}
  \includegraphics[width=0.19\textwidth]{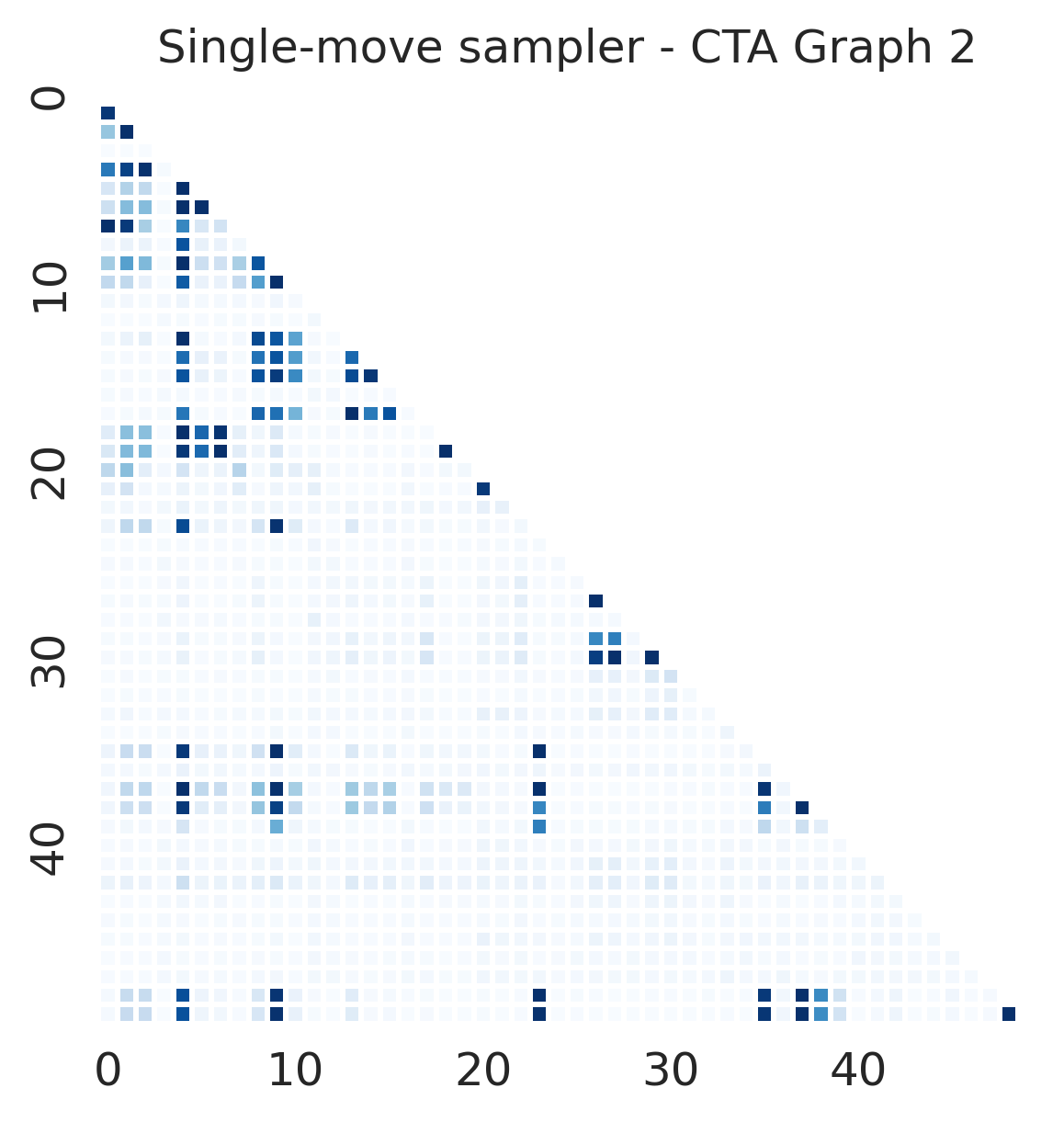}
  \includegraphics[width=0.19\textwidth]{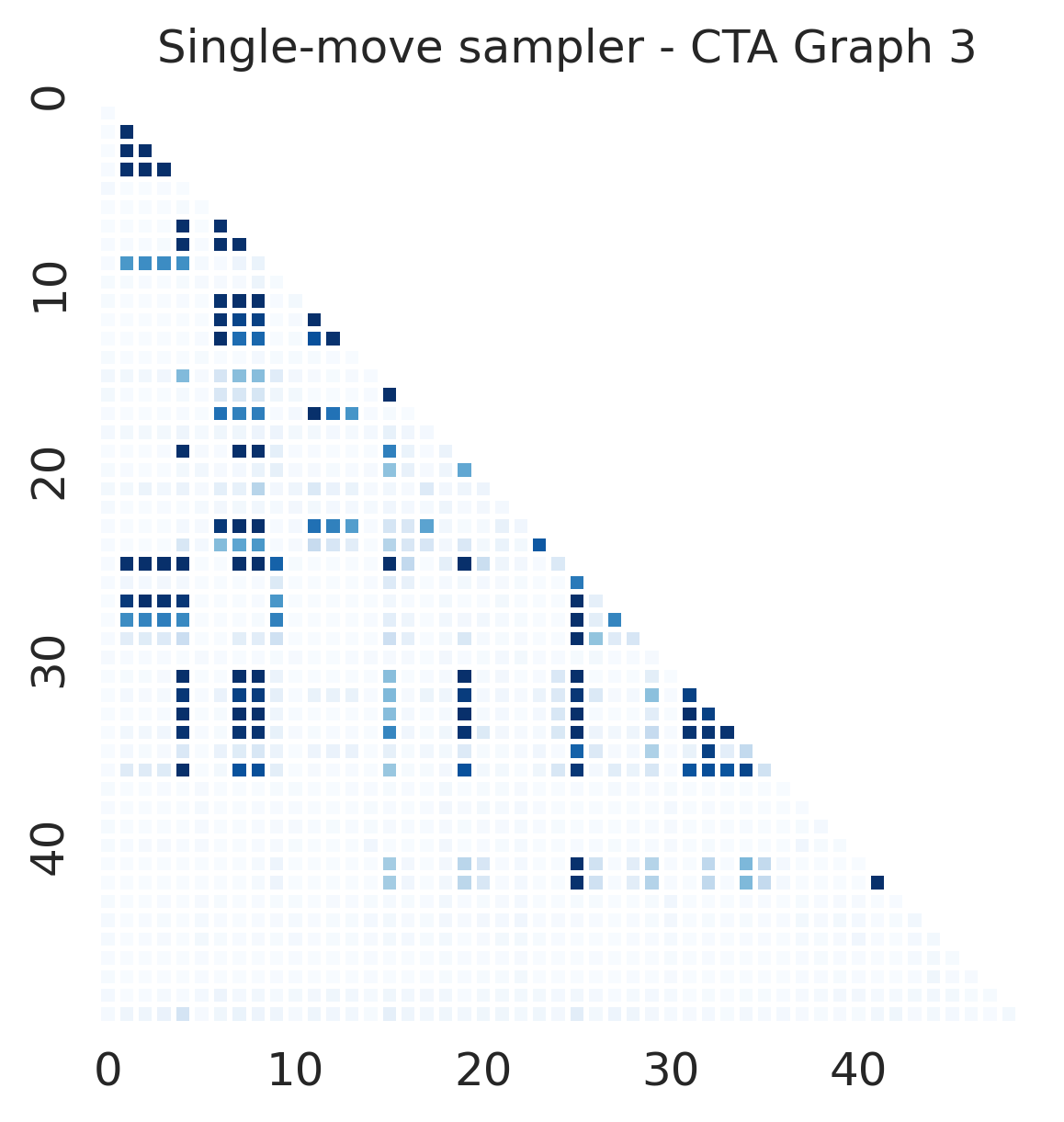}
  \includegraphics[width=0.19\textwidth]{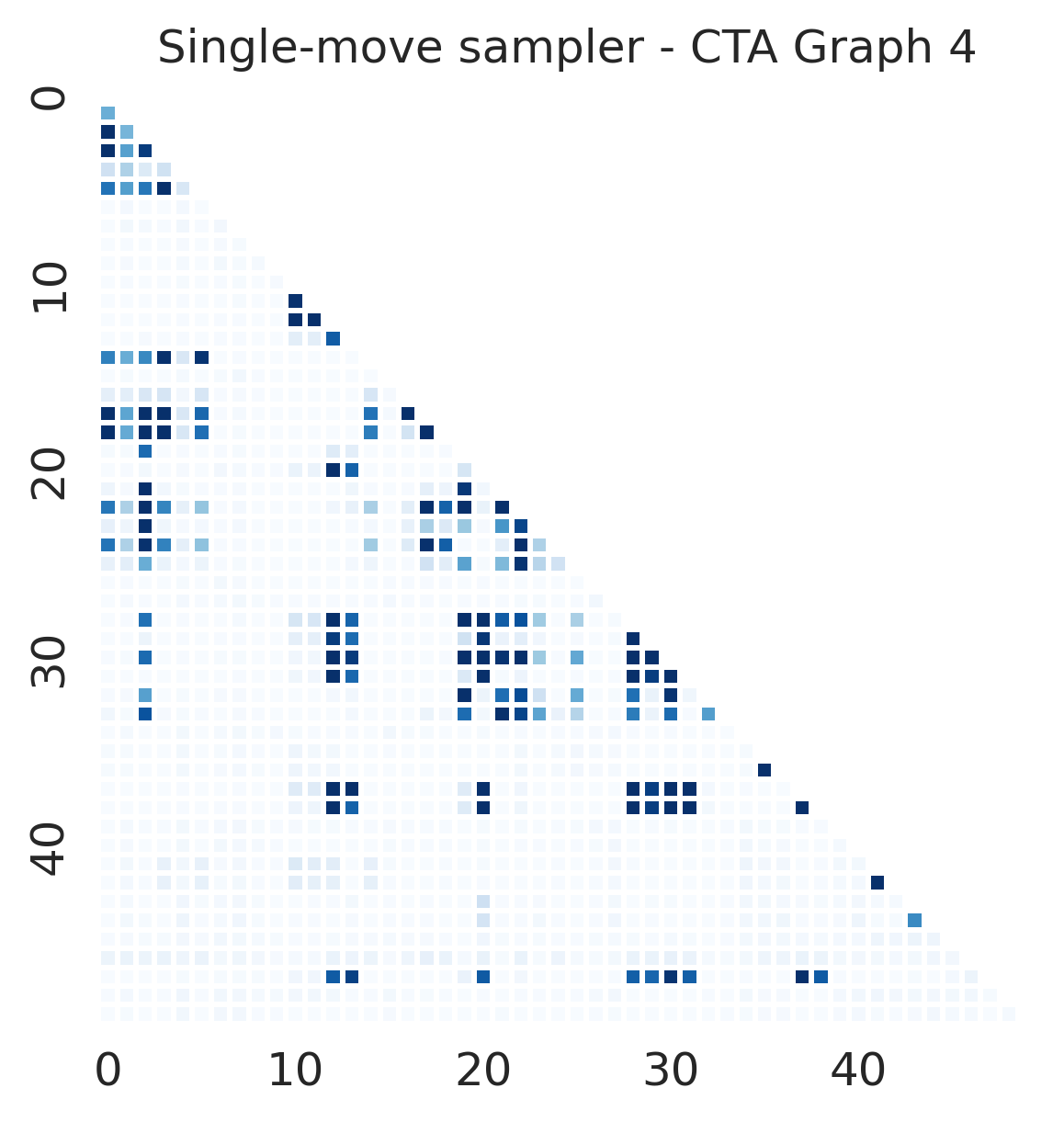}
  \includegraphics[width=0.19\textwidth]{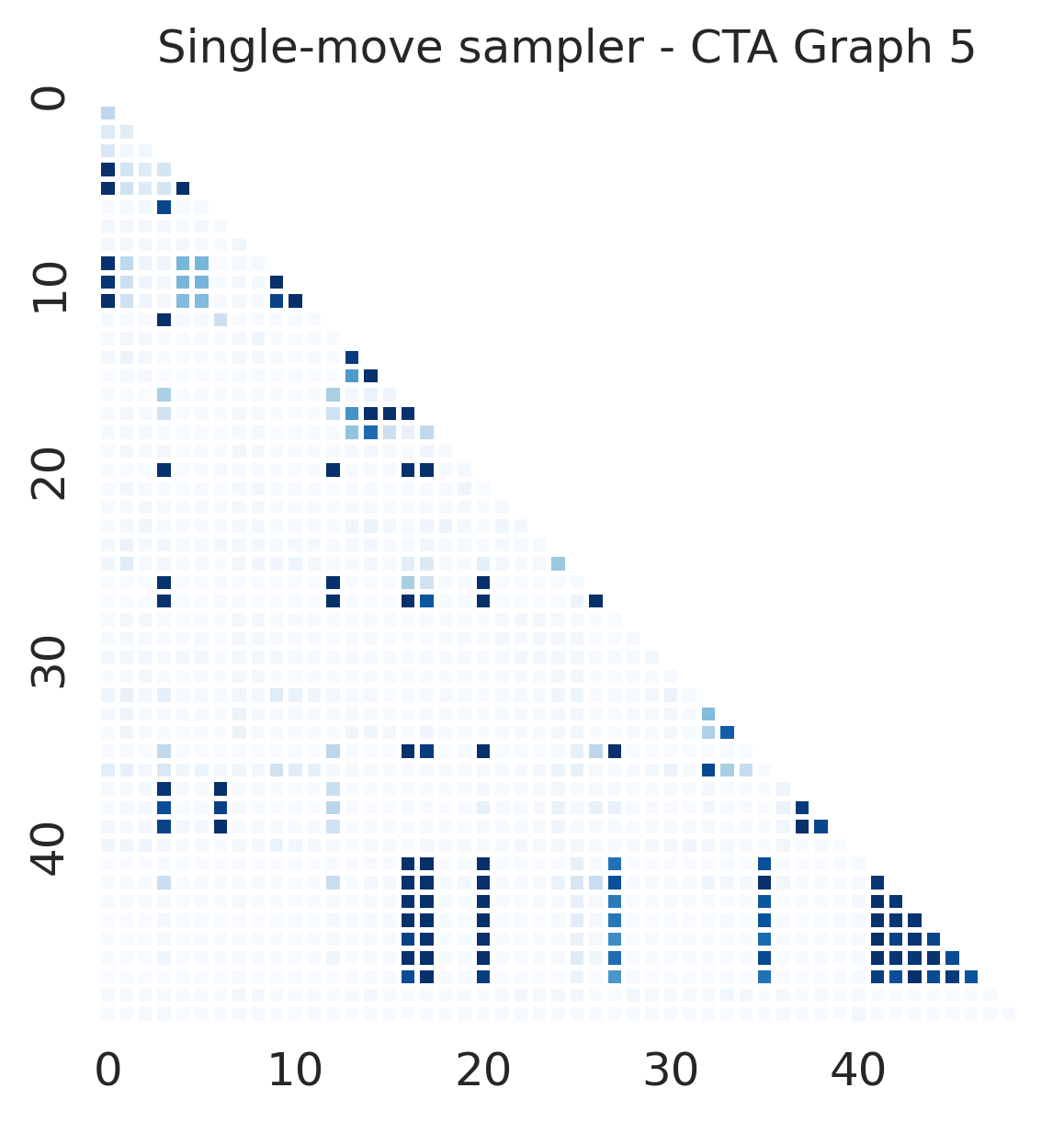}

  \includegraphics[width=0.19\textwidth]{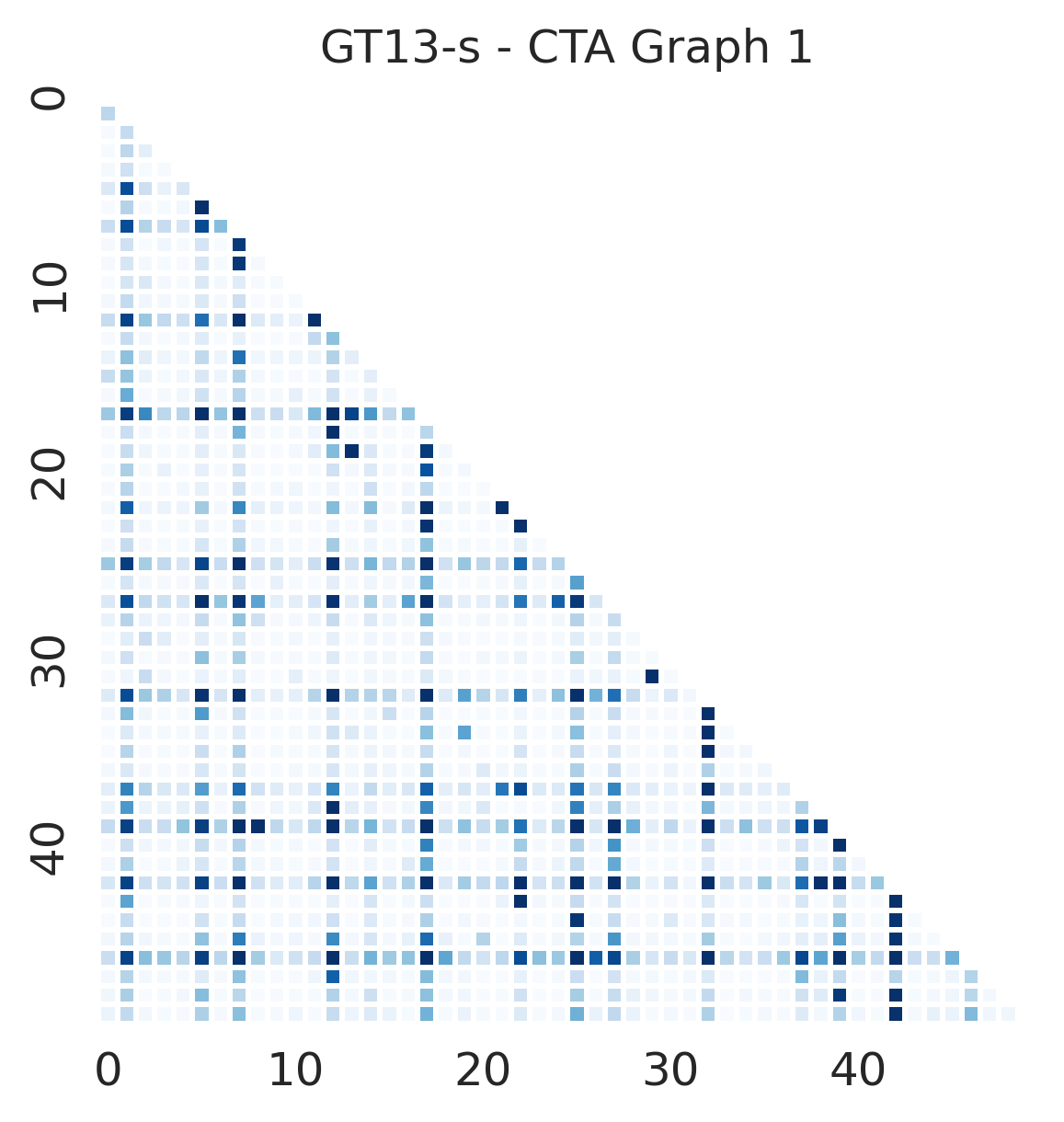}
  \includegraphics[width=0.19\textwidth]{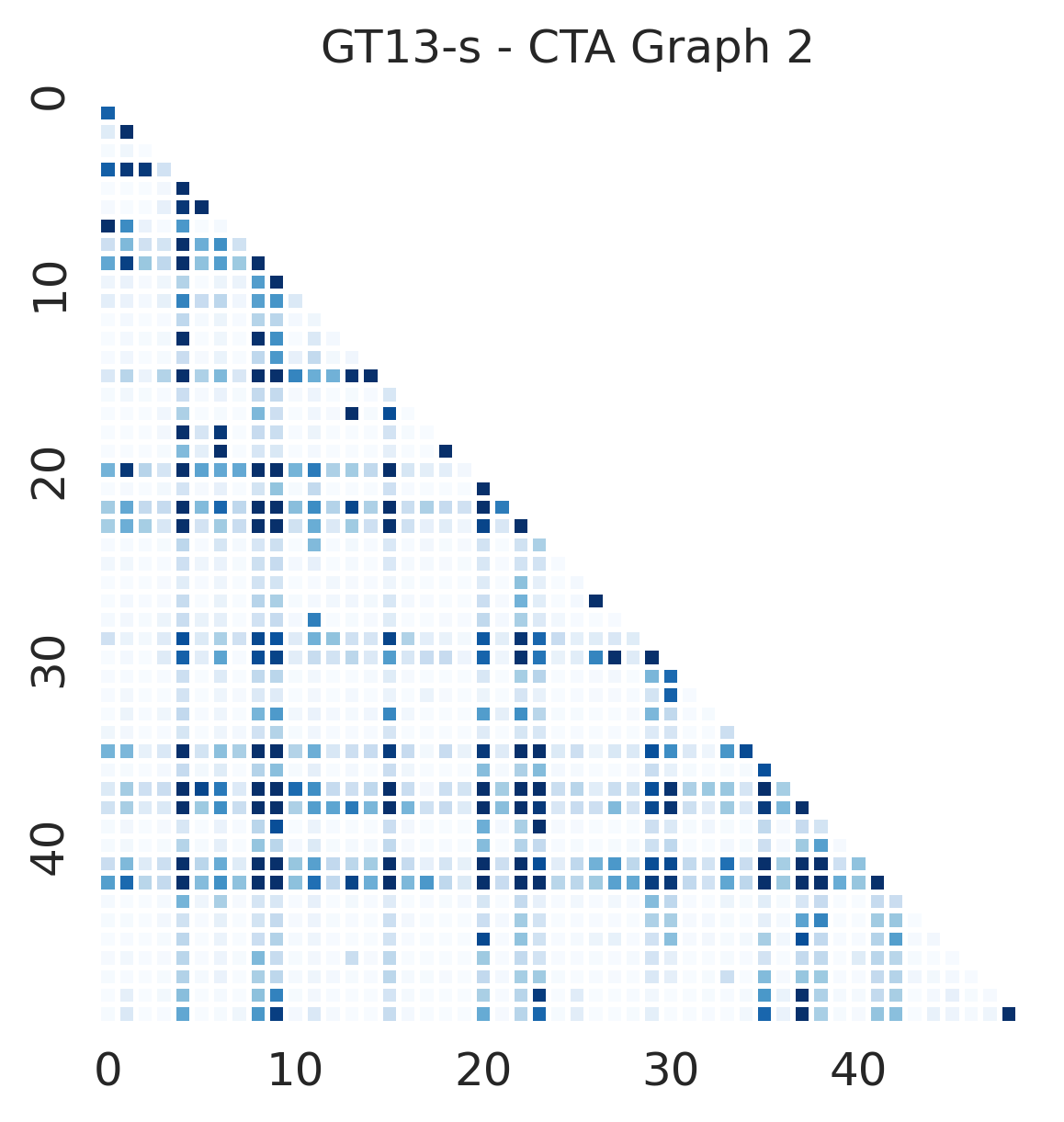}
  \includegraphics[width=0.19\textwidth]{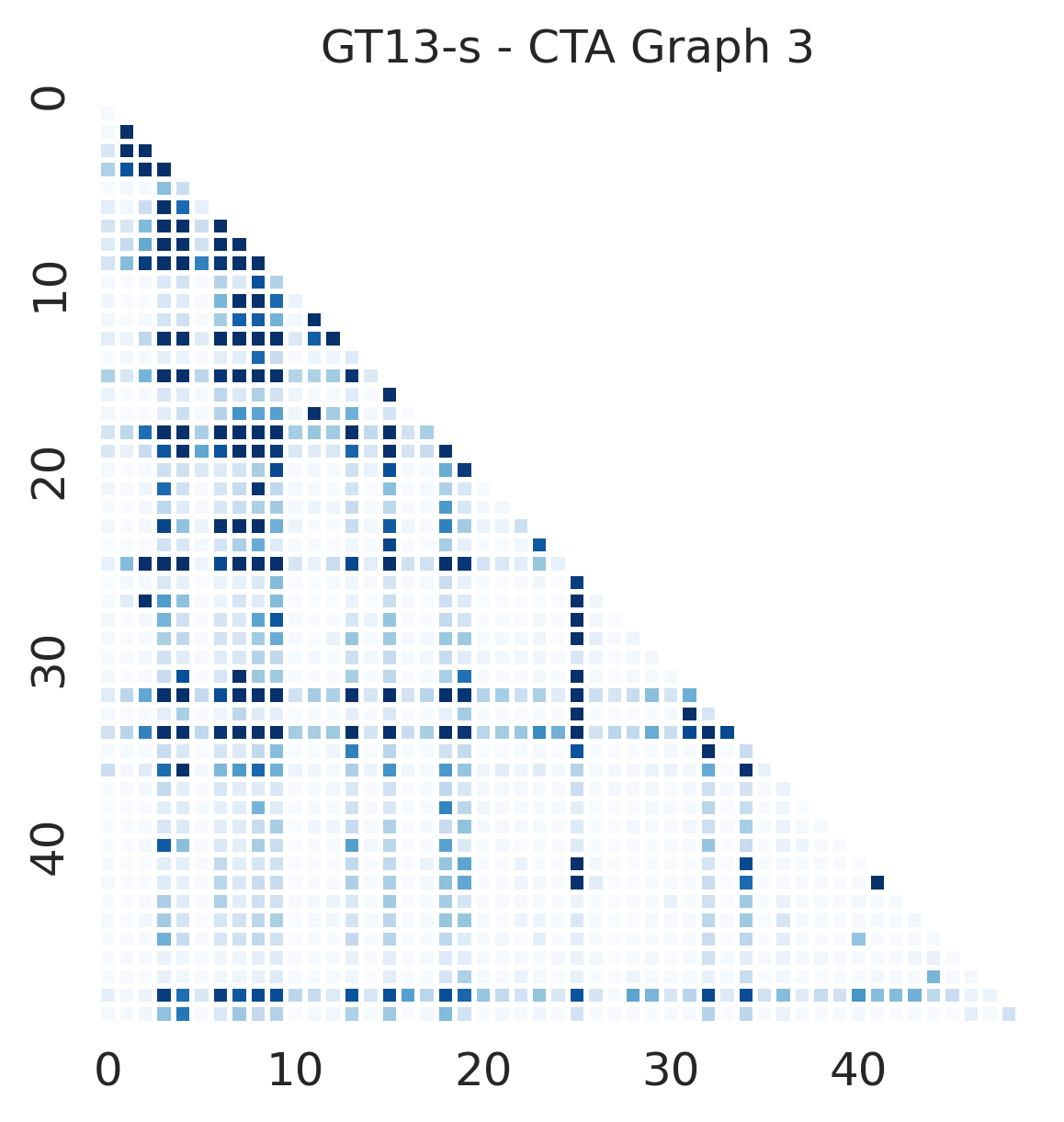}
  \includegraphics[width=0.19\textwidth]{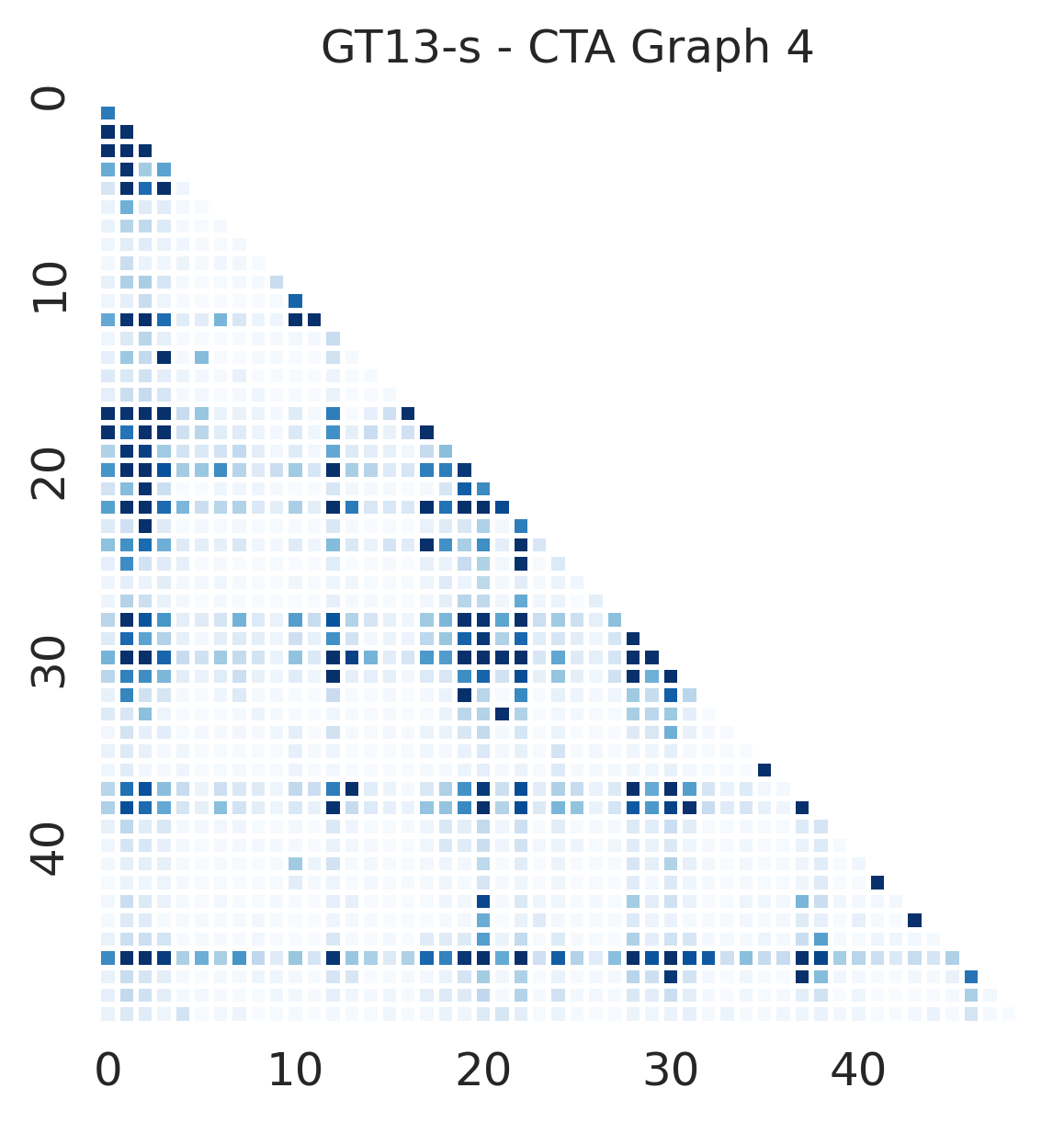}
  \includegraphics[width=0.19\textwidth]{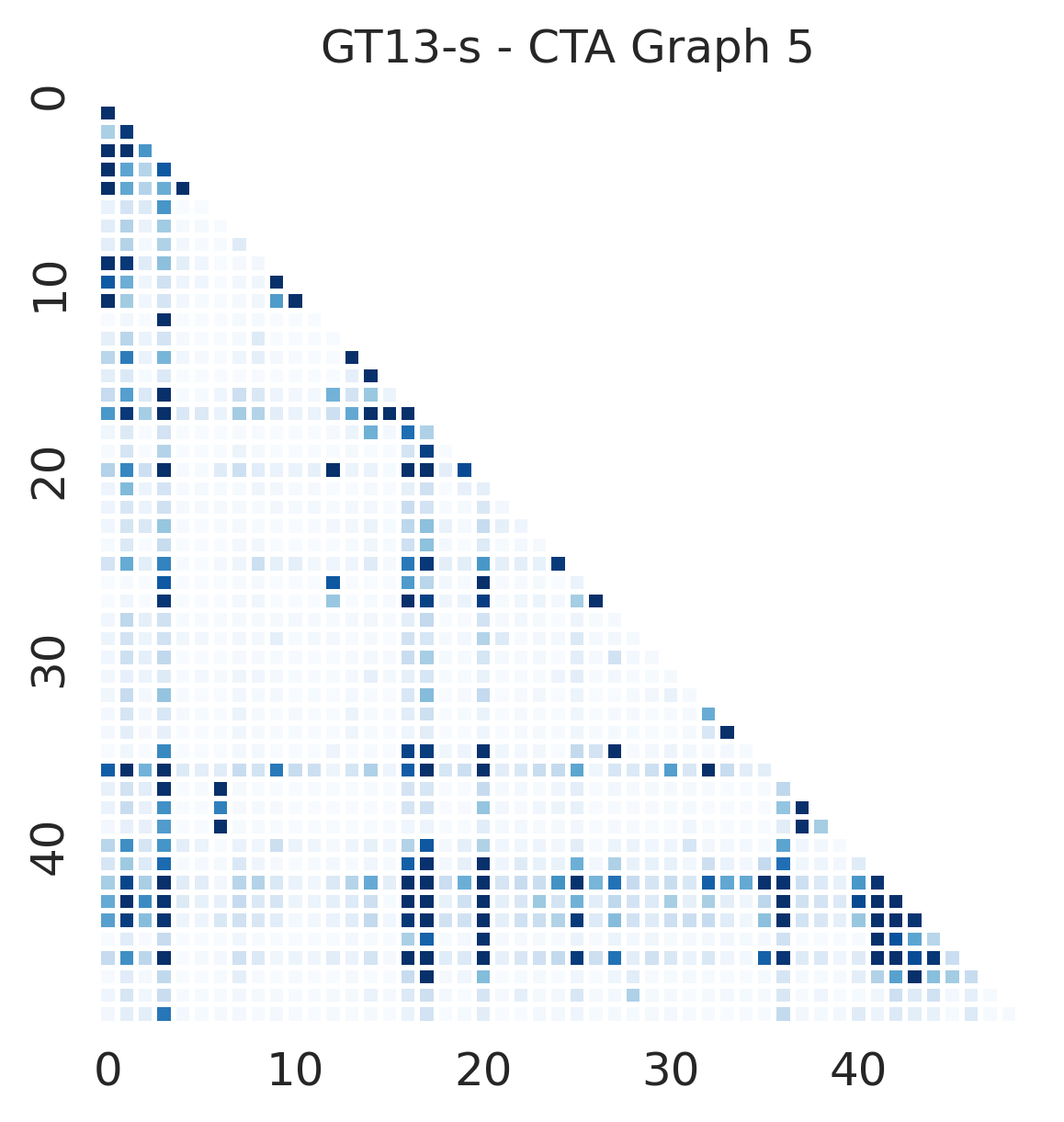}

  \includegraphics[width=0.19\textwidth]{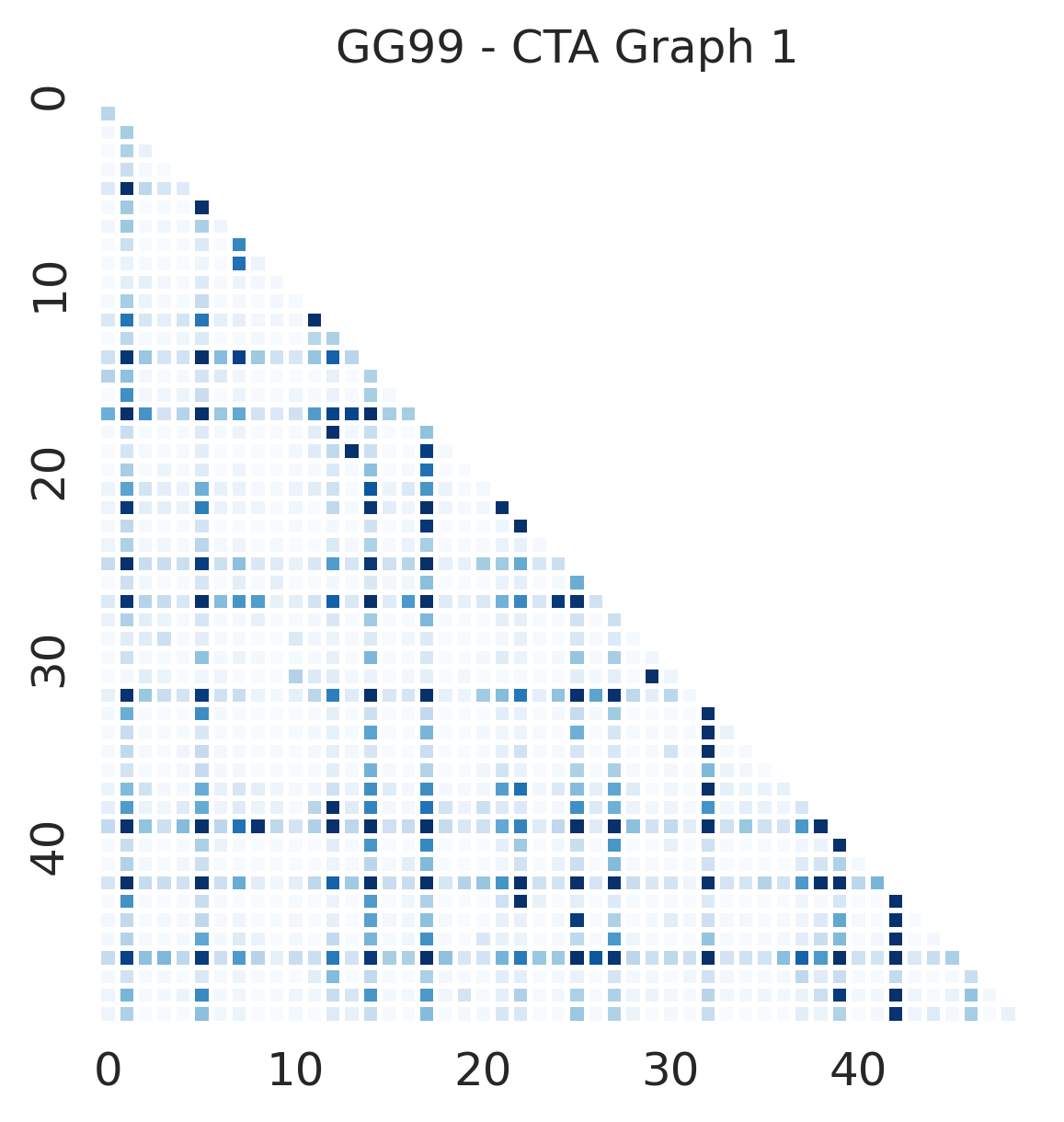}
  \includegraphics[width=0.19\textwidth]{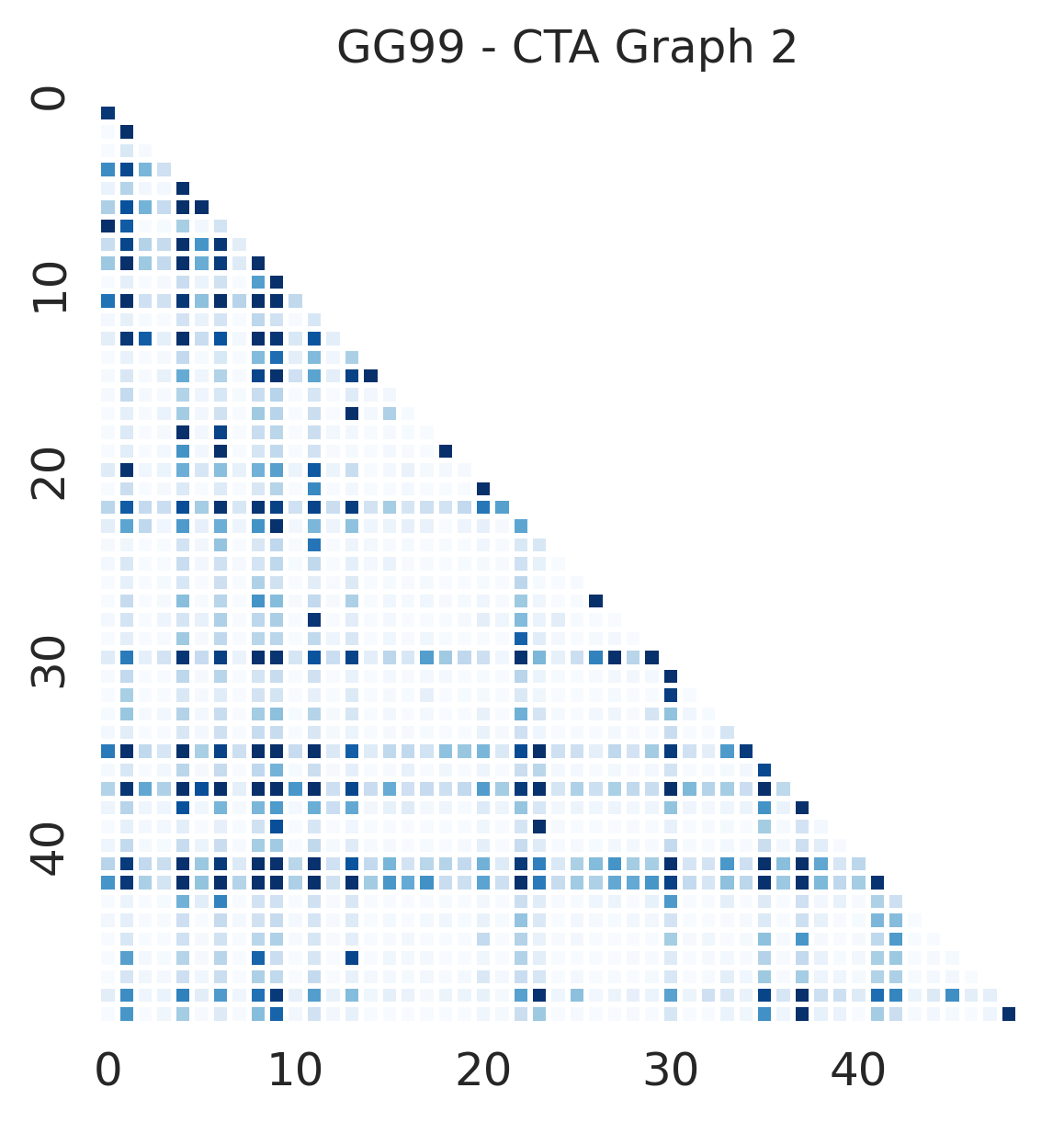}
  \includegraphics[width=0.19\textwidth]{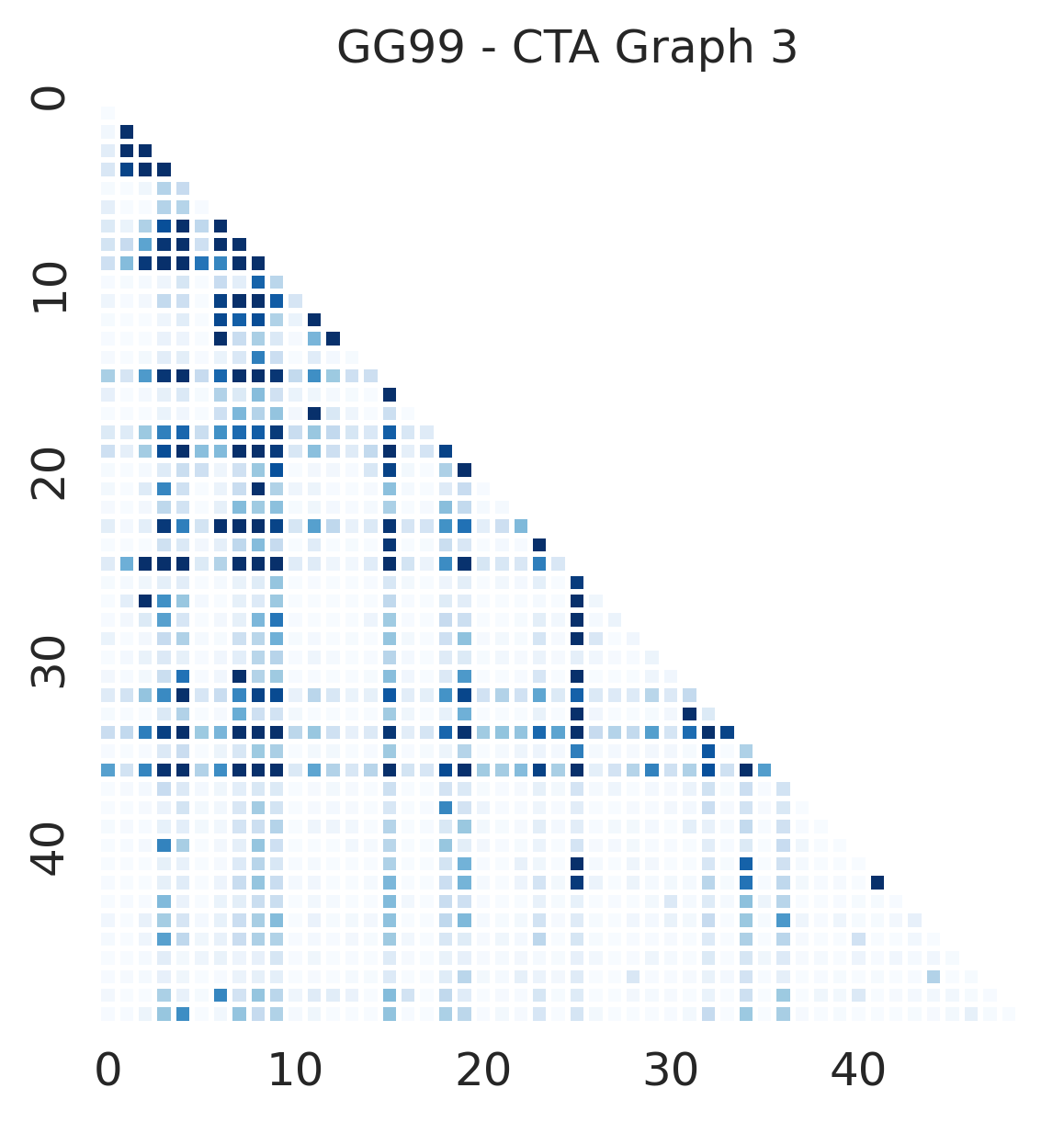}
  \includegraphics[width=0.19\textwidth]{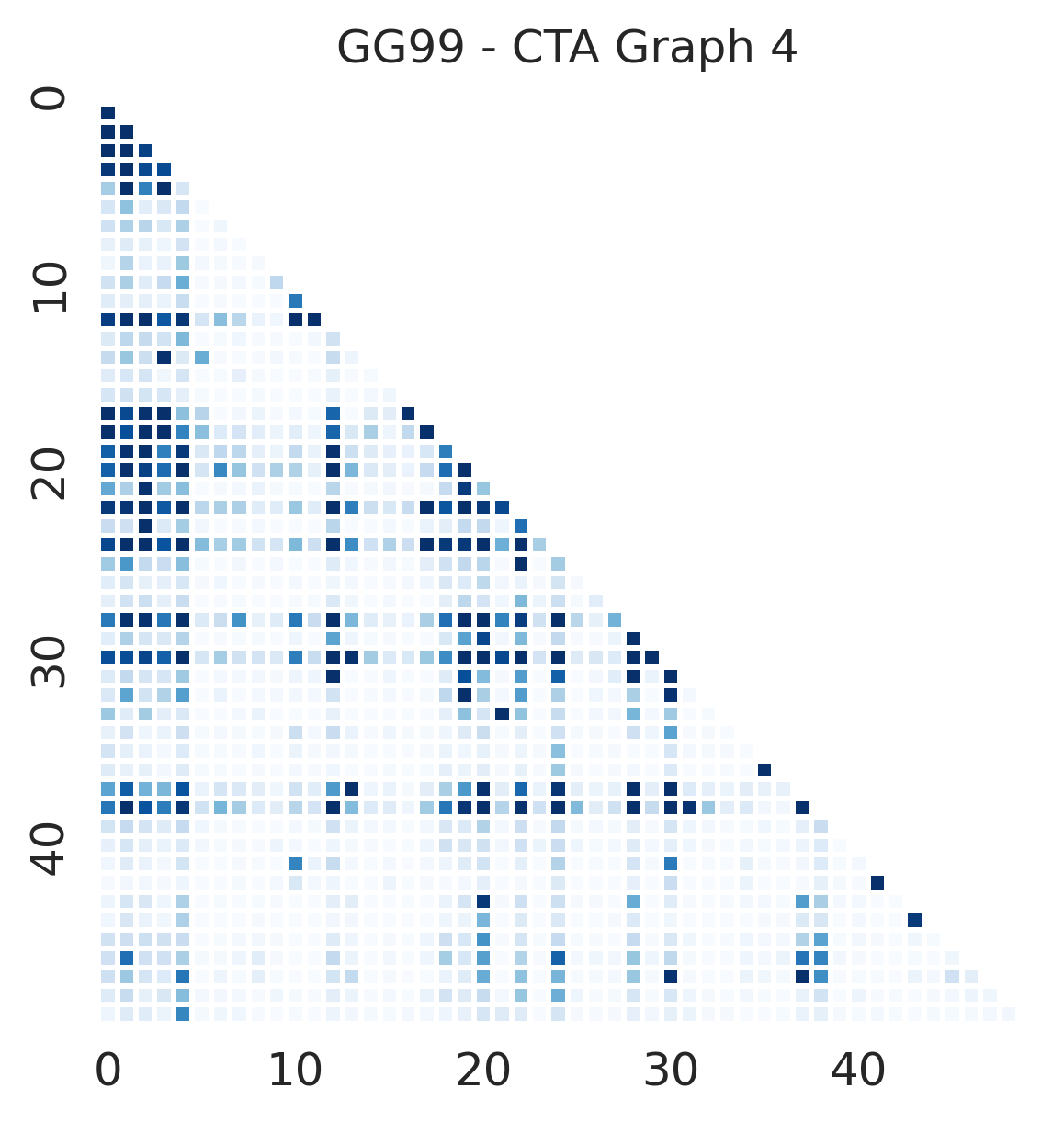}
  \includegraphics[width=0.19\textwidth]{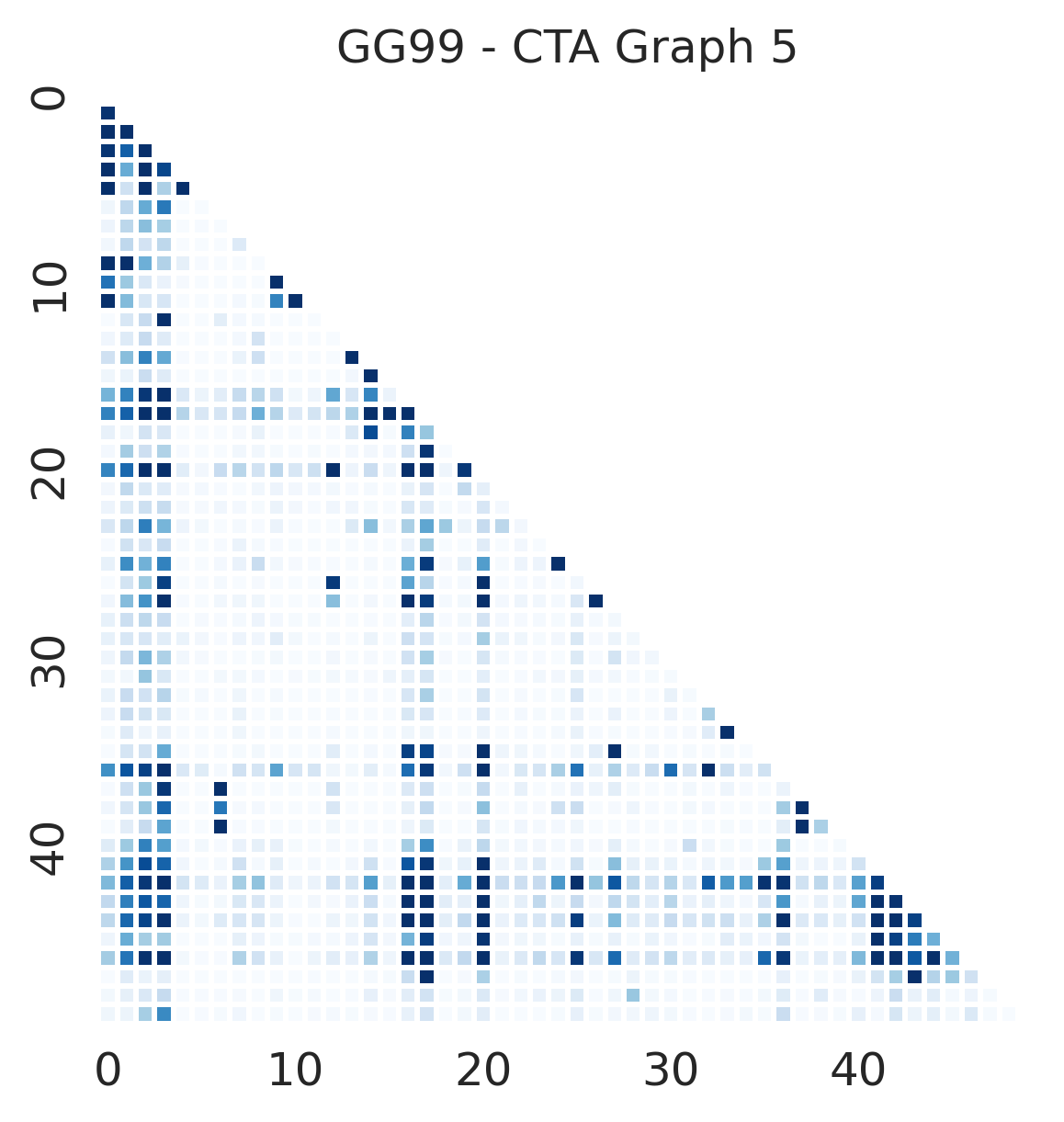}

  \includegraphics[width=0.19\textwidth]{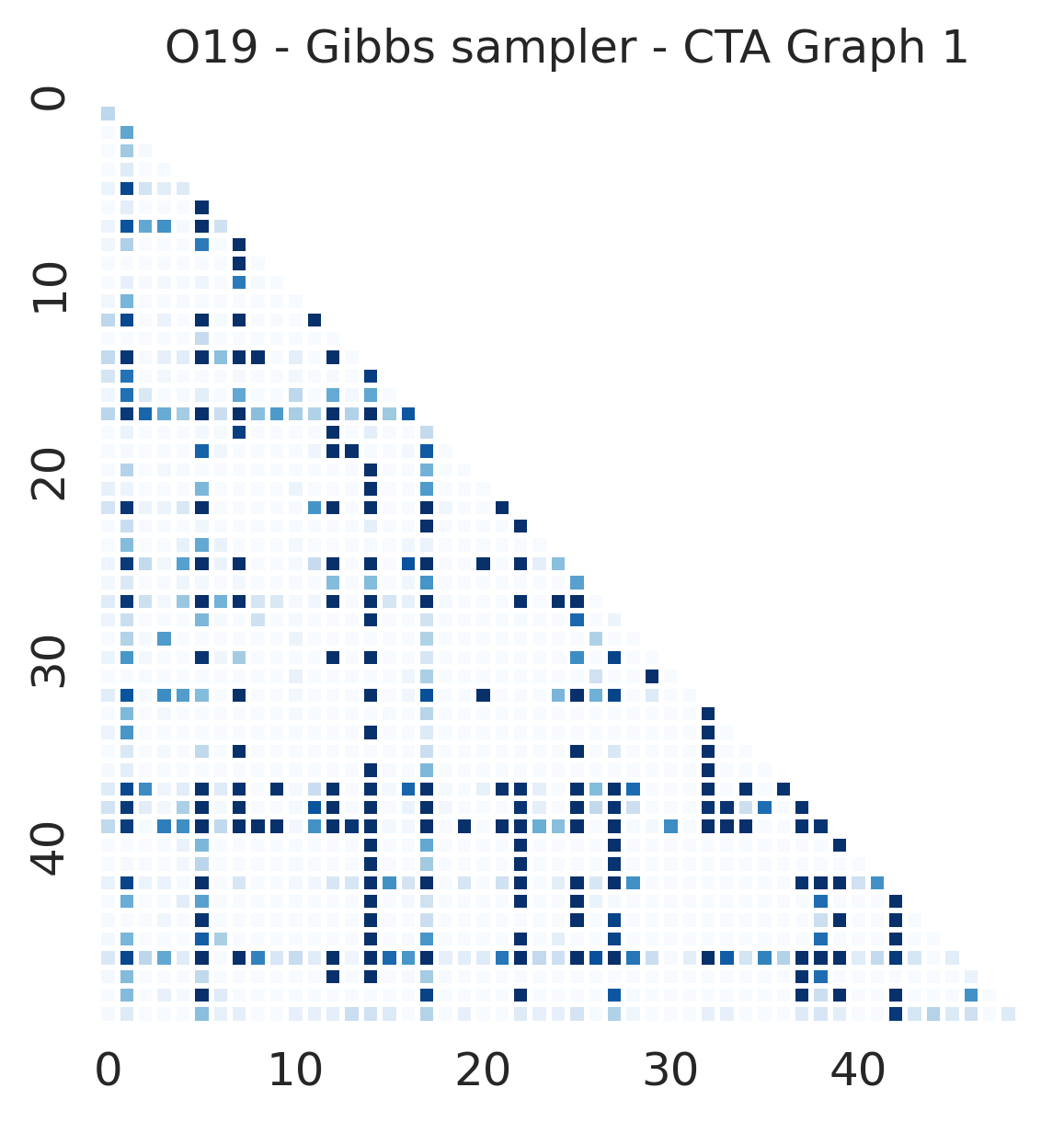}
  \includegraphics[width=0.19\textwidth]{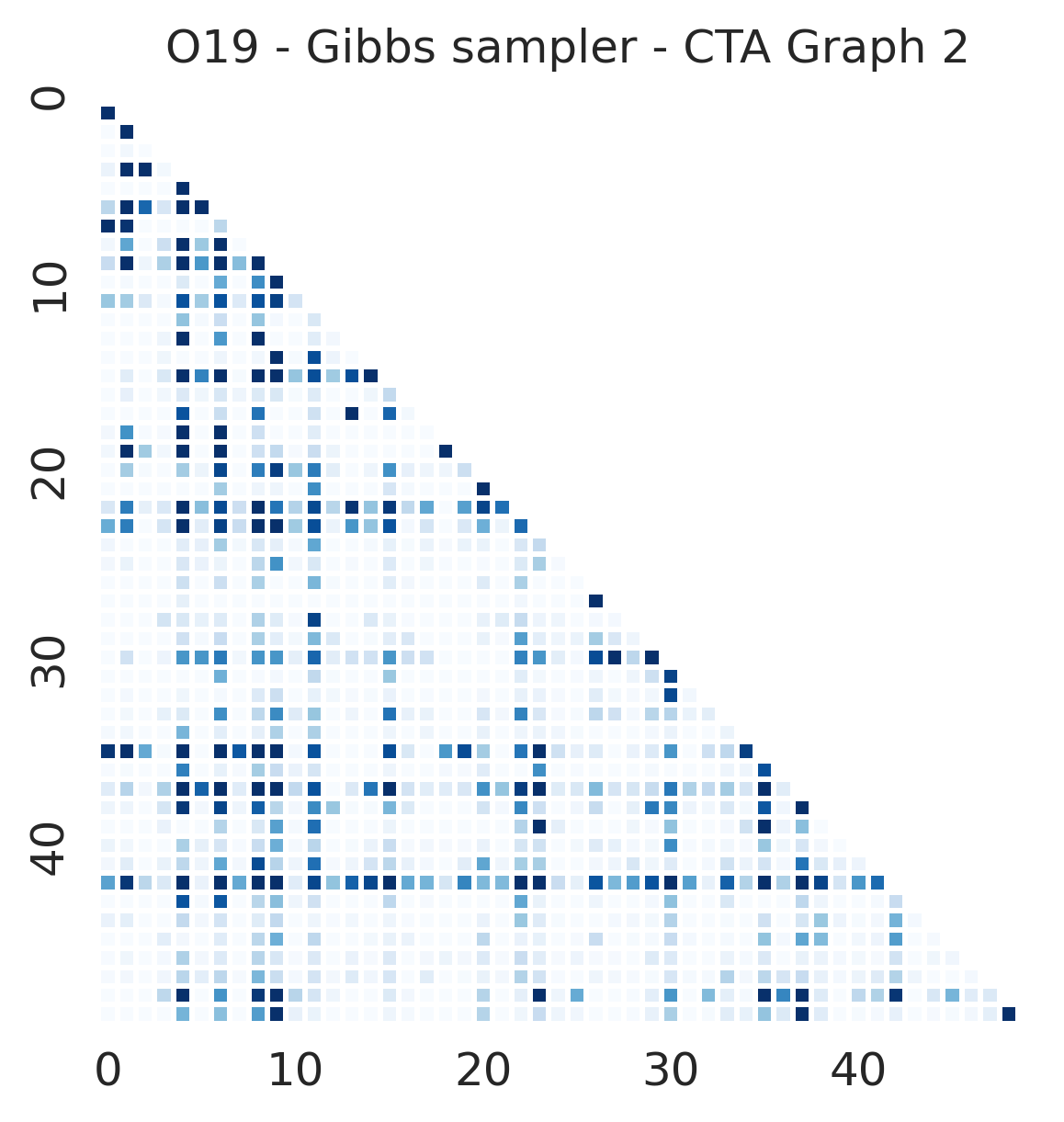}
  \includegraphics[width=0.19\textwidth]{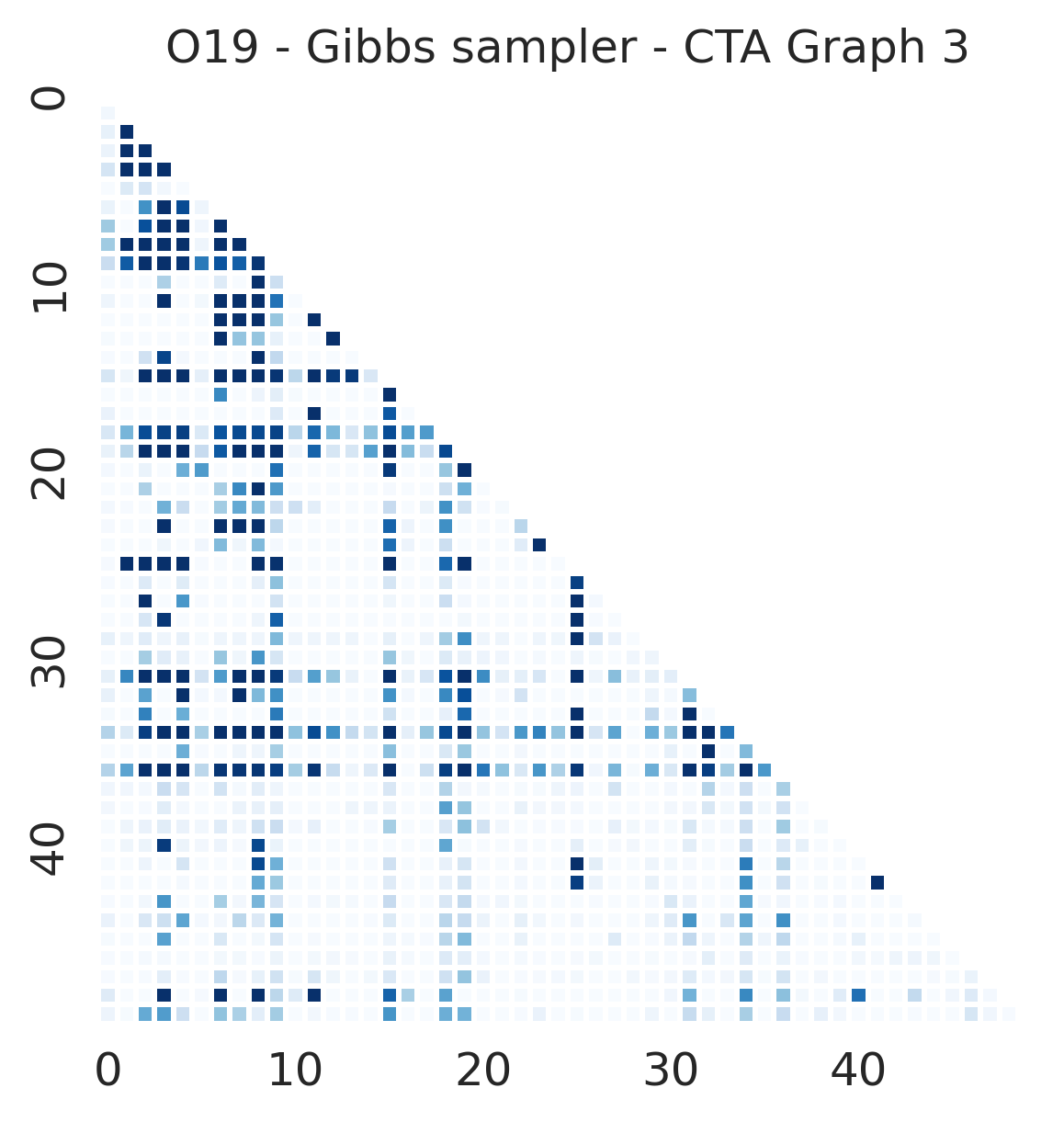}
  \includegraphics[width=0.19\textwidth]{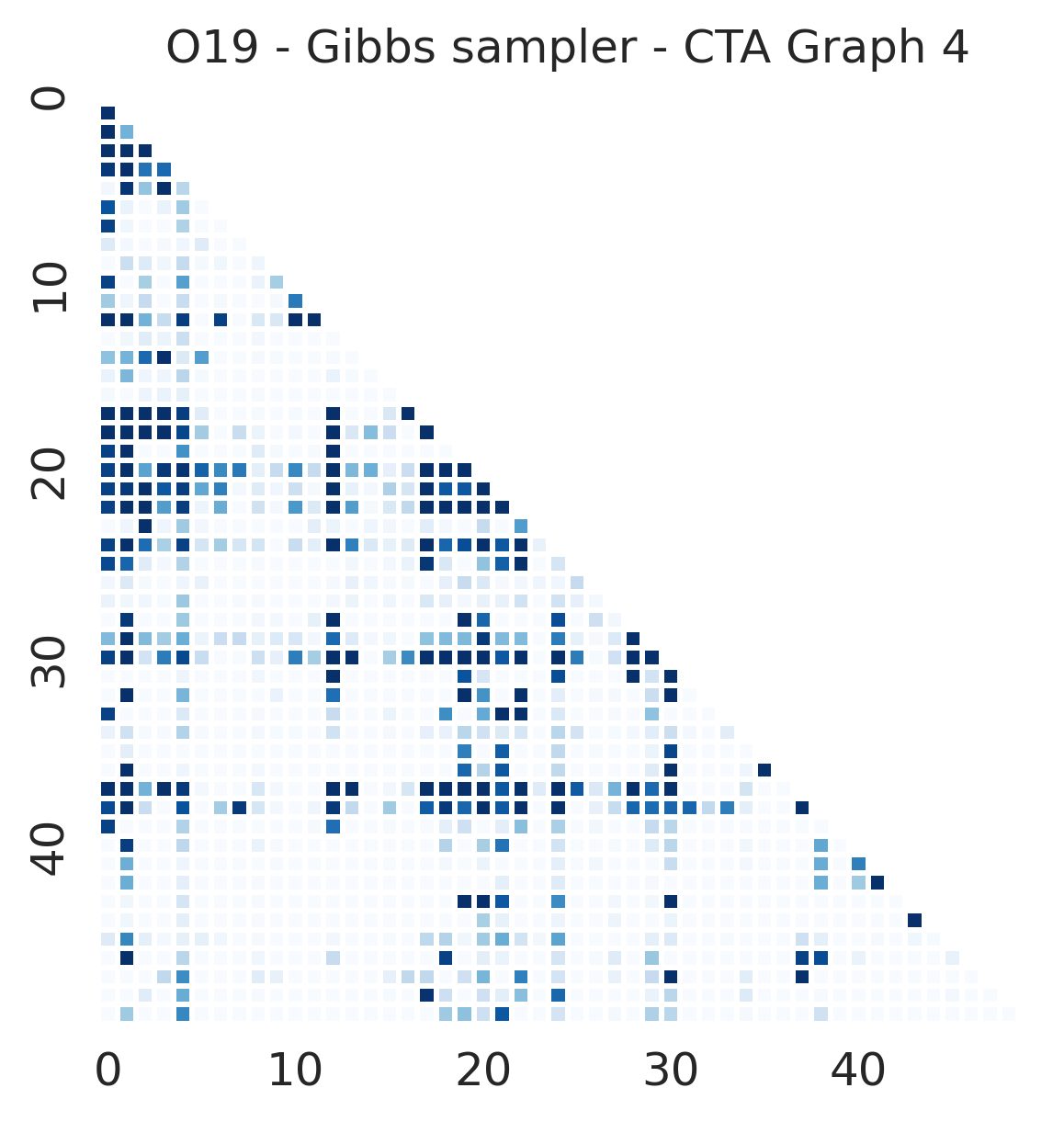}
  \includegraphics[width=0.19\textwidth]{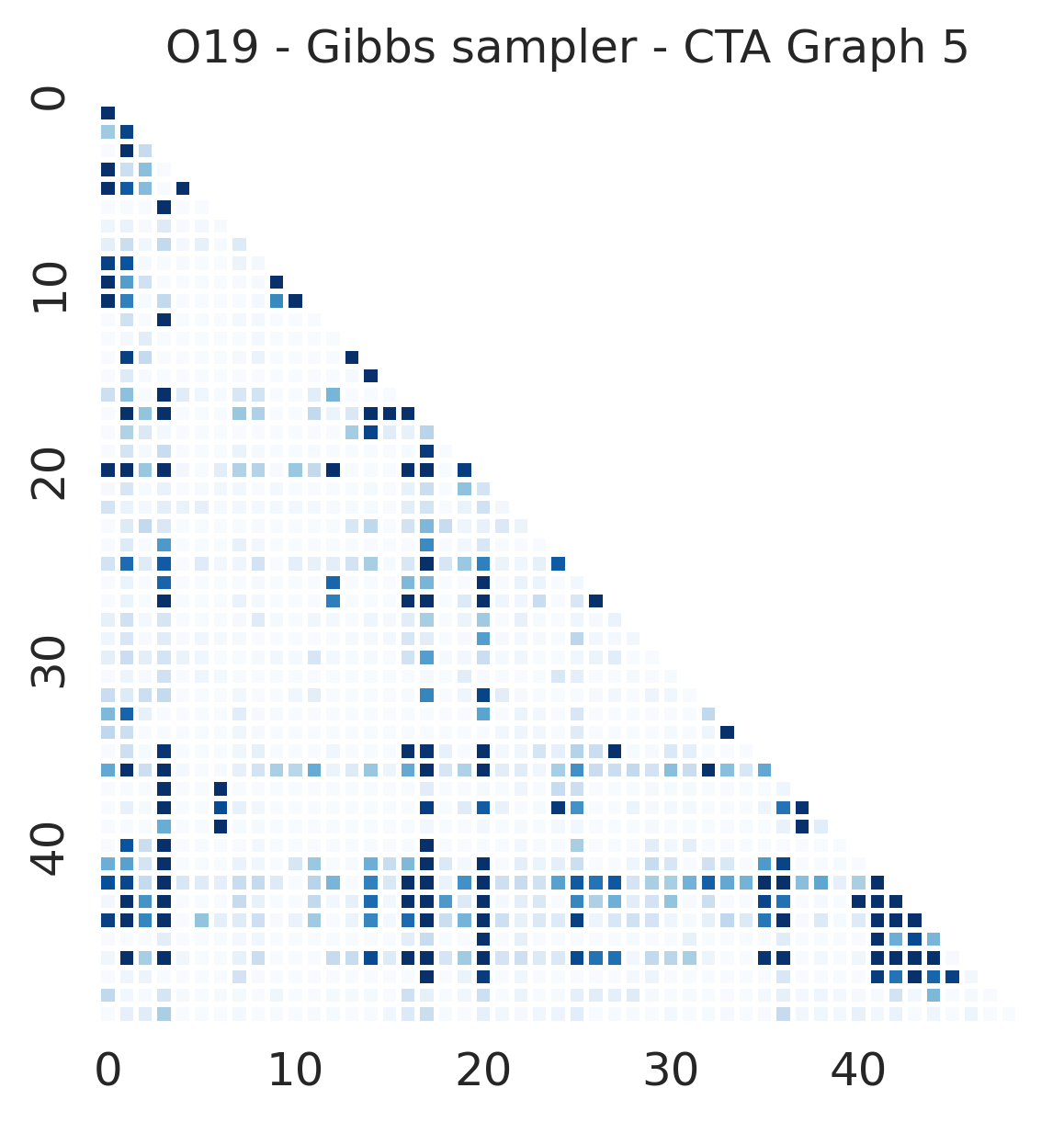}

  \caption{Heatmap derived from the final 300,000 steps of samplers in Section~\ref{sec:simul-setup-gauss}, covering replications 1 to 5 under the Christmas tree algorithm and $(n,p)=(100,50)$.}
  \label{app:fig:heatmap-cta-0}
\end{figure}

\begin{figure}[ht]
  \centering
  \includegraphics[width=0.19\textwidth]{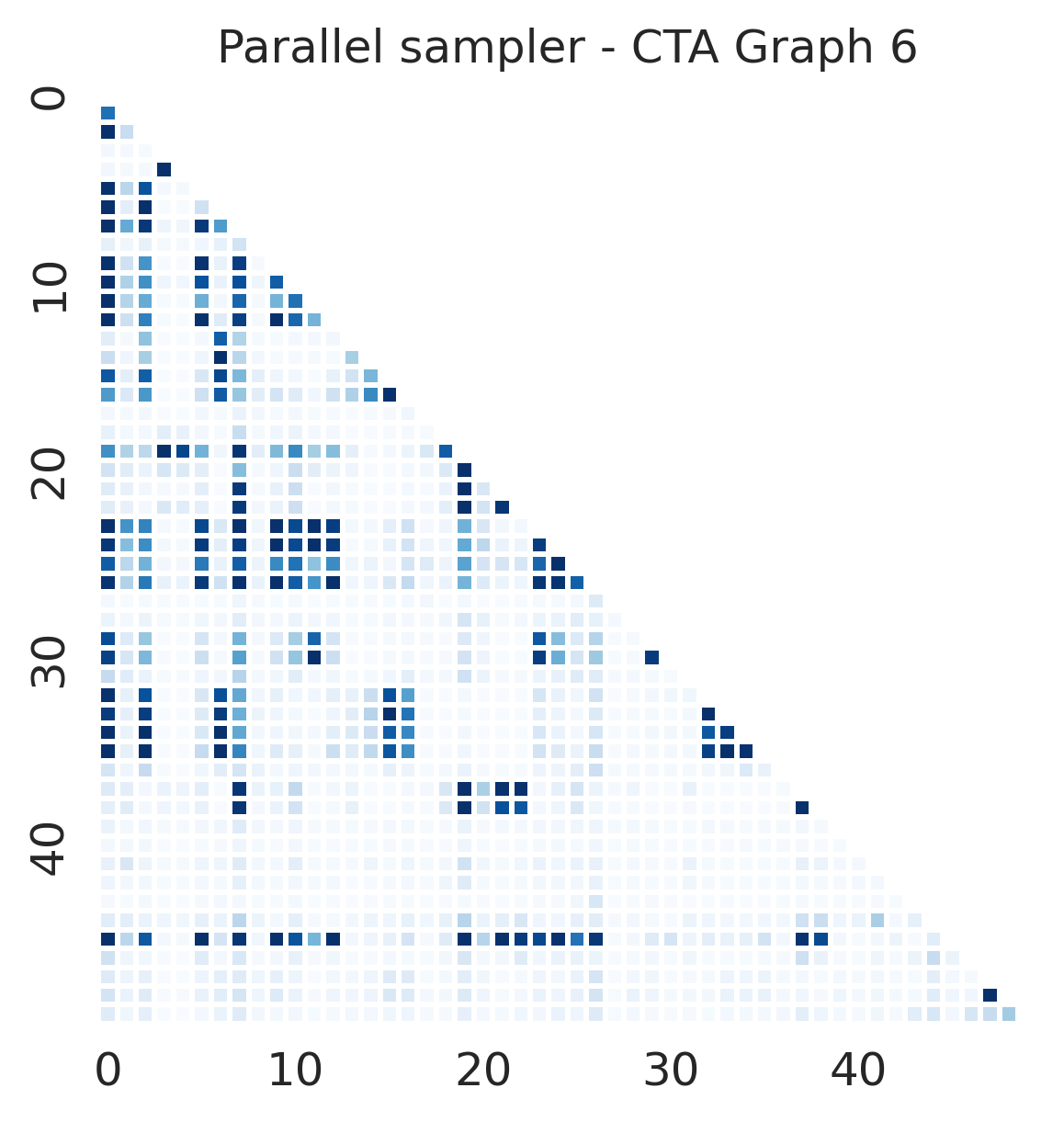}
  \includegraphics[width=0.19\textwidth]{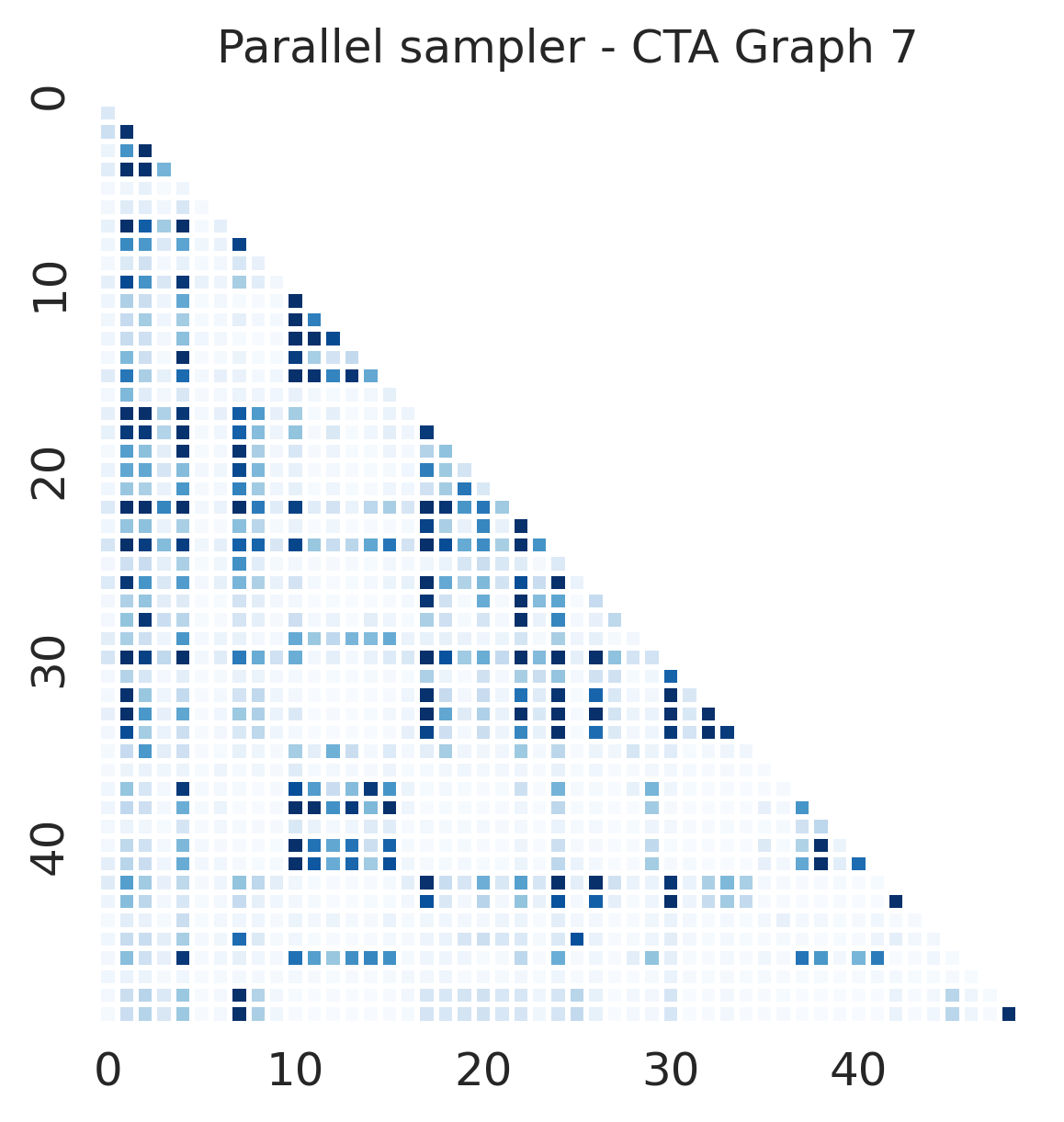}
  \includegraphics[width=0.19\textwidth]{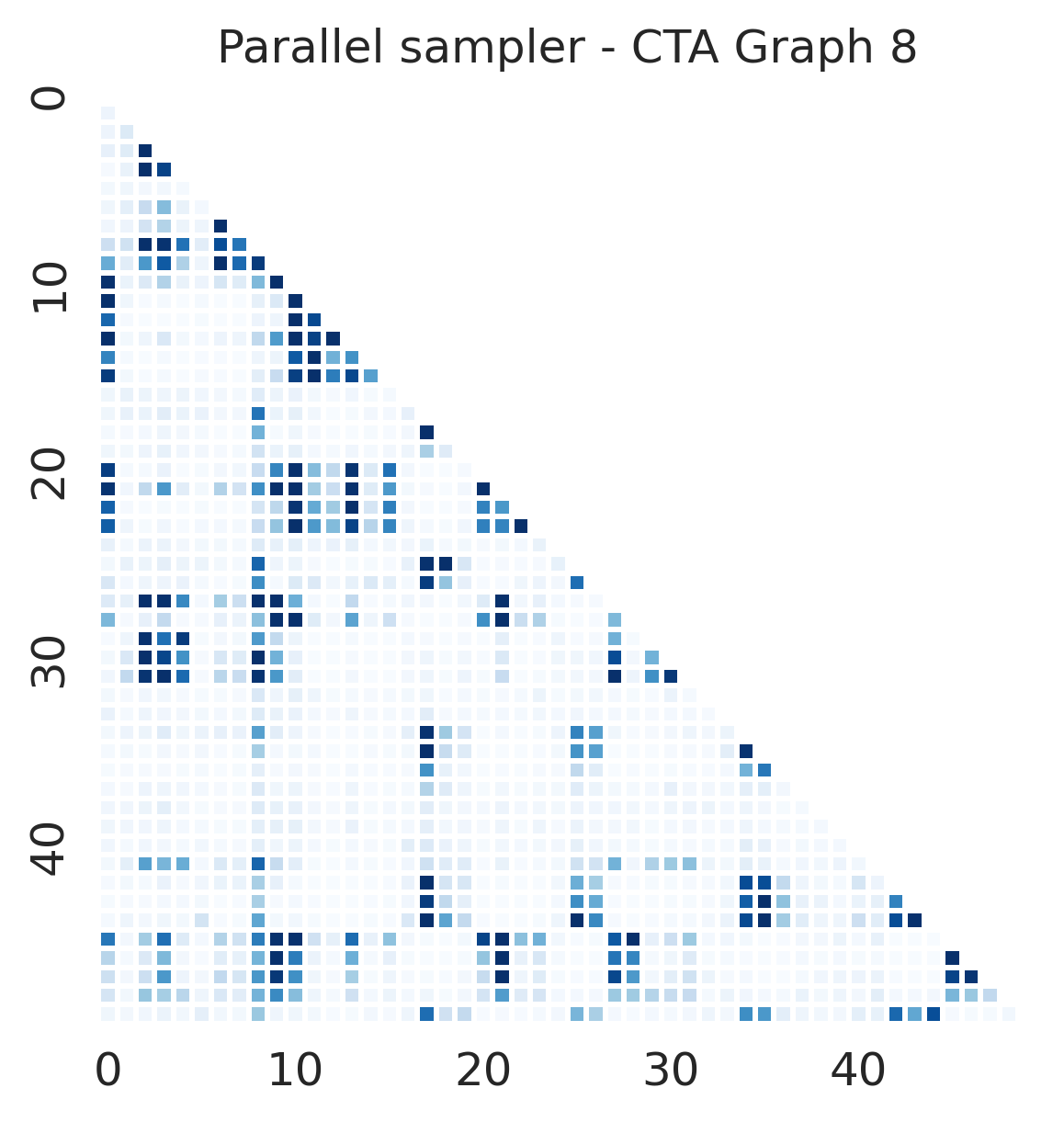}
  \includegraphics[width=0.19\textwidth]{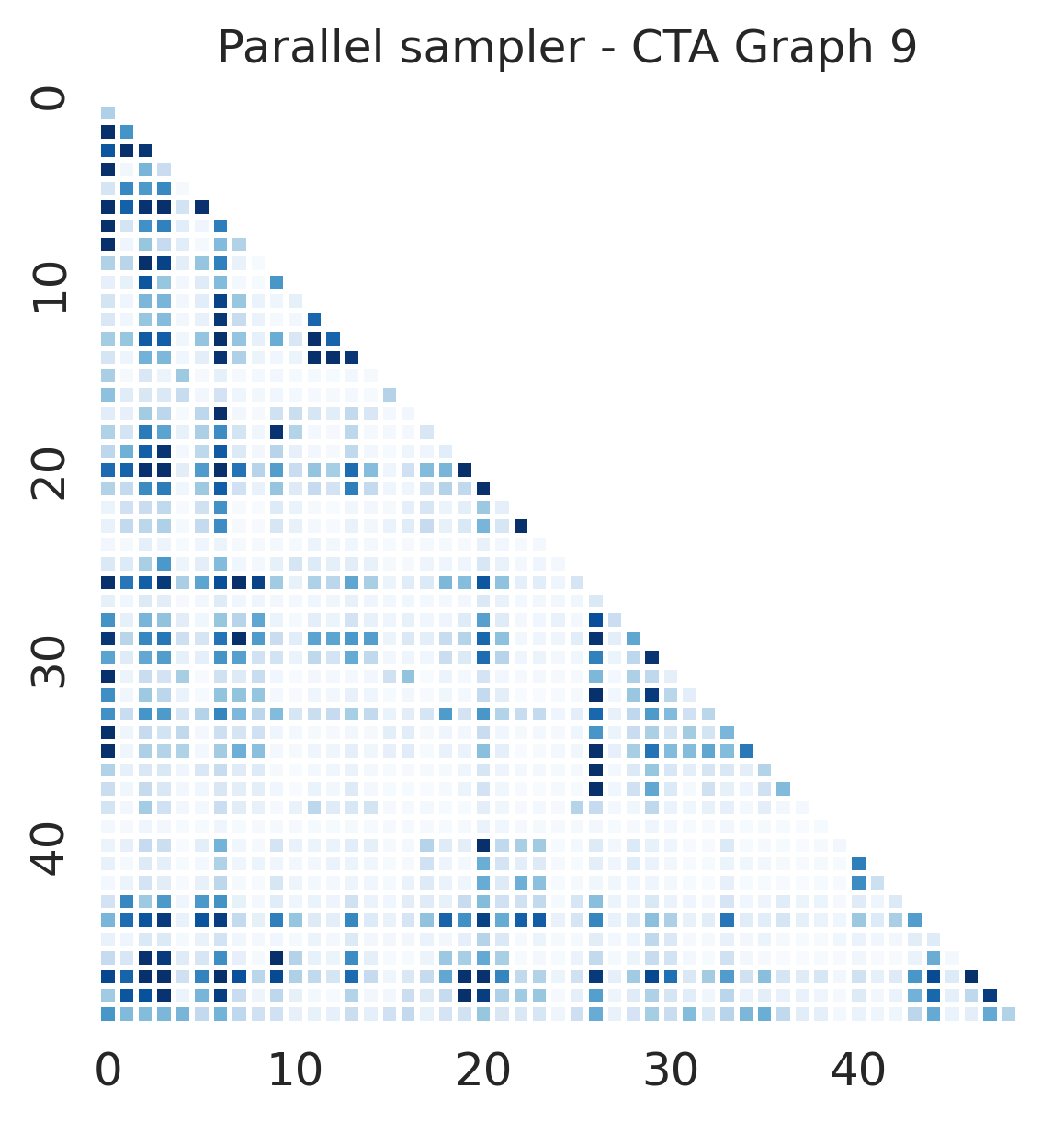}
  \includegraphics[width=0.19\textwidth]{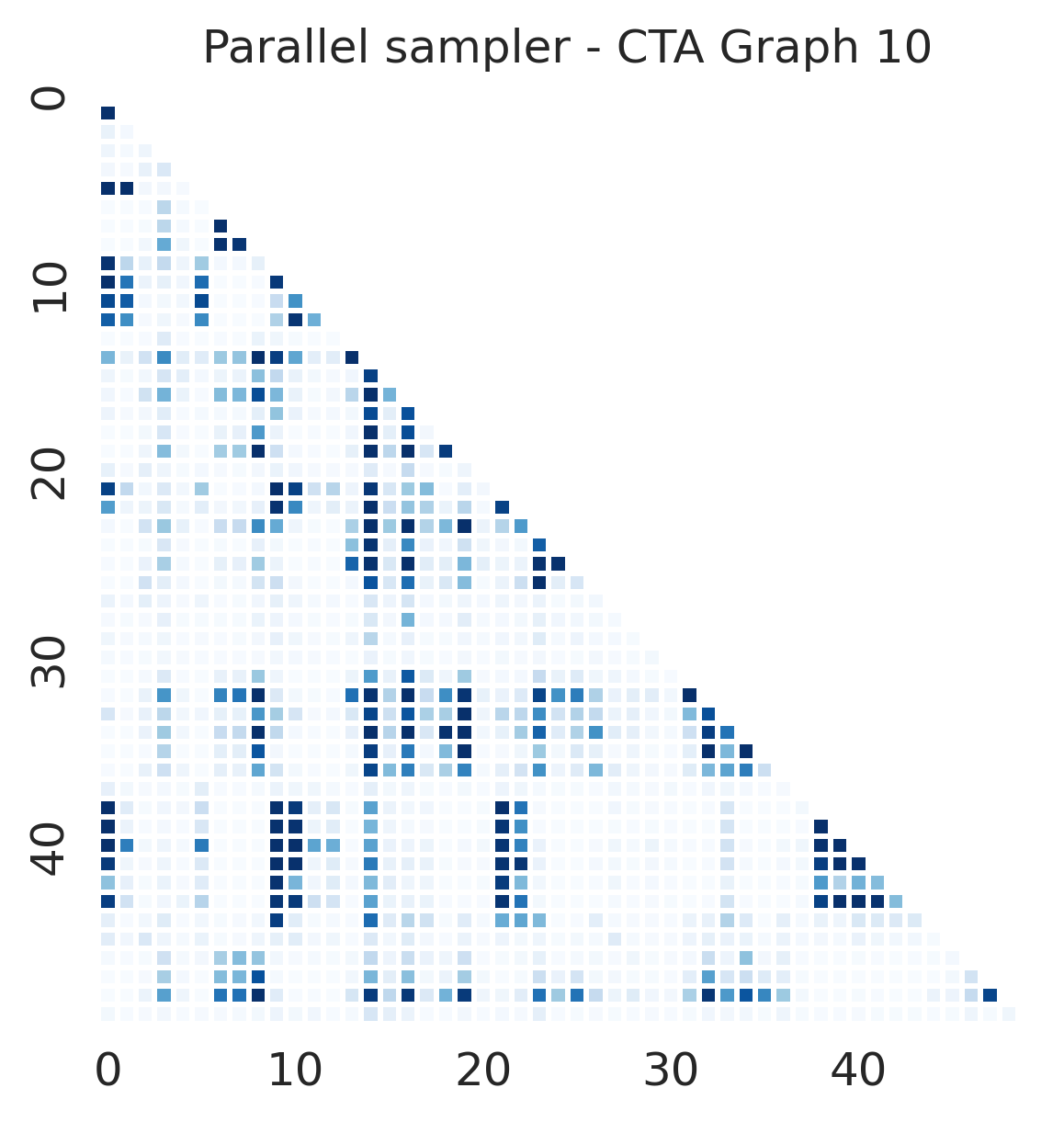}

  \includegraphics[width=0.19\textwidth]{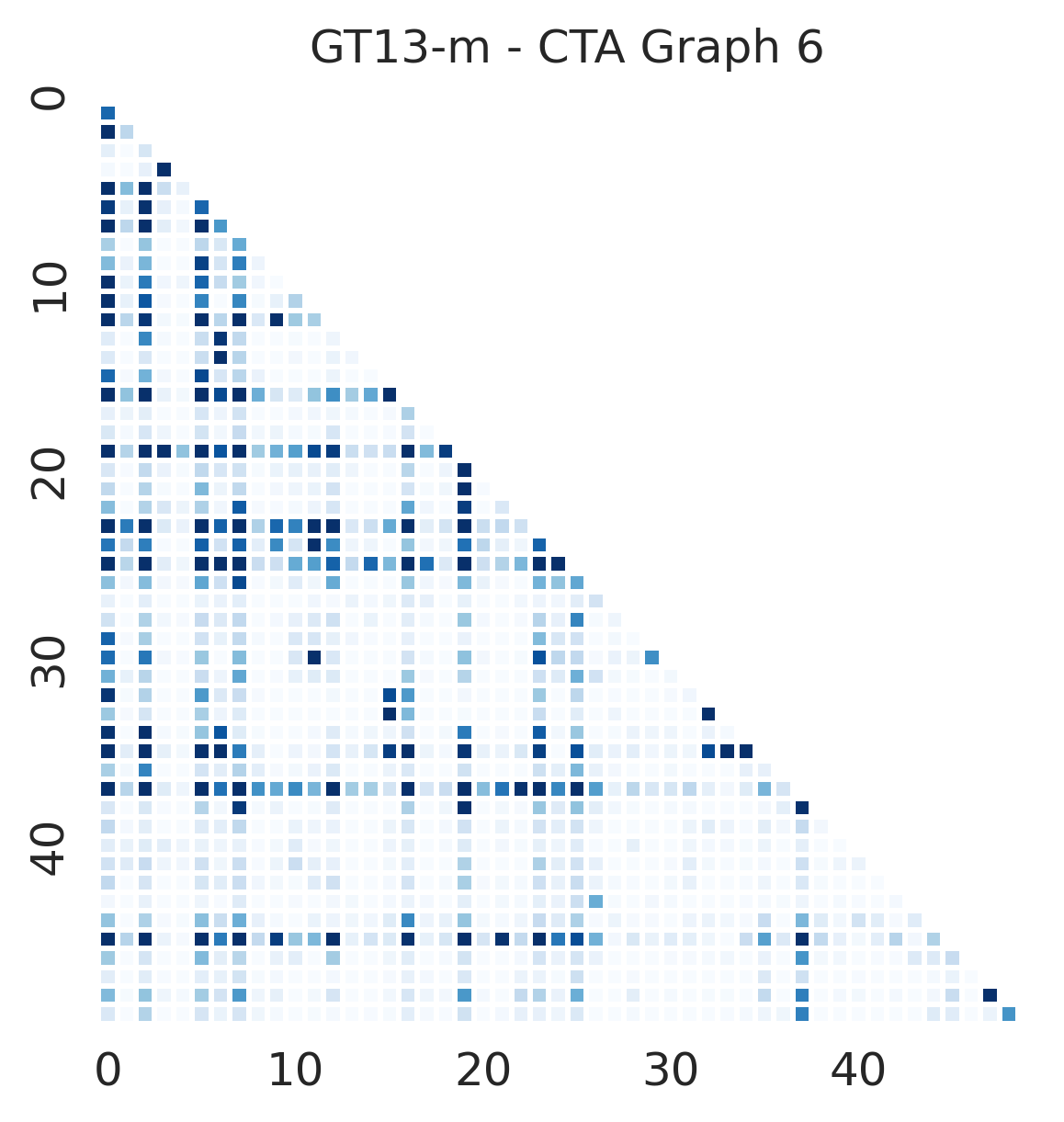}
  \includegraphics[width=0.19\textwidth]{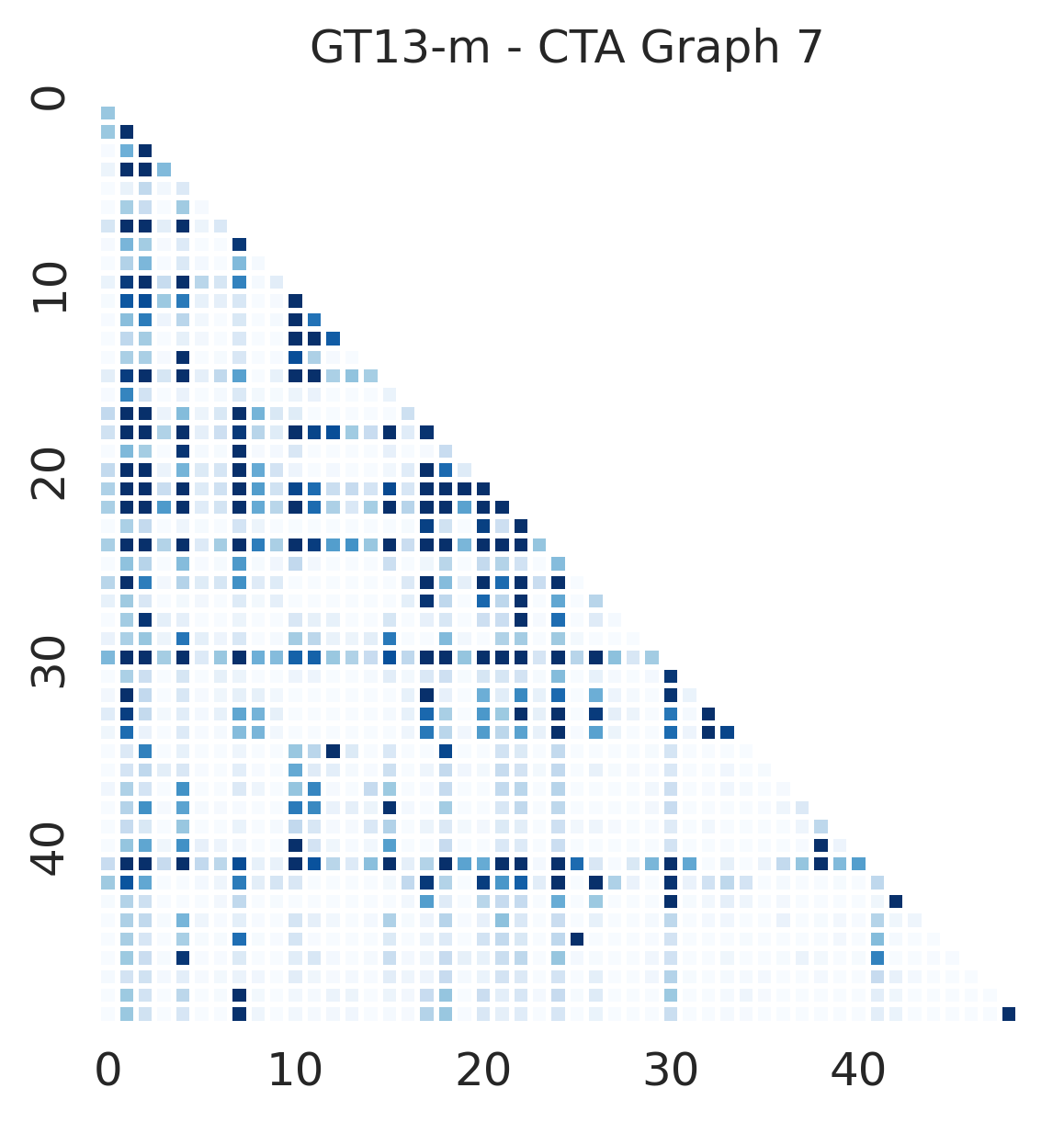}
  \includegraphics[width=0.19\textwidth]{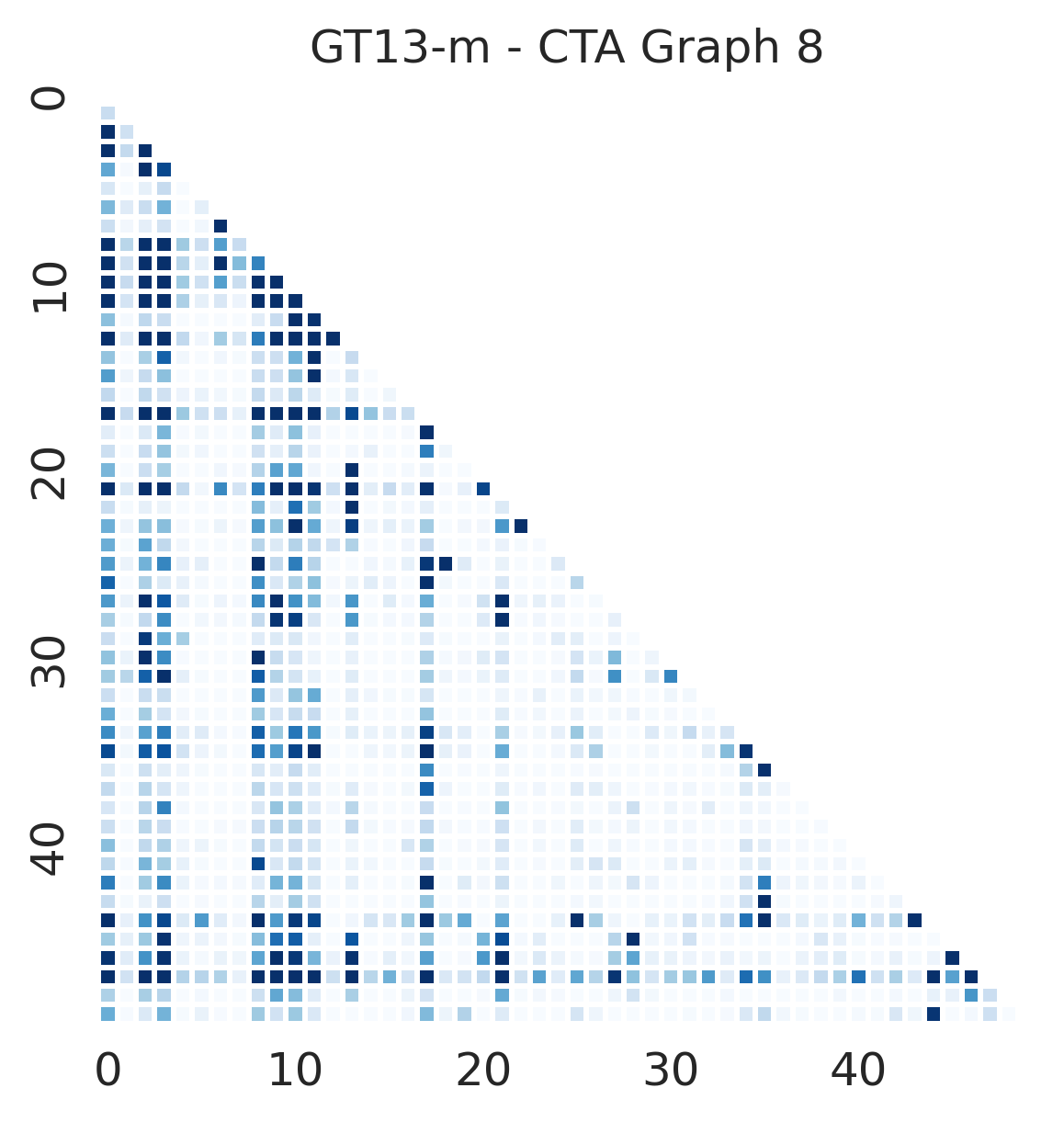}
  \includegraphics[width=0.19\textwidth]{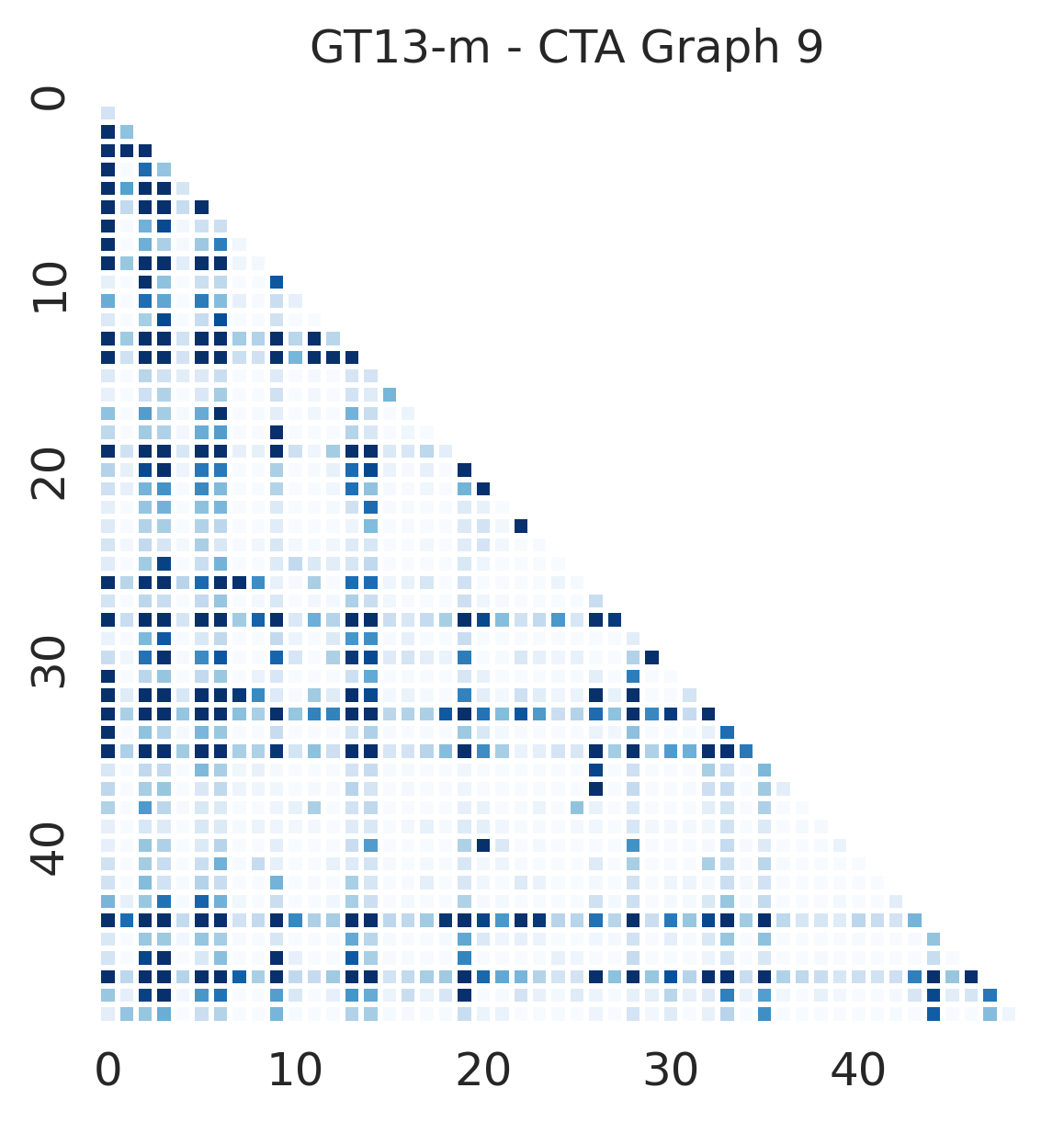}
  \includegraphics[width=0.19\textwidth]{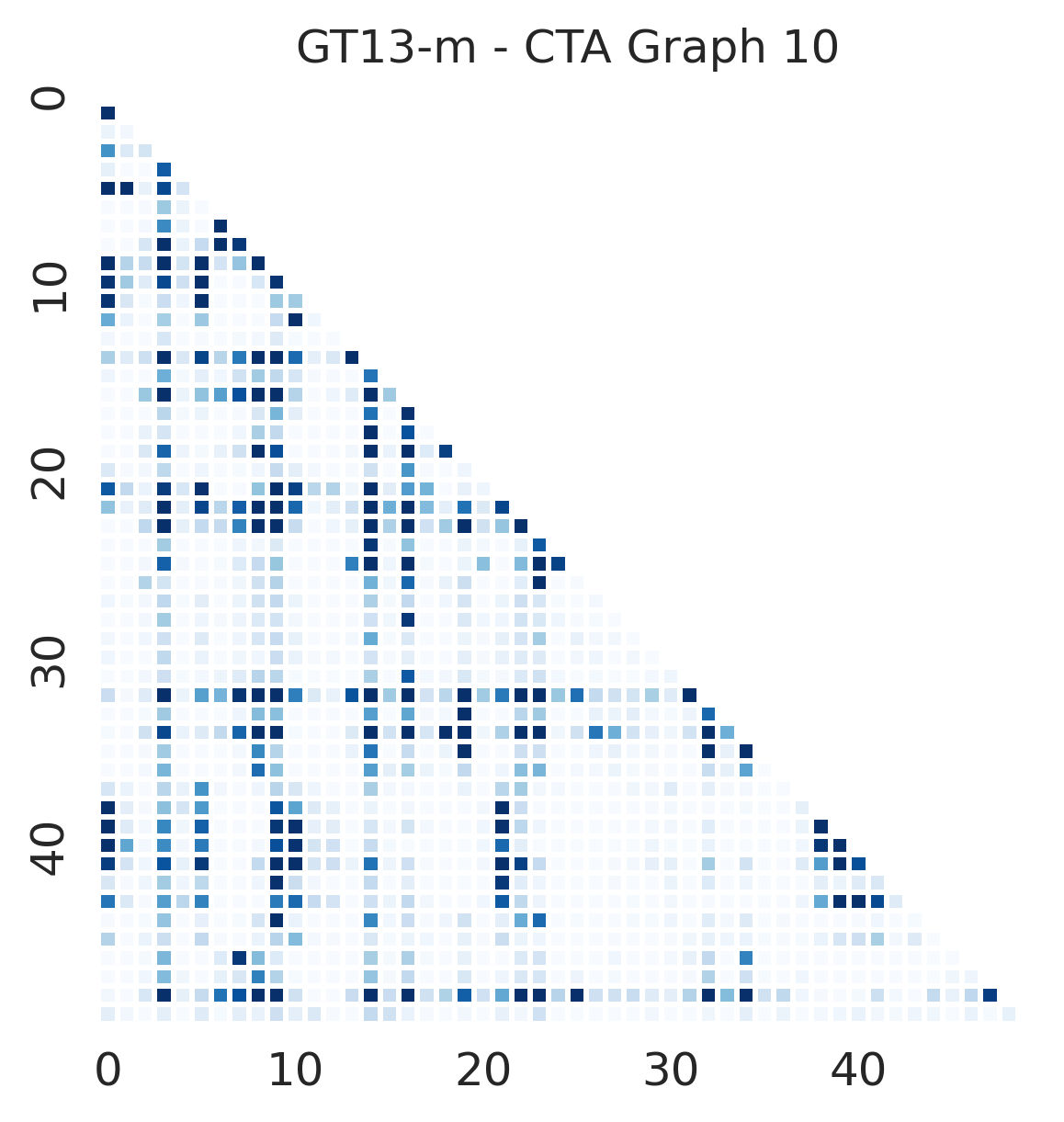}
  
  \includegraphics[width=0.19\textwidth]{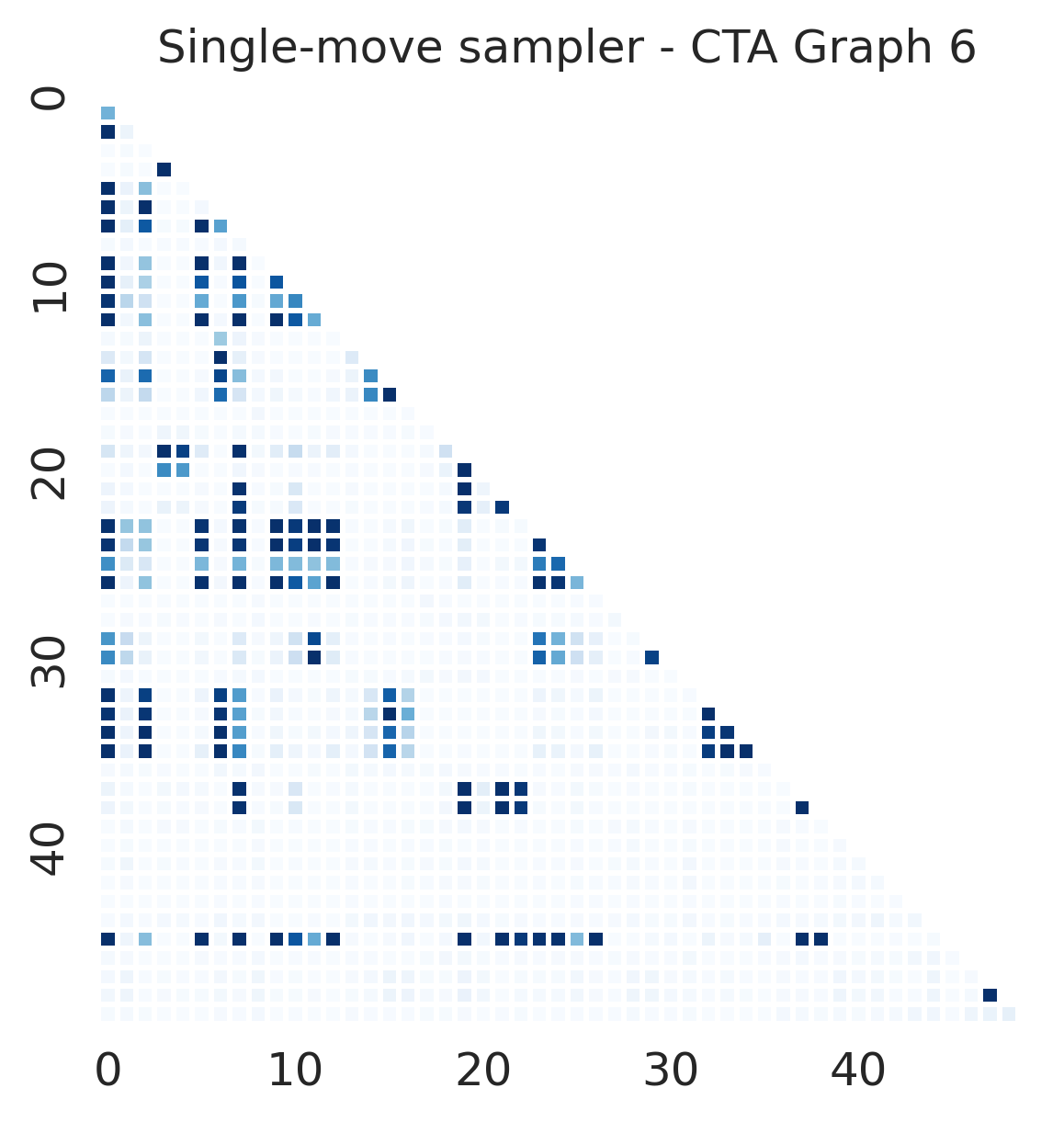}
  \includegraphics[width=0.19\textwidth]{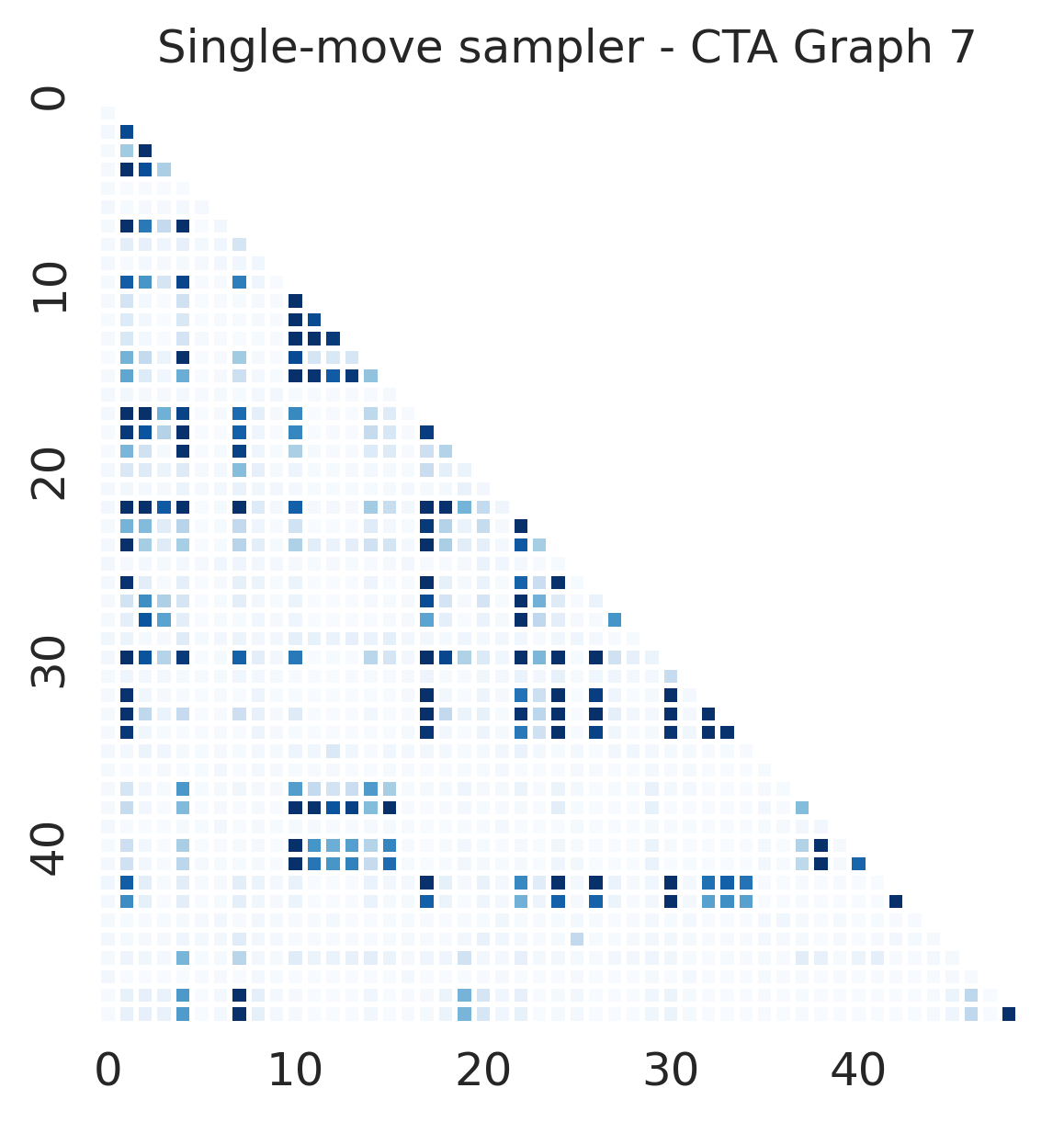}
  \includegraphics[width=0.19\textwidth]{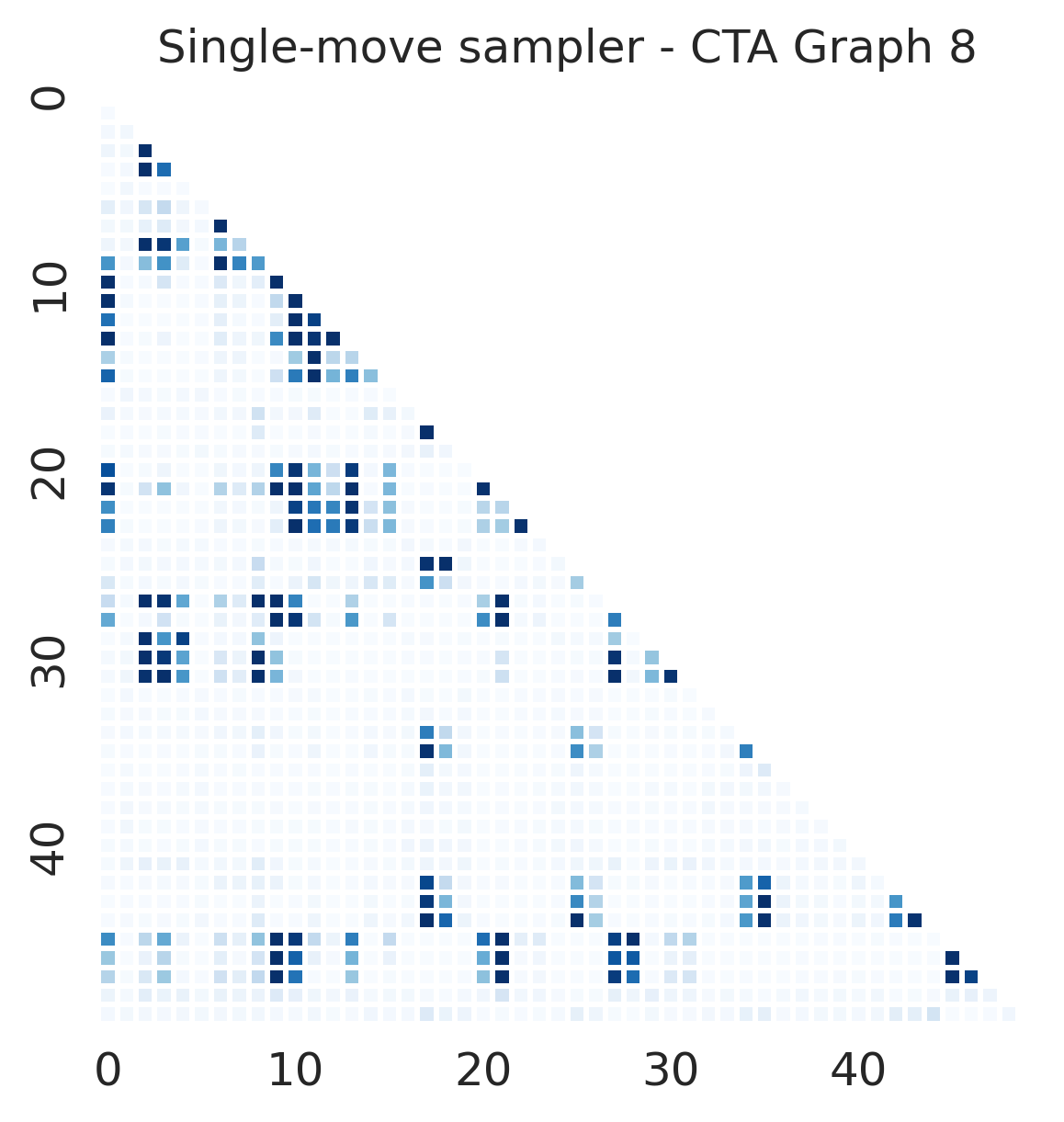}
  \includegraphics[width=0.19\textwidth]{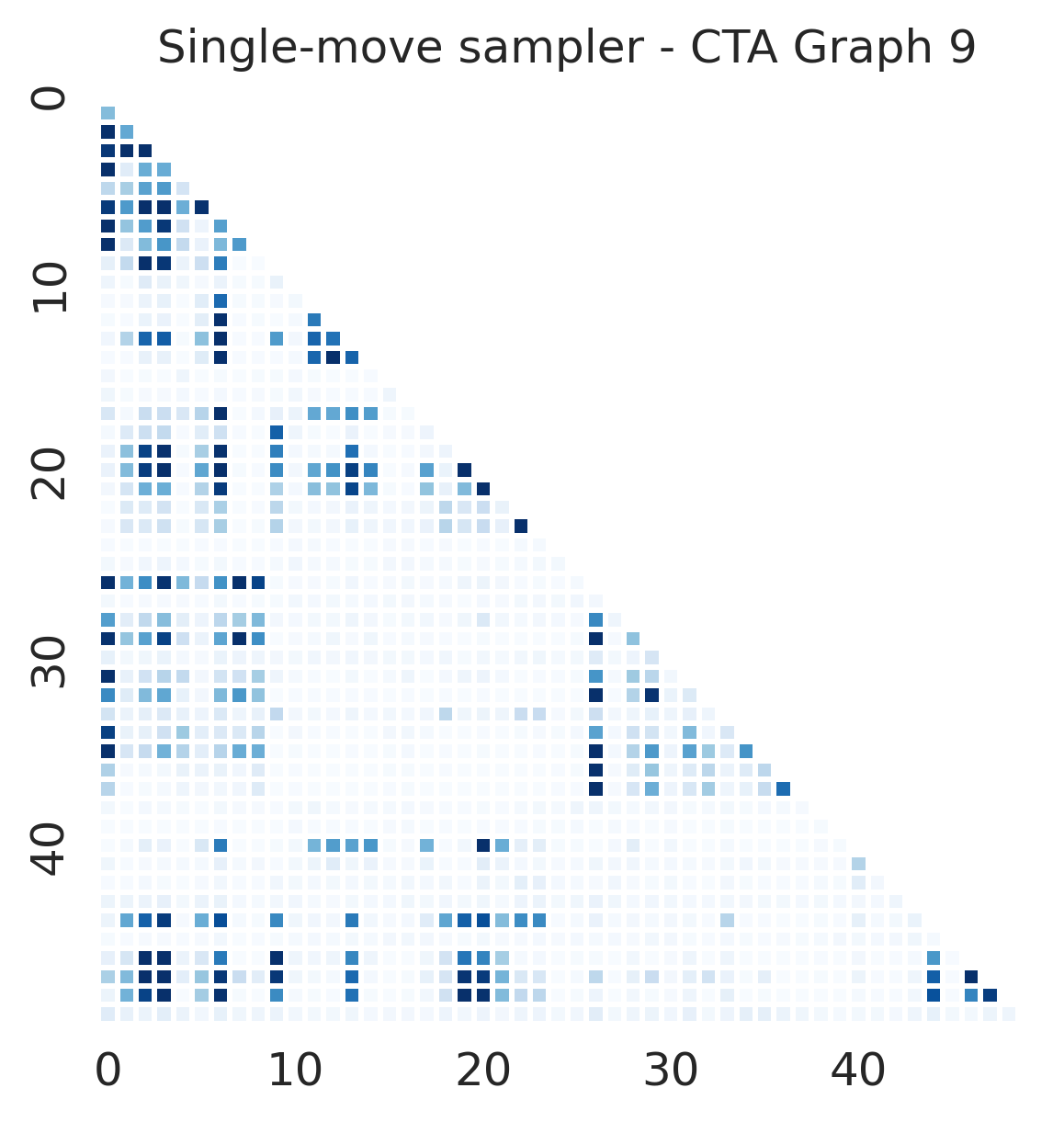}
  \includegraphics[width=0.19\textwidth]{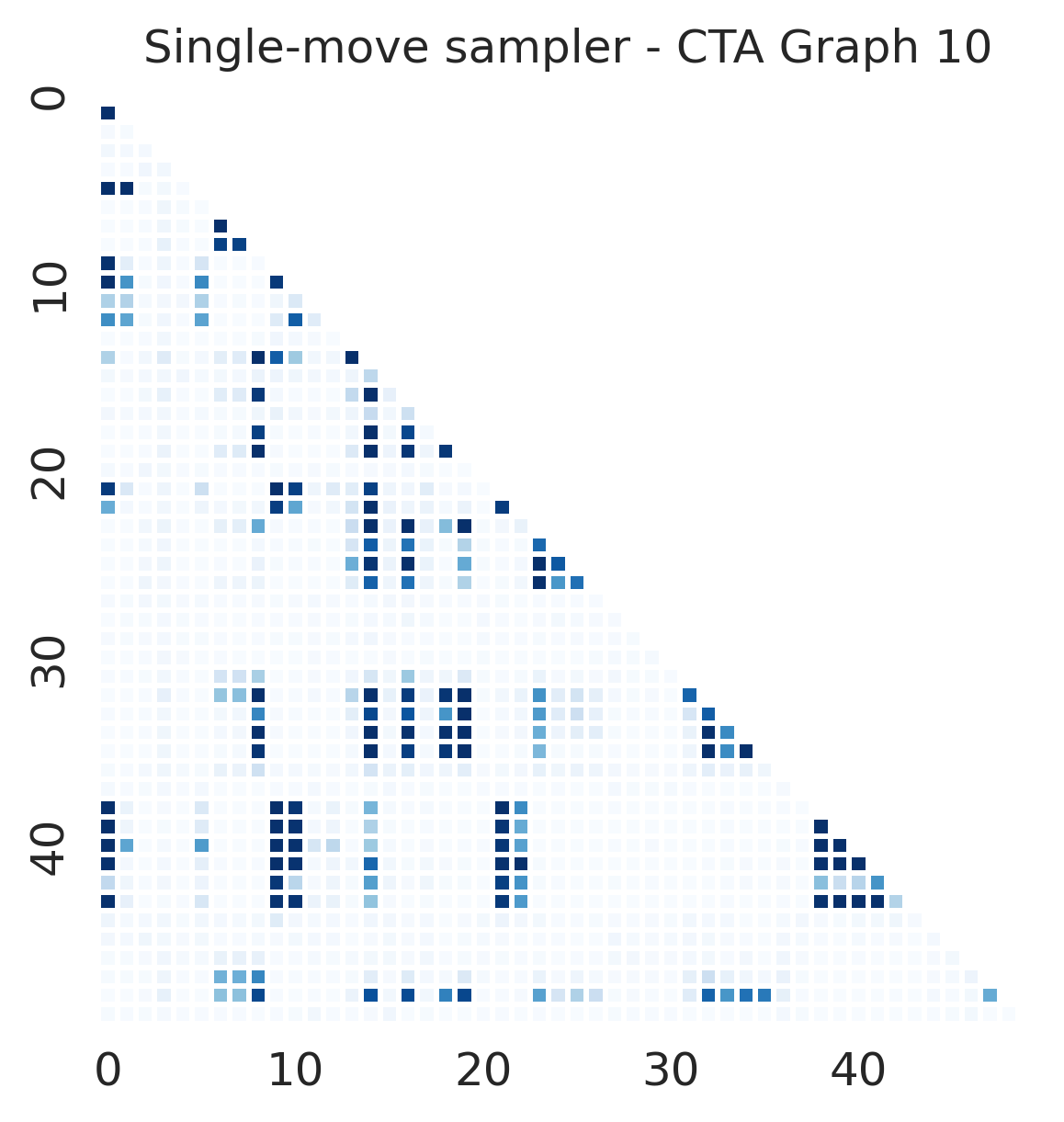}

  \includegraphics[width=0.19\textwidth]{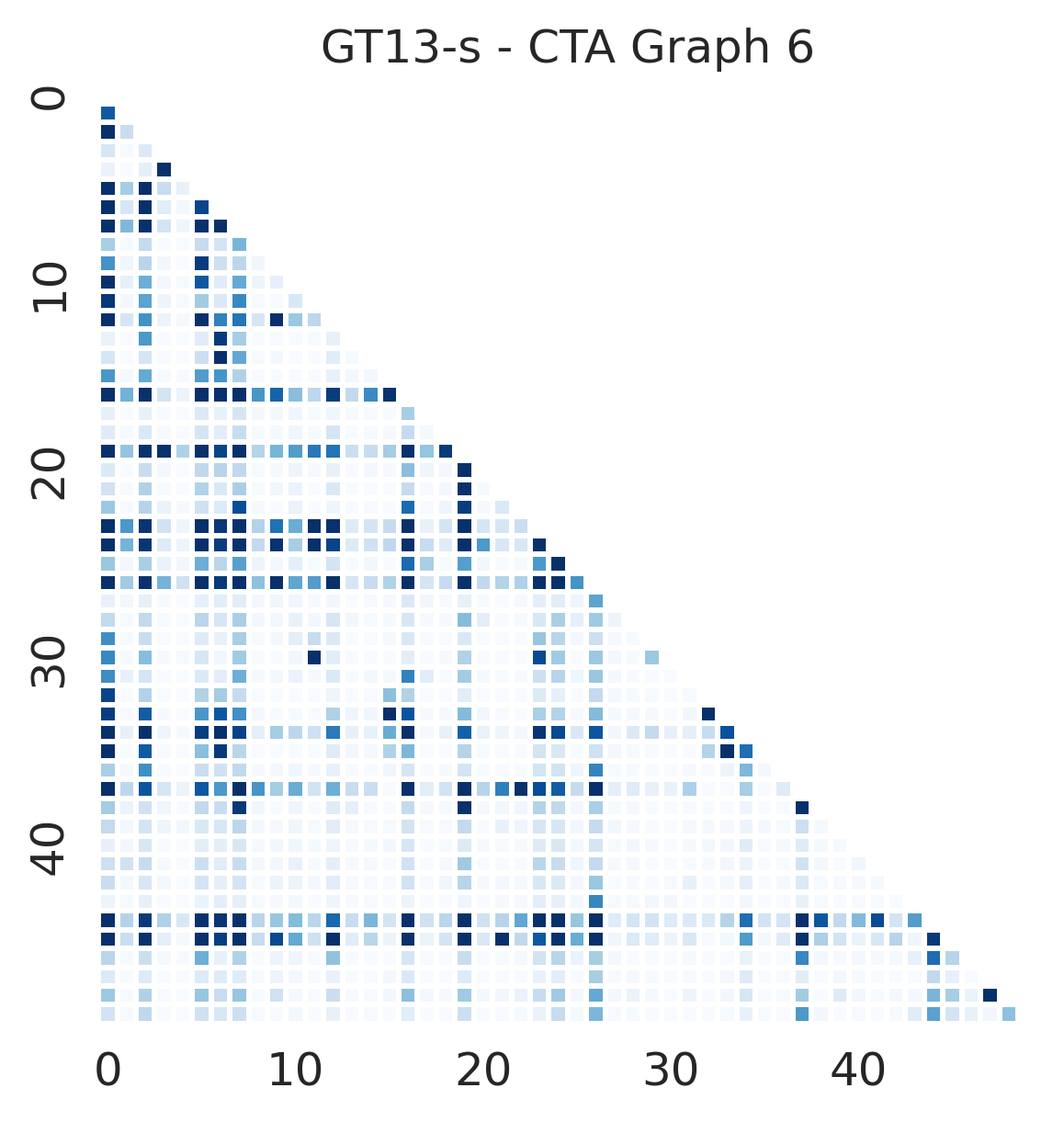}
  \includegraphics[width=0.19\textwidth]{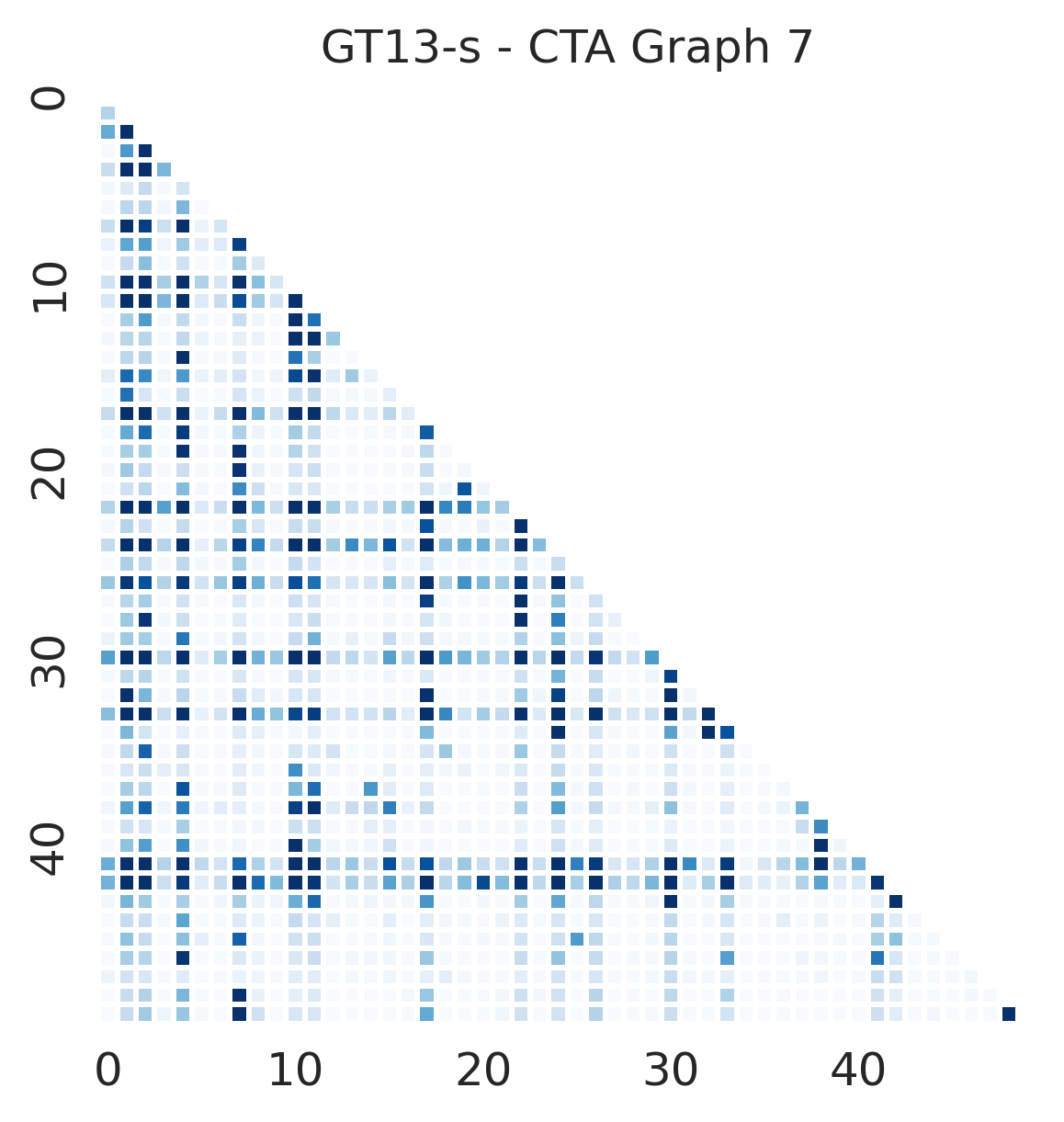}
  \includegraphics[width=0.19\textwidth]{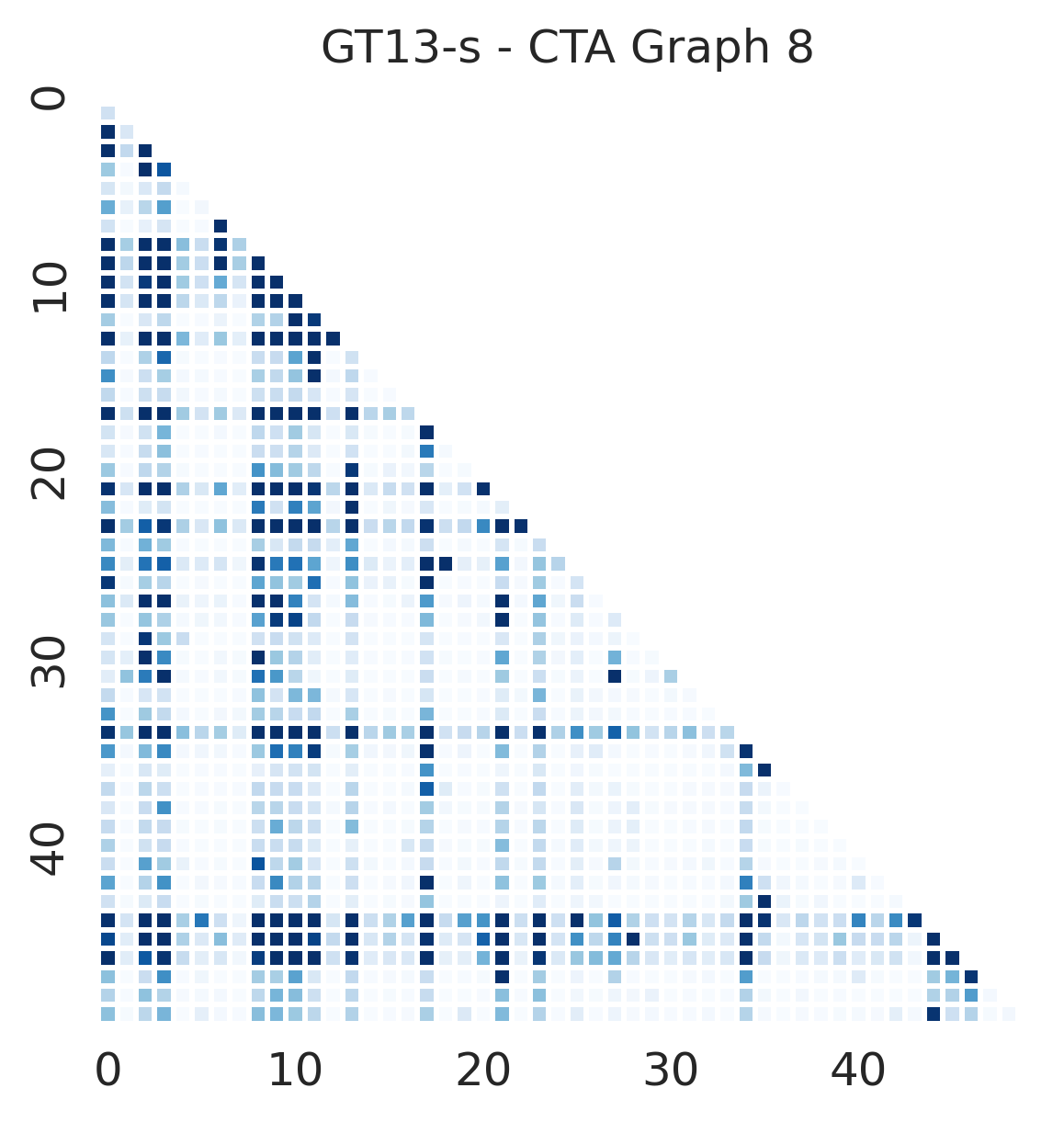}
  \includegraphics[width=0.19\textwidth]{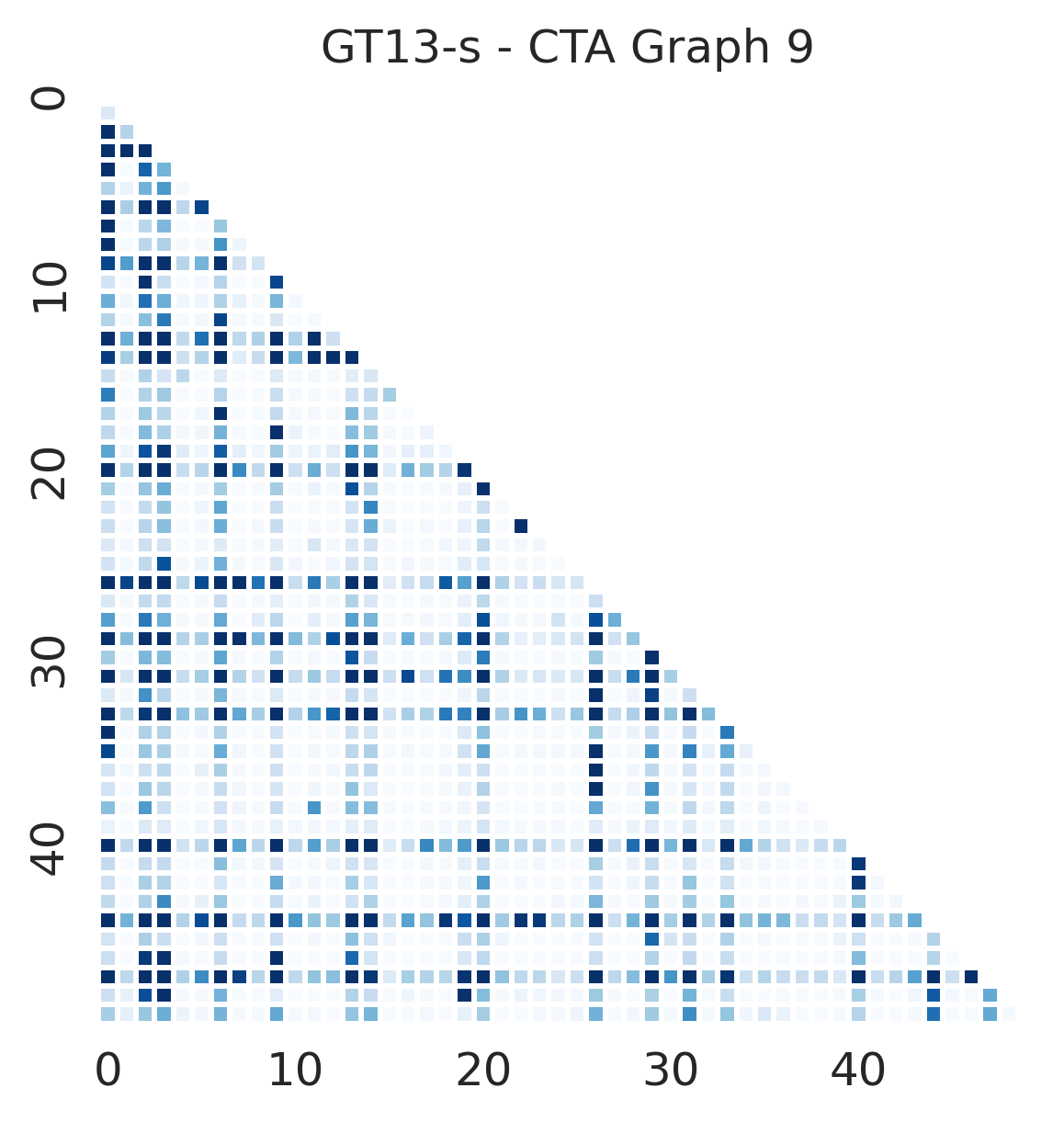}
  \includegraphics[width=0.19\textwidth]{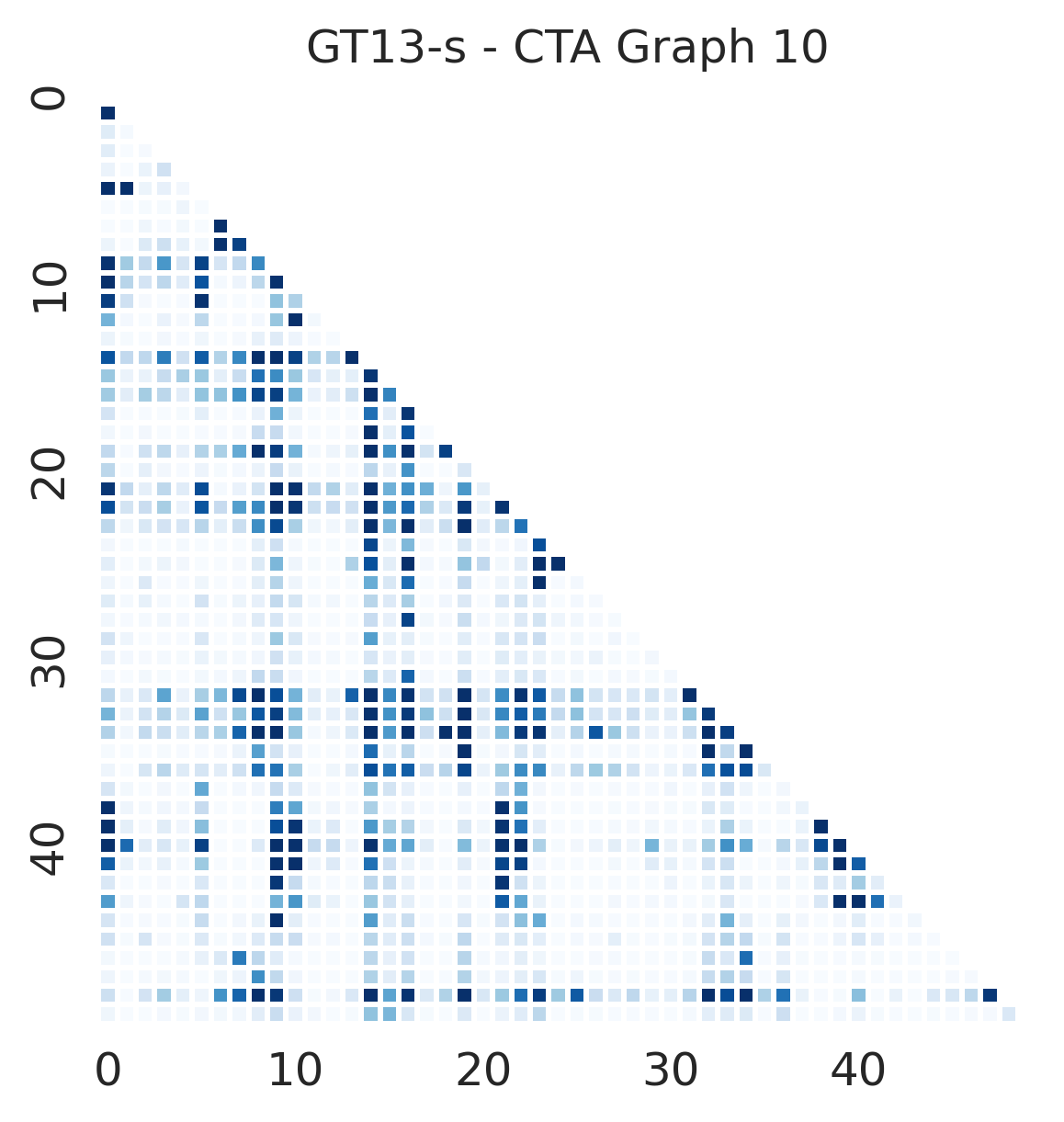}

  \includegraphics[width=0.19\textwidth]{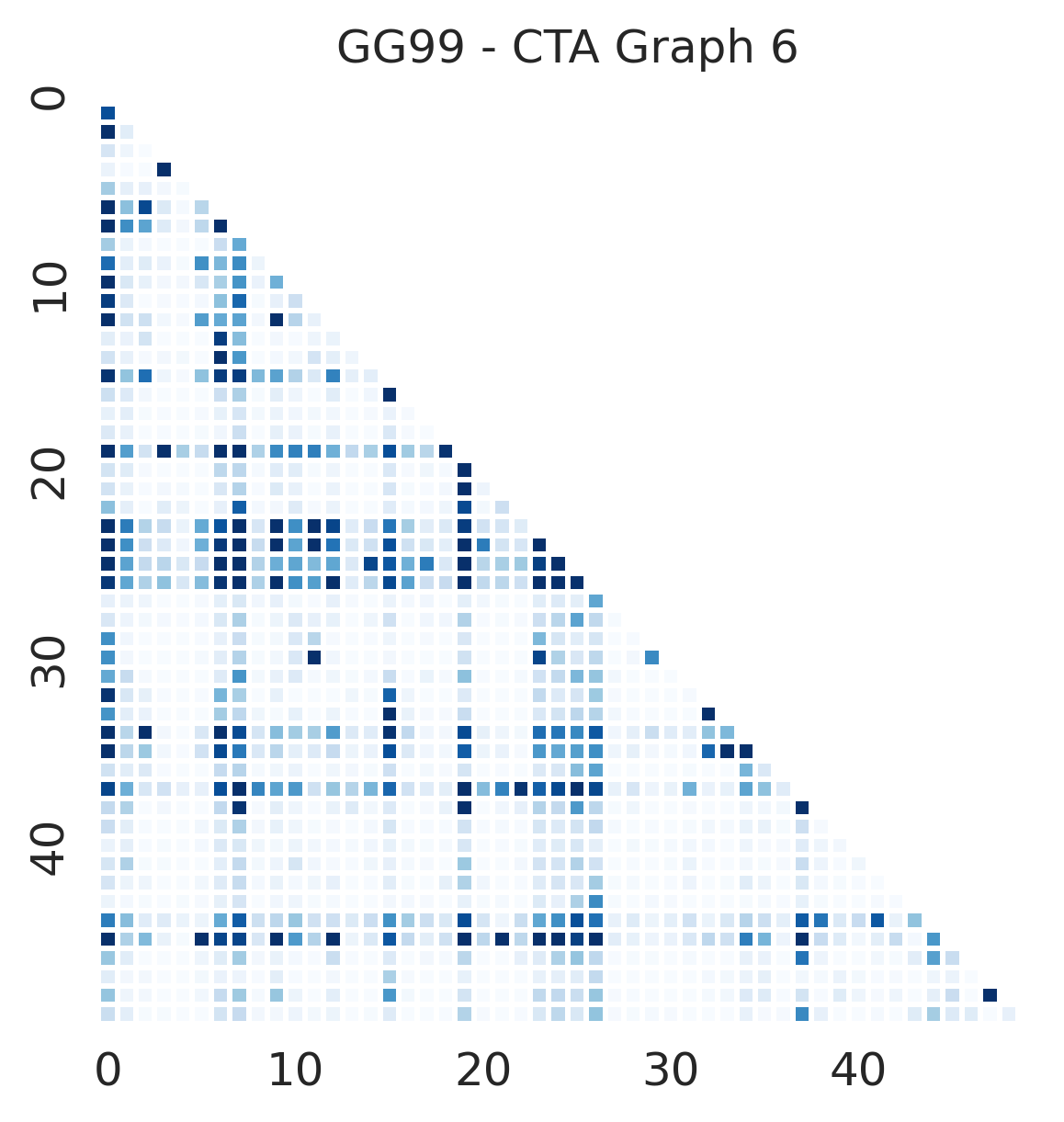}
  \includegraphics[width=0.19\textwidth]{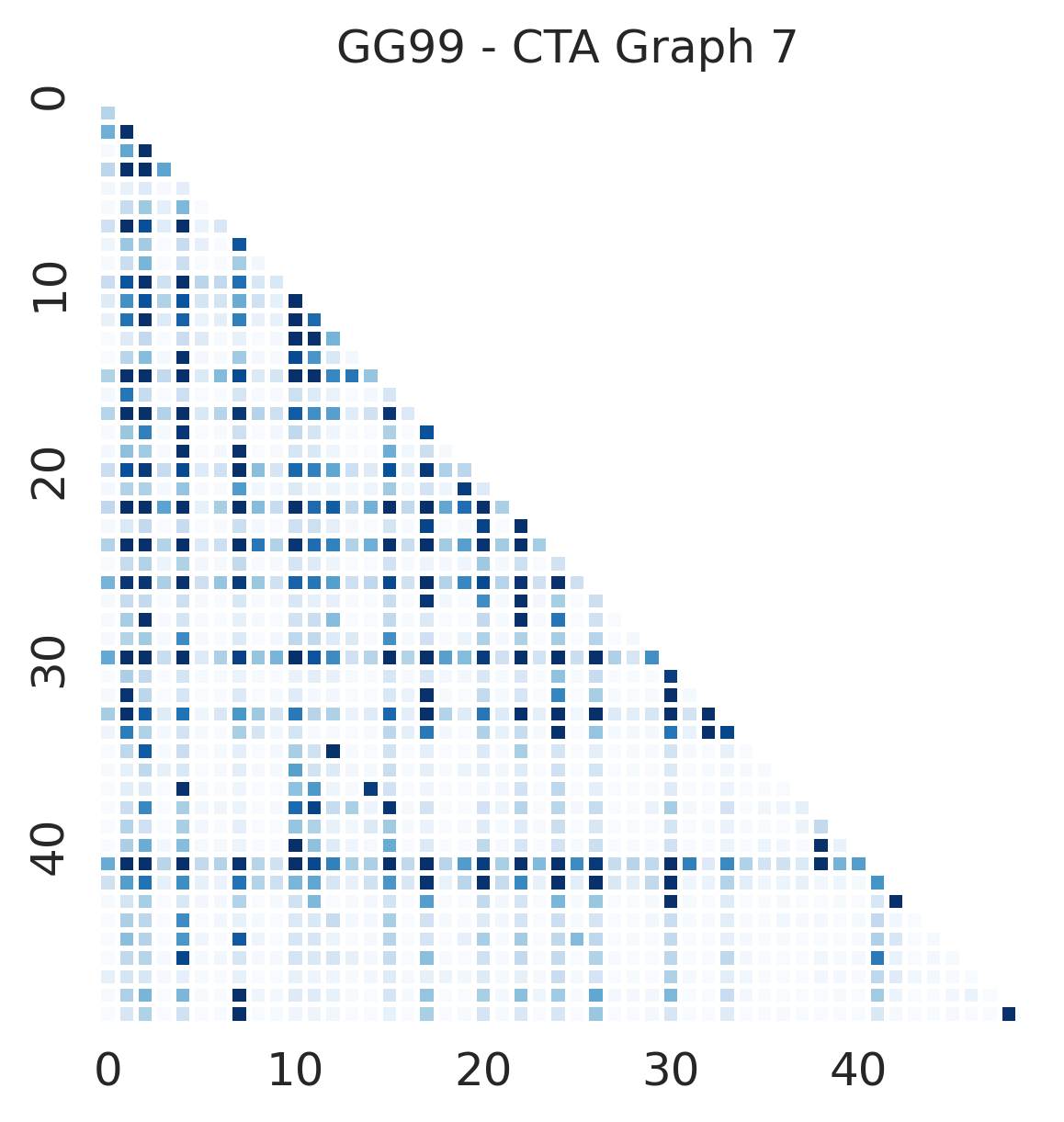}
  \includegraphics[width=0.19\textwidth]{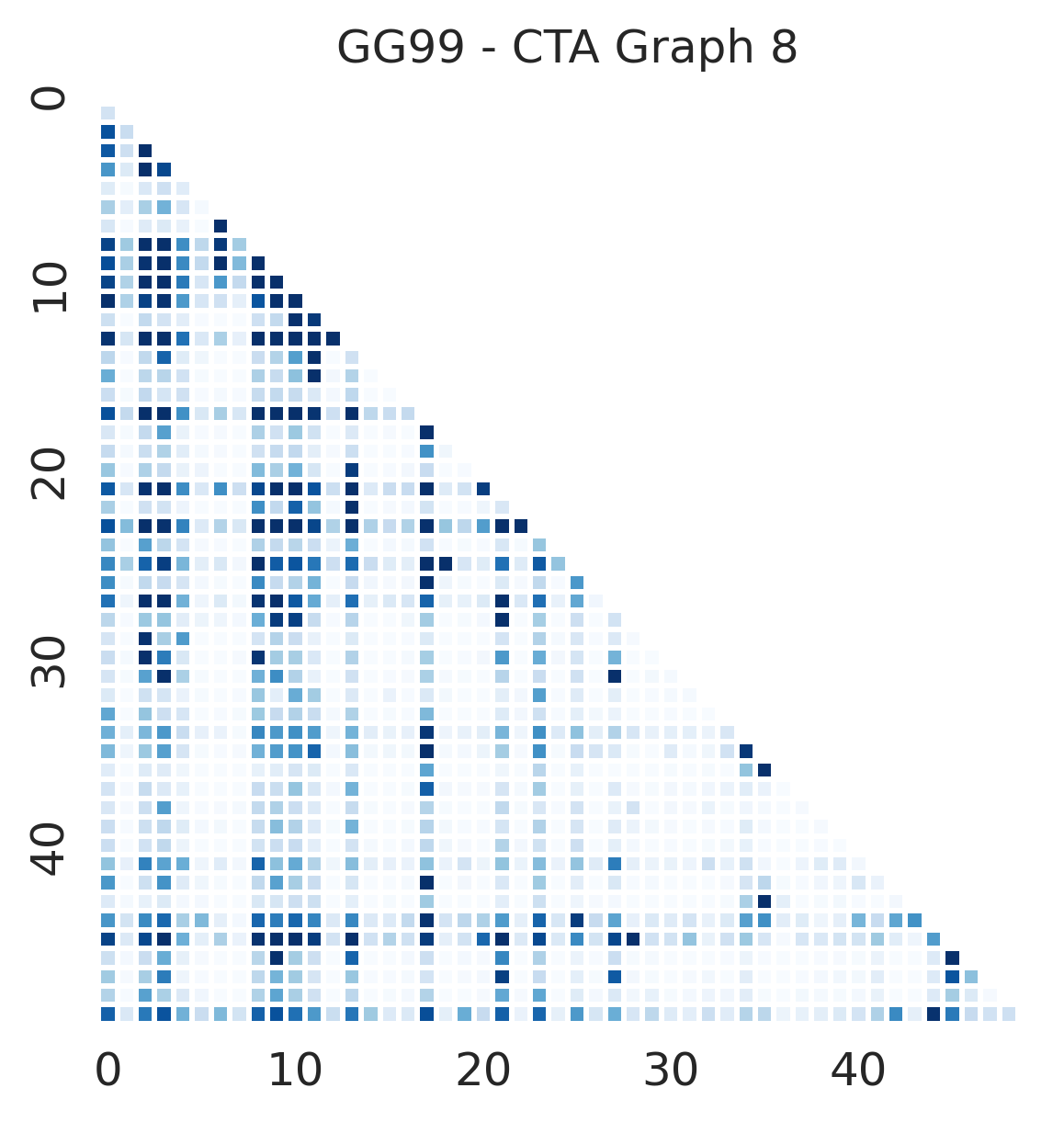}
  \includegraphics[width=0.19\textwidth]{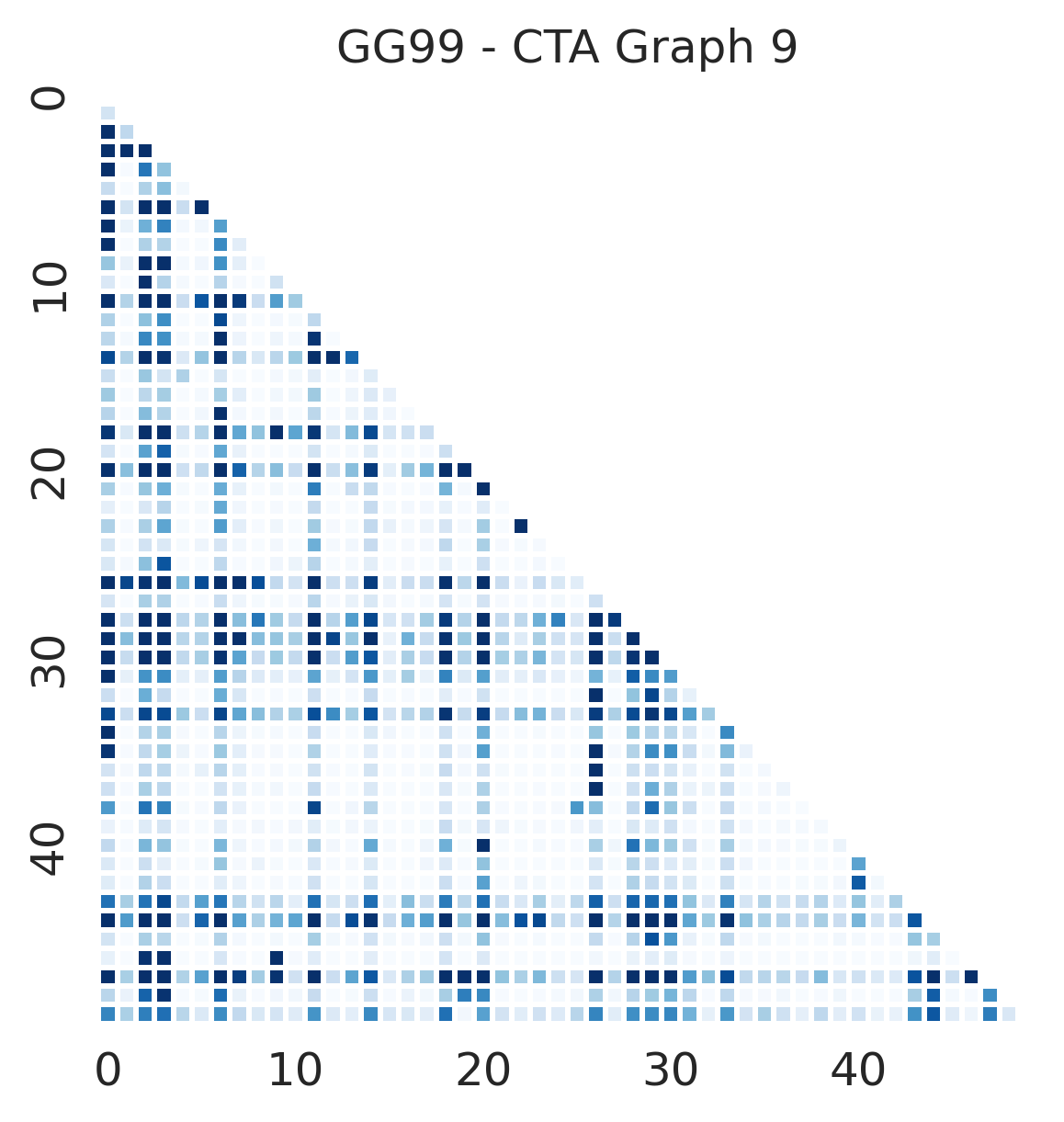}
  \includegraphics[width=0.19\textwidth]{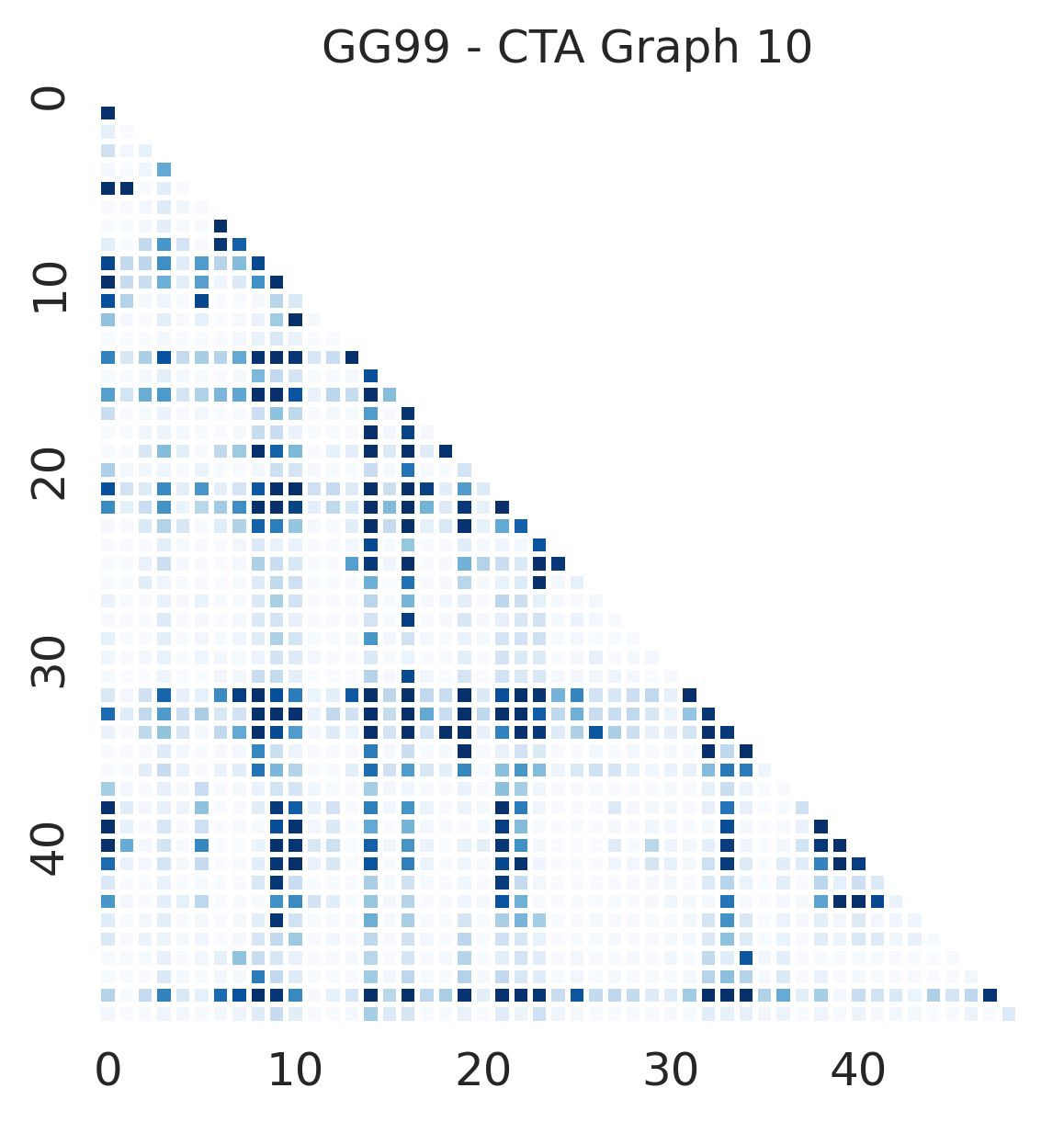}

  \includegraphics[width=0.19\textwidth]{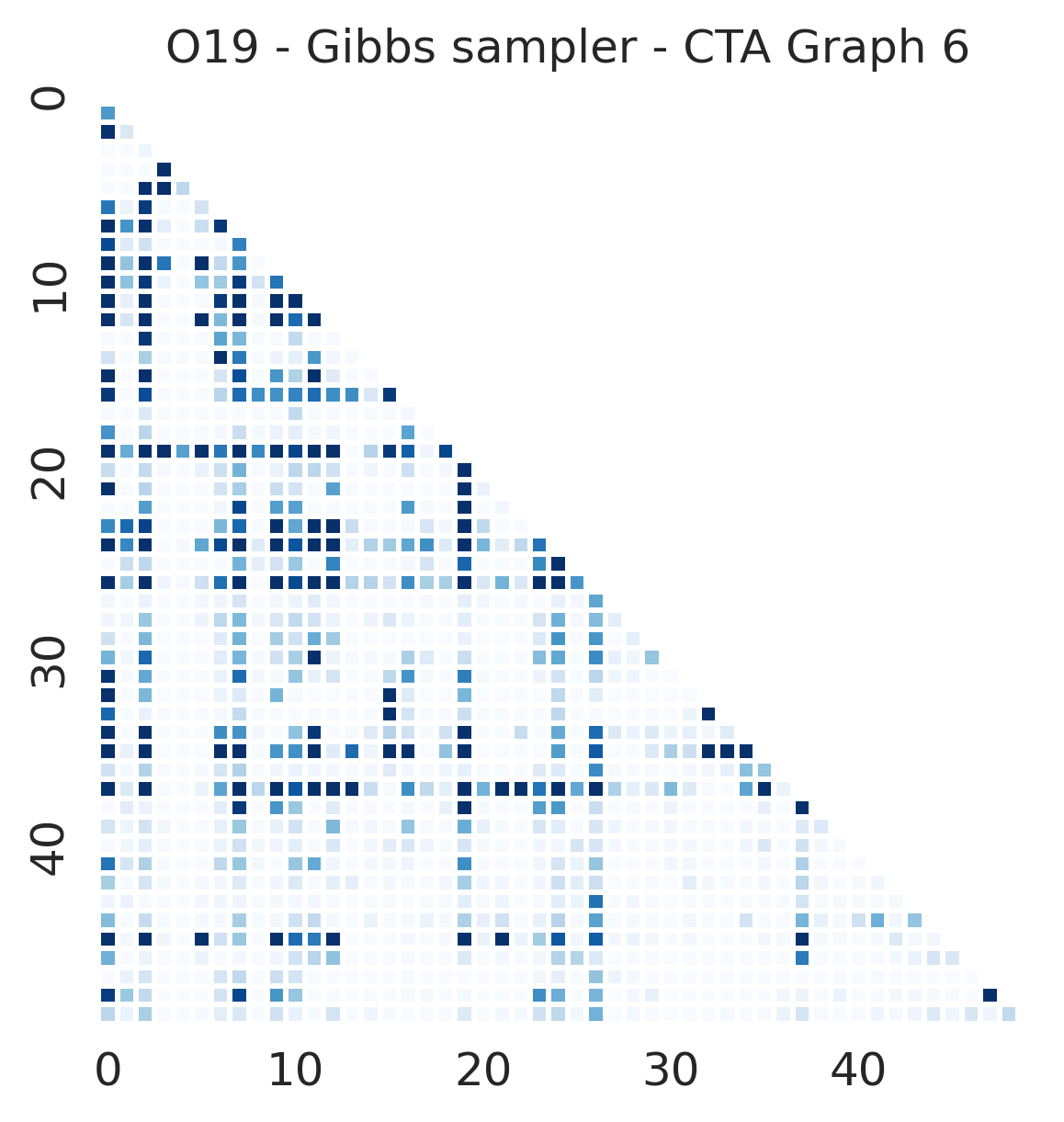}
  \includegraphics[width=0.19\textwidth]{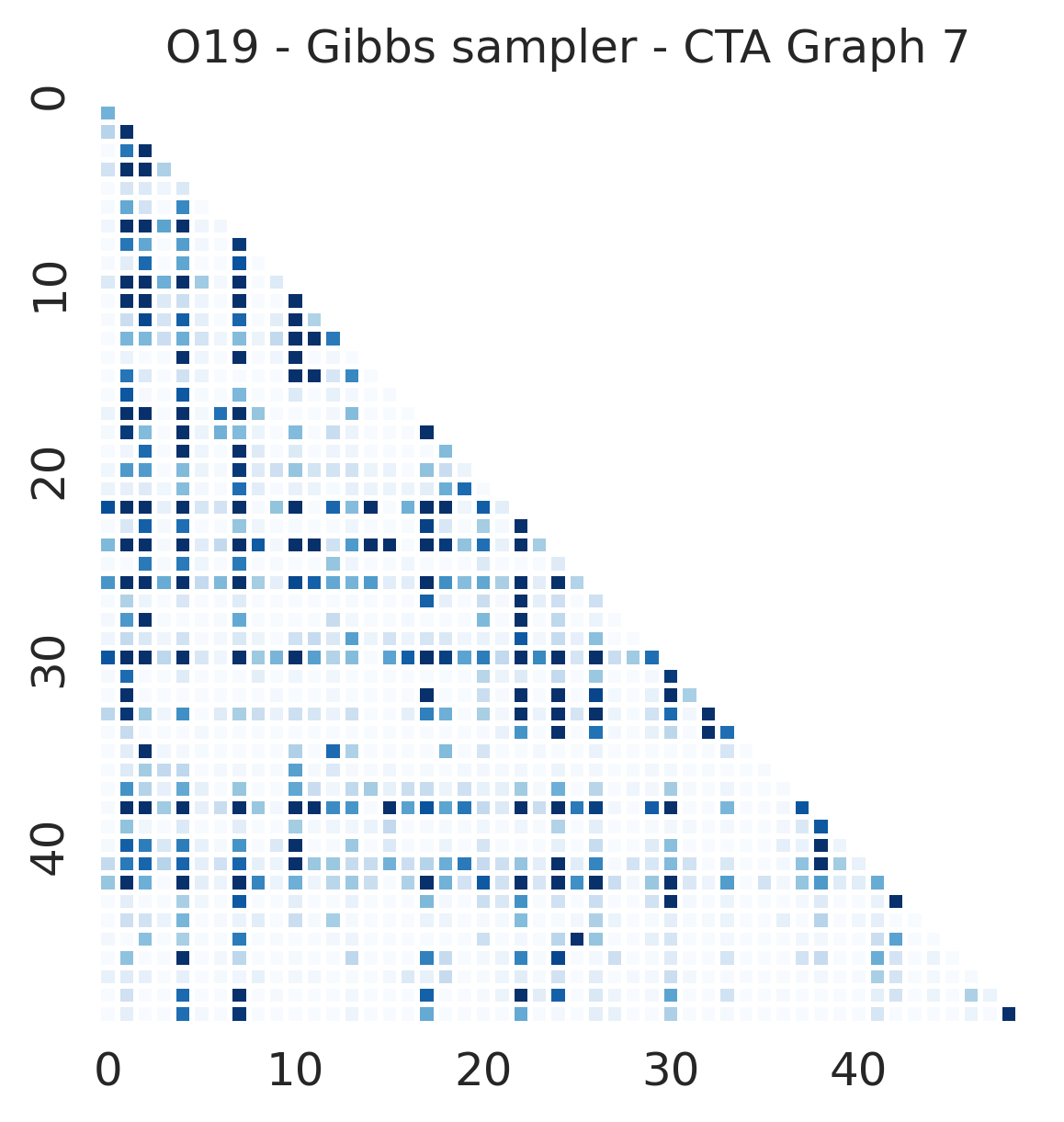}
  \includegraphics[width=0.19\textwidth]{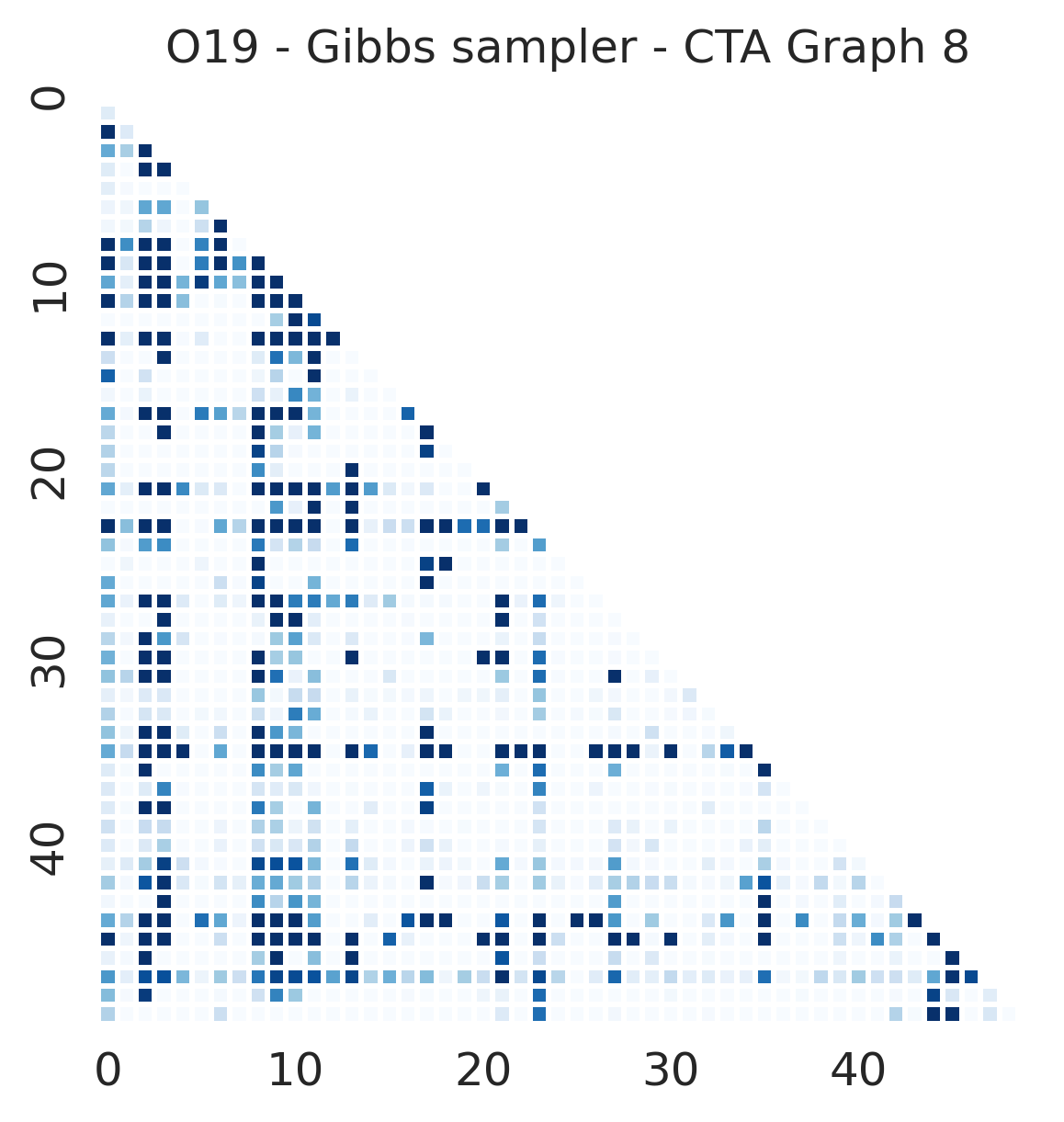}
  \includegraphics[width=0.19\textwidth]{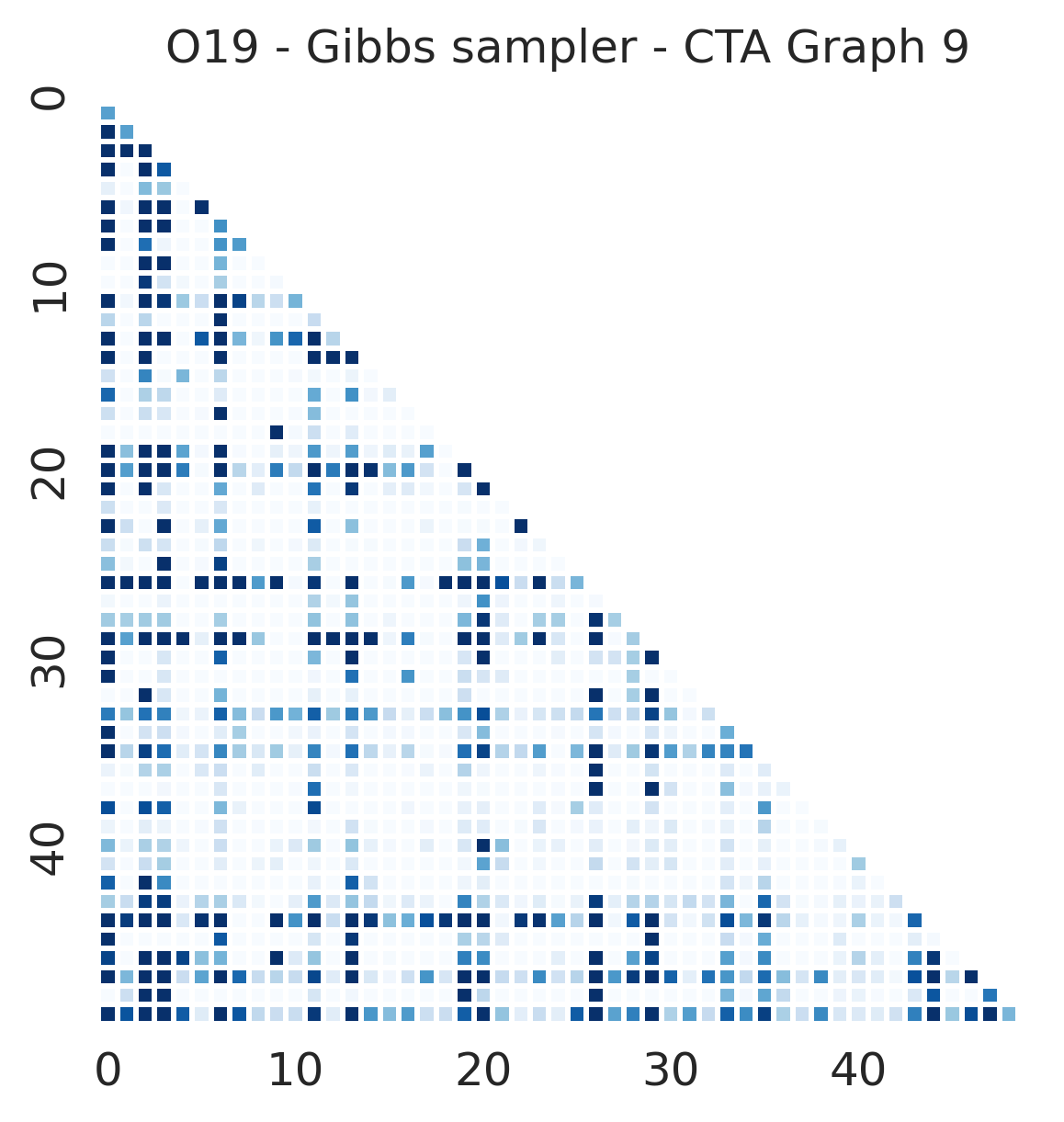}
  \includegraphics[width=0.19\textwidth]{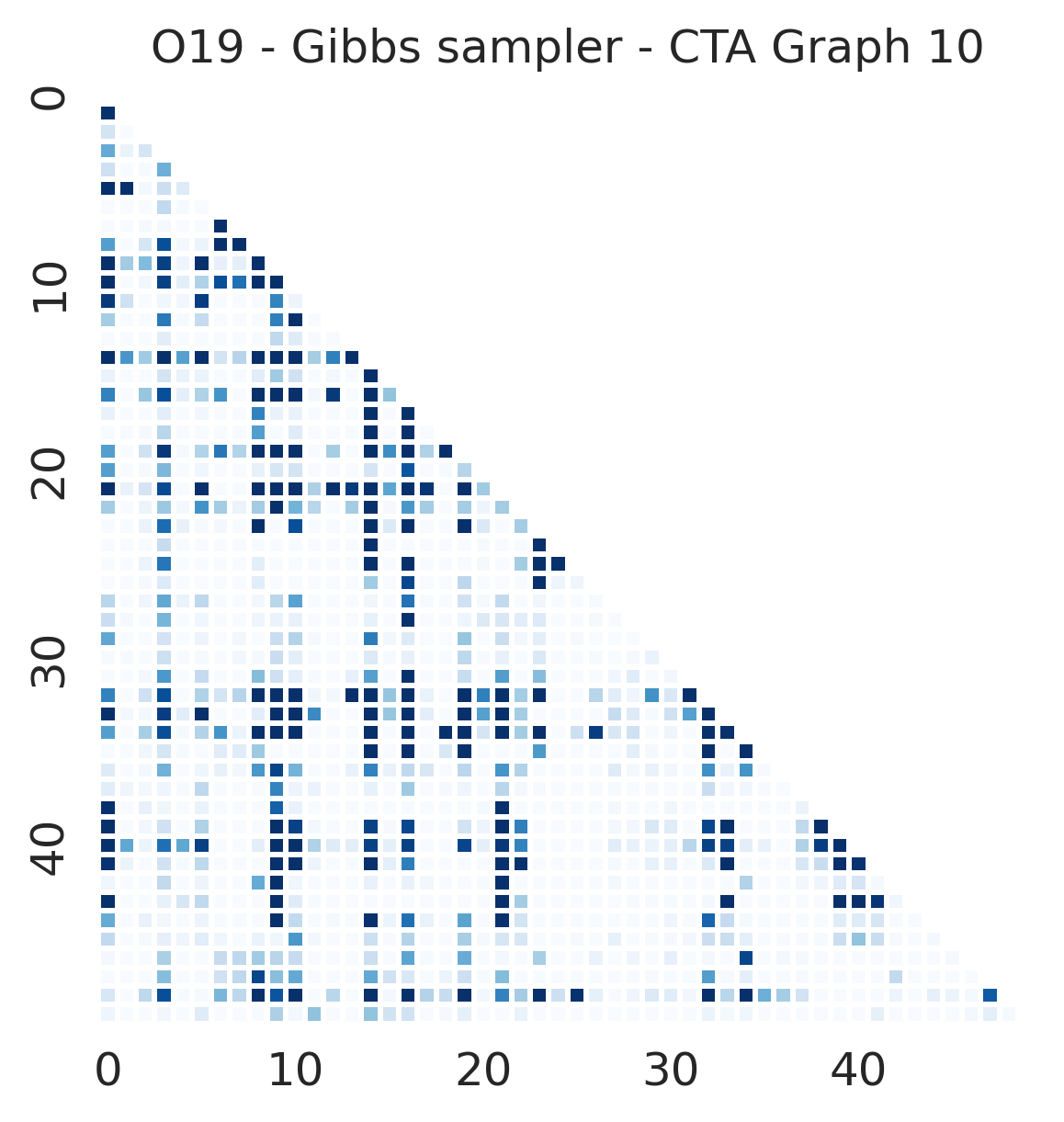}

  \caption{Heatmap derived from the final 300,000 steps of samplers in Section~\ref{sec:simul-setup-gauss}, covering replications 5 to 10 under the Christmas tree algorithm and $(n,p)=(100,50)$.}
  \label{app:fig:heatmap-cta-1}
\end{figure}

\begin{figure}[ht]
  \centering
  \includegraphics[width=0.19\textwidth]{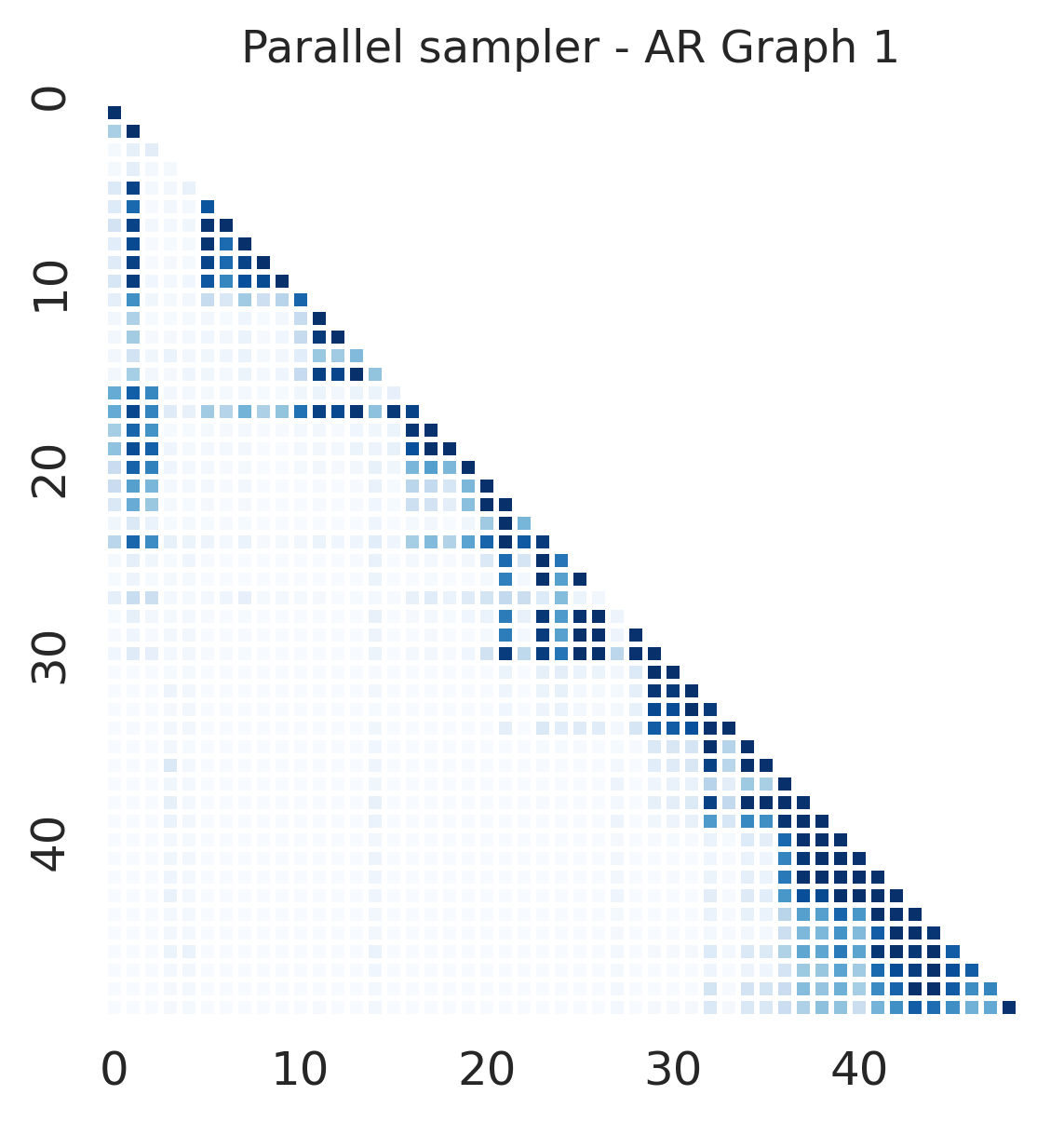}
  \includegraphics[width=0.19\textwidth]{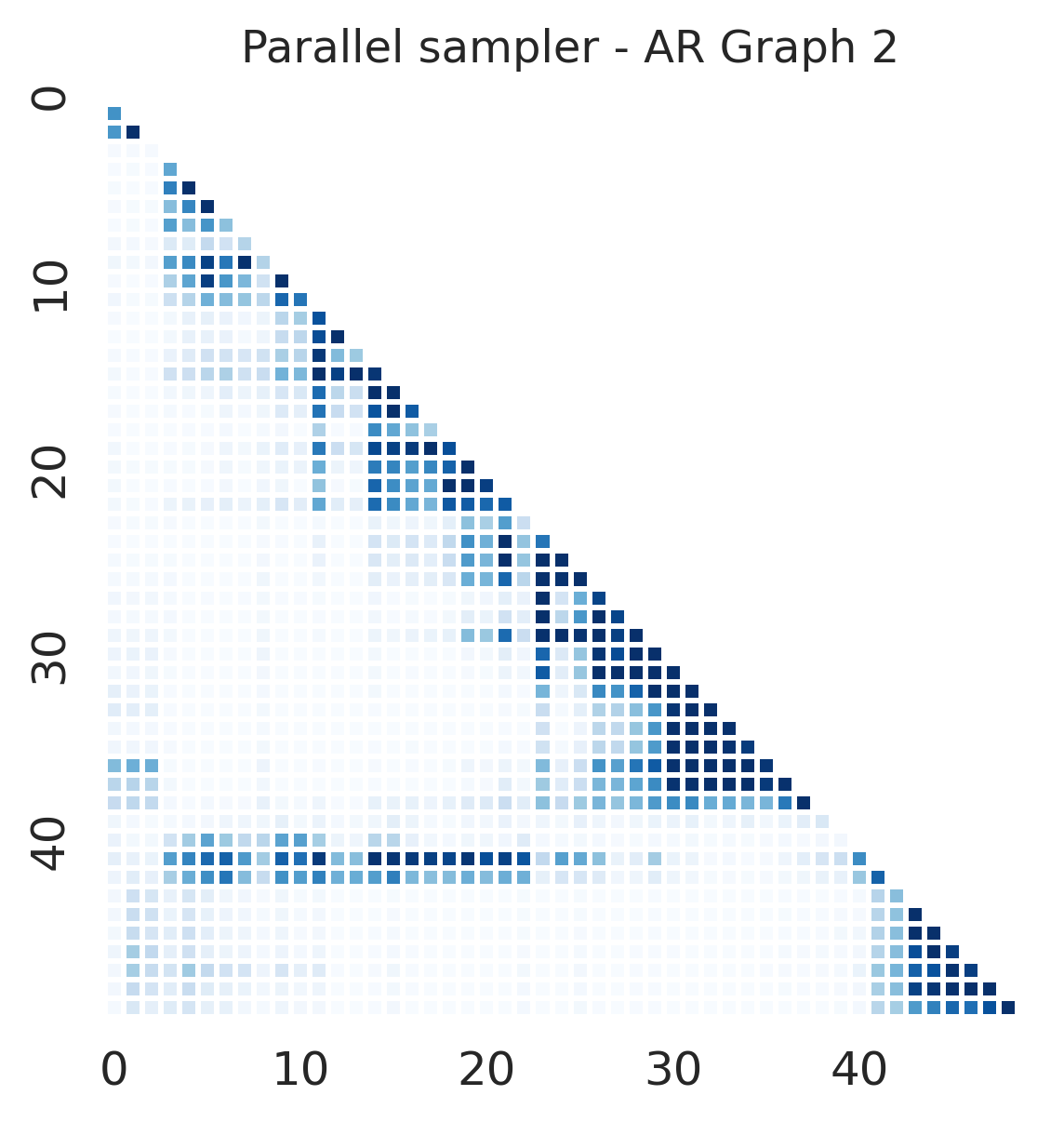}
  \includegraphics[width=0.19\textwidth]{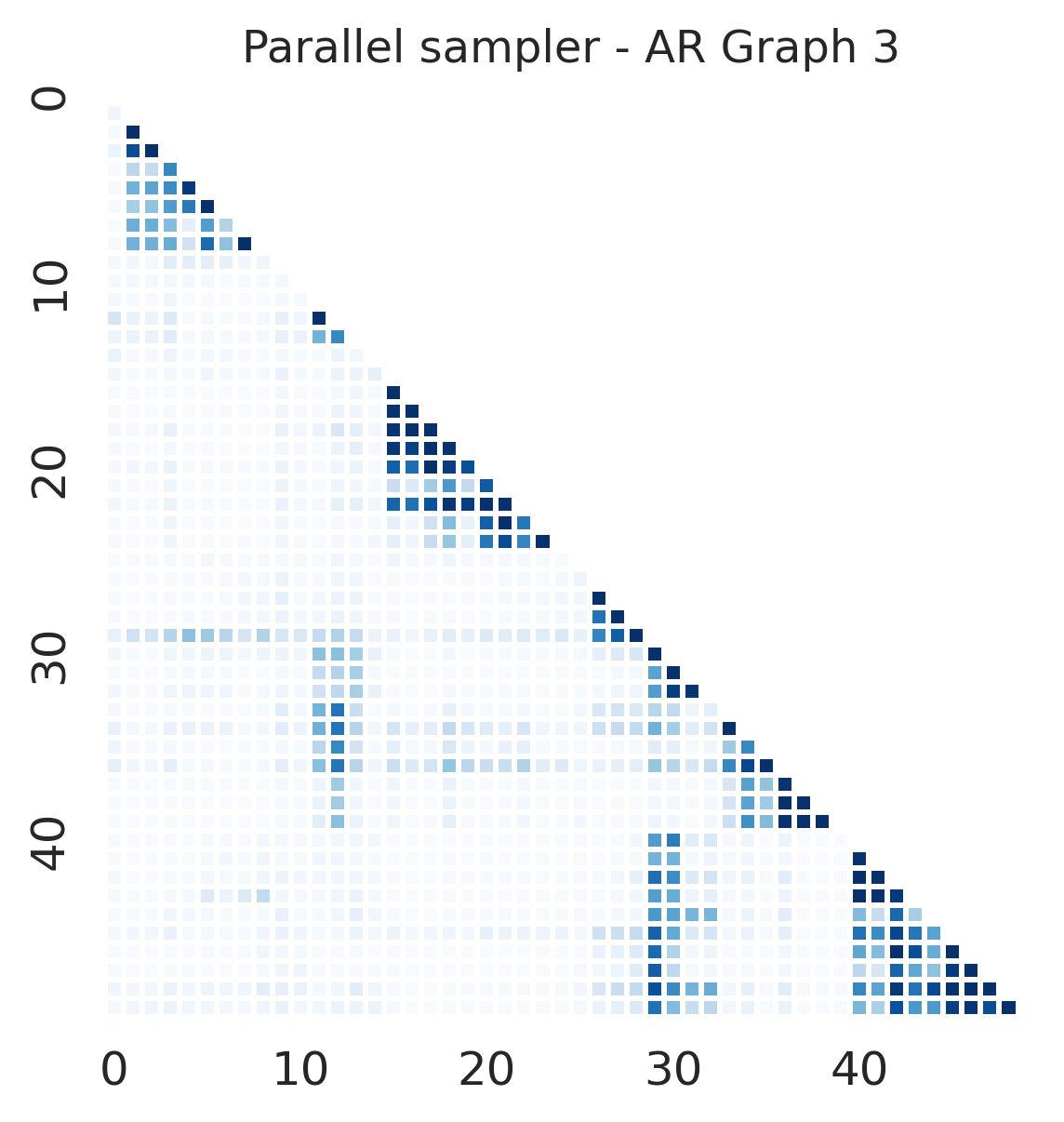}
  \includegraphics[width=0.19\textwidth]{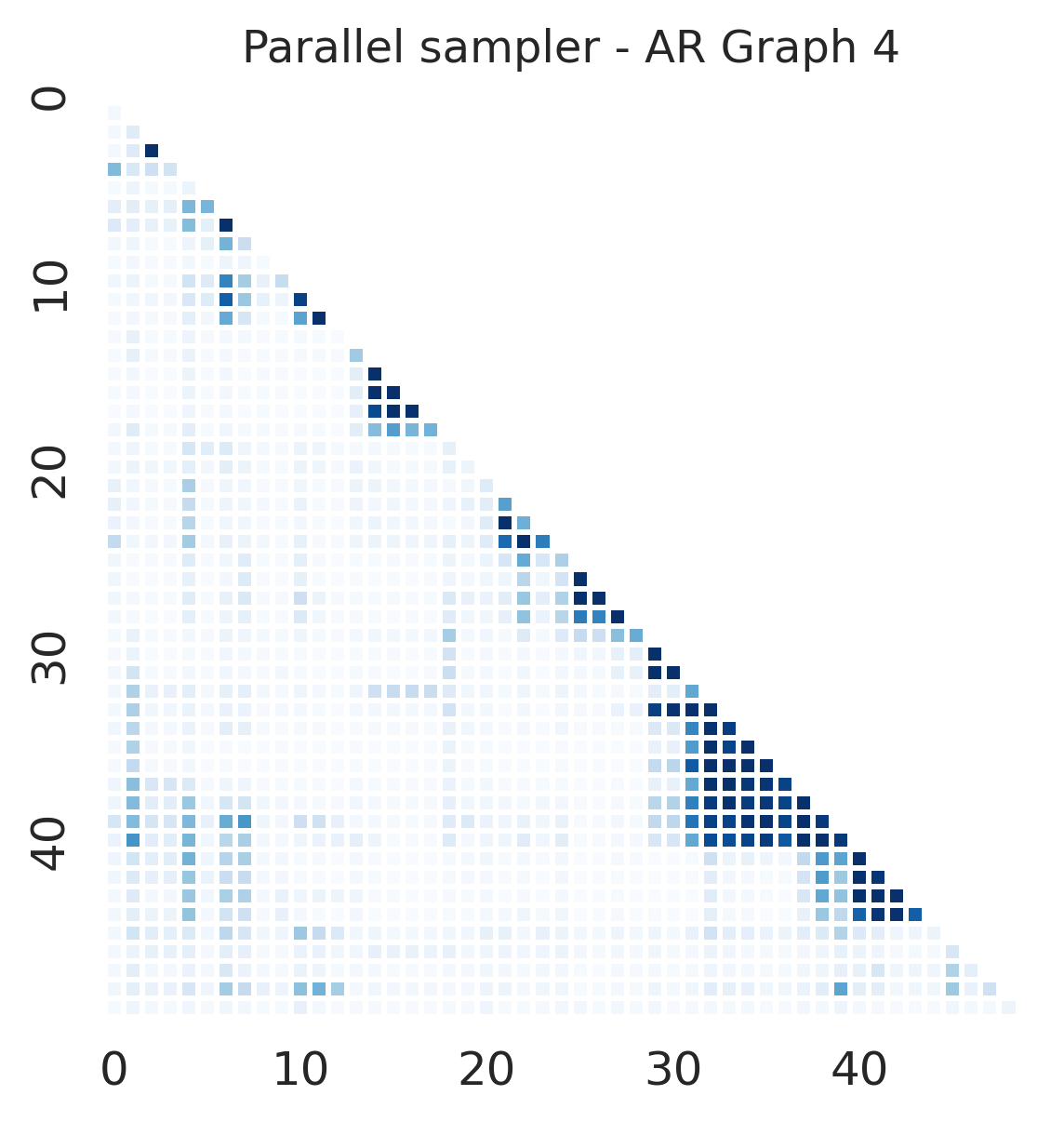}
  \includegraphics[width=0.19\textwidth]{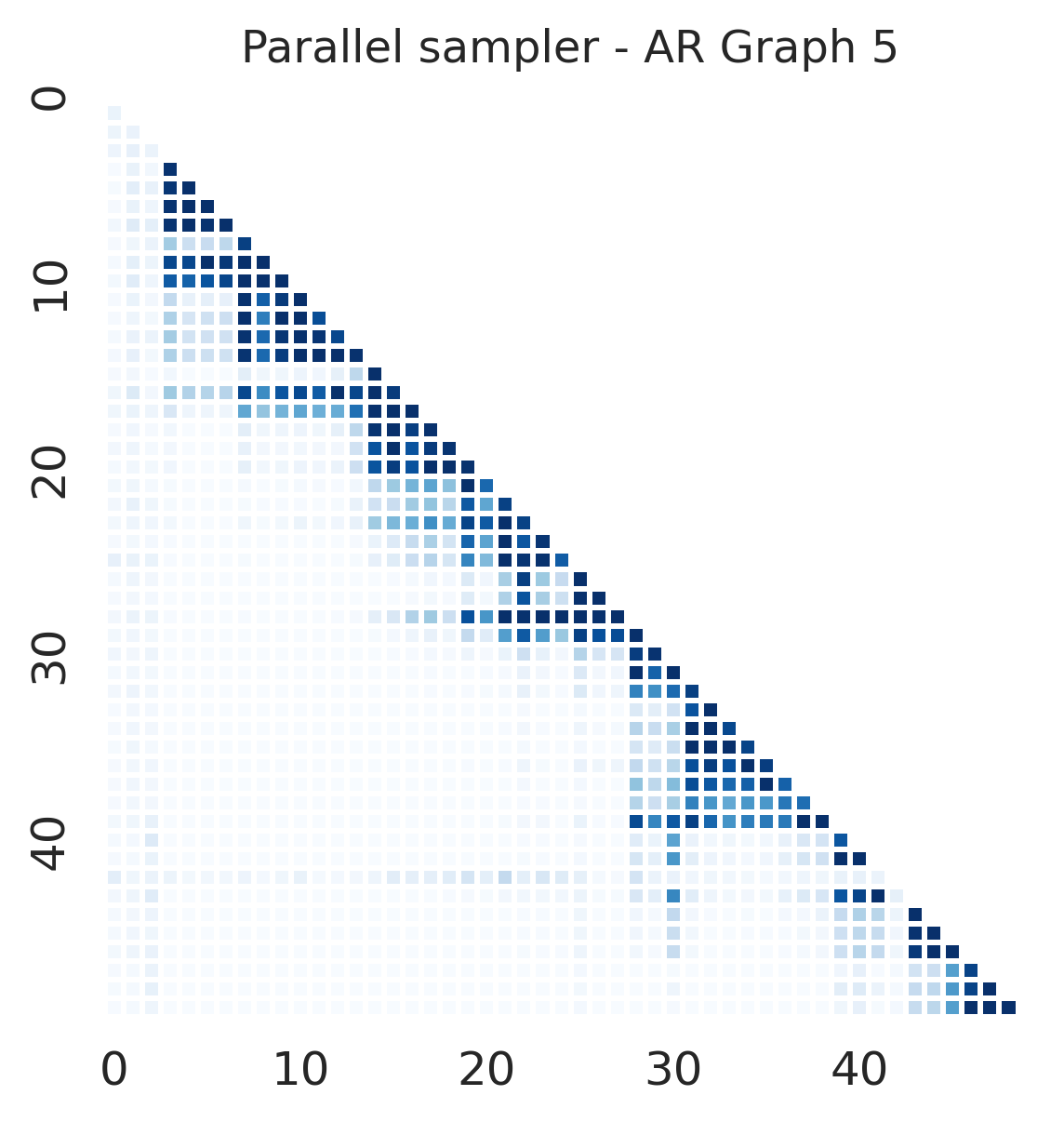}

  \includegraphics[width=0.19\textwidth]{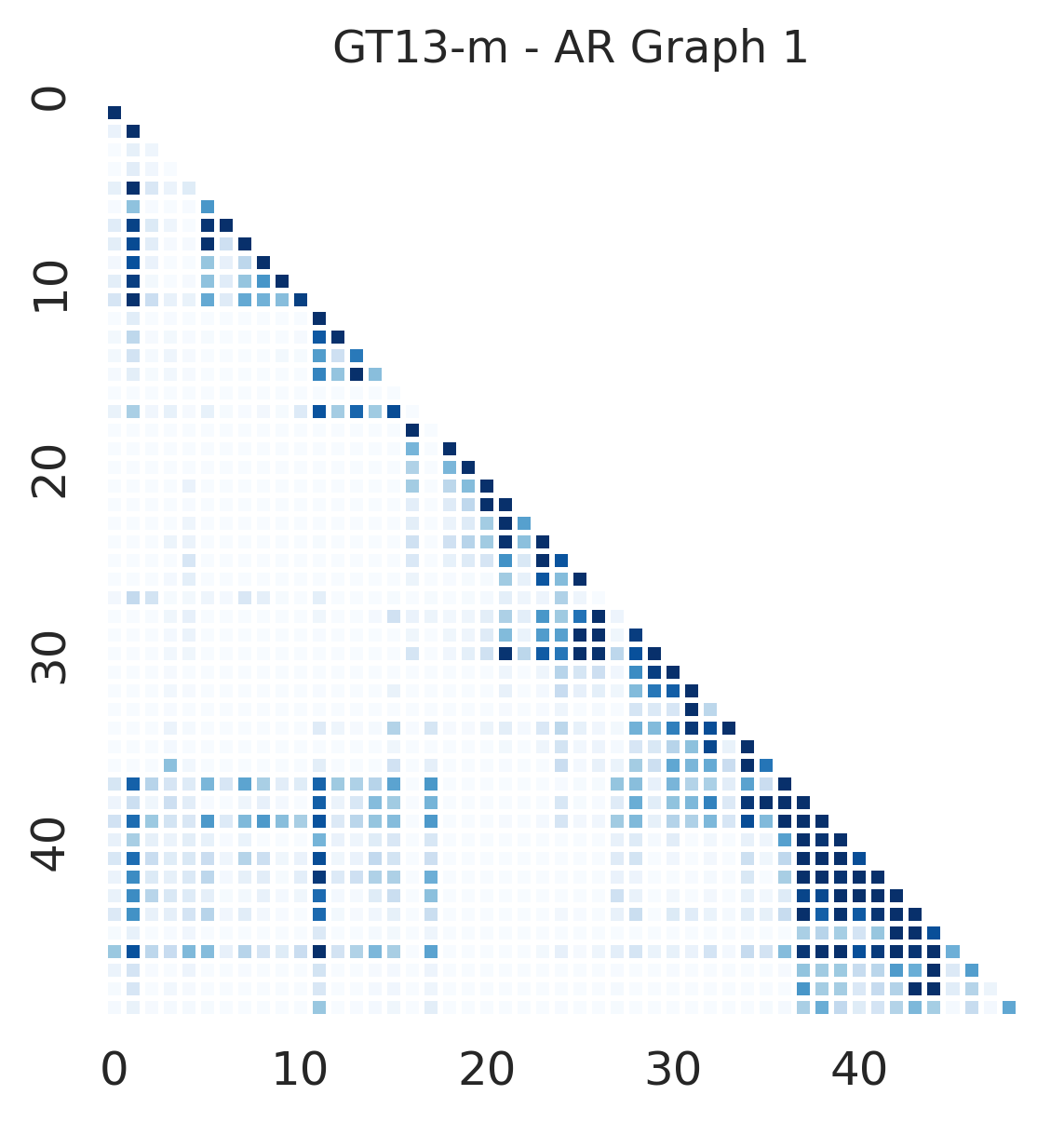}
  \includegraphics[width=0.19\textwidth]{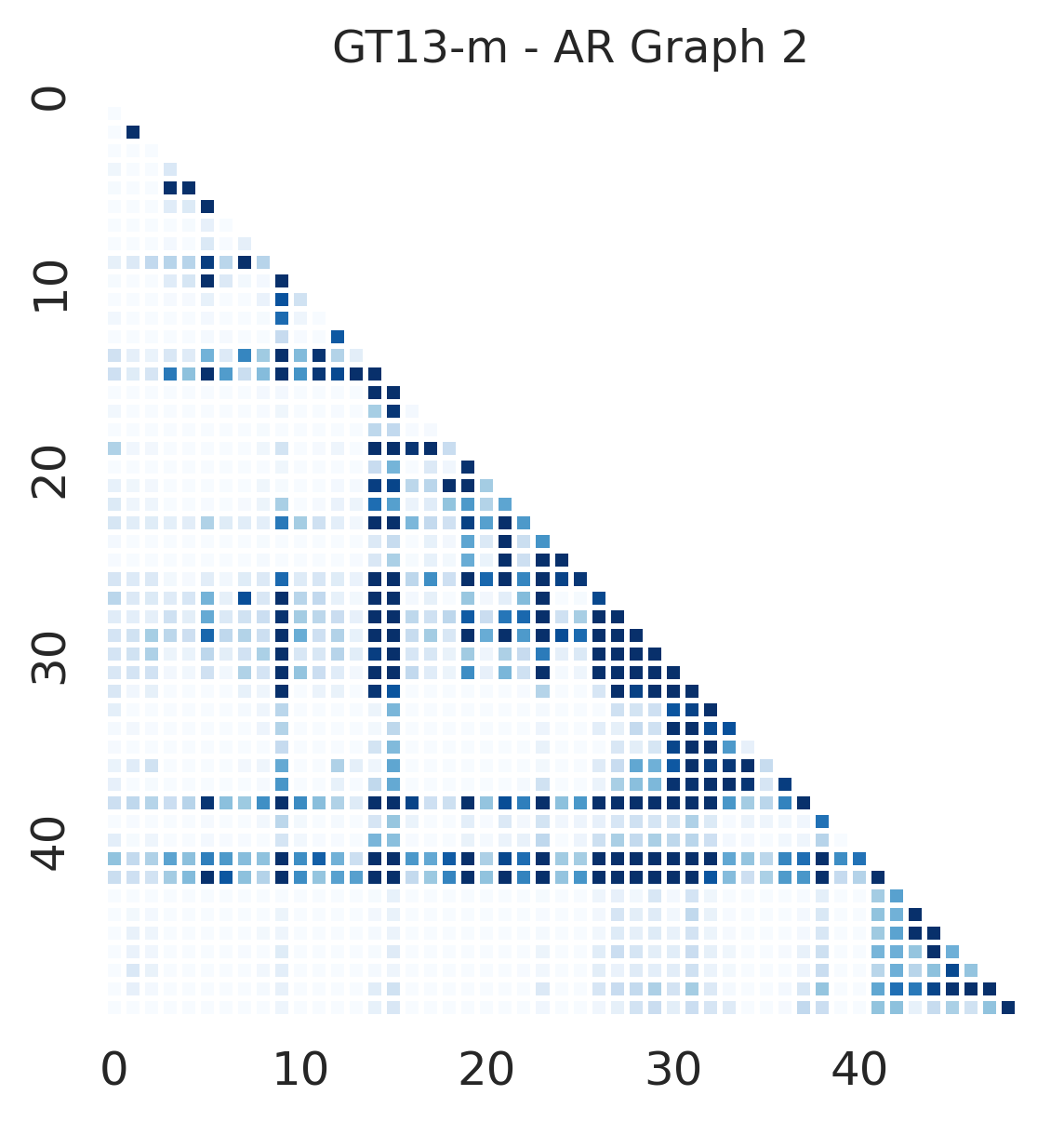}
  \includegraphics[width=0.19\textwidth]{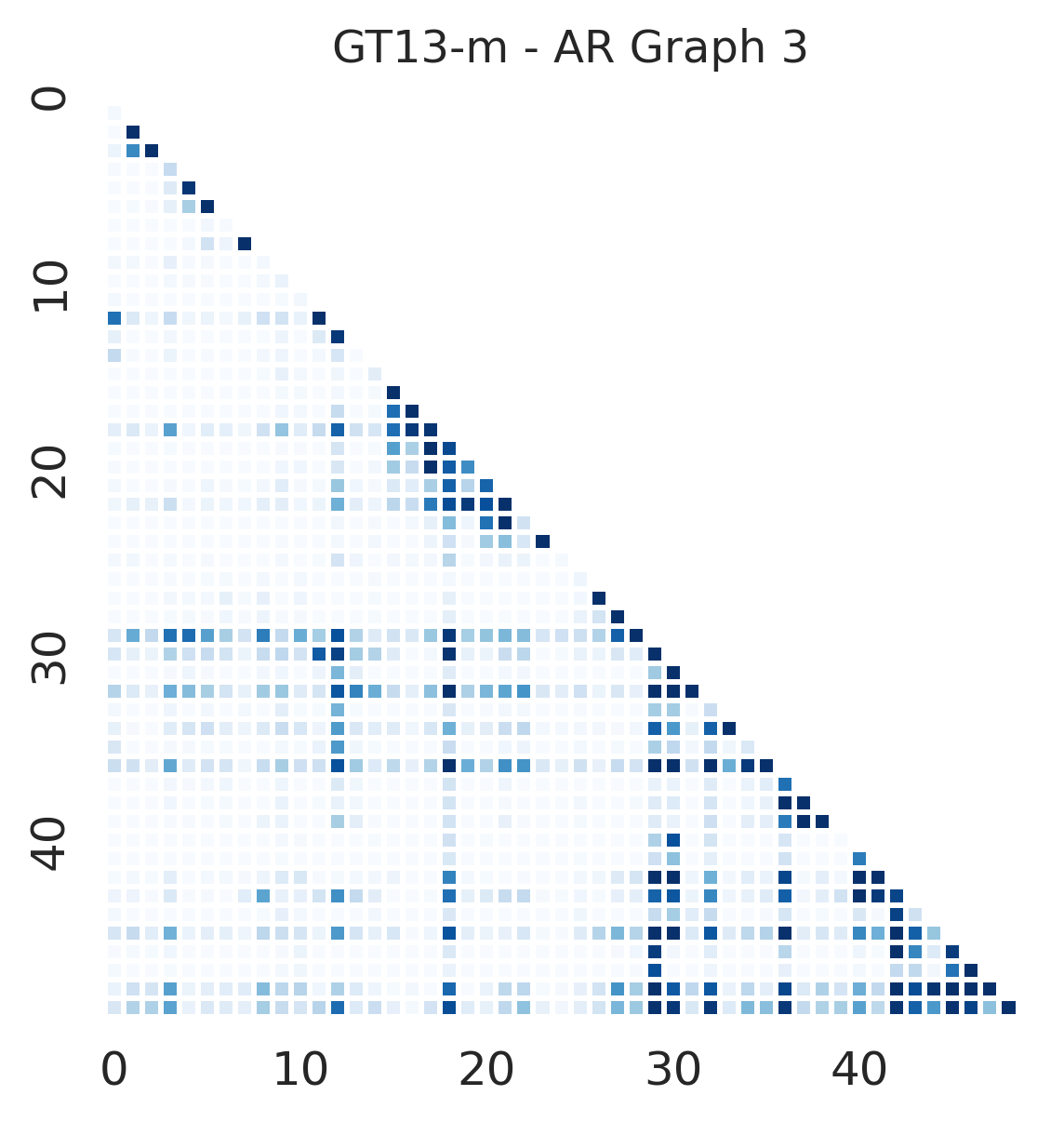}
  \includegraphics[width=0.19\textwidth]{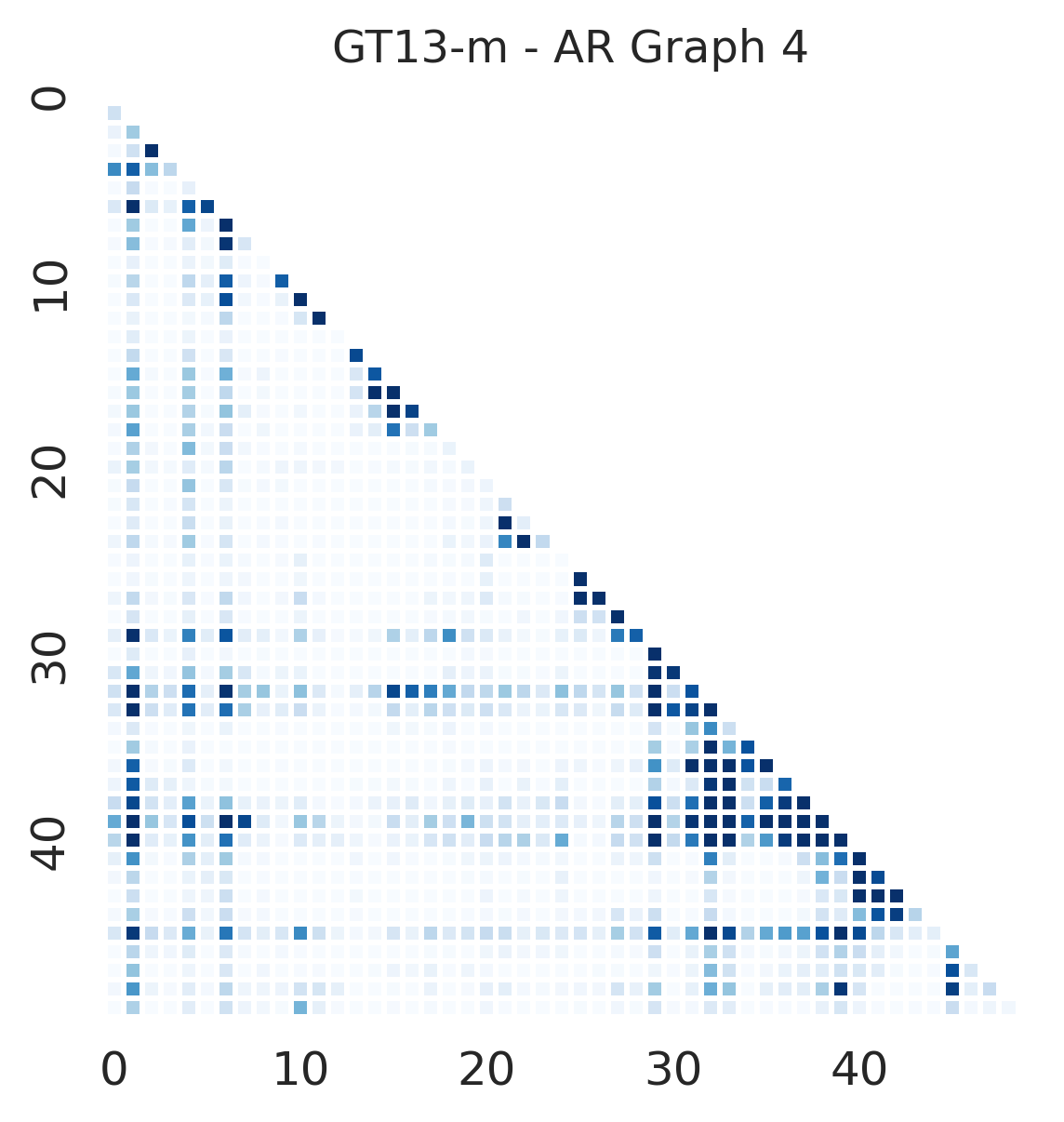}
  \includegraphics[width=0.19\textwidth]{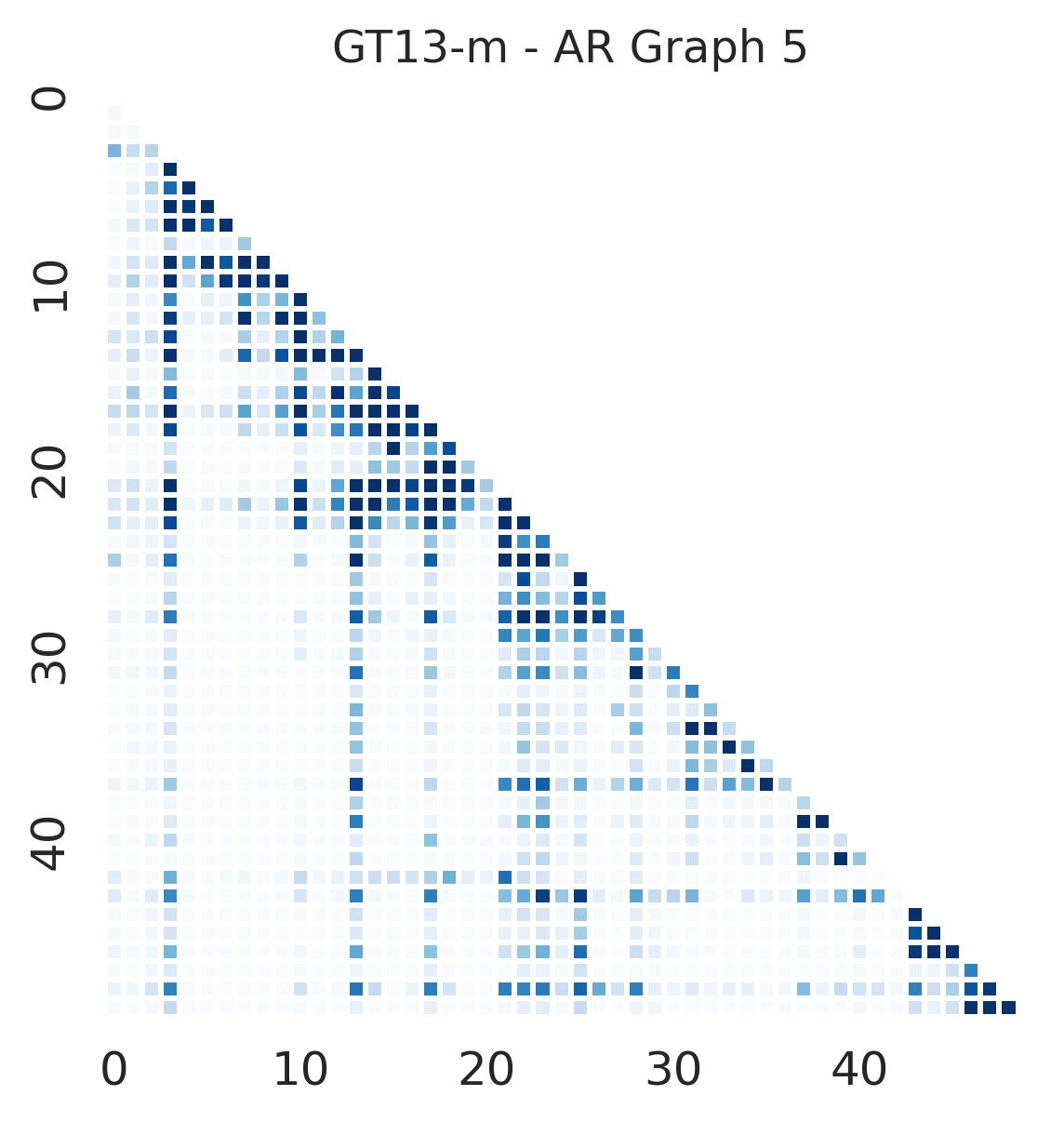}

  \includegraphics[width=0.19\textwidth]{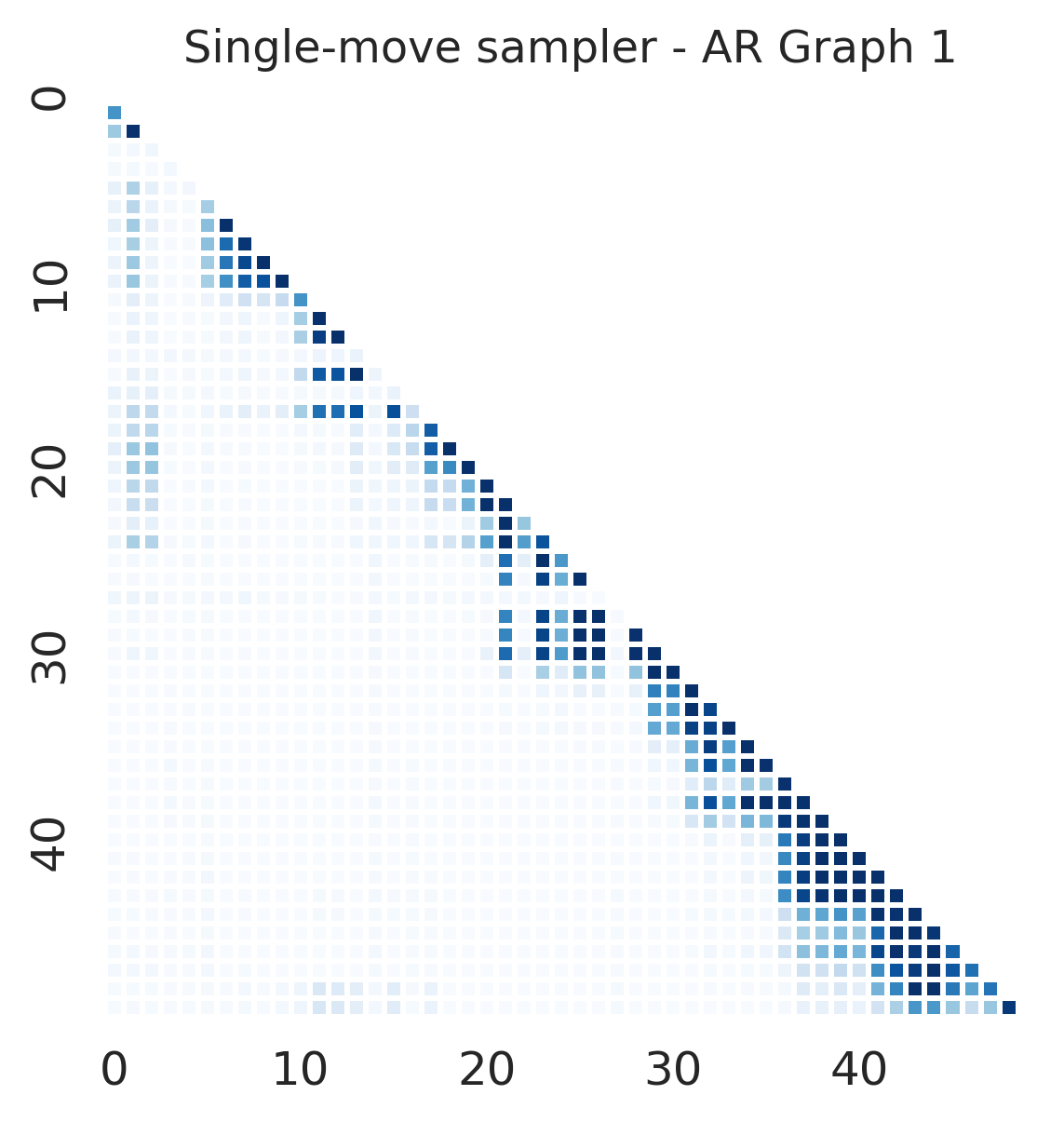}
  \includegraphics[width=0.19\textwidth]{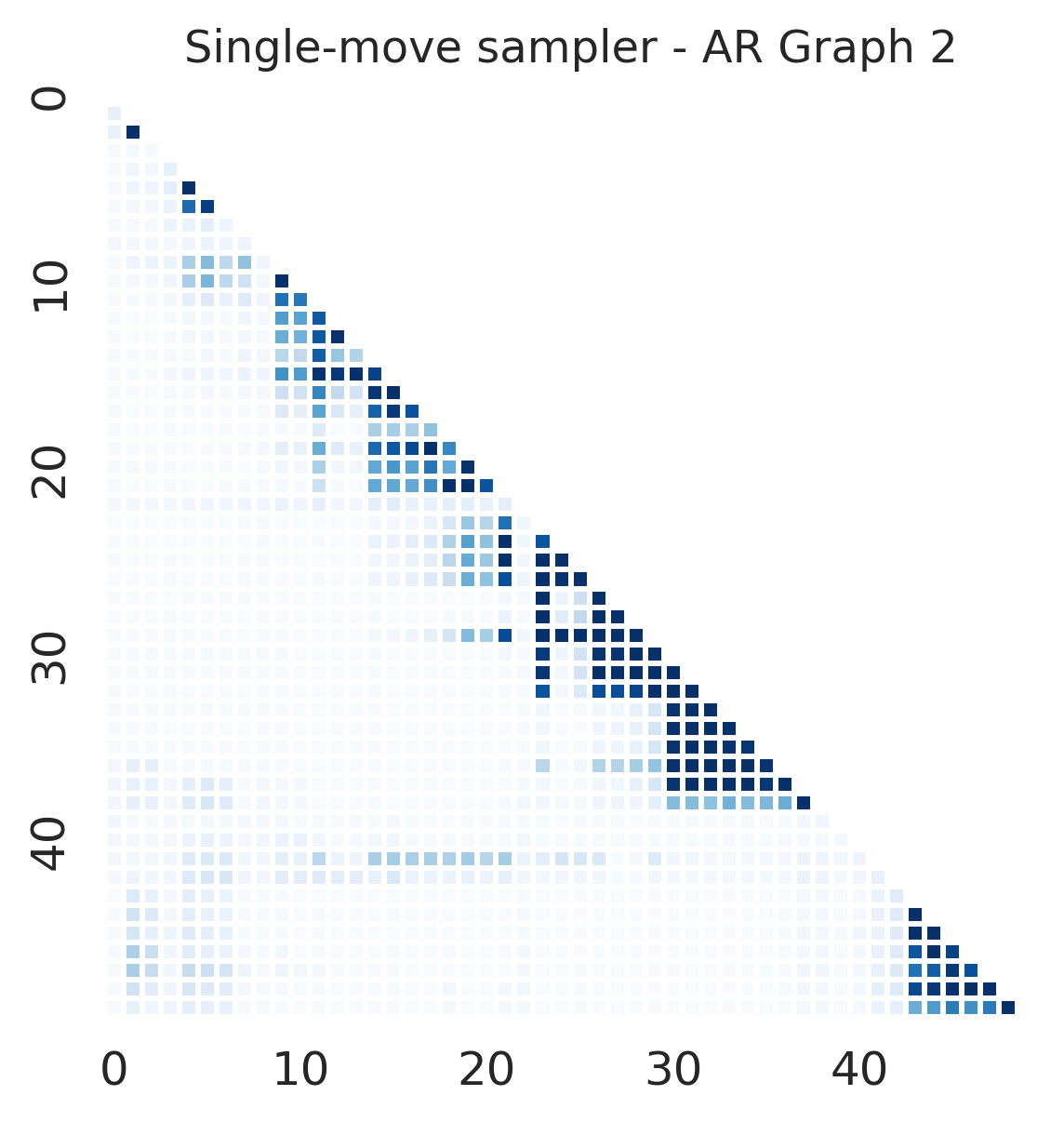}
  \includegraphics[width=0.19\textwidth]{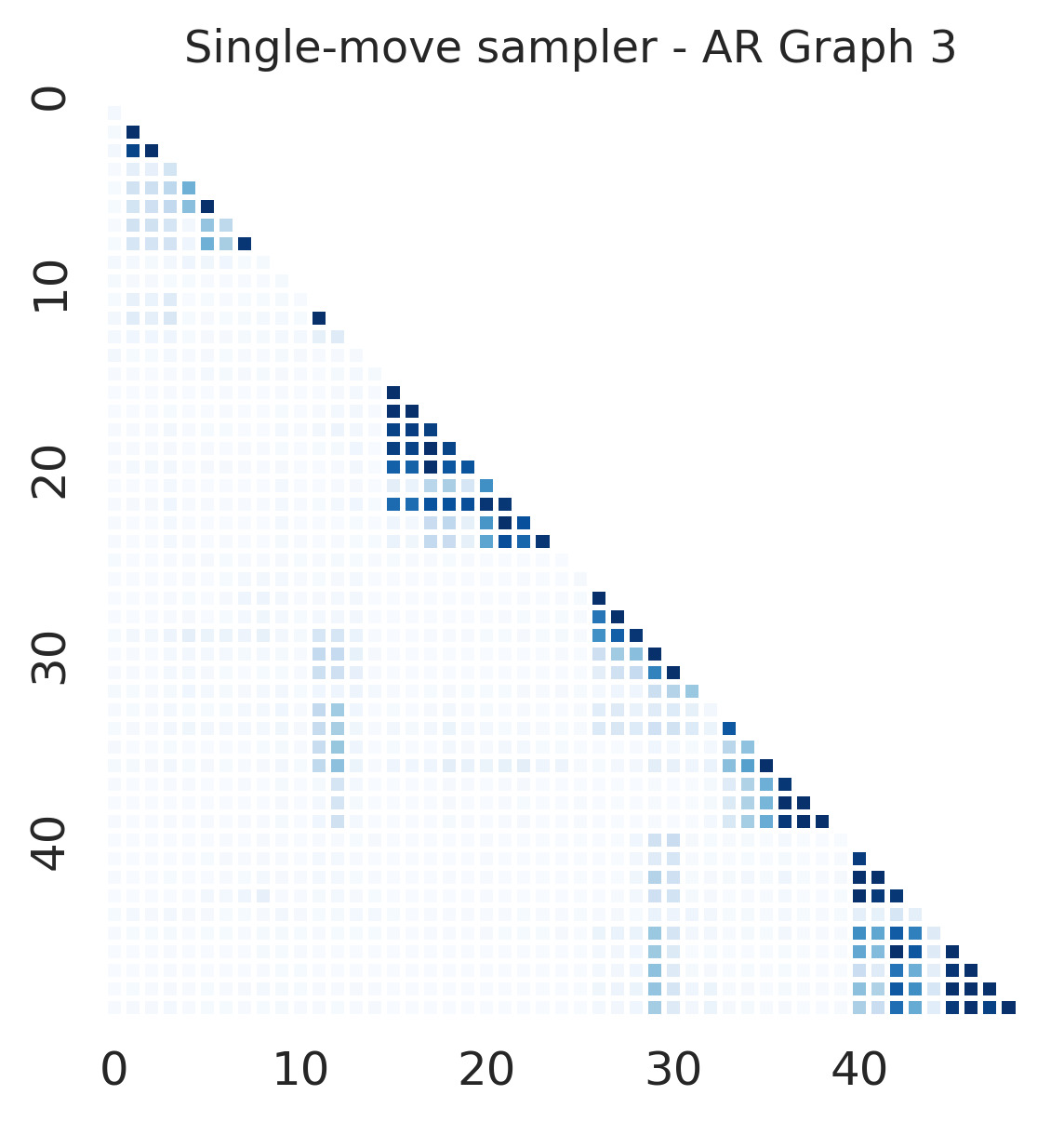}
  \includegraphics[width=0.19\textwidth]{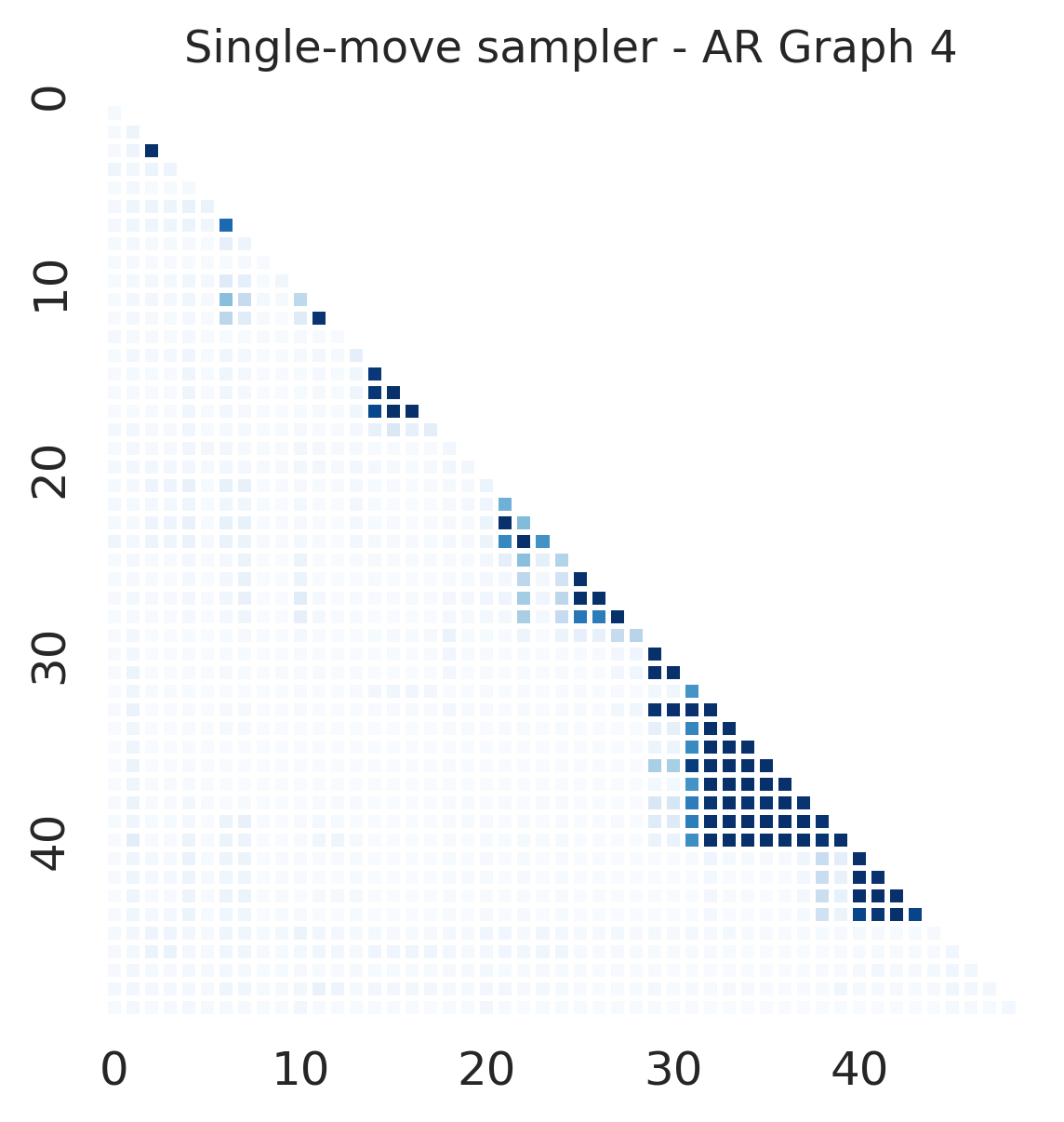}
  \includegraphics[width=0.19\textwidth]{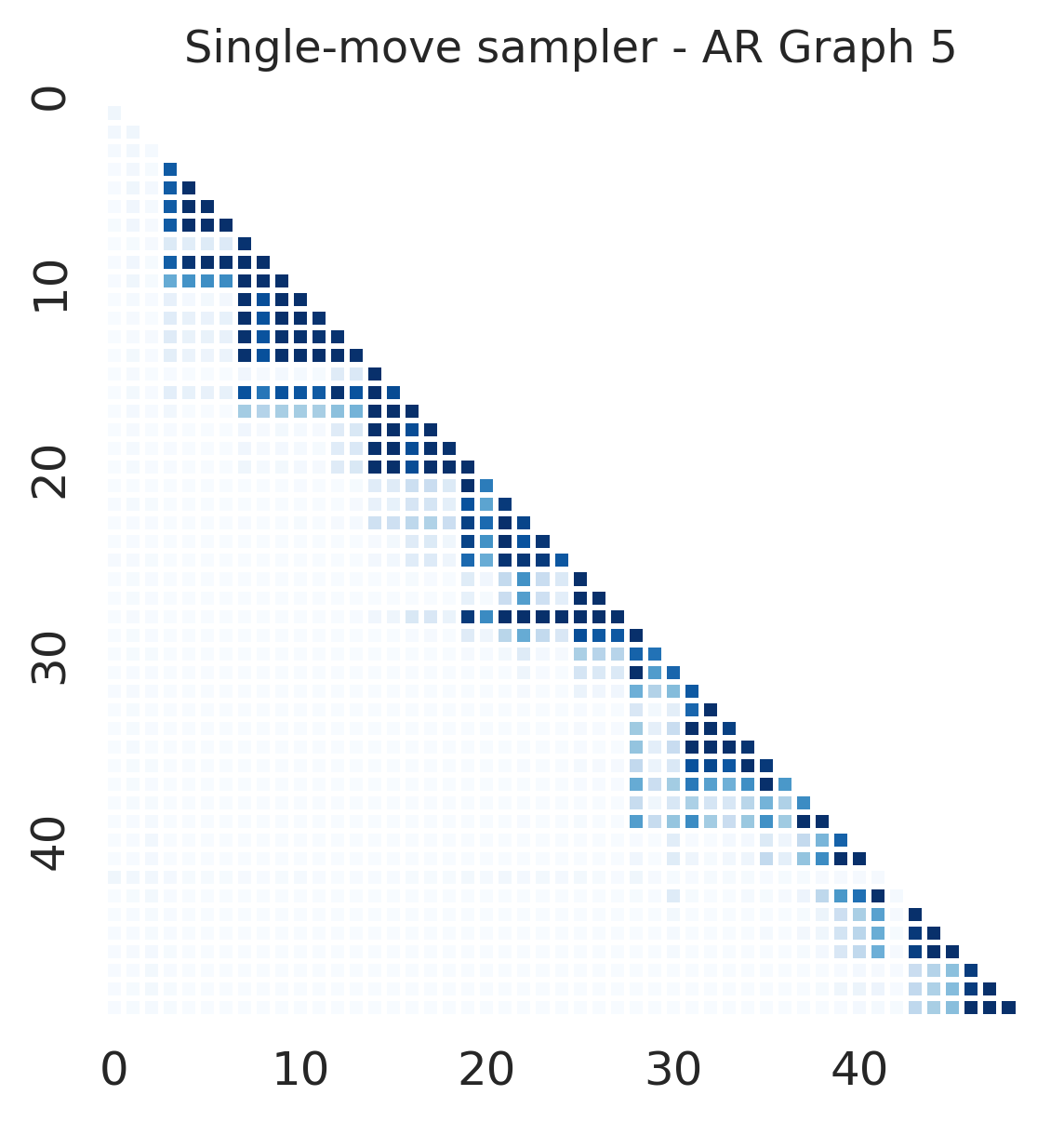}

  \includegraphics[width=0.19\textwidth]{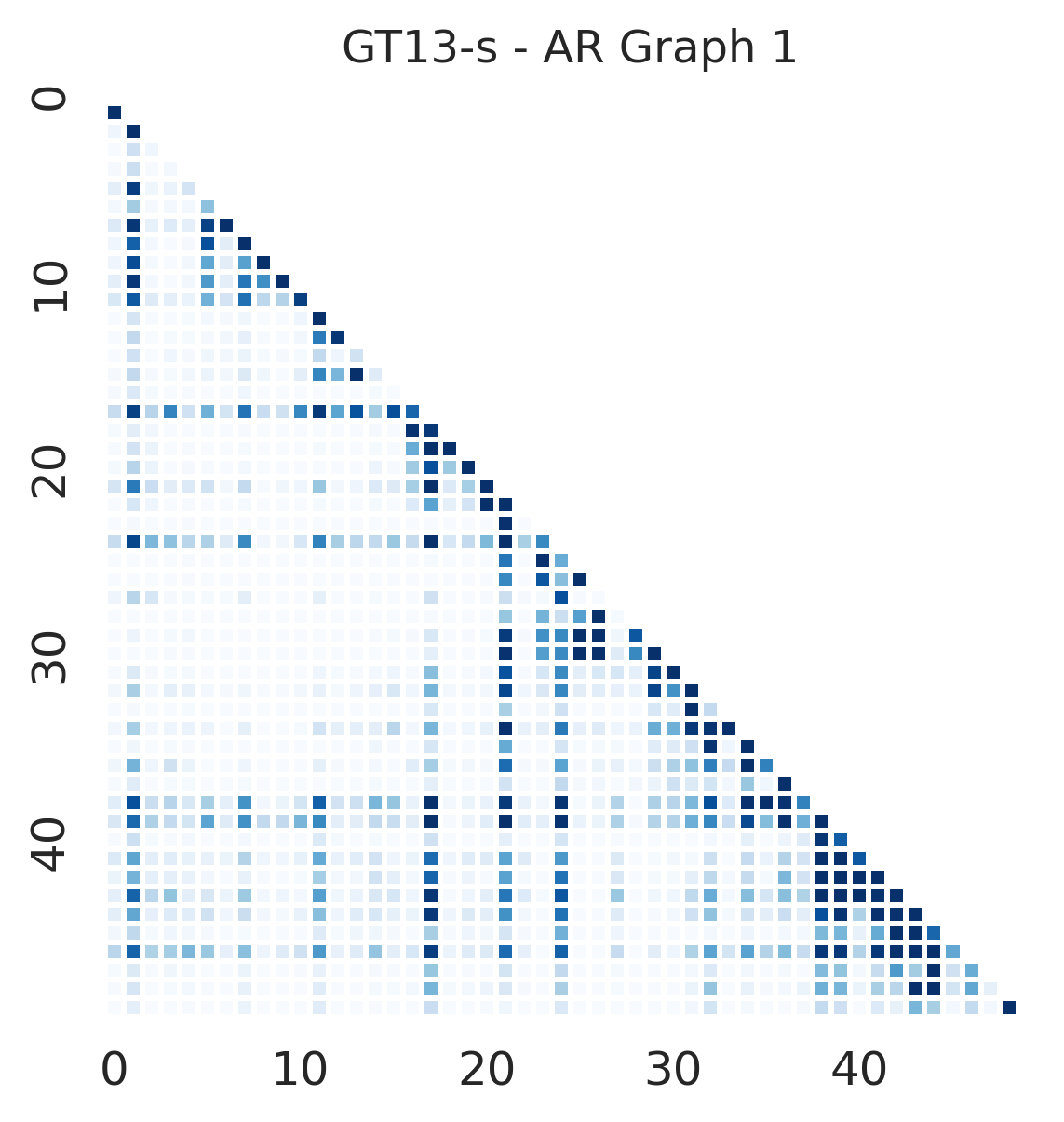}
  \includegraphics[width=0.19\textwidth]{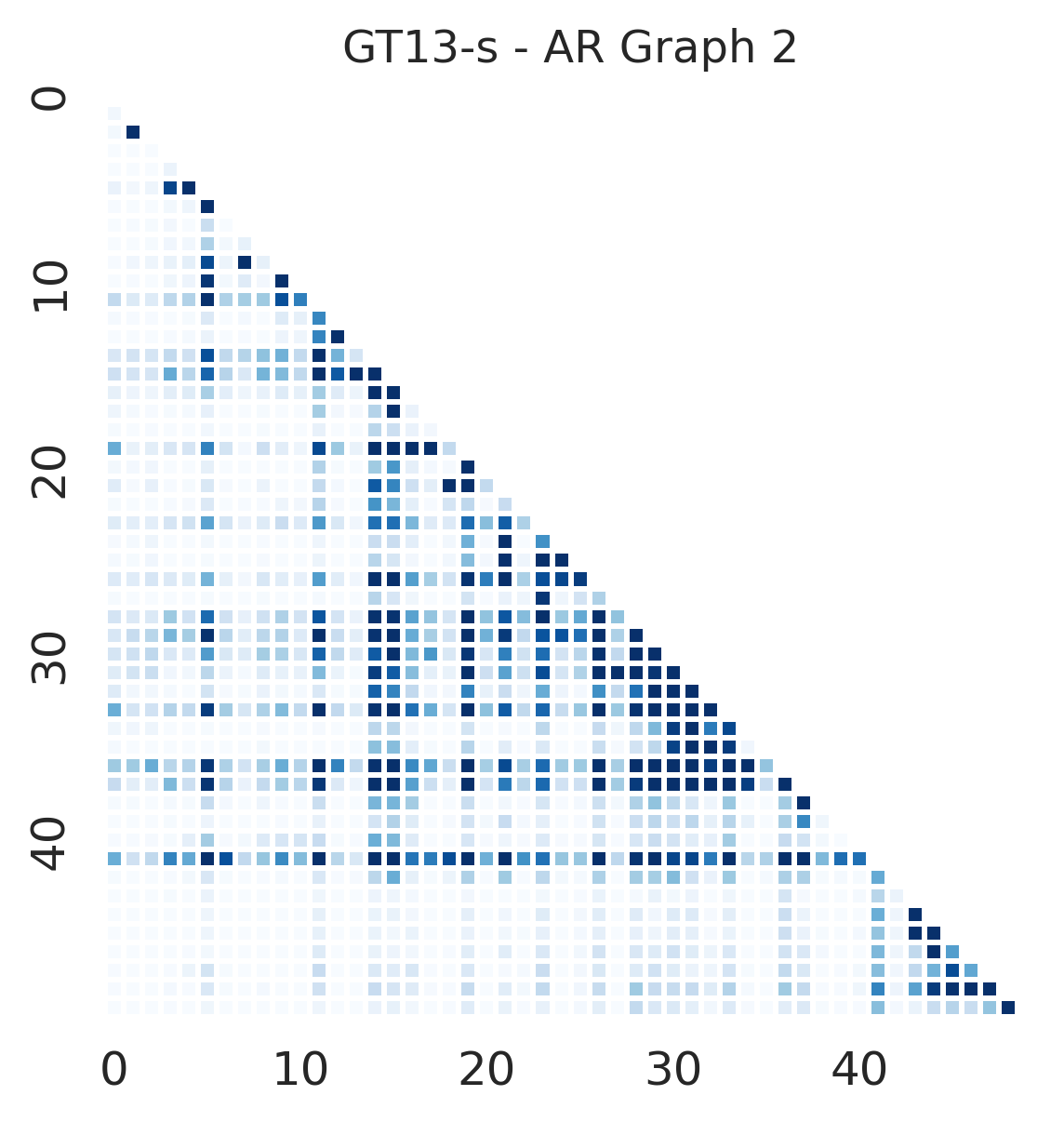}
  \includegraphics[width=0.19\textwidth]{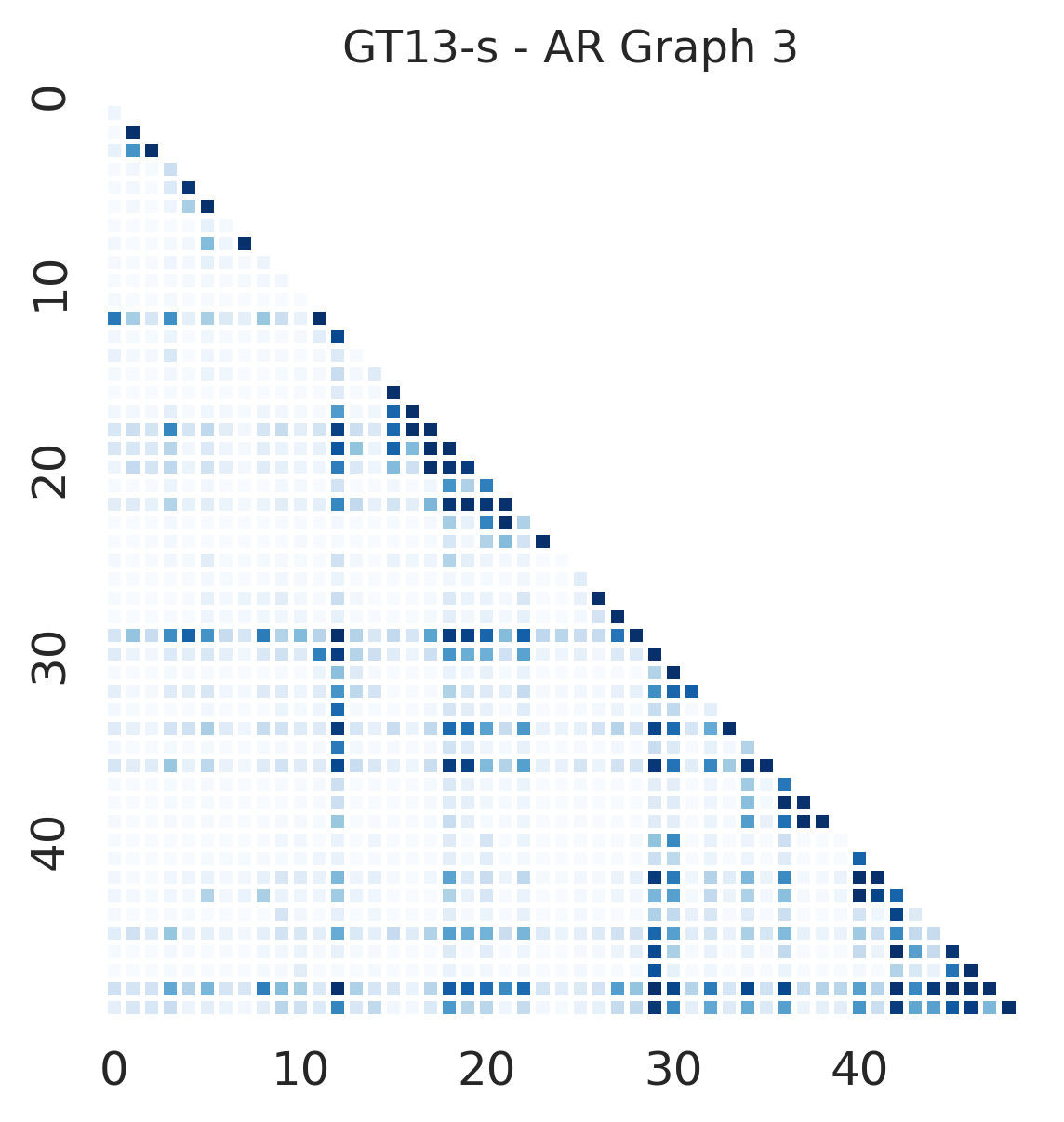}
  \includegraphics[width=0.19\textwidth]{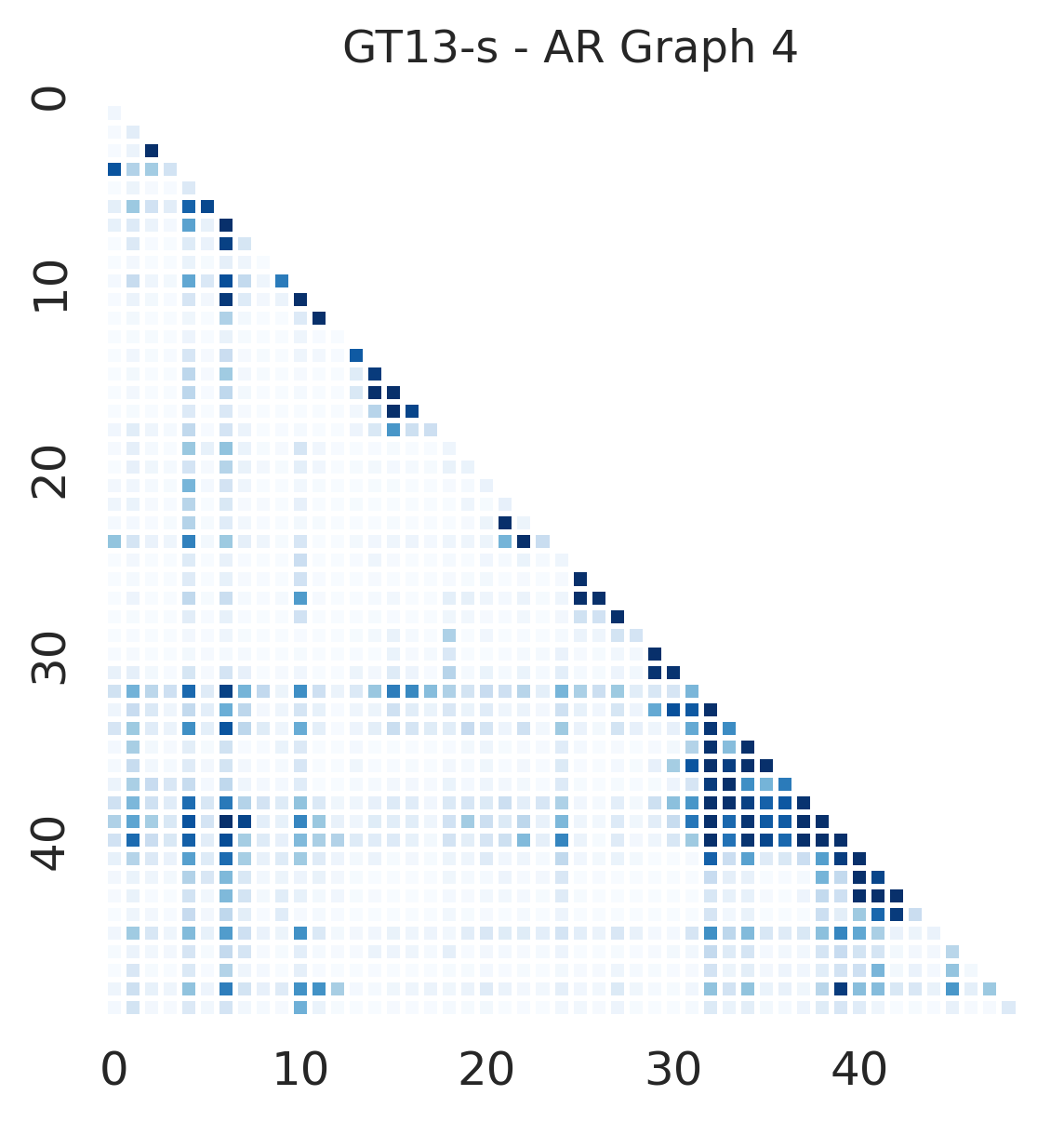}
  \includegraphics[width=0.19\textwidth]{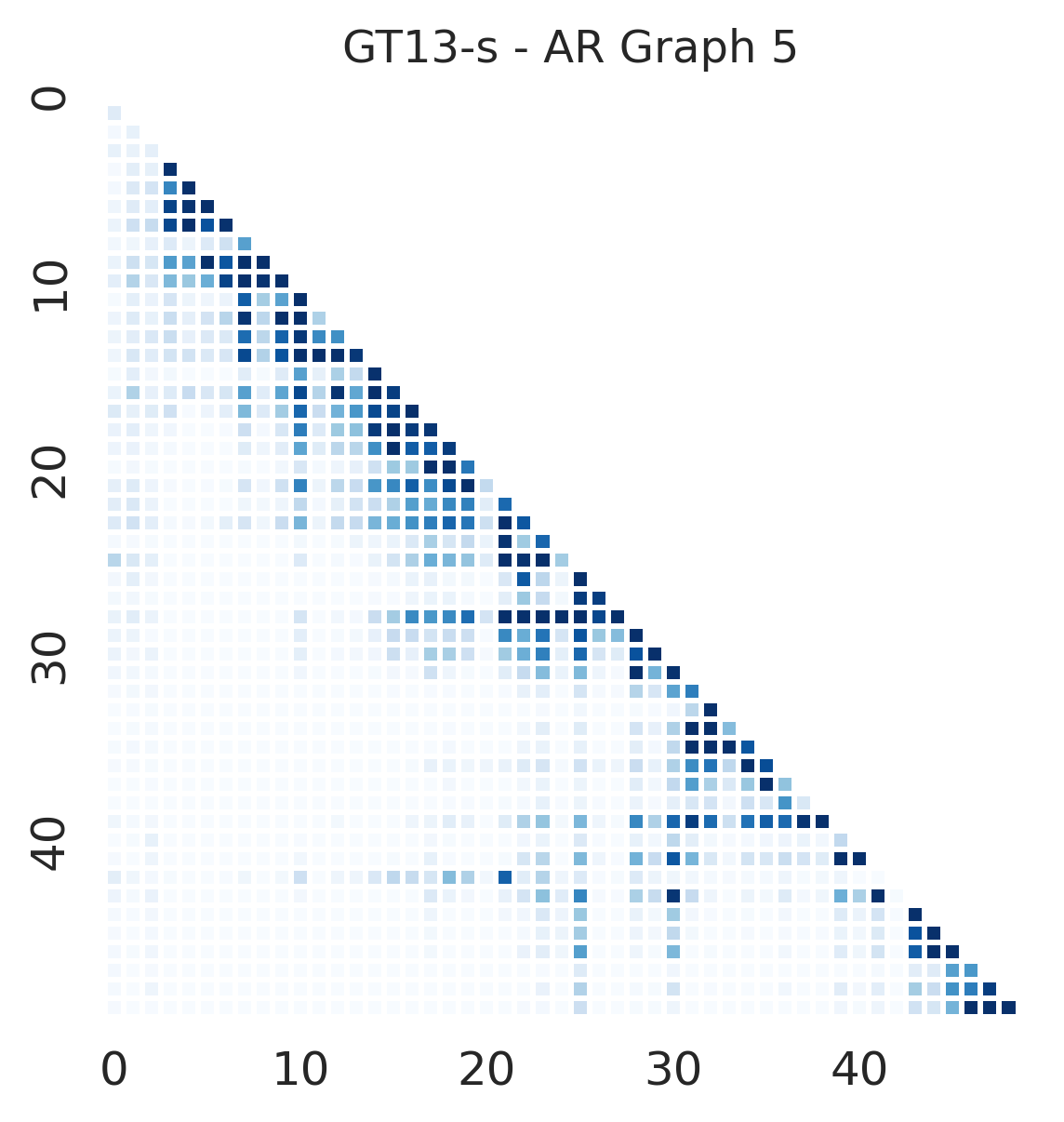}

  \includegraphics[width=0.19\textwidth]{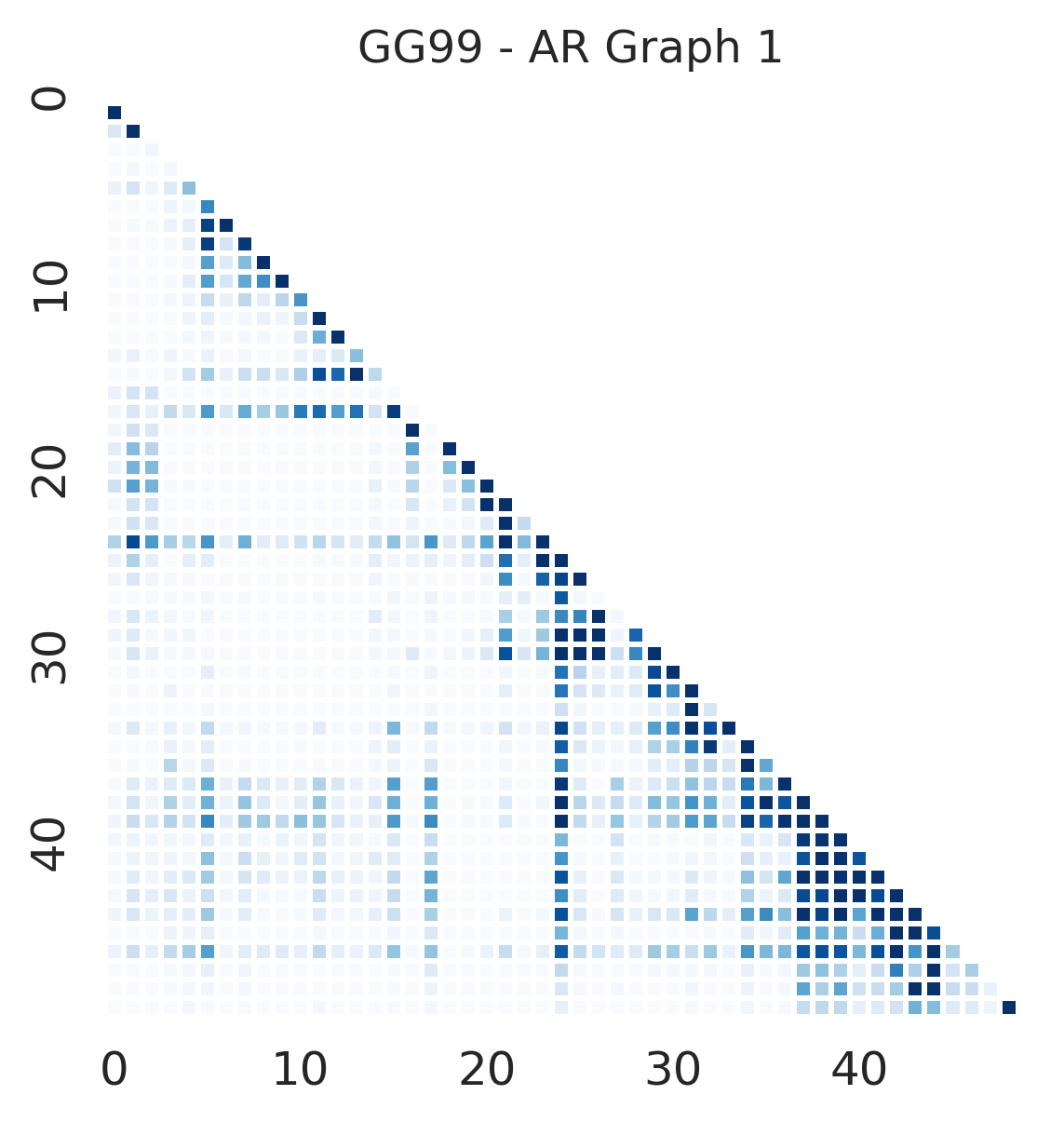}
  \includegraphics[width=0.19\textwidth]{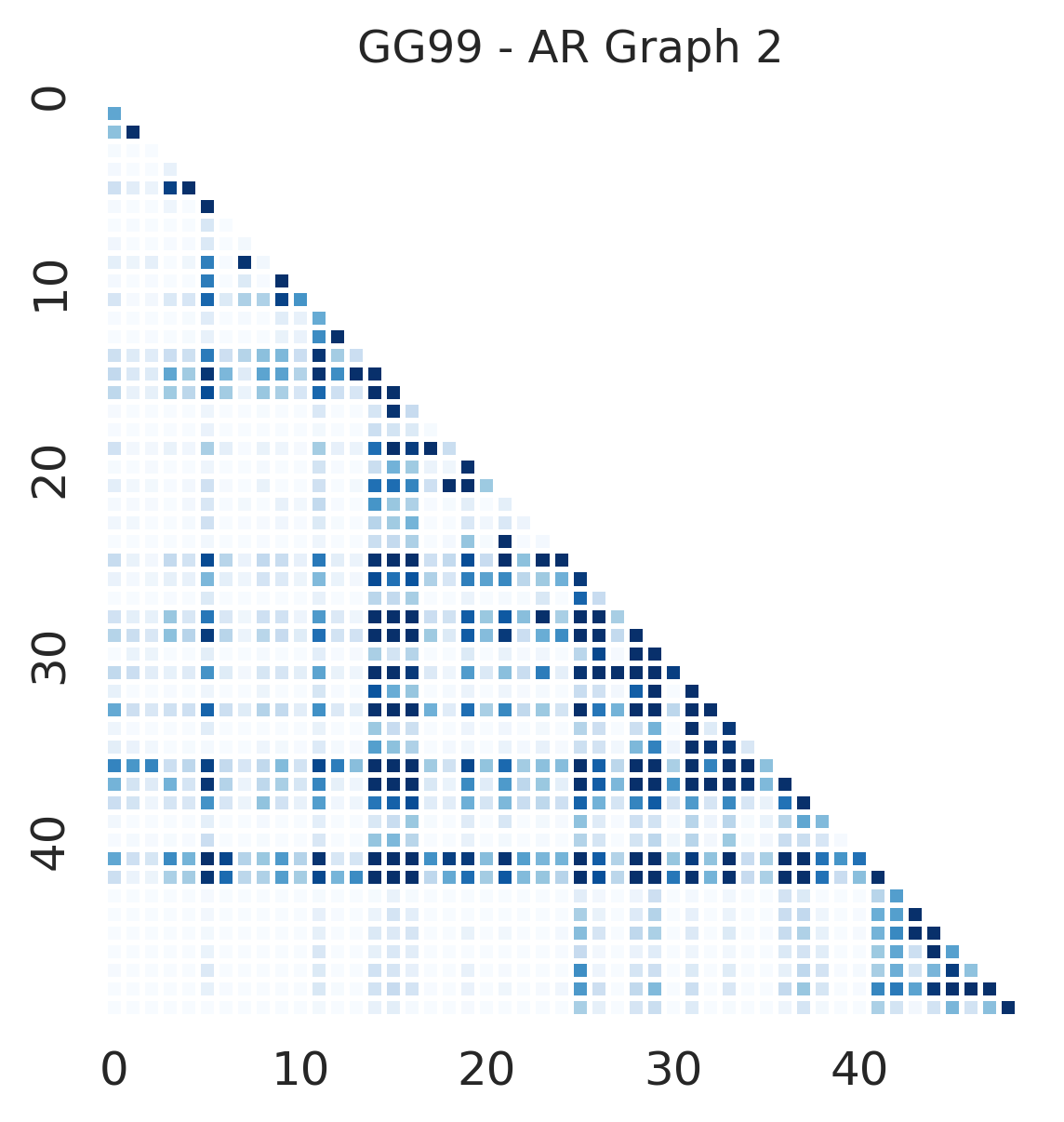}
  \includegraphics[width=0.19\textwidth]{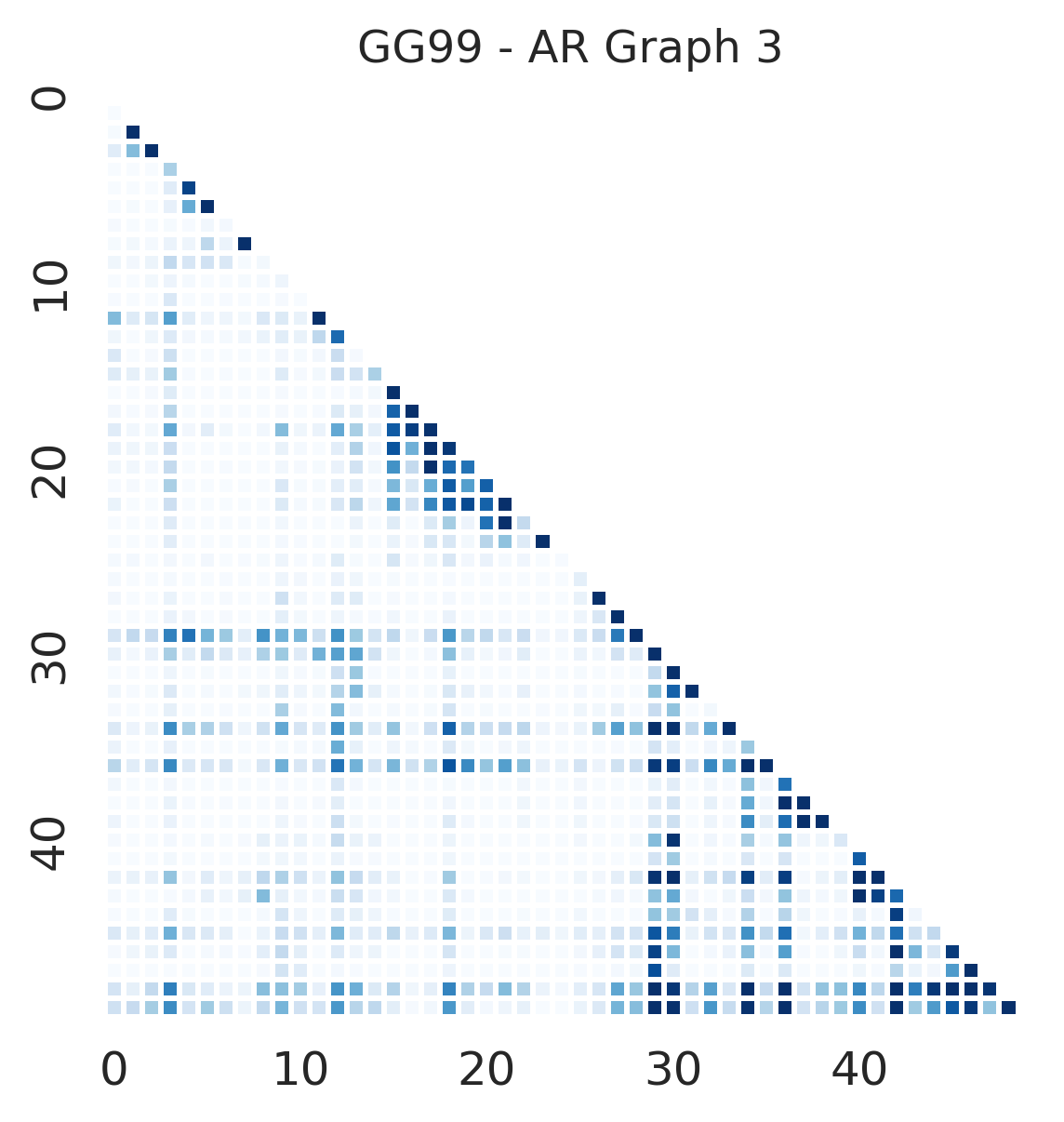}
  \includegraphics[width=0.19\textwidth]{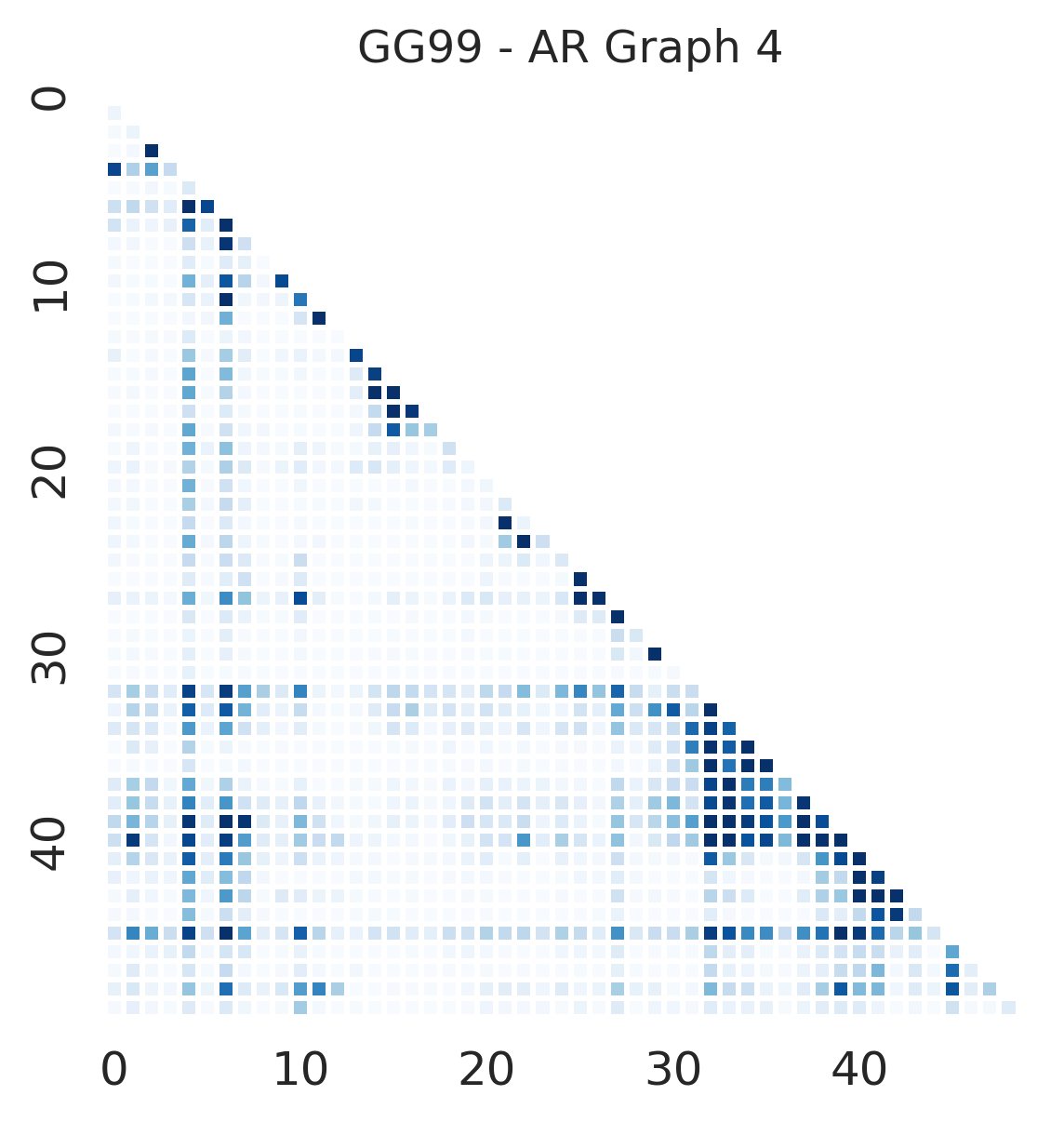}
  \includegraphics[width=0.19\textwidth]{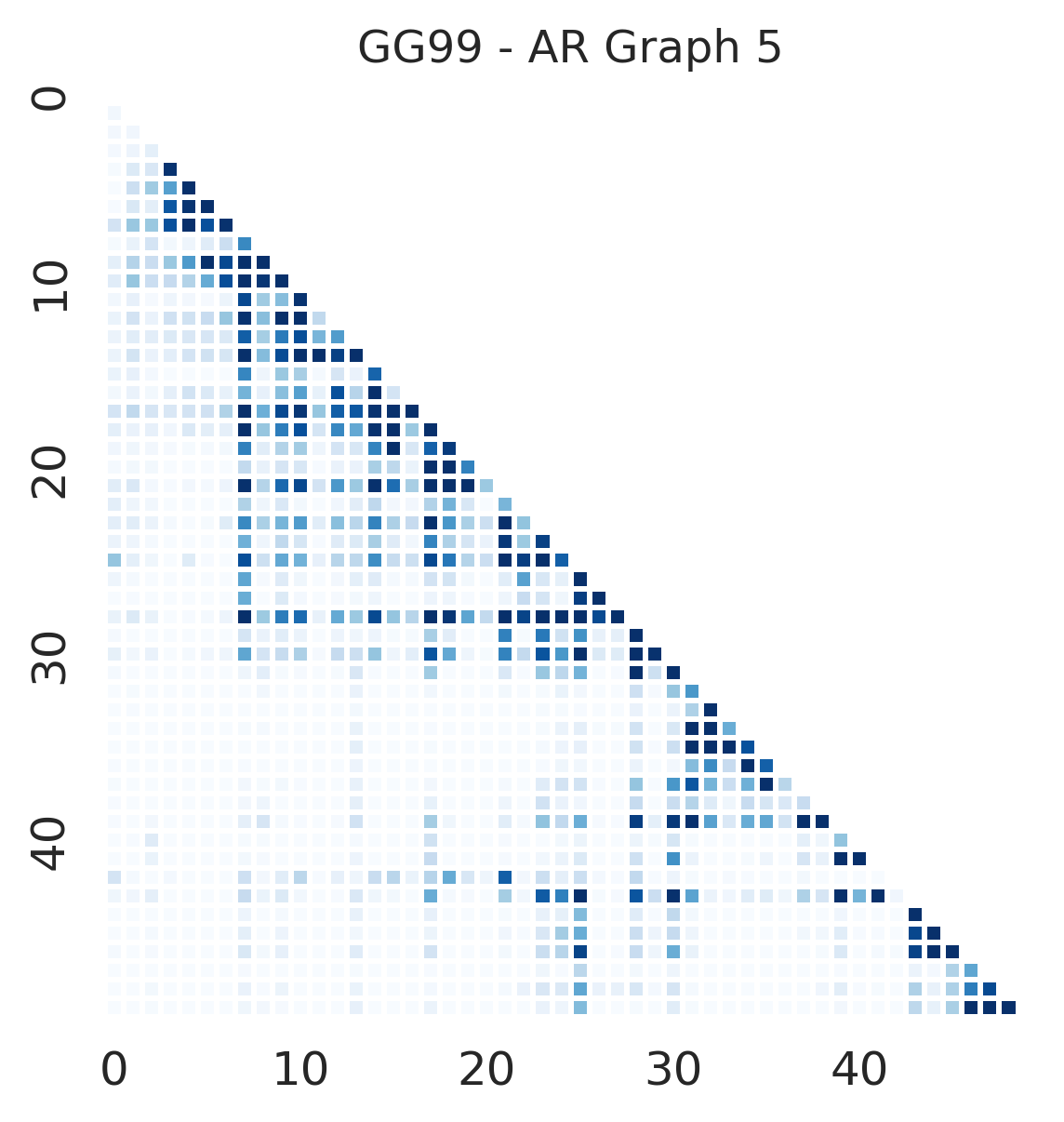}

  \includegraphics[width=0.19\textwidth]{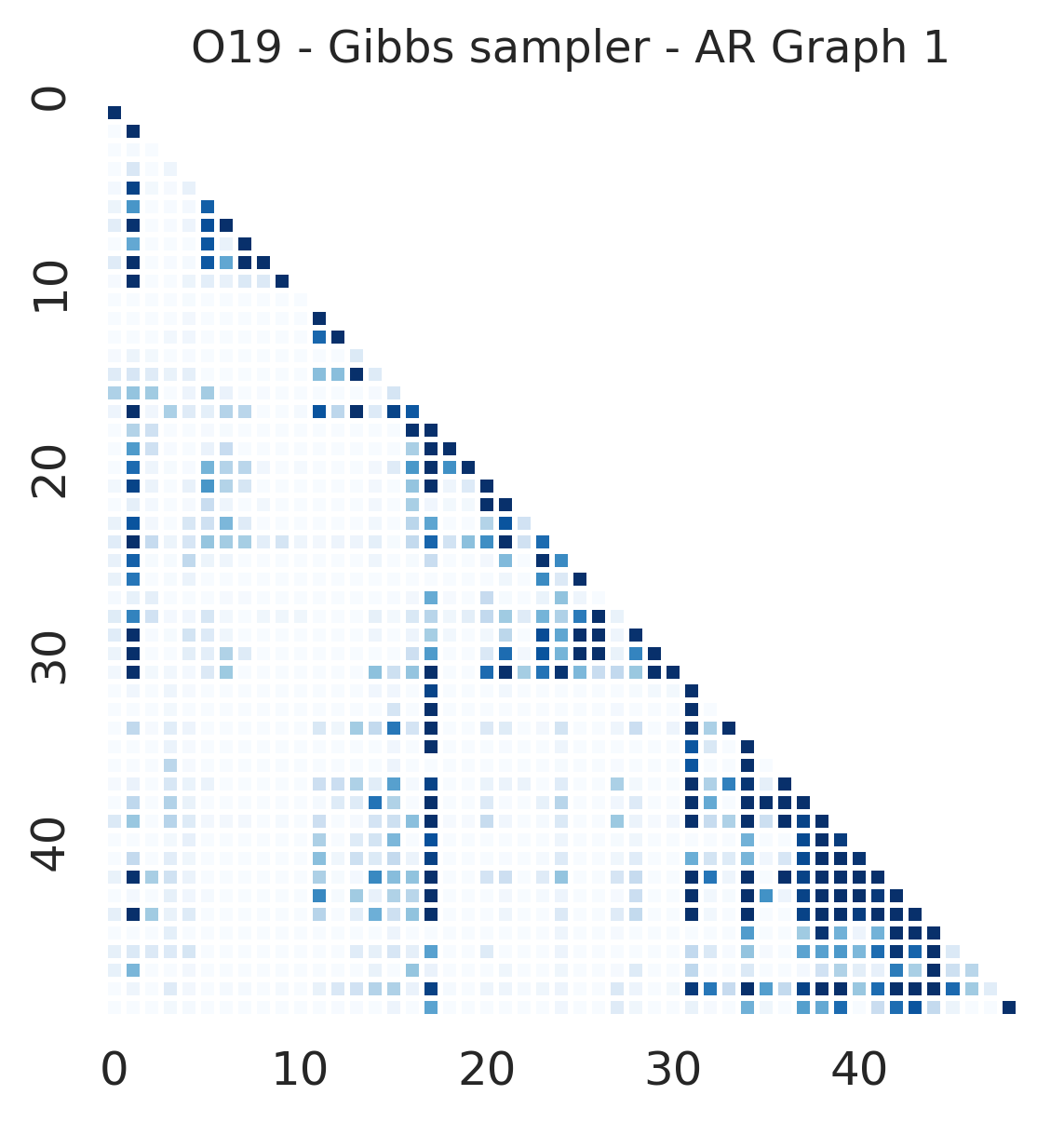}
  \includegraphics[width=0.19\textwidth]{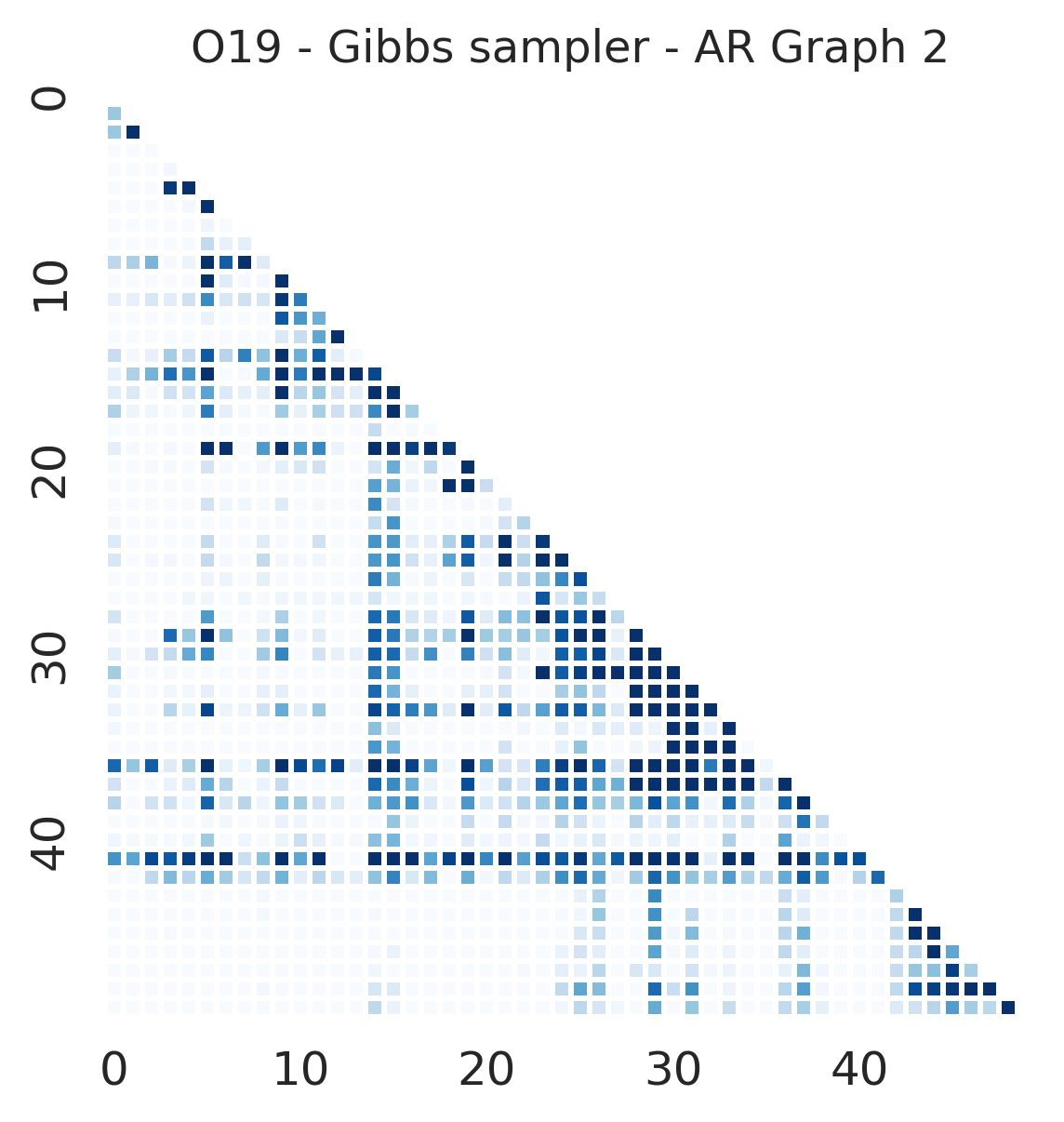}
  \includegraphics[width=0.19\textwidth]{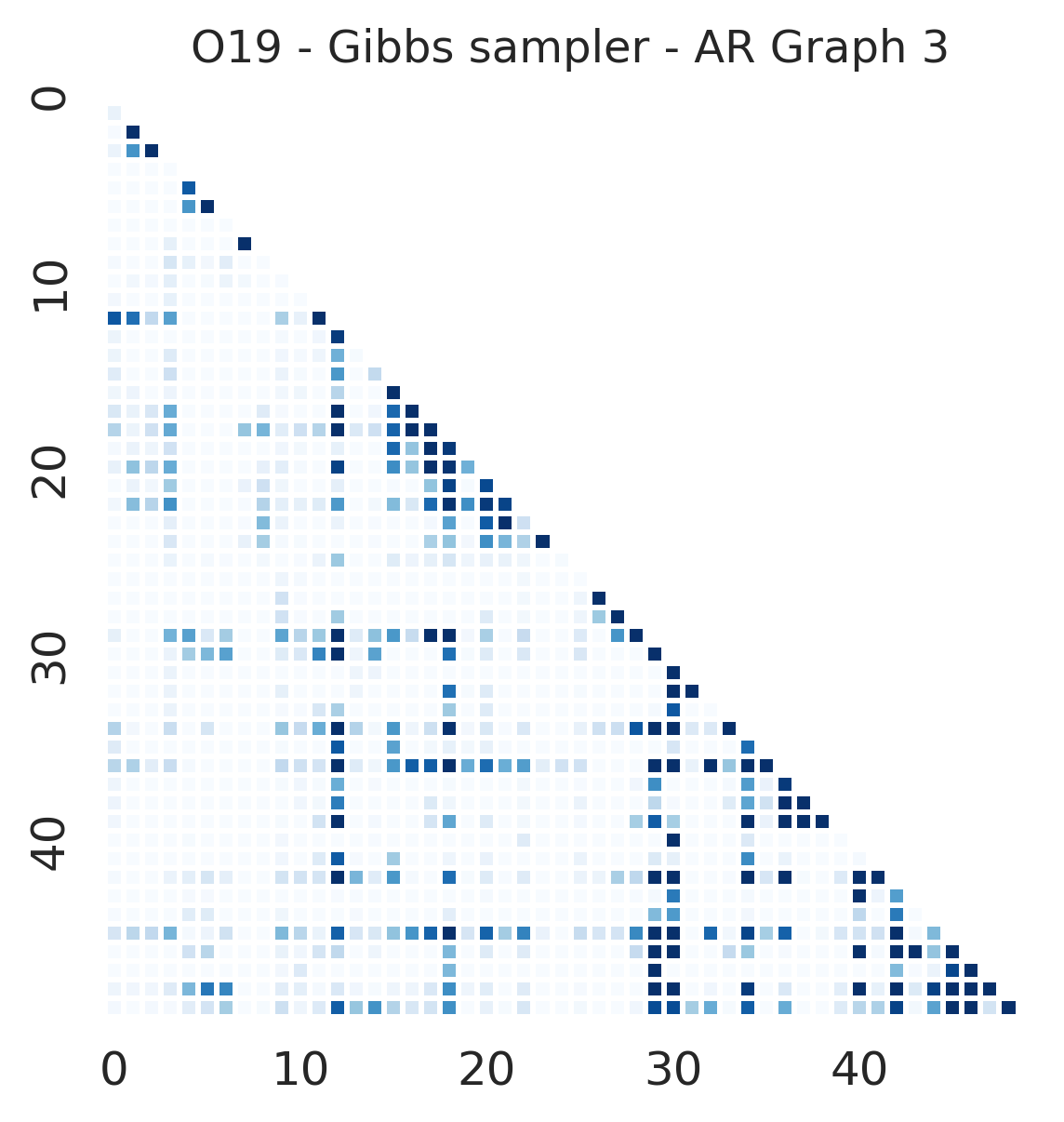}
  \includegraphics[width=0.19\textwidth]{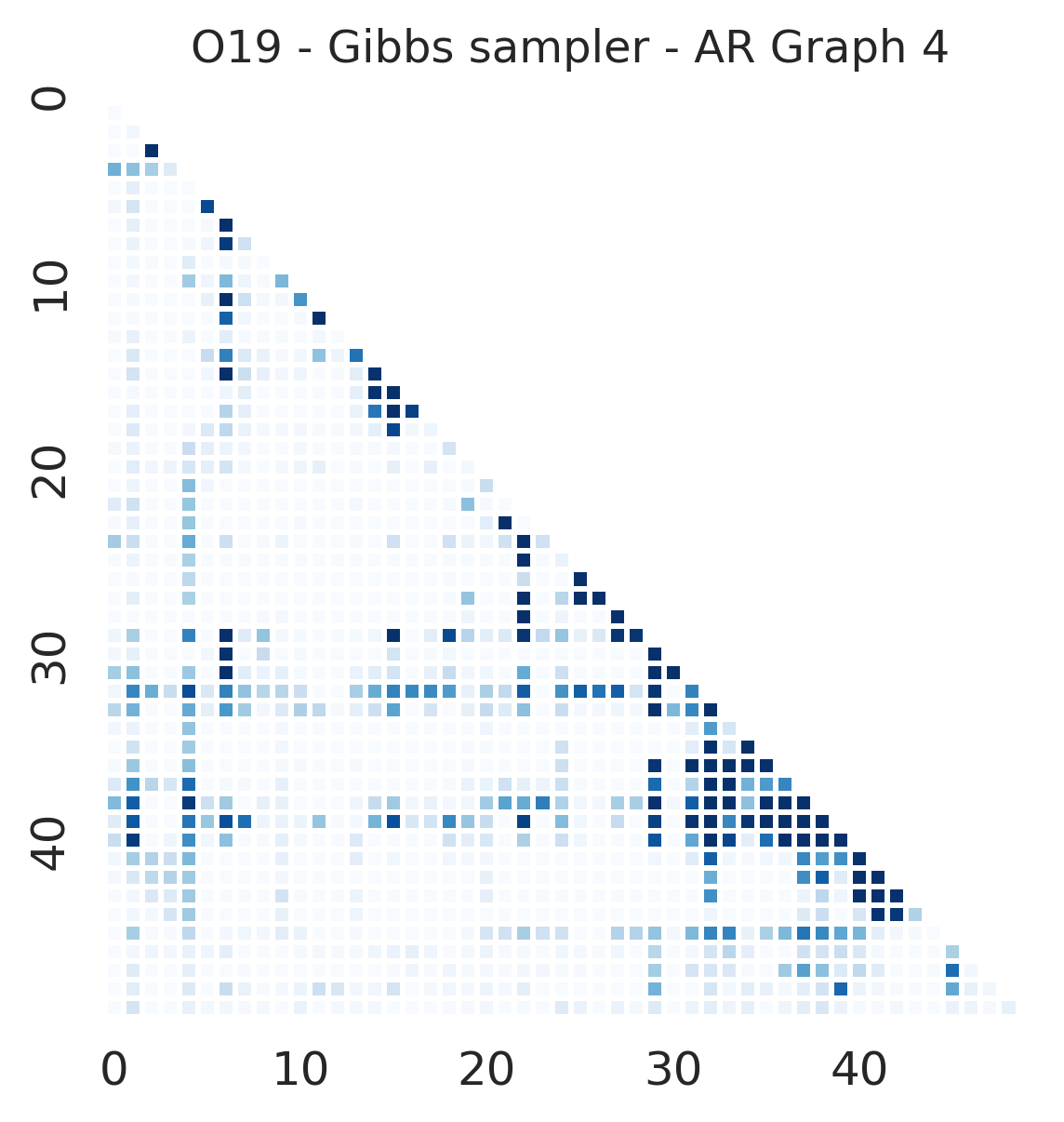}
  \includegraphics[width=0.19\textwidth]{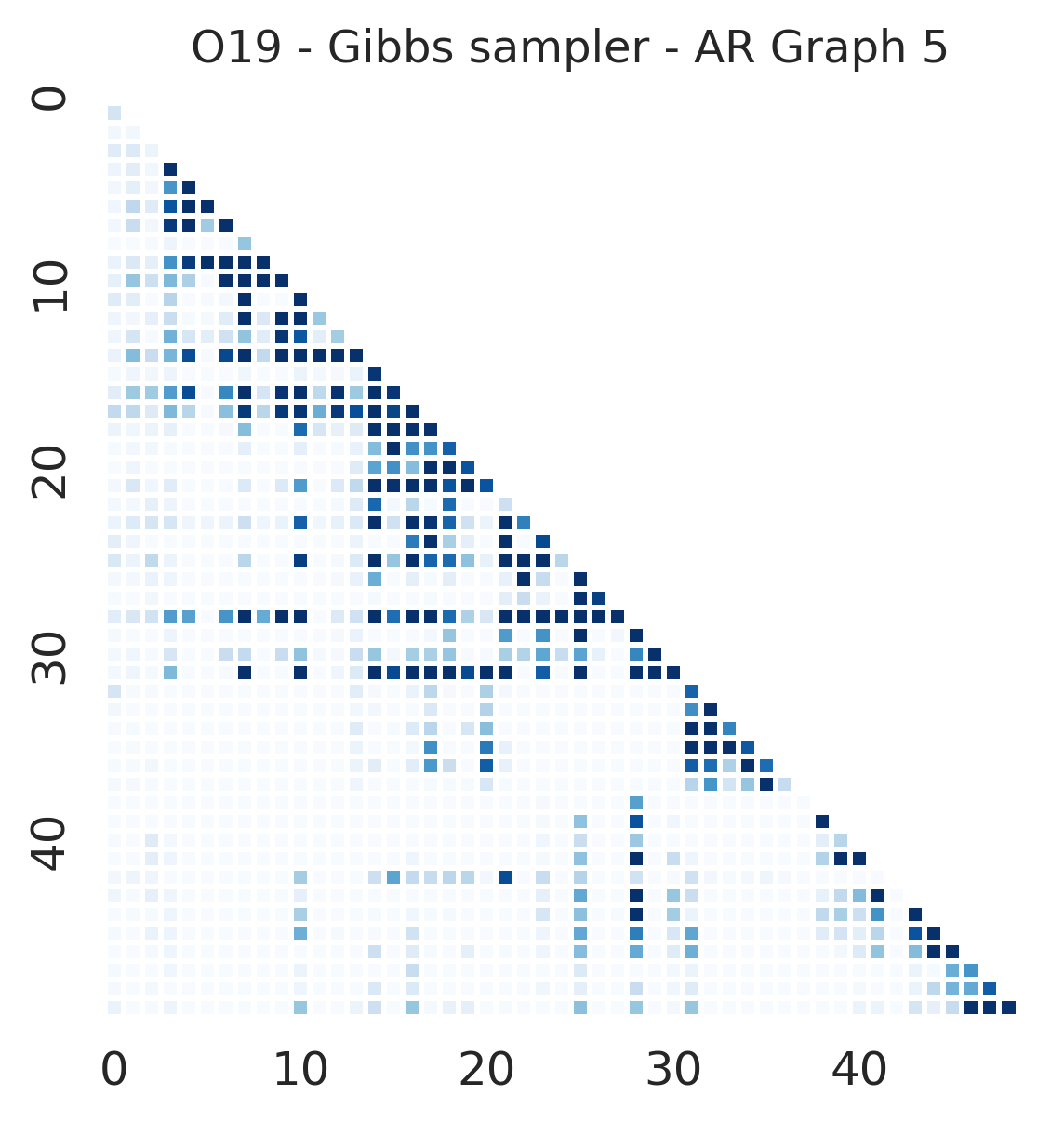}

  \caption{Heatmap derived from the final 300,000 steps of samplers in Section~\ref{sec:simul-setup-gauss}, covering replications 1 to 5 under the autoregressive structure and $(n,p)=(100,50)$.}

  \label{app:fig:heatmap-bandmat-0}
\end{figure}

\begin{figure}[ht]
  \centering
  \includegraphics[width=0.19\textwidth]{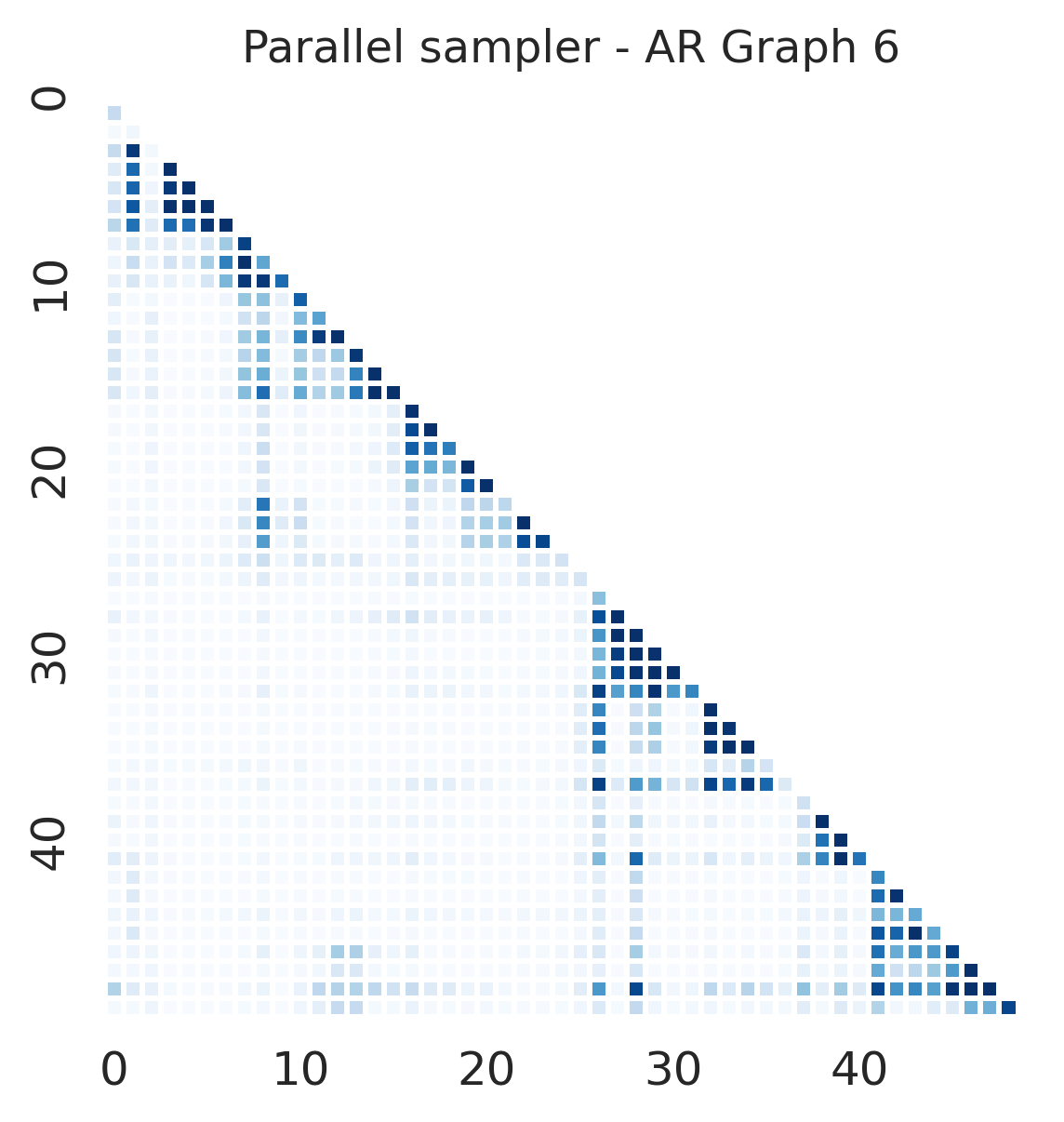}
  \includegraphics[width=0.19\textwidth]{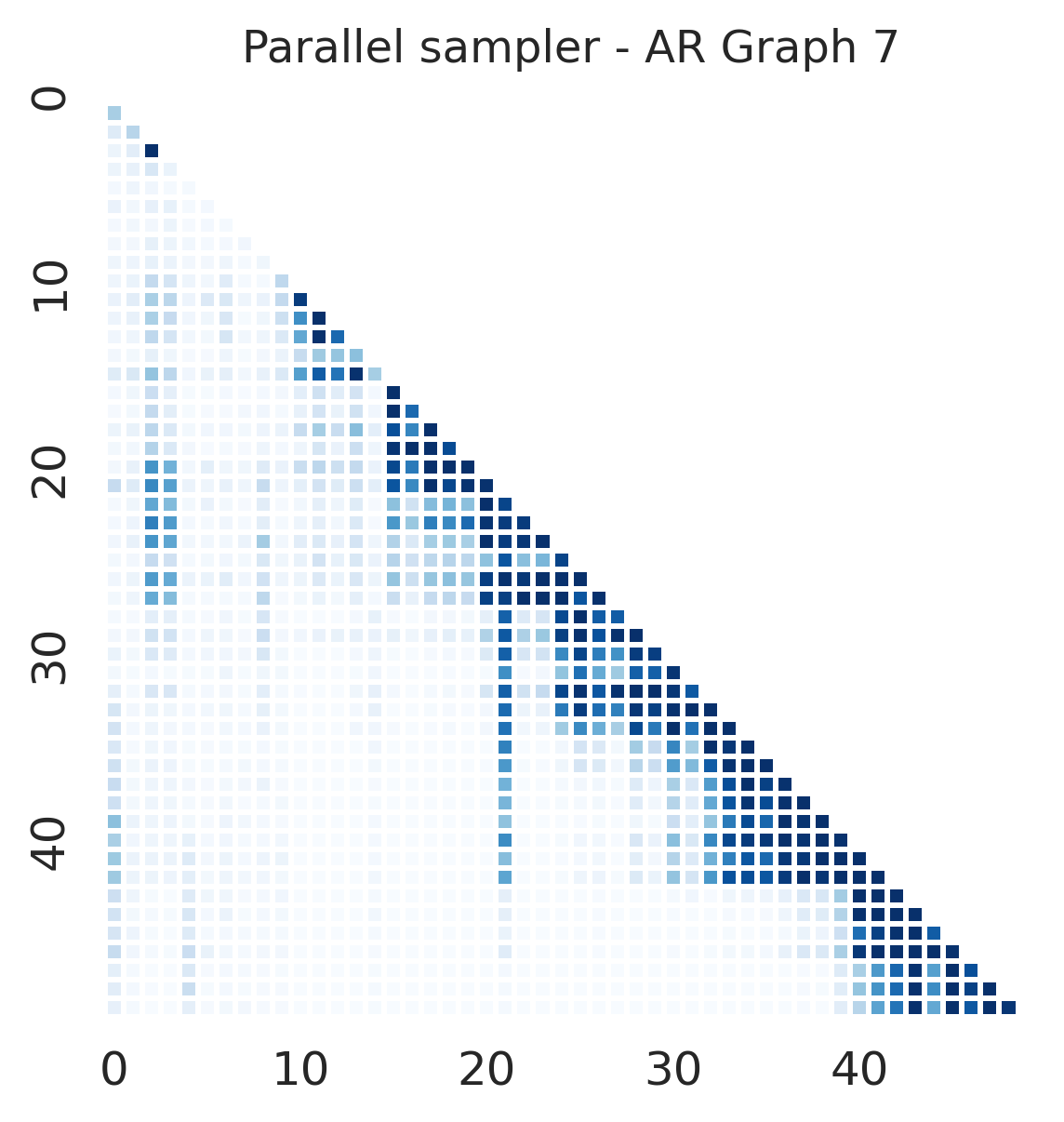}
  \includegraphics[width=0.19\textwidth]{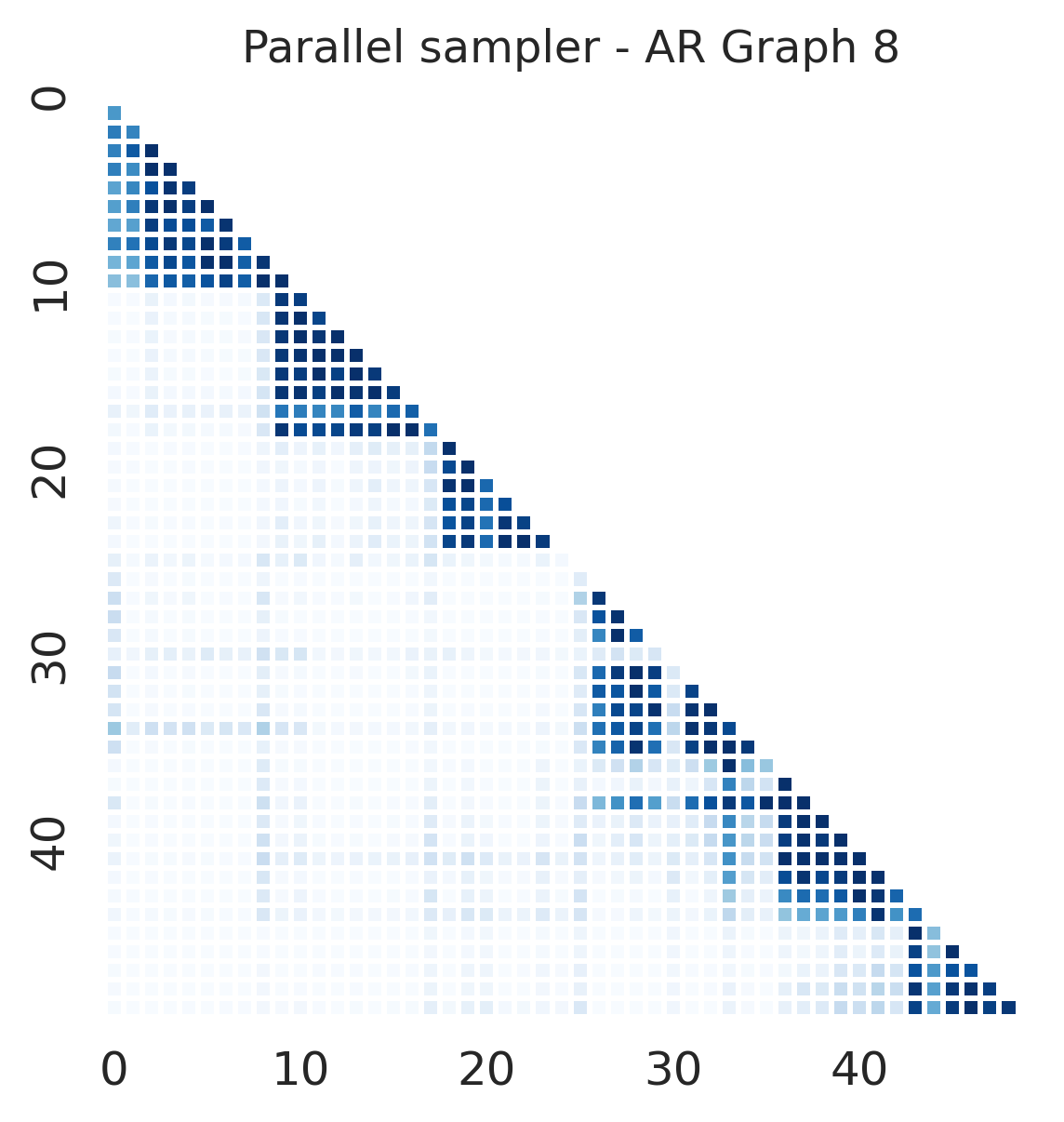}
  \includegraphics[width=0.19\textwidth]{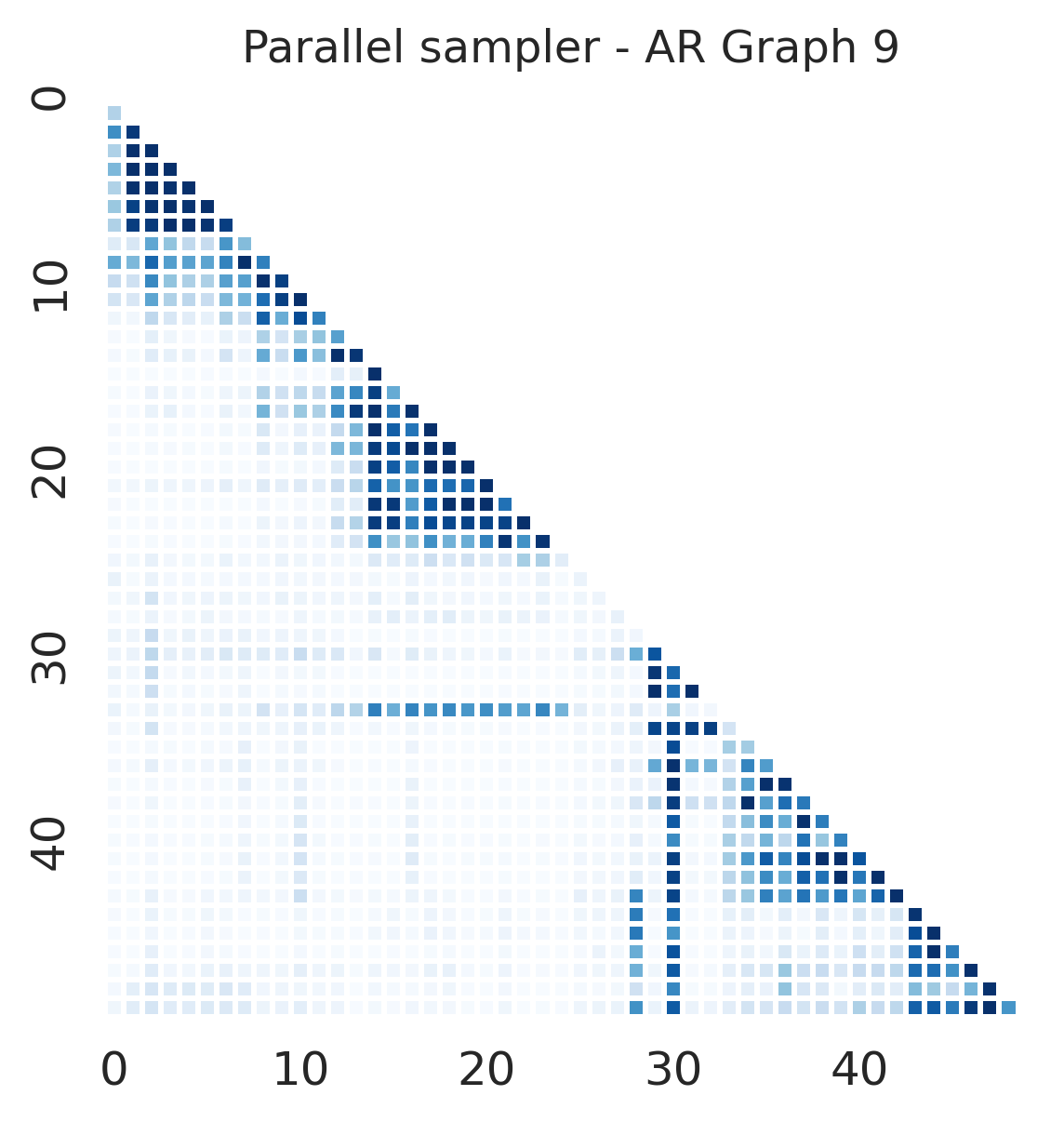}
  \includegraphics[width=0.19\textwidth]{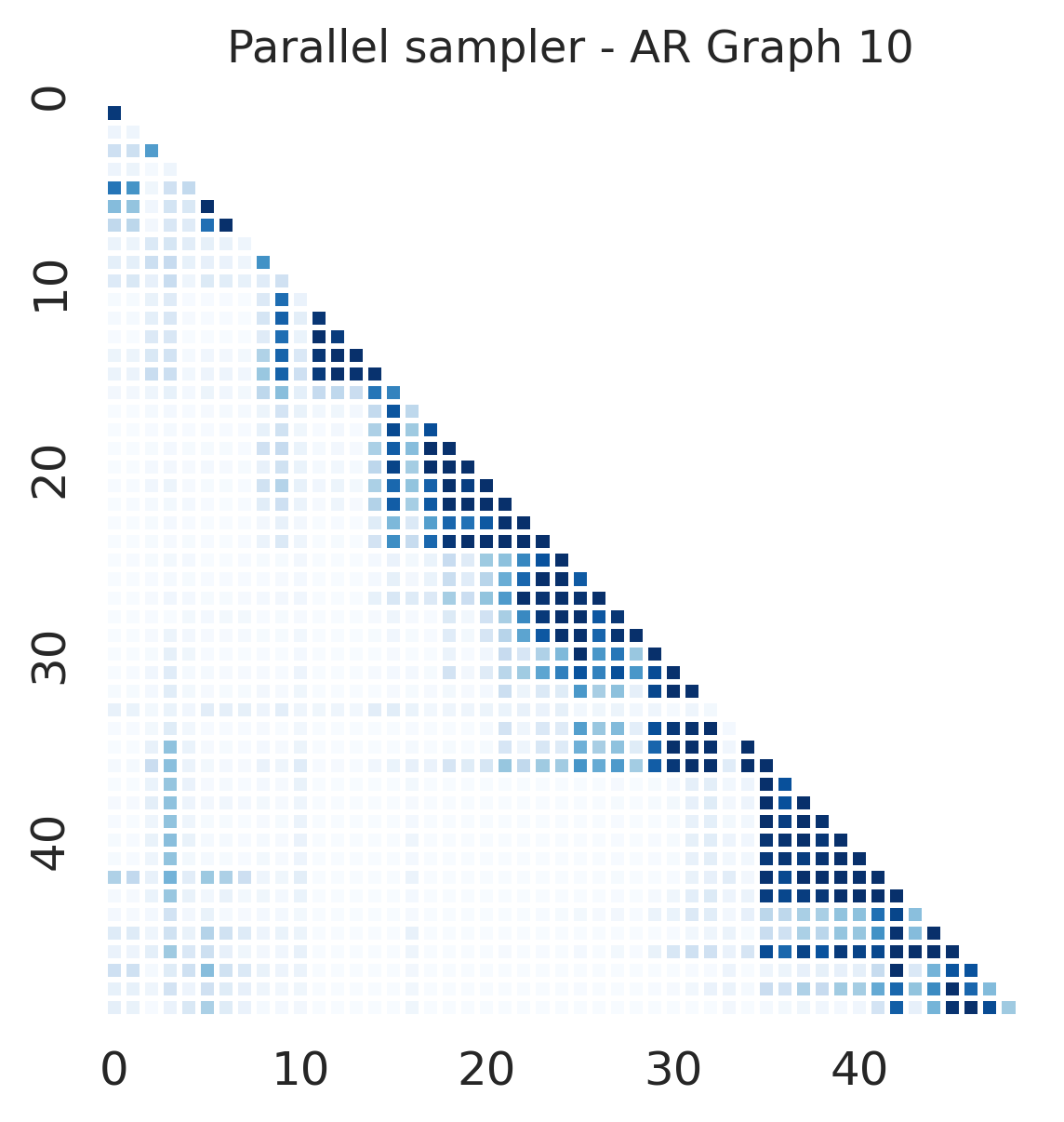}

  \includegraphics[width=0.19\textwidth]{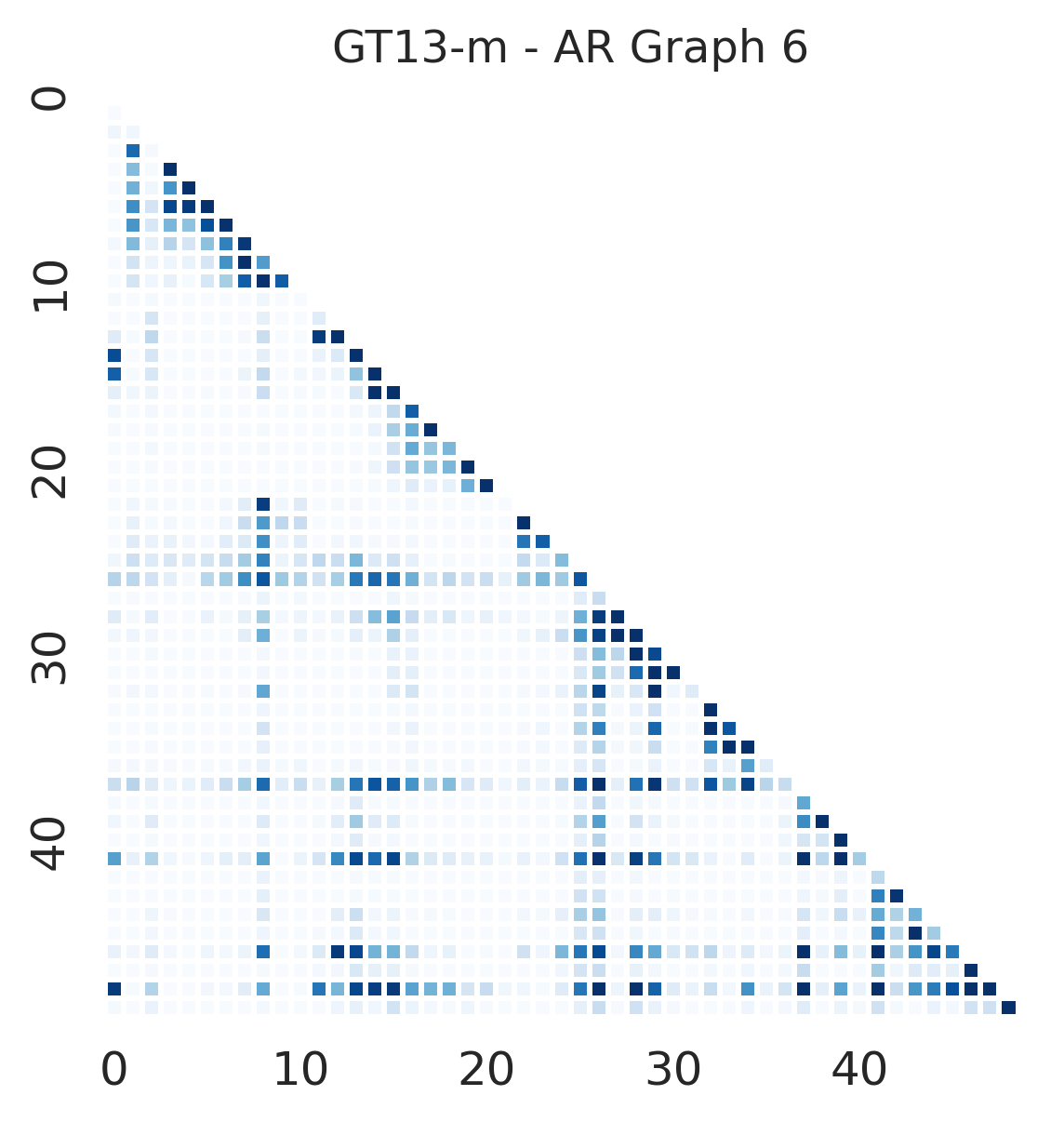}
  \includegraphics[width=0.19\textwidth]{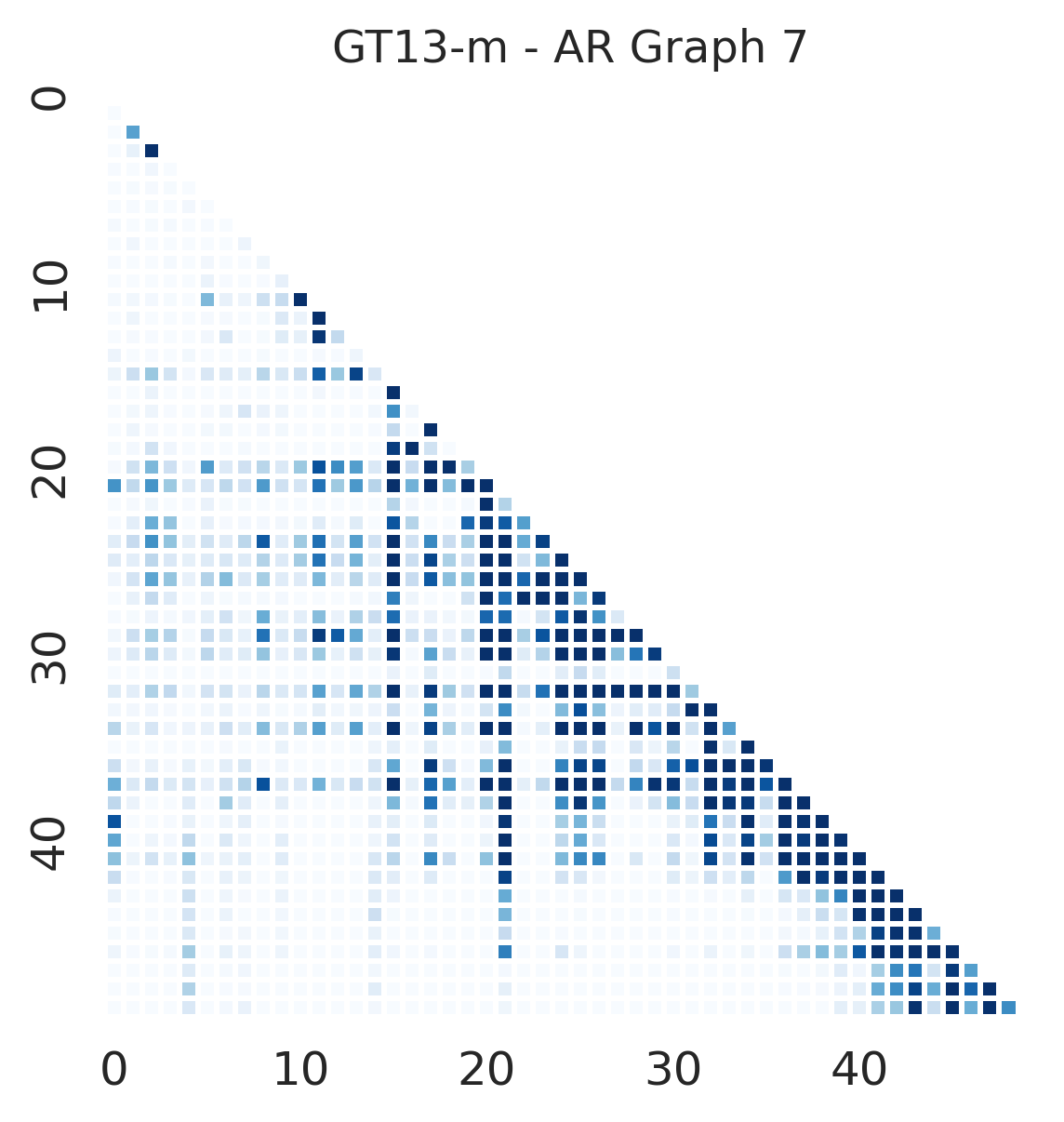}
  \includegraphics[width=0.19\textwidth]{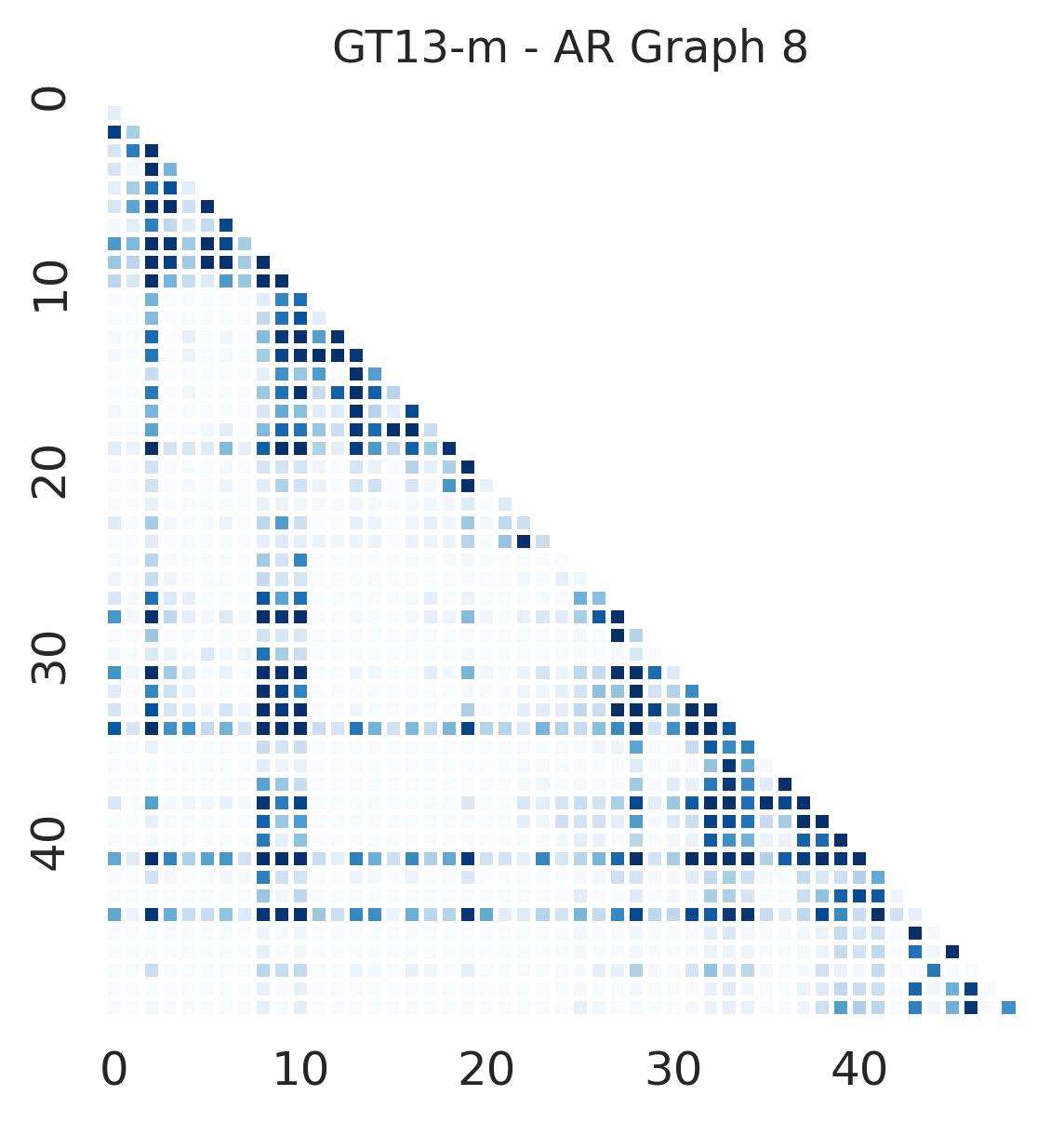}
  \includegraphics[width=0.19\textwidth]{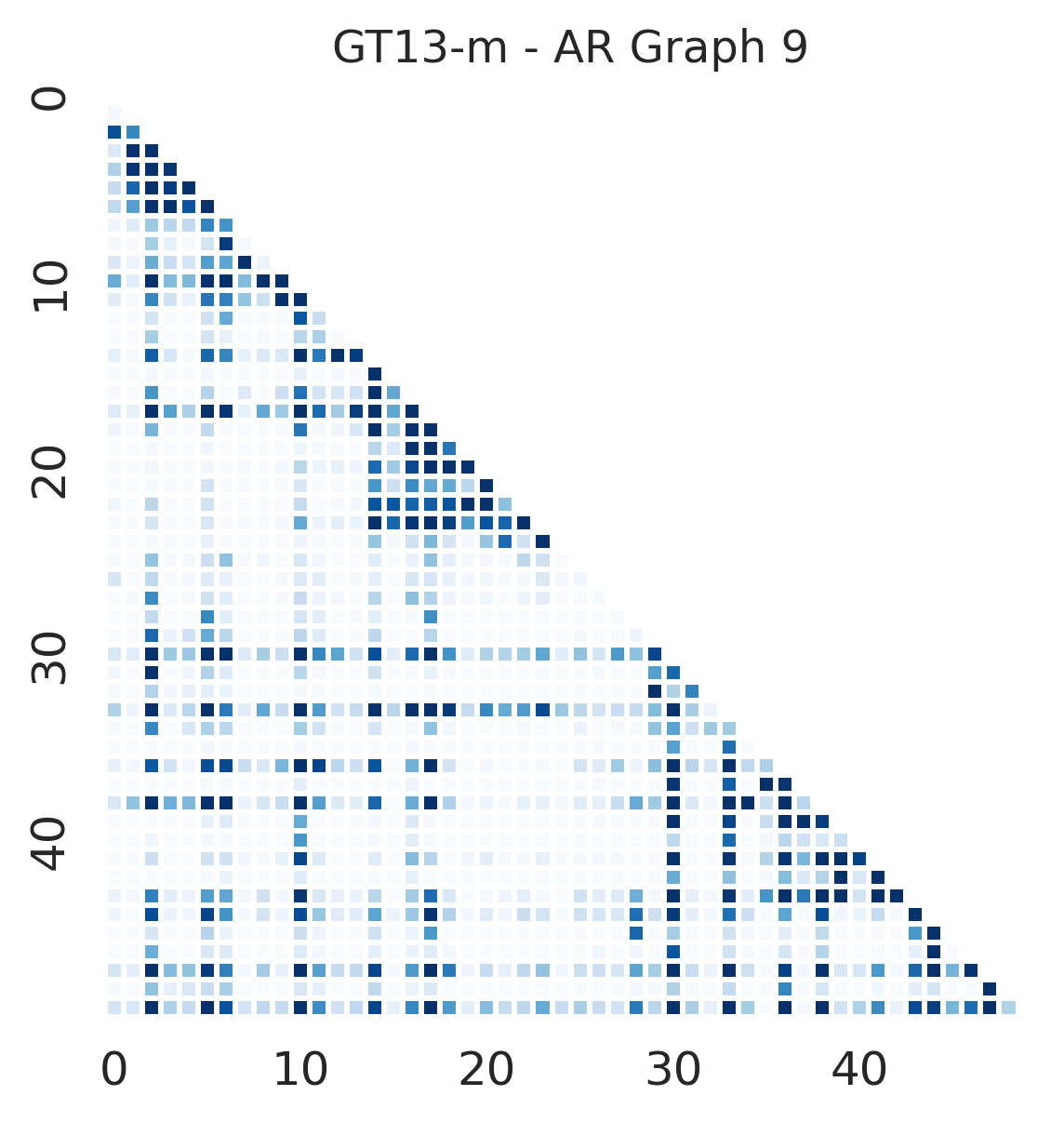}
  \includegraphics[width=0.19\textwidth]{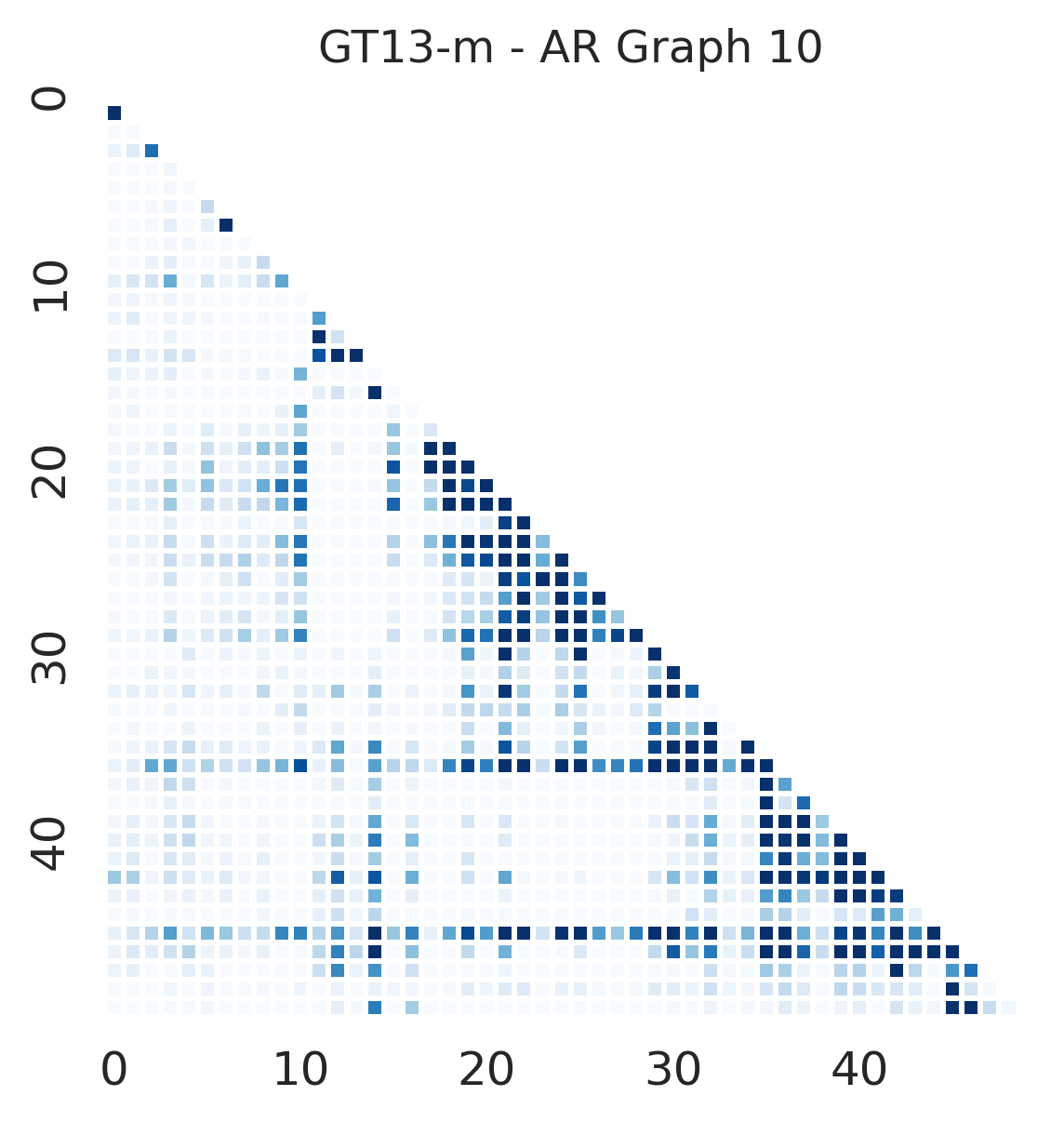}
  
  \includegraphics[width=0.19\textwidth]{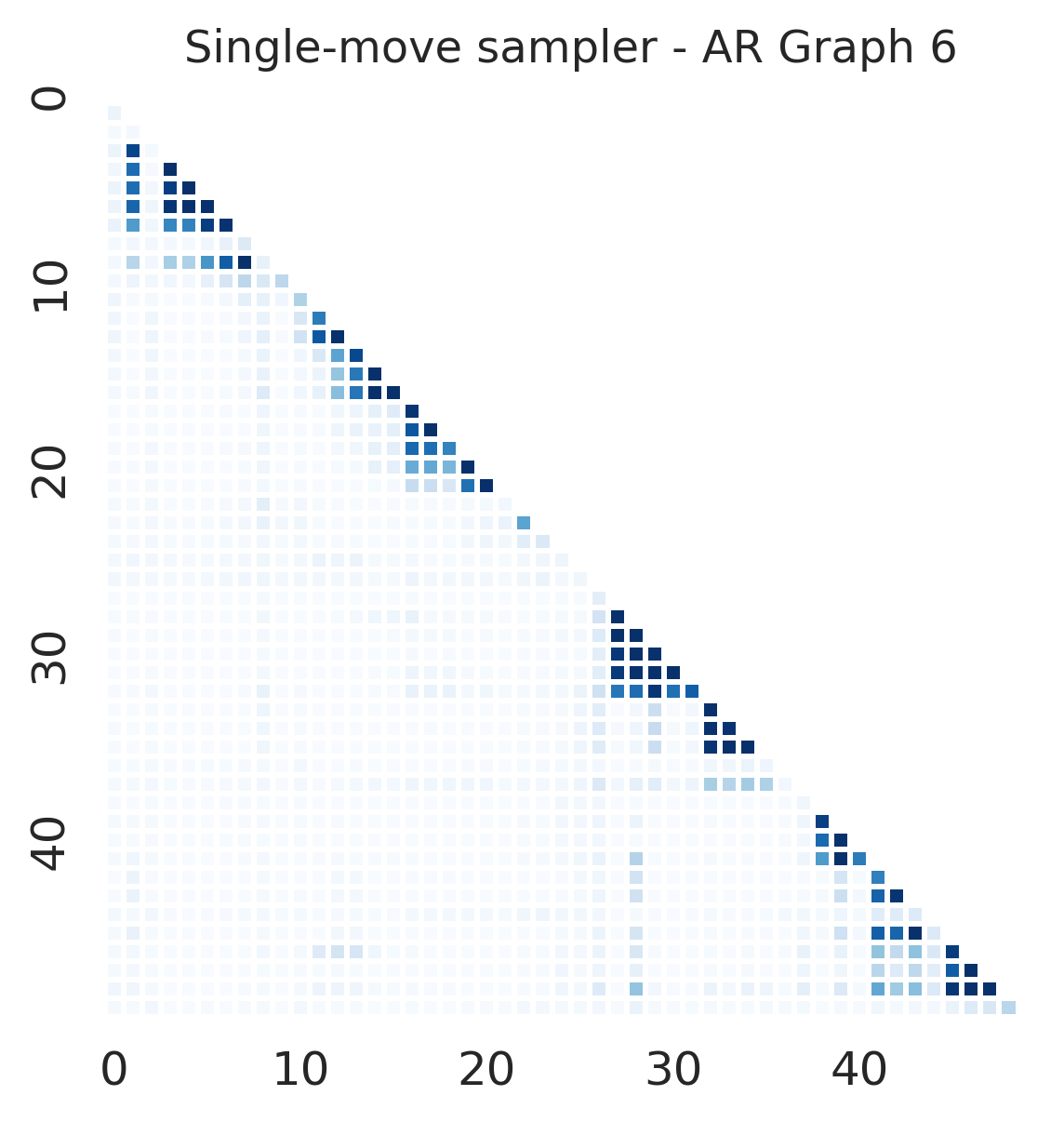}
  \includegraphics[width=0.19\textwidth]{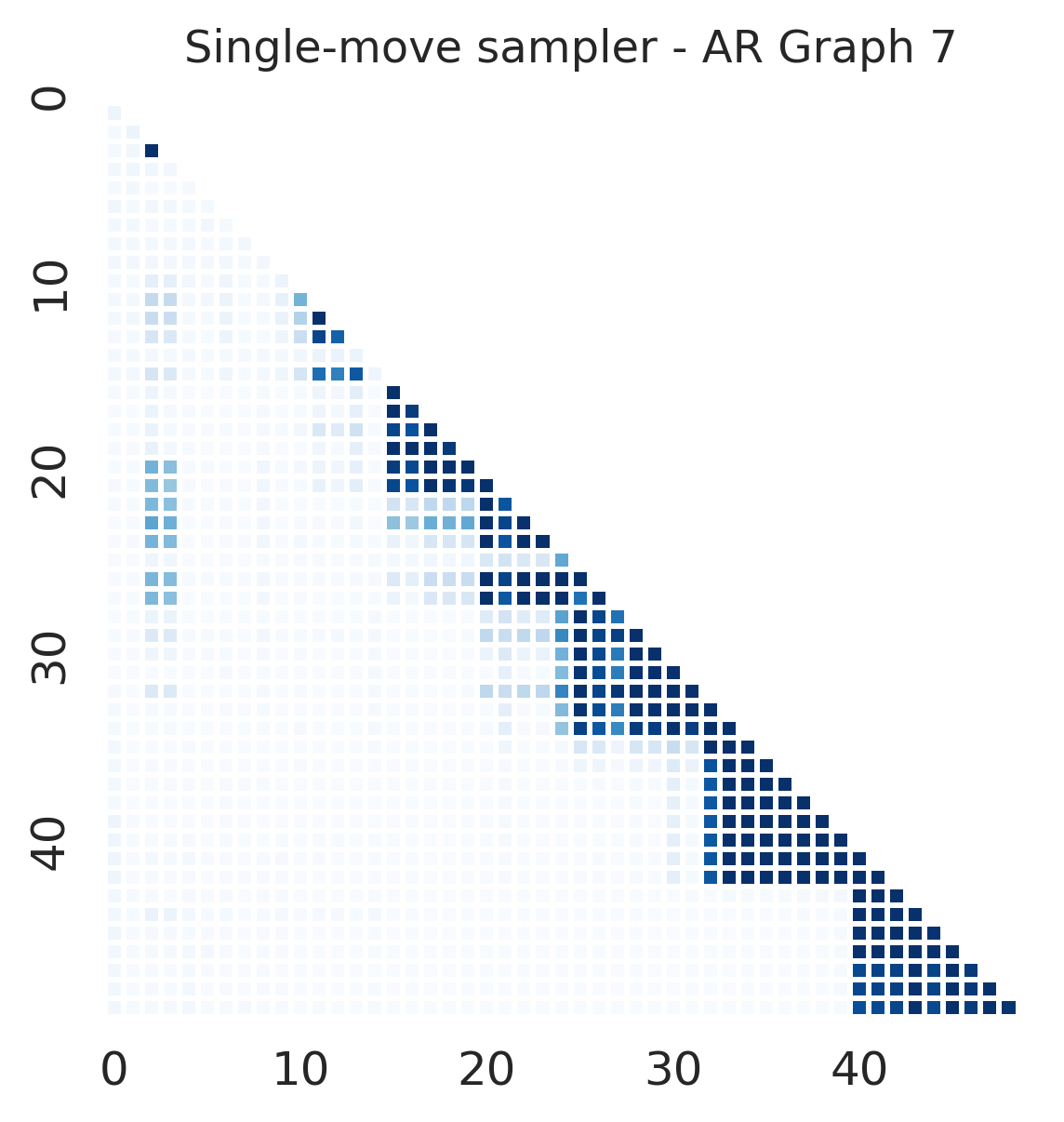}
  \includegraphics[width=0.19\textwidth]{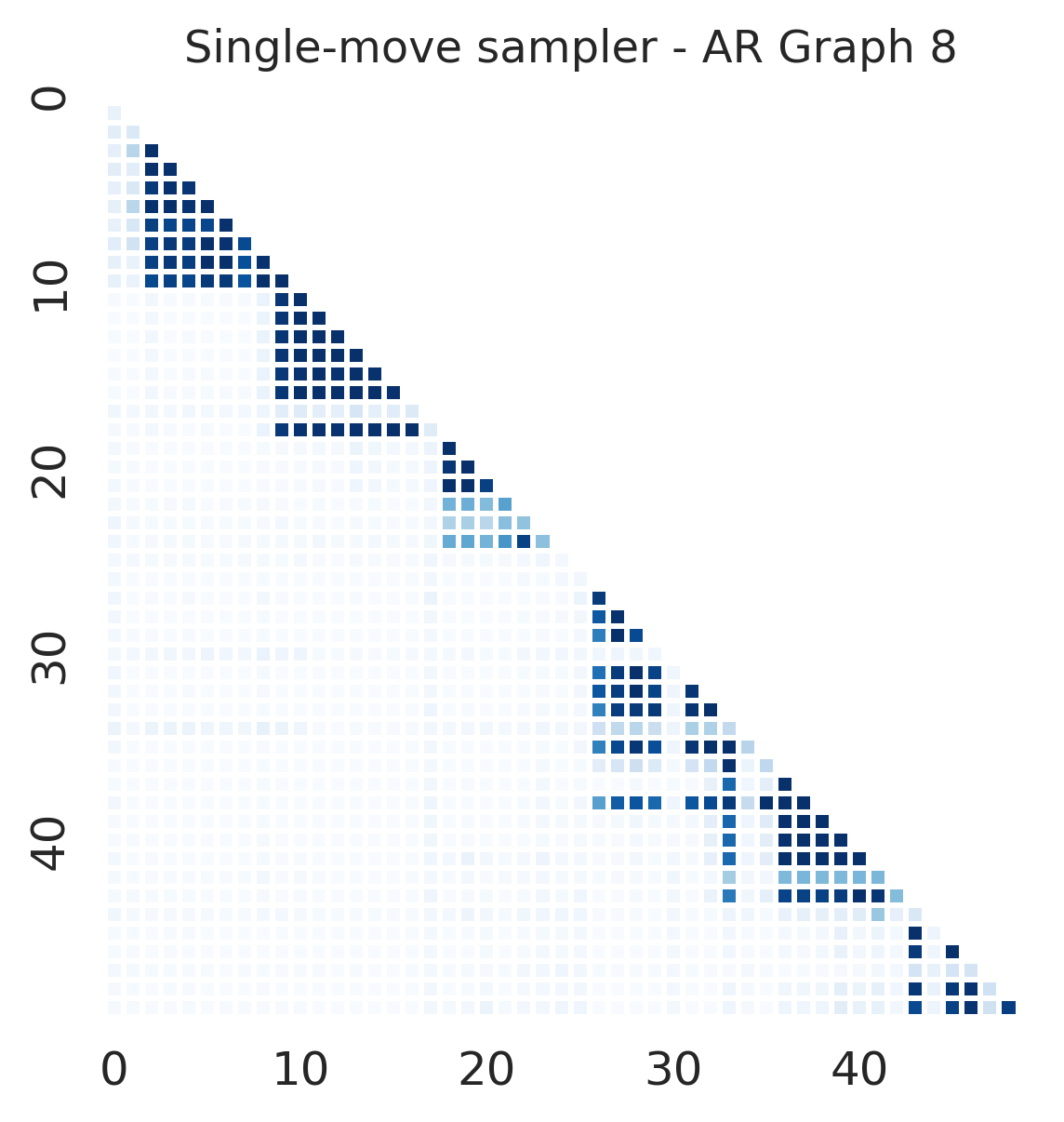}
  \includegraphics[width=0.19\textwidth]{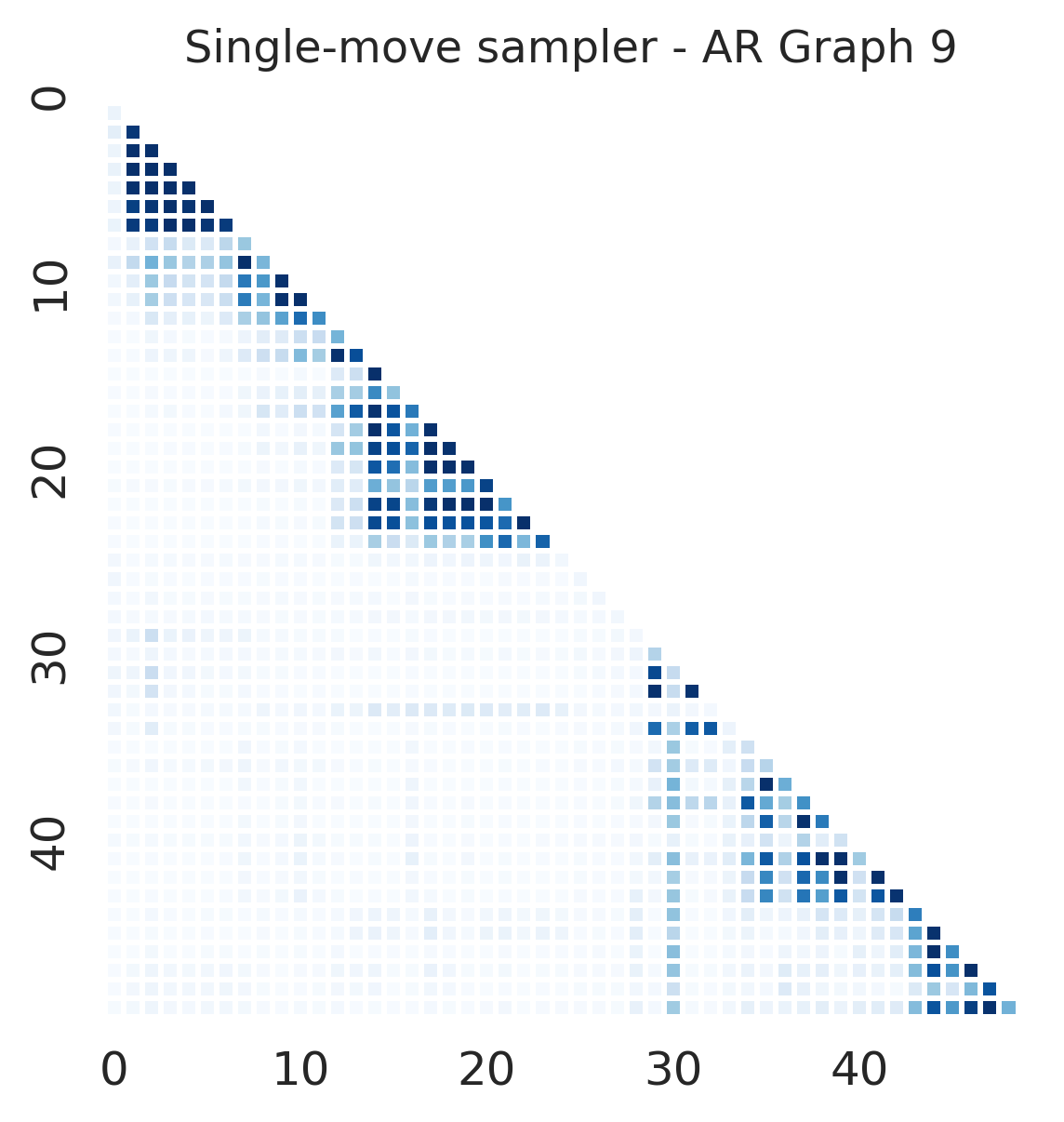}
  \includegraphics[width=0.19\textwidth]{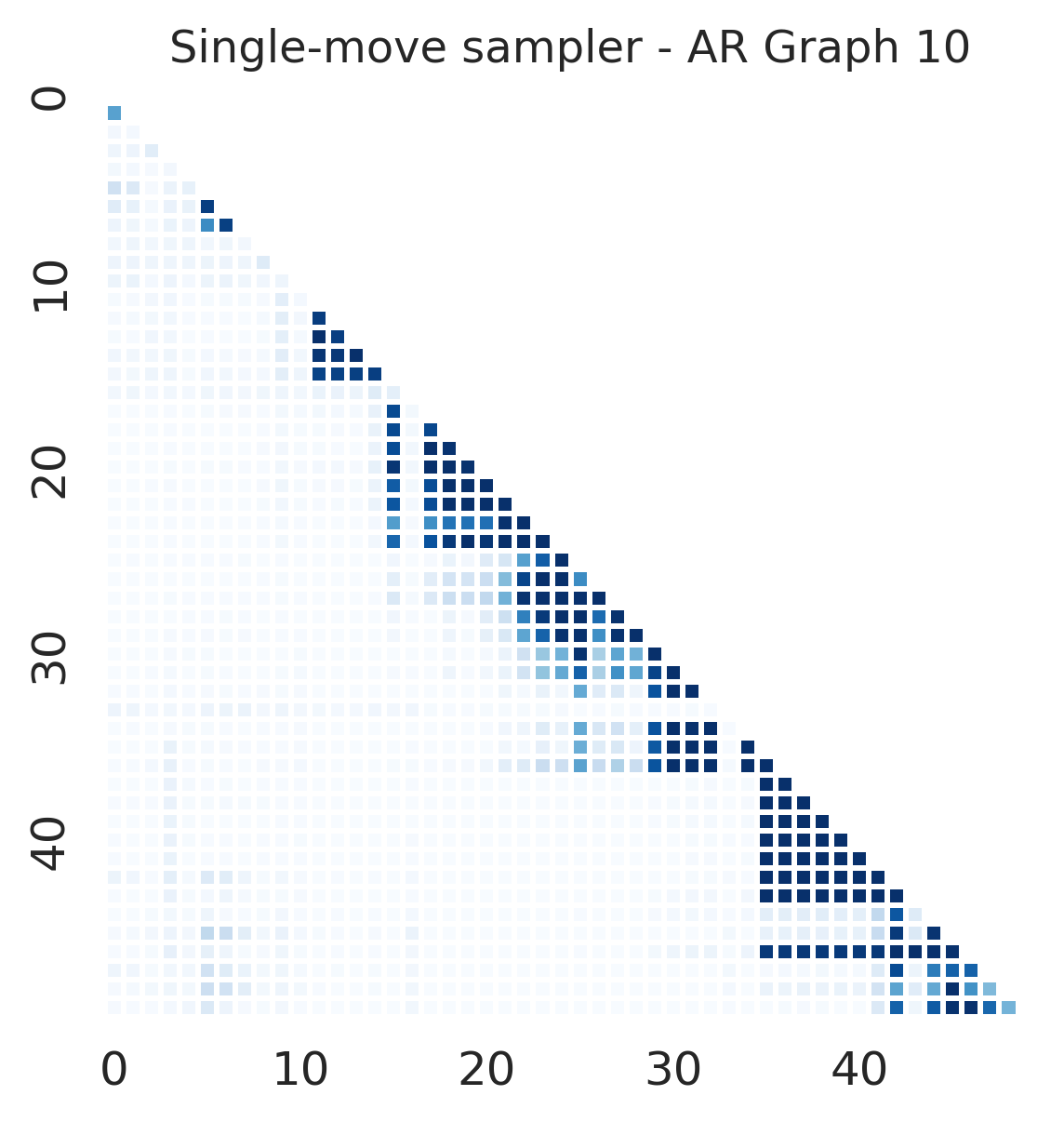}

  \includegraphics[width=0.19\textwidth]{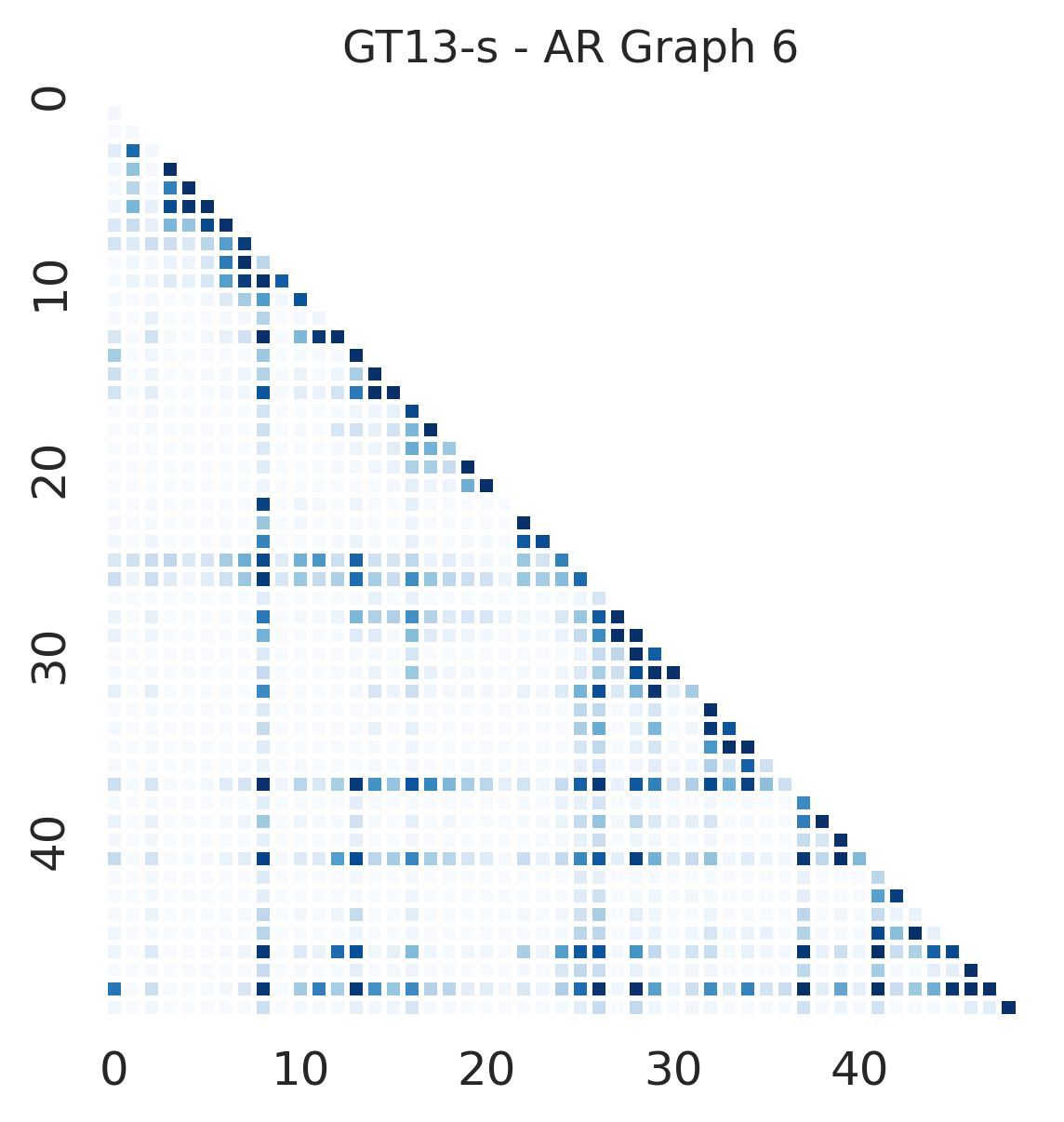}
  \includegraphics[width=0.19\textwidth]{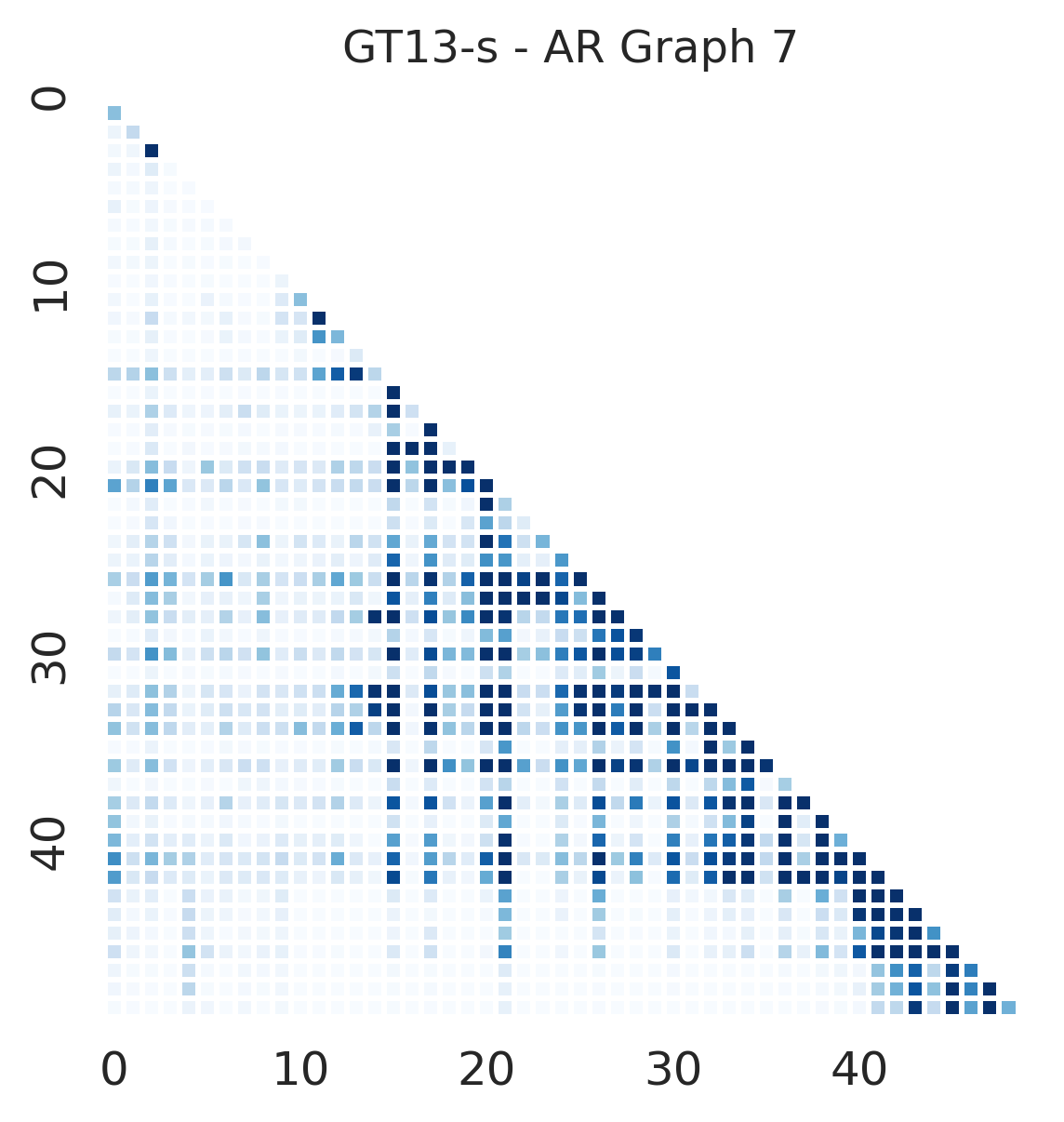}
  \includegraphics[width=0.19\textwidth]{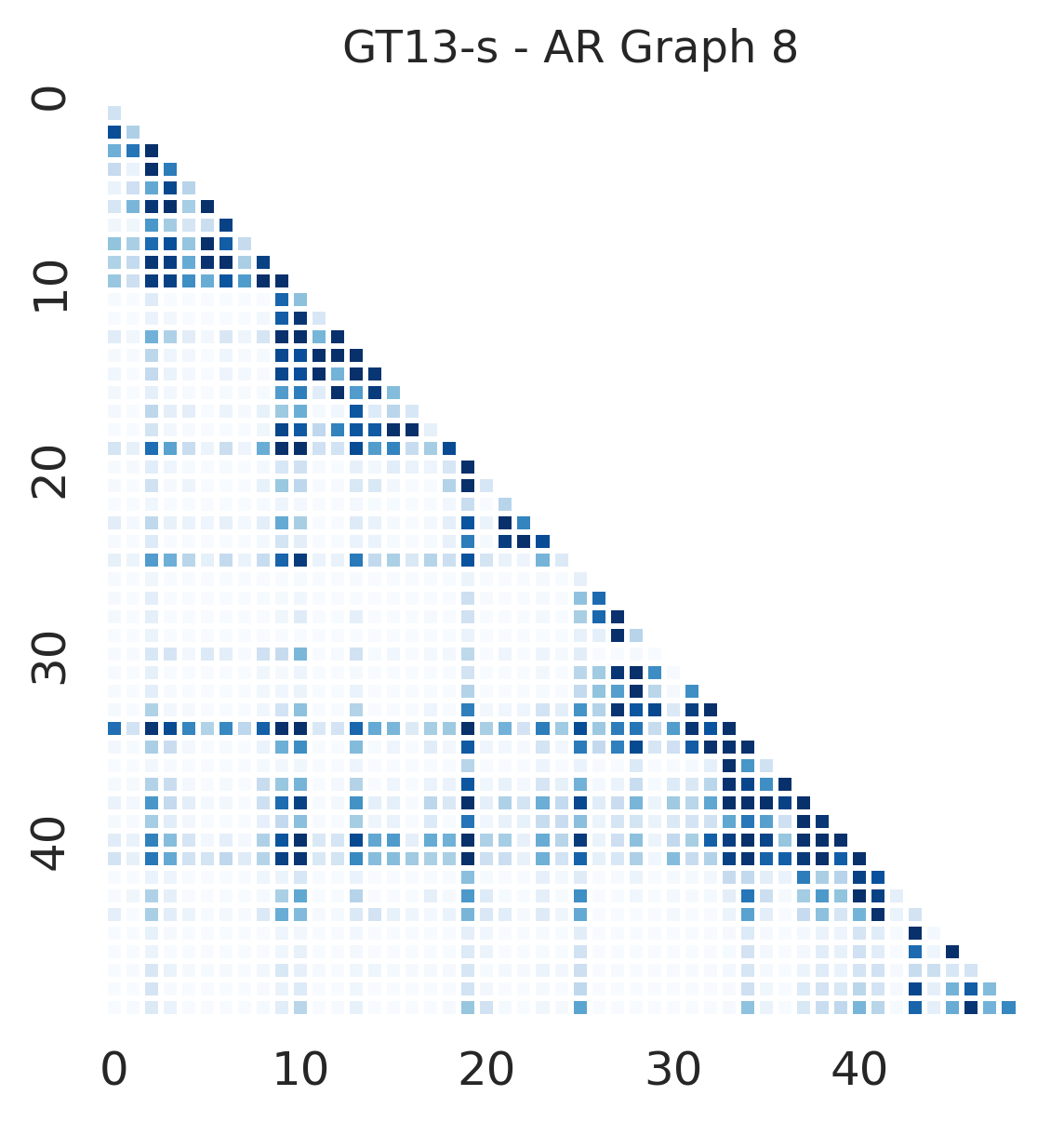}
  \includegraphics[width=0.19\textwidth]{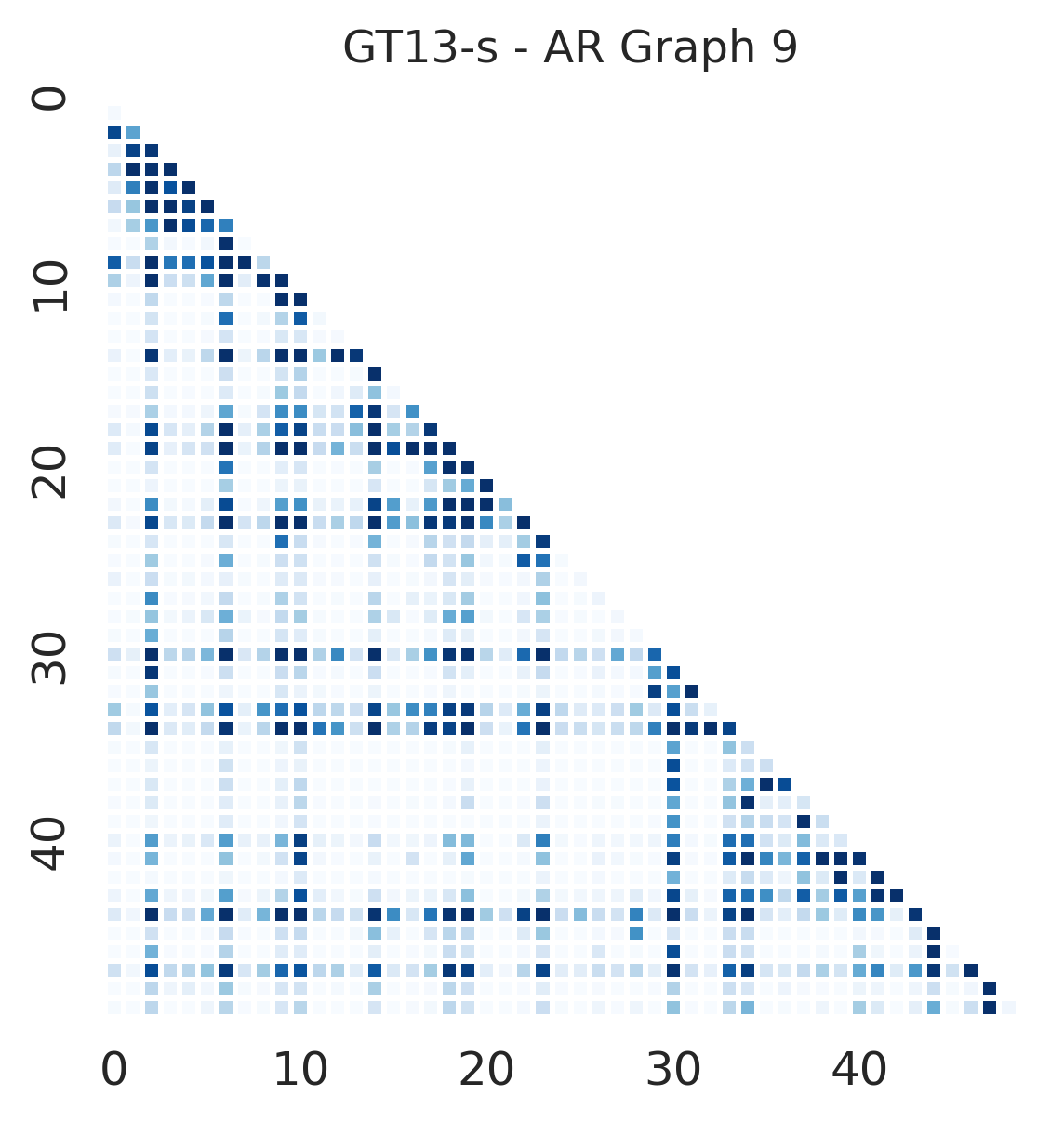}
  \includegraphics[width=0.19\textwidth]{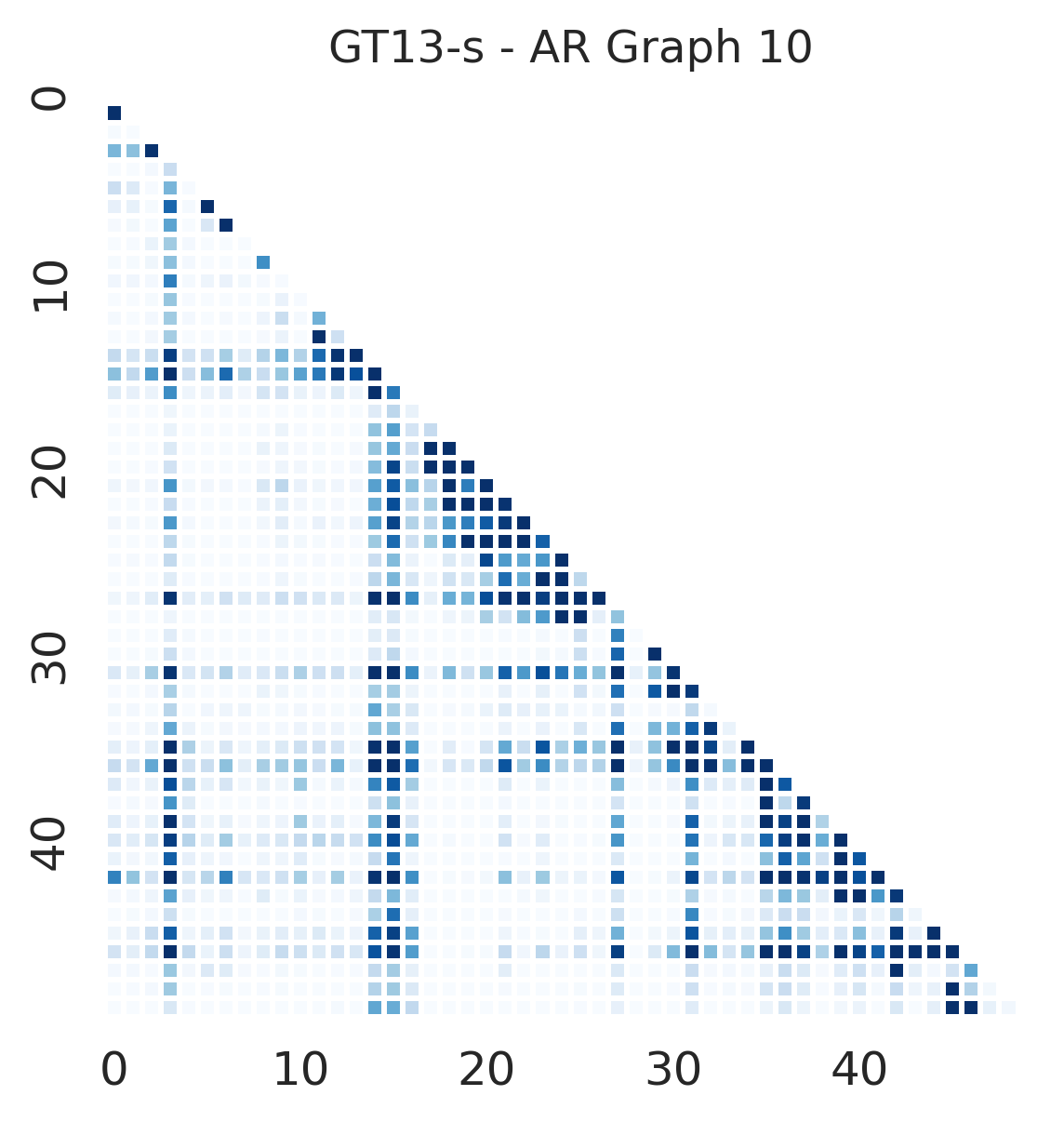}

  \includegraphics[width=0.19\textwidth]{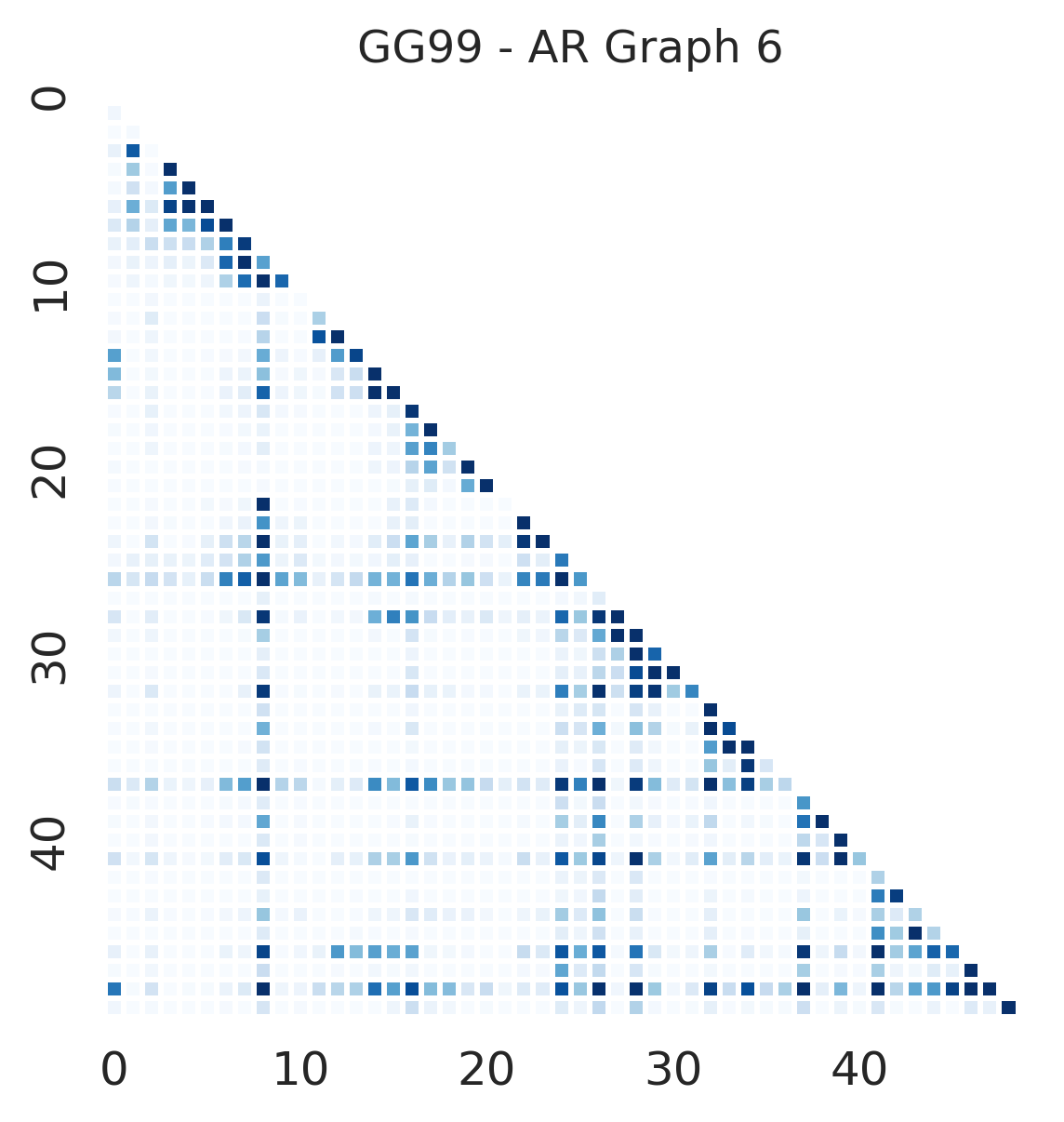}
  \includegraphics[width=0.19\textwidth]{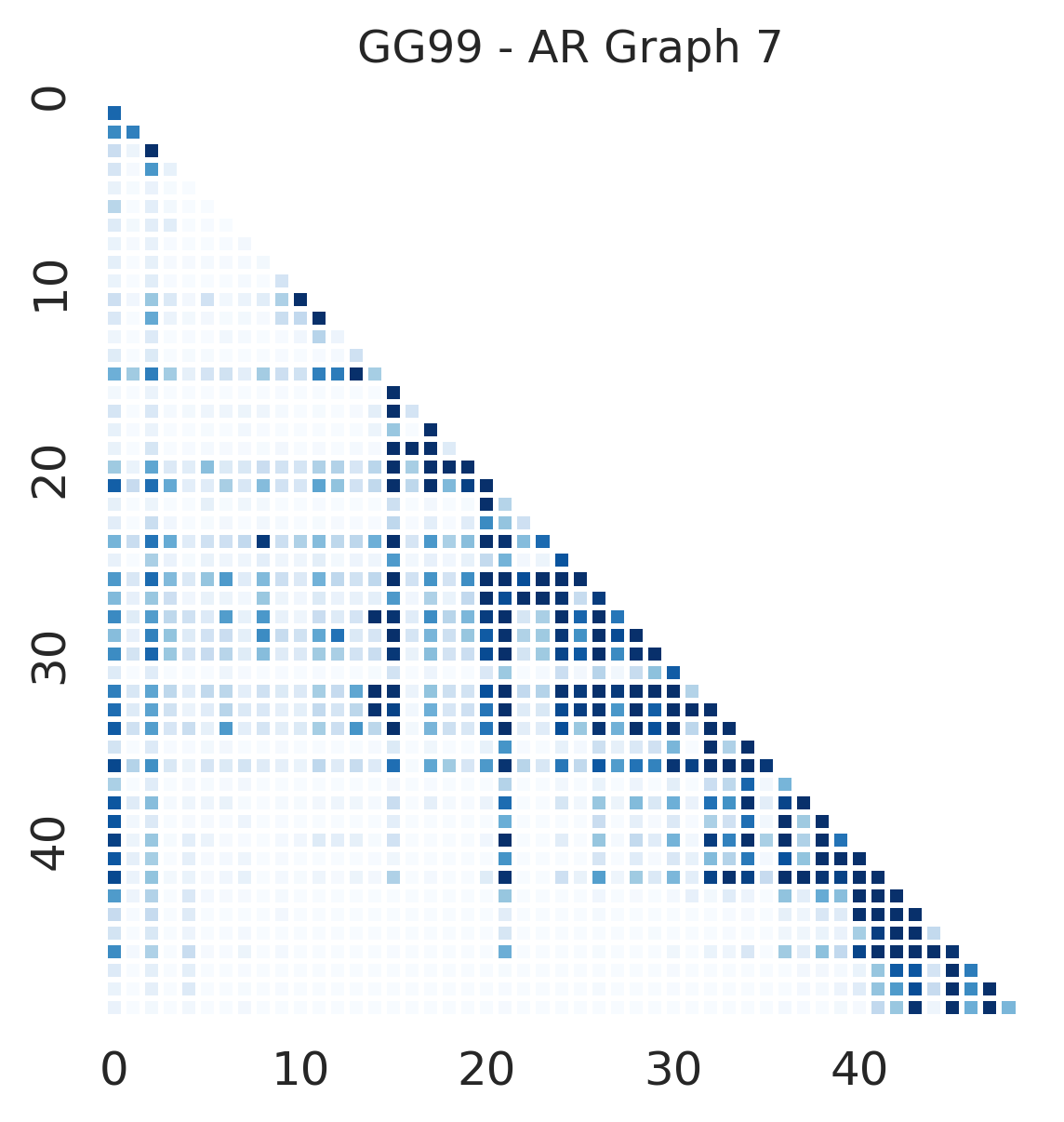}
  \includegraphics[width=0.19\textwidth]{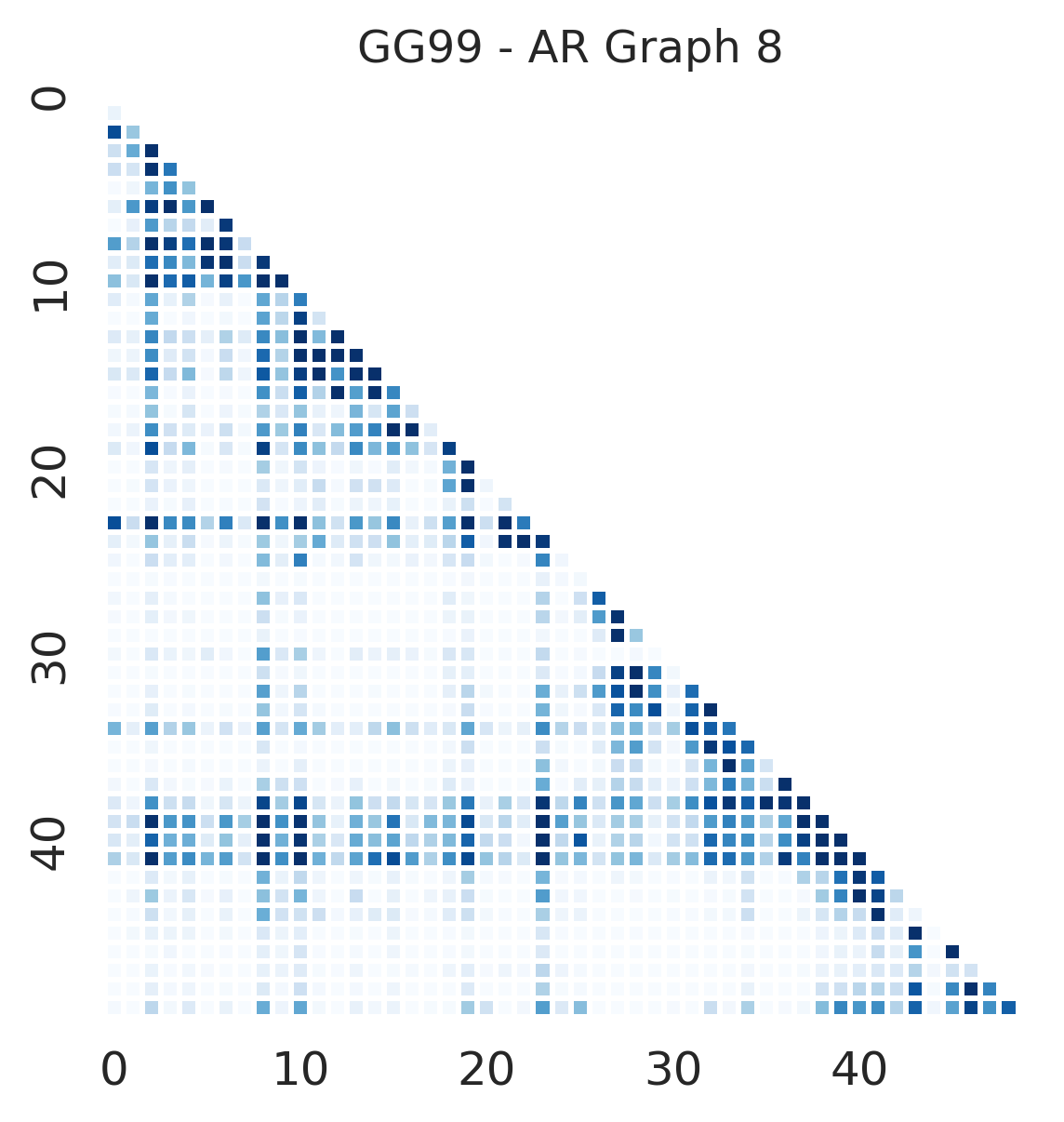}
  \includegraphics[width=0.19\textwidth]{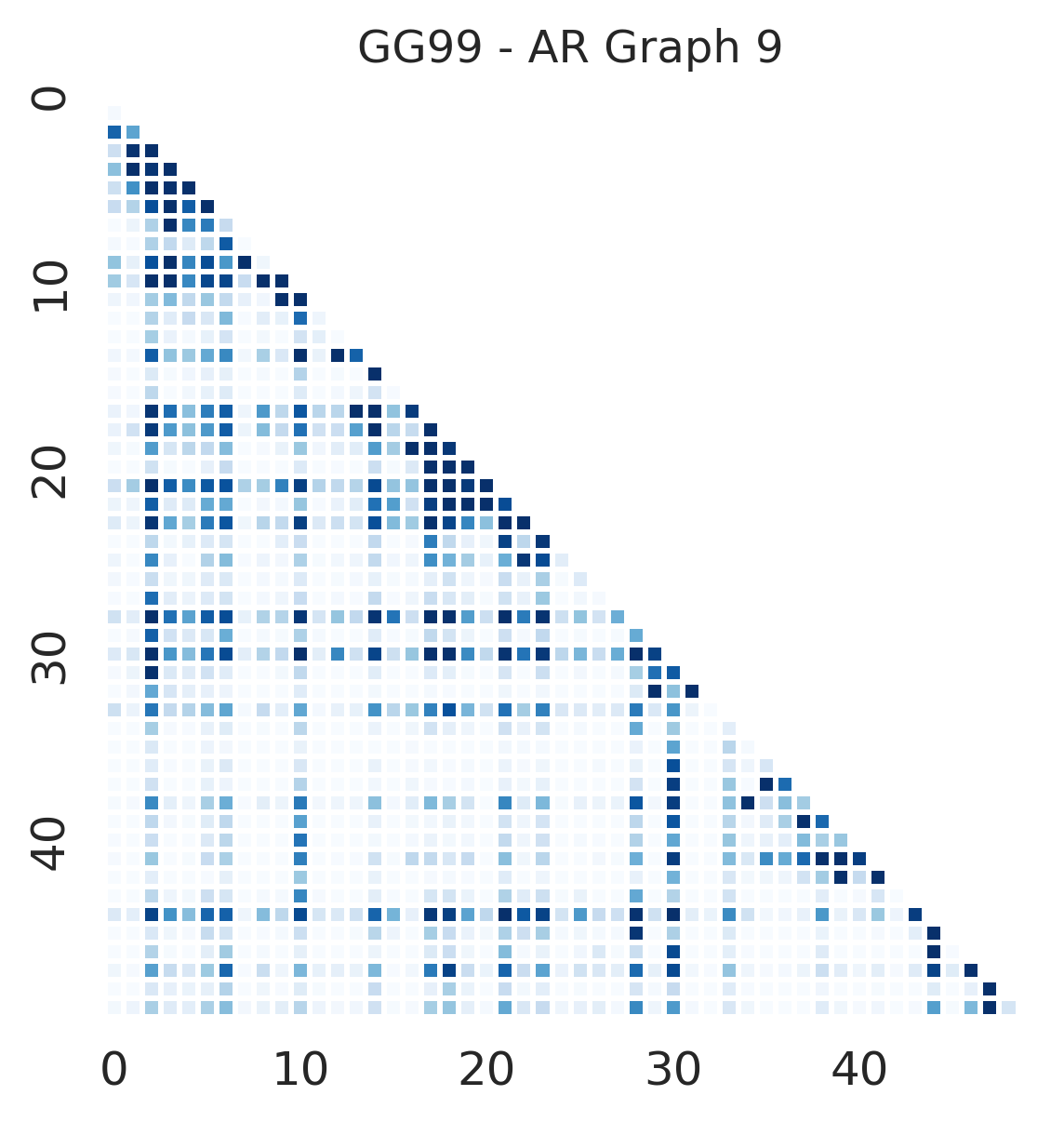}
  \includegraphics[width=0.19\textwidth]{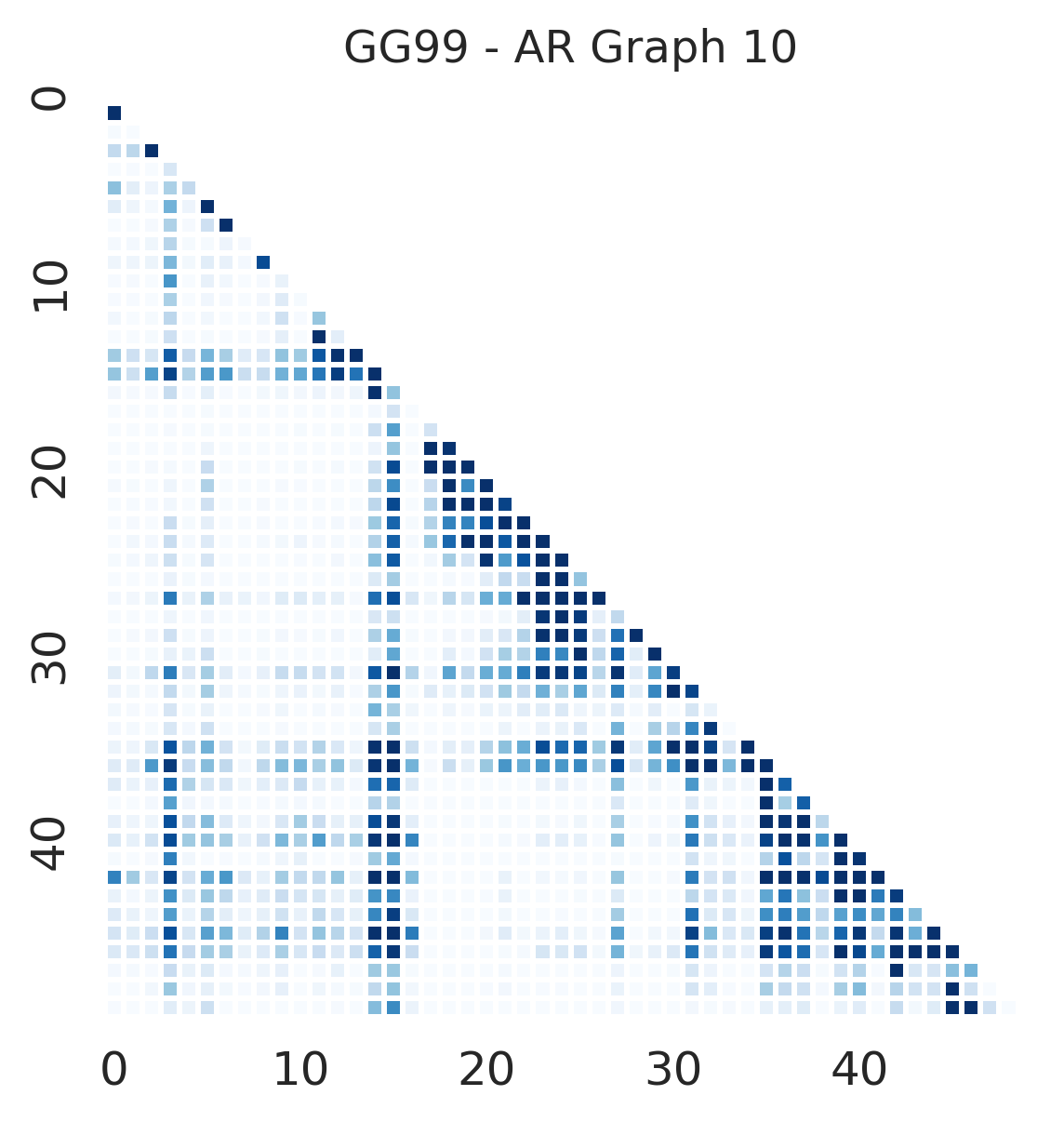}

  \includegraphics[width=0.19\textwidth]{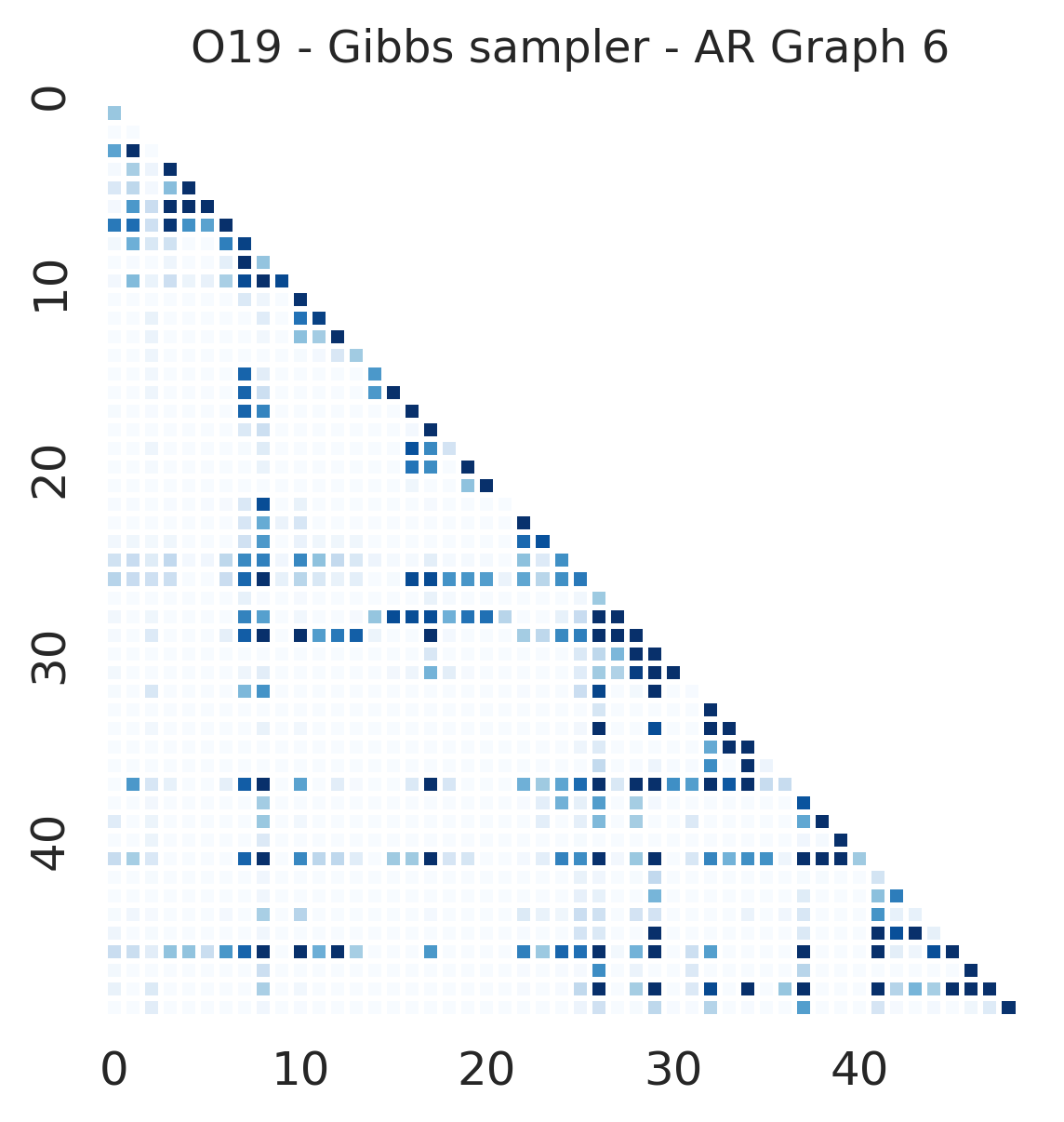}
  \includegraphics[width=0.19\textwidth]{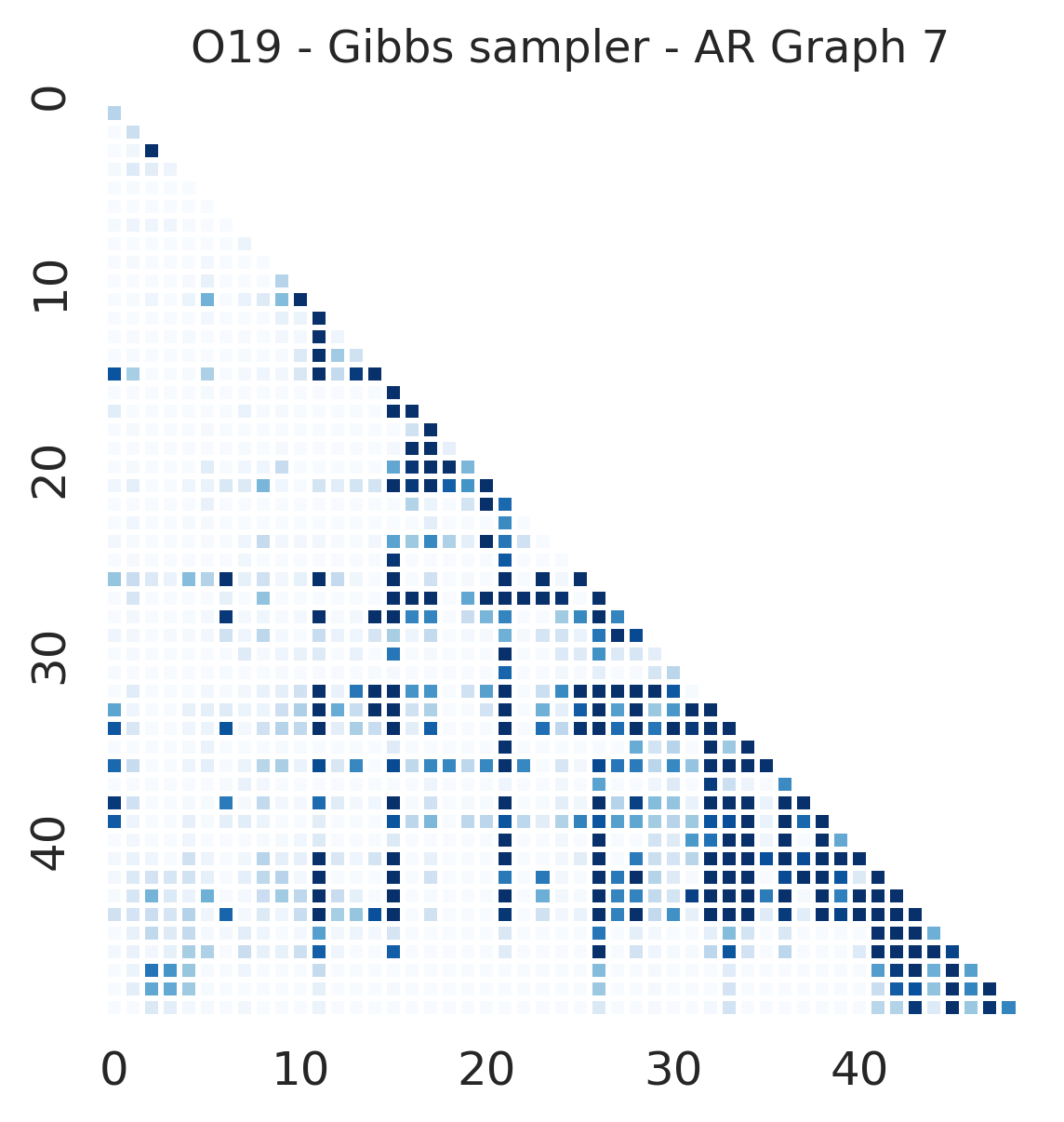}
  \includegraphics[width=0.19\textwidth]{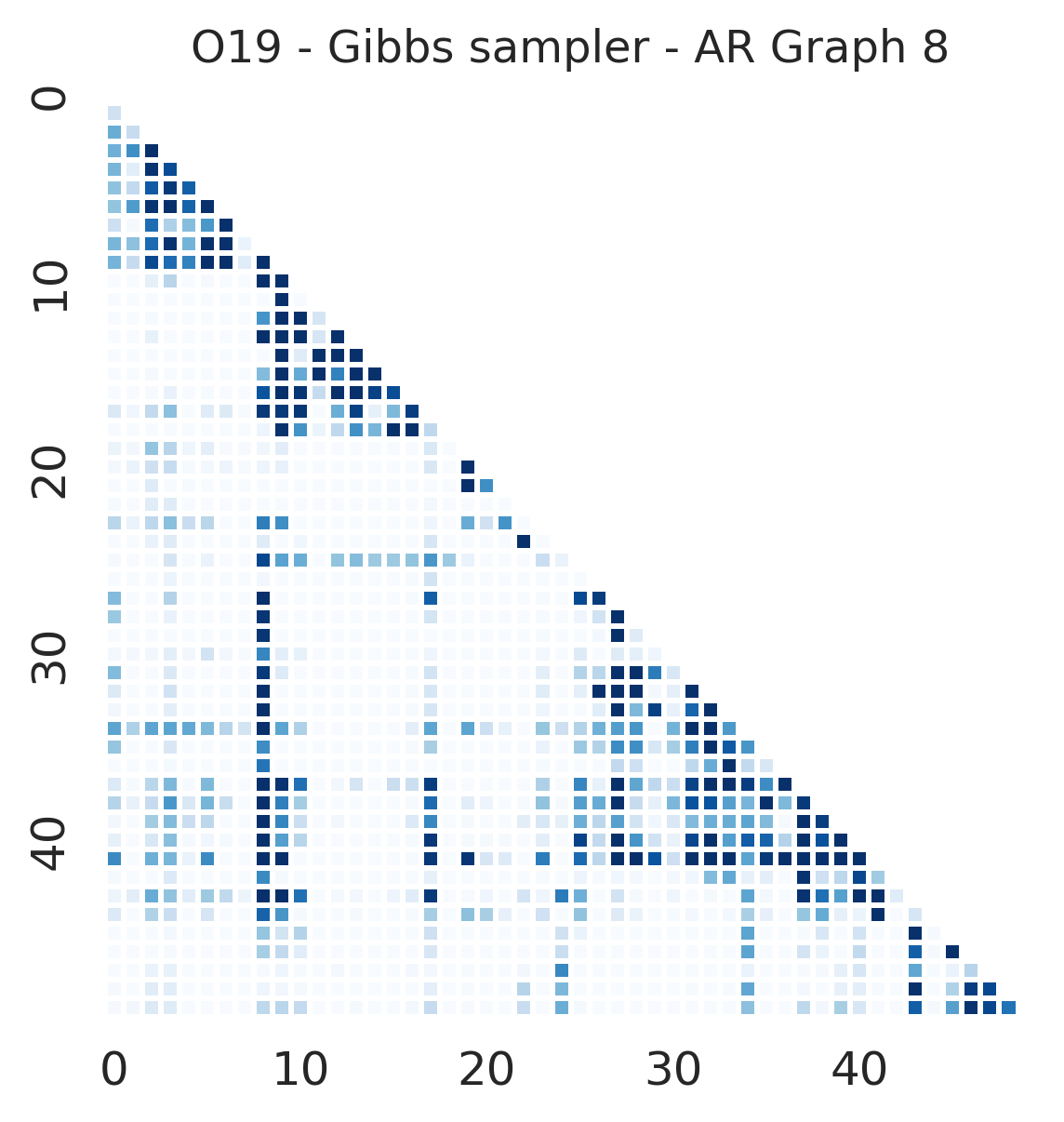}
  \includegraphics[width=0.19\textwidth]{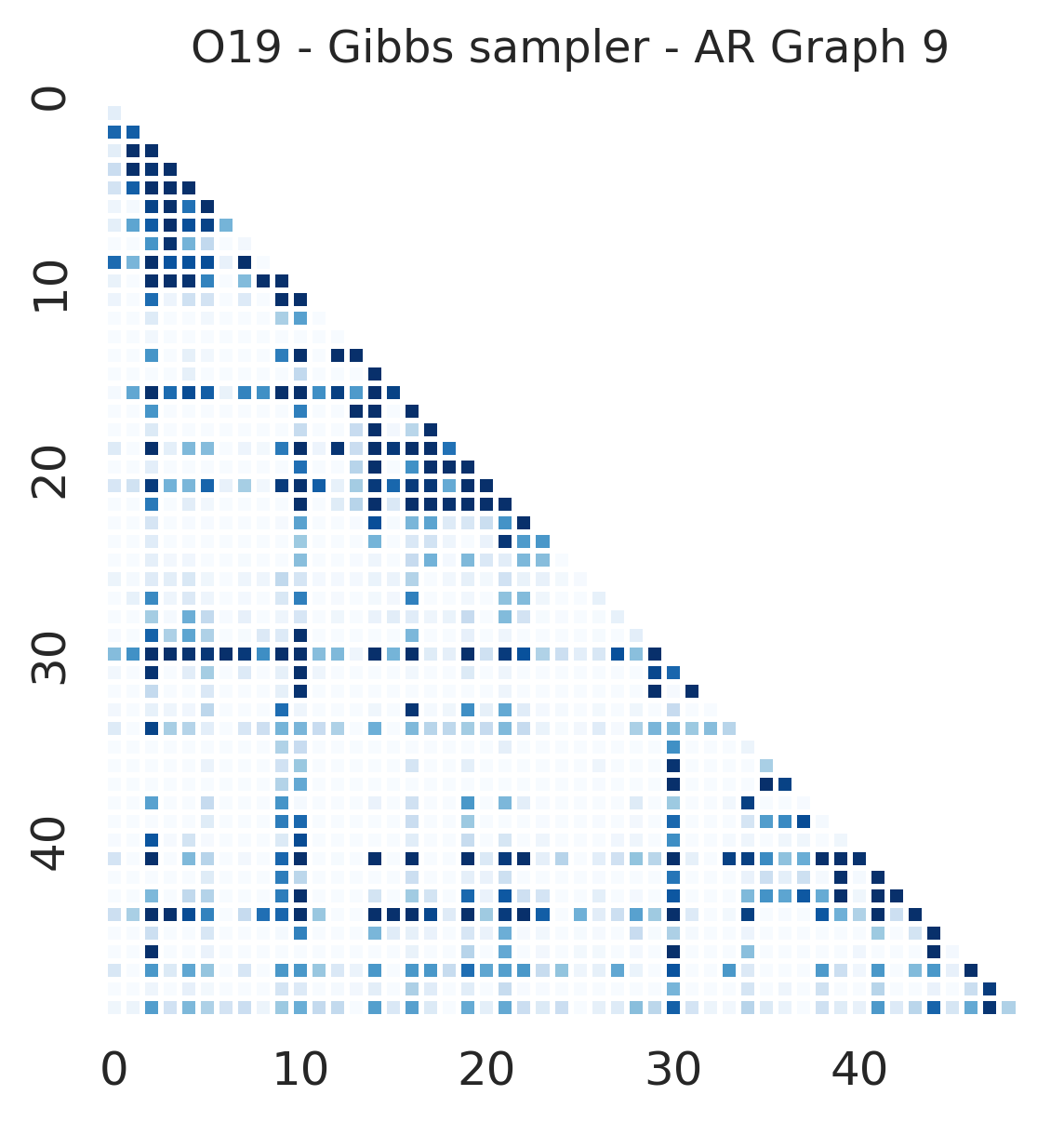}
  \includegraphics[width=0.19\textwidth]{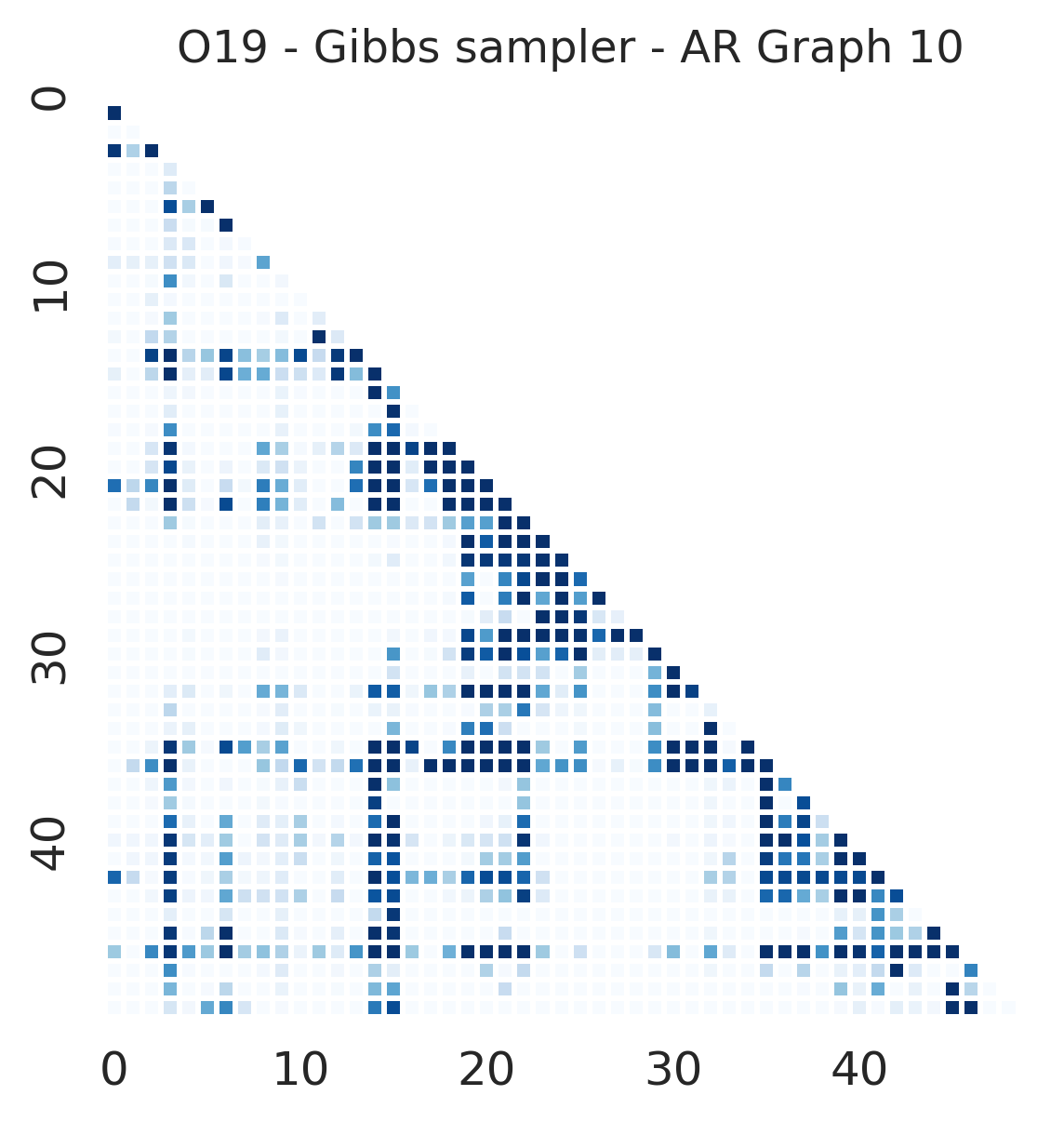}

  \caption{Heatmap derived from the final 300,000 steps of samplers in Section~\ref{sec:simul-setup-gauss}, covering replications 6 to 10 under the autoregressive structure for $(n,p)=(100,50)$.}
  \label{app:fig:heatmap-bandmat-1}
\end{figure}

\clearpage
\section{Extra simulation results for high-dimensional case} 
\label{app:sec:ggm:sparse}

\begin{figure}[ht!]
  \centering
  \includegraphics[width=0.7\textwidth]{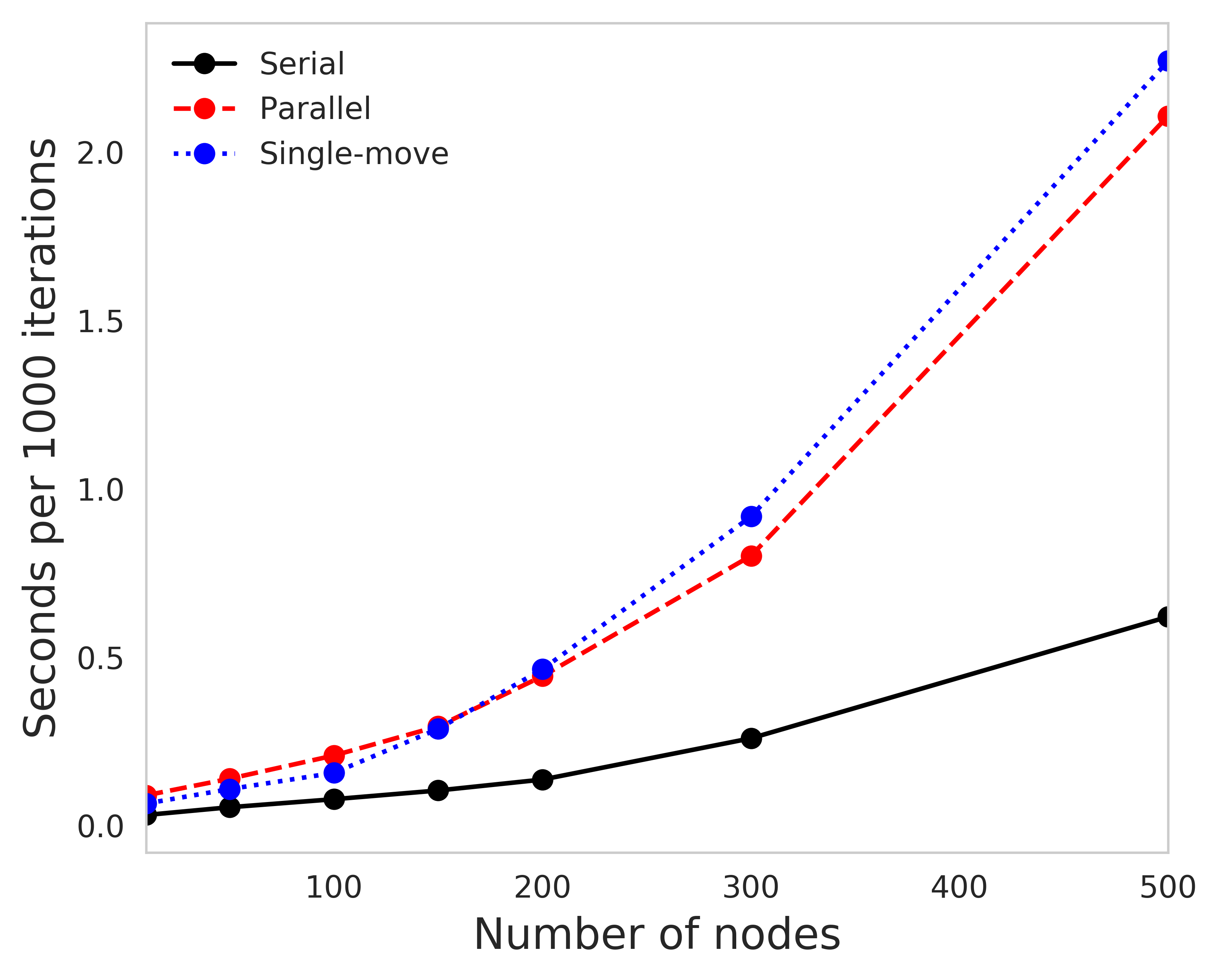}
  \caption{Average time (in seconds) for 1000 steps of the Gaussian decomposable graphical model with an autoregressive structure, comparing the parallel sampler, its serial variant, and the single-move sampler. The number of data samples equal the dimension.}
  \label{fig:time-samplers}
\end{figure}

\begin{figure}[ht!]
  \centering
  \includegraphics[width=0.32\textwidth]{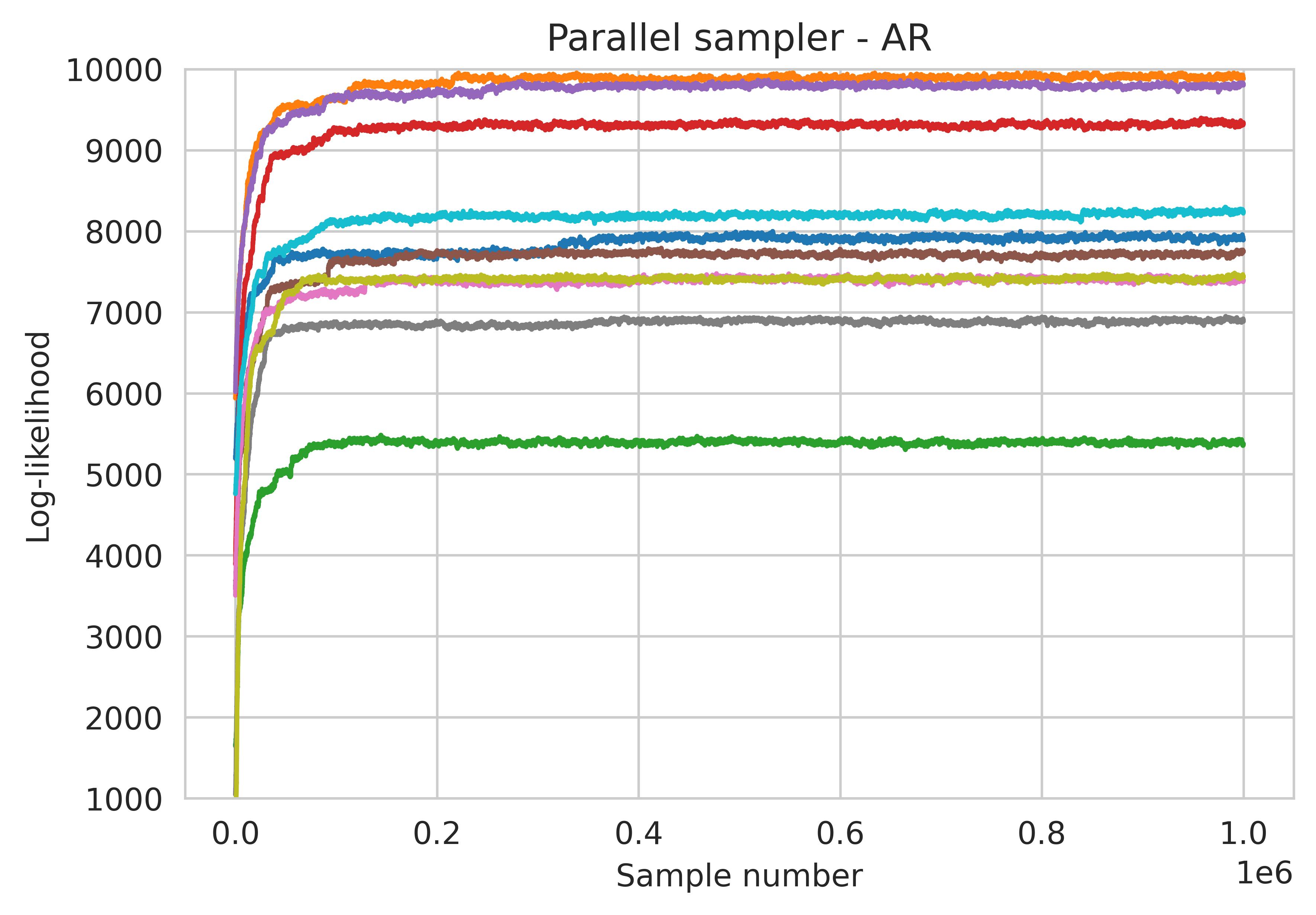}
  \includegraphics[width=0.32\textwidth]{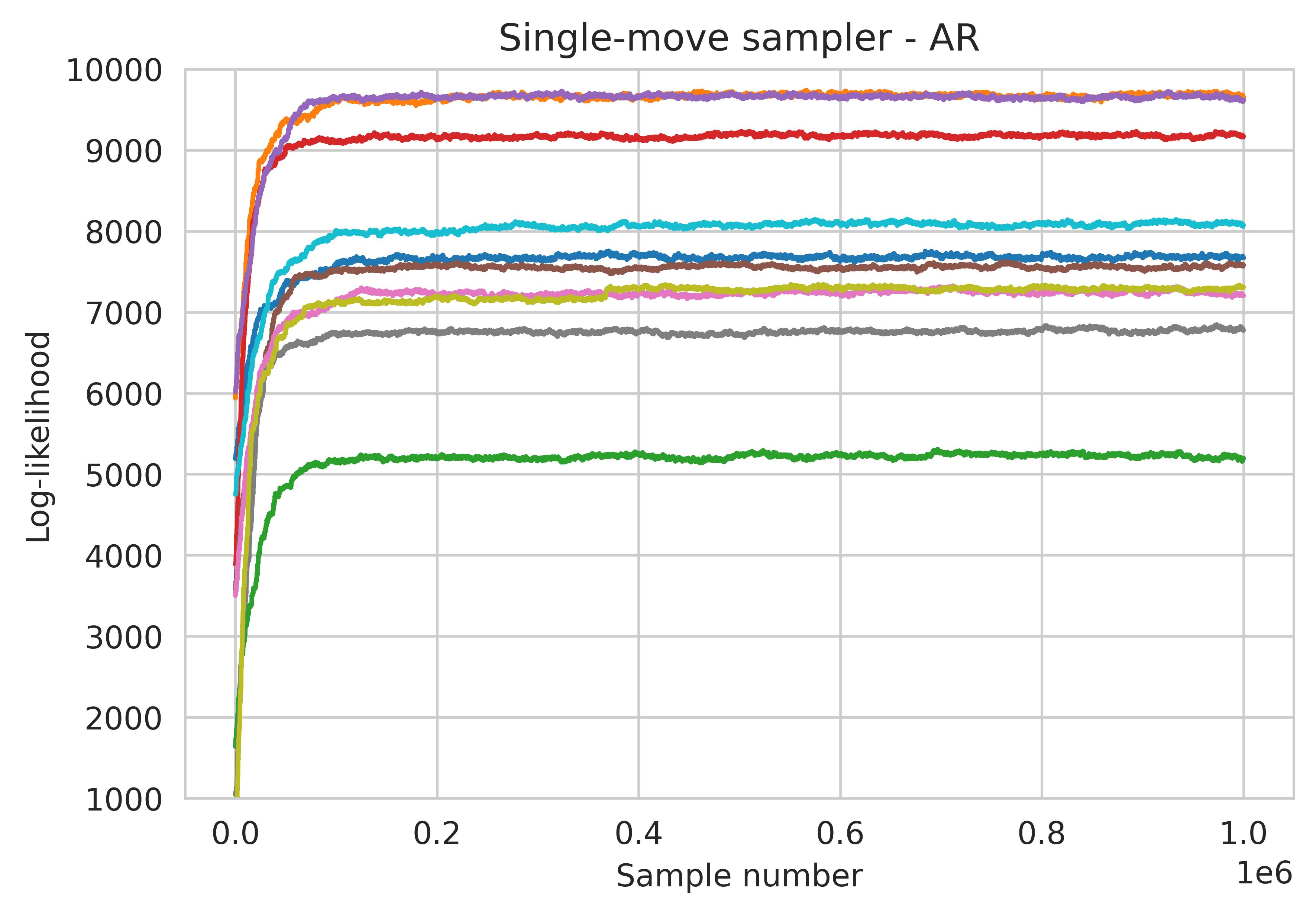}
  \includegraphics[width=0.32\textwidth]{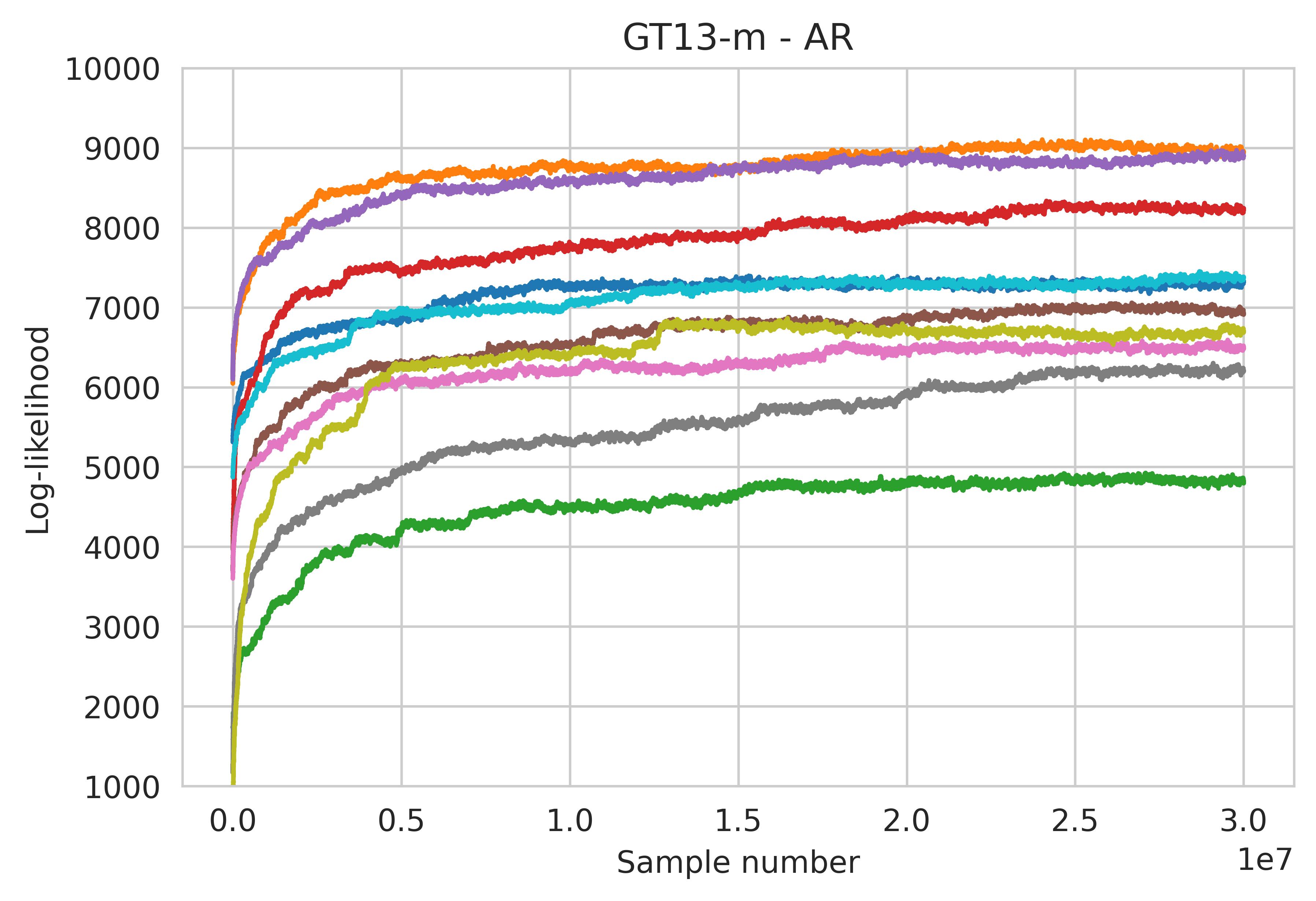}
  
  \includegraphics[width=0.32\textwidth]{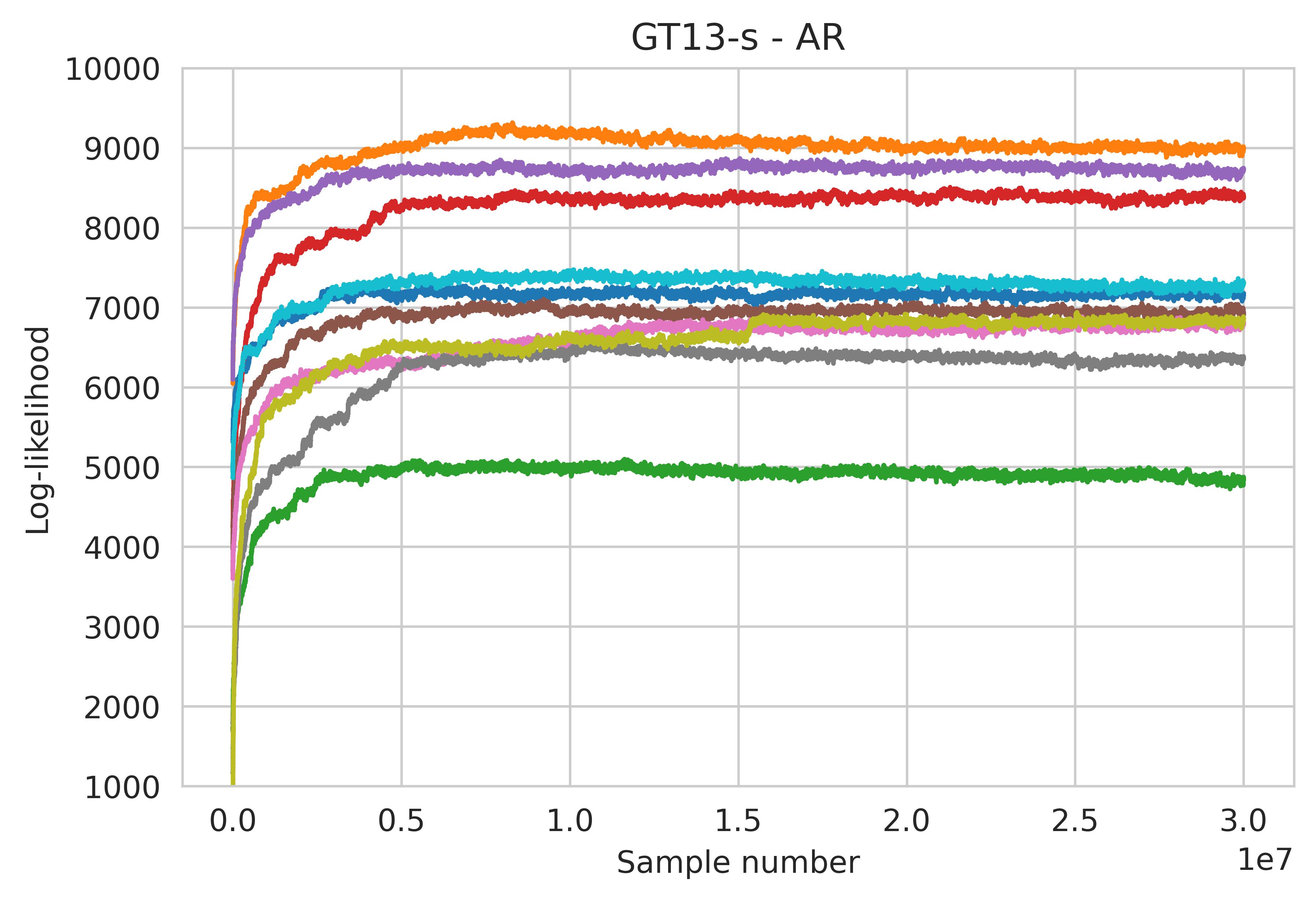}
  \includegraphics[width=0.32\textwidth]{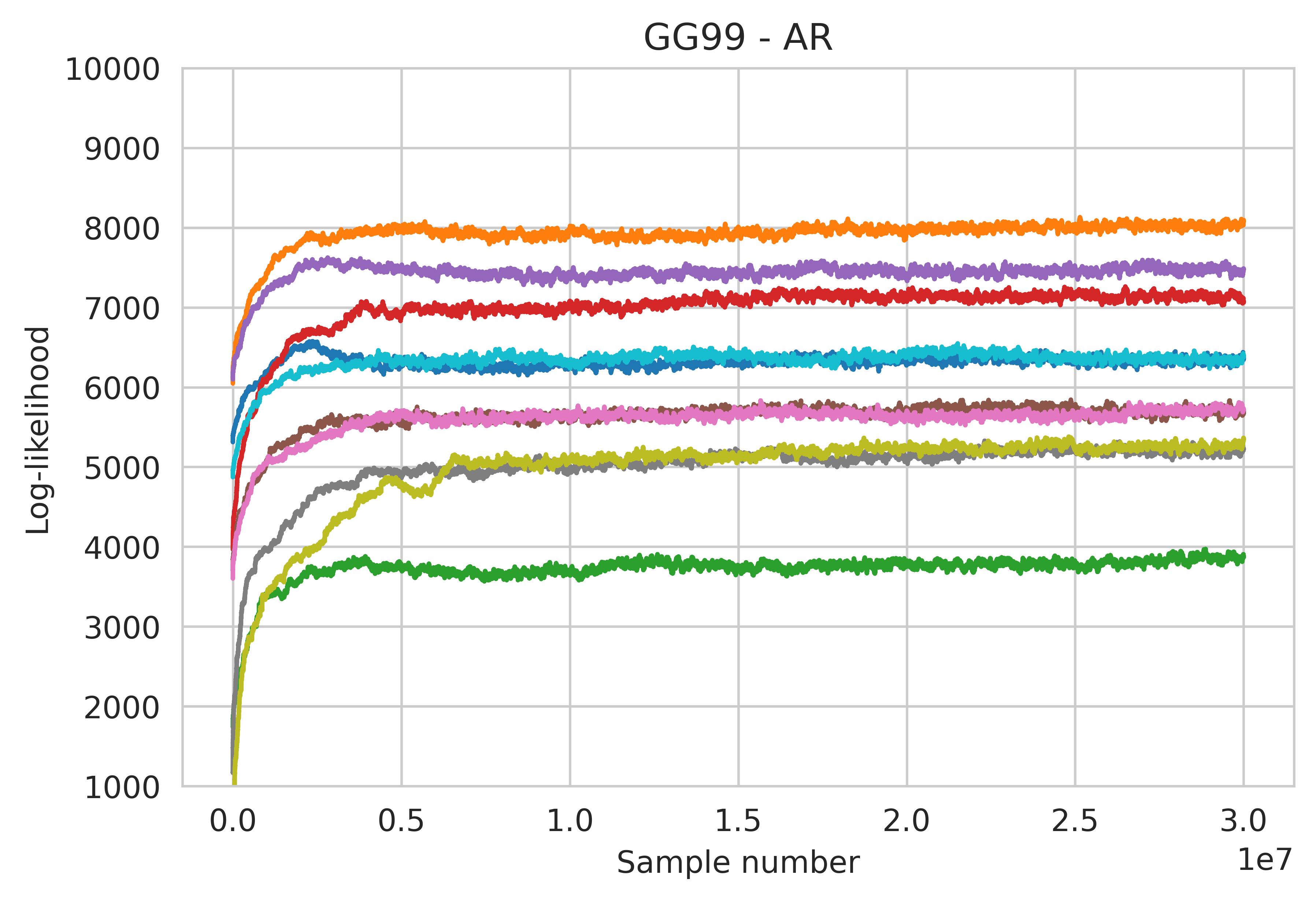}
  
  \includegraphics[width=0.32\textwidth]{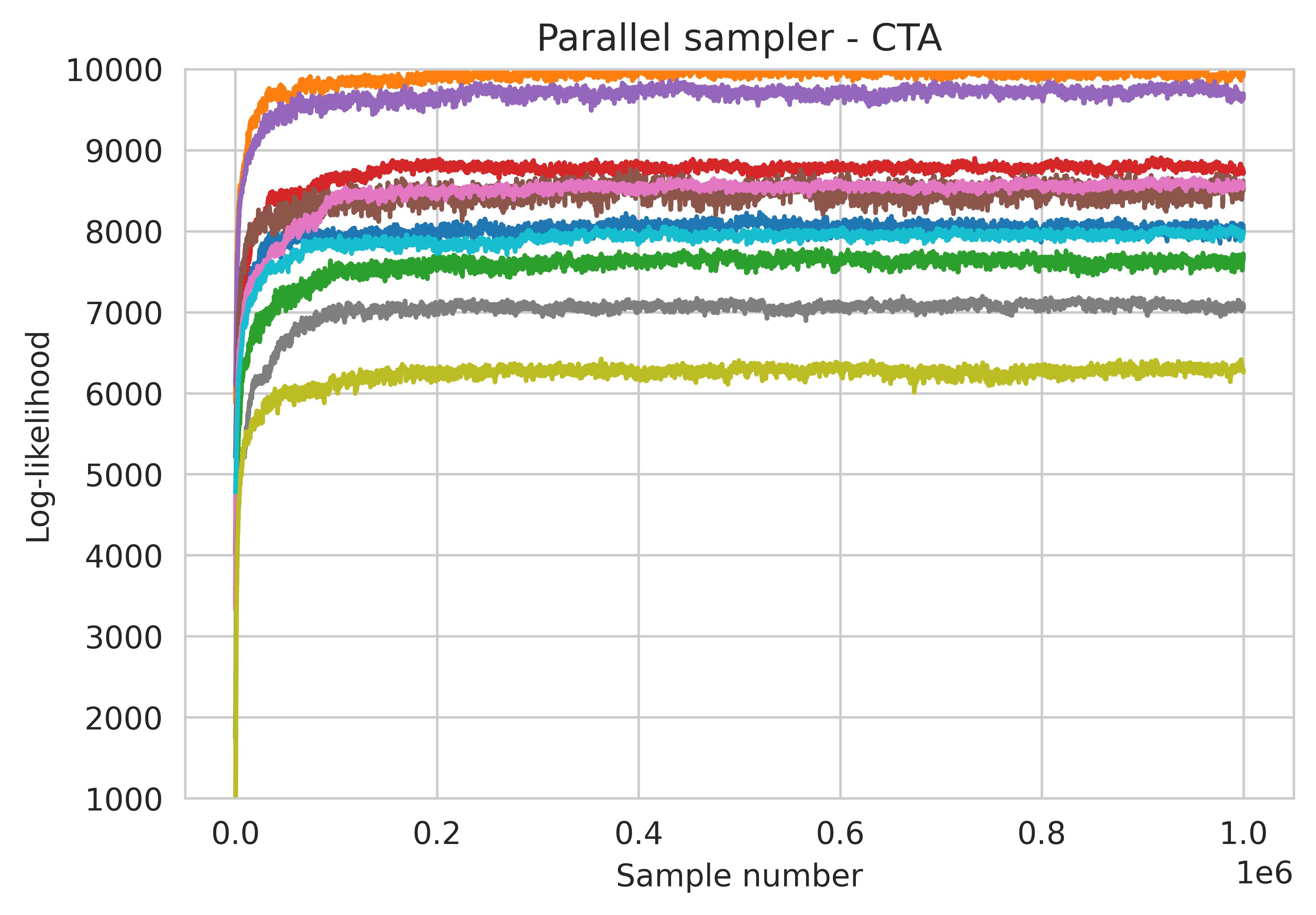}
  \includegraphics[width=0.32\textwidth]{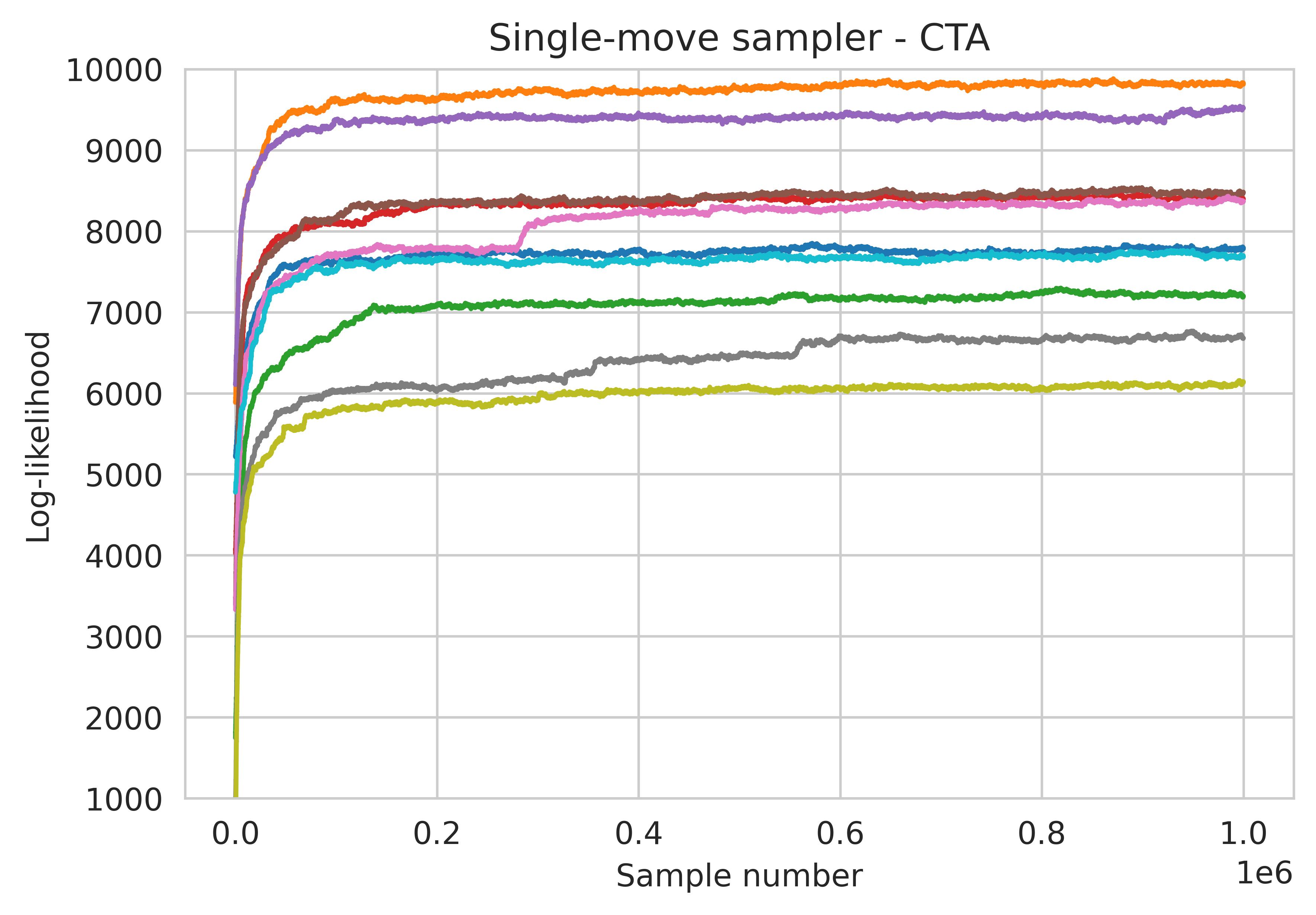}
  \includegraphics[width=0.32\textwidth]{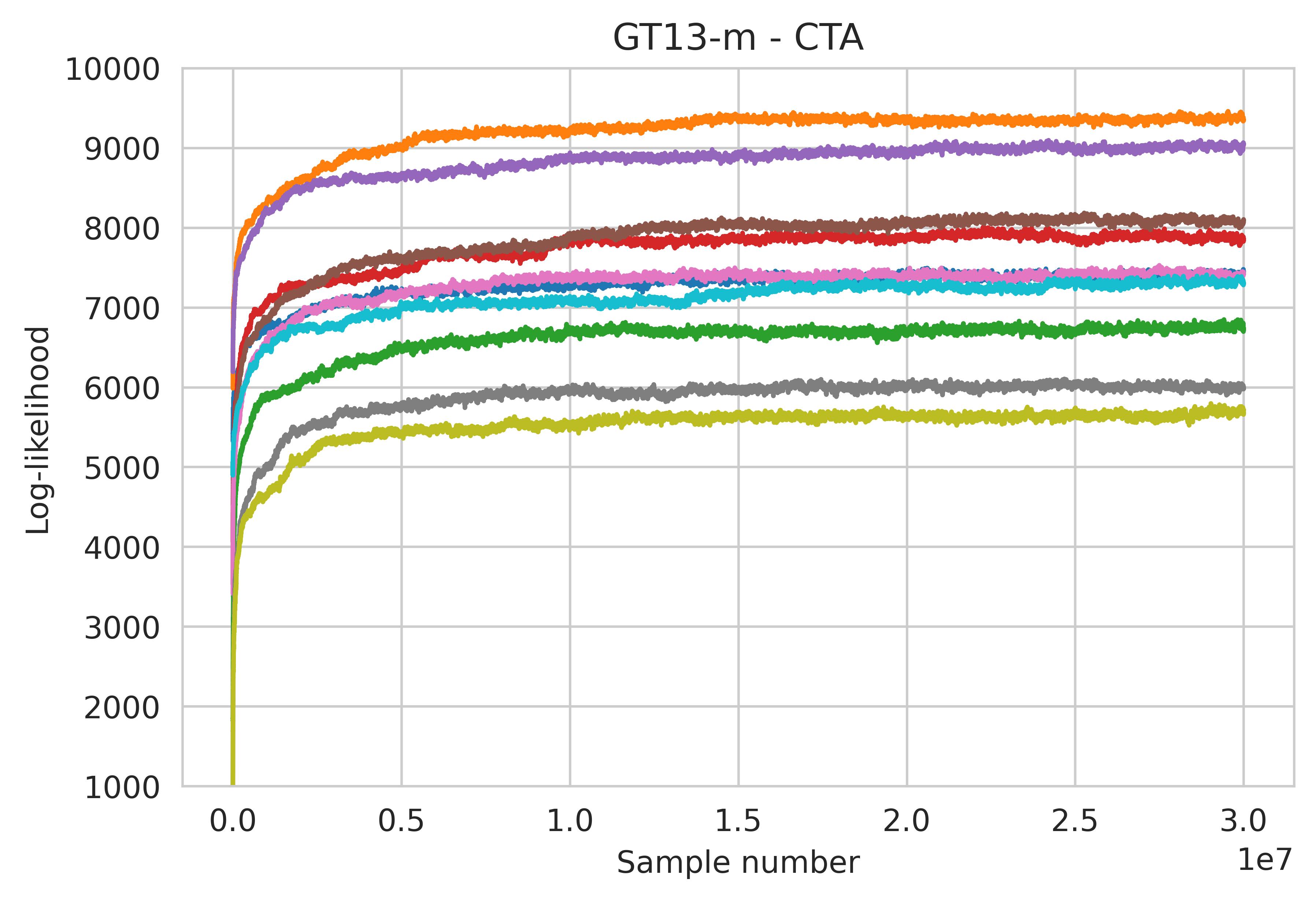}
  
  \includegraphics[width=0.32\textwidth]{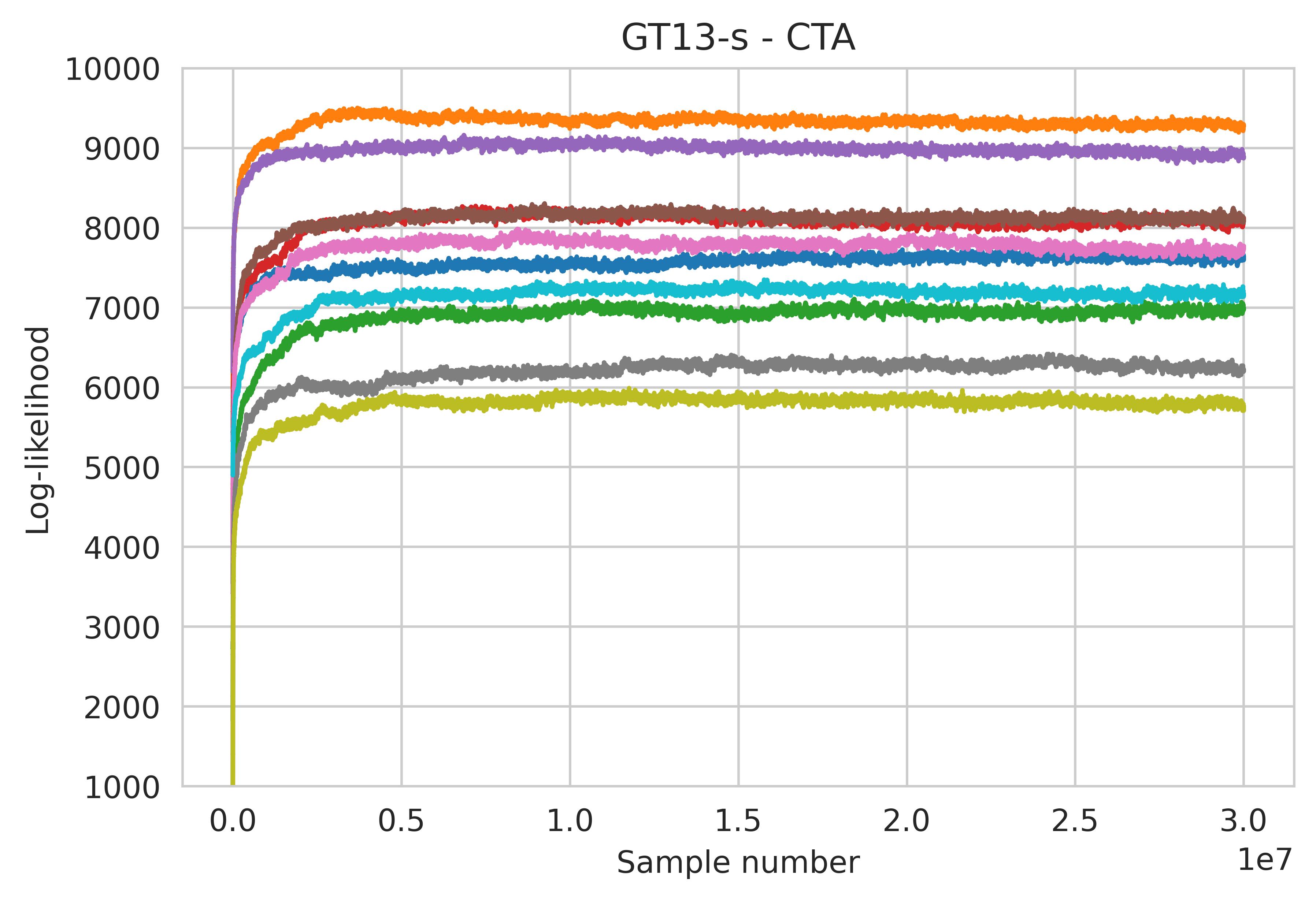}
  \includegraphics[width=0.32\textwidth]{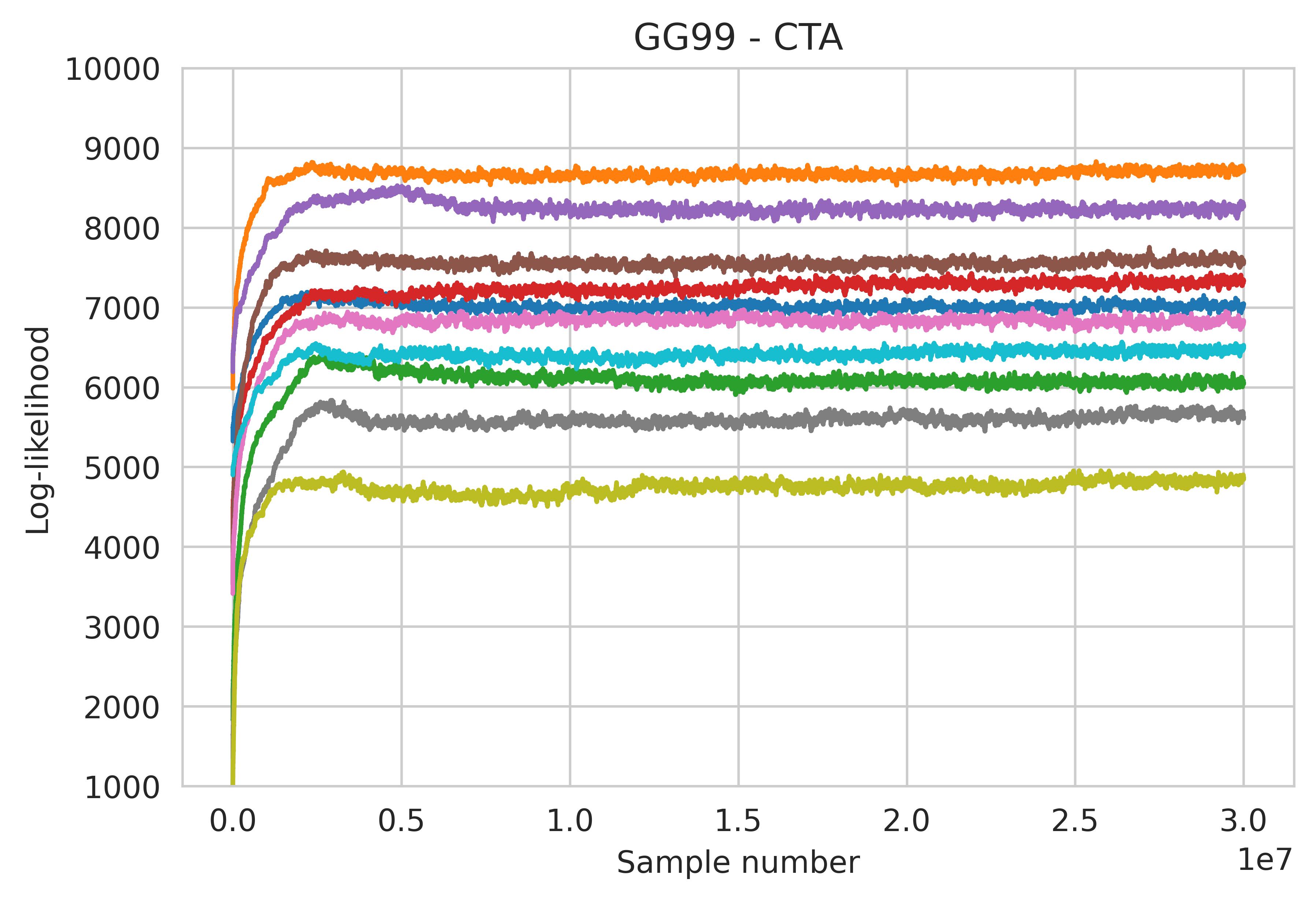}
  
\caption{Likelihood traceplots for 10 replications (color-matched) for samplers discussed in Section~\ref{sec:numerical}, under the autoregressive graph structure for $(n,p)=(100,200)$.}
   \label{app:fig:loglikeli-traceplots-sparse}
 \end{figure}

\begin{figure}[ht!]
  \centering

  \includegraphics[width=0.32\textwidth]{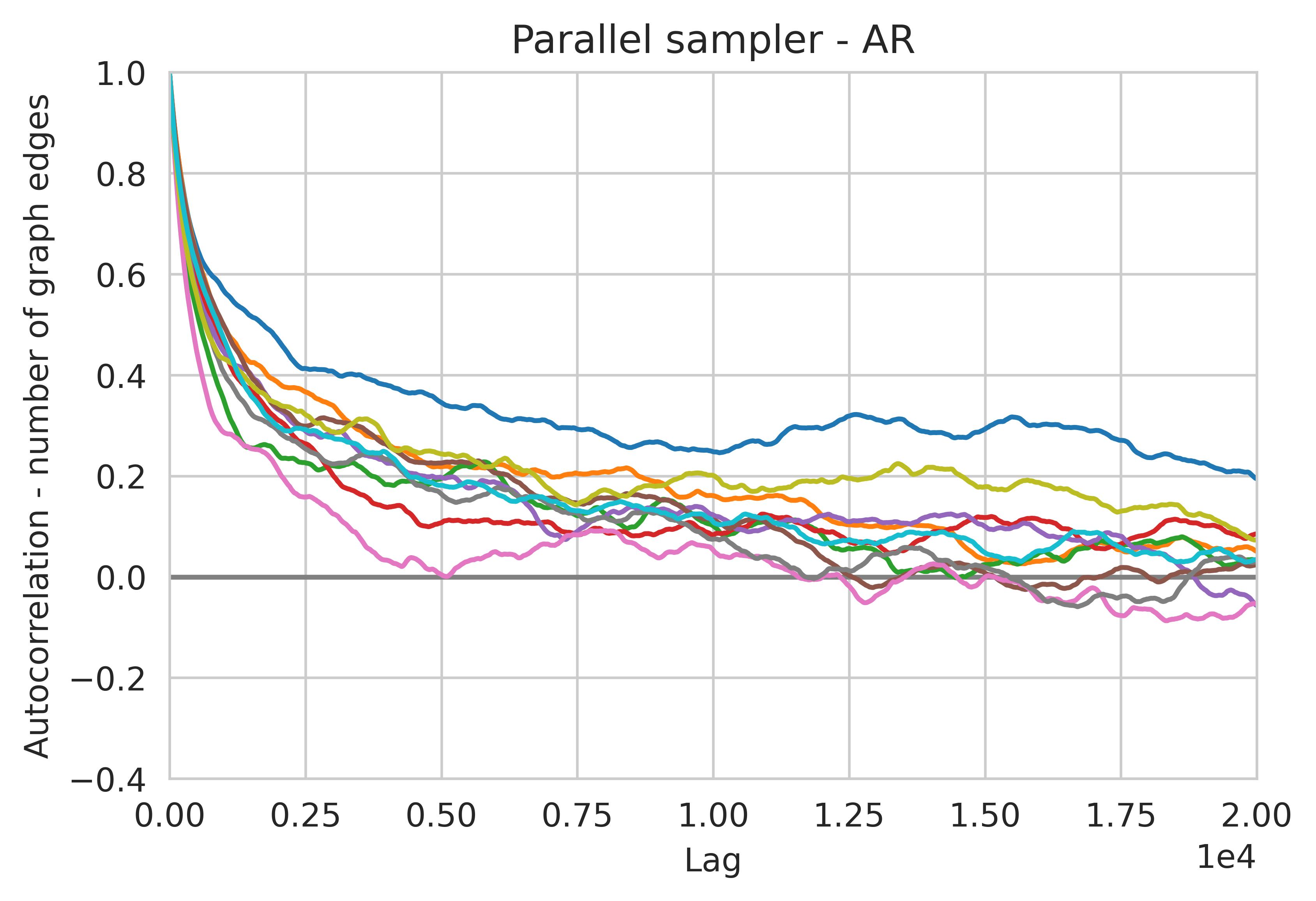}
  \includegraphics[width=0.32\textwidth]{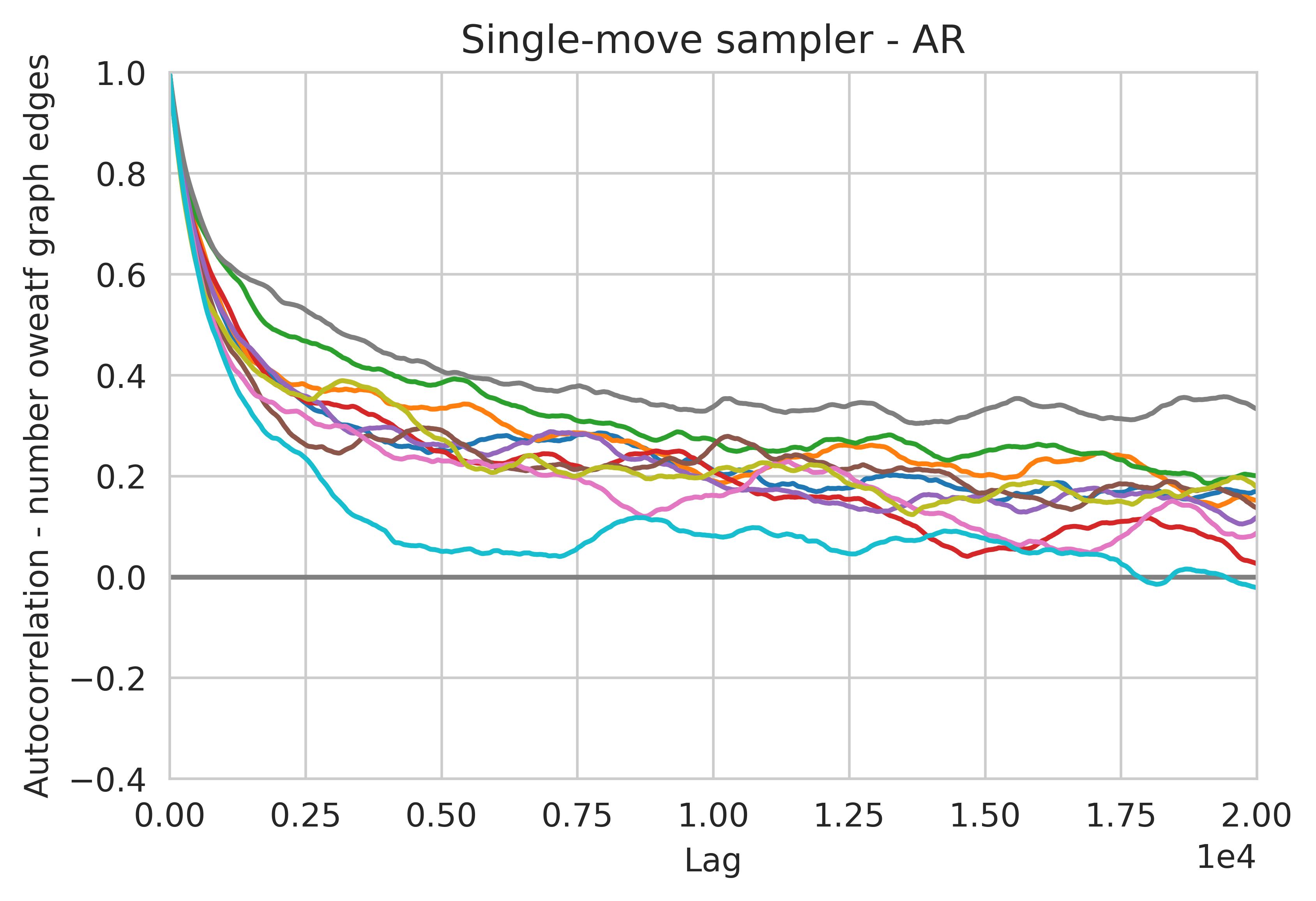}
    \includegraphics[width=0.32\textwidth]{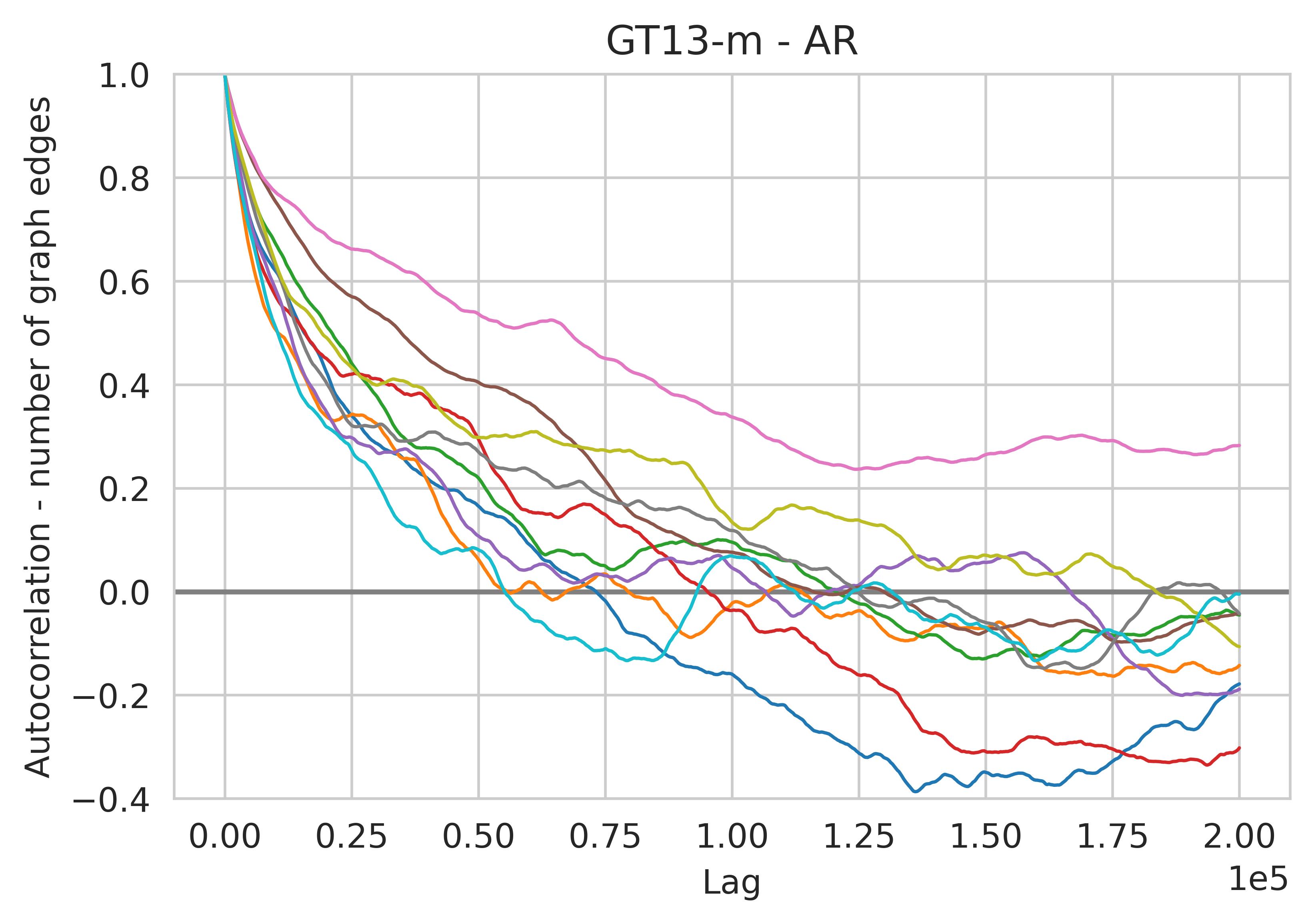}
  
    \includegraphics[width=0.32\textwidth]{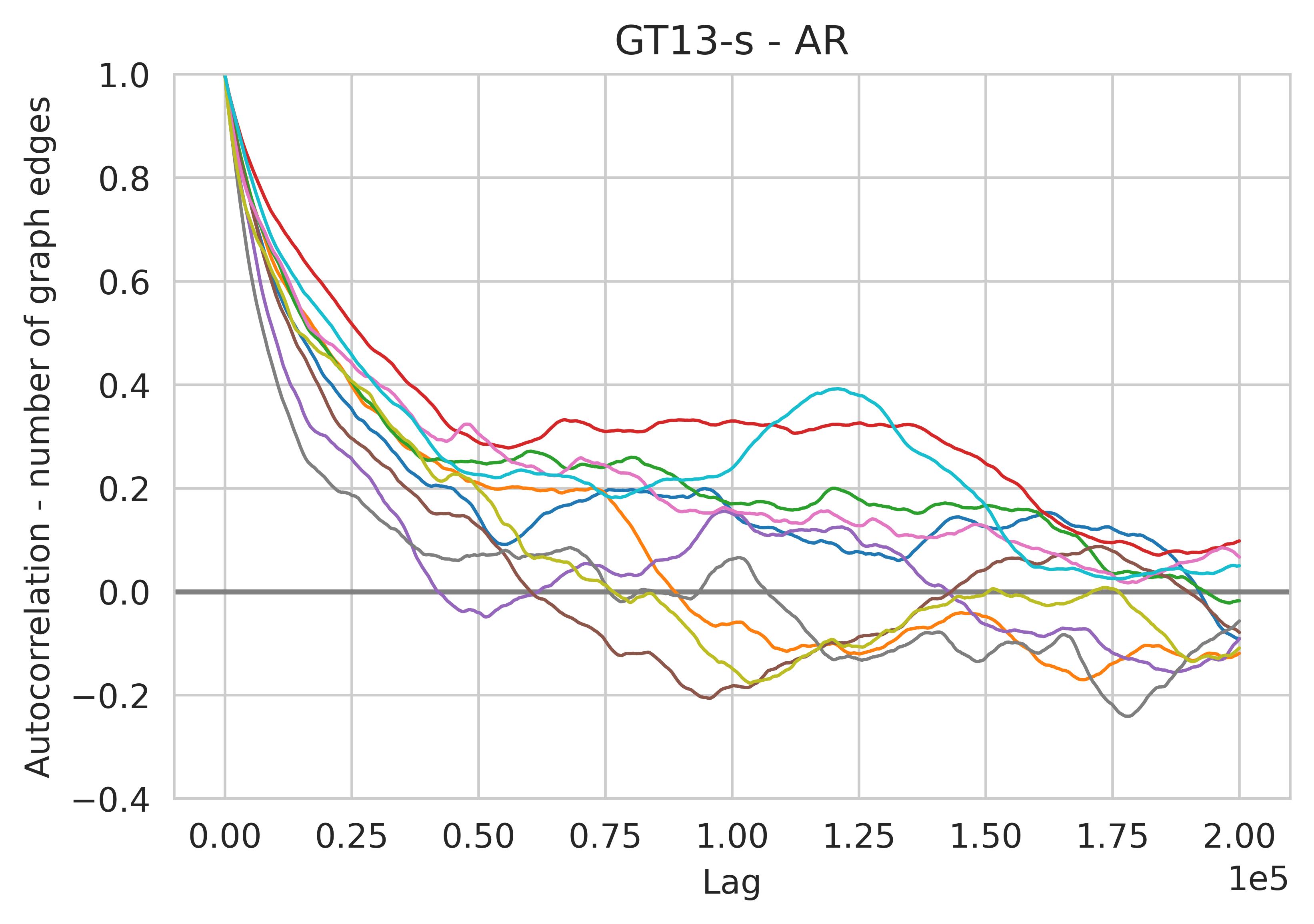}
  \includegraphics[width=0.32\textwidth]{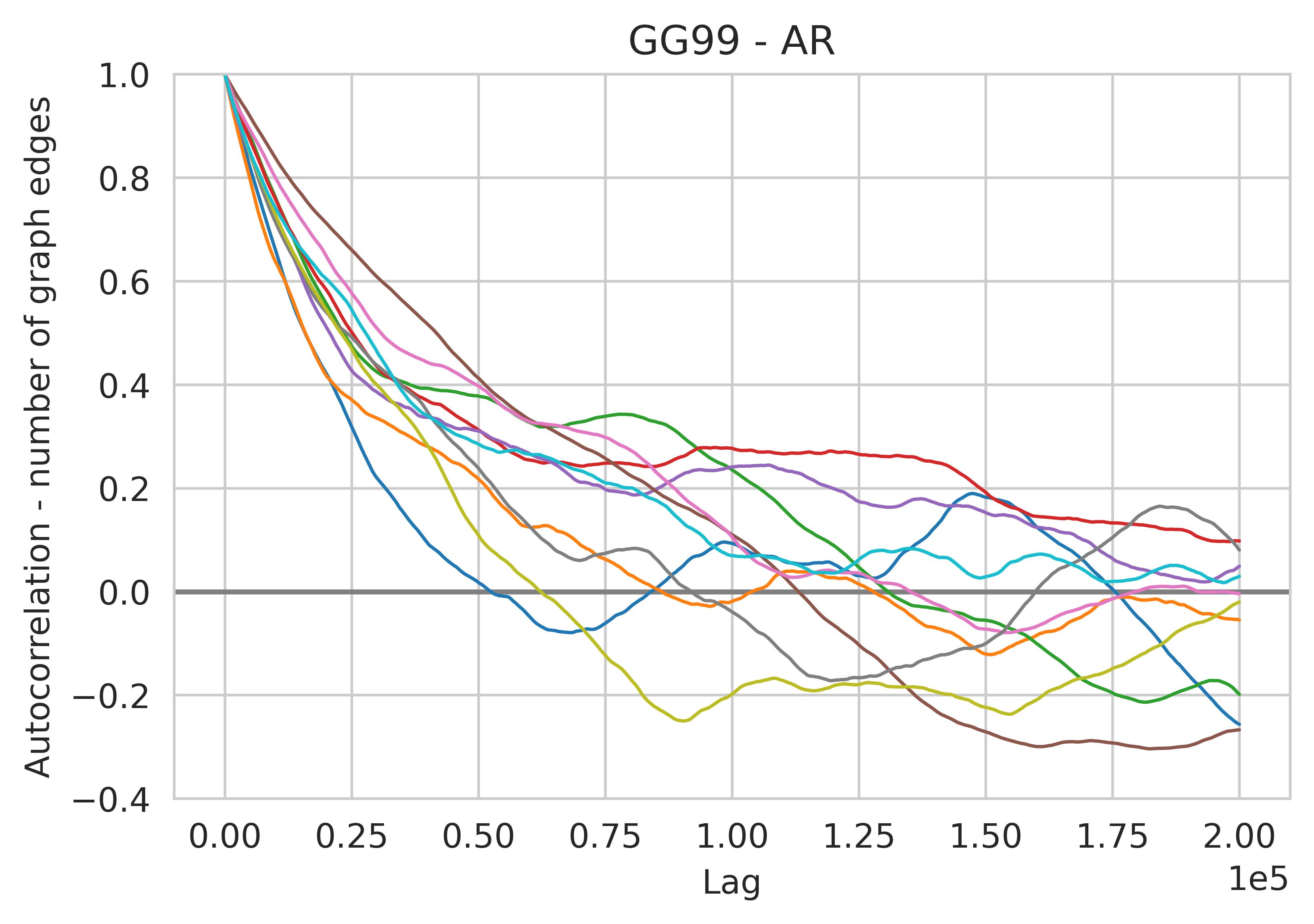}
  
   \includegraphics[width=0.32\textwidth]{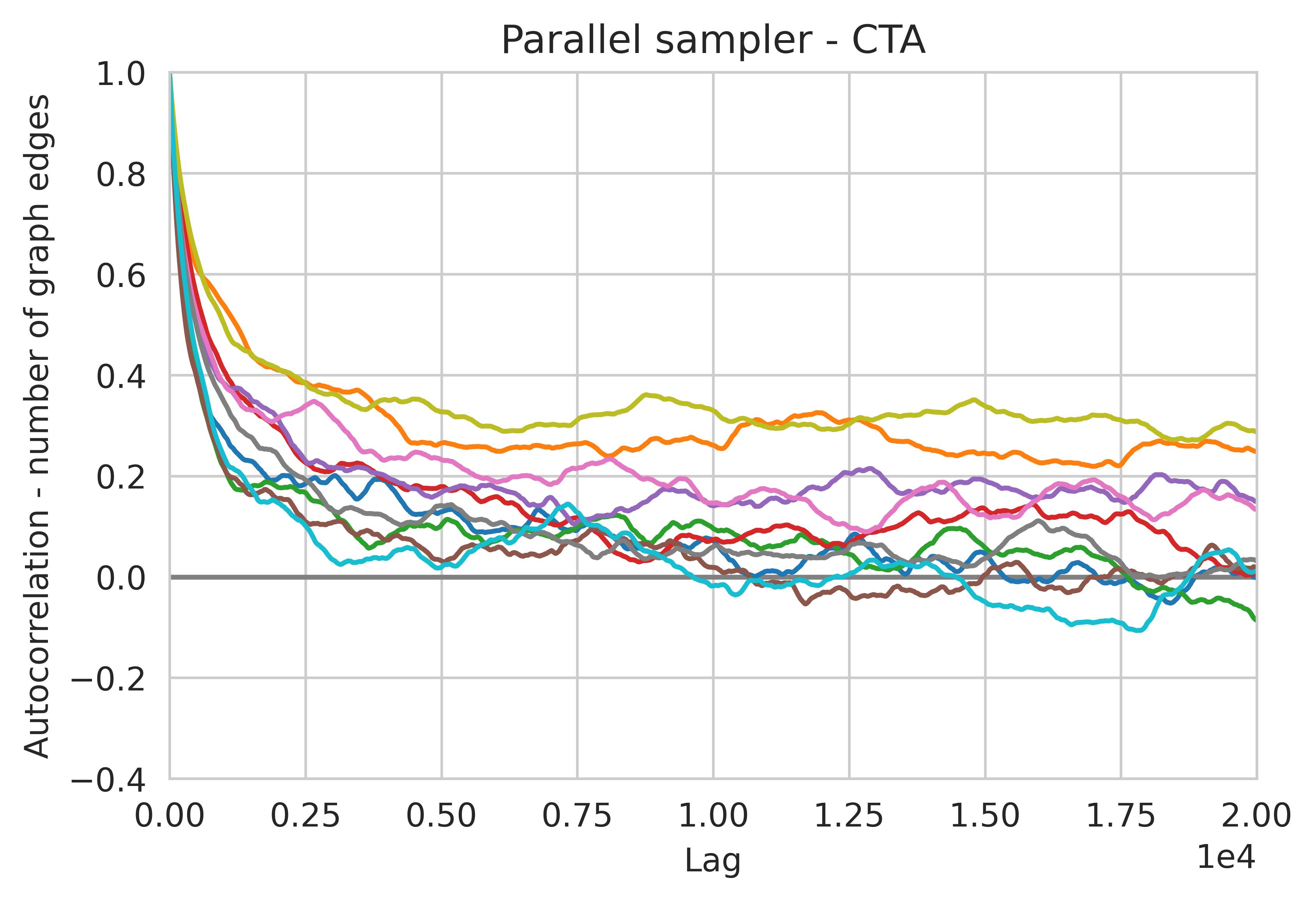}
   \includegraphics[width=0.32\textwidth]{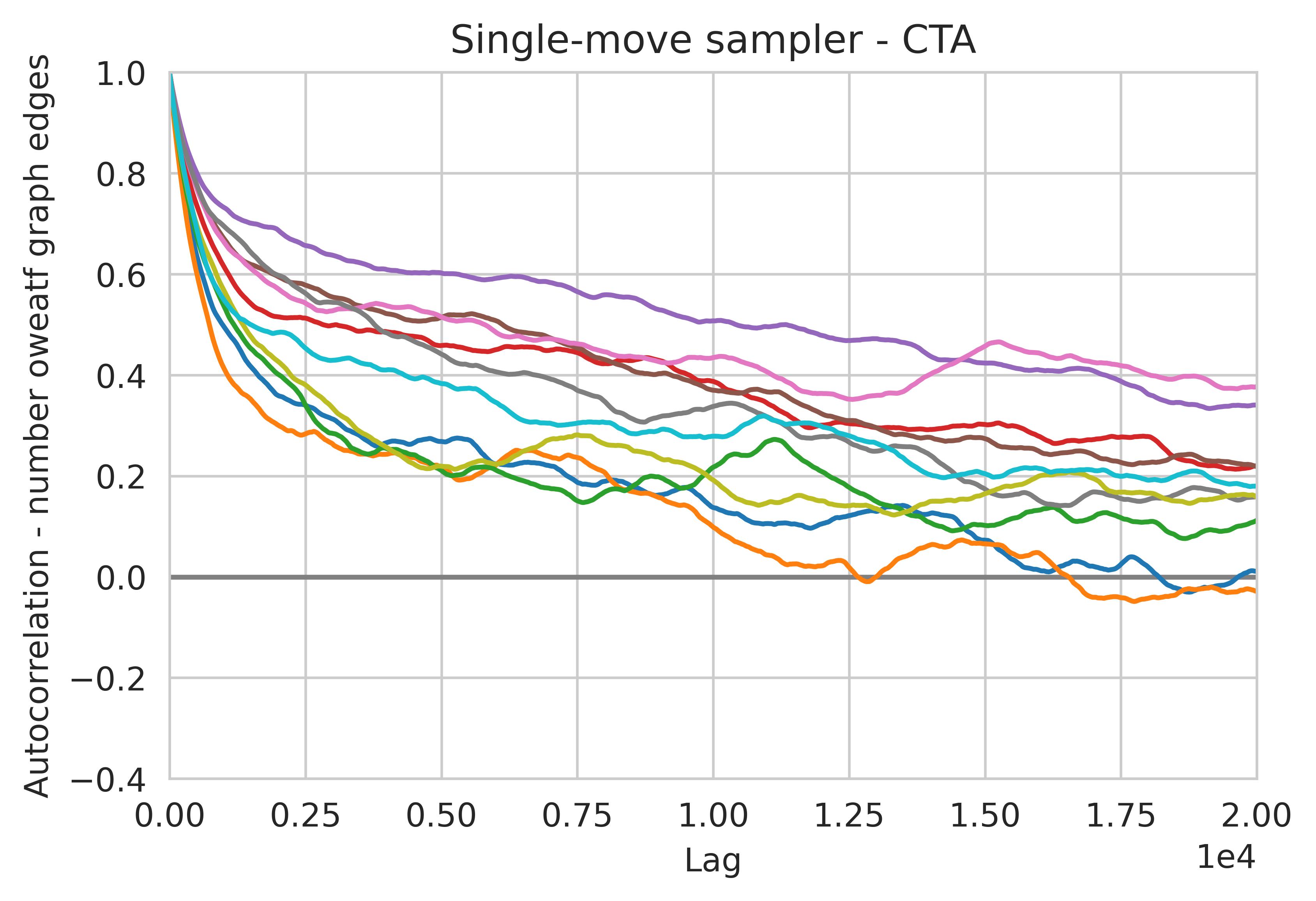}
    \includegraphics[width=0.32\textwidth]{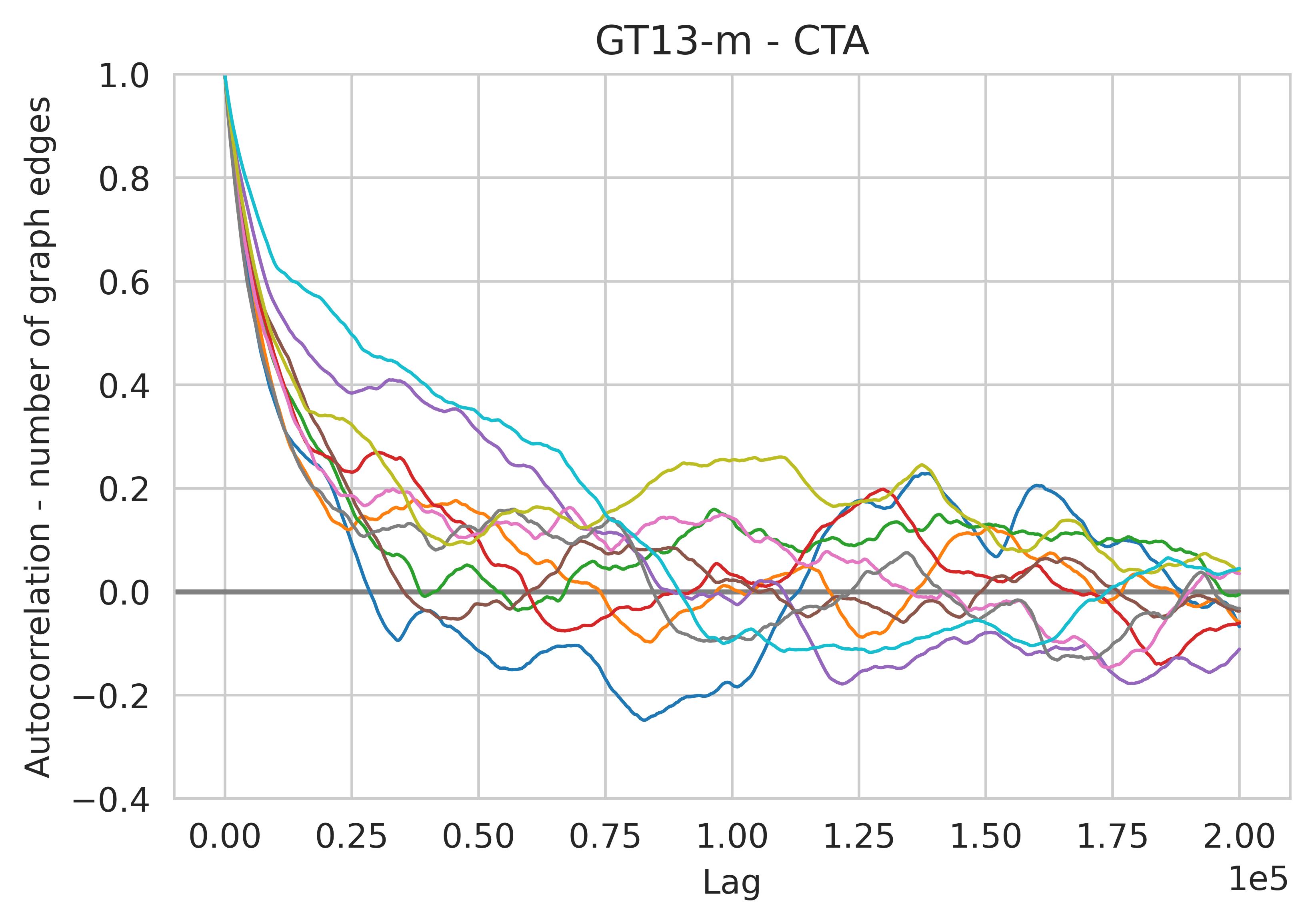}

    \includegraphics[width=0.32\textwidth]{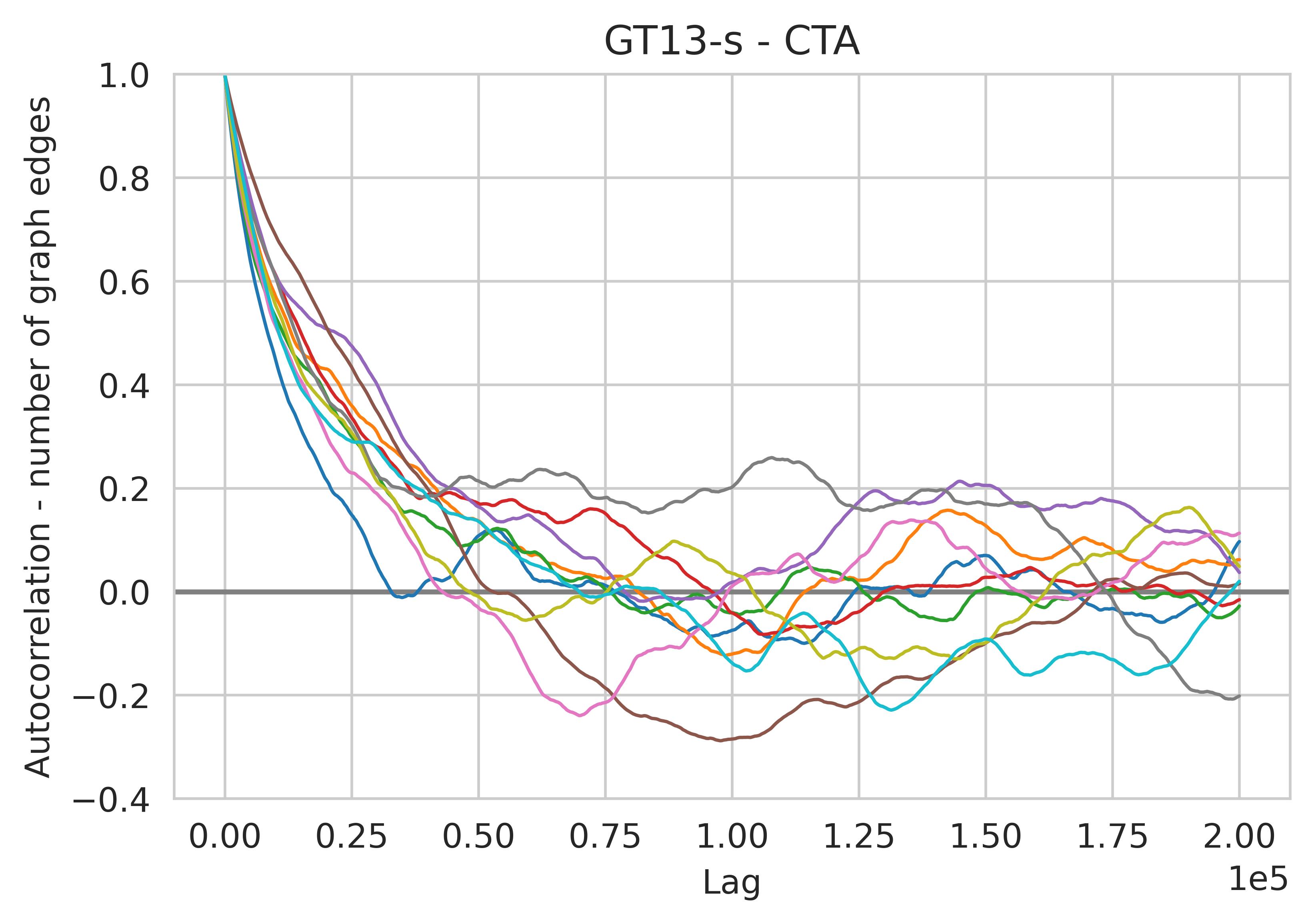}    
  \includegraphics[width=0.32\textwidth]{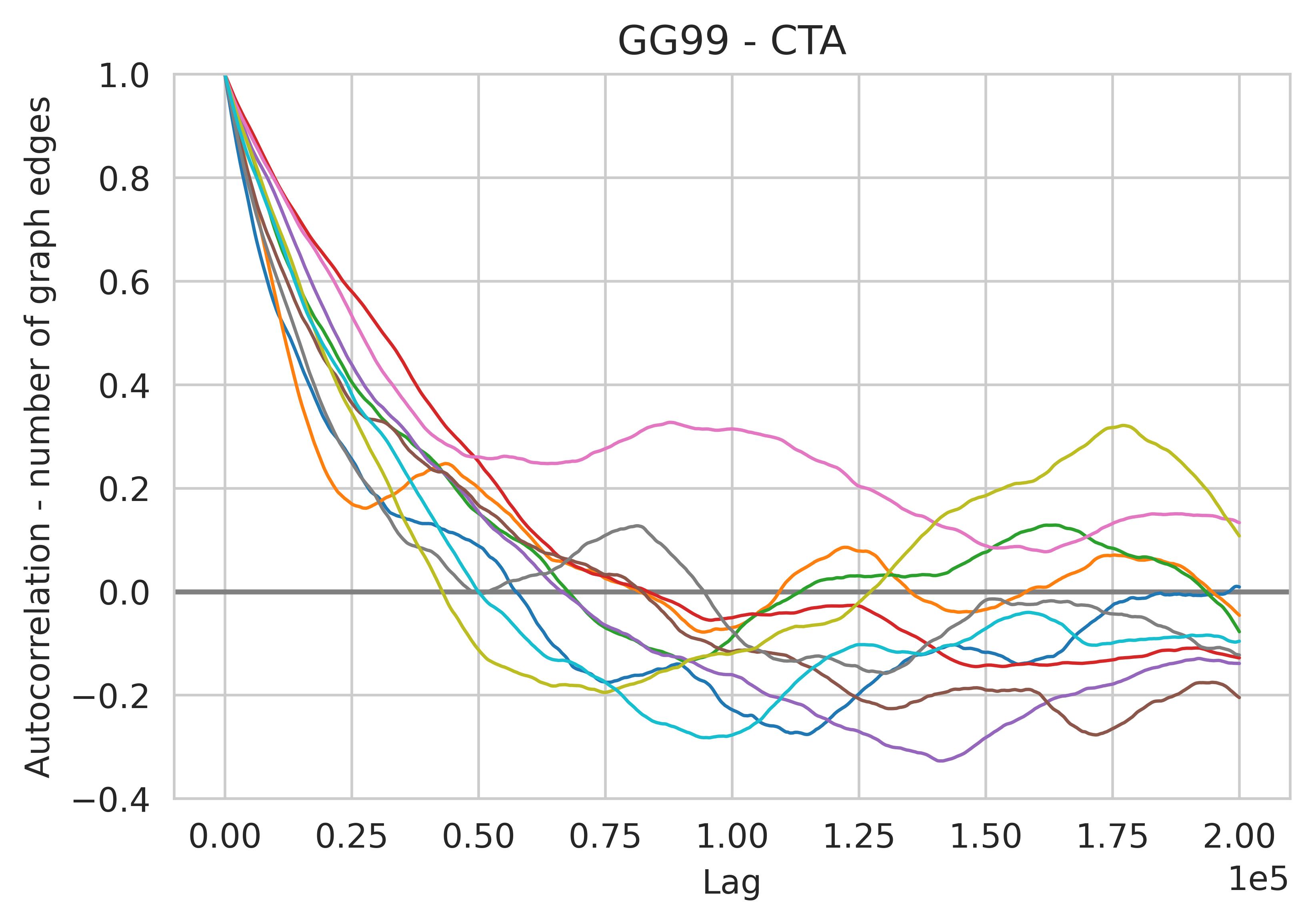}
\caption{Autocorrelation plots for 10 replications (color-matched) for samplers discussed in Section~\ref{sec:numerical}, under the autoregressive graph structure for $(n,p)=(100,200)$.}
   \label{app:fig:corr-traceplots-sparse}
 \end{figure}

\begin{figure}[ht]
  \centering
  \includegraphics[width=0.32\textwidth]{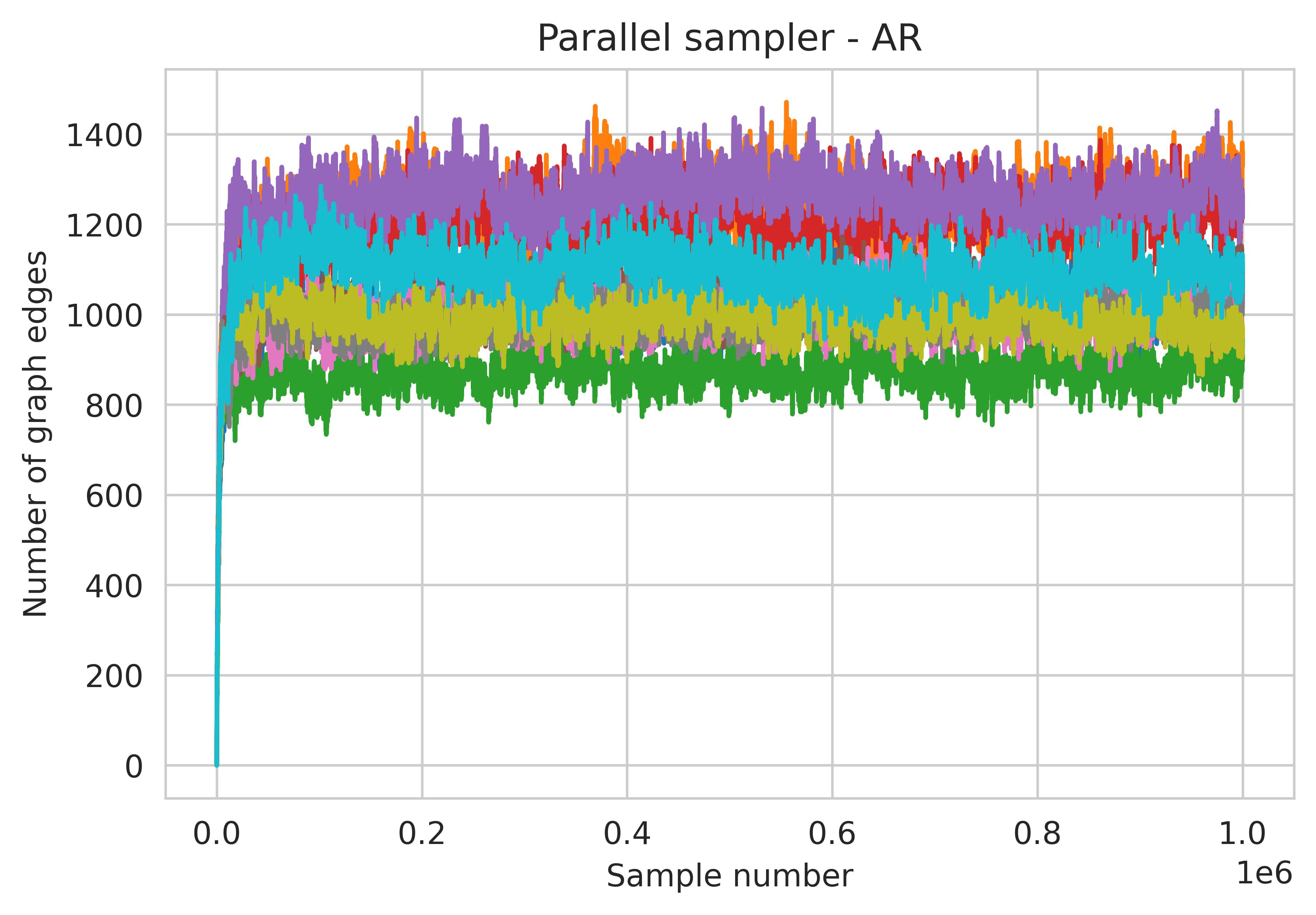}
  \includegraphics[width=0.32\textwidth]{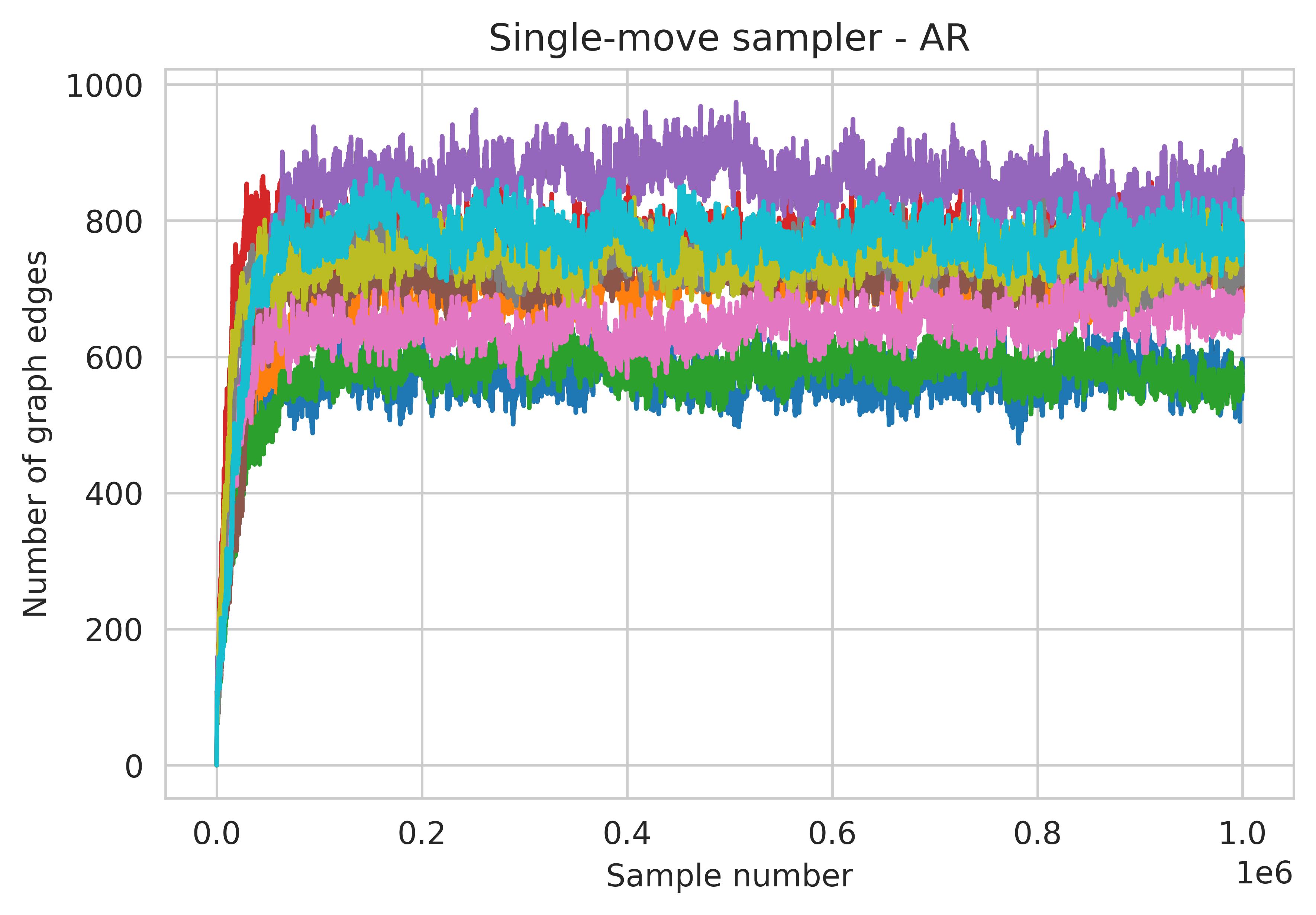}
  \includegraphics[width=0.32\textwidth]{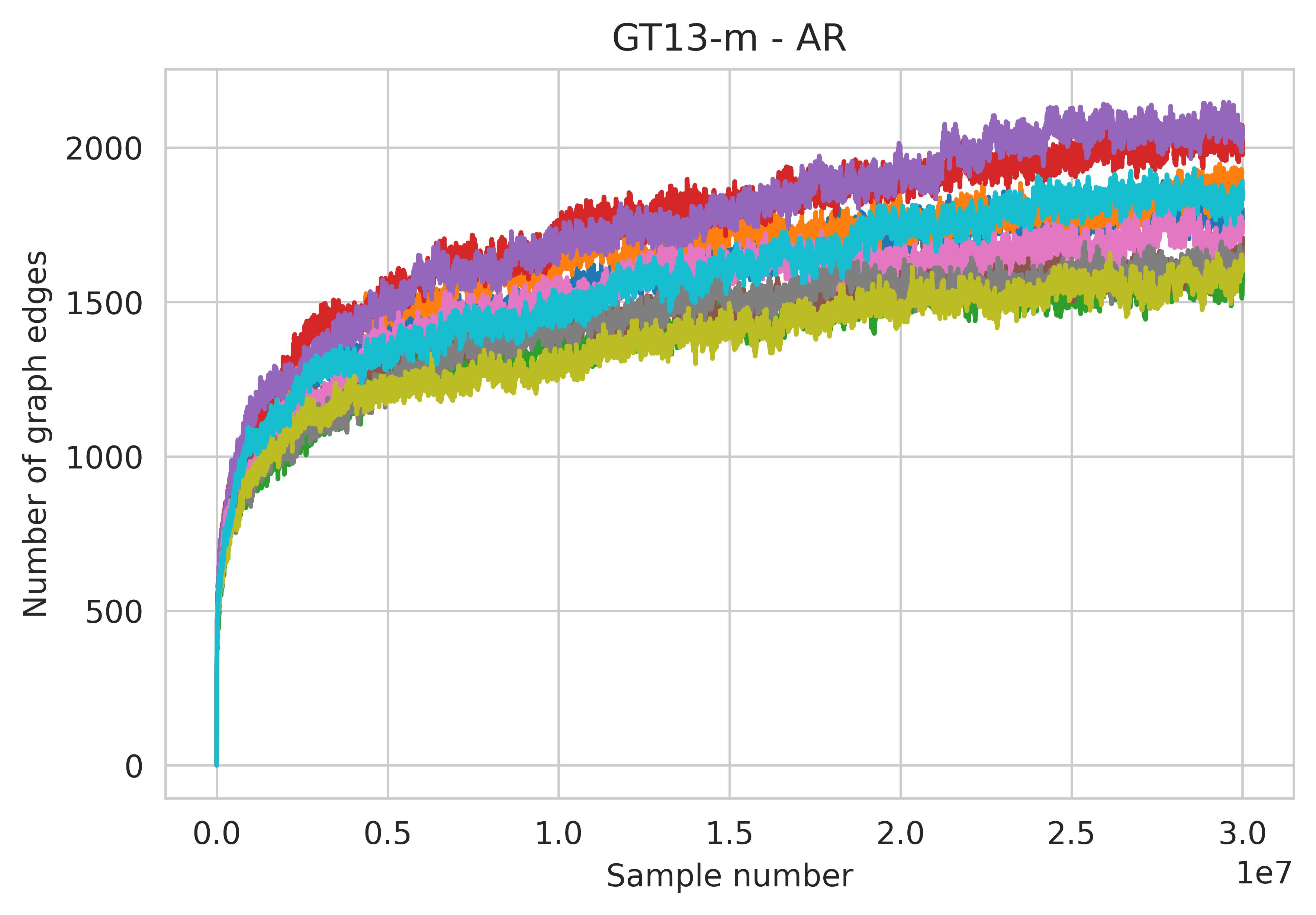}
  
  \includegraphics[width=0.32\textwidth]{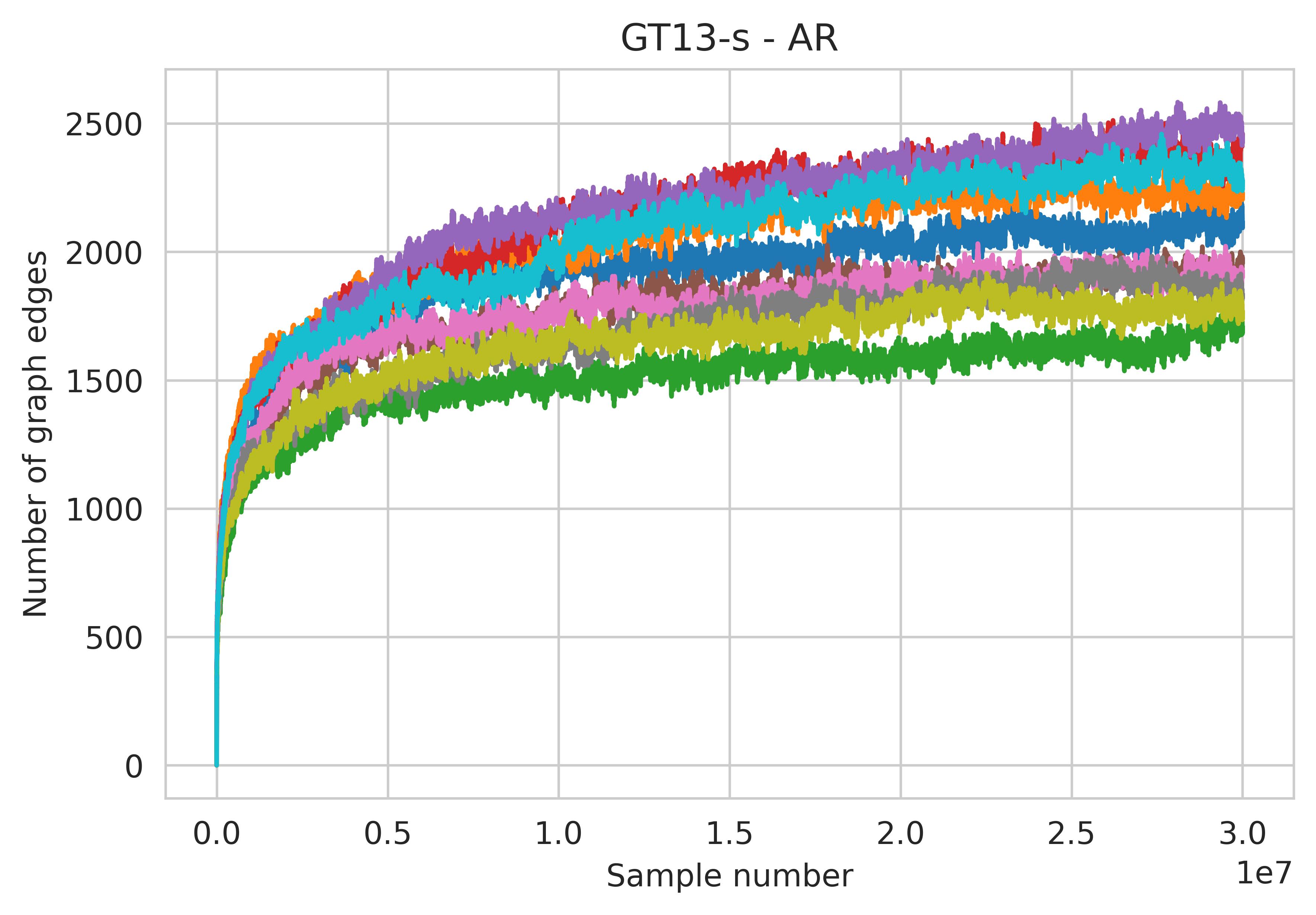}
  \includegraphics[width=0.32\textwidth]{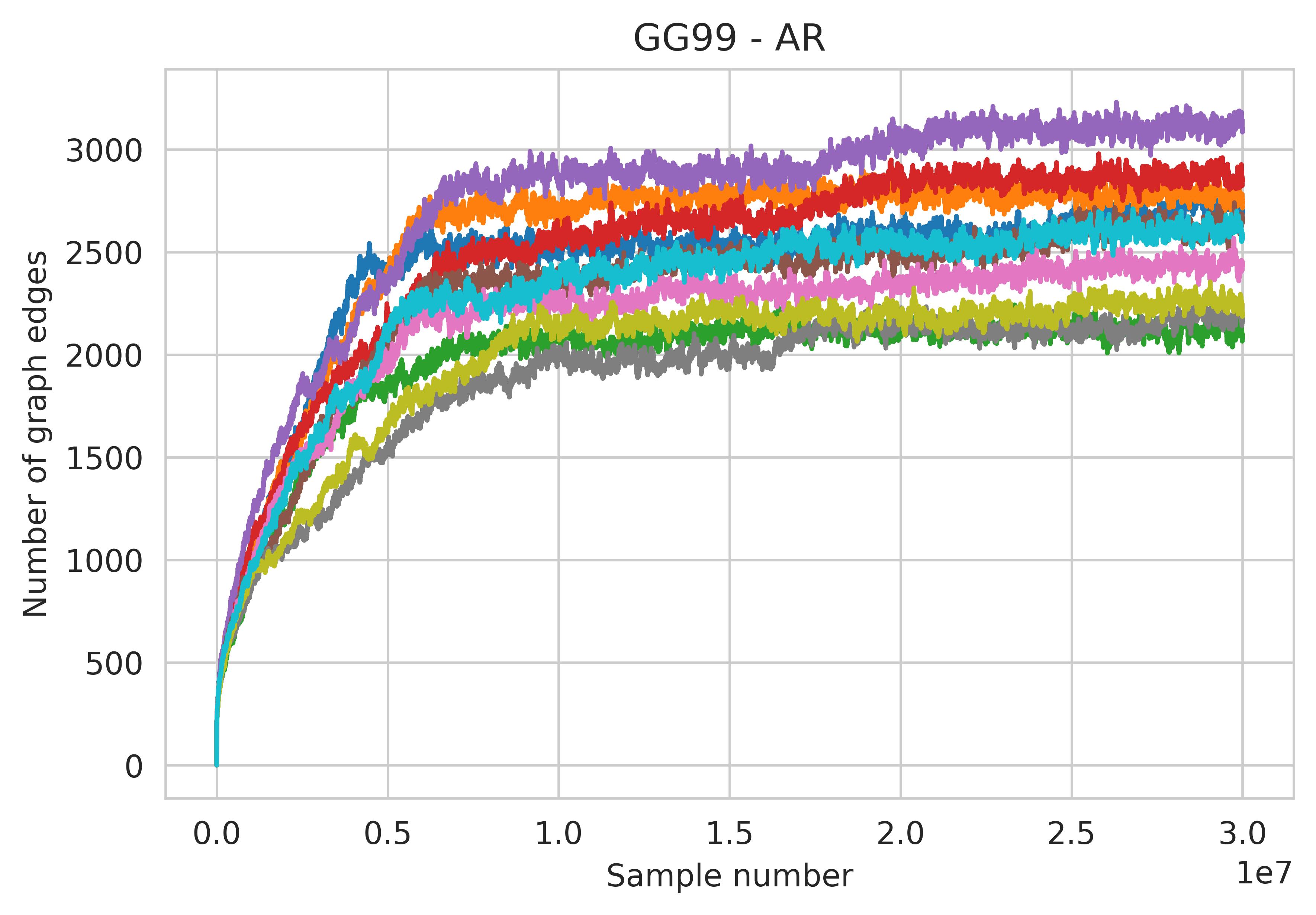}
  
  \includegraphics[width=0.32\textwidth]{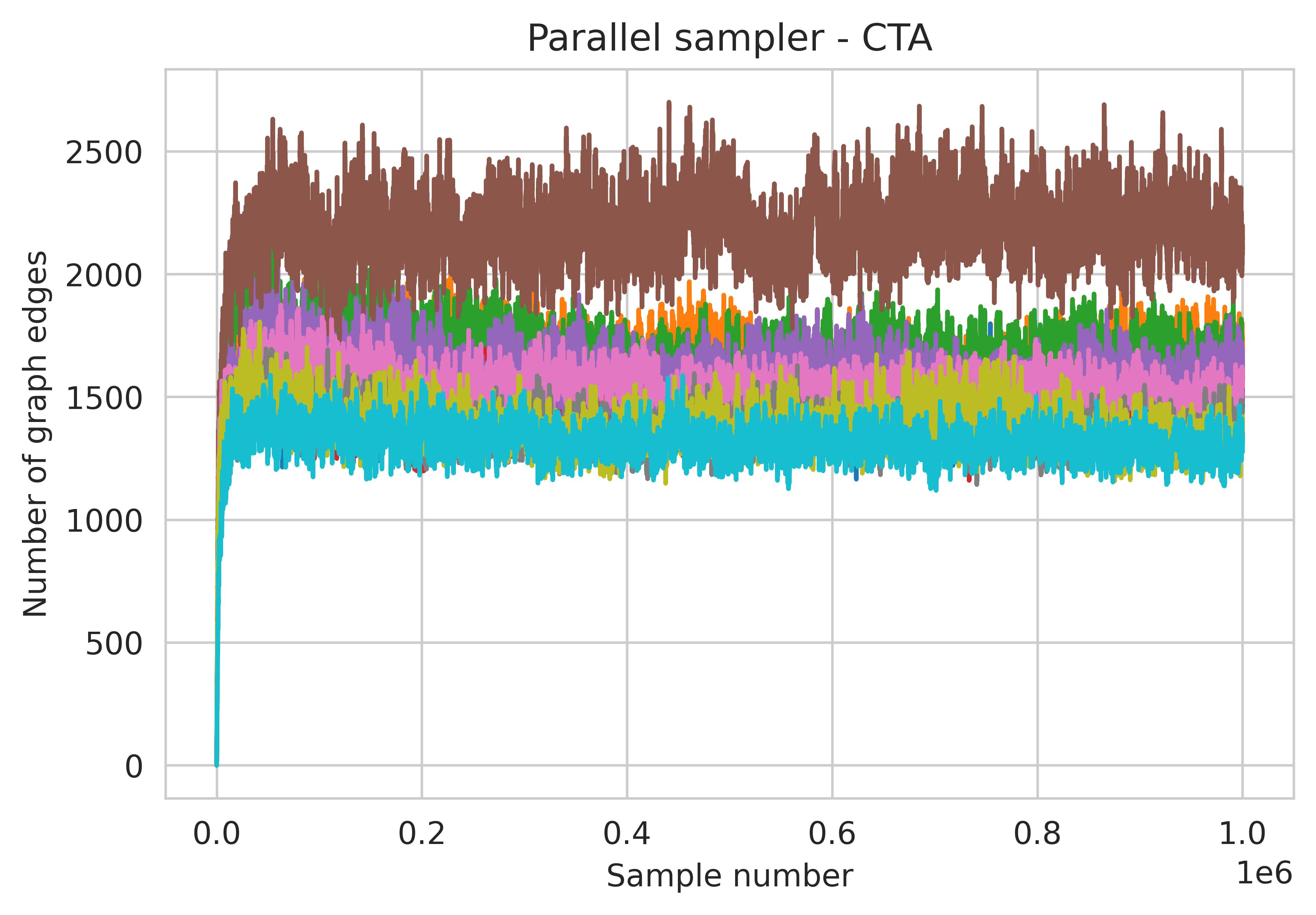}
  \includegraphics[width=0.32\textwidth]{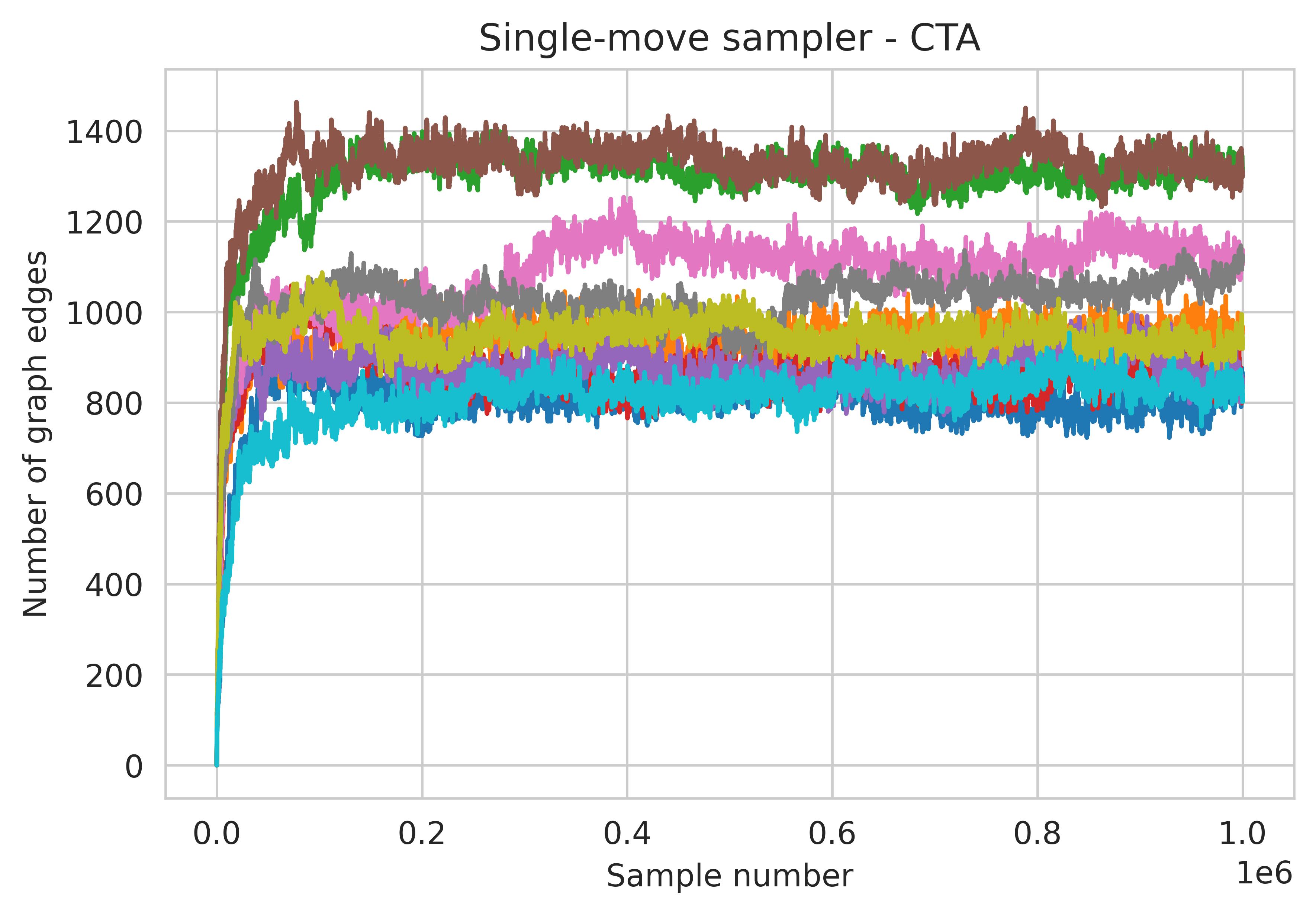}
  \includegraphics[width=0.32\textwidth]{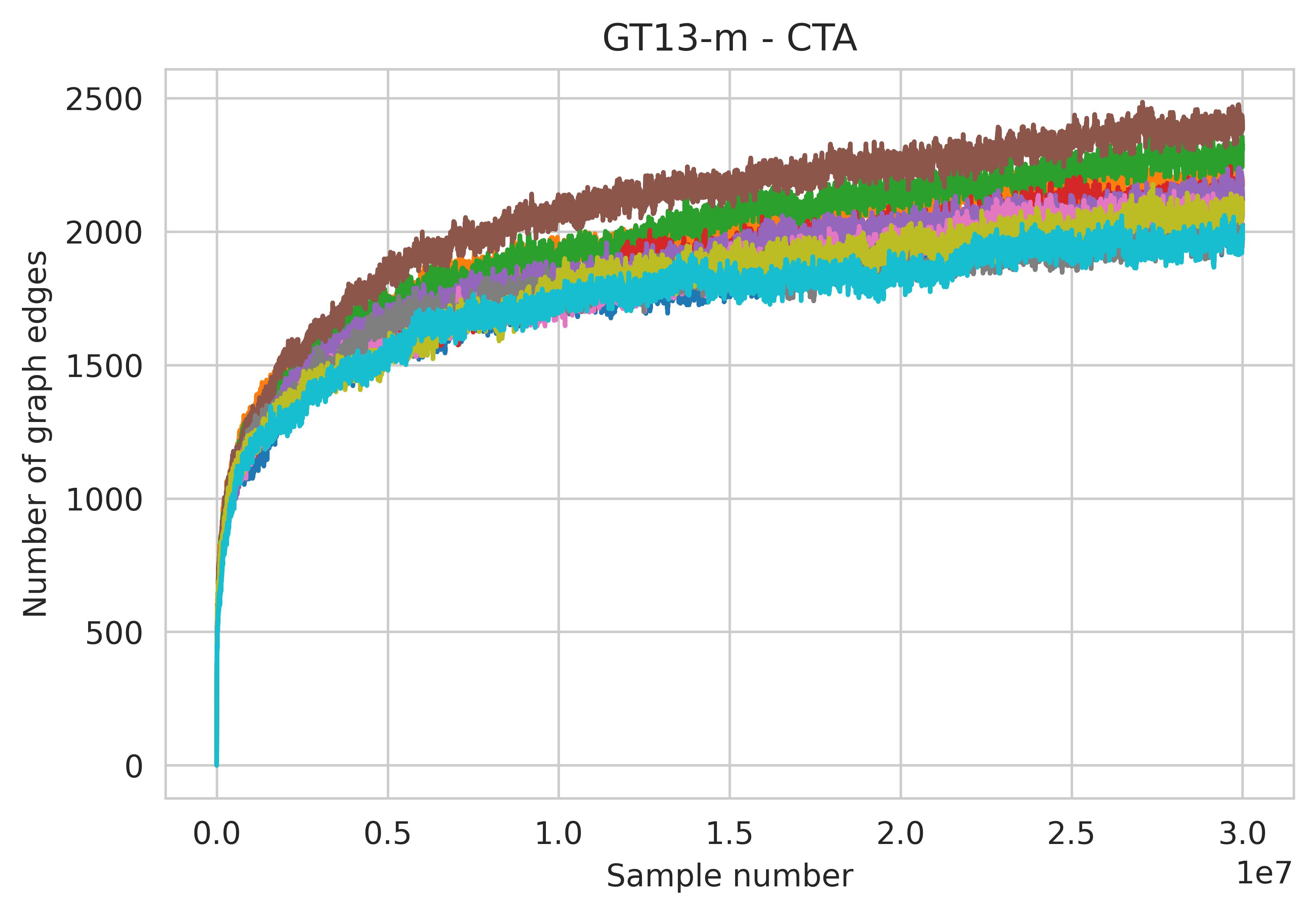}

  \includegraphics[width=0.32\textwidth]{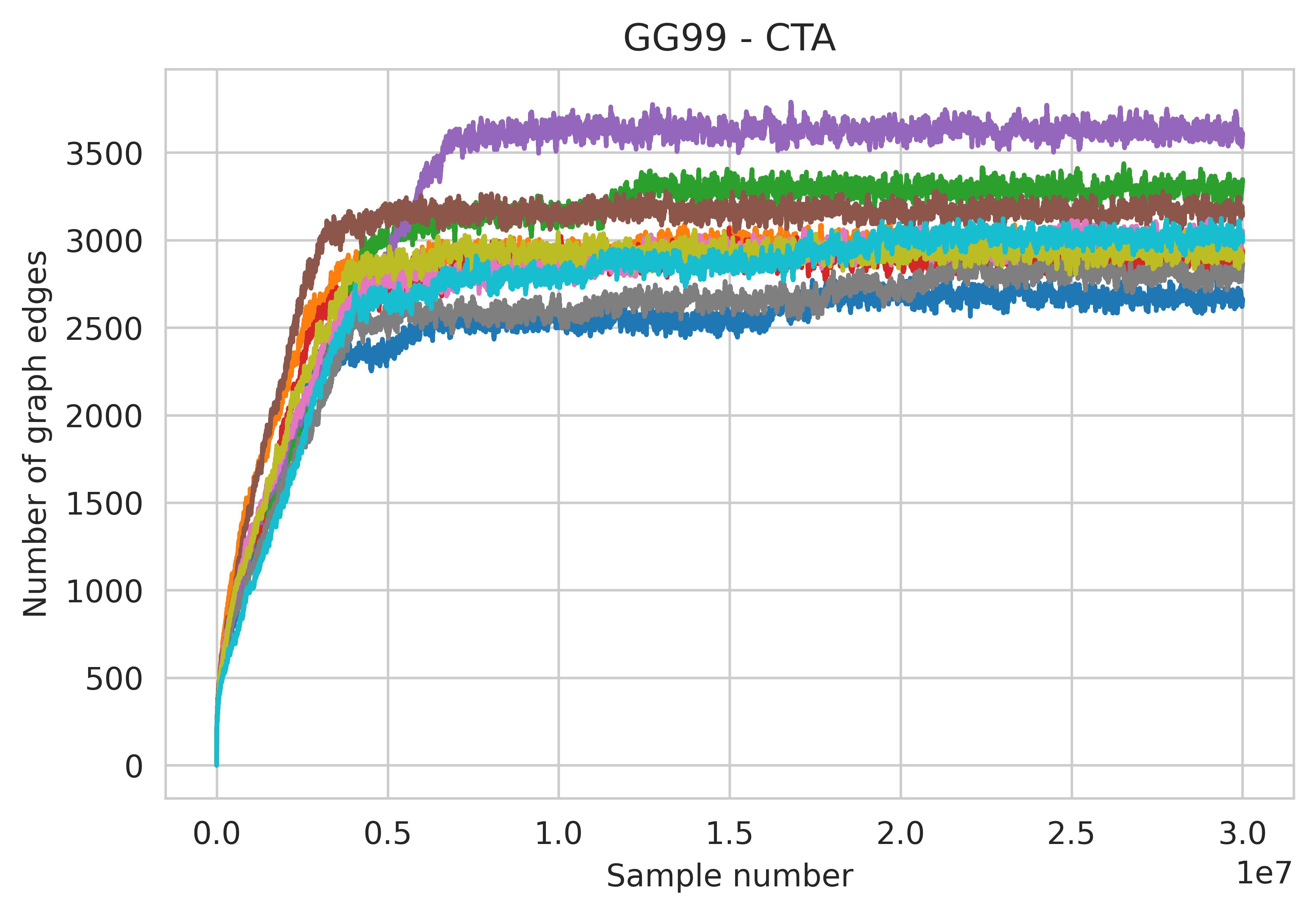}
  \includegraphics[width=0.32\textwidth]{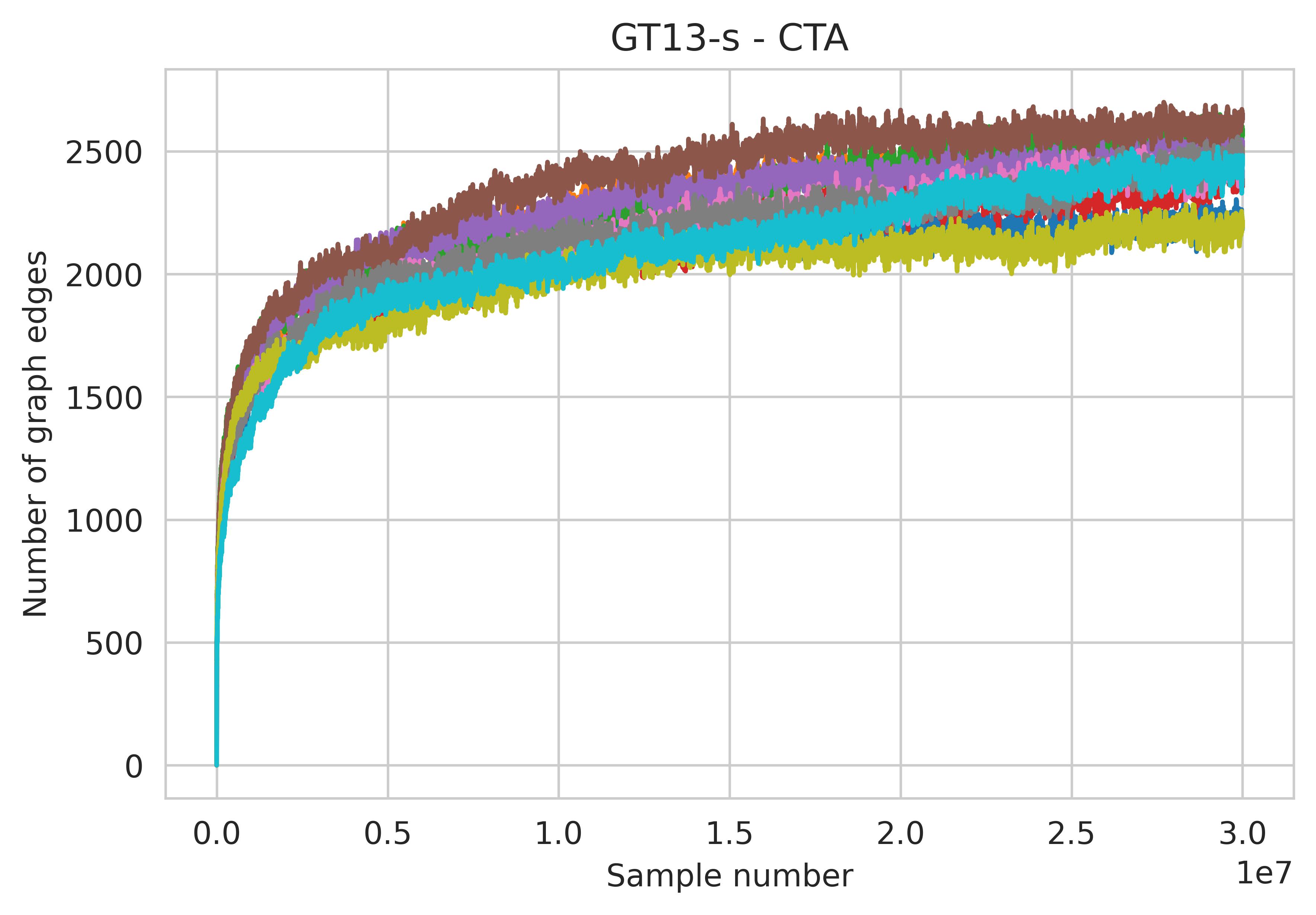}
  
 \caption{Traceplots showing the number of graph edges for 10 replications (color-matched), as outlined in Section~\ref{sec:simul-setup-gauss} for $(n,p)=(100,200)$.}
  \label{fig:gt_parallel_size_traceplot-sparse}
\end{figure}

\begin{figure}[!ht]
  \centering
  \includegraphics[width=0.53\textwidth]{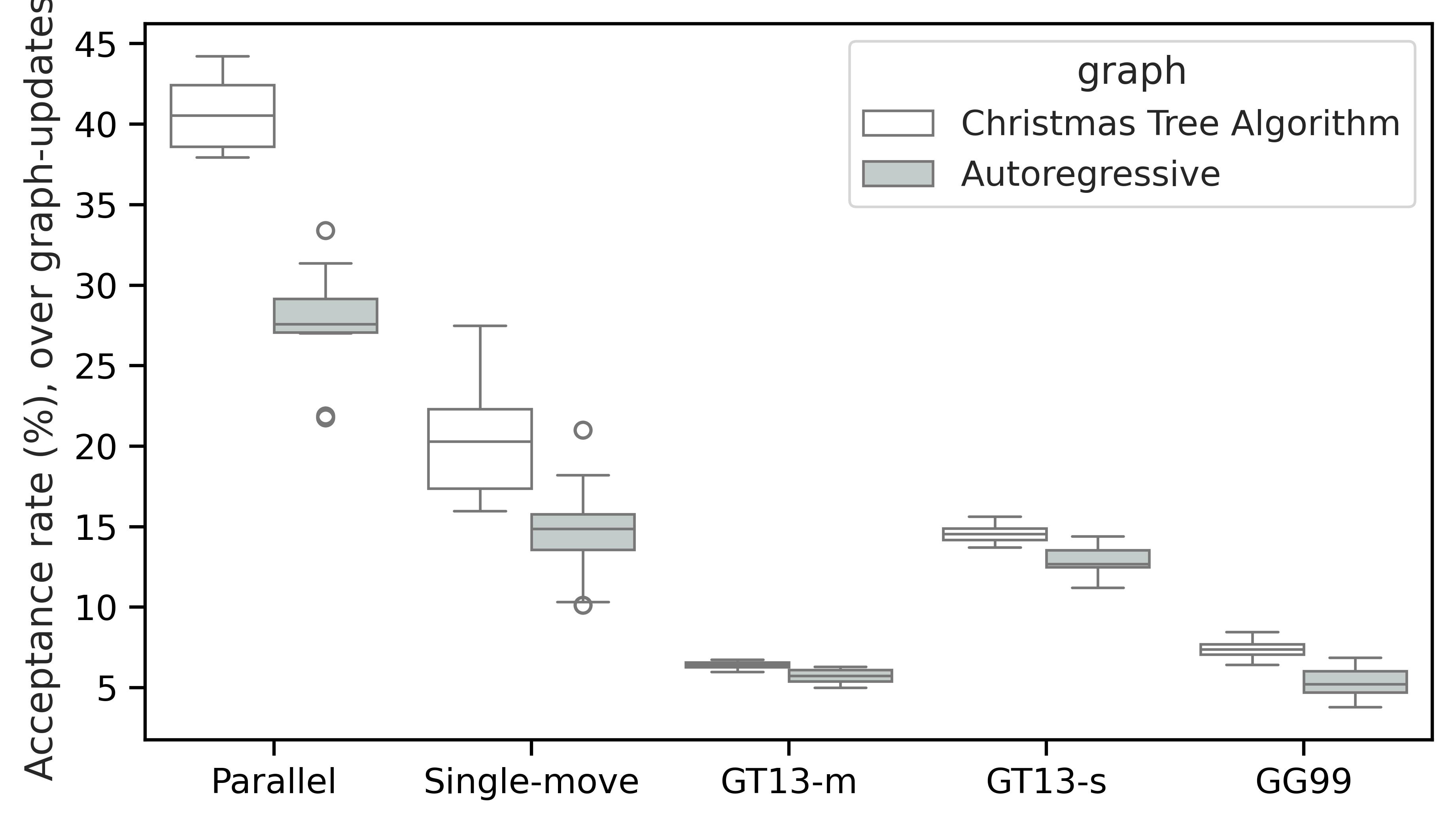}
  \includegraphics[width=0.44\textwidth]{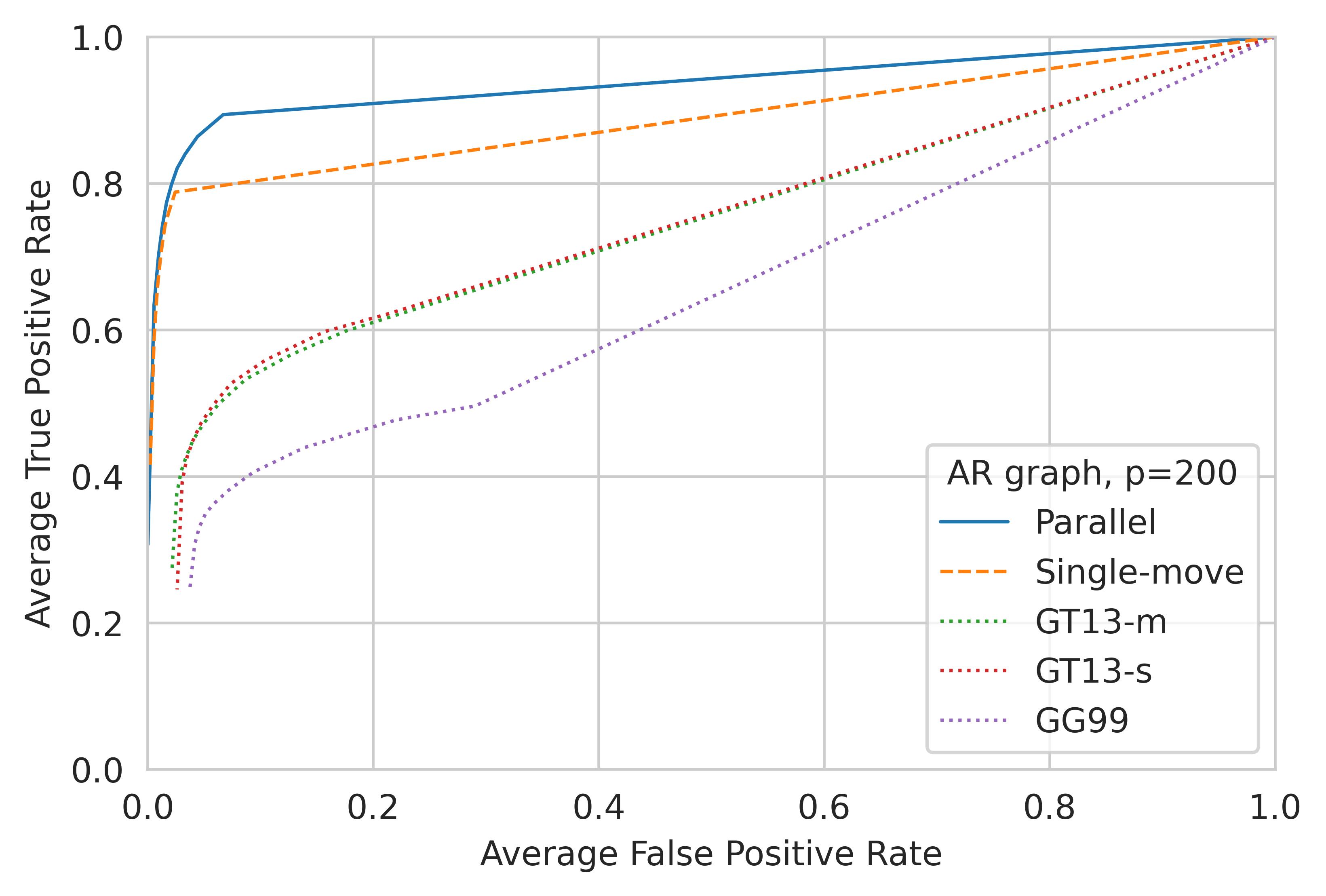}
  \caption{ (Left) Boxplot comparing acceptance rates over graph-updates between our parallel and single-move samplers and competitors, across two graph structures and 10 replications. (Right) ROC curves comparing the same samplers under the autoregressive graph structure only. Here $(n,p)=(100,200)$.}
  \label{fig:boxpot-roc-sparse}
\end{figure}

\clearpage
\section{Decomposable graphs over seven vertices}

\begin{figure}[ht]
  \centering
  \includegraphics[width=0.7\textwidth]{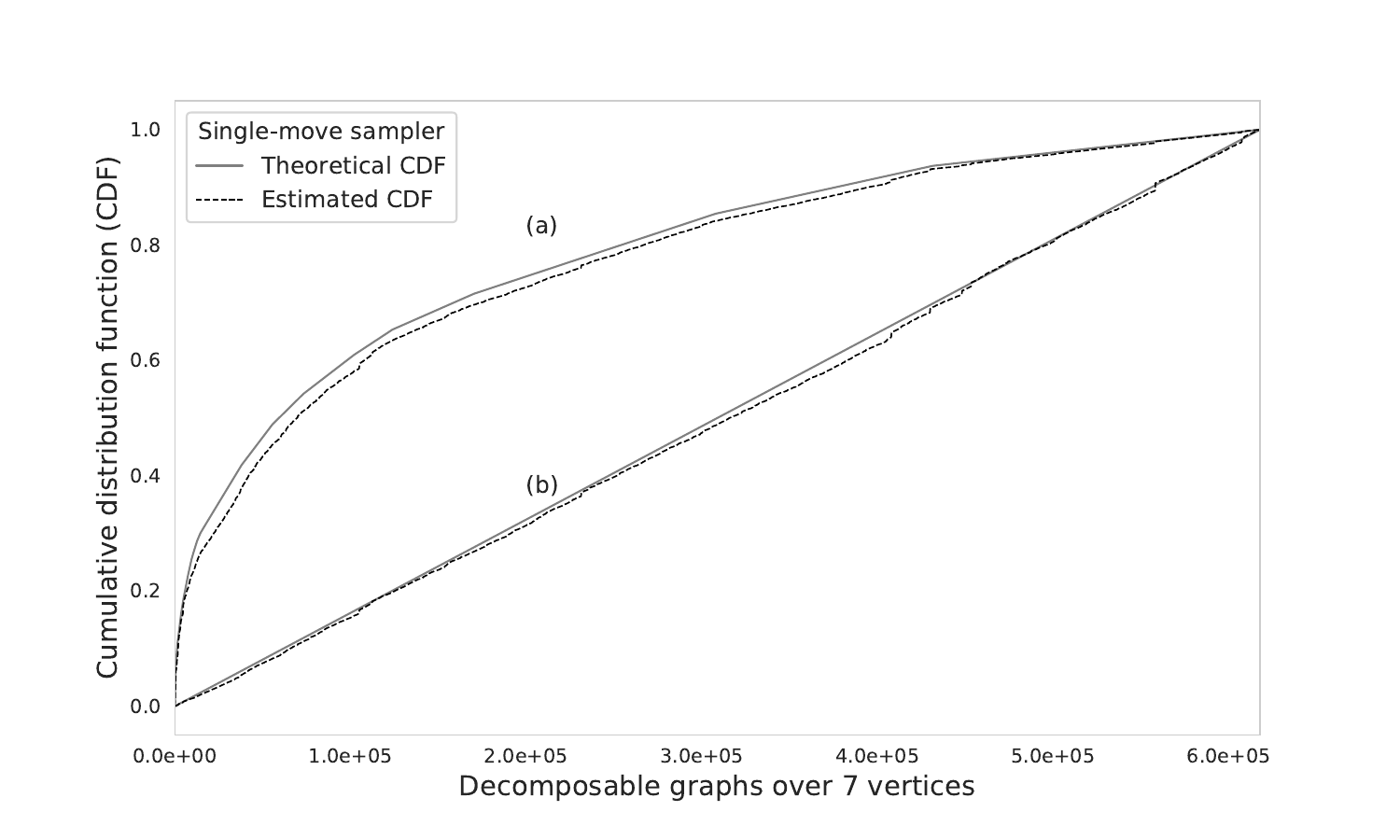}
  \caption{Cumulative distribution functions for decomposable graphs over seven vertices, estimated from samples drawn from the single-move sampler (a) with probability proportional to the number of junction tree representations, and (b) uniformly. The solid lines represent expected frequencies, and the dashed lines represent observed frequencies. The x-axis enumerates the graphs in decreasing order based on the number of reduced junction tree representations.}
  \label{app:fig:single-move-empirical-CDF}
\end{figure}
 \begin{figure}[ht]
  \centering
  \includegraphics[width=0.7\textwidth]{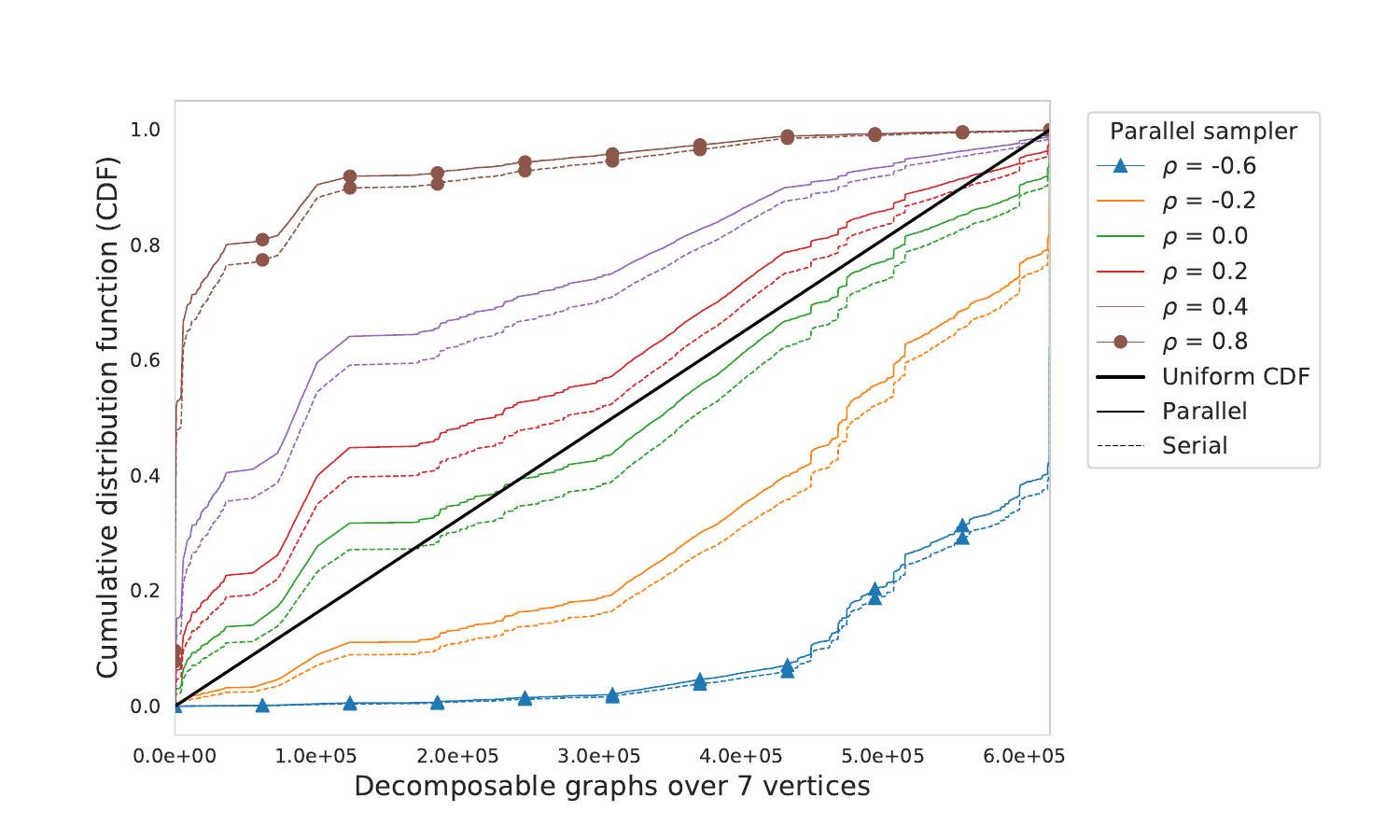}
  \caption
  {Cumulative distribution functions for decomposable graphs over seven vertices, estimated from samples drawn from the parallel sampler with proposal specified in~\eqref{eq:prior-update-type}. The x-axis enumerates the graphs in decreasing order based on the number of reduced junction tree representations.}
\label{app:fig:single-move-prior-sensitivity}
\end{figure}

\end{appendices}
\end{document}